\documentclass[12pt,a4paper]{book}
\pdfoutput=1 
\usepackage{cite} 
\usepackage{latexsym,amsfonts,amsmath,amssymb}
\usepackage[latin1]{inputenc}
\usepackage[dvips,pdftex]{graphicx}
\usepackage{color}
\usepackage{epsfig}
\usepackage{multirow}
\usepackage{epstopdf}
\usepackage[innercaption]{sidecap} 
\usepackage[usenames,dvipsnames]{xcolor}
\usepackage{eso-pic}

\usepackage[small,bf]{caption} 
\usepackage{tikz} 
\usetikzlibrary{decorations.pathmorphing,calc}
\usepackage[bookmarks=false,colorlinks=true,linkcolor=Black,citecolor=Black,urlcolor=Black]{hyperref} 

\advance \topmargin by -\headheight
\advance \topmargin by -\headsep

\evensidemargin \oddsidemargin
\newcommand{\reef}[1]{(\ref{#1})}

\def\a{\alpha}
\def\b{\beta}
\def\d{\delta}

\def\g{\gamma}
\def\G{\Gamma}
\def\h{\eta}

\def\l{\lambda}
\def\m{\mu}
\def\n{\nu}
\def\o{\omega}  \def\w{\omega}
\def\p{\pi}
\def\r{\rho}                                     
\def\s{\sigma}                                   

\newcommand{\td}{\mathrm{d}}

\newcommand{\be}{\begin{equation}}
\newcommand{\ee}{\end{equation}}
\newcommand{\beq}{\begin{equation}}
\newcommand{\eeq}{\end{equation}}
\newcommand{\beqn}{\begin{equation*}}
\newcommand{\eeqn}{\end{equation*}}
\newcommand{\bear}{\begin{eqnarray}}
\newcommand{\eear}{\end{eqnarray}}
\newcommand{\bal}{\begin{aligned}}
\newcommand{\eal}{\end{aligned}}
\newcommand{\nn}{\nonumber}

\newcommand{\cN}{{\cal N}}
\newcommand{\cO}{{\cal O}}
\newcommand{\cE}{{\mathcal E}}
\newcommand{\cP}{{\mathcal P}}

\newcommand{\cF}{{\cal F}}

\newcommand{\cD}{{\cal D}}

\def\IR {\mathbb{R}}
\def\IZ {\mathbb{Z}}

\def\ie{\hbox{\it i.e. }}
\def\eg{\hbox{\it e.g. }}

\def\CC{{\mathchoice
{\rm C\mkern-8mu\vrule height1.45ex depth-.05ex
width.05em\mkern9mu\kern-.05em}
{\rm C\mkern-8mu\vrule height1.45ex depth-.05ex
width.05em\mkern9mu\kern-.05em}
{\rm C\mkern-8mu\vrule height1ex depth-.07ex
width.035em\mkern9mu\kern-.035em}
{\rm C\mkern-8mu\vrule height.65ex depth-.1ex
width.025em\mkern8mu\kern-.025em}}}

\def\a{\alpha}
\def\b{\beta}
\def\d{\delta}
\def\e{\epsilon}           

\def\g{\gamma}
\def\h{\eta}

\def\l{\lambda}

\def\m{\mu}
\def\n{\nu}
\def\o{\omega}  \def\w{\omega}
\def\r{\rho}                                     
\def\s{\sigma}                                   

\def\6{\partial}

\newfont{\namefont}{cmr10}
\newfont{\addfont}{cmti7 scaled 1440}
\newfont{\boldmathfont}{cmbx10}
\newfont{\headfontb}{cmbx10 scaled 1728}

\newcommand{\cS}{\mathcal{S}}

\allowdisplaybreaks[1]

\numberwithin{equation}{section}

\begin{document}

\frontmatter 


\begin{titlepage}

\begin{center} 

\vskip.5cm

\centerline{\epsffile{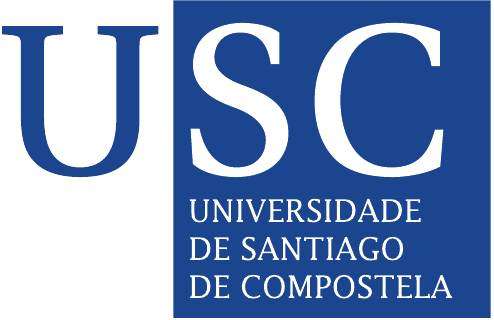}}

\vspace{.3cm}

\large Departamento de F\'isica de Part\'iculas 


\vspace{8cm}


\Huge \bfseries \ttfamily  LOVELOCK GRAVITY, BLACK HOLES\\ AND HOLOGRAPHY

\end{center}

\vspace{6.5cm}

\begin{center} 

{\sf\bf \large Xi\'an Otero Cama\~no}

\vspace{0.2cm}

\sf TESE DE DOUTORAMENTO

\end{center}

\end{titlepage}




\newpage
\thispagestyle{empty}
\mbox{}

\newpage


\thispagestyle{empty}

\begin{center} 

\large \sf  UNIVERSIDADE DE SANTIAGO DE COMPOSTELA

\vspace{.3cm}

\large Departamento de F\'isica de Part\'iculas 

\end{center}

\vspace{6cm}

\begin{center} 

\LARGE  \bf LOVELOCK GRAVITY, BLACK HOLES\quad {\bfseries \itshape \&} \  HOLOGRAPHY

\end{center}

\vspace{3cm}

\vspace{2cm}

\begin{center} 

{\sf\bf \large Xi\'an Otero Cama\~no}
\vskip1mm
{PhD supervisor: \bf Jos\'e D.~Edelstein Glaubach}
\vskip6mm
\sf Santiago de Compostela, May 2013\footnote{[Revised: minor corrections, updated references] Berlin, September 2015.}.

\vfill

\end{center}



\newpage
\thispagestyle{empty}
\mbox{}

\newpage































\setcounter{footnote}{0}






\

\vspace{6cm}


\begin{quotation}

{\it ``Unha serea que canta de noite polos tellados,
e un astronauta en bicicleta que no asomar das estrelas ven beijarlle as palmas das mans. 
Nese intre, no espello fr\'io da l\'ua 
acendes a noite,
e ardemos.
Eu creo que foi as\'i como naceu o Universo'' 

\hfill {\rm [Patrieira}, Big Bang{\rm ]}
}

\vspace{1.5cm}

\flushright
A todos os que confiaron en min, m\'ais do que eu mesmo.
\end{quotation}



















\newpage
\thispagestyle{empty}
\mbox{}

\newpage


\begin{quotation}
\flushright
{\it ``We are like dwarfs standing upon the shoulders of giants, \\ and so able to see more and see farther than the ancients.''}\\

\vspace{.5cm}

Bernard of Chartres
\end{quotation}


\vspace{4cm}

{\LARGE \bfseries\itshape {\centerline{Agradecementos}}}

\vspace{1cm}

\noindent \ldots \ e at\'e aqu\'i chegou o cami\~no, un ronsel entre tantos, fin de etapa, porto de abrigo. Tempo de ollar atr\'as antes do seguinte paso, a pr\'oxima traves\'ia. Porque un nunca est\'a s\'o na s\'ua viaxe, porque cada persoa que cruzou o noso carreiro, cada compa\~neiro de andaina, deixa pegada e leva un anaco de n\'os. {\it Ubuntu}, eu son porque n\'os somos. Un non \'e, non se pode entender, sen todas as persoas que vai atopando ao cami\~nar. 

Mari\~neiro son, coma meu bisav\'o, meu av\'o e meu pai; por\'en un non pode navegar s\'o. Non pode un sa\'ir ao mar sen tribo, sen porto, sen barco e mans amigas. 
\'E tempo de lembrar e adicar unha verba a todos os que fixeron que eu poida hoxe estar escribindo estas li\~nas. 

A todos os moitos e bos mestres que tiven. Eles ensin\'aronme que vivir \'e procurar o propio cami\~no. Hoxe que inventar novos cami\~nos \'e m\'ais importante ca nunca, hoxe que nos est\'an a retirar o enlousado de baixo os p\'es. \'E tempo de voltar cami\~nar sobre a herba. A eles por soprar as velas da mi\~na curiosidade insaci\'abel. 

A toda a xente de Compostela, xa a mi\~na segunda aldea, campo base. Aos compa\~neiros dos anos da carreira: Patxi, Patri, Edu, Gonza, Luc\'ia, Xe, Celes, Vane, Jesus, Lionel, Brais, Vero, Meri, Rub\'en, \'Angel, Gemma... e tantas e tantos outros con quen compart\'in conversas, ceas, troulas... e mesmo alg\'un escenario do QMF (que ousad\'ia!). Todos me vistes medrar para ser quen son hoxe. A todos os f\'isicos, Paolo, John, Jose, Ricardo, Josi\~no, Alfonso, Javier, Tarr\'io, Daniel... \'a FROGsS! Mesmo os meteor\'ologos. Todos me arroupastes nos meus primeiros pasos coma f\'isico? te\'orico? Literatos do m\'ais mi\'udo e o m\'ais grande, de todo o que escapa aos sentidos, a\'inda aos m\'ais sofisticados aparellos. Quen dixo que poderiamos sequera enxergar tales cousas? Pouco m\'ais do que simios ollando ao ceo, do cabo do mundo escoitar o universo. Que poder\'ia aportar eu?

Mais houbo quen confiou en min, deume unha palmadi\~na nas costas e dixo -- {\it ti podes}. Grazas Jose por ensinarme novos mares, na f\'isica e na vida, polo teu entusiasmo contaxioso, pola t\'ua paciencia e o teu apoio. Grazas por compartires as t\'uas grandes ideas e por escoitar sempre as mi\~nas.

\newpage

Moitas veces me levou o cami\~no lonxe da costa, Chamb\'ery, Cambridge, Porto, Waterloo, Amsterdam, Buenos Aires, Santiago de Chile ou Princeton. De todos estes lugares gardo lembranzas imborr\'abeis, amigos que a\'inda lonxe me acompa\~nar\'an sempre. A Sonya e Letizia, a Stephan, Savan, Pilar ou Valentin, a toda a familia do outro lado do mar. A todos mil milleiros de grazas, por acollerme cos brazos abertos, por cami\~nar comigo, por darme un empurr\'on no momento certo. Grazas tam\'en a todos os grand\'isimos f\'isicos cos que tiven a sorte de traballar, todos contribu\'iron enormemente a expandir os meus horizontes cient\'ificos. A Miguel, Alex, Rob, Jan, Juan e Sasha. E menci\'on especial para Gast\'on e Andy, por axudarme a co\~necer algunhas das moitas marabillas que o cono sul ten para ofrecer. Foi todo un pracer poder colaborar con todos v\'os e espero que poidamos seguir traballando xuntos no futuro. 

Deixo para o final a ra\'iz, o cerne, a madeira m\'ais dura, a que aguanta o peso. Sempre ser\'a pouco o que poida agradecer \'a familia e aos amigos, por confiar sempre en min m\'ais do que eu mesmo, pola paciencia e o apoio que sempre me brindaron. \'A mi\~na nai, pola s\'ua forza sosegada, o son dos seus pasos canda min, ao longo dun cami\~no que nen sempre soubo onde \'ia levar. 

E grazas a mares mi\~na Patrieira, meu rumbo, por cami\~nares comigo, por ollares ao futuro aos ollos. Grazas porque sen ti non dar\'ia chegado at\'e aqu\'i, sen o teu ollar sobre o mundo, sen o teu pulo creador de universos, mi\~na inspiraci\'on. 

\hfill Grazas a todos!



\newpage


\pagestyle{headings}


\tableofcontents


\chapter{\bfseries\itshape Foreword}
\chaptermark{Foreword}
\label{chp:intro}

\vspace{0.6cm}

\begin{quotation}
\flushright
{\it ``La science cherche le mouvement perp\'etuel. \\
Elle l'a trouv\'e; c'est elle meme.''}\\
\vspace{.3cm}
Victor Hugo
\end{quotation}

\vspace{3cm}


\noindent In recent years, there has been a revival of interest in higher curvature theories of gravity. Higher order corrections to the Einstein-Hilbert action appear in any sensible theory of quantum gravity, either in the context of Wilsonian approaches, as next-to-leading orders in the effective action of string theory or motivated by the possibility of higher dimensional spacetimes. In particular, Lovelock theories represent the most natural generalization of the Einstein-Hilbert action to dimensions larger than four. Moreover, its first non-trivial action, Lanczos-Gauss-Bonnet, also appears in {\it bona fide} realizations of string theory, with the advantage that it can be considered as a finite correction. Gravity theories of the Lovelock type, yielding two derivative equations of motion, avoid some of the problems of other  higher curvature gravities while capturing some of their characteristic features, namely the existence of several branches of solutions, more general black hole spacetimes and complex dynamics. This class of theories provides a particularly suitable playground to test our ideas about gravity in a much broader context. We will investigate the consequences that follow from the assumption that the model is the classical limit of a fundamental theory, not an effective one. This attitude is also the most efficient one to eventually uncover reasons to reject the assumption.

In the holographic context, the addition of higher curvature corrections allows for the description of more general field theories, \eg CFTs with different central charges in four dimensions. Doing this in a controlled way, we will uncover previously unsuspected connections between some central concepts in physics, such as causality or positivity of the energy. These correlations extend smoothly and meaningfully to any dimension and any Lovelock theory, thus supporting the possibility that AdS/CFT may be applicable beyond the framework of string theory, as long as there is a consistent theory of quantum gravity in AdS.

\begin{figure}
\centering
\includegraphics[width=\textwidth]{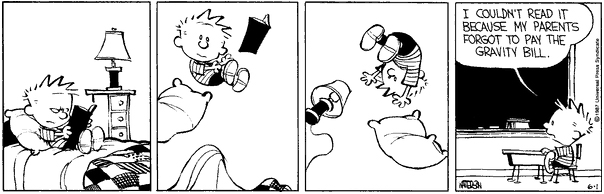}
\caption{Calvin's thoughts in modified gravity}
\end{figure}




\section*{Outline of the thesis}

This thesis is divided in two separate parts, the first concerned with gravitational aspects of Lovelock theories, the second with their holographic applications. Both are based on a series of papers \cite{Camanho2010,Camanho2010a,Camanho2010d,Camanho2011a,Camanho2012,Camanho2013a,Buchel2013b, Camanho2013b} and encompass some original work based on those developments and part of ongoing projects in collaboration with my supervisor Jos\'e Edelstein, Alexander Zhiboedov and professors Gast\'on Giribet, Andy Gomberoff and Juan Maldacena \cite{Camanho2013,Camanho2013c}. 

The first part is devoted to the analysis of static black holes and other solutions in the context of general Lovelock gravities. In chapter \ref{chp:LLgravity} a formal introduction to Lovelock theories of gravity is given; field equations,  structure of vacua and boundary terms, among other aspects, are discussed there. We also introduce the generalization of the junction conditions of Israel for the case of Lovelock. We focus mostly on the simplest cases of Lanczos-Gauss-Bonnet and third order Lovelock theory for concreteness, but most results are completely general. 

Chapter \ref{chp:LLbh} is concerned with black holes solutions in these theories, the different branches of solutions and horizon structure. We present a novel approach to deal with this class of static solutions in arbitrary Lovelock theories. This was introduced in \cite{Camanho2011a} for the case of uncharged black holes although some preliminary ideas were already present in previous works \cite{Camanho2010,Camanho2010a,Camanho2010d}. This method allows for the description and classification of these solutions regardless the dimensionality, the order and the type of branches of the Lovelock theory, for completely arbitrary values of its coupling constants. In the last sections of the chapter, we generalize this proposal to the case of charged and cosmological solution. Following this very same approach, chapter \ref{chp:thermo} deals with thermodynamic properties of Lovelock black holes \cite{Camanho2011a}. We start by deriving all the relevant thermodynamic variables for then analyzing the structure of the solutions, their stability and phase transitions. 

Chapter \ref{chp:bhstability} is also dedicated to the stability of black hole solutions, a different kind of stability though. Studying the equations of motion for perturbations about these backgrounds we find several situations in which static black holes become unstable. These instabilities appear to be related to the occurrence of naked singularities and other special solutions present in the Lovelock family of theories. In particular, stability seems to protect the cosmic censorship hypothesis and the third law of thermodynamics in this context, at least in some cases \cite{Camanho2013b}. Finally, closing the first part of the thesis, chapter \ref{genHP} presents a more general class of solutions in Lovelock gravity. These correspond to {\it bubbles} separating two regions of the spacetime corresponding to different branches of the Lovelock theory. These exist even though they do not carry any matter and have dynamics inherited from the junction conditions. These {\it bubble} configurations allow us to describe several interesting effects, namely phase transitions between branches. This new type of transitions was the subject of a series of recent papers \cite{Camanho2012,Camanho2013a,Camanho2013}. 

After a lengthy study of our gravitational theories of interest, we move on to the analysis of their r\^ole in the context of holography. The second part of the thesis starts with a conceptual and computational introduction to the AdS/CFT correspondence in chapter \ref{chp:AdSCFT}. Following \cite{Camanho2010d}, we calculate holographically the parameters that enter 2- and 3-point functions of the stress-energy tensor for Lovelock theories and discuss briefly the kinds of higher order terms that may enter in that computation \cite{Camanho2013c}.

Higher curvature corrections arise in the context of string theory as next to leading orders in the low energy effective action. As such, the corresponding corrections are necessarily small, this being also true for any string construction of the AdS/CFT duality. Nonetheless, Lovelock gravities are characterized by having second order field equations, they are thus consistent\footnote{\S\ In some situations having second order field equations will not be enough as has been shown in \cite{Camanho2013c}.} for finite values of the couplings. This will allow to explore the AdS/CFT correspondence in a broader setup, describing much more general CFTs than the Einstein-Hilbert (super)gravity approximation. In particular, this will allow for field theory duals with unequal central charges $a\neq c$ in four dimensions. 

As it will be discussed, some of the gravitational effects analyzed in the first part of the thesis turn out to have a nice field theory counterpart.  For instance, the Lovelock analogue of the instability found by Boulware and Deser in Laczos-Gauss-Bonnet gravity may be interpreted as non-unitarity of the dual CFT \cite{Camanho2010d}. In that context, the Hawking-Page phase transitions of chapter \ref{chp:thermo} correspond to confinement/deconfinement phase transitions of the field theory. This can be affected by hydrodynamic effects, such as cavitation, that may effectively shift the temperature at which the phase transition occurs \cite{Buchel2013b} (see Annex \ref{cavitation}). Chapters \ref{chp:LLcausality} and \ref{LLfate} are then devoted to dug further into this CFT/Lovelock duality.  In particular we show the exact equivalence between positivity of energy correlators and causality in the corresponding gravity dual. Chapter \ref{chp:LLcausality} is an account of the work done in \cite{Camanho2010,Camanho2010a} where we  also analyze the restrictions on the space of parameters imposed by these causality/positivity constraints. These also impose restrictions in other variables of the theory such as the shear viscosity to entropy density of the dual plasma. The exploration of the possible values of this quantity and the existence of bounds will be extensively treated in chapter \ref{LLfate} following closely the discussion of \cite{Camanho2010d}.  The aim of the work is to deeper scrutinize in the amazing relations between gravity and gauge theories, relations that seem to go beyond the framework of
string theory.

We end up by a small summary of the work done in chapter \ref{summary} where we also draw  some final conclusions. 

\vskip3mm\noindent 
\textbf{Note added:} Some comments (indicated by \S) have been included following the publication of \cite{Camanho2013c}. The original text of the thesis remains otherwise unchanged.

\section*{Preliminaries {\bfseries \itshape \&} notation}
\label{notation}

Before getting our hands dirty, let me summarize the notations and conventions that will be assumed to hold throughout this thesis. Some of these are quite standard but I prefer to gather them here instead of being scattered along the text. In particular we will assume $\hbar=c=k_B=1$ units and avoid these constants in all the expressions except when relevant for the discussion. 

Rather than working with tensors, in most of this thesis we will make extensive use of differential forms and the exterior algebra (see for instance \cite{Eguchi1980, Willison2004}). Instead of the metric and affine connection we will be referring to orthonormal frames (or {\it vielbein}) and spin connection (or connection 1-form). This formalism will make our expressions more compact and many  manipulations much easier as it will become clear in the next chapters.

We will be working in general in a $d$-dimensional spacetime with  $(-1,1,\ldots,1)$ signature. The vielbein is a non-coordinate basis which provides an orthonormal basis for the tangent space at each point on the manifold,
\beq
g_{\m\n}\, d x^\m \otimes dx^\n = \h_{ab}\, e^a \otimes e^b~, 
\eeq
where $\h_{ab}$ is the $d$-dimensional Minkowski metric. The latin indices $\{a,b,\ldots\}$ are called \textit{flat} or \textit{tangent space indices}, while the Greek ones $\{\m,\n,\ldots\}$ are called \textit{curved} or \textit{spacetime indices}. In some cases we will also distinguish spacelike indices $\{i,j,\ldots\}$ from timelike ones and, in the presence of hypersurfaces, we will use capital letters $\{A,B,\ldots\}$ for the vielbeine adapted to the hypersurface. The vielbein are $d$ 1-forms, 
\beq
e^{a}=e^{a}_{\m}\,dx^{\m}~, 
\eeq
that we may use in order to rewrite the metric as
\beq
g_{\m\n}=\eta_{ab}\,e^{a}_{\m}\,e^{b}_{\n}~.
\eeq
We also introduce the metric compatible (antisymmetric) connection 1-form that is necessary in order to deal with tensor valued differential forms. In addition to the usual exterior derivative, $\td$, we define the {\it covariant} exterior derivative, $D$, that reduces to the former when applied to a scalar valued form. For a general $(p,q)$-tensor valued form,
\beq
DV^{a_1\cdots a_p}_{b_1 \cdots b_q} \equiv \td V^{a_1\cdots a_p}_{b_1 \cdots b_q}+\sum_{i=1}^p \w^{a_i}_{\ c}\wedge V^{a_1\cdots c \cdots a_p}_{b_1 \cdots b_q}-\sum_{j=1}^q \w^{d}_{\ b_j}\wedge V^{a_1 \cdots a_p}_{b_1 \cdots d \cdots b_q}~.
\eeq
We can in this way define the torsion and curvature 2-forms as derivatives of the vielbein,
\bear
De^a &=& \td e^a+\w^{a}_{\  b}\wedge e^b \equiv T^a~, \\
DDe^a &=& \left(\td\w^{a}_{\ b}+\w^a_{\ c}\wedge\w^c_{\ b}\right)\wedge e^b\equiv R^{a}_{\ b}\wedge e^b~,
\eear
or equivalently,
\bear
T^a &=& De^a~, \\
R^{ab} &=& \td\w^{ab}+\w^{a}_{\ c}\wedge \w^{cb}=\frac{1}{2} R_{~b\mu\nu}^{a}\; dx^{\mu} \wedge dx^{\nu}~,
\eear
expressions known as the Cartan structure equations. Due to the nilpotency of the exterior derivative, $\td^2=0$, the covariant derivative of Cartan's equations give the Bianchi identities,
\bear
DT^a &=& R^a_{\ b}\wedge e^b~, \\
DR^{ab} &=& 0~.
\eear
In the absence of torsion the spin connection is not independent from the metric and coincides with the Levi-Civita connection,
\beq
\w^\mu_{\ \n}=\G^{\m}_{\ \n\r}\,dx^{\r}~.
\eeq
In GR the torsion tensor is constrained to vanish. When this constraint is not imposed, we have the Einstein-Cartan theories. These are very important when considering spinor fields as these generally source the torsion. 

Other notations that will be used extensively in this thesis are
\bear
R^{a_1 a_2 \ldots a_{2n}} &\equiv& R^{a_1 a_2}\wedge\ldots \wedge R^{a_{2n-1} a_{2n}}~,\\ e^{a_1 \ldots a_n} &\equiv& e^{a_1}\wedge \ldots \wedge e^{a_n}~.
\eear
We will also use the antisymmetric tensor $\e_{a_1 a_2 \cdots a_d}$ when writing down and manipulating the Lovelock lagrangian and the derived equations of motion.
It is antisymmetric on any pair of indices with $\e_{123\ldots d}=+1$. Some times, in order to write more compact expressions we will even write scalars constructed with the antisymmetric tensor as
\beq
\e\!\left(\psi\right)=\e_{a_1\ldots a_d} \psi^{a_1\ldots a_d}   ~,
\eeq
\eg when writing down the order $k$ Lovelock term, we may write
\beq
\mathcal{L}_{k}= \e\!\left(R^k e^{d-2k}\right)= \e_{a_1 a_2\ldots a_d}R^{a_1 \ldots a_{2k}}\wedge e^{a_{2k+1}\ldots a_d}      ~,
\eeq
where wedge product inside the brackets is understood.

Generically, working in flat indices is much easier than doing it in curved ones. An excellent implementation of Cartan's formalism for \textit{Mathematica}, developed by Prof.~Bonanos, can be found at \url{http://www.inp.demokritos.gr/~sbonano/EDC/}.




\thispagestyle{empty}


\mainmatter 

\part{LOVELOCK THEORIES {\bfseries\itshape \&} BLACK HOLES}

\chapter{\bfseries\itshape Lovelock theories of gravity}
\chaptermark{Lovelock theories of gravity}
\label{chp:LLgravity}

\vspace{.6cm}

\begin{quotation}
\flushright
{\it ``Imagination will often carry us to worlds that never were.\\ But without it we go nowhere''}\\

\vspace{.3cm}

Carl Sagan
\end{quotation}

\vspace{3cm}


\noindent The general theory of relativity \cite{Einstein1916} is one of the greatest scientific accomplishments of the XX$^{\text{th}}$ century. It was born from the need to reconcile the Newtonian laws of the gravitational interaction with the new paradigm of the special theory of relativity \cite{Einstein1905}. It was independently pursued, at the same time, by two of the greatest minds of that time, Albert Einstein and David Hilbert, reason of the name of the action of the theory. 

Two basic ideas stand behind this extraordinary mathematical construction, the special theory of relativity and the principle of equivalence. On its weakest version the latter is just the observation of the exact equivalence between inertial and gravitational mass, two very different concepts with exactly the same value. This equality allows, at any point of spacetime, to choose a locally inertial reference frame such that the effect of the gravitational force is completely screened at that point by an equal and opposite acceleration. This in turn motivated the {\it strong equivalence principle} that moreover asserts that in a small enough neighbourhood of that point the laws of nature, not just those of dynamics, take the well known form  of special relativity without gravity. In other words it is impossible to tell the difference {\it locally} between an experiment in the presence of gravitational forces and the same experiment in an accelerated laboratory. Gravity cannot be avoided {\it globally} in this way as we would need to give different accelerations to different points. 

The view of spacetime that arises is that of a curved manifold whose metric parametrizes the gravitational interaction. The spacetime is no longer the inert scene for all physical phenomena to become the dynamical fabric of the universe. The set of transformations under which the laws of physics must be invariant is enlarged to general changes of coordinates that include, of course, Lorentz transformations as a particular example. In a sense the strong equivalence principle states that {\it the laws of physics are independent of the coordinates chosen to describe them}, whether they correspond to inertial or non-inertial observers. Moreover, the source of the gravitational field is the matter content of the spacetime or, more specifically, the induced stress energy tensor. The mass that entered the Newtonian theory, equivalent to energy through the celebrated $E=mc^2$, is just one of its components.

Once the field carrying the gravitational force and its sources have been identified, the final piece of information needed for a complete description of the {\it classical} theory is the choice of action that encodes the dynamics of the interaction. There is \`a priori a plethora of lagrangians that realize the requirement of general covariance and are therefore viable candidates. Nonetheless if we restrict the possibilities to those yielding second order equations of motion the choice becomes almost unique in four dimensions. In particular we require the field equations to be of the form 
\begin{equation}
\mathcal{G}_{\mu\nu}(g_{\alpha\beta},g_{\a\b,\g},g_{\a\b,\g\l})=T_{\mu\nu} ~,
\end{equation}
where the left hand side is a tensor valued local functional of its local arguments, symmetric and conserved,
\beq
\mathcal{G}^{\m\n}_{\ \ ;\n}=0 ~,
\eeq
in agreement with the analogous property for the stress energy tensor. Then Lovelock's theorem \cite{Lovelock1971} states that the possible equations reduce to
\begin{equation}
R_{\m\n}-\frac{1}{2}g_{\m\n}R+\hat{\Lambda} g_{\m\n}= 8\p G_N T_{\m\n}~,
\end{equation}
where the constant of proportionality is chosen in order to reproduce the correct Newtonian limit. These equations of motion arise from the Einstein-Hilbert (EH) action with cosmological constant coupled to matter, 
\begin{equation}
\mathcal{I}_{d=4}=\frac{1}{16\p G_N}\int{\sqrt{-g}\left(R-2\hat\Lambda\right)}  \,+\, \mathcal{I}_{mat} ~ .
\label{EHaction}
\end{equation}
The cosmological constant, $\hat\Lambda$, was first introduced by Einstein \cite{Einstein1917} in order to describe a stationary universe. He later referred to this episode as his {\it greatest blunder}, once the observation of the Hubble redshift made clear the Universe is actually expanding. The actual value of $\hat\Lambda$ in the observed universe is not zero though, and a number of observations, including the discovery of cosmic acceleration, have revived the cosmological constant. The $\Lambda$CDM model of the Universe, the most accepted modern cosmological model to date, asserts that $\hat\Lambda$ is positive, although negligible even by the scale of our galaxy, the Milky Way. In a much more general context, this constant will also play a very important r\^ole in our discussion although the most interesting case for us will be that of negative $\hat\Lambda$. 

Another possible characterization of the EH lagrangian, valid as well in higher dimensions, is that of the corresponding field equations being {\it linear} in second order derivatives of the metric \cite{Vermeil,Weyl,Cartan1922}. This restriction together with the above requirements singles out,  in any dimension, the action of general relativity. In dimensions greater than four, there are however other tensors admissible if this linearity condition is relaxed, the Lovelock lagrangians. The modified equations of motion will then just be {\it quasi-linear} (see \cite{Deruelle2003} for a detailed definition), quasi-linearity implying the absence of squared or higher order terms in second derivatives of the metric with respect to a given direction. This is important in order to have a well-defined initial value problem for gravity. The coefficient of this second derivatives can depend however on first derivatives of the metric and may vanish for this reason, leaving the second derivative in question indetermined. In \cite{Deruelle2003} some of the problems which may arise because of the quasi-linearity of the Lovelock equations are discussed. 

Lanczos \cite{Lanczos1932,Lanczos1938} found in 1932 a generalization of the EH lagrangian quadratic in the Riemann tensor and whose equations of motion are symmetric, conserved and second order in the metric. Yet another property of the EH lagrangian is that it is a pure divergence in two dimensions and the Einstein tensor vanishes identically in one and two dimensions. Similarly, the Lanczos, or Lanczos-Gauss-Bonnet (LGB), lagrangian is a pure divergence in four dimensions and the corresponding equations are identically zero in four or less. Also the LGB term is the Euler density appearing in the Gauss-Bonnet theorem \cite{Eguchi1980} in four dimensions. 

Lovelock \cite{Lovelock1971} generalized these results in 1971 and obtained, for any dimension, a formal expression for the most general, symmetric and conserved tensor which is quasi-linear in the second derivatives of the metric without any higher derivatives. He also found the lagrangian from which that tensor is derived: in $d$-dimensions it corresponds to a linear combination of the\footnote{Square brackets denote here the integer part of a number.} $\left[\frac{d-1}{2}\right]$ dimensionally continued Euler densities. In dimensions 5 and 6, the explicit form of the Lovelock lagrangian reduces to a linear combination of the EH and LGB lagrangians (with the possible addition of the cosmological constant). The Lovelock lagrangians are, due to their properties, the most natural generalization of that of Einstein and Hilbert to describe pure gravity in dimensions larger than four.

In physics, actions are built based on general principles, such as symmetry, causality and other consistency requirements. All terms satisfying these and built from the appropriate fields should then be included in the lagrangian. In this sense, there is no \`a priori reason\footnote{Even for general covariant actions, the issue of causality in gravity is a non-trivial one, as we will analyze in the second part of this thesis (\S\ see also \cite{Camanho2013c}). The same happens for non-gravitational theories \cite{Adams2006} where covariance does not imply causality in relativistic quantum field theories with higher dimensional terms.}, why higher order Lovelock terms should be excluded from the action. The dimensionful couplings of the theory increase their length dimension with the order in curvatures in such a way that higher order contributions become important at shorter distances (or higher energies) while solutions of Lovelock gravity reduce to those of general relativity asymptotically.

In this thesis we will be mainly concerned with gravity theories of the Lovelock family. As mentioned before, these only contribute to the gravitational dynamics in dimensions five or higher\footnote{In some cases they may contribute to the equations of motion when coupled to other fields in lower dimensions.} so that we need to first answer a more pressing question. Why should we be interested in spacetimes with dimensions different from the four known to our experience? The idea of higher dimensional spacetimes goes back to the groundbreaking papers of Kaluza \cite{Kaluza1921} and Klein \cite{Klein1926} but most of the present renewed interest comes from the advent of string theory. Inspired by the physics of strings and other motivations, much effort has been devoted in the last quarter of a century in high energy physics dealing with scenarios involving higher dimensional gravity, and it is fair to say that at present it is unclear if gravity is a truly four-dimensional interaction. 

On the other hand, the inclusion of terms non-linear in the curvature modifying the EH lagrangian is an idea first proposed by Weyl \cite{Weyl} and Eddington \cite{Eddington1924}. Of course,  these extra terms introduce contributions with derivatives of the metric up to the fourth. In the seventies and early eighties such quadratic lagrangians were exploited in view of renormalizing the linearized version of general relativity (see {\it e.g.} \cite{BoulwareQGR} for a review of that period) as well as to renormalize the stress-energy tensor of quantized matter fields in classical, curved, backgrounds; see \cite{Birrell1982} and references therein. They made their most forceful entrance however when it was shown that they should arise as next-to-leading corrections to the low energy limit of string theory. In particular, the simplest Lovelock lagrangian, the LGB term, has been explored to a large extent, mainly as a consequence of its appearance in this context \cite{Zwiebach1985}.

Since their inception, a steady attention has been devoted to scrutinize the main properties of Lovelock theories of gravity, their vacuum structure, induced cosmologies, Hamiltonian formalism, dimensional reduction, wormhole configurations and, most importantly, their black hole solutions, including their formation, stability and thermodynamics. In spite of the abundant literature on the subject, most articles deal with particular cases of the general Lovelock formalism due to the intricacy endowed by the increasing number of coupling constants: there are $[\frac{d-3}{2}]$ dimensionful quantities (alongside the Newton and cosmological constants) in a $d$-dimensional theory. For this reason, many investigations on black hole solutions of Lovelock gravities are restricted to one-parameter (zero measure) subspaces in the space of couplings. It is the aim of the first three chapters of this thesis to tackle the existence and main features of Lovelock black holes for arbitrary values of the full set of gravitational couplings. We will be dealing with arbitrary orders in the Lovelock action and arbitrary dimensions, most of our results being completely general. We will just focus in specific examples for illustrative purposes or when the intricacy of the equations so requires. 

Despite its debatable phenomenological interest, Lovelock gravities provide an interesting framework from a theoretical point of view for several reasons. As higher dimensional members of Einstein's general relativity family, they allow to explore several conceptual issues of gravity in depth in a broader setup. Among these, we can include features of black holes such as their existence and uniqueness theorems, their thermodynamics, the definition of their mass and entropy, etc. Lovelock theories are perfect toy models to contrast our ideas about gravity. 

A final piece of motivation comes from the theoretical framework proposed by Juan Maldacena \cite{Maldacena1998}. This will be the object of the second part of the thesis. The AdS/CFT correspondence establishes a holographic identification between conformal field theories and quantum gravities in higher dimensional AdS spaces. Besides its original maximally supersymmetric formulation, the correspondence seems robust enough to survive its generalization to less supersymmetric scenarios \cite{Klebanov1998}, and even non-supersymmetric \cite{Polyakov1999}, as well as non-stringy realizations \cite{Strominger:1997eq} (see also the seminal paper \cite{Brown1986b}). In particular, even if some caution remarks should be quoted at this point, the AdS/CFT correspondence seems to apply in higher dimensions too.

We know very little about non-trivial conformal field theories in higher dimensions (see \cite{Witten2007c} for a recent discussion). The interest of the AdS/CFT correspondence in this context is twofold. It provides an effective definition of higher dimensional CFTs from the gravitational side, whereas, in the reverse direction it opens new perspectives in the dynamics of gravity and its quantization. In the particular case of Lovelock theories this effective approach has yielded some unsuspected surprises in the form of very interesting connections. These will be reviewed in the second part of the thesis and involve some central concepts in physics, such as positivity of energy and causality \cite{Boer2009,Camanho2010,Buchel2010a,Camanho2010a,Boer2009a}.  These results have also motivated the discovery of new relations between unitarity and causality in CFTs \cite{Kulaxizi2011}.

Applications of AdS/CFT towards the understanding of the hydrodynamics of CFT plasmas in arbitrary dimensions demand a proper understanding of Lovelock black holes in AdS. This provides the final bit of motivation to pursue the present investigation. Regardless of the phenomenological dimensionality required by these applications, it is customarily the case that pushing some ideas to their extremes, besides verifying their robustness, allows to uncover novel features that are hidden in the somehow simpler original formulation (see, for instance, \cite{Paulos2011} for a beautiful recent example of this statement).

\section{Lovelock gravity}

Lovelock theories of gravity are the most general second order gravity theories in higher-dimensional spacetimes. They have the same degrees of freedom as general relativity and it is free of higher derivative ghosts \cite{Lovelock1971,Zumino1986}. The bulk action has a very complicated form in terms of the Riemann tensor and its contractions, nonetheless it can also be written very simply in terms of differential forms as
\begin{equation}
\mathcal{I} =\frac{1}{16\pi G_N (d-3)!}\, \sum_{k=0}^{K} {\frac{c_k}{d-2k}} \mathop\int \mathcal{L}_{k} ~,
\label{LLaction}
\end{equation}
$G_N$ being the Newton constant in $d$ spacetime dimensions. In some cases we will omit the overall normalization factor and simply set $16\pi G_N (d-3)!=1$ $\{c_k\}$ is a set of couplings with length dimensions $L^{2(k-1)}$, $L$ being a length scale related to the cosmological constant, while $K$ is a positive integer,
\begin{equation}
K\leq \left[\frac{d-1}{2}\right] ~,
\label{maximalK}
\end{equation}
labelling the highest non-vanishing coefficient, {\it i.e.}, $c_{k>K} = 0$.  $\mathcal{L}_{k}$ is the exterior product of $k$ curvature 2-forms with the required number of vielbein to construct a $d$-form,
\begin{equation}
\mathcal{L}_{k} = \e\!\left(R^k e^{d-2k}\right)=\epsilon_{f_1 \cdots f_{d}}\; R^{f_1 f_2\ldots f_{2k-1} f_{2k}} \wedge e^{f_{2k+1}\ldots f_d} ~.
\end{equation}
The zeroth and first term in (\ref{LLaction}) correspond, respectively, to the cosmological term and the Einstein-Hilbert action. It is fairly easy to see that $c_0 = L^{-2}$ and $c_1 = 1$ correspond to the usual normalization of these terms, the cosmological constant having the customary value $2 \hat\Lambda = - (d-1)(d-2)/L^2$. Either a negative ($c_0 = -L^{-2}$) or a vanishing ($c_0 = 0$) cosmological constant can be easily incorporated as well. The first non-trivial Lovelock term contributes just for dimensions larger than four and corresponds to the LGB coupling $c_2=\lambda L^2$.

The Lovelock action written in this way has the advantage that it can be equivalently considered in first order formalism, \ie we can consider the vielbein and the spin connection as independent variables. We then have  two equations of motion, one for each {\it field}. First, varying the action with respect to the connection 1-form the resulting equation is proportional to the torsion. We may use all the technology of exterior algebra and treat exterior covariant derivatives as normal derivatives inside the brackets. We can then integrate by parts to show,
\bear
\label{spineq}
\delta_{\w}\mathcal{L}_{k}&=& k\,\e\!\!\left[D(\delta\w)R^{k-1}e^{d-2k}\right]\\ 
&=&k\,\td\!\!\left[\e(\d\w R^{k-1} e^{d-2k})\right]-k(d-2k)\,\e(\d\w T R^{k-1}e^{d-2k-1}) \nonumber
\eear
where we have used that $\delta_{\w}R^{ab}=D(\delta\w^{ab})$ and the Bianchi identity $DR^{ab}=0$. The first term in the above variation is a total derivative and does not contribute to the equations of motion whereas the second is proportional to the torsion. In most cases we may safely restrict to the torsionless sector as usual, allowing us to compare our results with those coming from the tensorial formalism based on the metric.

The second equation is obtained by varying the action with respect to the vielbein. It can be cast into the form
\begin{equation}
\mathcal{E}_a \equiv \epsilon_{a f_1 \cdots f_{d-1}}\;c_K\, \mathcal{F}_{(1)}^{f_1 f_2} \wedge \cdots \wedge \mathcal{F}_{(K)}^{f_{2K-1} f_{2K}} \wedge e^{f_{2K+1}\ldots f_{d-1}} = 0 ~,
\label{eqlambda}
\end{equation}
where $\mathcal{F}_{(i)}^{a b} \equiv R^{a b} - \Lambda_i\, e^a \wedge e^b$. This expression involves just the curvature 2-form and no extra covariant derivatives, making explicit the two derivative character of the Lovelock equations of motion. Also, for the critical dimension $d=2k$, the $k^{th}$ term contribution to the equations vanishes. In our approach this is simply due to the absence of vielbeine in the corresponding action term, thus yielding zero upon variation. More generally, the integral of that term becomes a topological invariant, the Euler number for that particular dimension. We will comment more on this in the next section. In dimensions lower than the critical one the corresponding Lovelock term exactly vanishes and we are led to the restriction (\ref{maximalK}).

Besides, (\ref{eqlambda}) makes manifest that, in principle, this theory admits $K$ constant curvature {\it vacuum} solutions,
\begin{equation}
\mathcal{F}_{(i)}^{a b} = R^{a b} - \Lambda_i\, e^a \wedge e^b = 0 ~.
\end{equation}
Inserting $R^{ab}=\Lambda\,e^a \wedge e^b$ in (\ref{eqlambda}), one finds that the $K$ different {\it cosmological constants} are the solutions of the $K^{th}$ order {\it characteristic} polynomial
\begin{equation}
\Upsilon[\Lambda] \equiv \sum_{k=0}^{K} c_k\, \Lambda^k = c_K \prod_{i=1}^K \left( \Lambda - \Lambda_i\right) =0 ~,
\label{cc-algebraic}
\end{equation} 
each one corresponding to a different {\it vacuum}, positive, negative or zero for dS, AdS and flat spacetimes. The effective cosmological constants correspond to the possible radii of these (A)dS spaces and should not be confused with the bare cosmological constant, $\hat\Lambda$, appearing in the action. The theory will have degenerate behavior whenever two or more effective cosmological constants coincide. This is captured by the discriminant,
\begin{equation}
\Delta = \prod_{i < j}^K (\Lambda_i - \Lambda_j)^2 ~,
\label{discriminant}
\end{equation}
that vanishes in a certain locus of the parameter space of  Lovelock couplings where some special features arise. The discriminant can be written as well in terms of the first derivative of the Lovelock polynomial, $\Upsilon$, as
\be
\Delta=\frac{1}{c_K^K}\prod_{i=1}^K\Upsilon'[\Lambda_i] ~.
\ee
As we move forward through the text it will become clear the preeminent r\^ole played by this polynomial in the most diverse situations. 


For the sake of clarity let us briefly consider the $K=2$ case. In LGB gravity there are two possible values of the effective cosmological constant
\begin{equation}
\Lambda_\pm = - \frac{1 \pm \sqrt{1 - 4\lambda}}{2 \lambda L^2} ~,
\end{equation}
and they agree when the discriminant 
\begin{equation}
\Delta \equiv (\Lambda_+ - \Lambda_-)^2 = \frac{1 - 4\,\lambda}{\lambda^2\, L^4} = 0 \qquad {\rm for} \qquad \lambda = \frac{1}{4} ~,
\label{discr}
\end{equation}
vanishes. This implies that, for $1 - 4\lambda > 0$, there are two (A)dS vacua around which we can define our theory. If $1 - 4\lambda < 0$, there is no constant curvature vacuum. For the exact value $4\lambda = 1$, the theory displays a degenerate behavior due to symmetry enhancement. In the particular case of $d=5$, the symmetry enhances to the full $SO(4,2)$ group and the expression (\ref{LLaction}) gives nothing but the Chern-Simons lagrangian for the AdS group \cite{Chamseddine1989e} (see also \cite{Zanelli2005}). It is a well-known fact in LGB gravity that one of the vacua, the one with the $+$ sign in front of the square root, leads to spacetimes with a naked singularity that signals the instability of the vacuum \cite{Boulware1985a}. We are thus led to the remaining branch of solutions, so-called EH branch as it is continuously connected to the solution of general relativity as $\lambda$ is taken to zero.


Another property of any degenerate vacuum is the absence of linearised gravitational degrees of freedom about it. The equations of motion for a metric perturbation, $h^{ab}$, around a given vacuum, $\Lambda_1$, are easily obtained from the perturbation of the curvature
\beq
R^{ab}=\Lambda_1 e^{ab}+\delta_{\!g}R^{ab}
\eeq
yielding
\beq
\mathcal{E}_a=\Upsilon'[\Lambda_1]\,\e_{a f_1\cdots f_{d-1}}\delta \mathcal{F}^{f_1 f_2}_1\,e^{f_3\cdots f_{d-1}}
\eeq
to the linear level, thus exactly zero as the first derivative of $\Upsilon$ vanishes for a degenerate vacua, 
\beq
\Upsilon'[\Lambda_1]=c_K\prod_{i\neq 1}(\Lambda_1-\Lambda_i)=0 ~.
\eeq
Moreover, it is easy to verify that the equations of motion around a non-degenerate vacuum are exactly the same as for Einstein-Hilbert gravity multiplied by a global factor proportional to $\Upsilon'[\Lambda_i]$. The propagator of the graviton corresponding to the vacuum $\Lambda_i$ is then proportional to $\Upsilon'[\Lambda_i]$ in such a way that when $\Upsilon'[\Lambda_i]<0$ it has the opposite sign with respect to the Einstein-Hilbert case and thus the graviton becomes a ghost. This generalizes the observation first done by Boulware and Deser \cite{Boulware1985a} in the context of LGB gravity. Thus, a given vacuum of Lovelock gravity, $\Lambda_1$, must satisfy
\be
\Upsilon'[\Lambda_1]>0
\ee
in order to correspond to a vacuum that hosts gravitons propagating with the right
sign of the kinetic term. See \cite{Charmousis2008a} for a recent discussion on the subject. In the non-degenerate case the number of degrees of freedom about any of these vacua is exactly the same as in general relativity.

Curiously enough, most of the studies in the context of Lovelock theory have been performed within the degenerate locus, $\Delta=0$. In this thesis however we will aim at making the complementary effort of digging into the non-degenerate case $\Upsilon[\Lambda_k]=0$, $\Upsilon'[\Lambda_k]\neq 0$, where $\Lambda_k$ is the {\it vacuum} under consideration. We will eventually see that, among the branches of solutions of (\ref{cc-algebraic}), only one would end up being physically relevant, say $\Lambda=\Lambda_\star$. Degeneracies that do not involve $\Lambda_\star$ are harmless, our analysis being thus valid for the whole parameter space, except the zero measure set $\Upsilon[\Lambda_\star]=\Upsilon'[\Lambda_\star]=0$.


As the simplest examples of the Lovelock family, we will later focus in the LGB and third order Lovelock lagrangians. Let us discuss in some detail the cubic case. The lowest dimensionality where this term arises is 7d reducing in lower dimensions to  LGB gravity. Consider the following action,
\begin{equation}
\mathcal{I} = \frac{1}{16\pi G_N}\, \mathop\int d^{d}x \, \sqrt{-g}\, \left[ R - 2 \hat\Lambda + \frac{(d-5)!}{(d-3)!}\; \lambda\, L^2\; \mathcal{L}_{2} + \frac{(d-7)!}{(d-3)!}\; \frac{\mu}{3}\, L^4\; \mathcal{L}_{3} \right] ~,
\label{trdaction}
\end{equation}
where the quadratic and cubic lagrangians are
\begin{eqnarray}
\mathcal{L}_{2} \!\!\!\!&=&\!\!\!\! R^2 - 4 R_{\mu\nu} R^{\mu\nu} + R_{\mu\nu\rho\sigma} R^{\mu\nu\rho\sigma} ~, \label{LGB} \\ [1em]
\mathcal{L}_{3} \!\!\!\!&=&\!\!\!\! R^3 + 3 R R^{\mu\nu\alpha\beta} R_{\alpha\beta\mu\nu} - 12 R R^{\mu\nu} R_{\mu\nu} + 24 R^{\mu\nu\alpha\beta} R_{\alpha\mu} R_{\beta\nu} + 16 R^{\mu\nu} R_{\nu\alpha} R_{\mu}^{~\alpha} \nonumber \\ [0.6em]
\!\!\!\!& + &\!\!\!\! 24 R^{\mu\nu\alpha\beta} R_{\alpha\beta\nu\rho} R_{\mu}^{~\rho} + 8 R_{~~\alpha\rho}^{\mu\nu} R_{~~\nu\sigma}^{\alpha\beta} R_{~~\mu\beta}^{\rho\sigma} + 2 R_{\alpha\beta\rho\sigma} R^{\mu\nu\alpha\beta} R_{~~\mu\nu}^{\rho\sigma} ~.
\label{LL3}
\end{eqnarray}
Up to an overall constant, this complicated tensorial expression can be cast very simply in the language of our previous discussion as ($c_2 = \lambda L^2$ and $c_3 = \frac{\mu}{3} L^4$)
\begin{eqnarray}
\mathcal{I} & = & \frac{\mu\, L^4}{3 (d-6)} \mathop\int \epsilon_{abcdef{g_1}\cdots{g_{d-6}}}\,\left( R^{abcdef} \ \ + \ \ \frac{d-6}{d-4}\; \frac{3 \lambda}{\mu\, L^2}\; R^{abcd} \wedge e^{ef} \right. \nonumber \\ [1em]
&+ & \left. \, \frac{d-6}{d-2}\; \frac{3}{\mu\, L^4}\; R^{ab} \wedge e^{cdef}
+ \frac{d-6}{d}\; \frac{3}{\mu\, L^6}\; e^{abcdef} \right) \wedge e^{{g_1}\cdots{g_{d-6}}} ~,
\end{eqnarray}
%
\begin{figure}\centering
\includegraphics[width=0.67\textwidth]{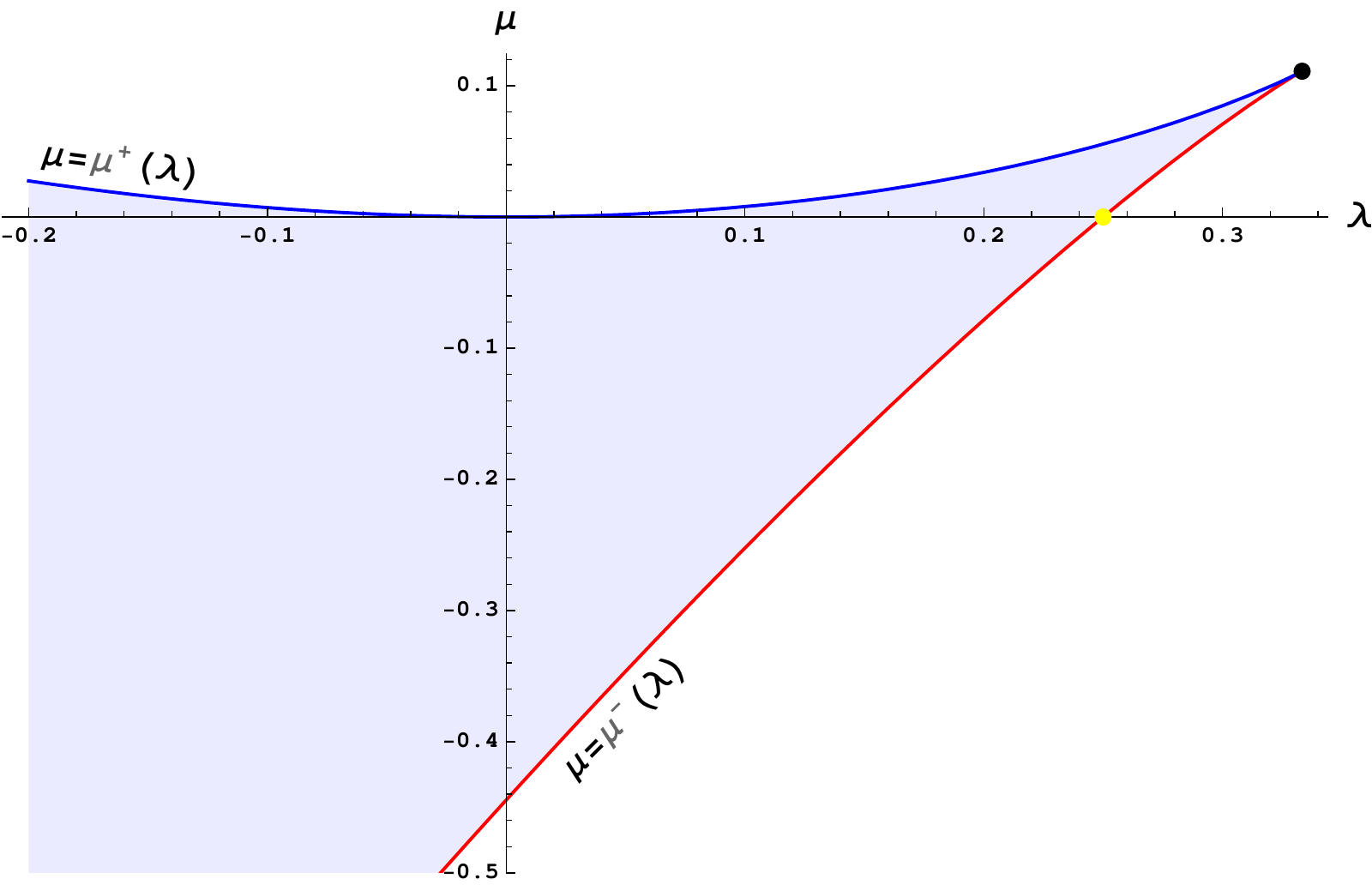}\caption{Singular locus for cubic Lovelock gravity. The shaded region is the domain where $\Delta > 0$. At the (dotted) vertex, $\lambda = 1/3$ and $\mu^+(1/3) = \mu^-(1/3) = \mu = 1/9$, the theory becoming maximally degenerated (in seven dimensions the symmetry enhances at that point to the full AdS group). The (dotted yellow) point, belonging to the locus, is nothing but the singular locus of LGB gravity, $\lambda = 1/4$. The point $\lambda = \mu = 0$ naturally belongs to the curve: it corresponds to Einstein's gravity.}
\label{locus-three}
\end{figure}
whose equations of motion, once the torsion is again set to zero, can be written as
\begin{equation}
\epsilon_{a \cdots f{g_1}\cdots{g_{d-6}}}\, \left( R^{ab} - \Lambda_1\; e^{ab} \right) \wedge \left( R^{cd} - \Lambda_2\; e^{cd} \right) \wedge \left( R^{ef} - \Lambda_3\; e^{ef} \right) \wedge e^{{g_1}\cdots{g_{d-7}}} = 0 ~.
\end{equation}
The values of the effective cosmological constants are complicated functions of the couplings that are not really important (see section \ref{LL3bh} for specific formulas). They must satisfy,
\begin{equation}
\sum_{i} \Lambda_i = - \frac{3 \lambda}{\mu\, L^2} ~, \qquad \sum_{i<j} \Lambda_i\, \Lambda_j = \frac{3}{\mu\, L^4} ~, \qquad \sum_{i<j<k} \Lambda_i\, \Lambda_j\, \Lambda_k = - \frac{3}{\mu\, L^6} ~,
\label{cosmconst}
\end{equation}
or, equivalently,
\begin{equation}
L^2\,\Upsilon[\Lambda]=\frac{\mu}{3}\, (L^2\,\Lambda)^3 + \lambda\, (L^2\,\Lambda)^2 + (L^2\,\Lambda) + 1 = 0 ~.
\label{polcosmo}
\end{equation}
There is always ({\it i.e.} for any $\lambda$ and $\mu\neq 0$) at least one real cosmological constant. This is to be contrasted with the case of LGB gravity where $1 - 4\lambda$ should be positive for the theory to have a real maximally symmetric vacuum. The theory will have degenerate behavior whenever the discriminant (\ref{discriminant}), which can be nicely written in terms of the couplings of the action,
\begin{equation}
\Delta = - \frac{1}{3}\left( 12\, \lambda^3 - 3\,\lambda^2 - 18\,\lambda\mu + \mu (9\,\mu + 4) \right) ~,
\label{disLLcubic}
\end{equation}
vanishes. The singular locus is $\Gamma: \mu(\lambda) \equiv \mu^+(\lambda) \cup \mu^-(\lambda)$, with
\begin{equation}
\mu^{\pm}(\lambda) \equiv \lambda - \frac{2}{9}  \pm \frac{2}{9}\, (1 - 3 \lambda)^{3/2} ~.
\end{equation}
It is convenient, for later use, to write the discriminant as
\begin{equation}
\Delta = - 3 \left( \mu - \mu^+(\lambda) \right) \left( \mu - \mu^-(\lambda) \right) ~,
\label{discLLcubic}
\end{equation}
since it makes clear that $\Delta < 0$ if $\mu > \mu^+(\lambda)$ or $\mu < \mu^-(\lambda)$, $\Delta = 0$ if $\mu = \mu^+(\lambda)$ or $\mu = \mu^-(\lambda)$, and $\Delta > 0$, elsewhere. We will see in what follows that the singular locus $\Gamma$ plays an important r\^ole. Notice that it does not depend in the spacetime dimensionality (for $d \geq 7$). A similar analysis can be readily performed for higher-order Lovelock theories.

\section{Gauss-Bonnet theorem and boundary terms}
\label{bdyterms}

The action \reef{EHaction} defines a variational principle from which the equations of motion of general relativity can be derived. However, in the event that the manifold has a boundary, the action should be supplemented by a boundary term so that this variational principle is well defined. This can be understood by means of a simple 1-dimensional example.

Consider the free particle lagrangian\footnote{We thank Andr\'es Gomberoff for pointing this very pedagogical example to us.}
\begin{equation}\label{I}
I=\frac{1}{2}\int_{t_1}^{t_2} \dot{x}^2dt ~.
\end{equation}
This action principle is designed in order to fix the position $x$ at initial and final times, $t_{1,2}$, \ie $\delta x(t_{1,2})=0$. Indeed,
\begin{eqnarray}\label{di}
\delta I &=& \int_{t_1}^{t_2} \dot{x}\delta \dot{x}  \ dt=   \int_{t_1}^{t_2} \left[ - \ddot{x} \delta x +  \partial_t (\dot{x} \delta x )\right] dt \\
&=&  \int_{t_1}^{t_2} (-\ddot{x})\delta x \ dt  + \dot{x}(t_2) \delta x(t_2) -   \dot{x}(t_1) \delta x(t_1)~. \nonumber 
\end{eqnarray}
Therefore for a solution of the equations, $\ddot{x}=0$, of motion the action is minimal, $\delta I=0$.

We could equivalently have started off with a different action,
\beq\label{II}
I'= - \frac{1}{2}\int_{t_1}^{t_2} x\ddot{x}dt ~,
\eeq
yielding the same equations of motion. This action principle is analogous to EH in the sense that it is linear on second derivatives of $x$. In the same way as for EH, it is not possible to fix just $x$ at the borders. In this case we get
\begin{eqnarray} \label{dip}
\delta I'&=&  \int_{t_1}^{t_2} (-\ddot{x})\delta x \ dt   \\ &-&\frac{1}{2} {x}(t_2) \delta \dot{x}(t_2) + \frac{1}{2} {x}(t_1) \delta \dot{x}(t_1)
+ \frac{1}{2} {\dot x}(t_2) \delta x (t_2) - \frac{1}{2} {\dot x}(t_1) \delta x (t_1)\ \ . \nonumber
\end{eqnarray}
Obviously, fixing $x$ at $t_{1,2}$ is not enough for solutions of the equations of motion to minimize the action. We need to add a boundary term, the analogue of Gibbons-Hawking term in order to fix the metric in GR. In this case it is easy to identify the missing term. The second action is related to the first by integration by parts, the difference being
\beq
\frac{1}{2}\left( x(t_2) \dot{x}(t_2)  -  x(t_1) \dot{x}(t_1) \right) ~.
\eeq
The variation of this term cancels the first two terms in (\ref{dip}) and adds up to the other two such that we get the original result (\ref{di}). Again the action with the boundary term is obviously devised to fix $x$ at the boundaries, it is exactly the original action. Although trivial in this context, conceptually the same  happens with the EH action. We need to supplement it with the Gibbons-Hawking term so that the variational principle is well defined. The difference is that in the context of gravity we do not have available an analogue of the action \reef{I}.


The same logic carries over to any gravity theory and in particular to the Lovelock class. The need for boundary terms can be seen explicitly from the variation of the action with respect to the spin connection \reef{spineq}. The first (boundary) term in the right hand side is analogous to the unwanted terms in \reef{dip}. Even though these terms do not change the equations of motion, they contribute to the variation of the action in such a way that the latter is not minimized for solutions. We want  to fix just the metric on the boundary, $\delta e^a=0$, not its normal derivative, $\delta\w^{ab}\neq 0$. Thus in order to cancel this boundary contribution we must add a boundary term analogous to the Gibbons-Hawking term of GR \cite{York1972,Gibbons1977},
\begin{equation}
\mathcal{I}_{GH}=\pm\frac{1}{8\pi G_N}\int_{\partial \cal M}\!\!\!\!d^{d-1}x\,\sqrt{\pm h}K ~,
\end{equation}
where $h_{\mu\nu}$ is the boundary metric and $K=K^\m_{\ \m}$ the trace of the extrinsic curvature, the plus (minus) sign applying to a spacelike (timelike) boundary.

The well-posedness of the variational problem in the context of Lovelock gravity was first analyzed in \cite{Myers1987} and the explicit expression for the boundary terms was found. In the same way as the $k^{th}$ Lovelock term is the dimensional continuation of the $2k$-dimensional Euler characteristic for closed manifolds, the corresponding boundary terms appear in the generalization of the Gauss-Bonnet theorem to manifolds with boundaries. For completeness we give account of some basic formul\ae\ connected to the Gauss-Bonnet theorem below. We basically follow the discussion of \cite{Willison2004}.

The {\it Euler number} or Euler-Poincar\'e characteristic, $\chi(\cal M)$, is a topological invariant of any manifold $\cal M$ on even dimensions, being zero if the dimension is odd. It is preserved under homeomorphisms, \ie one-to-one maps from one manifold to another in such a way that the topology is preserved. For instance, a coffee cup and a doughnut share the same Euler number, $\chi=0$. In general the Euler number is related to the genus, $g$, of the manifold, the number of {\it handles} on it, as
\beq
\chi=2-2g ~.
\eeq

The Gauss-Bonnet theorem \cite{Eguchi1980} allows to calculate the Euler number of a manifold of dimension $2k$ as the integral of a density constructed solely from the curvature 2-form. It can be written as 
\beq
\chi({\cal M})=\frac{(-1)^k}{(4\pi)\,k!}\int_{\cal M} \e(R,\ldots R)~.
\eeq
A very important property of the function function $\mathcal{L}(\w)=\e(R^k)$ is that under a continuous change of the connection, $\w\rightarrow \w'$, $\mathcal{L}(\w)$ changes by an exact form. This can be easily seen as follows. We omit indices for simplicity and define an interpolating connection
\beq
\w_t=t\w+(1-t)\w' ~,
\eeq
calling
\beq
\theta=\frac{d}{dt}\w_t=\w-\w' ~.
\eeq
We also define the curvature associated with this new connection,
\beq
R_t=\td\w_t+\w_t\wedge\w_t ~,
\eeq
and notice that
\beq
\frac{d}{dt}R_t=D_t\theta ~,
\eeq
where $D_t$ is the covariant derivative associated with $\w_t$. Then, we may write the variation of the Euler density as
\bear
\e(R,\ldots R)- \e(R', \ldots R')&=&\int_0^1 dt \frac{d}{dt}\e(R_t,\dots R_t)
= k\int_0^1\e\!\left(\frac{d}{dt}R_t, R_t,\ldots R_t\right)\nonumber\\
&=& k\int_0^1\e(D_t\theta, R_t,\ldots R_t) =k\int_0^1\td\e(\theta, R_t,\ldots R_t) ~,
\eear
where symmetry and linearity of $\e(\ldots)$ has been used, as well as the Bianchi identity, $D_tR_t=0$. If we define 
\beq
\mathcal{Q}(\w,\w')=k\int^1_0 dt\, \e(\w-\w',R_t,\ldots R_t)~,
\eeq
then
\beq
\mathcal{L}(\w)=\mathcal{L}(\w')+\td\mathcal{Q}(\w,\w')~,
\eeq
in such a way that, in case the manifold $\cal M$ is compact (or non-compact, $\theta$ vanishing fast enough asymptotically),
\beq
\int_{\cal M} \mathcal{L}(\w)
\eeq
does not depend on $\w$, and obviously does not depend on the vielbein. Hence, the integral above is a topological invariant as implied by the Gauss-Bonnet theorem. In the case of a manifold with boundary, $\partial \cal M$, we can slightly modify the argument very easily in order to construct another invariant quantity. We may introduce a third (reference) connection, $\w_0$, in such a way that
\beq
\mathcal{L}(\w)-\mathcal{L}(\w')=\left[\mathcal{L}(\w)-\mathcal{L}(\w_0)\right]-\left[\mathcal{L}(\w')-\mathcal{L}(\w_0)\right]=\td\mathcal{Q}(\w,\w_0)-\td\mathcal{Q}(\w',\w_0)
\eeq
and we have demonstrated
\beq
\mathcal{L}(\w)-\td\mathcal{Q}(\w,\w_0)=\mathcal{L}(\w')-\td\mathcal{Q}(\w',\w_0)
\eeq
Therefore the new quantity 
\beq
\int_{\cal M} \mathcal{L}(\w)-\int_{\partial \cal M} \mathcal{Q}(\w,\w_0)
\eeq
is invariant under a continuous change of connection. This is
a poor man's way to show that the Euler class is a topological invariant, the
real work is to prove that the actual value of the integral is related to $\chi$.

For vanishing torsion the above properties carry on trivially to any of the Lovelock terms. Once the corresponding boundary terms have been added, the variation of the full action with respect to the spin connection is exactly zero without any boundary contribution. The variation of the generalized Gibbons-Hawking term with respect to $\w$ cancels the unwanted term coming from the bulk action, and the variational problem is thus well posed. The reference connection, $\w_0$, is chosen to depend on the boundary metric, therefore it is kept fixed when extracting the equations. 

In fact, even though the above argument is independent of $\w_0$, this reference connection is usually taken to be the spin connection for a product metric that agrees with the original one on the boundary. In particular if we choose a set of adapted coordinates such that the boundary is $y=0$, $\w_0$ would be the connection associated with the metric
\begin{equation}
ds^2_0=n^2 dy^2+ds^2_{y=0}~,
\label{prodds}
\end{equation}
where $n^a$ is the unit normal vector to the boundary. It is clear that all the components aligned along the normal direction of this intrinsic spin connection are zero in the same way as it happens for the corresponding intrinsic curvature, 
\beq
R_0=\td \w_0+\w_0\wedge \w_0 ~. 
\label{intR}
\eeq
In this way, the spin connection difference with respect to this reference metric,
\beq
\theta=\w-\w_0 ~,
\label{secfund}
\eeq
has a very neat geometric interpretation, becoming the {\it second fundamental form} of the boundary surface. As we will explain below, the second fundamental form is related to the extrinsic curvature of the surface and it is zero for purely boundary components. 

The boundary terms appearing in the Gauss-Bonnet theorem, once dimensionally continued, are the natural generalization of the Gibbons-Hawking term. The latter corresponds to the simplest $k=1$ case that can also be written as
\begin{equation}
\mathcal{Q}_{1}=\theta^{a_1 a_2}\wedge e^{a_3\cdots a_d}\, \epsilon_{a_1\cdots a_d} ~,
\end{equation}
in terms of differential forms, and the analogous term in LGB gravity, so-called Myers term \cite{Myers1987},
\begin{equation}
\mathcal{Q}_2=2\theta^{a_1 a_2}\wedge (R^{a_3 a_4}-\frac{2}{3}\theta^{a_3}_{\ \;c}\wedge\theta^{c a_4}) \wedge e^{a_5\cdots a_d}\, \epsilon_{a_1\cdots a_d}~.
\label{Mterm}
\end{equation}

For the reasons outlined above, any general Lovelock action \reef{LLaction}, defined by a set coupling constants $\{c_k\}$, has to be supplemented by a boundary term 
\begin{equation}
\mathcal{I}_{\partial} = \sum_{k=1}^{K} {\frac{c_k}{d-2k}\int_{\partial\cal M}\mathcal{Q}_k} ~,
\end{equation}
the individual boundary densities being simply,
\begin{equation}
\mathcal{Q}_k =  k \int_0^1 d\xi \, \epsilon_{a_1\cdots a_d}\,\theta^{a_1 a_2}\wedge \mathcal{R}_\xi^{a_3 a_4}\wedge\ldots\wedge \mathcal{R}_\xi^{a_{2k-1} a_{2k}} e^{a_{2k+1}\cdots a_d} ~,
\label{genGH0}
\end{equation}
where 
\be
\mathcal{R}_\xi^{ab}= \xi R^{ab}+(1-\xi)R_0^{ab}-\xi(1-\xi)(\theta\wedge \theta)^{ab}    ~.
\ee
We have omitted the overall normalization factor $1/16\pi G_N (d-3)!$. The total action is in this way $\widetilde{\mathcal{I}}=\mathcal{I}-\mathcal{I}_\partial$ and can be written as a sum of dimensionally continued Euler densities. 

In order to get simplified expressions involving just the curvature $R$ and $\theta$ in the spirit of \reef{Mterm} we may make use of
\be
R^{ab}= R_0^{ab}+D_0\theta^{ab}+(\theta\wedge\theta)^{ab}
\ee
and also take into account that $D_0\theta^{ab}=d\theta^{ab}+(\omega_0\wedge\theta)^{ab}+(\theta\wedge\omega_0)^{ab}$ is zero unless either one of the indices is in the normal direction. As the whole expression is multiplied by the second fundamental form that necessarily picks one index in the normal direction in order not to vanish, this term does not contribute to \reef{genGH0} and we can write it in terms of just the full curvature and the second fundamental form as
\begin{equation}
\mathcal{Q}_k = k\int_0^1 d\xi \, \theta^{a_1 a_2}\wedge \mathfrak{R}^{a_3 a_4}(\xi)\wedge\ldots\wedge\mathfrak{R}^{a_{2k-1} a_{2k}}(\xi)\wedge e^{a_{2k+1}\cdots a_d} \epsilon_{a_1\cdots a_d}~,
\label{genGH}
\end{equation}
where $\mathfrak{R}^{ab}=R^{ab}+(\xi^2-1)(\theta\wedge\theta)^{ab}$. The Gibbons-Hawking and Myers terms as written above are the individual contributions for $k=1,2$. 

Sometimes it is more natural and useful to write \reef{genGH} in terms of the extrinsic and intrinsic curvatures of the boundary. In chapter \ref{genHP} we will make extensive use of this form of the boundary terms. The extrinsic curvature is related to the second fundamental form as
\begin{equation}
\theta^{ab}=n^2(n^aK^b-n^bK^a) ~,
\label{extrinsiK}
\end{equation}
where again $n^{a}$ is the normal vector to the surface  and $K^A=K^A_{\ \ B} e^B$ is the extrinsic curvature 1-form. The extrinsic curvature can in turn be calculated as covariant derivative of the normal vector as 
\begin{equation}
K_{AB}=e_A^\m\nabla_{e_B}n_\m=-n_\m\nabla_{e_B}e_A^\m ~,
\end{equation}
where we used the fact that $n^a$ is normal to the vielbein basis induced on the surface.  Taking into account \reef{extrinsiK} and
\begin{equation}
\theta^a_{\ b}\wedge\theta^{b}_{\ c}=-n^2\,K^a\wedge K_c ~
\end{equation}
we finally get
\begin{equation}
\mathcal{Q}_k = -2k\int_0^1 d\xi \, K^{A_1}\wedge \mathfrak{R}_0^{A_2 A_3}(\xi)\wedge\ldots\wedge\mathfrak{R}_0^{A_{2k-2} A_{2k-1}}(\xi)\; e^{A_{2k}\cdots A_{d-1}} \epsilon_{A_1\cdots A_{d-1}}~,
\label{intrinsicaction}
\end{equation}
where $\mathfrak{R}_0^{AB}=R_0^{AB}-\xi^2\,n^2\,K^A \wedge K^B$ and everything has been expressed in terms of the vielbein basis adapted to the surface such that 
\beq
n^a\e_{ab\ldots} \sim -n^2\e_{B\ldots}
\eeq
This will be consistent with our conventions in chapter \ref{genHP}.

Another important r\^ole of the boundary terms in general relativity is that they give rise to the so-called Israel junction conditions that govern the dynamics of shells separating different bulk domains \cite{Israel1967}
By performing an integration of Einstein's equations across the shell and taking the thickless limit it is possible to show that the jump in the extrinsic curvature is related to the surface stress-energy tensor, $S_{AB}$,
\beq
K^+_{AB}-K^-_{AB}=8\pi G_N\left(S_{AB}-\frac12\eta_{AB}S\right)~.
\label{EHjunc}
\eeq
It can be shown that Israel's method for singular hypersurfaces is equivalent to an action principle with boundary terms at the hypersurface, the variation of the latter yielding the junction conditions \cite{Hayward1990}. 

The Lovelock action in the presence of singular hypersurfaces can also be written in terms of smooth bulk integrals plus boundary terms \cite{Gravanis2003a,Maeda2004,Gravanis2003}, that in the case of $n = 2$ LGB theory was first written down by Davis \cite{Davis2003}. One may consider the spacetime manifold as the union of two submanifolds with a common boundary $\Sigma$ in such a way that in addition to the boundary term at infinity we get two extra surface terms at the boundary between the two. The action in this case would be written as
\begin{equation}
\mathcal{I}_{tot}=(\mathcal{I}_- -\mathcal{I}_{\partial}^-) + (\mathcal{I}_+ +\mathcal{I}_{\partial}^+)-\mathcal{I}_\partial~,
\end{equation}
where $-$ ($+$) denote the inner (outer) region. Remark that the form of the surface terms at $\Sigma$ is the same as that of the boundary term at infinity,
\begin{equation}
\mathcal{I}_{\Sigma}=-\mathcal{I}_{\partial}^-
+\mathcal{I}_{\partial}^+  ~,
\end{equation}
the plus sign in the second term coming from the fact that the bulk regions induce opposite orientations on the common boundary\footnote{One can also construct solutions with the same orientation on both sides leading in turn to wormholes and Randrall-Sundrum-like models \cite{wormholes,Garraffo2008b}}. In case the spin connection is continuous at $\Sigma$, $\mathcal{I}_\Sigma$ vanishes and we recover the usual Lovelock action.

This way of writing the action is useful in order to find solutions where the spin connection is discontinuous across some codimension one hypersurface (the metric being continuous). The variation of the bulk terms on each side yield the usual Lovelock equations of motion while the junction conditions arise from the variation of the boundary terms with respect to the induced vielbein field or equivalently the pullback of the metric on $\Sigma$. We will comment more on this in section \ref{genHP} where we will be interested in this kind of distributional solutions. The issue of finding equations with singular sources is a non-trivial one in non-linear gravity theories as many operations with distributions are not unambiguously defined.


In order to take the variation of the action written in this way we also have to vary with respect to the spin connection induced in the intermediate surface. This variation is again zero as it cancels the boundary term coming from the bulk integrals (this is the reason we introduced these terms to begin with) whereas the variation with respect to the intrinsic metric is proportional to the canonical momenta in such a way that the variation of each term may be written as
\be
\delta \mathcal{I}_\partial=-\int_{\partial M}\!\!\! d^{d-1} x\, \pi^{AB}\delta h_{AB}~.
\ee
Therefore, the junction conditions at the surface $\Sigma$ amount precisely to continuity of the canonical momenta \cite{Teitelboim1987,Banados1994a,Miskovic2007}. The analogous term from the boundary does not contribute as we keep the boundary metric fixed, $\delta h=0$. The canonical momenta in Lovelock gravity can be expressed as 
\bear
\pi^B_{\ C}&=&-\frac12\frac{\delta \mathcal{I}_\partial}{\delta e^C}\wedge e^B \label{momenta}\\&=& \sum_{k=1}^{K}{k\, c_k} \int_0^1 d\xi \, K^{A_1}\wedge \mathfrak{R}_0^{A_2 A_3}(\xi)\wedge\ldots\wedge\mathfrak{R}_0^{A_{2k-2} A_{2k-1}}(\xi)\; e^{A_{2k}\cdots A_{d-2}B} \epsilon_{A_1\cdots A_{d-2}C}~,\nonumber
\eear
The generalization of the Israel junction conditions to Lovelock gravity being 
\be
\pi^+_{AB}-\pi^-_{AB}=8\pi G_N\, S_{AB}
\label{genjunc}
\ee
that reduces to \reef{EHjunc} in the case of EH gravity.

Junction conditions can also be seen to arise in our 1-dimensional example. First of all, the same kind of boundary term appears if we split the action, or the interval of integration, in two. The variation of each term in that case is 
\be
\delta I^-=\int_{t_1}^{t_\star}{ dt \left[\frac{\partial L}{\partial q}-\frac{d}{dt}\left(\frac{\partial L}{\partial \dot{q}}\right)\right]\delta q}+\left.\frac{\partial L}{\partial\dot{q}}\right|_{t_\star}\delta q_\star
\ee
The first term vanishes because of the equations of motion but the second does not as the position is not fixed for $t=t_\star$. That term combines from an analogous one coming from the other part of the action, $I^+$, to yield the {\it junction condition}
\be
p^--p^+=0 ~,
\ee
\ie continuity of the canonical momentum. For univalued momentum such as in \reef{I} this in turn implies the continuity of the velocity. The free particle trajectory is necessarily continuous and smooth without any need of imposing any of these conditions \`a priori. There is no non-smooth solution of $\ddot{x}=0$. In the same way as for EH gravity, in order to add a discontinuity in the velocity we need to include new {\it source} terms. In the particle example the analogue of a {\it dust shell} would be localized at a given time, $t=0$ for simplicity. 
\be
\tilde{I}=I+\int_{t_1}^{t_2}{dt \lambda x\delta(t) }=I+\lambda x(0) ~.
\ee
When varying the action, fixing $x$ in the borders, $x(t_{1,2})=x_{1,2}$ we get the equation
\be
\frac{d}{dt}\left(\frac{\partial L}{\partial \dot{x}}\right)-\frac{\partial L}{\partial x}=\lambda \delta(t)~.
\ee
This equation includes {\it junction conditions} that can be found by integrating in infinitesimal region around $t=0$, between $t=0_-$ and $t=0_+$. In this way we find
\be
p^+-p^-=\lambda ~,
\ee
where $p^{\pm}=p(0_\pm)$. We could also have started with the splitted action in which case the junction conditions would arise from the variation on the {\it shell}, in this case the variation of $x(0)$ which is not fixed by the boundary conditions.

In order to keep the discussion as general as possible we will consider the action written in terms of the canonical variables $q$ and $p$ instead of the velocity,
\be
L(q,p)=p\dot{q}-H(q,p)~.
\label{Lag1}
\ee
In this way the lagrangian can be varied independently with respect to the two canonical variables, $q$ and $p$, yielding the well known Hamilton equations,
\be
\dot{p}=-\frac{\partial H}{\partial q}\qquad ; \quad \dot{q}=\frac{\partial H}{\partial p}~.
\ee
In the same way as before this action is prepared to fix the value of the position at the extrema, $\delta q=0$. However we can also use a different lagrangian,
\be
L'(q,p)=-\dot{p}q-H(q,p)~,
\label{Lag2}
\ee
that in turn is prepared to fix the momenta $p$ instead. This can be easily understood as a result of the transformation $q\leftrightarrow p$, $\dot{q}\leftrightarrow -\dot{p}$ between the two lagrangians. However we can supplement the latter \reef{Lag2} with a boundary term in such a way that it is equivalent to \reef{Lag1},
\be
L(q,p)=L'(q,p)+\frac{d}{dt}(pq)
\ee
Obviously the right hand side is obtained from the original lagrangian just by integrating by parts. The nice thing about this way of writing the action is that now we can split a given interval in two pieces and vary the action not just with respect to the {\it bulk} variables but also with respect to the ones at the {\it shell}, $t=t_\star$,
\be
I'=\int_{t_1}^{t^\star}{dt\,L'(q,p)}+(p^--p^+)q_\star+\int_{t^\star}^{t_2}{dt\,L'(q,p)}~.
\ee
The variation with respect to the momentum on the {\it shell} again cancels the contribution of the boundary term, whereas inside the intervals $(t_1,t_\star)$ and $(t_\star,t_2)$ it also vanishes due to the equations of motion. The only contribution to the variation thus comes from the variation of the boundary term with respect to the {\it shell} position, 
\be
\delta I'=(p^--p^+)\delta q_\star ~,
\ee
again implying the continuity of the canonical momentum across the {\it discontinuity}. We may also add source terms that induce jumps in the canonical momentum, similar to singular matter distributions in GR. 

As we mentioned above, for univalued momentum, its continuity implies that the the velocity is continuous as well. In more general cases however, the momentum may be multivalued this becoming a non-trivial equation. The velocity may jump as long as the canonical momentum is conserved. We will comment more on this on chapter \ref{genHP}, with a specific example.

\chapter{\bfseries\itshape Lovelock black holes}
\chaptermark{Lovelock black holes}
\label{chp:LLbh}

\vspace{.6cm}

\begin{quotation}
\flushright
{\it ``To myself I am only a child playing on the beach,\\ while vast oceans of truth lie undiscovered before me.''}\\

\vspace{.3cm}

Isaac Newton
\end{quotation}

\vspace{3cm}

\noindent The concept of singularity is central to general relativity. Due to the attractive and universal nature of the gravitational interaction, the theory predicts that these kind of objects inevitably form, either in the form of a black hole or as a cosmological singularity such as the {\it Big Bang}.

The first to describe a singular solution of the equations of general relativity was Karl Schwarzschild \cite{Schwarzschild1916} already in 1916, soon after the publication of Einstein's original paper. It was the first exact solution of Einstein's equations besides the trivial Minkowski metric. Unfortunately, Schwarzschild had contracted a disease while serving in the German army during World War I and died shortly after his paper was published. The singular character of the solution he found was at first considered just as a mathematical curiosity, of none physical relevance, until it was realized quite a long time afterwards that such objects actually do generally form from the collapse of matter \cite{Chandrasekhar1931,Oppenheimer1939} such as that of a dying star. The density of a physical object of radius $R$ is bounded, in such a way that if $R$ dips below the Schwarzschild radius the system will undergo gravitational collapse and become a black hole. However it was not until the sixties, with the advent of the singularity theorems of Hawking and Penrose \cite{Penrose1965,Hawking1965}, that the debate was definitely settled. In short,  a black hole is a self-gravitating object so densely packed that nothing, not even light, can scape its gravitational attraction. Nowadays black holes are thought to be quite common objects in the Universe being generally present at the center of galaxies such as the Milky Way. They cannot be directly seen but their presence is detected through the trajectories of stars on their vicinity or radiation coming from their accretion disks.

The fact that the gravitational field can affect the trajectory of light rays is well known. In fact it was the way Eddington proved Einstein theory right in his famous 1919 expedition to Africa. The idea of an object from which not even light can scape is much older though. We can trace it back as far as 1783, to a letter \cite{Michell1784} John Michell sent to Henry Cavendish, his fellow at the Royal Society of London. In that letter, using just Newtonian gravity, Michell describes the hypothetical case of a {\it heavenly object} so massive that not even light could escape its gravitational pull. Michell speculated, were the escape velocity at the surface of a star equal or greater than the speed of light, the generated light would be gravitationally trapped, and the star invisible to a distant observer. He named his discovery {\it dark star}, the precursor of black holes.

For an object leaving the surface of a dark star of mass $M$ with some speed $v\geq v_s$ to reach infinity we need the sum of its kinetic and gravitational energy to be equal or greater than zero,
\be
\frac{1}{2}mv^2-\frac{G_N M m}{R}\geq 0 \qquad\Rightarrow \quad v_s^2=\frac{2 M G_N}{R}
\ee
in such a way that the radius of the dark star has to be smaller than
\be
R\leq R_{\rm S} \equiv \frac{2 M G_N}{c^2} 
\ee
which, curiously enough, is independent of the mass of the object and actually coincides with the Schwarzschild radius of general relativity, $r=2M$ in geometric units. In that context this particular radial position is named {\it event horizon}.

The concept of black hole is quite different from that of a dark star. Nothing {\it sent} from the dark star can reach infinity but it can leave the {\it star} and even reach infinity if we furnish some extra acceleration. The black hole however is provided with an event horizon that acts as a one way membrane. Objects can get into the horizon but they cannot get back out. More precisely, consider the form of the Schwarzschild metric in general relativity
\be
ds^2=-\left(1-\frac{2M}{r}\right)dt^2+\frac{dr^2}{\left(1-\frac{2M}{r}\right)}+r^2d\Omega^2~.
\ee
The time and radial variables exchange their r\^oles beyond $r=2M$ in such a way that the $r$ coordinate becomes timelike and $t$ spacelike. Being timelike, $r$ has to decrease along any timelike trajectory in the same way as the time ticks forward outside the black hole. In fact, any object falling through the event horizon will reach the central singularity in finite proper time, it would be inevitably driven there. 

The event horizon plays yet another very important r\^ole. As it prevents anything from leaving the black hole, it effectively divides the spacetime in two. Nothing happening inside the horizon can ever influence the dynamics of the exterior region. This is essential in order to have a well-defined initial value problem in the presence of a singularity. The singularity represents a break down of the theory, in a sense, it is the place where general relativity shows its failure. It is also the place where quantum effects become dramatically important so that we would need a quantum theory of gravity to disclose the dynamics at the singularity. The existence of the event horizon protects the exterior region from this unknown dynamics, the exterior evolution being always well defined.  In 1969 Penrose made this idea precise and conjectured that, in the context of general relativity,  there can be no singularity visible from future null infinity. In other words, singularities need to be hidden from any observer at infinity by the event horizon of a black hole. This is known as the {\it weak cosmic censorship hypothesis} \cite{Penrose1969}.

We will study the analogous solution to that of Schwarzschild in the context of general Lovelock theories of gravity. Finding an analytic black hole solution
requires to explicitly solve a polynomial equation and we are certainly restricted by
the implications of Galois theory; meaningly, quartic is the highest order polynomial
equation that can be generically solved by radicals (Abel-Ruffini theorem). However,
an implicit but exact solution can be found, and we develop some tools to extract all
relevant information, mainly their horizon structure and thermodynamics. We will devote this chapter to present our proposal to deal with generic black holes in Lovelock theory, focusing in the case of LGB and cubic Lovelock for a detailed description. In the next chapter we perform a classification of all possible black hole solutions, including the case of asymptotically dS solutions, and all possible horizon topologies within maximally symmetric configurations.

The analysis of these solutions for the Lovelock family may also provide some useful information about the dynamics of black holes in more general gravity theories. This is specially important due to the high nonlinearity of the field equations that makes very difficult finding nontrivial analytical solutions of Einstein's equation with higher derivative terms. In most cases, one has to adopt some approximation methods or find solutions numerically. In the last few months there were some papers constructing gravitational theories that share some compelling properties with Lovelock lagrangians \cite{Oliva2010, Oliva2010a}. In particular, these are lower dimensional theories displaying black hole solutions whose profile precisely correspond to Lovelock black holes \cite{Myers2010c,Oliva2011}.  In particular these theories allow for the addition of an extra term of degree $K=\frac{d+1}{2}$ in odd dimensions \cite{Oliva2011}, contributing in every way as the corresponding Lovelock term in higher dimensions. Some other higher order terms may be also added that do not change the form of the black hole solution. Some of the results of this thesis are therefore of direct application to those cases as well. This is particularly interesting due to the fact that quasi-topological gravities are higher curvature theories in dimensions lower than their corresponding Lovelock cousins, thus the results are of interest in more `physical' setups of AdS/CFT \cite{Myers2010d}.

\section{Black holes in Lovelock gravity}

It has been shown in \cite{Aros2001} that Lovelock theories admit asymptotically (A)dS solutions with non-trivial horizon topologies. We can consider for instance solutions with a planar or hyperbolic symmetry as a straightforward generalization of the usual spherically symmetric ansatz,
\begin{equation}
ds^2 = - A(t,r)\, dt^2 + \frac{dr^2}{B(t,r)} + \frac{r^2}{L^2}\, d\Sigma_{d-2,\sigma}^2 ~,
\end{equation}
where
\begin{equation}
d\Sigma_{d-2,\sigma}^2 = \frac{d\rho^2}{1 - \sigma \rho^2/L^2}+\rho^2 d\Omega^2_{d-3} ~,
\end{equation}
is the metric of a $(d-2)$-dimensional manifold of negative, zero or positive constant curvature ($\sigma =-1, 0, 1$ parametrizing the different horizon topologies), and $d\Omega^2_{d-3}$ is the metric of the unit $(d-3)$-sphere. This does not imply that the horizon is just spherical or non-compact. By means of the Killing-Hopf theorem \cite{Wolf2011}, any complete connected Riemannian manifold of Euclidean signature and constant curvature $\sigma$ can be written as a quotient space $\Sigma_{d-2,\sigma}/\Gamma$, where $\Gamma$ is a discrete subgroup of the isometry group of $\Sigma_{d-2,\sigma}$. Thus, even in (what we shall call) the {\it spherical} case, we have non-spherical possibilities; for example, one may take the horizon to be a lens space. Besides, planar or hyperbolic horizons can be made compact in this way.

It has been proven in \cite{Zegers2005} that these black holes admit a version of Birkhoff's theorem, in such a way that in addition to the $SO(d-1)$, $E_{d-2}$ or $SO(1,d-2)$ isometry groups, these spacetimes admit  an extra timelike killing vector (for $A,B>0$). This means that these solutions of the field equations are locally isometric to their corresponding static counterparts, which can be found by means of the ansatz
\begin{equation}
ds^2 = - f(r)\, dt^2 + \frac{dr^2}{f(r)} + \frac{r^2}{L^2}\, d\Sigma_{d-2,\sigma}^2 ~.
\label{bhansatz}
\end{equation}
There are extra solutions with different functions in the timelike and radial direction but they are just valid for degenerate values of the cosmological constant \cite{Charmousis2002a}. In that case, the most general solution is
\begin{equation}
ds^2 = - f(r)\, dt^2 + \frac{dr^2}{(\sigma-\Lambda\, r^2)} + \frac{r^2}{L^2}\, d\Sigma_{d-2,\sigma}^2 ~,
\label{lifshitzbh}
\end{equation}
for any function $f(r)$. This allows in particular Lifshitz-like solutions $f(r)\sim r^{2z}$ for any value of the critical exponent $z$.

These black hole solutions are all three asymptotic to a maximally symmetric space. Thus, when considering the same curvature for all of them they are locally asymptotically equivalent, but globally different. They are often referred to as topological black holes for this reason. Indeed, there are global changes of coordinates that relate the sets of coordinates corresponding to the three topologically different vacuum solutions associated with a given $\Lambda$ \cite{Emparan1999b}. Each set  covers a different patch of AdS and has a different time coordinate. Thus, we can also look at the different topologies as static black holes for different classes of observers.

Using the natural frame,
\begin{equation}
e^0 = \sqrt{f(r)}\, dt ~, \qquad e^1 = \frac{1}{\sqrt{f(r)}}\, dr ~, \qquad e^a = \frac{r}{L}\, \tilde e^a ~,
\label{vierbh}
\end{equation}
where $a = 2, \ldots, d-1$, and $\tilde R^{ab} = \sigma\, \tilde e^a \wedge \tilde e^b$. The Riemann 2-form reads
\begin{eqnarray}
& & R^{01} = - \frac12\, f''(r)\; e^0 \wedge e^1 ~, \quad\qquad R^{0a} = - \frac{f'(r)}{2 r}\; e^0 \wedge e^a ~, \nonumber \\ [1em]
& & R^{1a} = - \frac{f'(r)}{2 r}\; e^1 \wedge e^a ~, \qquad\qquad R^{ab} = - \frac{f(r) - \sigma}{r^2}\; e^a \wedge e^b ~.
\label{riemannbh}
\end{eqnarray}
If we insert these expressions into the equations of motion, we get
\begin{equation}
\left[ \frac{d~}{d\log r} + (d-1) \right]\, \left( \sum_{k=0}^{K} c_k\, g^k \right) = 0 ~,
\end{equation}
where $g = (\sigma- f)/r^2$. This can be readily solved as
\begin{equation}
\Upsilon[g]=\sum_{k=0}^{K} c_k\,  g^k = \frac{\kappa}{r^{d-1}} ~,
\label{eqg}
\end{equation} 
where $\kappa$ is an integration constant related to the mass of the spacetime \cite{Kastor2010,Kastor2011},
\begin{equation}
M=\frac{(d-2)V_{d-2}}{16\pi G_N}\, \kappa ~,
\label{mmass}
\end{equation} 
$V_{d-2}$ being the volume of the unit $(d-2)$-dimensional horizon. Notice that the polynomial giving the implicit black hole solution is the same as the one defining the possible vacua of the theory. This is not surprising as maximally symmetric spaces appear as {\it massless} solutions, $M=0$. This can also be understood as follows. If there is actually a mass source for the gravitational equations of motion, $\rho=M \delta^{(d-1)}(r)$, therefore
\begin{equation}
\mathcal{E}_0\wedge e^0 \sim T^0_0 \qquad\quad \Rightarrow \qquad\qquad \left[\frac{d}{d\log r}+(d-1)\right]\frac{\kappa}{r^{d-1}}\sim \rho ~,
\label{handy}
\end{equation}
and, as the right hand side of the equation does not depend on the Lovelock theory we are considering, the left hand side cannot either. Thus, the relation between $\kappa$ and the mass must be the same as in Einstein-Hilbert gravity (\ref{mmass}). 
This assertion can be made precise using the Hamiltonian formalism \cite{Kastor2011}.

For arbitrary dimension, the spherically symmetric solutions where found in \cite{Boulware1985a,Wheeler1986,Wheeler1986a} and their extension to planar and hyperbolic symmetry was given in \cite{Cai1999,Aros2001}. 

\section{Branches}

Notice that (\ref{eqg}) leads to $K$ different roots for every value of the radius and, thus, to $K$ different branches associated to each of the cosmological constants (\ref{cc-algebraic}) (some of them may be imaginary, though), in such a way that $g_i(r\rightarrow\infty)=\Lambda_i$. For instance, in LGB gravity there are two branches that read 
\begin{equation}
g_{(\pm)} = -\frac{1}{2 L^2 \lambda} \left( 1 \pm \sqrt{1 - 4 \lambda \left( 1 - \frac{\kappa L^2}{r^{d-1}} \right)} \right) ~,
\label{GBbranches}
\end{equation}
each one associated with a different cosmological constant. As for the corresponding vacua we need $\lambda\leq1/4$ in order to have real solutions, otherwise the argument of the square root may become negative at some finite radius. Only one of the solutions, $g_{(-)}$, is {\it connected} to the standard Einstein-Hilbert gravity, in the sense that it reduces to it when $\lambda\rightarrow 0$,
\begin{equation}
g_{(-)} \approx -\frac{1}{2 L^2 \lambda} \left( 1 - \left[ 1 - 2 \lambda \left( 1 - \frac{\kappa L^2}{r^{d-1}} \right) \right] \right) = -\frac{1}{L^2} \left( 1 - \frac{\kappa L^2}{r^{d-1}} \right) ~,
\end{equation}
while $g_{(+)}$ blows up in that limit. It will be referred to as the EH branch. It can be seen that this is the branch corresponding to the intersection of $\Upsilon[g]$ with the vertical axis, $g = 0$. The $K$ different branches of (\ref{eqg}) are continuous functions of the radial coordinate, as long as the roots of a polynomial equation depend continuously on its coefficients \cite{Marden1949}, and $r$ enters monotonically in the zeroth order coefficient $ \tilde c_0(r)\equiv c_0 -\kappa/r^{d-1}$. When $r\to\infty$, (\ref{eqg}) is nothing but the expression leading to the $K$ cosmological constants.

\begin{figure}
\centering
\includegraphics[width=0.54\textwidth]{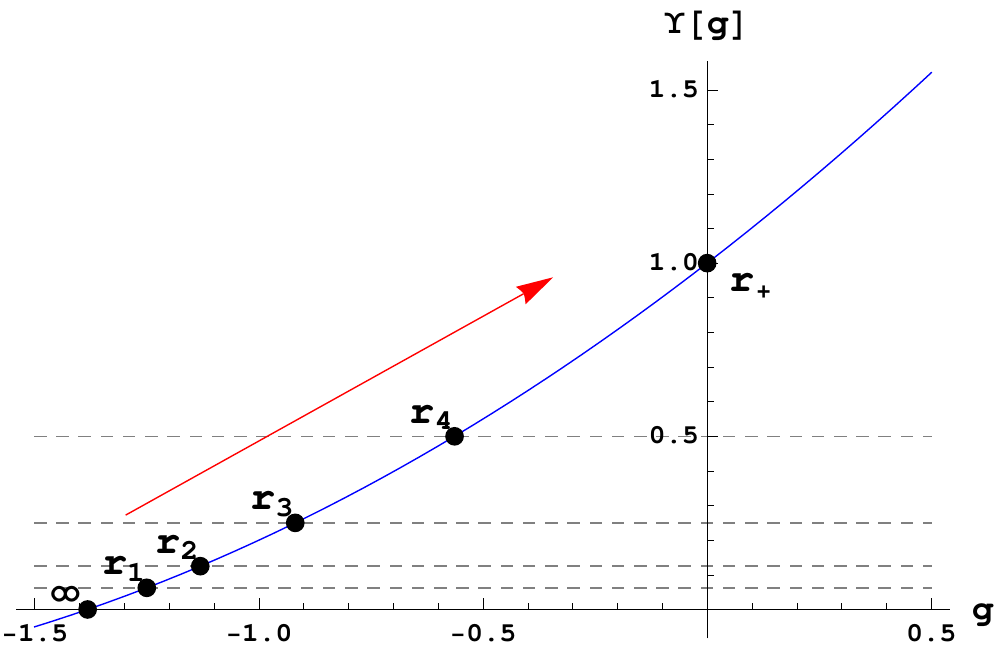}
\caption{A branch of the polynomial $\Upsilon[g]$ for the case $K=2$, {\it i.e.}, GB theory (with $\lambda = 0.2$ and $L = 1$), in arbitrary spacetime dimension, for different values of the radius ranging from $\infty$ to $r_+$, $r_1>r_2>\ldots>r_+$. The projection of the depicted points give $g(r_i)$ for the EH branch in the planar case ($\sigma=0$).}
\label{polynomial-root}
\end{figure}

The different Lovelock couplings $c_{k>1}$ fix the shape of the polynomial $\Upsilon[g]$. While varying $r$ from $\infty$ to $r_+$ (see figure \ref{polynomial-root}), the function $g(r)$ is given by the implicit solution of equation (\ref{eqg}) that graphically corresponds to climbing up (down for negative masses) a given monotonic part of the curve $\Upsilon[g]$ starting from one of its roots (tantamount of a given cosmological constant). 
The metric function $g$ is a monotonic function of $r$ since $\tilde c_0(r)$ is so and the remaining coefficients are frozen. Then each branch can be identified with a monotonic section of the polynomial $\Upsilon[g]$, and can easily be visualized graphically.

As discussed above, the propagator of the graviton corresponding to the vacuum $\Lambda_i$ is proportional to $\Upsilon'[\Lambda_i]$ in such a way that we will restrict to positive values of that derivative. It is zero just for degenerate vacua where there are actually no linearized degrees of freedom. In the present context the restriction $\Upsilon'[\Lambda_i]>0$ is just verified by positive slope branches thus we will only consider those in the future. All the {\it relevant} or {\it BD-stable} branches correspond then to positive slope sections of the polynomial and, therefore, $g$ will be considered a monotonically decreasing function of $r$.

For positive $\kappa$ the solution runs over the points with positive value for $\Upsilon[g]$ while for negative mass it is the other way around. Either way, every branch always encounters a maximum/minimum, or it grows unboundedly.

For the sake of clarity and the ease of reading, let us first classify the different types of branches that one may encounter when dealing with a Lovelock theory of gravity. The appearance of a given type of branch will depend, in general, on the specific theory considered and on the values of the different coupling constants. On the one hand, we may classify the branches by their asymptotics: AdS, flat or dS branches. In the particular case we are considering, with $c_0=L^{-2}$, there are no asymptotically flat branches. The sign of the cosmological constant corresponding to the EH branch (when real) is the opposite to $c_0$ (or, equivalently, the same as the explicit cosmological constant, as in standard Einstein-Hilbert gravity); thus, the EH branch is asymptotically AdS. Due to the particular features and relevance of this branch, we will consider it separately.

Some of the branches (monotonic sections of the polynomial) may also be associated to complex values of $\Lambda$. Therefore, they do not correspond to real metrics and should be disregarded as unphysical. We will refer to these as excluded branches, and to the sector of the parameter space where the EH branch is excluded as the {\it excluded region}. 

We will then exhaustively classify branches on (non-EH) AdS ({\it i.e.}, not crossing $g=0$), EH, dS and excluded branches. The latter, being unphysical, do not need further discussion. The AdS branches must end at a maximum of the polynomial in order not to cross $g=0$. The other two cases may end at a maximum or, else, continue all the way up to $g\rightarrow\infty$. We will then consider two subclasses of branches: those (a) continuing all the way to infinity or (b) ending at a maximum. For the AdS branches we will also consider two subclasses: (a) positive mass and (b) negative mass.

\section{Singularities and horizons}
\label{singhor}

Where are the singularities of these spacetimes located? The simplest way to answer this question is to calculate the curvature scalar and see where it diverges. As it depends on the metric and its derivatives, these divergences can be traced back to those of the first derivative of $g$,
\begin{equation}
g' = - \frac{(d-1)\kappa}{r^d} \Upsilon'[g]^{-1} ~.
\label{singularity}
\end{equation} 
Then, the metric is regular everywhere except at $r=0$ and at points where $\Upsilon'[g]=0$; that is, whenever the branch we are looking at coincides with any other. In such case, 
\begin{equation}
\Upsilon'[g]=\sum_{k=1}^{K} k\,c_k\, g^{k-1} = 0 ~.
\end{equation}
These are precisely the maxima/minima at which all branches end, except those growing unboundedly (that also approach asymptotically to a singularity located at $r = 0$). The values of $r$ where this happens exhibit a curvature singularity that prevents from entering a region where the metric becomes complex. Type (a) branches correspond to solutions with a singularity at $r=0$ whereas those of (b) type display the singularity at finite radius. 

It can be easily seen that the mass parameter $\kappa$ must be positive in the planar case ($\sigma=0$) in order for the spacetime to have a well-defined horizon.  We can actually rewrite equation (\ref{eqg}) as
\begin{equation}
\sum_{k=1}^{K} c_k\,  g^k = \frac{\kappa}{r^{d-1}} - \frac{1}{L^2} ~,
\end{equation} 
and realize that the equation admits a vanishing $g$ only when $r = r_+ \equiv (\kappa L^2)^\frac{1}{d-1}$. In the planar case, furthermore, only one branch has a horizon at $r_+$ and all the rest display naked singularities. This is so since the polynomial root $g=0$ has multiplicity one at $r=r_+$ (higher multiplicity would require a vanishing coefficient of the Einstein-Hilbert term). In the case of LGB theory, for instance, we can see from (\ref{GBbranches}) that it is $g_{(-)}$. This is the above mentioned EH branch, a deformation of the solution to pure Einstein-Hilbert theory, and the only branch that remains when turning all the extra couplings off. Since $\Upsilon'[0] > 0$, $g(r)$ in the EH branch is decreasing close to $r_+$ and, thus, it is a relevant branch.

For non-planar horizons the situation is more complicated and, in principle, some of the branches admit horizonful black hole solutions even for negative values of $\kappa$. The physical mass of the black hole has to match the one of the matter contained in that region of spacetime. Thus, $\kappa$ will be considered a positive quantity, except for hyperbolic black holes for which some comments on negative mass solutions shall be made. In \cite{Mann1997a}, indeed, the formation by collapse of black holes with negative mass has been considered. We shall see that spherical or planar black holes always exhibit a naked singularity in the case of negative mass. The only horizon that may arise for those solutions is a cosmological one. This is the case for negative mass asymptotically dS branches even for hyperbolic topology.

Taking into account that the value of $g$ at the event horizon, $r = r_+$, reads
\begin{equation}
g_+\equiv\frac{\sigma}{r_+^2} ~,
\label{gplus}
\end{equation}
we can write $\kappa = r_+^{d-1}\Upsilon[\sigma/r^2_+]$. The other way around, the radii of the location of the horizons are given by solutions of the previous equation for any given value of $\kappa$. This leads to a more handy formula for the mass
\begin{equation}
M = \frac{(d-2)V_{d-2}}{16\pi G_N}\, r_{+}^{d-1}\;\Upsilon\left[\frac{\sigma}{r_+^2}\right] ~.
\label{mass}
\end{equation}
Following the argument used to derive (\ref{handy}), and taking into account that Einstein-Hilbert gravity has a positive energy theorem, we are tempted to conjecture that the same should apply to any Lovelock theory though this has not been proven so far. This conjectured positivity would not in principle rule out the negative mass solutions mentioned before as it may happen that the {\it positive mass} corresponds to the difference of the mass previously defined with the extremal one \cite{Horowitz1998}. 

We can recast the equation for the horizon, by means of (\ref{eqg}) and (\ref{gplus}), in such a way that it can be plotted in the $(g,\Upsilon[g])$-plane,
\begin{equation}
\Upsilon[g_+] = \kappa\left(\sqrt{\frac{g_+}{\sigma}}\right)^{d-1} ~,
\label{rheq}
\end{equation}
where the right hand side is just defined for positive/negative values of $g_+$, for $\sigma = \pm 1$, while for $\sigma=0$ the expression is not strictly valid since, in that case, $g_+$ exactly vanishes. Notice that in the high mass limit, $\kappa\rightarrow\infty$, the curve (\ref{rheq}) approaches the vertical axis --the planar black hole-- regardless of the value of $\sigma$.

\begin{figure}
\centering
\includegraphics[width=0.57\textwidth]{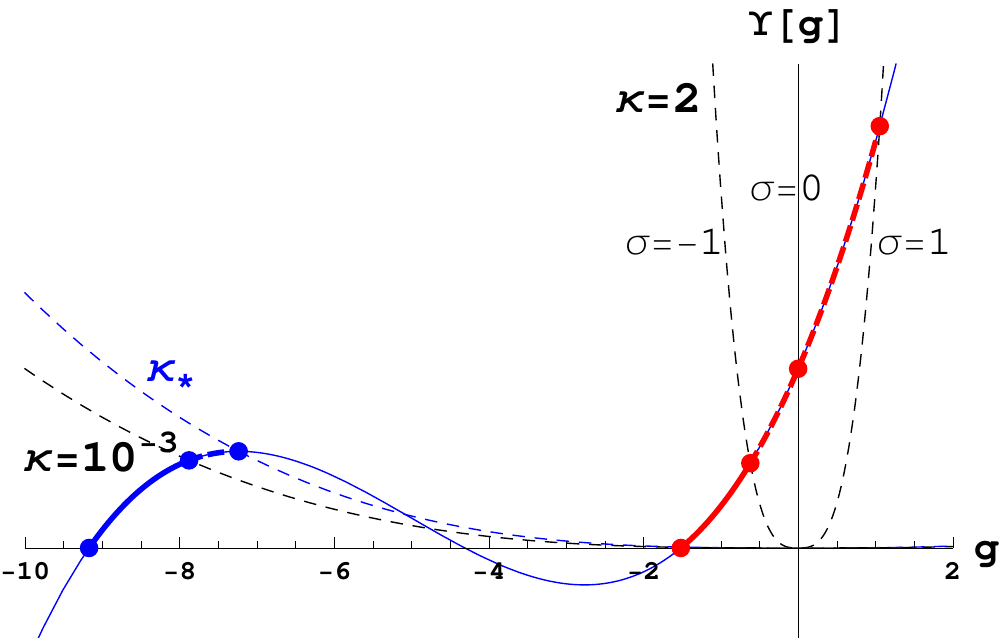}
\caption{Seven dimensional cubic Lovelock theory for $\lambda=1/4$ and $\mu=1/20$ ($L=1$) possesses two hyperbolic black holes for sufficiently low positive values of the mass. The dashed lines are plots of (\ref{rheq}) for the indicated values of $\sigma$ and $\kappa$ (in units of $L$). The crossing of these lines with the polynomial give the possible values for $g$ at the horizon and then of $r_+$. For $\sigma=-1$ and large enough values of $\kappa$, $\kappa > \kappa_\star$, the blue branch has a naked singularity (as it always has for $\sigma=0, 1$).}
\label{horizons}
\end{figure}

It is interesting to note that monotonicity of the function $g$ implies that every branch of black holes, for (positive mass) hyperbolic or planar topology, can have just one horizon. For $\sigma=0$ we recover just $g_+=0$, but for $\sigma\neq0$ we can actually have several possible values for $g_+$ (see figure \ref{horizons}).
Nonetheless, the right hand side of (\ref{rheq}) is monotonic in $g_+$ and each branch corresponds to a monotonic part of the polynomial $\Upsilon[g]$. We observe that, contrary to what happens for planar topology, there exists the possibility of having several branches with a horizon for $\sigma\neq0$. Some of them may be discarded by means of Boulware-Deser-like instabilities, while for some other branches horizons will appear or disappear depending on the actual values of the different couplings and $\kappa$.

For the case of hyperbolic horizons, as the slopes of both sides have opposite signs, there can just be at most one horizon per branch. In the positive curvature case, the determination of the number of horizons is, however, a non trivial matter. As the slope in both sides of the equation are positive we can even conceive the possibility of them crossing each other several times. We will illustrate this phenomenon below.

\begin{figure}
\centering
\includegraphics[width=0.7\textwidth]{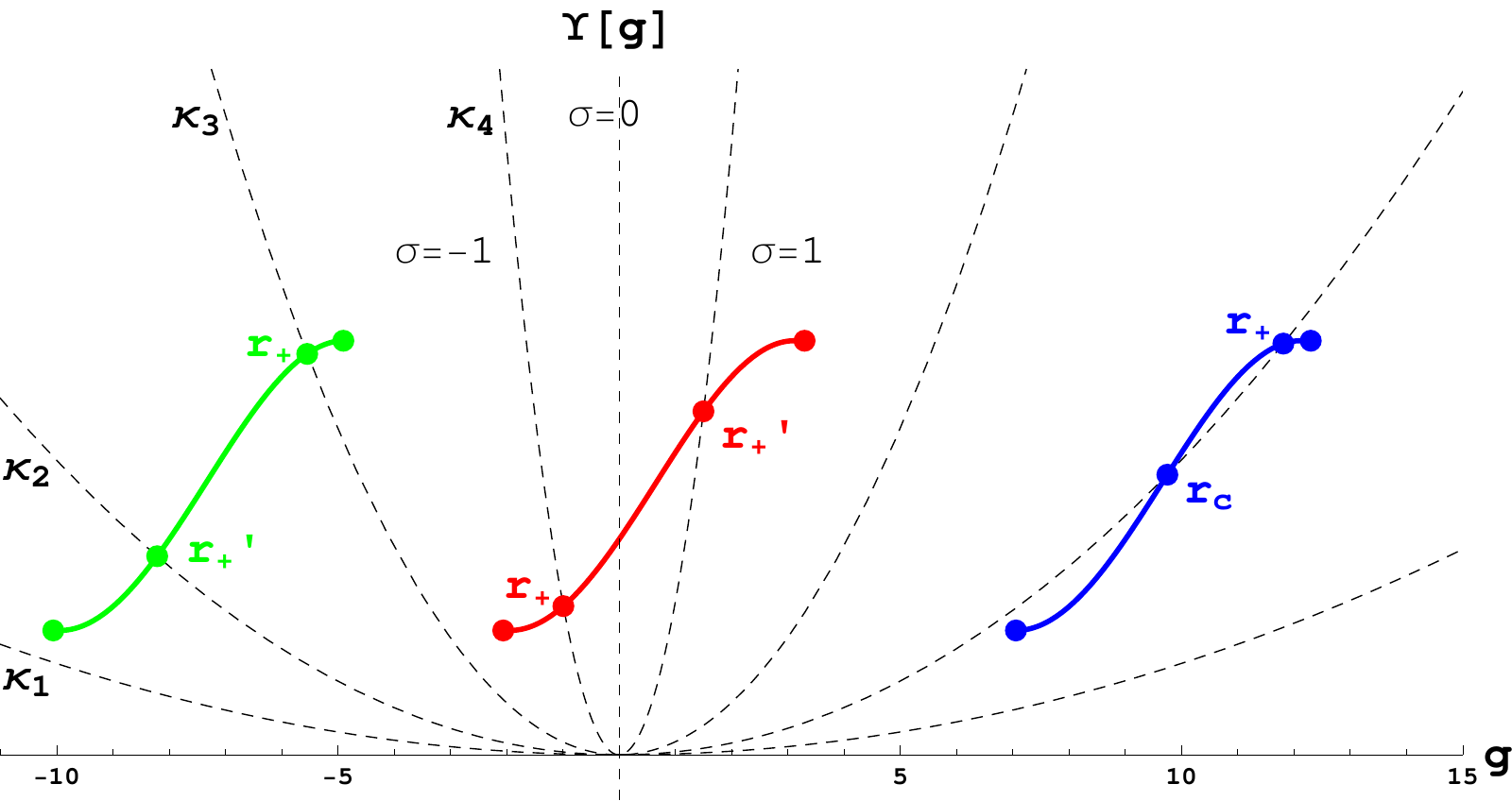}
\caption{Three examples of excluded branches running over positive, negative and positive, and just negative values of $g$ respectively. We also plot the values of the horizons for several values of the mass $\kappa_1<\kappa_2<\kappa_3<\kappa_4$ and all the topologies. The one crossing $g=0$ corresponds to the (excluded) EH branch. The blue branch describes a well-defined spacetime for some values of the mass with both singularities hidden behind the black hole and the cosmological horizons.}
\label{excluded}
\end{figure}

Depending on the couplings of Lovelock theory, it may happen that certain branches do not correspond to a proper vacuum. These coefficients fix the shape of the polynomial and, as they vary, some branches can become pathological in reason of their cosmological constant becoming imaginary. This happens whenever a monotonic part of the polynomial ends (towards the left) at a minimum without ever touching the $g$-axis (see figure \ref{excluded}). We refer to them as {\it excluded branches}. When the EH branch is excluded we say that we are in the excluded region of the parameter space.
These spacetimes have two singularities, one for small values of the radial coordinate at the maximum, and another one for large values of $r$ at the minimum. In the cases where we can just have one horizon, the nakedness of the singularity associated with the minimum cannot be avoided. In the $\sigma=1$ case we may have two (or more) horizons, each of them hiding a singularity and describing a regular spacetime in between.

At this point it should be clear that several different kinds of branches may generically arise in Lovelock theory, depending on the topology of the spacetime slicing, the coupling constants and the relevant AdS/dS vacuum. These are schematically summarized in Table \ref{cases}.
\begin{table*}
\centering
\begin{tabular}{c||c|c|c|}
asympt. & $\sigma=-1$ & $\sigma=0$ & $\sigma=1$ \\
\hline
\raisebox{9.3ex}{AdS}  & \includegraphics[width=0.26\textwidth]{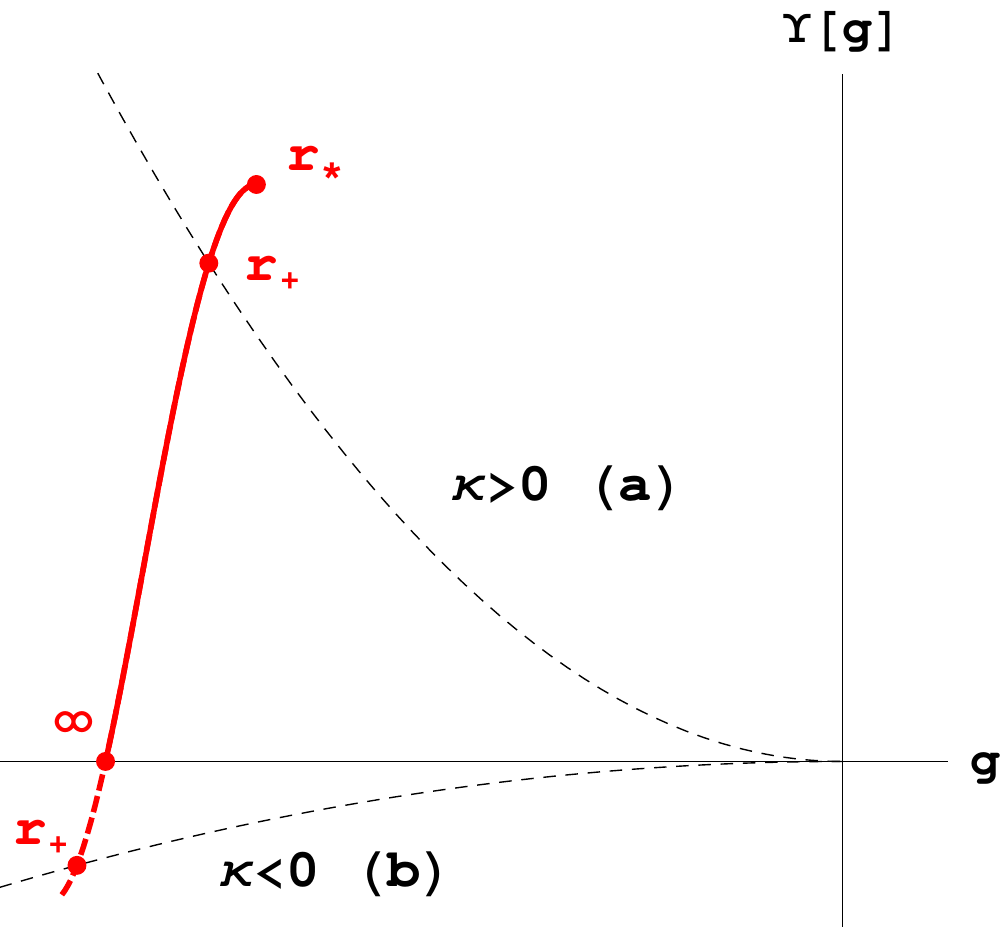} & \includegraphics[width=0.26\textwidth]{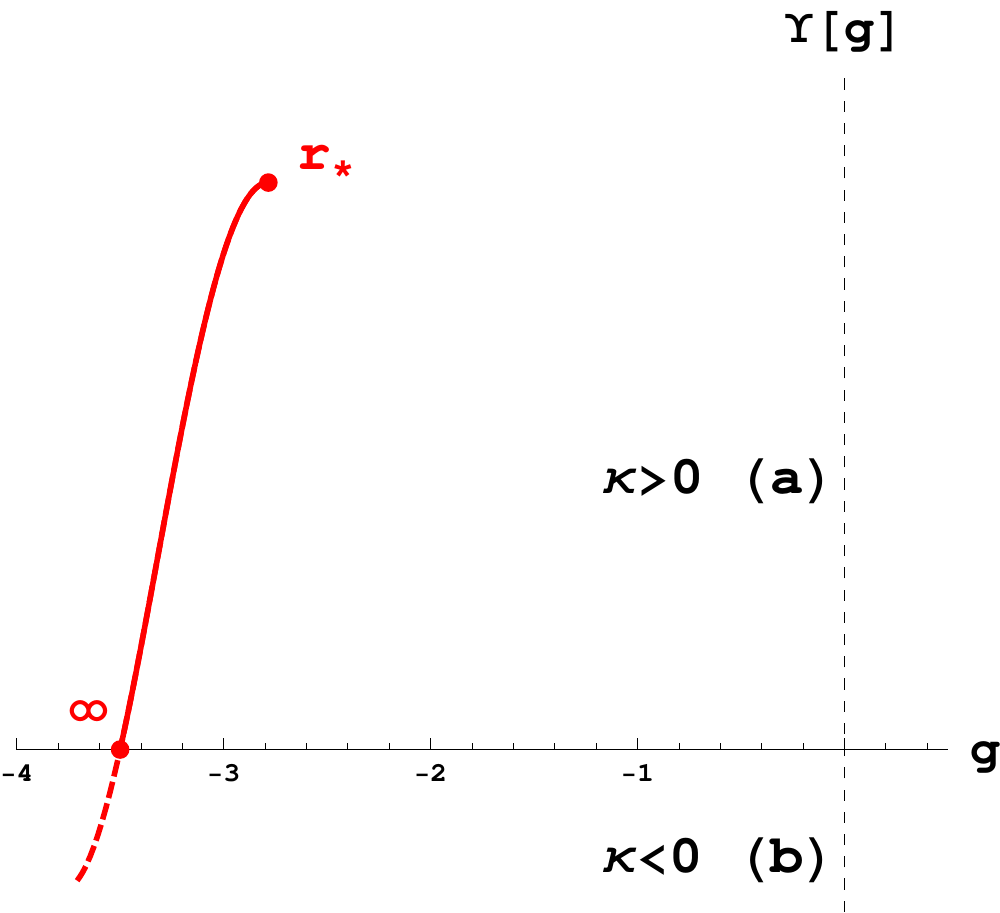} & \includegraphics[width=0.26\textwidth]{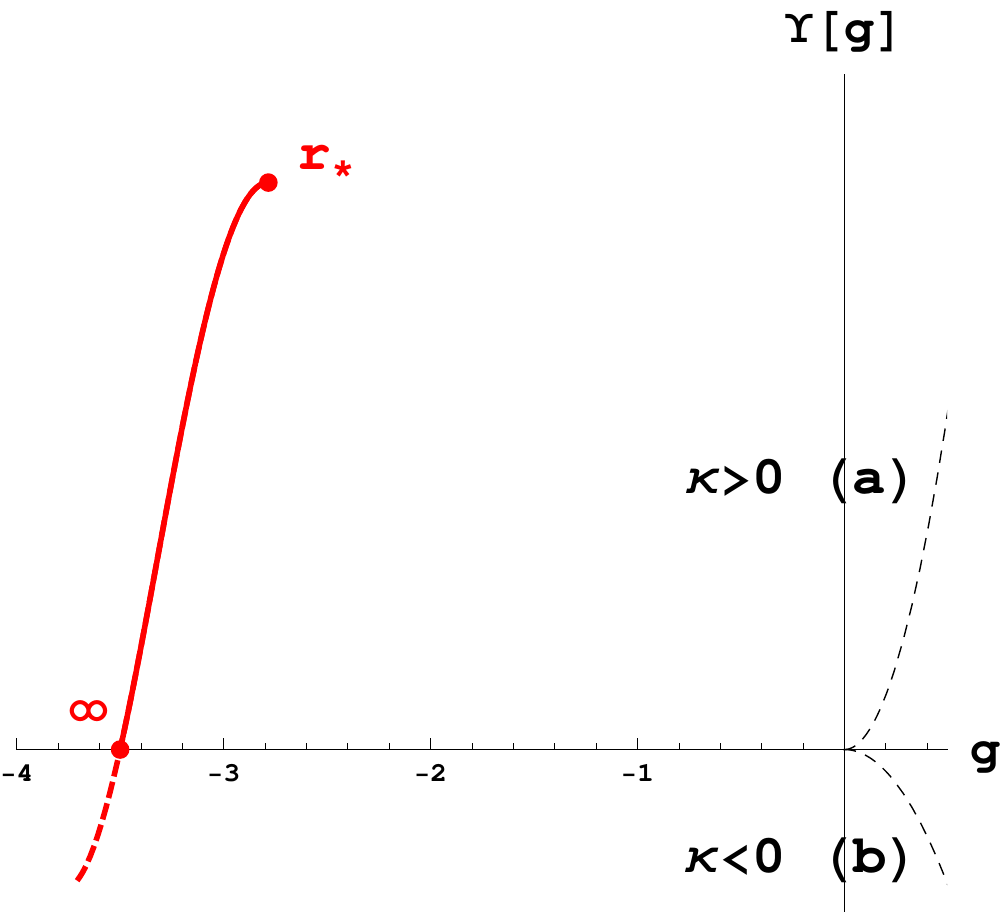} \\
\hline
\raisebox{9.3ex}{EH}  & \includegraphics[width=0.26\textwidth]{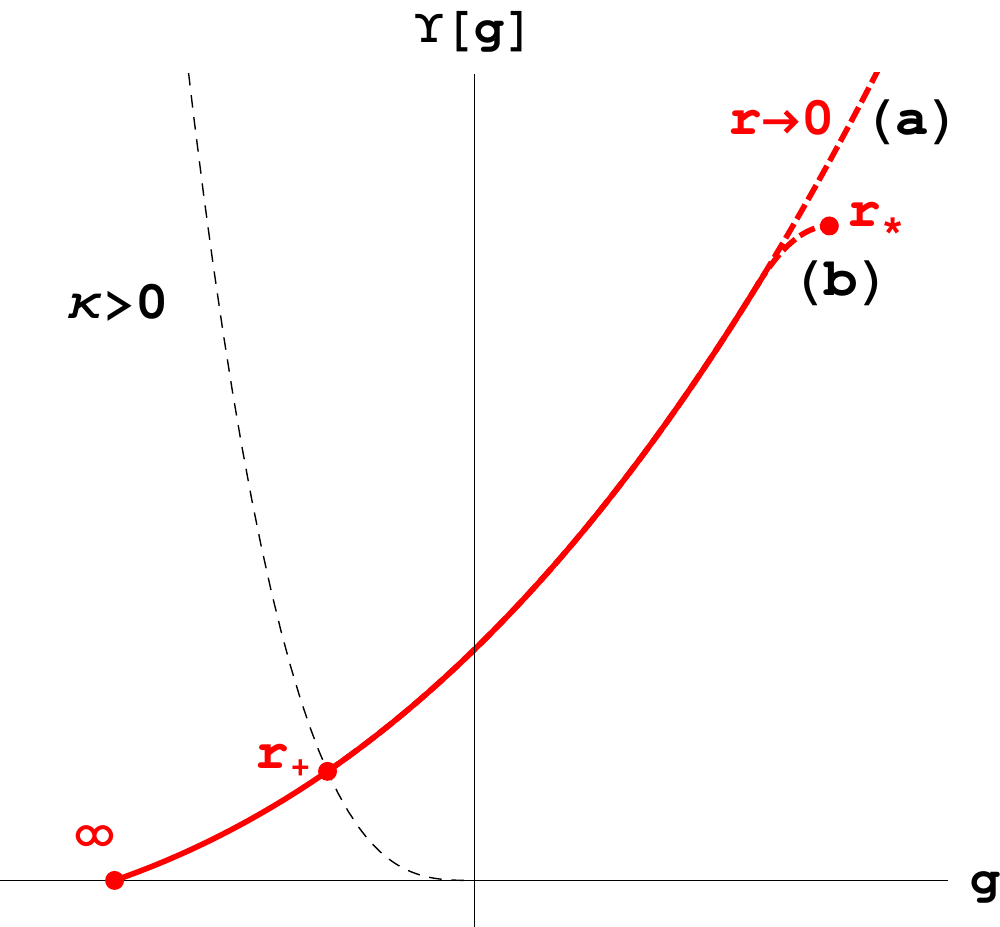} & \includegraphics[width=0.26\textwidth]{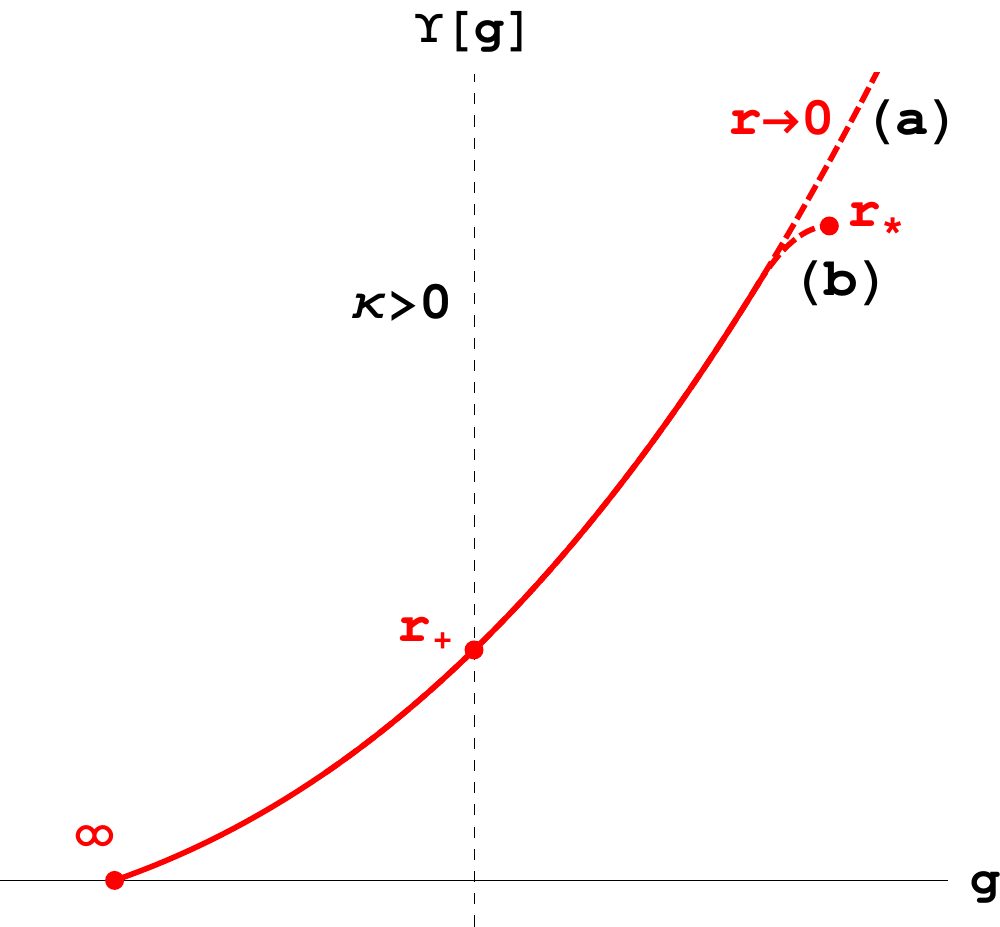} & \includegraphics[width=0.26\textwidth]{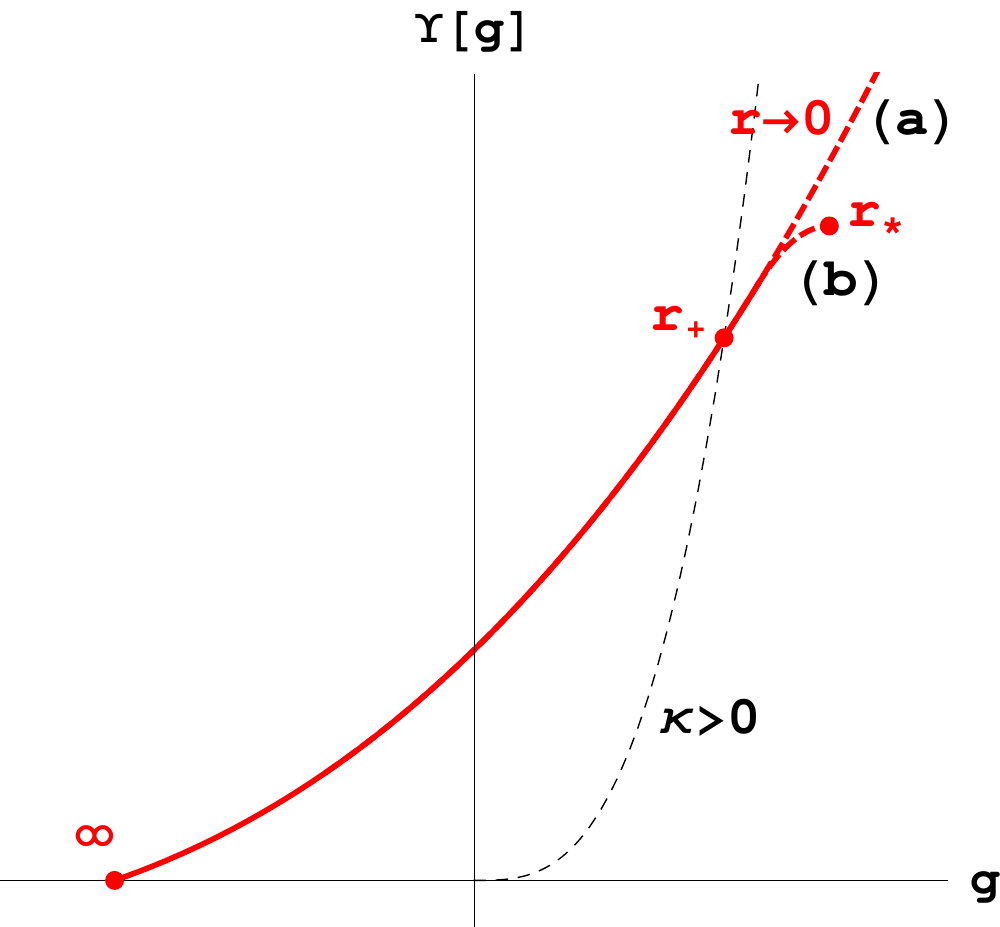} \\
\hline
\raisebox{9.3ex}{dS}  & \includegraphics[width=0.26\textwidth]{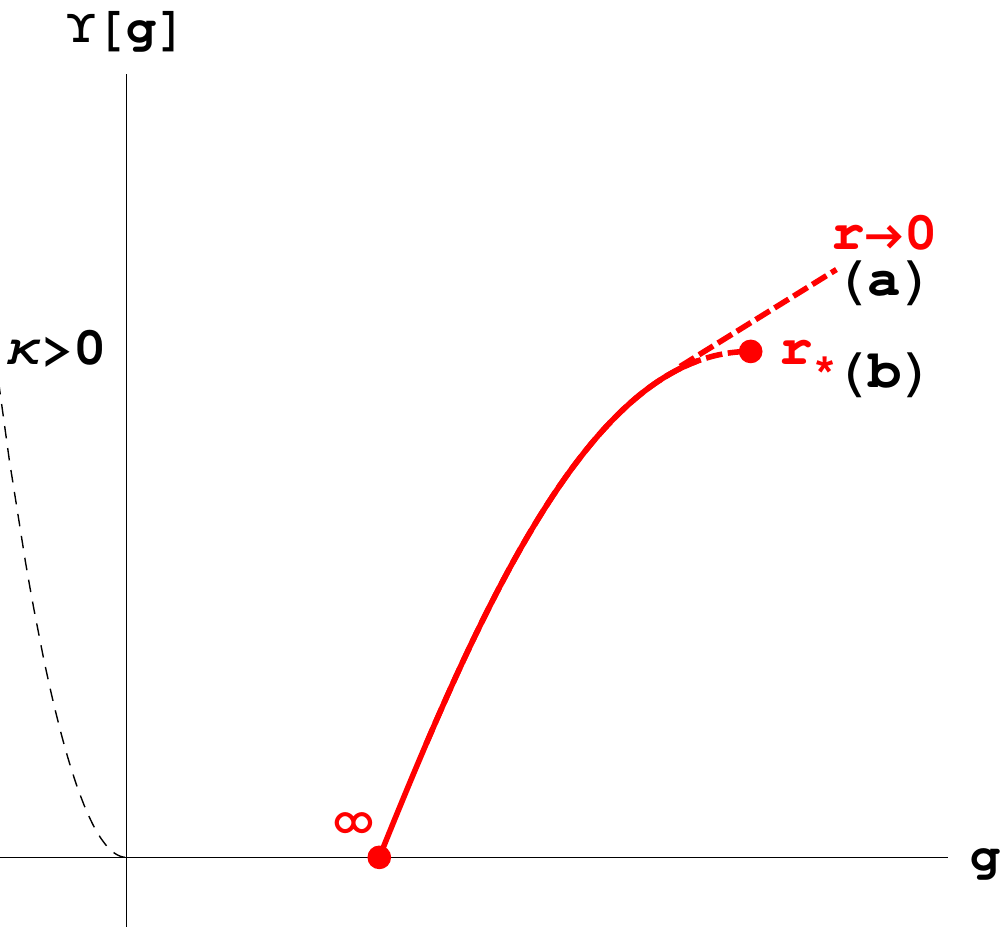} & \includegraphics[width=0.26\textwidth]{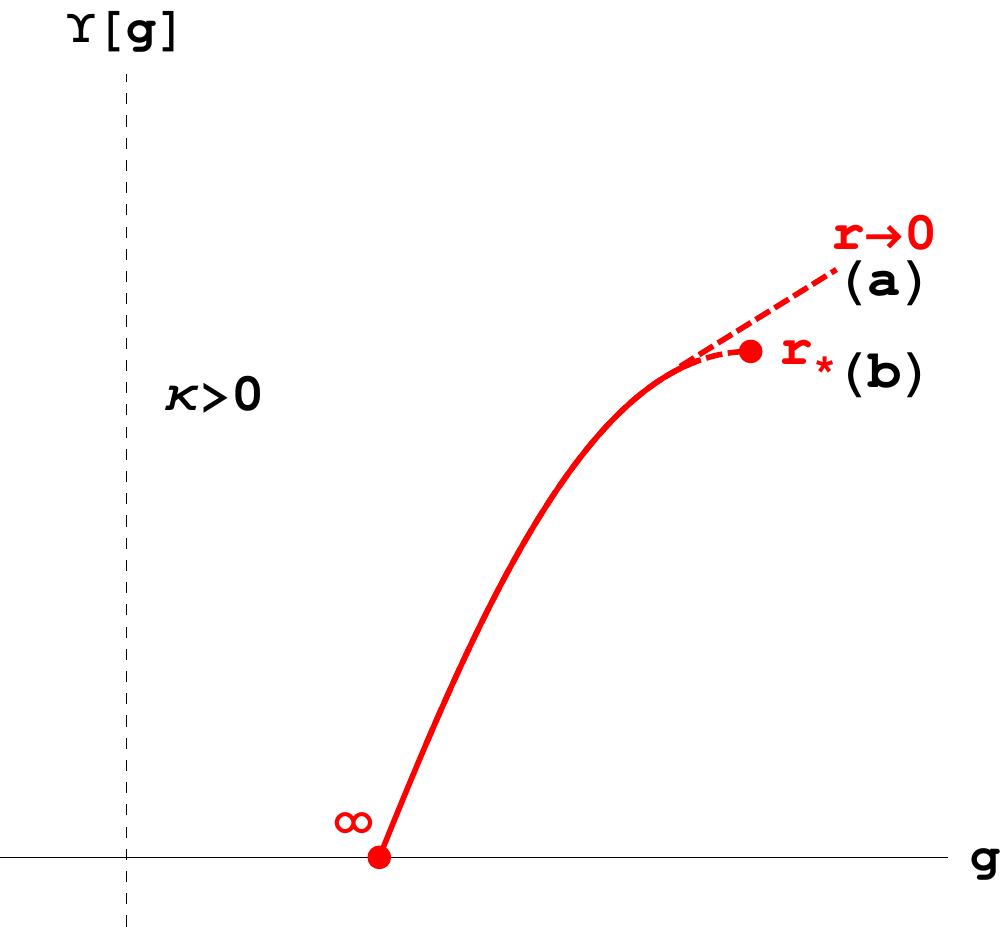} & \includegraphics[width=0.26\textwidth]{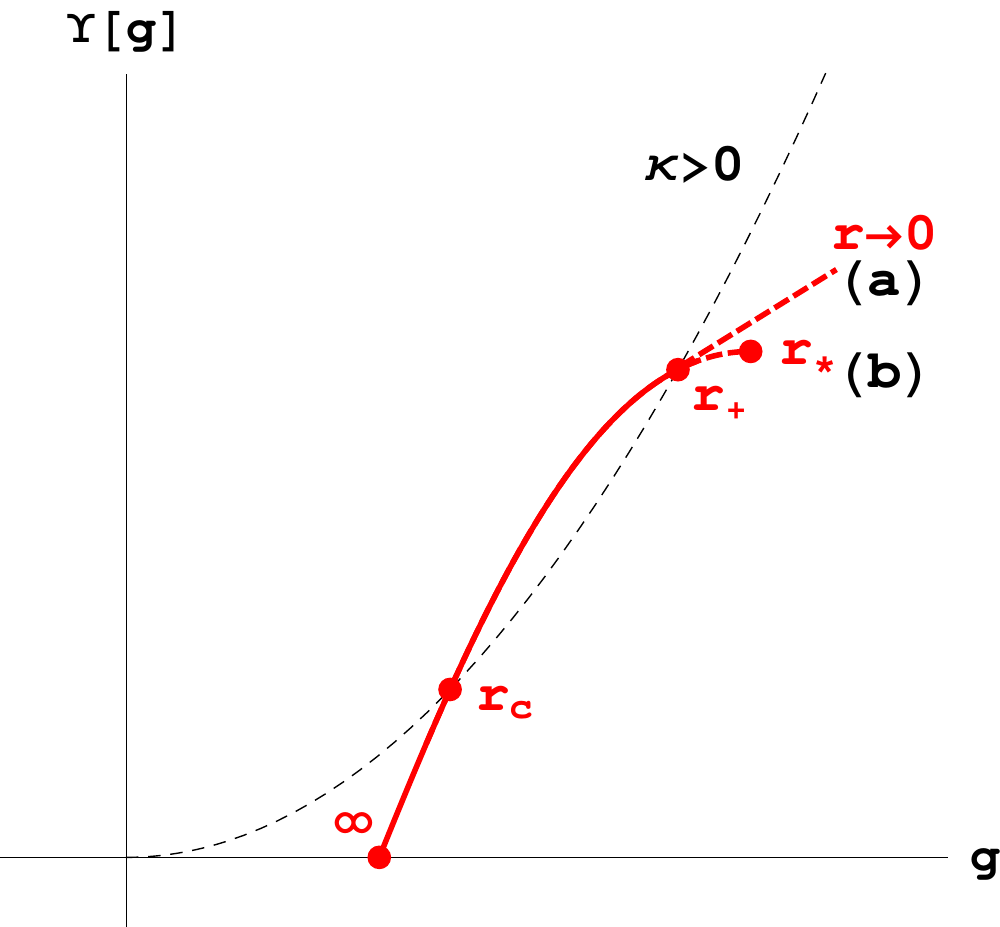} \\
\hline
\end{tabular}
\caption{Classification of (non-excluded) branches attending their asymptotics and topology. Black hole solutions exist in those cases where the given branch (in red) intersects the dashed curve: hyperbolic for an AdS branch (top left), spherical for a dS branch (bottom right), while only for the EH branch supports all possible horizon topologies (second row). Spherical black holes in dS branches exhibit, in addition to the event horizon, a cosmological horizon, $r_c$. We distinguish those branches ending up at extrema of $\Upsilon[g]$, called type (b), from those extending all the way to $r = 0$, named type (a).}
\label{cases}
\end{table*}

The existence of at least one horizon fixes hyperbolic topology as the only possible one for AdS branches (see the first row in the table), as well as it sets an upper bound on the mass of such spacetimes (corresponding to $r_\star$ in such plot). It also sets a lower bound if we consider the possibility of negative mass black holes in those branches. Also this sets a lower bound for the spherical black hole in the EH branches that end up at a maximum, that we call type (b) (or even in those extending all the way to $r = 0$, named type (a), in the critical case, $d=2K+1$).

For dS branches this requirement also fixes the only possible topology admitting an event horizon as spherical at the same time as it imposes a double bound, upper and lower, as will be discussed further later on. The {\it physical} or untrapped region of the spacetime ($f>0$) is that located to the left of the dashed line in all figures appearing in the table. The region to the right corresponds to the inside of the would be horizon or trapped region ($f<0$).

\begin{figure}
\centering
\includegraphics[width=0.43\textwidth]{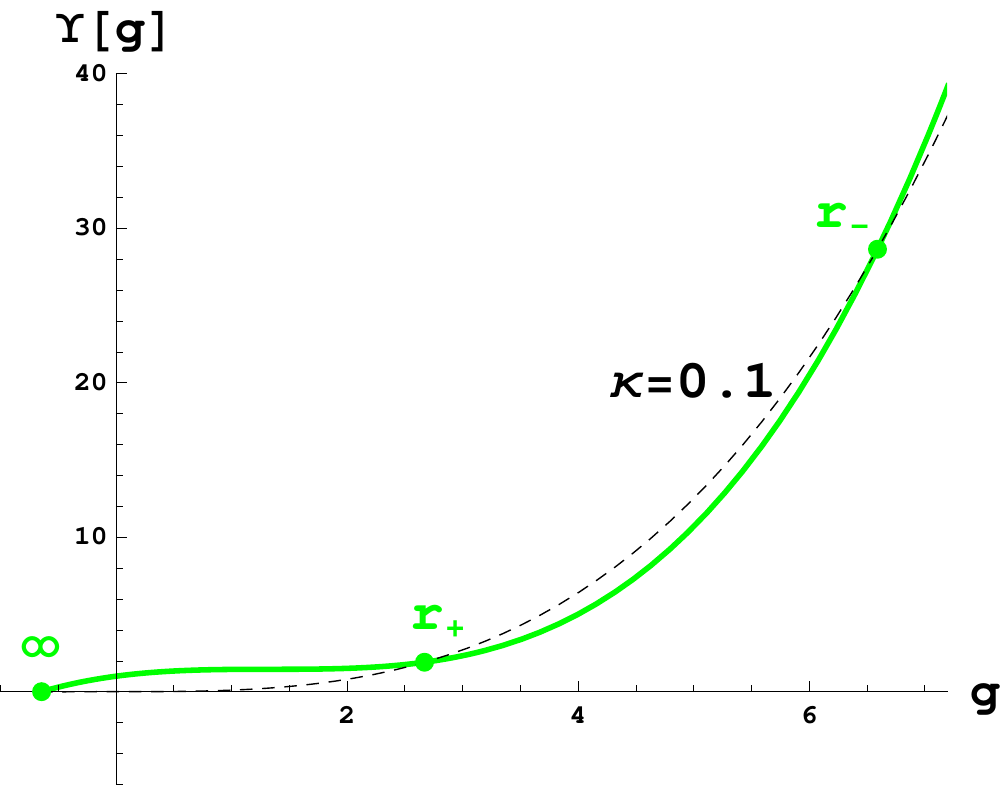}
\caption{Seven dimensional cubic Lovelock theory for $\lambda=-0.746$ and $\mu=0.56$. In the green branch we observe the occurrence of two (outer and inner, respectively) horizons, $r_+$ and $r_-$. For $d=8$ we find a similar behavior with one further inner horizon, three in total.}
\label{mhorizons}
\end{figure}

We will also have more untrapped regions inside the black hole in the presence of several black hole horizons. We already mentioned, indeed, the possibility of black hole spacetimes with multiple horizons. This is for instance the case for some regions of the parameter space with (either EH or dS) branches displaying inflection points. The simplest situation where this can be observed is therefore the cubic theory, as shown in figure \ref{mhorizons}.
This very same behavior will be found in general for some region of the parameter space in the critical Lovelock theory; for $d=2K+1$ this can be easily understood as we can always construct a polynomial 
\begin{equation}
\Upsilon[g]=\prod_{i=1}^{L\leq K} {\left(1-\frac{g}{g_i}\right)} + \alpha\,g^K ~,
\end{equation}
for large enough $\alpha$ and suitable coefficients, $g_i > \Lambda_\star$, in order to make the slope everywhere positive for the EH or dS branches with cosmological constant $\Lambda_\star$. Then, for $\kappa=\alpha$, the polynomial equation has all $g_i$ as solutions. Different (positive) $g_i$ correspond to different spherical horizons, each degenerate $g_i$ giving rise to a degenerate horizon. From this value, varying the mass of the solution the number of horizons will in general change. For the EH branch, the number of black hole horizons has to be always one for high enough mass, and it is so as well in the low mass regime for $d>2K+1$ as well for the EH as for any dS branch. Thus, we have in general couples of horizons appearing and disappearing depending on the values of the mass. We will always refer as $r_+$ to the outermost horizon of the black hole, {\it i.e.}, the biggest root of (\ref{rheq}) besides the cosmological horizon, if present. The very same logic applies to the case of negative mass black holes with hyperbolic horizon. We will comment more on this later.

The algebraic reasoning presented in previous paragraphs leads us to an alternative formulation: we can visualize each branch as a segment corresponding to the continuous running of the roots, $g_i$ (in the complex plane) while $r$ goes from the boundary to the horizon. We will focus on the planar case where the horizon corresponds simply to $g=0$. This will also be the most relevant case in view of the holographic applications described in the second part of this thesis. As discussed before just the EH branch displays a horizon in that case so that it is the only relevant solution. Rescaling the coefficients $c_k = a_k\, L^{2k-2}$ and calling $x \equiv L^2  g$,
\begin{equation}
p[x;r] = \left( 1 - \frac{r_+^{d-1}}{r^{d-1}} \right) + x + \sum_{k=2}^{K} a_k\,x^{k} = 0 ~,
\end{equation}
where, of course, $x = x(r)$.

\begin{figure}
\centering
\includegraphics[width=0.63\textwidth]{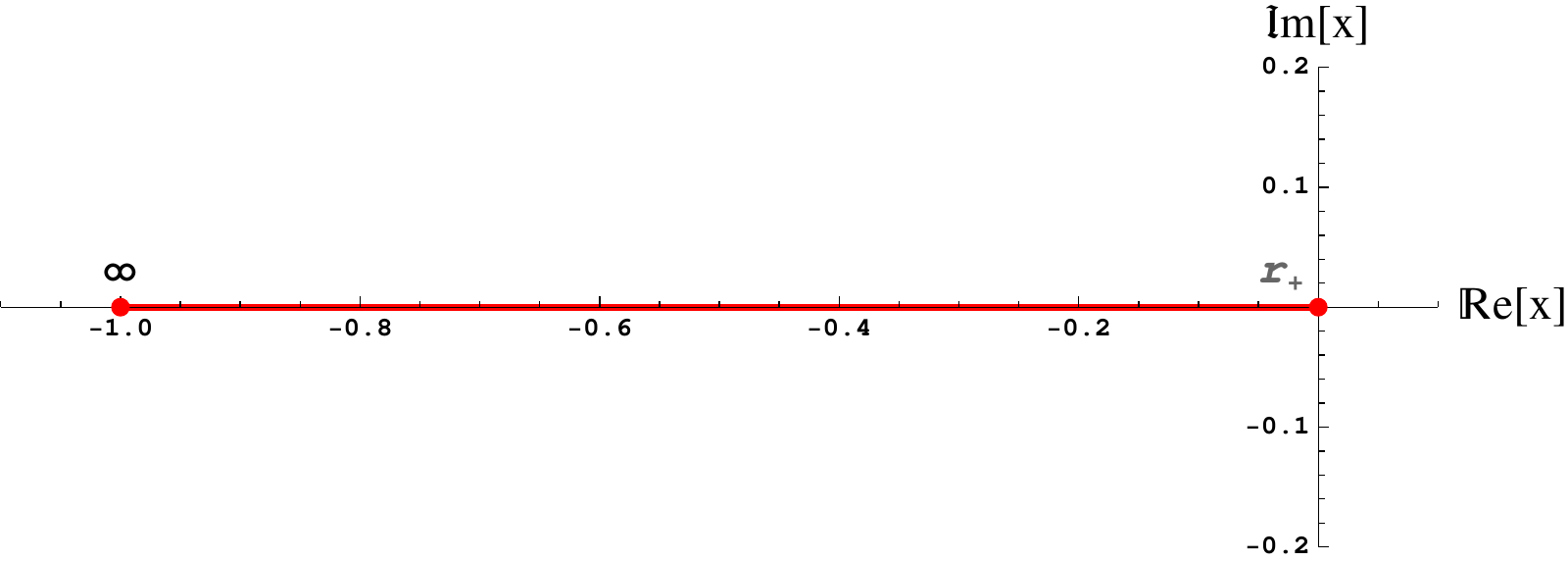}
\caption{The AdS black hole solution of the Einstein-Hilbert theory.}
\label{einstein-root}
\end{figure}

For instance, the solution of the standard Einstein-Hilbert gravity with negative cosmological constant is described by the root of the linear polynomial (see figure \ref{einstein-root}),
%
\begin{equation}
p_{\scriptstyle{\rm EH}}[x;r] = \left( 1 - \frac{r_+^{d-1}}{r^{d-1}} \right) + x = 0 ~.
\end{equation}
The LGB case, instead, presents a richer structure. It is given by the roots of the quadratic polynomial
\begin{equation}
p_{\scriptstyle{\rm LGB}}[x;r] = \left( 1 - \frac{r_+^{d-1}}{r^{d-1}} \right) + x + \lambda\,x^2 = 0 ~.
\end{equation}
The discriminant is $\Delta(r) = 1 - 4\,a_0(r)\,\lambda$, which immediately suggests that there are two different regions.\hskip-1mm
\begin{figure}
\centering
\includegraphics[width=0.63\textwidth]{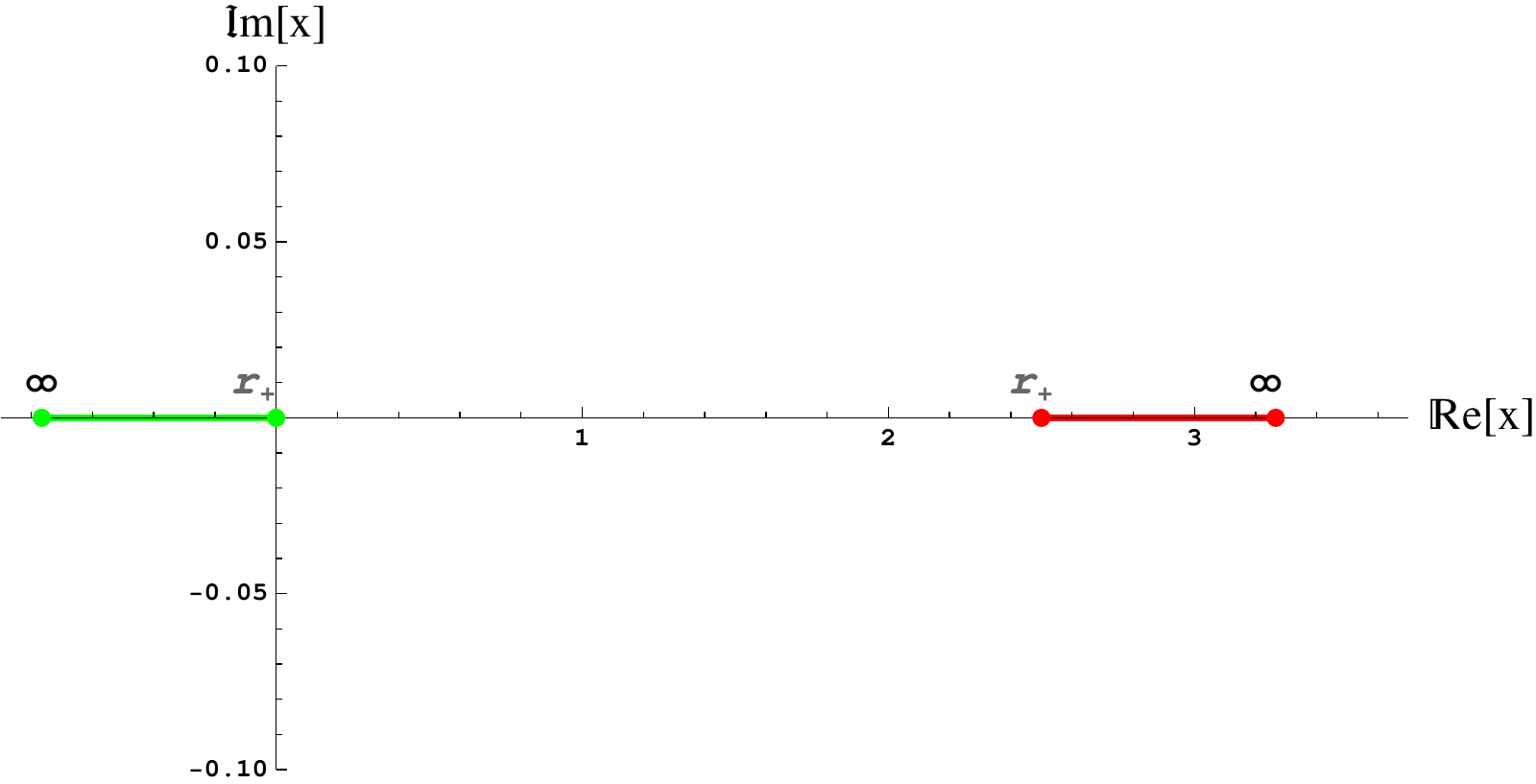}
\caption{The AdS black hole solution of the LGB theory for $\lambda = -0.4$. There are two real branches. The green one is the branch connected to the Einstein-Hilbert AdS black hole, the one we named $f_-$ in (\ref{GBbranches}).}
\label{gb-roots-1}
\end{figure}
When $\lambda < 1/4$, $\Delta(r) > 0$ $\forall r \in [r_+,\infty)$, while for $\lambda > 1/4$, there is a value $r_\star \in [r_+,\infty)$ where $\Delta(r_\star) = 0$ (see figures \ref{gb-roots-1} and \ref{gb-roots-2}).\hskip-1mm
\begin{figure}
\centering
\includegraphics[width=0.63\textwidth]{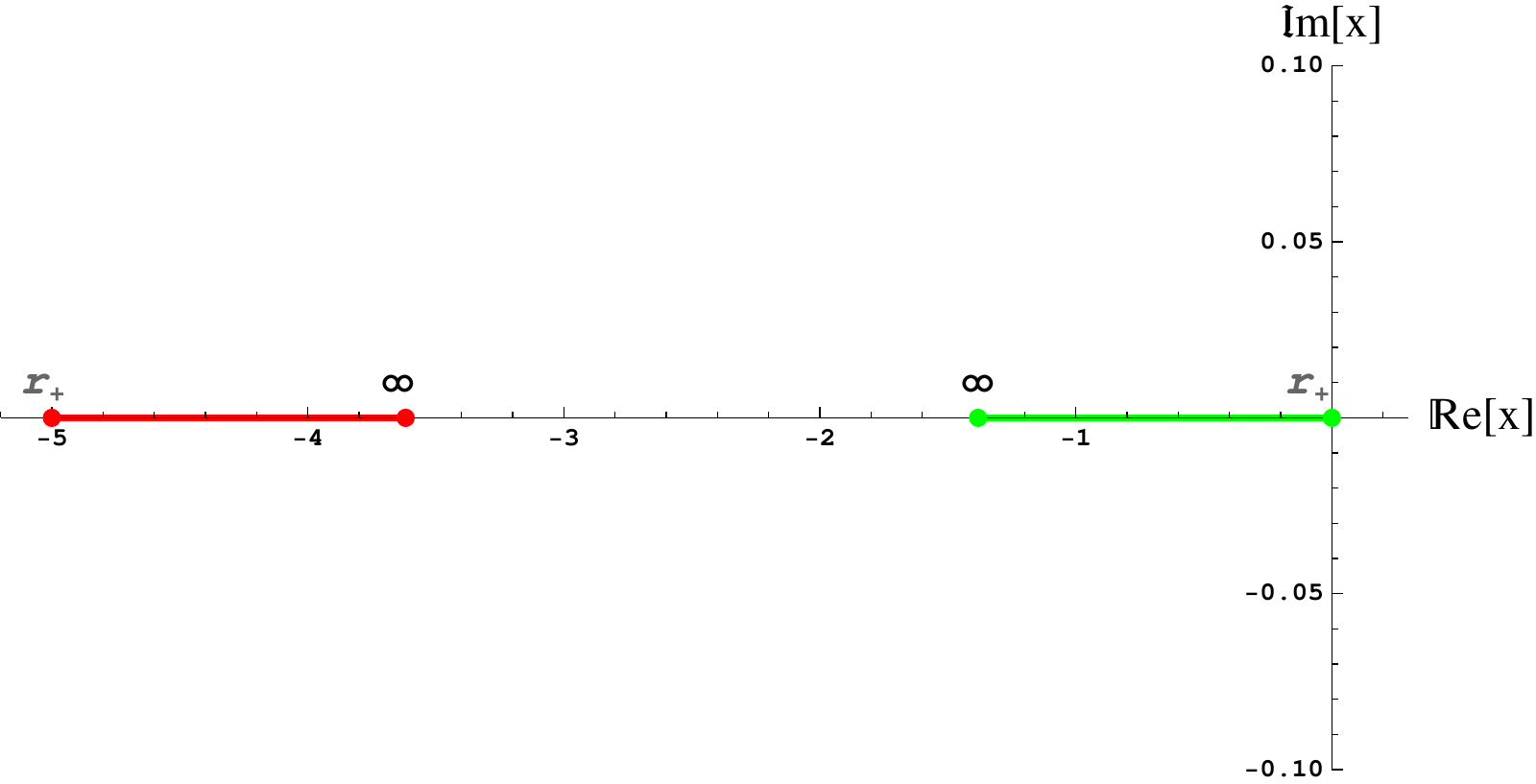}
\caption{The AdS black hole solution of the Gauss-Bonnet theory for $\lambda = 0.2$. For $\lambda > 0$, the (still real) red branch jumps to the left of the green one, without flipping its orientation. For bigger values of $\lambda$, it slides to the right, approaching the green branch. Needless to say, their infinities collide at $\lambda = 1/4$.}
\label{gb-roots-2}
\end{figure}
For $r > r_\star$, $\Delta(r) < 0$ and the solutions become complex (see Figure \ref{gb-roots-3}). The critical value $\lambda = 1/4$, which is the singular locus of the LGB theory, leads to $r_\star = \infty$. In the previous figures we have plotted the different solutions $x(r)$ as a function of $a_0(r)$ between one, that correspond to $r=\infty$, and zero, corresponding to the horizon in the EH branch. For all other solutions $r_+$ does not represent a horizon.

 This quite simple behavior for different values of the LGB coupling serves as a preliminary exercise to clarify the algebraic approach in higher order theories.\hskip-1mm
\begin{figure}
\centering
\includegraphics[width=0.63\textwidth]{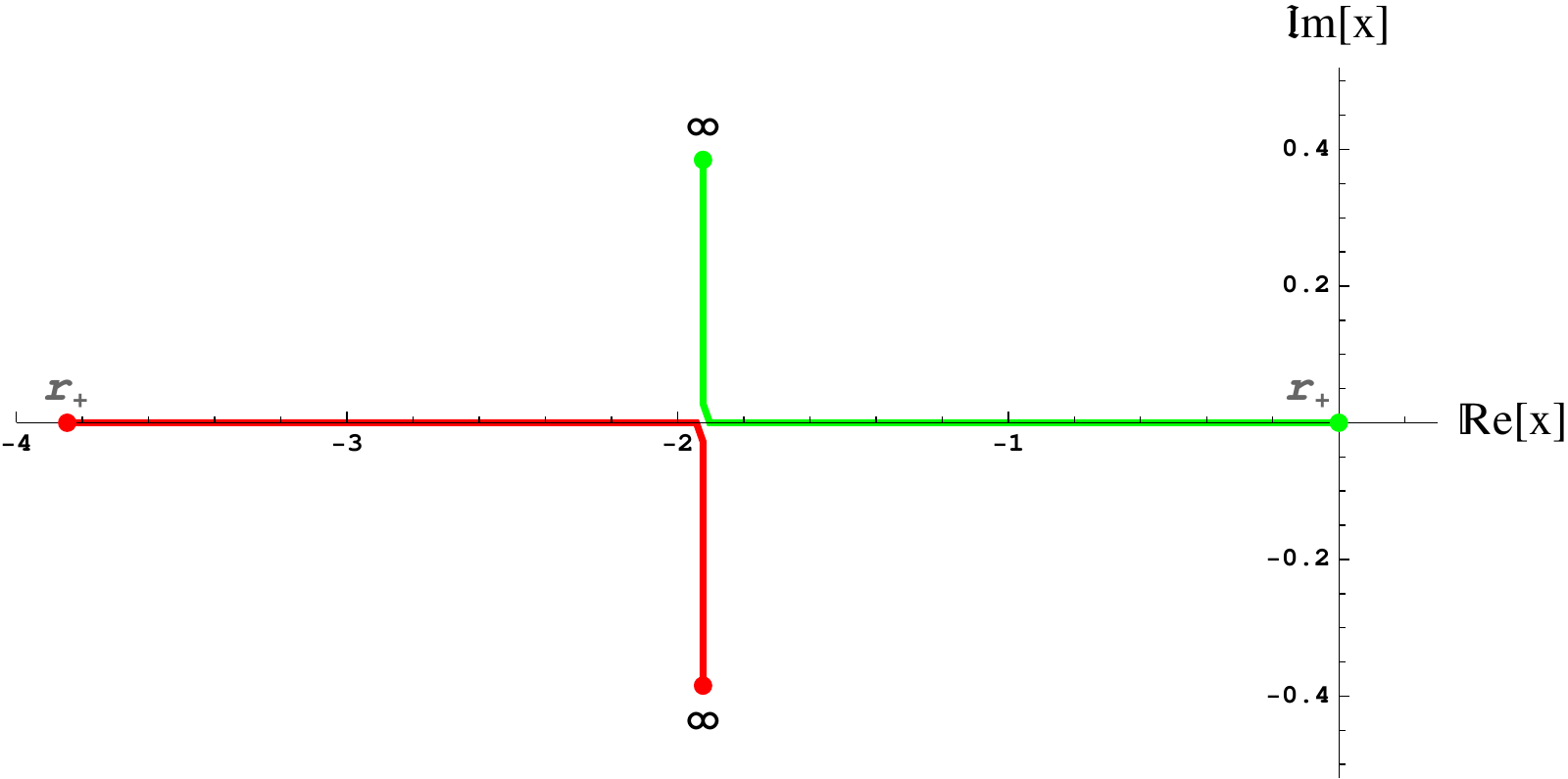}
\caption{The AdS black hole solution of the LGB theory for $\lambda = 0.26$. For $\lambda > 1/4$ the branches become complex conjugates to each other. The critical value, indeed, is the singular locus of the theory (see figure \ref{locus-three}).}
\label{gb-roots-3}
\end{figure}
The situation, indeed, gets more involved in the third order Lovelock theory. The relevant polynomial is
\begin{equation}
p_{\scriptstyle{\rm LL}_3}[x;r] = \left( 1 - \frac{r_+^{d-1}}{r^{d-1}} \right) + x + \lambda\,x^2 + \frac{\mu}{3}\,x^3 = 0 ~,
\label{LL3polynomial}
\end{equation}
whose discriminant is $\Delta(r) = a_0(r)\,\lambda \left (6 \mu - 4 \lambda^2 \right) - 3 a_0^2(r)\, \mu^2 + \lambda ^2 - 4 \mu/3$. As a function of the radius, $\Delta(r) = 0$ spans a family of curves in the $(\lambda,\mu)$-plane that go from $\mu = \frac{3}{4}\,\lambda^2$ (namely, $\Delta(r_+) = 0$) to the singular locus $\Gamma$ depicted in figure \ref{locus-three}. It is not difficult to see that for intermediate values of $r$, the curves look roughly like $\Gamma$, the singular vertex sliding up the parabola $\mu = \lambda^2$ (see figure \ref{cubic-regions}).\hskip-1mm
\begin{figure}
\centering
\includegraphics[width=0.58\textwidth]{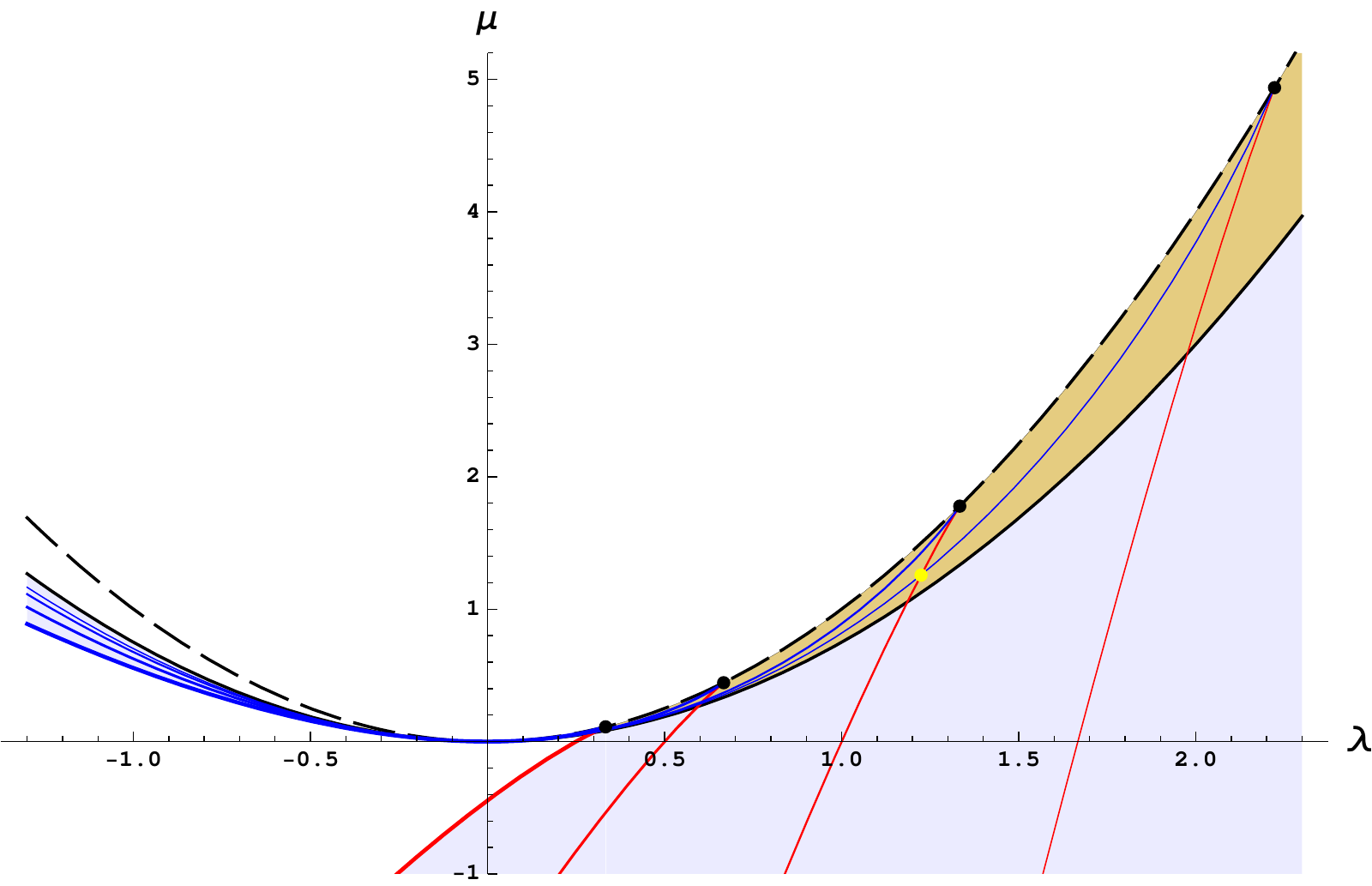}~~\includegraphics[width=0.39\textwidth]{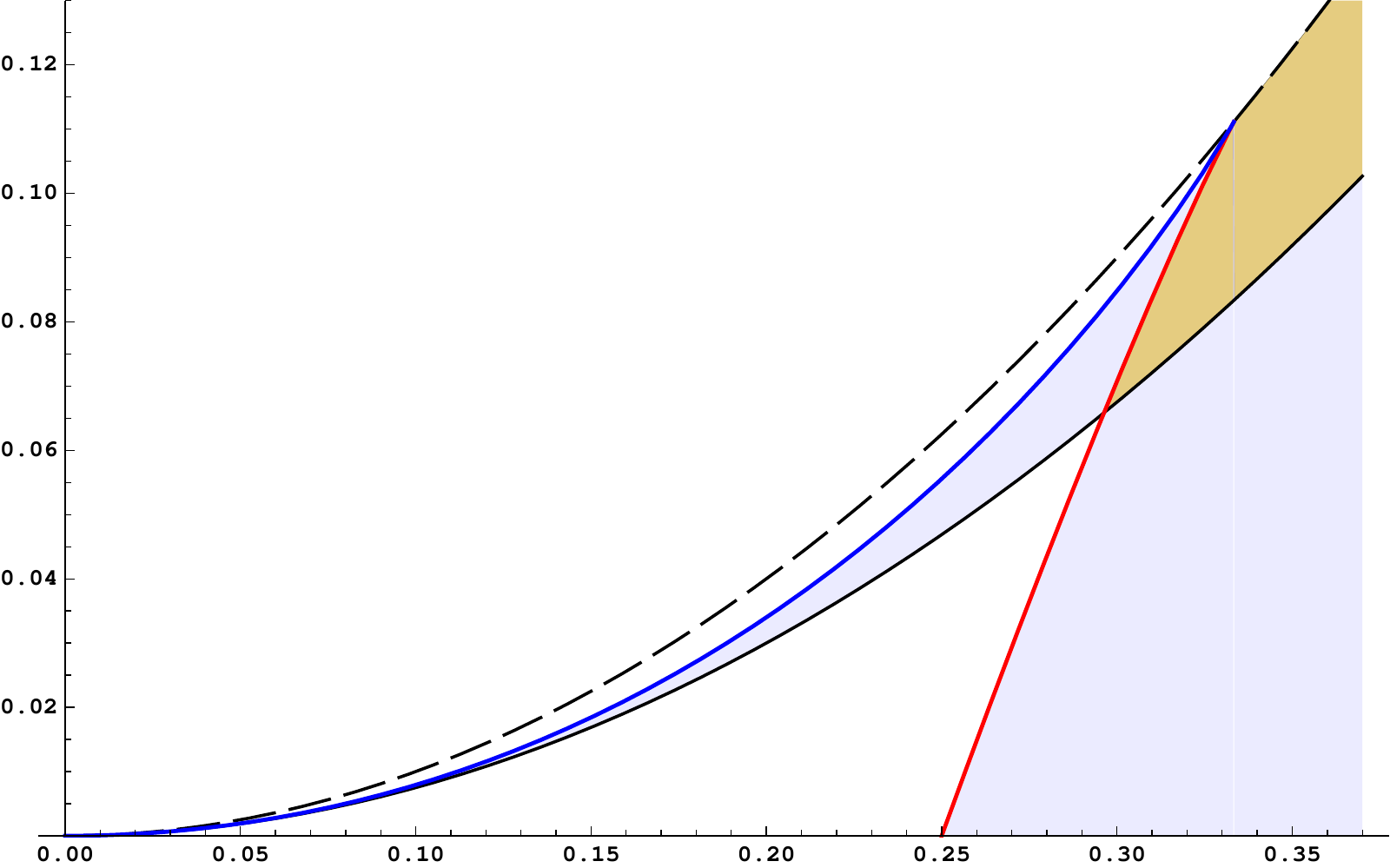}
\caption{Vanishing locus of $\Delta(r)$ for different values of $r$, spanning from $\infty$ (thickest red and blue curve in the left) to $r = (20/17)^{1/(d-1)}\,r_+$ (thinnest red and blue curve in the right). The vertex of the curve slides up along the parabola $\mu = \lambda^2$ (dashed curve). Points in the brown region (as the one depicted in yellow) belong to two curves. The black curve is $\mu = \frac{3}{4} \lambda^2$ (that corresponds to $\Delta(r_+) = 0$), which bounds the regions coloured in light blue, where $\Delta(r)$ only vanishes for one value of the radius between infinity and the horizon. The zoom in the right allows to better understand the structure where the different regions merge.}
\label{cubic-regions}
\end{figure}
Thus, the $(\lambda,\mu)$-plane has two regions, $\mathcal{M}^{(\pm)}$, where there is a value $r_\star \in [r_+,\infty)$ such that $\Delta(r_\star) = 0$,
\begin{eqnarray}
\mathcal{M}^{(+)} & = & \bigg\{ \lambda \leq 0\;, ~\mu^+(\lambda) \leq \mu \leq \frac{3}{4} \lambda^2 \bigg\} ~\cup~ \bigg\{ 0 < \lambda < \frac{8}{27}\;, ~\frac{3}{4} \lambda^2 \leq \mu \leq \mu^+(\lambda) \bigg\} \nonumber \\ [0.5em]
& & \qquad ~\cup \bigg\{ \frac{8}{27} \leq \lambda \leq \frac{1}{3}\;, ~\mu^-(\lambda) < \mu \leq \mu^+(\lambda) \bigg\} ~, \\ [0.7em]
\mathcal{M}^{(-)} & = & \bigg\{ \lambda < \frac{8}{27}\;, ~\mu \leq \mu^-(\lambda) \bigg\} ~\cup~ \bigg\{ \lambda \geq \frac{8}{27}\;, ~\mu < \frac{3}{4} \lambda^2 \bigg\} ~,
\end{eqnarray}
and one region, $\mathcal{M}^{(2)}$, where there are two values $r^\pm_\star \in [r_+,\infty)$ where $\Delta(r^\pm_\star) = 0$,
\begin{equation}
\mathcal{M}^{(2)} = \bigg\{ \frac{8}{27} < \lambda \leq \frac{1}{3}\;, ~\frac{3}{4} \lambda^2 \leq \mu \leq \mu^-(\lambda) \bigg\} ~\cup~ \bigg\{ \lambda > \frac{1}{3}\;, ~\frac{3}{4} \lambda^2 \leq \mu < \lambda^2 \bigg\} ~.
\end{equation}
Everywhere else, $\Delta(r)$ does not vanish for real values of $r$ (unless they are hidden by the horizon). The previous analysis suggests that the space of parameters is divided into different regions that should be treated separately. Depending on the sign of the discriminant we will have one or three real solutions (cosmological constants). As we shall see, for $\Delta > 0$ we have three real solutions while for $\Delta < 0$ we will have just one.

The algebraic approach portrayed in this section can be extended to arbitrary higher order Lovelock theory in higher dimensions. It can be surely dealt with analytically up to the fourth order gravity while higher order cases may require numerical analysis since the quartic is the highest order polynomial equation that can be generically solved by radicals (Abel-Ruffini theorem). Despite these difficulties most of the relevant information about black holes in a given Lovelock theory can be extracted from the implicit solution \reef{eqg}, as we will see in the next chapter. This effective approach will allow the discussion of completely general Lovelock theories, namely their thermodynamic properties. \hskip-1mm
\begin{figure}
\centering
\includegraphics[width=0.4\textwidth]{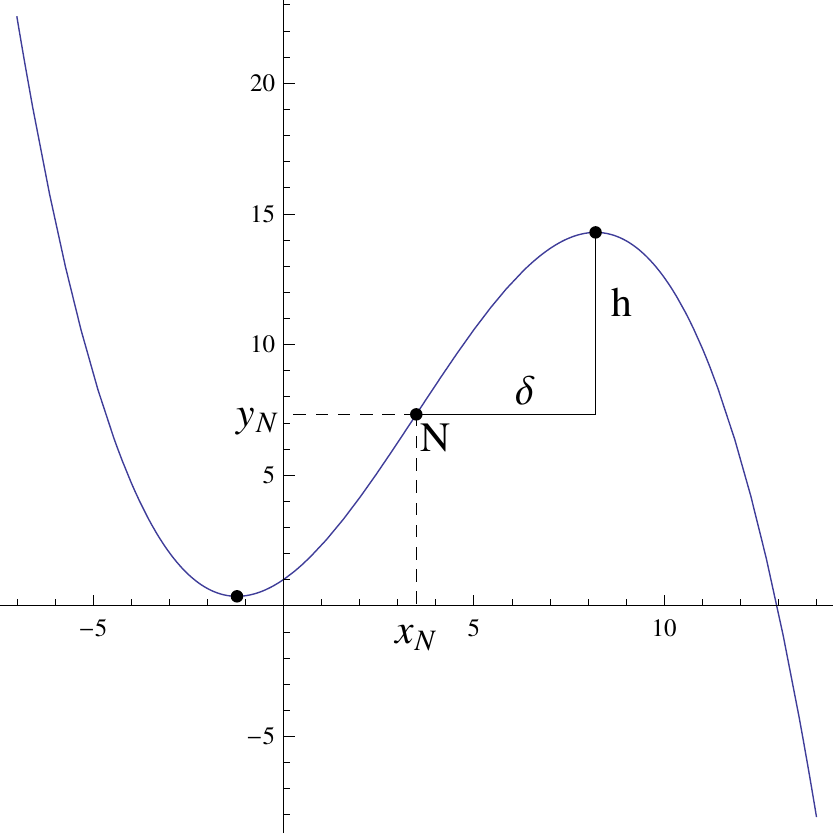}
\caption{Geometrical meaning of the parameters of the cubic. Recall also that the cubic is symmetric with respect to the point of inflexion. }
\label{Nickalls}
\end{figure}
We will focus from here on in the case of cubic Lovelock theory. This particular case has taught us that it is a subtle issue to determine which is the branch of solutions connected to Einstein-Hilbert and, thus, presumably stable. From the point of view of a black hole background, the relevant question is whether one can find an asymptotically AdS black hole with a well-defined horizon. This is a delicate problem that can be amusingly cast in terms of a purely algebraic setup.

\section{Black holes in third order Lovelock theory}
\label{LL3bh}

As discussed in the preceding paragraphs, most of the information needed to clarify the existence of black hole solutions for different values of the Lovelock couplings relies in the behavior of a cubic polynomial. There is a very convenient way of parametrizing the cubic, suitable for analyzing the EH branch of the black solution in terms of the geometry of the polynomial. Following the discussion of \cite{Nickalls1993}, we can characterize a cubic in terms of four fundamental parameters, $\delta,\,h,\,x_N$ and $y_N$ (see figure \ref{Nickalls}). 

If we start from our cubic polynomial, $p_{\scriptstyle{\rm LL}_3}[x;r]=\frac{\mu}{3} x^3 +\lambda x^2 + x + a_0(r)$, a shift on the variable $x = z + x_N$, where $x_N=-\frac{\lambda}{\mu}$, leads to a polynomial, $p_{\scriptstyle{\rm LL}_3}[z;r]$, that is known as the reduced cubic. $N$ is the point of inflexion, and it can be shown that $h$ is a simple function of $\delta$, namely
\begin{equation}
h = - \frac{2}{3} \mu\,\delta^3 ~, \qquad {\rm where} \qquad \delta^2=\frac{\lambda^2-\mu}{\mu^2} ~.
\end{equation}
Thus, the shape of the cubic is completely characterized by the parameter $\delta$. Either the maximum and minimum are different ($\delta^2 > 0$), or they coincide at N ($\delta^2 = 0$), or there are no turning points ($\delta^2 < 0$). We choose the sign of $\delta$ so that $\delta > 0$ corresponds to the situation depicted in the figure ($\mu<0$) and $h$ (when $\delta$ is real) is a positive quantity,
\begin{equation}
\delta=-\frac{1}{\mu}\sqrt{\lambda^2-\mu} \qquad \Rightarrow \qquad h=\frac{2}{3\mu^2}\left(\lambda^2-\mu\right)^{3/2} ~.
\end{equation}
Consider the usual form of the cubic equation $p_{\scriptstyle{\rm LL}_3}[x;r]=0$, with roots $\alpha$, $\beta$ and $\gamma$, and obtain the reduced form by the substitution $x = x_N + z$. The equation has now the form,
\begin{equation}
\frac{\mu}{3} z^3 - \mu\,\delta^2\, z + y_N(r) = 0 ~,
\end{equation}
with roots $\alpha-x_N$, $\beta-x_N$ and $\gamma-x_N$. The parameter $y_N(r)$ is obviously the only one that depends on the radial coordinate through $a_0(r)$,
\begin{equation}
y_N(r) = a_0(r) + \frac{\lambda}{3\mu^2}\left(2\lambda^2-3\mu\right) ~.
\end{equation}
This form allows us to use the identity
\begin{equation}
(p+q)^3-3p\,q(p+q)-(p^3+q^3) = 0 ~.
\end{equation}
Thus $z=p+q$ is a solution where 
\begin{equation}
p\,q=\delta^2  \qquad \text{and} \qquad p^3+q^3=-\frac{3\,y_N}{\mu} ~.
\end{equation}
Solving these equations by cubing the first and substituting into the second, and solving the resulting quadratic in $p^3$ gives
\begin{equation}
p^3=\frac{3}{2\mu}\left(-y_N\pm\sqrt{y_N^2-h^2}\right) ~.
\end{equation}
The discriminant of the polynomial reads 
\begin{equation}
\Delta(r)=-3\mu^2\left(y_N^2-h^2\right) = -3\mu^2\left(y_+\,y_-\right) ~,
\end{equation}
where $y_\pm=y_N\pm h$ are the $y$-coordinates of the maximum and the minimum respectively. $\Delta(r)$ acquires a neat geometrical meaning in the light of this expression. We can see that the sign of the discriminant is determined by the position of the maximum and the minimum with respect to the $x$-axis. When $h \in \mathbb{C}$, $y_+,\,y_-\in \mathbb{C}$, this corresponding to the case $\mu>\lambda^2$. The singular locus $\Delta=0$ corresponds to $y_\pm=0$ at $r=\infty$ (see figure \ref{locus-three}). 

We can treat separately the different regions. For $y_N^2>h^2$ ($y_+y_->0$), there are just one real root that can be easily obtained as\footnote{The complex branches being,
\begin{eqnarray}
& & \beta = x_N + \frac{-1 + i \sqrt{3}}{2}\; \sqrt[3]{-y_N+\sqrt{y_N^2-h^2}}+ \frac{-1 - i \sqrt{3}}{2}\; \sqrt[3]{-y_N-\sqrt{y_N^2-h^2}} ~,\nonumber \\ [0.5em]
& & \gamma=x_N + \frac{-1 - i \sqrt{3}}{2}\; \sqrt[3]{-y_N+\sqrt{y_N^2-h^2}} + \frac{-1 + i \sqrt{3}}{2}\; \sqrt[3]{-y_N-\sqrt{y_N^2-h^2}} ~. \nonumber
\end{eqnarray}
}
\begin{equation}
\alpha=x_N+\sqrt[3]{-y_N+\sqrt{y_N^2-h^2}}+\sqrt[3]{-y_N-\sqrt{y_N^2-h^2}} ~.
\label{realroot}
\end{equation}
For $y_N^2=h^2$ ($y_+\,y_-=0$), there are three real roots two of them equal. The roots are $\alpha=x_N+2\tilde{\delta}$  and $\beta=\gamma=x_N-\tilde{\delta}$, where the sign of $\tilde{\delta}$ depends on the sign of $y_N$ and has to be determined from 
\begin{equation}
\tilde{\delta}=\sqrt[3]{\frac{-y_N}{2a}}=\pm \sqrt[3]{\frac{-h}{2a}}=\pm \delta
\end{equation}
If $y_N=h=0$, then $\delta=0$, in which case there are three equal roots at $x=x_N$.

For $y_N^2<h^2$ ($y_+y_-<0$), all three roots are real and distinct. The easiest way to proceed, without having to find the cube root of a complex number, is to use trigonometry to solve the reduced form with the substitution $z=2\delta \cos\theta$, that gives
\begin{equation}
\cos3\theta=\frac{y_N}{h} ~.
\end{equation}
The three roots are therefore given by 
\begin{eqnarray}
\alpha &=& x_N + 2\delta\; cos\theta ~, \nonumber \\ [0.5em]
\beta  &=& x_N + 2\delta\; cos(\theta+2\pi/3) ~, \\ [0.5em]
\gamma &=& x_N + 2\delta\; cos(\theta+4\pi/3) ~. \nonumber
\end{eqnarray}
The important point to notice here is that, as $y_N$ varies (or equivalently $r$ since $y_N(r)$ is monotonic), the angle $\theta$  can run from zero (where $\beta=\gamma$ and $\alpha$ is the {\it real} branch parametrized by (\ref{realroot})) and $\pi/3$ (where $\alpha=\gamma$ and $\beta$ is the {\it real} branch parametrized by (\ref{realroot})). We can follow the real root(s) as $y_N$ changes. For $y_N>h$ we have just (\ref{realroot}) as a real root. For $y_N=h$ we have $\alpha=x_N+2\delta$ and $\beta=\gamma=x_N-\delta$. For $-h<y_N<h$, the angle $\theta$ is monotonic with $y_N$ going from $\theta=0$ to $\theta=\pi/3$. Beyond that point, two roots become complex again but not the same two that were imaginary for $y_N>h$. Thus we have $\beta=x_N-2\delta$ and $\alpha=\gamma=x_N+\delta$ for $y_N=-h$ the latter two roots becoming complex for $y_N<-h$ where $\beta$ correspond to (\ref{realroot}). Each of this solutions is continuous with $y_N$ and the same can be checked with the other parameters of interest, $\mu$ and $\lambda$. The only presumably singular point is $\mu=0$ where the degree of the equation changes, but we can take a well-defined limit where one of the solutions diverges and the other two coincide with those of the quadratic polynomial. These solutions are continuous, but the `always real branch' as usually parametrized in (\ref{realroot}) is discontinuous since there is an interval of values of $x$ that cannot be taken by the $\alpha$ nor the $\beta$ branches (see figure \ref{roots2}). From now on we will refer to as $\alpha$ ($\beta$) the branch real for $y_N\rightarrow\infty$ ($-\infty$).

Just by analyzing the shape of the cubic we can understand which one is the branch connected to the horizon ($x=0$ for $r=r_+$ ($a_0=0$)). We have given values of $a_1$ and $a_0$ and so the polynomial at the origin and its first derivative there. The first derivative of the polynomial at $x=0$ is always 1, and so the polynomial is growing at that point. Also, we know the sign of $x_N$ when $\lambda$ and $\mu$ are given. We can distinguish several different cases. We know from our previous analysis that we can take a representative point in each of the relevant regions and analyze the behavior of the solutions. 

For $\mu\geq\lambda^2$ ($\delta^2<0$) there are no turning points (or they coincide at the inflexion point) and the polynomial is monotonically growing. There is just one real branch of solutions for all values of $r$ and it has then no discontinuity. The same happens in the region contained in between the curve $\mu=3/4\lambda^2$ and $\mu=\lambda^2$ for $\lambda<0$ and in between $\mu=\mu^+(\lambda)$ and $\mu=\lambda^2$ for $\lambda>0$. In this whole region we have $\Delta(r)<0$ ~$\forall r\in [r_+,\infty)$. This happens for any point $(\lambda,\mu)$ lying in the upper white part of figure \ref{cubic-regions}. Let us call this region $\mathcal{M}_-^{(0)}$,
\begin{figure}
\centering
\includegraphics[width=0.63\textwidth]{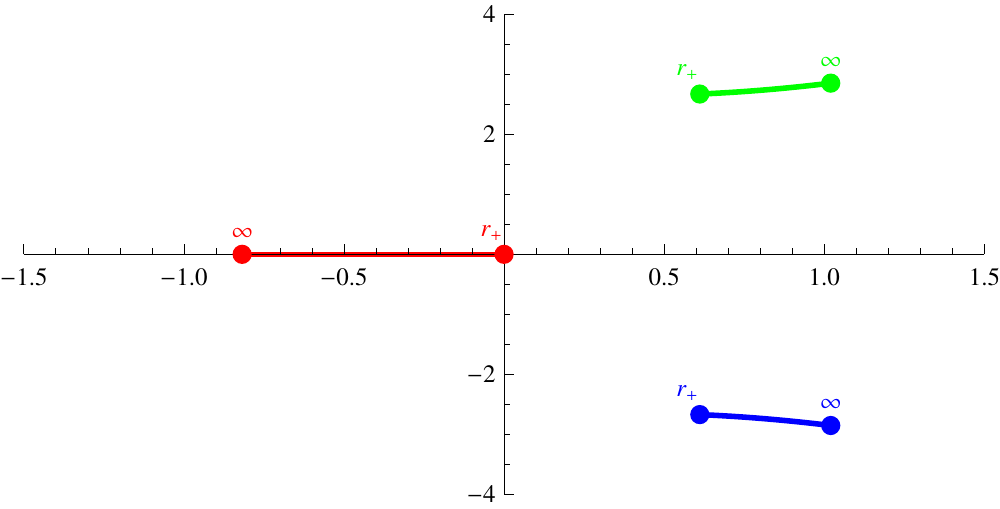}
\caption{$\Delta(r)<0$ $\forall r\in [r_+,\infty)$ corresponding to $\mathcal{M}_-^{(0)}$. There is a single branch which is real. As we will see later, it is the EH branch for the black hole solution.}
\label{roots0}
\end{figure}
%
\begin{equation}
\mathcal{M}_-^{(0)} = \bigg\{ \lambda \leq 0\;, ~ \mu > \frac{3}{4} \lambda^2 \bigg\} ~\cup~ \bigg\{ 0 < \lambda \leq \frac{1}{3}\;, \mu > \mu^+(\lambda) \bigg\} ~\cup~ \bigg\{ \lambda > \frac{1}{3}\;, ~\mu > \lambda^2 \bigg\} ~.
\end{equation}
In this case, the roots of (\ref{LL3polynomial}) behave the same all along the radial {\it flow}; there is a single real branch which is the relevant one for asymptotically AdS black hole solutions with a well-defined horizon, and two complex conjugate unphysical branches (see figure \ref{roots0}). The $\mu=\lambda^2$ line for $\lambda>1/3$ is excluded from $\mathcal{M}^{(0)}_-$ since in this case the inflection point is situated above the $x$-axis
\be
y_N(\infty)=\frac{1}{\lambda}\left(\lambda-\frac13\right)>0
\ee
with $x_N<0$, and then, when the inflection point crosses the $x$-axis ($r=r_\star$ where $\Delta(r_\star)=0$) a (naked) singularity shows up even if the spacetime is perfectly regular for $r\neq r_\star$.
The lower white part of figure \ref{cubic-regions}, instead, has $\Delta(r)>0$ ~$\forall r\in [r_+,\infty)$. Let us call this region $\mathcal{M}_+^{(0)}$,
\begin{figure}
\centering
\includegraphics[width=0.63\textwidth]{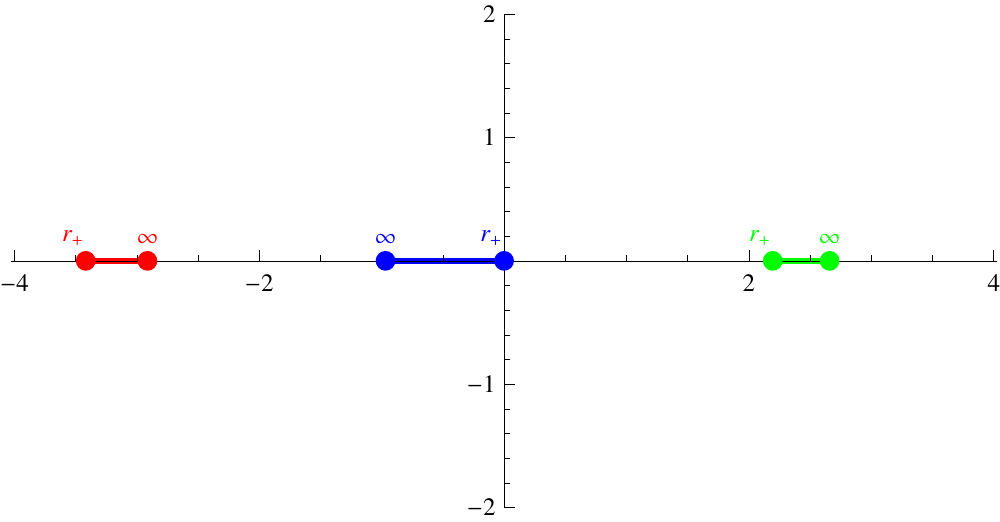}
\caption{$\Delta(r)>0$ $\forall r\in [r_+,\infty)$ corresponding to $\mathcal{M}_+^{(0)}$. This situation is very similar to LGB gravity. One of the solutions diverges as $\mu \to 0$}
\label{roots8}
\end{figure}
%
\begin{equation}
\mathcal{M}_+^{(0)} = \bigg\{ \lambda \leq 0\;, ~\mu^-(\lambda) < \mu < \mu^+(\lambda) \bigg\} ~\cup~ \bigg\{ 0 < \lambda < \frac{8}{27}\;, ~\mu^-(\lambda) < \mu < \frac{3}{4} \lambda^2 \bigg\} ~.
\end{equation}
In this case, again, the roots of (\ref{LL3polynomial}) behave the same all along the radial {\it flow} (see figure \ref{roots8}); there are three real branches but only one is relevant for asymptotically AdS black hole solutions with a well-defined horizon (see figure \ref{roots8}), the EH one. This is reminiscent of the LGB case where there are two real solutions for $\lambda < 1/4$. One of the branches diverge as $\mu \to 0$.

\begin{figure}
\centering
\includegraphics[width=0.63\textwidth]{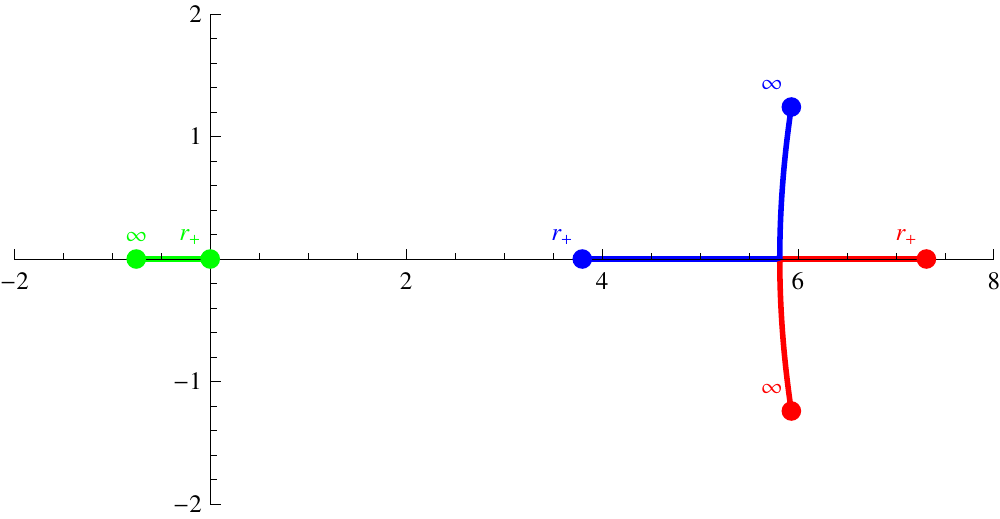}
\caption{$\Delta_\infty <0$ and $\Delta(r_+)>0$ corresponding to $\mathcal{M}^{(+)}$. }
\label{roots6}
\end{figure}

For $\mu<\lambda^2$ we have in general two values of $r$ (or $y_N$) for which $\Delta(r)=0$, but they can be included or not in the interval $r\in [r_+,\infty)$ (in the previously analyzed regions these values lie outside this interval). For $\mu>0$ ($\delta<0$) $\lambda<0$ ($x_N>0$) maximum and minimum are located at positive values of $x$ and so the branch connected to the horizon has no problems of reality or continuity even if eventually (in the subregion contained in $\mathcal{M}^{(+)}$) $\Delta(r_{*})=0$ for some value $r_{*}\in[r_+,\infty)$ (see figure \ref{roots6}).

\begin{figure}
\centering
\includegraphics[width=0.63\textwidth]{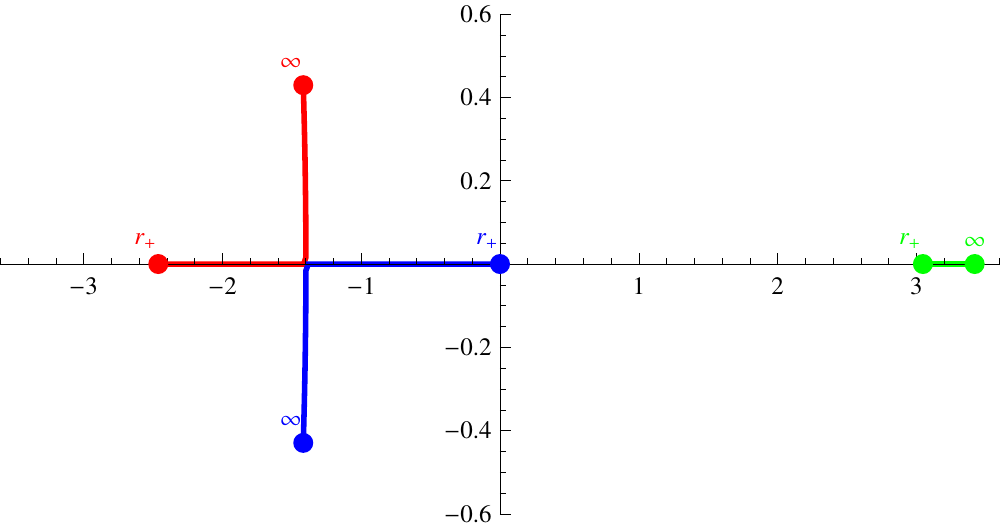}\caption{$\Delta_\infty <0$ and $\Delta(r_+)>0$ corresponding to $\mathcal{M}^{(-)}$. Regarding the behavior of $\Delta(r)$ this region seems similar to the previous one but in this case one of the degenerate branches when $\Delta(r)=0$ is the one connected to the horizon. Thus, there is a real cosmological constant but it does not correspond to the branch with horizon.}
\label{roots3}
\end{figure}

For $\mu<0$ ($\delta>0$) the only growing part of the polynomial is the one in between the minima and the maxima, corresponding to the branch $\gamma$. The maximum is located to the right of the origin $x=0$ and does not pose any problem regarding the EH branch, but the minimum is located in negative values of $x$ and then we must have $y_-(\infty)<0$ ($y_N(\infty)<h$) in order to have a real cosmological constant. The singular curve $y_-=0$ correspond in this region to $\mu=\mu^-(\lambda)$. The region with real cosmological constant correspond to $\mathcal{M}^{(0)}_+$ while the region with complex cosmological constant correspond to the region $\mathcal{M}^{(-)}$ (see figure \ref{roots3}).

The $\gamma$ branch for $\mu<0$ is continuously deformed into $\alpha$ or $\beta$ when crossing the $\mu=0$ line. For $\lambda<0$ the $\alpha$ branch diverges and $\gamma$ and $\beta$ interchange their r\^oles. For $\lambda>0$ is $\beta$ the diverging branch and $\gamma$ is deformed into $\alpha$. For $\mu=0$ we can identify the remaining branches as the solutions for the LGB case. 

\begin{figure}
\centering
\includegraphics[width=0.63\textwidth]{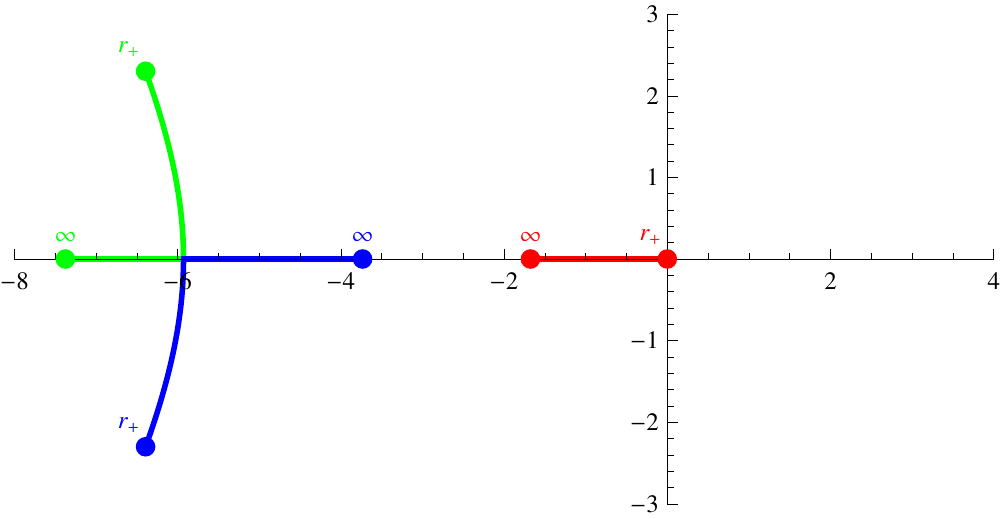}
\caption{$\Delta_\infty >0$ and $\Delta(r_+)<0$ corresponding to $\mathcal{M}^{(+)}$. }
\label{roots1}
\end{figure}

\begin{figure}
\centering
\includegraphics[width=0.63\textwidth]{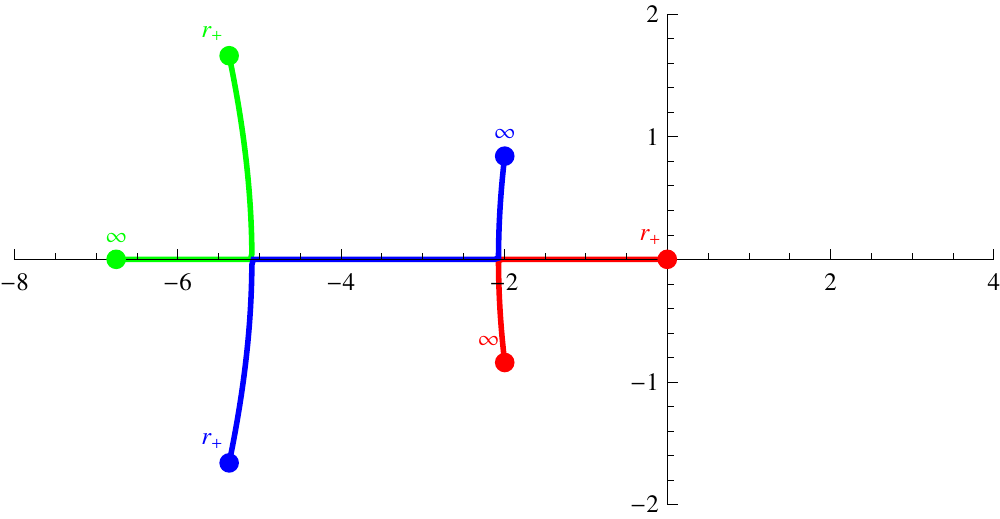}
\caption{$\Delta(r)<0$ $\forall r\in [r_+,r_\star^-)\cup(r_\star^+,\infty)$ corresponding to $\mathcal{M}^{(2)}$. In this case as in the figure \ref{roots3} there is no way of continuously connect the real cosmological constant to the horizon }
\label{roots2}
\end{figure}

The remaining case to be discussed is $\mu>0$ ($\delta<0$) and $\lambda>0$ ($x_N<0$). In this region (except as already discussed for $\mu\geq\lambda^2$) the two critical points are located at negative values of $x$, and they can have positive or negative $y$-values. For the subregion contained in $\mathcal{M}^{(+)}$, see figure \ref{roots1}. As in the previous case, in order to have a real value for the relevant cosmological constant the value $y_-$ for the minimum must be negative. Again, the limiting case ($y_-=0$) corresponds to $\mu=\mu^-(\lambda)$ as can be seen in figures \ref{roots2} and \ref{roots3}  

In any case the necessary and sufficient condition for the cosmological constant to be real is $y_-(\infty)<0$ or $y_N(\infty)<h$. Therefore we have an excluded region of parameters below $\mu=\mu^-(\lambda)$. The other curve $\mu=\mu^+(\lambda)$ does not affect the qualitative behavior of the EH solution since it is just indicating the appearance of two new real cosmological constants. Thus, in most of the space of parameters the existence of two extra solutions does not qualitatively affect the solution with horizon, except in the excluded region where the cosmological constant connected to the horizon is not real. This excluded region reduces to $\lambda>1/4$ in the LGB limit, as it should be expected. The remaining boundary of the excluded region, for $\lambda>1/3$, is the $\mu=\lambda^2$ line. 

Notice that the well behaved solution is regular when crossing the curve $\mu=\mu^+(\lambda)$; the would be symmetry enhancement affects the other two cosmological constants. In other words, the two cosmological constants that agree over this curve are not those connected to the horizon, and then the theory has propagating linear perturbations when expanded about the EH vacuum. Symmetry enhancement for this vacuum occurs at $\mu=\mu^-(\lambda)$ and this is the boundary of the excluded region.

The usual parametrization of the black hole solution \cite{Dehghani2009,Ge2009a} shall make manifest the different properties portrayed in the previous section. In fact, the usual parametrization of the three, in general complex, solutions for the function $f$ in third order Lovelock gravity is
\begin{equation}
f_i = \frac{r^2}{L^2} \frac{\lambda}{\mu} \left[ 1 + \alpha_i \left( J(r) + \sqrt{\Omega(r)} \right)^{1/3} + \bar{\alpha}_i \left( J(r) - \sqrt{\Omega(r)} \right)^{1/3} \right] ~,
\label{BHsolution}
\end{equation}
where
\begin{equation}
\Omega(r) = J(r)^2 + \Gamma^3 ~,
\label{Deltar}
\end{equation}
with
\begin{eqnarray}
& & J(r) = 1 - \frac{3\,\mu}{2\,\lambda^2} + \frac{3\,\mu^2}{2\,\lambda^3}\, \left( 1 - \frac{r_+^6}{r^6} \right) = \frac{3\,\mu^2}{2\,\lambda^3}\; y_N(r) ~, \\ [0.5em]
& & \Gamma \equiv \left( \frac{\mu}{\lambda^2} - 1 \right) \qquad \Rightarrow \qquad \Gamma^3 = \frac{3\,\mu^2}{2\,\lambda^3}\; h^2 ~.
\label{JGamma}
\end{eqnarray}
$\alpha_i$ are the three cubic roots of unity, $\alpha_0 = 1$, $\alpha_\pm = -\frac{1 \pm i\sqrt{3}}{2}$, and the bar indicates complex conjugation. Each of these solutions is associated with one possible value of the cosmological constant, and so, {\it fixing the value of the cosmological constant fixes the function}. Notice that $f$ is directly related to our previous parametrization as
\begin{equation}
f = - \frac{r^2}{L^2}\; x ~,
\end{equation}
where $x$ is the relevant branch ($\alpha$, $\beta$ or $\gamma$) in each region.

\section{Charged and rotating solutions}
\label{sec:chargedBH}

One obvious extension of Lovelock gravity, probably the simplest one, is that of Lovelock-Maxwell theory. Solutions charged under both Maxwell and Born-Infeld electrodynamics have been known for a long time \cite{Wiltshire1986,Wiltshire1988}, and were reconsidered recently  \cite{Lidsey2002,Cvetic2002b,Aiello2004,Banados2004,Aiello2005,Torii2005a,Maeda2009,Li2011,Ohta2012,Dehghani2003,Dehghani2005a,Dehghani2006b,Dehghani2008,Takahashi2011,Takahashi2012}\footnote{The analogous solutions in quasi-topological gravity have been considered in \cite{Brenna2012}}. Most of these efforts have been devoted to the simpler LGB case and a complete classification of all possible black hole solutions in Lovelock theories is still missing. Even though Lovelock-Maxwell solutions have in general a more complex structure, it is quite straightforward  to modify the general approach outlined throughout this chapter to the charged case. 

For the action principle we just need to include the Maxwell action as part of the matter action
\begin{equation}
\mathcal{I} =\frac{1}{16\pi G_N}\left( \frac{1}{(d-3)!}\sum_{k=0}^{K} {\frac{c_k}{d-2k}} \mathop\int \mathcal{L}_{k}\;+\;F\wedge\star F \right)~,
\label{LMaction}
\end{equation}
where the normalization of the Maxwell term has been chosen for later convenience. The equations of motion for the electromagnetic field in absence of currents reduce to
\beq
\td \star F=0 \qquad ; \quad \td F=0
\eeq
where the second equation corresponds to the Bianchi identity, consequence of the field strength arising from a Maxwell 1-form as $F=\td A$. The black hole ansatz will be the same assumed for the uncharged black hole \reef{bhansatz} whereas the field strength $F$ takes the form
\beq
F=\frac{q}{r^{d-2}}e^0\wedge e^1 
\eeq
In 4-dimensions we can also add the magnetic charge but this would not change the form of the black hole metric. 

\begin{figure}
\centering
\includegraphics[width=0.67\textwidth]{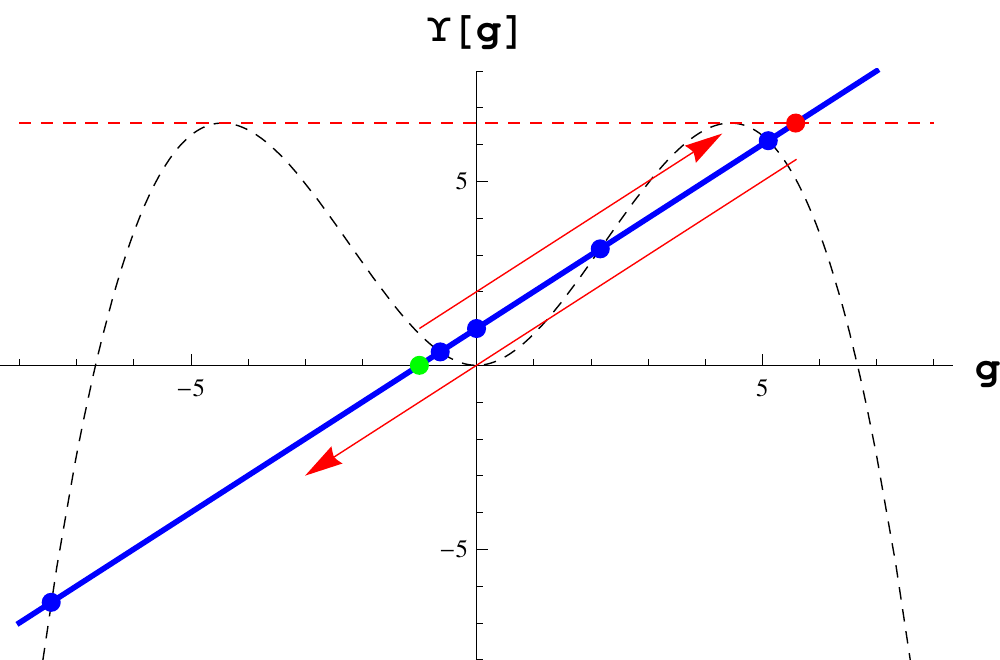}
\caption{Einstein-Hilbert gravity in five dimensions for $L=1$, $\kappa=1$ and $Q^2=0.15$. The dashed lines are plots of \reef{rheqCH} for the indicated values of the mass and charge parameters and $\sigma=\pm 1$, for positive and negative values of $g$ respectively. The line corresponding to $\sigma=0$ is just the vertical line $g=0$. The crossing of these lines with the polynomial give the possible values for $g$ (in blue) at the horizon and then of $r_+$. Contrary to the uncharged case, the solution climbs up the polynomial up to a maximal value given by the maximum of \reef{rheqCH} (in red) and then goes back down. This happens for all three topologies in such a way that for planar black holes $g=0$ is actually a double root corresponding to two different horizon radius. }
\label{chargedBH}
\end{figure}

This expression solves the above equations and sources the gravitational interaction in such a way that the $tt$-component of the equations of motion now becomes
\begin{equation}
\left[ \frac{d~}{d\log r} + (d-1) \right]\, \Upsilon[g] = \frac{(d-3)\,Q^2}{r^{2(d-2)}} ~,
\end{equation}
$Q$ being related to the electric charge $q$ (in geometric units),
\be
Q^2=\frac{q^2}{(d-2)(d-3)}
\ee
whereas the remaining field equations are either equal or follow from energy conservation. The final, implicit but exact, solution is very similar to the uncharged case, having just  an extra term in the right hand side of the polynomial equation, changing its radial dependence,
\begin{equation}
\Upsilon[g] =\frac{\kappa}{r^{d-1}} -\frac{Q^2}{r^{2(d-2)}} ~,
\label{eqgCH}
\end{equation}
Instead of just having a monotonically decreasing piece, the right hand side grows at first, it reaches a maximum and then decreases. The two usual horizons appearing in the general relativistic case, each correspond to one of the monotonic pieces. In order to find the horizons in general, we may perform the change $g_+=\sigma/r_+^2$ as to get an equation for $g_+$ that we can analyze in very much the same way as in the uncharged case. See figure \ref{chargedBH} for a graphical visualization of the resulting equation,
\be
\Upsilon[g_+] =\kappa\left(\frac{g_+}{\sigma}\right)^{\frac{d-1}{2}}\!\!-Q^2\, \left(\frac{g_+}{\sigma}\right)^{d-2} \label{rheqCH} 
\ee
%

As for the neutral case, the different Lovelock couplings fix the shape of the polynomial $\Upsilon[g]$. While varying $r$ from infinity to zero, the function $g(r)$ is given by the implicit solution of equation \reef{eqgCH} that graphically corresponds to climbing up (for positive mass) a given monotonic part of the curve $\Upsilon[g]$ starting from one of its roots (tantamount of a given cosmological constant). This is illustrated in the figure by just Einstein-Hilbert gravity but, except for the position of the singularities, everything holds for any branch of solutions in Lovelock in qualitatively the same way. The metric function $g$ is monotonic in $r$ up to a given radial position $r_\star$ corresponding to the maximum value of the right hand side of \reef{eqgCH}. Beyond that point the solution slope changes sign and $g$ is again monotonic in the opposite direction. Despite the differences, each branch can be again identified with a monotonic section of the polynomial $\Upsilon[g]$, and can easily be visualized graphically.

For each of the two monotonic sections of $g$ we can analyze the existence of horizons in the same way as we did in previous sections. Indeed planar black holes may just have horizons for the EH branch as before. As in the uncharged case $g=0$ is just crossed by this particular branch. Nevertheless, the addition of charge implies that we need to satisfy an extra requirement for $g=0$ to actually be a horizon.  We need the maximum of the right hand side of \reef{eqgCH}, let us call it $\Upsilon_\star$, be bigger than $\Upsilon[0]$ so that we get two crossings of the polynomial under radial {\it flow}. In fact, $\Upsilon[0]=\Upsilon_\star$ corresponds to the extremal limit in the planar case. Once this condition is fulfilled, the solution will always display two horizons for type (a) solutions whereas those of (b) type may have one or two depending on $\Upsilon_\star$ being smaller or bigger than the value of $\Upsilon[g]$ at the maximum that indicates the position of the singularity. In the latter case, we run into the singularity as we climb up our branch of interest, radial positions beyond that point being unphysical. Notice that the negative parts of the polynomial become relevant in this case, even though for regular solutions these regions lie behind the horizon. They are explored in general as we approach $r=0$, in such a way that zero radius singularities happen now for branches extending all the way to $g=-\infty$ whereas the minima indicate the position of finite size singularities. 

For non-planar topology the discussion is a bit more involved but follows in similar terms. For spherical black holes we may have several horizons as we climb up the branch, as in the uncharged case, but just one when going back down. As before the existence of horizons in the second monotonic section of $g$ is precluded in type (b) branches when $\Upsilon_\star$ is bigger than the value of the polynomial at the maximum, we encounter the singularity before we get to the turning point $\Upsilon[g]=\Upsilon_\star$. For hyperbolic branches the maximal number of horizons is one for the first section of $g$ (going up) but it may be bigger than one in the second (going down), generalizing the observation made for negative mass hyperbolic solutions. Actually, for negative masses, there is just one section of $g$, this function being completely monotonic.

Considering masses for which the black hole has a horizon in the uncharged case, there will be in general an upper bound in the charges one may consider, corresponding the bound itself to an extremal black hole. Even though the explicit expression for the radius of the extremal solution in terms of the charge and mass is complicated, we may find its implicit form combining the horizon equation \reef{rheqCH} and its first derivative
\be
\Upsilon'[g_+] =\frac{d-1}{2\sigma}\kappa\left(\frac{g_+}{\sigma}\right)^{\frac{d-3}{2}}-\frac{d-2}{\sigma}\,Q^2\, \left(\frac{g_+}{\sigma}\right)^{d-3}
\ee
the horizon being degenerate when both equations are satidfied. This system can be solved in parametric form as
\bear
\kappa&=&\frac{2 r_+^{d-1}}{d-3}\left((d-2)\Upsilon[g_+]-g_+\Upsilon'[g_+]\right)\\
Q^2&=&\frac{r_+^{2(d-2)}}{d-3}\left((d-1)\Upsilon[g_+]-2g_+\Upsilon'[g_+]\right)
\eear
this expressions parametrizing a curve that divides the $(\kappa,Q^2)$ space in two pieces, one corresponding to black holes with at least one horizon and the other to naked singularities. At this point, it would be interesting to generalize the arguments of \cite{Wang1998} to charged solutions in Lovelock gravity, in order to elucidate the possibility of turning non-extremal into extremal black holes through any physical process. However, the existence of naked singularities in the uncharged case suggests this is not straightforward. In chapter \ref{chp:bhstability} we will explore an alternative route to address this problem, although just for the neutral case. Besides, for type (b) branches the extremality curve ends at the value of $g_+$ corresponding to the curve, giving the minimal mass (and charge) extremal solution. Beyond that point we may go from a black hole to a naked singularity without ever becoming extremal.  

\begin{figure}
\centering
\includegraphics[width=0.4\textwidth]{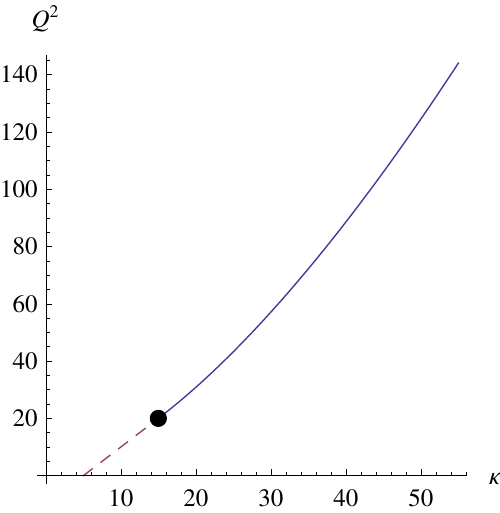}~~\qquad \includegraphics[width=0.4\textwidth]{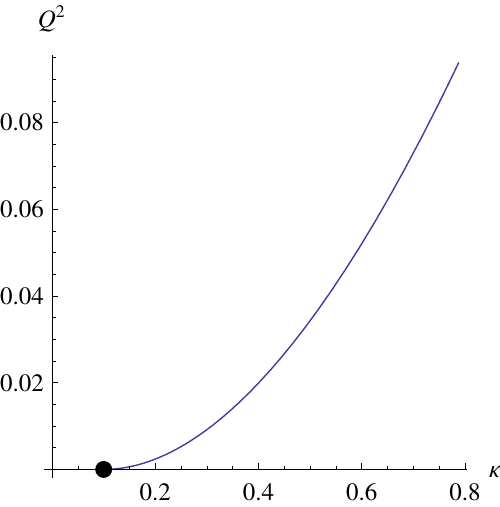}
\caption{LGB extremality curves in the $(\kappa,Q^2)$ parameter space in 5 dimensions for $\lambda=-1$ and $\lambda=0.1$ respectively in units of $L$. The dot indicates where the curve ends. For negative $\lambda$ the extremality curve ceases to exist for masses below the one for which the degenerate horizon coincides with the singularity. The dashed line indicates the boundary between black holes and naked singularities in that case.}
\label{extcurve}
\end{figure}

Two prototypical examples in the case of LGB gravity are shown in figure \ref{extcurve}. For negative $\lambda$, the extremality curve stops at the value of $g_+$ corresponding to the maximum. Beyond that point the separation between black holes and naked singularities corresponds to those solutions where the (non-degenerate) outer horizon coincides with the singularity. Instead of zero,  this solutions have diverging temperature, as we will see in the next chapter. For positive $\lambda$ the curve extends all the way to $Q^2=0$. In five dimensions the extremal value of $\kappa$ at zero $Q^2$ is finite, corresponding to the mass gap $\kappa=\lambda$ found in the neutral case, whereas for dimensions greater than five the intersect is zero. 

For non-linear electrodynamic sources, such as for instance Born-Infeld charges, the $r$-dependence of the solution will be a bit different but in general lines the same logic applies.

Before concluding this section let us say something about the addition of angular momentum. Rotating solutions which would generalize the Kerr or Myers-Perry rotating black holes of general relativity are hard to find in general Lovelock gravity. Despite the many  attempts in that direction this remains an unsolved problem. Recently, it was proven in \cite{Deruelle2008} that the Kerr-Schild ansatz does not yield any solution in Lovelock theory (except for the very special cases of Einstein-Hilbert and Chern-Simons gravity), making manifest the intricacy of Lovelock equations in this case. 

Nonetheless, some advances in this area were recently achieved. In \cite{Deruelle2008} an exact analytic rotating solution was found for Chern-Simons gravity in five dimensions. This LGB solution, however, does not present a horizon, and thus it does not represent a black hole. Moreover the numerical analysis of \cite{Brihaye2008} supports the idea of the existence of rotating solutions, in the same way as the approximated analytic solutions found in \cite{Kim2008}. Other solutions are known which represent rotating flat branes, a simple extension of topological black holes with $\sigma = 0$. 

Despite the difficulties, the linear approximation to the ({\it slowly}) rotating solution is quite simple, and can be easily generalized to any order in curvatures. We just need  to modify the spherically symmetric ansatz as 
\bear
e^0 = \sqrt{f(r)}\, (dt-a \sin^2\theta d\phi^2) ~, & \qquad e^1 = \frac{1}{\sqrt{f(r)}}\, dr ~, &\qquad  e^2=rd\theta ~, \nonumber\\
e^3= \frac{1}{r}\sin\theta\left(r^2 d\phi+a(\Lambda r^2-1)dt\right)&\qquad e^a = r\, cos\theta\tilde e^a ~, &
\label{vierot}
\eear
where the $\tilde e^a$ are the vielbein basis of the $(d-4)$-sphere, yielding a very simple form for the curvature 2-form
\begin{eqnarray}
& & R^{01} = - \frac12\, f''(r)\; e^0 \wedge e^1 +a g'(r) \cos\theta\,e^2\wedge e^3 ~, \nonumber \\ [1em]
& & R^{02} = - \frac{f'(r)}{2 r}\; e^0 \wedge e^2 + \frac12 a g'(r)\cos\theta\, e^1\wedge e^3 ~, \nonumber \\ [1em]
& & R^{12} = - \frac{f'(r)}{2 r}\; e^1 \wedge e^2 + \frac12 a g'(r)\cos\theta\, e^0\wedge e^3 ~,  \\ [1em]
& & R^{03} = - \frac{f'(r)}{2 r}\; e^0 \wedge e^3 - \frac12 a g'(r)\cos\theta\, e^1\wedge e^2~, \nonumber \\ [1em]
& & R^{13} = - \frac{f'(r)}{2 r}\; e^1 \wedge e^3 - \frac12 a g'(r)\cos\theta\, e^0\wedge e^2 ~, \nonumber \\ [1em]
& & R^{0a} = - \frac{f'(r)}{2 r}\; e^0 \wedge e^a ~, \qquad\qquad
 R^{1a} = - \frac{f'(r)}{2 r}\; e^1 \wedge e^a ~, \nonumber \\ [1em]
& & R^{23} = g(r)\; e^2 \wedge e^3 -a g'(r) \cos\theta \, e^0\wedge e^1 ~, \qquad\qquad R^{ab} = g(r)\; e^a \wedge e^b ~, \nonumber
\label{riemannrot}
\end{eqnarray}
for $f(r)=1-r^2 g(r)$. In this way, it is simple to see that the extra terms appearing in the curvature do not contribute to the equations of motion to first order in $a$. The two terms on any component of the curvature 2-form are {\it orthogonal}, they have components in 2-planes with no common directions. Therefore, the first correction to the field equation is order $a^2$. The same happens for the mass or any of the thermodynamic quantities, they do not get corrected either. The only thermodynamic variables that are modified to linear order are the angular velocity and the angular momentum,
\beq
J=\frac{2 a M}{d-2}~.
\eeq

The obtained metric becomes
\beq
ds^2=-f(r)dt^2+\frac{dr^2}{f(r)}+r^2 d\Omega_{d-1}^2- 2a r^2\sin^2\theta (g(r)-\Lambda)dtd\phi 
\eeq
with an undetermined constant $\Lambda$ related to the angular velocity at infinity. We may choose $\Lambda$ to be the effective cosmological constant of the branch under study, $g(\infty)=\Lambda$, in such a way that this angular velocity vanishes,
\beq
\Omega_\infty=\frac{g_{t\phi}}{g_{\phi\phi}}=a r^2 (g(r)-\Lambda)\;\rightarrow\; 0~.
\eeq
At the horizon however the angular velocity is nonzero,
\beq
\Omega_H=a (1-r_+^2\Lambda)~.
\eeq
%

\section{Lovelock cosmologies}
\label{LLcosmo}

Another particularly simple solution in the context of Lovelock gravities are FRW cosmologies. We may still use the same graphical techniques to describe and analyze them. We consider the usual homogeneous FRW ansatz
\beq
ds^2=-dt^2+a(t)^2 d\Sigma^2_{d-1,\sigma}~,
\eeq
whose natural vielbein basis and resulting spin connection correspond to
\bear
e^0 = dt ~, \qquad e^a = a(t)\tilde e^a ~,\\
\w^a_{\ 0}= \frac{\dot{a}}{a} e^a ~, \qquad \w^{ab}=\tilde \w^{ab}~,
\label{cosmo}
\eear
where $(\tilde{e}^a,\tilde{\omega}^{ab})$ are the vielbein and spin connection of the maximally symmetric slices. This choice yields a particularly simple form for the curvature 2-form,
\begin{eqnarray}
& & R^{0a} = \frac{\ddot{a}}{a}\; e^0 \wedge e^a ~,  \qquad\qquad R^{ab} =  \frac{\sigma + \dot{a}^2}{a^2}\; e^a \wedge e^b ~.
\label{riemanncosmo}
\end{eqnarray}
The equations of motion are also very simple, diagonal, the $tt$-component being proportional to the characteristic polynomial of the Lovelock theory, $\Upsilon[G]$,
the argument function $G$ being in this case $G=\frac{\sigma + \dot{a}^2}{a^2}$ (the r\^ole of the $f$ function of the black hole being played by $-\dot{a}^2$). The rest of the components are fixed by the Bianchi identity to
\beq
\mathcal{E}_b\sim \left[\frac{d}{d\log a}+(d-1)\right]\Upsilon[G]~.
\eeq
The vacuum solutions are in this way exactly the same ones we found using the black hole ansatz, \ie maximally symmetric solutions with
\beq
\frac{\sigma + \dot{a}^2}{a^2} = \frac{\ddot{a}}{a} = \Lambda~.
\eeq
Moreover, if we source these equations by a homogeneous isotropic perfect fluid filling the whole universe, the polynomial is basically the energy density, $\rho$ and the second equation of motion defines the corresponding pressure, $p$, subject to the energy conservation constraint. This system of equations can be equivalently written as
\bear
\Upsilon[G]=\frac{8\pi G_N}{d-1}\rho ~,\label{conserv}\\
\label{constraint}
\dot{\rho}+(d-1)\frac{\dot{a}}{a}(\rho+p)=0 ~.
\label{cosmocons}
\eear
We can even introduce a third equation for $G$
\be
\left[\frac{d }{d\log a}+2\right]G=2\frac{\ddot{a}}{a}
\label{accel}
\ee
in such a way that also $\sigma$ is set by the initial conditions as an integration constant. We thus have a set of three differential equations that fully determine the dynamics of our system, the only missing piece is the fluid equation of state. Consider for simplicity a {\it single species} fluid with a linear equation of state, $p=(\omega-1)\rho$. Zero $\w$ is equivalent to the cosmological constant and can be reabsorbed into $\Upsilon$. We will not consider this case and assume $0<\w\leq 2$, which satisfies the dominant energy condition, $\rho>\|p\|$. Then, we can use the conservation equation together with the equation of state to get the density, $\rho(a)$, in our case
\beq
\rho=\frac{\rho_0}{a^{(d-1)\w}}~,
\eeq
effectively reducing the number of equations to two; one {\it conservation equation} \reef{conserv} and one {\it acceleration equation} \reef{accel}. We can make an analogy with a one-dimensional system of a particle on a potential, although with a non-canonical kinetic term. The conservation equation is in general of the form
\be
E(\dot{a}^2,a)=0~,
\ee
whereas the acceleration equation comes from the derivative with respect to time of the first one,
\be
\frac{\partial E}{\partial (\dot{a}^2)}\ddot{a}=-\frac12\frac{\partial E}{\partial a}~.
\ee
These are analogous to the $\frac12\dot{x}^2+V(x)=E_0$ and $\ddot{x}=-V'(x)$ equations of the particle. Due to the higher order form of the equations of motion, the expression of these is more complicated but the logic remains the same. The dynamics of the system can be described using only the conservation equation. In our case the {\it conservation} and {\it acceleration} equations are independent because we have considered an extra variable $G$. Still we may integrate \reef{accel} in order to get $G=\frac{\sigma+\dot{a}^2}{a^2}$ and everything will reduce to one conservation equation
\be
\Upsilon\left[\frac{\sigma+\dot{a}^2}{a^2}\right]=\frac{8\pi G_N}{d-1}\rho(a)~.
\ee
We just introduced the {\it acceleration equation} \reef{accel} in order to make clear that $\sigma$ may be considered an integration constant instead of something of our choice. Even though we can rescale this constant to the three $0,\pm1$ usual values, it may take in principle any real value. $\sigma$ plays the same r\^ole as the energy, $E_0$, of the one-dimensional particle. We have effectively changed from a system in {\it real} space $(t,a(t))$ to a problem in {\it phase} space $(a,\dot{a}(a))$. We can even write the original metric in those variables as
\be
ds^2=-\frac{da^2}{\dot{a}(a)^2}+a^2 d\Sigma^2_{d-1}~,
\ee
making the analogy with the black hole solution much more transparent. This parallelism is even more explicit taking into account that $R_{AH}^2=\frac{1}{G(a,\dot{a})}$ is the radius of the {\it apparent horizon} of that spacetime \cite{Faraoni2011}, with associated thermodynamic variables in very much the same way as the black hole. 

Once the conservation equation \reef{conserv} has been stablished the dynamics of the system is completely determined by two initial conditions
\be
a(0)=a_0 \qquad\qquad ; \qquad \quad \dot{a}(0)=v_0
\ee
and the choice of a given branch of the polynomial. 

In the general relativistic case we would just plug $\rho(a)$ into the constraint in order to find $a=a(t)$. In this more general case this cannot always be done and even when we can the equation for the different branches would be very complicated. Nevertheless we can still use graphical techniques in order to analyze the qualitative behavior of the solutions.

 We can determine the position of the turning points as we did for the black hole positions in the black hole case. This is even clearer taking into account that the r\^ole of $f$ is now played by $-\dot{a}^2$, thus {\it horizons}, $f=0$, are now turning points, $\dot{a}=0$. There is a crucial difference however as the cosmological evolution corresponds to $\dot{a}^2\geq 0$, hence the trapped region in black hole language. The regions of interest are exactly complementary to those analyzed in the black hole case. The turning points $G_+=\sigma/a^2$, are the roots of the following equation
\beq
\Upsilon[G_+]=\frac{8\pi G_N}{d-1}\rho\left(\sqrt{\sigma/G_+}\right)
\label{turning}
\eeq
For pressureless matter ($\omega=1$) this reduces to exactly the same equation as for black hole horizons with $\kappa=\frac{8\pi G_N}{d-1}\rho_0$. Remark that considering positive energy, $\rho_0>0$ density the behavior of $\rho(a)$ is monotonic even for several species. This follows from conservation of the energy and the energy conditions. Notice also that we recover the vacuum solutions in the limit of infinite expansion, $a\rightarrow\infty$, at least for dS branches. For AdS branches the vacuum is in the untrapped region and thus we always encounter a turning point before reaching that point.  Remarkably, in the presence of radiation ($\omega=2$) the right hand side of \reef{turning} grows at least as $G_+^{d-1}$ whereas the polynomial grows at most as $G_+^\frac{d-1}{2}$. The Big Bang singularity $a=0$ is then always in the $\dot{a}^2>0$ physical ({\it trapped}) region and for low enough densities there is also always at least one turning point. 

As we have seen, the analogy with the black hole solution is very useful, vacua and branches are the same, horizons map to turning points and $f(r)$ to $-\dot{a}^2$. In black holes we usually restrict our analysis to the untrapped region $f>0$ whereas in this case the physically relevant region is the complement, $\dot{a}^2>0$. Cosmological stories lie to the right of the $\rho\left(\sqrt{\sigma/G_+]}\right)$ curve. 

The singularities are also the same as in the black hole case and occur either as $a\rightarrow 0$ ({\it Big Bang}) or at points $\Upsilon'[G]=0$,  where the acceleration (or equivalently $\dot{G}$) is not determined by the equations of motion. 
In this case however, contrary to the black hole case this singularity is {\it traversable}. There is no reason \`a priori for the energy density to be monotonous with $G$. We just need to change the affine parameter describing the trajectory to the {\it length} in phase space, $ds=\sqrt{da^2+d\dot{a}^2}$ and verify that the motion is then regular and the time spent finite.  An analogous phenomenon has also been observed in the context of braneworlds in LGB gravity \cite{Maeda2012}. 

The curvature singularity appears just because the potential is multivalued and has a degenerate point as $\Upsilon'[G]=0$. The prototypical example of this is a potential with two branches of the form 
\be
V_\pm(a)=\pm\sqrt{\frac{a_\star^k}{a^k}-1}
\ee
that degenerates at $a=a_\star$ and becomes imaginary beyond that point. The acceleration is not well defined at the degenerate point
\be
\ddot{a}=\frac{k}2\frac{a_\star^k}{a^{k+1}V(a)}
\ee
diverging in opposite directions depending on whether we approach $a_\star$ trough the $+$ or $-$ branch. On the contrary, if we perform the change of variable mentioned above we get
\bear
\frac{da}{ds}&=&\frac{1}{\sqrt{1+\left(\frac{\ddot{a}}{\dot{a}}\right)^2}}\nonumber\\
\frac{d\dot{a}}{ds}&=&\pm\sqrt{1-\left(\frac{da}{ds}\right)^2}
\eear
well-defined evolution equations for $a$ and $\dot{a}$. As we approach the singular point, $a=a_\star$, $da/ds$ vanishes and then changes sign whereas $d\dot{a}/ds$ goes to plus or minus one, depending on the branch. In other words we have a turning point with finite $\dot{a}$! Our {\it cosmological} particle does not go back the way it came, it just follows the potential through the singularity changing from one branch to the other. The same happens for maxima and minima of the Lovelock polynomial in this context. If we do not encounter a turning point the system may start for $a\to\infty$ in a given dS vacuum, pass through the singularity and end up in a different dS space after enough time\footnote{Notice that the final vacuum will be BD unstable if we started from a stable one.}. This would be the case, for instance, for LGB gravity with two dS branches (\ie $c_0<0$ and $c_2<0$). 

We can take the analogy with the one particle system a bit further. 
The existence of turning points will imply in some cases the existence of forbidden regions separating possible cosmological trajectories. Quantum-mechanically we may calculate the tunneling probability by performing a Wick rotation and computing the Euclidean action of the resulting trajectory with the same energy. In the gravitational context we may think of doing the analogous thing, this would amount to the computation of the Euclidean on-shell action, $\widehat{\mathcal{I}}$, of our cosmological solution while going through the barrier. The tunneling probability will be proportional to $e^{-\widehat{\mathcal{I}}}$. As we will see in the next chapter, the Euclidean section has very important applications also in the context of black holes.  

We can now take any of the figures corresponding to black hole solutions from previous sections and reinterpret their trapped regions as possible cosmological solutions for a pressureless fluid. As an example we may analyze graphically the case of Einstein-Hilbert gravity with cosmological constant, our Lovelock cosmologies lie to the right of the black dashed lines.

\begin{figure}
\centering
\includegraphics[width=0.58\textwidth]{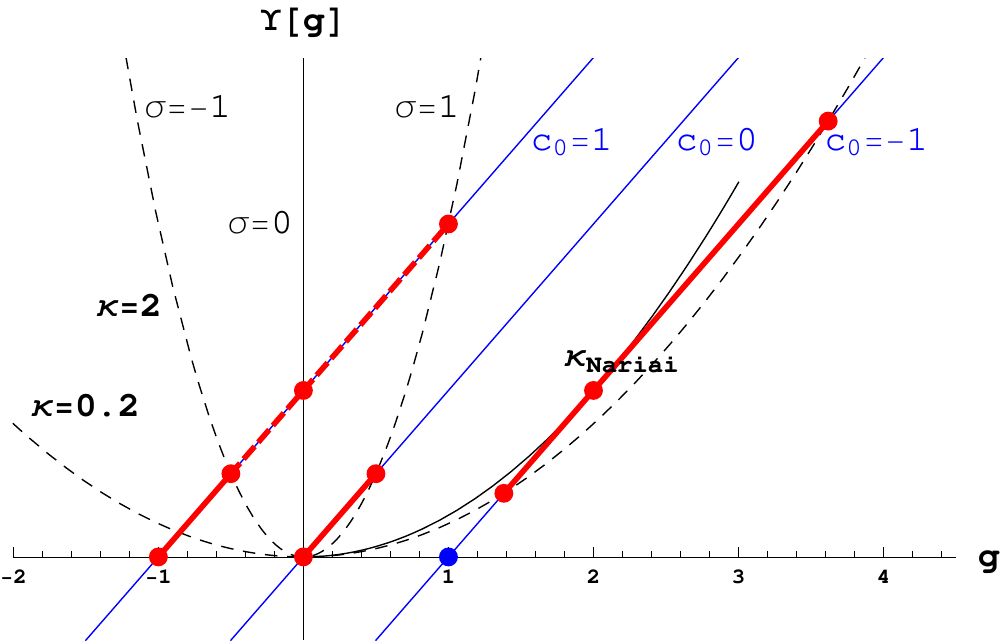} 
\caption{Linear polynomial corresponding to the usual EH branch for negative ($c_0=1$), zero ($c_0=0$) and positive ($c_0=-1$) cosmological constants ($L=1$). The dashed lines are just $\kappa\left(g/\sigma\right)^{\frac{d-1}{2}}$ for $d=5$, corresponding to pressureless matter. The crossing of these lines with the polynomial gives the turning points. The solid black line corresponds to the critical value of the mass, $\kappa_{\rm Nariai}=1/4$, that in this case corresponds to the critical mass for the existence of a potential barrier in the dS case.  For $\sigma=1$ and $\kappa>\kappa_{\rm Nariai}$ (or $r_+>r_{\rm Nariai}$), starting from the Big Bang singularity at $a=0$ (equivalently $\Upsilon=\infty$) the asymptotically dS branch describes an always expanding spacetime whereas for the other cases and topologies the turning point always exist. For the dS branch for $\kappa<\kappa_{\rm Nariai}$  there is also a second type of solution that describes a spacetime that collapses from vacuum and then reexpands. This solution can be connected to the Big Bang by tunneling. }
\label{cosmoEH}
\end{figure}

\chapter{\bfseries\itshape Black hole thermodynamics}
\chaptermark{Black hole thermodynamics}
\label{chp:thermo}

\vspace{.6cm}

\begin{quotation}
\flushright
{\it ``Pour atteindre la v\'erit\'e,\\ il faut une fois dans la vie se d\'efaire de toutes les opinions qu'on a re\c{c}ues,\\ et reconstruire de nouveau tout le syst\`eme de ses connaissances.''}\\

\vspace{.3cm}

Ren\'e Descartes
\end{quotation}

\vspace{3cm}

\noindent  
In the early 1970s, Bekenstein \cite{Bekenstein1973} argued, using simple thermodynamic arguments, that black holes should be assigned finite entropy, even though this seemed to contradict the fact that, as classical objects, they have zero temperature.  
Bekenstein was worried about the possibility of collapsing  matter into a black hole -- classically, a very simple object -- hiding any trace of the original entropy. The second law of thermodynamics, which asserts that the entropy of any closed system can never decrease, would be violated by this process. However, the mass increase due to the matter falling in also implies a growth of the black hole size, which led to the suggestion that the entropy of a black hole is given by its surface. This intuition became conviction when, in 1972, Hawking proved that the area of a black hole event horizon can never decrease.

The similarity between black holes and thermodynamic systems was considerably strengthened when Bardeen, Carter, and Hawking \cite{Bardeen1973} showed that  these enigmatic objects obey a complete set of rules that parallel precisely the laws of thermodynamics, 
\begin{itemize}
\item  The surface gravity $\kappa$ is constant over the event horizon ({\it zeroth law}).
\item For any two stationary black holes differing only by small variations in the mass $M$, angular momentum $J$, and charge $Q$,
\be
\delta M=\frac{\kappa}{8\pi G_N}\delta A+\Omega_H\delta J+\Phi_H \delta Q
\ee
where $\Omega_H$ and $\Phi_H$ are the angular velocity and the electromagnetic potential at the horizon respectively ({\it first law}).
\item The area of the event horizon of a black hole never decreases ({\it second law}),
\be
\delta A\geq 0
\ee
\item It is impossible by any procedure to reduce the surface gravity
$\kappa$ to zero in a finite number of steps ({\it third law}).
\end{itemize}
These laws were later shown to apply in much more diverse situations than the particular 4-dimensional setting where they were discovered. In particular, the first law holds for much more general gravitational actions, for which the entropy can be understood as a Noether charge \cite{Wald1993}.

In these laws the r\^ole of the temperature is played by the surface gravity but Bardeen, Carter and Hawking, in their original paper, added: {\it ``It should however be emphasized that $\kappa/8\pi G_N$ and $A$ are distinct from the temperature and the entropy of the black hole. In fact the effective temperature of a black hole is absolute zero. In this sense a black hole can be said to transcend the second law of thermodynamics.''}

At the time, there seemed to be a fundamental contradiction between the  hypothesis of black holes' entropy being nonzero and them being perfectly absorptive entities. In contact with a heat bath they would take in energy while giving off nothing, thus behaving as zero temperature objects. Couple of years later, Hawking himself came up with the resolution of this apparent paradox, the problem being solved by quantum theory \cite{Hawking1974}. In the presence of a horizon quantum fields were shown to be emitted with a thermal spectrum of temperature given roughly by the surface gravity of the hole, $T_H=\kappa/2\pi$. This temperature has not yet been directly observed, otherwise it would presumably grant Hawking the Nobel prize! For typical black hole masses, Hawking's temperature is several orders of magnitude smaller than the cosmic microwave background, thus impossible to measure. To have a Hawking temperature larger than 2.7 K (and actually be able to evaporate), a black hole needs to have less mass than the Moon. Such a black hole would be smaller than a needle's eye\footnote{Data from \href{http://www.einstein-online.info/elementary/quantum/evaporating_bh/?set_language=en}{\it Einstein online}}.

All the accumulating evidence indicate that the similarities between the laws of black hole mechanics and the laws of thermodynamics are not a simple coincidence. Indeed, they seem to be hints of some very deep physics, intertwining classical and quantum properties of these objects. Hawking established the equivalence of surface gravity and temperature of a black hole, whereas mass in this latter context is matched by the energy of the thermodynamic system. Moreover, this connection also requires identifying the black hole area with its entropy, as suggested by Bekenstein. More specifically, Bekenstein's entropy is proportional to the area of the event horizon of the black hole. The {\it Generalized} Second Law of thermodynamics then conjectures that the sum of the entropy of the matter outside a black hole and its own never decreases. Moreover, black holes, as isolated objects, can be described as thermodynamic ensembles to which all the usual machinery of thermodynamics can be applied, allowing the description of new and interesting phenomena, including phase transitions. 

The phase structure of general relativity is quite well understood at present.  As we increase the dimensionality however this phase structure gets increasingly intricate and diverse. In dimensions greater than four the metric has many more degrees of freedom and, as a result, the spectrum of the theory gets richer. We may have extended black objects, like strings or branes; but also more rotation planes and extra dimensions that may be compactified {\it \`a la} Kaluza-Klein. The analysis of this phase structure is very interesting for a wide variety of reasons. First of all it may help to elucidate which features of general relativity are universal and which are, on the contrary, $d$-dependent. This particular question is exacerbated in the context of Lovelock, or more general gravity theories, as we also increase the number of tunable couplings as we increase the dimension. In this context even the properties of static solutions are poorly understood and seemingly pathological features have appeared. The detailed analysis of the thermodynamics of these solutions is crucial for the understanding of the consistency of the theory.

On the other hand, the proposed study may uncover the existence of critical dimensions where properties of black holes change dramatically. It will yield information about stability and phase transitions, and the possibility of thermodynamic and perturbative stability being correlated. Also, the information about static and stationary solutions may provide some clues about the endpoint of instabilities or even about time dependent trajectories connecting different phases, many interesting questions that deserve further investigation.

The thermodynamics of Lovelock theories is not as well understood as the general relativistic case, not even at the level of the basic principles. The zeroth law has been shown to hold also in this context for matter respecting the dominant energy condition \cite{Sarkar2012}, exactly as in the Einstein-Hilbert case. The first law is also verified, as expected in general grounds, and has been used to extract the expression for the entropy \cite{Myers1988,Jacobson1993} of Lovelock black holes, that no longer coincides with the area of the event horizon. In particular, it has been shown that the entropy may become negative in some cases, in conflict with any microscopic interpretation. We will comment more on this below. An extended version of the first law has also been proposed \cite{Kastor2010} where the Lovelock couplings play the r\^ole of extra thermodynamic variables. 

The second law of thermodynamics has a crucial r\^ole in the thermodynamic picture of black hole dynamics as it enforces the irreversibility intrinsic to  thermodynamic processes. It has been shown that the second law also holds for general Lovelock theories in a number of cases \cite{Akbar2009,Akbar2009a,Sarkar2010}, namely a physical process version has been proven on \cite{Kolekar2012}. Some words on the third law can be also found in \cite{Torii2006}.

In the next sections we will take   the usual thermodynamic interpretation inherited from general relativity for granted, discussing its implications for the stability and phase transitions of static black holes in Lovelock gravity. Some seemingly pathological features will appear, its resolution, when available, being proposed and discussed. We will start with some clarifying description of the connection between gravity and thermodynamics and the presentation of the basic thermodynamics formul\ae\ for this chapter.

\section{The path-integral approach to Quantum Gravity}
\label{EQG}

The first attempts to the quantization of gravity were based naively on the usual techniques of Quantum Field Theory, combined with general considerations about the nature of the gravitational interaction itself. Even though these theories were not completely consistent they yield some partial results compelling enough to be trusted as a good first approximation to the problem. These results are related to the Hawking effect, the connection between black holes and thermodynamics and other semi-classical features. The most useful approach to make  this connection manifest is what is usually called {\em Euclidean Quantum Gravity}. 
		
The starting point for this approach is the idea that one can represent the amplitude to go from a state with metric $g_1$ and matter fields $\phi_1$, on a spacelike hypersurface $S_1$, to a state with a metric $g_2$ and matter fields $\phi_2$, on a hypersurface $S_2$, as a path integral over all field configurations $g$ and $\phi$ which take the given boundary values on $S_1$ and $S_2$ \cite{EQG1}. More precisely,

\be
\label{pathint}		
\left\langle g_2, \phi_2, S_2 \right. \left| g_1, \phi_1, S_1 \right\rangle = \int \cD[g,\phi]\, e^{i\mathcal{I}[g,\phi]}~.
\ee 
This is the usual way a path integral is defined in any quantum-mechanical system, where $\cD[g,\phi]$ is a measure in the space of all field configurations $g$ and $\phi$, and $\mathcal{I}[g,\phi]$ is the action of the fields. Not all the components of the metrics $g_1$ and $g_2$ are physically relevant. We need to specify only the three-dimensional induced metric $h$ on $S_1$ and $S_2$, up to diffeomorphisms which map this two surfaces into themselves. 

Consider now an intermediate surface $S_0$ between the two boundary surfaces. One would expect that

\be
\left\langle h_2, \phi_2, S_2 \right. \left| g_1, \phi_1, S_1 \right\rangle = \sum_{h_0, \phi_0}{\left\langle h_2, \phi_2, S_2 \right. \left| h_0, \phi_0, S_0 \right\rangle \left\langle h_0, \phi_0, S_0 \right. \left| g_1, \phi_1, S_1 \right\rangle }~.
\ee 
This is a general property of amplitudes in quantum mechanics that one would want the theory to verify. The amplitude to go from the initial state to the final state should be obtained also by summing over all possible configurations on the intermediate surface. The usual Lovelock bulk action alone does not verify this property, it needs to be supplemented with the boundary terms discussed in section \ref{bdyterms}\cite{EQG1}. This is an alternative motivation for its inclusion, now in the quantum framework.

\subsection{Spacetime complexification and thermodynamics}

For real Lorentzian metrics $g$ and real matter fields the action $\mathcal{I}[g,\phi]$ will be real. The path integral (\ref{pathint}) will then oscillate and it is not clear whether it converges or not. In quantum field theory one usually deals with this difficulty by a Wick rotation in the complex time variable, {\it i.e.}, $t=-i\tau$. This idea applied to our general case leads to a path integral of the form

\be
\label{path}
Z=\int{\cD[g,\phi]\, e^{-\widehat{\mathcal{I}}[g,\phi]}}~,
\ee 
where $\hat{\mathcal{I}}=-i\mathcal{I}$ is called the {\it Euclidean} action, and $g$ and $h$ are now positive-definite. For Lovelock gravity we may write 
\be
\widehat{\mathcal{I}}[g,\phi]= -\frac{1}{16\pi G_N (d-3)!}\, \sum_{k=0}^{K} {\frac{c_k}{d-2k}} \left(\int_{\mathcal{M}}\!\! \mathcal{L}_{k} -\int_{\partial \mathcal{M}}\!\!\!\mathcal{Q}_k\right)+\mathcal{I}_{mat}[\phi]~.
\label{Eucaction}
\ee

A very important use of the Euclidean section is to construct the canonical ensemble for a field theory. Consider a field $\phi$. The amplitude to propagate from a configuration $\phi_1$ at time $t_1$ to a configuration $\phi_2$ at time $t_2$ is given by the path integral

\be
\left\langle \phi_2, t_2\right|\left.\phi_1, t_1\right\rangle=\int{\cD[\phi]\, e^{i\mathcal{I}[\phi]}}~,
\ee
with the given boundary values. Using the Schr\"odinger picture we can also write this amplitude as 

\be
\left\langle \phi_2\right|\, e^{-iH(t_2-t_1)}\left|\phi_1\right\rangle=\int{\cD[\phi]\, e^{i\mathcal{I}[\phi]}}~,
\ee
where $H$ is the hamiltonian driving the time evolution of the system. By going  to periodic complex time via a Wick rotation with $t_2-t_1=-i\b$ and $\phi_2=\phi_1$, and summing over a complete orthonormal basis of configurations, we get the partition function,

\be
\mathcal{Z}=\sum_{E_n}e^{-\b E_n}=\int{\cD[\phi]\,  e^{-\widehat{\mathcal{I}}[\phi]}}~,
\ee 
for the field $\phi$ at temperature $T=\b^{-1}$, where $E_n$ is the energy of the $n$th eigenstate. In this last expression the path integral is taken over all fields $\phi$ which are real in the Euclidean section and periodic in imaginary time with period $\b$.

The same idea can be applied directly to any gravitational system making the connection with thermodynamics explicit. The canonical partition function associated with a gravitational system at temperature $T$ would be defined as the path integral (\ref{path}) extended to all configurations with given boundary values and identified in with period $\b$ in Euclidean time. Henceforth, we are going to consider only the gravitational part of the action setting all matter fields to zero. 

Once the partition function of our theory has been computed, it can be used to extract information about the system. In particular we can derive all the thermodynamic magnitudes of interest using the usual relations from statistical mechanics, namely the Helmholtz free energy, relevant thermodynamic potential for the canonical ensemble,

\be
F=M-TS=-T\log \mathcal{Z}~,
\ee
which, at equilibrium, sits at a minimum. The average energy and the entropy can be calculated from it as

\bear
\label{entropyenergy}
\left\langle E\right\rangle &=& \frac{1}{\mathcal{Z}}\sum_{E_n}{E_n e^{-\b E_n}}=-\frac{\partial}{\partial\b} \log \mathcal{Z}~,\label{Emass}\\
S&=&-\left( \frac{\partial F}{\partial T}\right)=\b \left\langle E\right\rangle+\log 
\mathcal{Z}\label{Eentropy}~.
\eear
In the canonical ensemble, the temperature is an external constraint applied over the system and cannot be determined using this framework. In the gravitational context the temperature of a black hole solution is computed from the required regularity of the Euclidean section of the solution. 

Finally, if one is to recover the classical gravitational theory we started with in the limit of macroscopic objects, one expects that the dominant contribution to the partition function will come from metrics which are an extremum of the action, \ie solutions of the classical equations of motion. 
This is known as the {\it stationary-phase} or {\it saddle point} approximation. As a extreme version of this we can estimate the partition function by the single contribution of the metric with least action,

\be
\log \mathcal{Z}\approx -\widehat{\mathcal{I}}[g_\text{min}]~,
\ee
and the free energy will coincide essentially with it, $\hat{\mathcal{I}}=\beta F=\beta M-S$. As we will see in the next section, the free energy is in general divergent due to infinite volume of the spacetime. In order to regularize the Euclidean action we will then subtract the contribution of some reference background that customarily is taken to be the maximally symmetric vacuum for the considered asymptotics. 

\subsection{AdS spacetime and the canonical ensemble}

The canonical ensemble describes a system in thermal equilibrium with an infinite heat reservoir so that the temperature of the system is fixed. This ensemble is ill-defined in asymptotically flat spacetimes \cite{Gross1982} because of the attractive nature of gravity and the possibility of having a black hole. To illustrate this, let us consider four dimensional flat space at temperature $T$. In this hypothetical situation every region of spacetime would be filled with homogeneous thermal energy density, $\rho\sim T^4$.  Due to the infinite volume of the spacetime, the total energy would be thus also infinite and the configuration, albeit simple, inconsistent. The backreaction of this infinite mass would lead the spacetime to warp, and it would be no longer flat. Another, argument from the classical point of view, is that thermal perturbations of wavelength larger than the Jeans' length scale\footnote{Critical radius of a cloud (typically a cloud of interstellar dust) where thermal energy, which causes the cloud to expand, is counteracted by gravity, which causes the cloud to collapse [{\it cf.} Wikipedia].}  grow exponentially and collapse to form a black hole \cite{Gross1982}. Thermal flat space is thus unstable to the formation of black holes and these black holes are also unstable, in this case thermodynamically. They cannot be in equilibrium with their environment, either decaying into pure thermal radiation or engulfing the whole spacetime.

This problem is naturally solved if we consider an anti-de Sitter space instead. In AdS the locally measured temperature
\beq
T_\text{loc}=\frac{T}{\sqrt{-g_{tt}(r)}}
\label{locT}
\eeq
from any origin undergoes an infinite redshift as $r\rightarrow\infty$, and the thermal energy density decreases fast enough to ensure that the mass of the AdS space at any temperature $T$ is finite. In addition to this, the higher mass branch of black hole solutions is stable and, as we will show in what follows, this determines a well-defined partition function. In fact, although AdS space at finite temperature is stable a black hole will be formed above some critical energy density, the action for the hole becoming lower to that of thermal AdS. The black hole metric is the dominant saddle in that case and the change from one phase to the other is known as the Hawking-Page (HP) phase transition \cite{HP}. Even though the previous argument is based in general considerations from general relativity, something similar will happen also for Lovelock theories. 

We proceed now to make some of the above considerations concrete. We will study (static) asymptotically AdS metrics identified periodically in $\tau$ with period $\b=T^{-1}$. Pure AdS spacetime periodic in imaginary time is one of the solutions to be considered and we take it to correspond to the zero of action and energy. The analogue of the Schwarzschild-AdS black hole solution is probably the only other non-singular positive-definite solution for the classical equations that satisfies the periodic boundary conditions. Its Euclidean section is
\be
ds^2=f(r)\, d\tau^2+\frac{dr^2}{f(r)}+r^2d\Sigma^2_{d-2,\sigma}   ~.
\ee
The metric inside the horizon is not positive definite, and the manifold should then end regularly at $r=r_{+}$. Near the horizon the metric in the ($\tau$,$r$)-plane looks like $\IR^2$ in polar coordinates,
\be
ds^2\approx\frac{4}{f'(r_{+})}\left(d\rho^2+\rho^2\frac{d\tau^2}{4/f'(r_{+})^2}\right)+r_{+}^2d\Sigma_{d-2,\sigma}^2~,
\ee
where $\rho^2=r-r_{+}$ and $'$ denotes derivation with respect to the radial coordinate $r$. Then, in order to avoid a conical singularity at $\rho=0$, the {\it angular} variable
\be
\frac{\tau}{2/f'(r_{+})}\equiv\varphi~,
\ee 
must be identified with period $2\pi$. We have a smooth Euclidean section that cannot be continued inside $r=r_{+}$ only with period in complex time
\be
\b=\frac{4\pi}{f'(r_{+})}~,
\ee
the solution being singular also for naked singularities. By the previous analysis this periodicity corresponds to the theory being at finite temperature $T=\b^{-1}$, exactly the same result we would get via the surface gravity, $\kappa=f'(r_+)/2$. As we will see in what follows, for any given possible asymptotics of the Lovelock theory, the black hole solutions under scrutiny can be divided into different branches, depending on the monotonicity properties of the temperature as a function of the mass (or horizon radius). Some of them, as the lower mass branch of GR, have negative specific heat this implying that the canonical ensemble is unstable when this kind of black holes are present. However, some of the branches, in particular the higher mass one of the EH solutions, have positive heat capacity and this allows us to define the partition function in the path integral approach. Moreover, we will see that, despite the non-compact nature of asymptotically AdS spaces, the action can be regularized yielding a finite value.

\section{Lovelock black holes and thermodynamics}

Some aspects of Lovelock black holes thermodynamics have been considered earlier in \cite{Cai2004}. In the present section we will proceed to a complete analysis including all possible cases and branches.

As discussed previously, Lovelock solutions displaying an event horizon have a well-defined temperature
\begin{equation}
T = \frac{f'(r_+)}{4\pi} = \frac{r_+}{4\pi}\,\left[(d-1)\,\frac{\Upsilon[g_+]}{\Upsilon'[g_+]}-2\,g_+ \right] ~.
\label{Temp}
\end{equation}
From this expression, however, it is not clear what the sign of the temperature is. The first term is always positive, for positive slope branches and mass. The sign of the second term depends on $\sigma$ in such a way that it is negative for spherical topology. Then the temperature is trivially positive for positive mass black holes with hyperbolic or planar topology. It is also positive for negative mass AdS black holes as will be explained shortly. This simply derives from the fact that the temperature can just change sign if it vanishes at some intermediate $r_+$, {\it i.e.}, if the black hole is extremal and the event horizon degenerates ($f'=0$). This very same logic will ensure the positivity of the temperature of spherical black hole horizons. This will be explained later following a different approach.  

We will also have negative temperature horizons but this is also a common feature of Einstein-Hilbert gravity. For instance, the temperature of the cosmological horizon of pure dS space is negative in our conventions, as it is negative for the inner horizon of a Reissner-N\"ordstrom black hole, regardless of the asymptotics of the solution. The same will happen here.

Following the approach outlined above we will compute the on-shell Euclidean action and from that all the relevant thermodynamic variables of the system. The contribution to the partition function from the black hole solution can be calculated as the difference between its Euclidean action and that of pure AdS space (the maximally symmetric vacuum with the chosen asymptotics) identified with the same period in imaginary time. The region inside the horizon is excluded since it is not positive-definite. Then, 
\begin{eqnarray}
\widehat{\mathcal{I}}&=&-\frac{V_{d-2}}{16\pi G_N}\sum_{k=0}^K \frac{a_k (d-2)}{d-2k}\left[ \int_{r_+}^R dr \int_0^{\beta} \partial_r^2\left(r^d g^k\right)-\int_0^R dr \int_0^{\beta'}\partial_r^2\left(r^d \Lambda^k\right)\right]\nonumber\\
&=& \beta \frac{V_{d-2}}{16\pi G_N}\partial_r\left[r^d_+\, \sum_{k=0}^K \frac{a_k (d-2)}{d-2k}g^k\right]
\label{Eaction}
\end{eqnarray}
There is a subtlety, as we need to impose a slightly different periods, $\b$ and $\b'$ (equal as $R\to \infty$), in each solution so that the geometry of the cut-off surface at any finite $R$ is the same in both cases. This is achieved by imposing the same physical radius in the $\tau$ circle,
\be
\b'\sqrt{\sigma-\Lambda R^2}=\b\sqrt{f(R)}~,
\ee 
otherwise it would appear an extra constant piece coming from the subleading behavior at infinity. With the above relation this constant piece becomes 
\be
\frac{\beta\kappa}{2\Lambda\Upsilon'[\Lambda]}\left[d\widetilde\Upsilon[\Lambda]-2\Lambda\widetilde\Upsilon'[\Lambda]\right]=\frac{\beta\kappa}{2\Lambda\Upsilon'[\Lambda]}\Upsilon[\Lambda]=0
\ee
where we introduced a new polynomial 
\be
\widetilde\Upsilon[g]=\sum_{k=0}^K{\frac{c_k}{d-2k}\,g^k}~,
\ee
related to the one used in previous sections, \reef{eqg}, as
\begin{equation}
\Upsilon[g]=\left.\partial_x \left[x^d\widetilde\Upsilon[g/x^2]\right]\right|_{x=1}
\end{equation}
or more precisely
\begin{eqnarray}
\Upsilon[g]&=&d\widetilde\Upsilon[g]-2g\widetilde\Upsilon'[g]\label{polynomials}\\
\Upsilon'[g]&=&(d-2)\widetilde\Upsilon'[g]-2g\widetilde\Upsilon''[g]\nonumber
\label{twopoly}
\end{eqnarray}
The characteristic polynomial used so far was related to the equations of motion whereas the one introduced above is directly related to the coefficients appearing in the action. In order to distinguish them in the text, they may be sometimes referred to as the {\it black hole} and {\it action} polynomials respectively.

In the computation of the on-shell Euclidean action we have not considered the contribution coming from the boundary terms appearing in \reef{Eucaction}. It can be easily verified that these terms yield actually no contribution. Their  subleading piece decays too fast for the integral to be nonzero. In chapter \ref{genHP} we will analyze the structure of these kind or terms in more detail. 

Using the new polynomial the expression for the free energy can be written in a much more compact way,
\be
F=\frac{(d-2) V_{d-2}}{16\pi G_N}\,\partial_r\!\left[r_+^d\widetilde\Upsilon[g_+]\right]
\ee
where the derivative is with respect to the radial variable not to the horizon radius, we have to derive before evaluating at $r=r_+$. The same simplicity will be inherited by other quantities such as the entropy and will ease some of the manipulations that follow. The free energy can also be written in a much more suggestive way as
\be
F=\frac{(d-2)V_{d-2}}{16\pi G_N}\left[\kappa-f'(r_+)\,r_+^{d-2}\widetilde\Upsilon'[g_+]\right]
\label{free-energy}
\ee
where we have taken into account the expression \reef{mmass} of the mass in terms of $r_+$.

This magnitude is relevant for processes at constant temperature to analyze the global stability of the solutions. If we express everything in terms of $g_+$ (see \reef{Temp} for the expression of $f'(r_+)$), the free energy has degree $2K-1$ in the numerator. Hence, that is the maximal number of zeros that may eventually correspond to Hawking-Page-like phase transitions (see \cite{Neupane2004} for a concrete example in the LGB theory). This expression also diverges at any extremum of the polynomial. We will comment further about this below. 

From the expression \reef{free-energy} of the free energy it is now very easy to calculate the rest of the thermodynamic variables. We can easily determine the expression of the entropy following \reef{Eentropy} yielding
\bear
S &\equiv& \frac{(d-2) V_{d-2}}{4\, G_N }\,r_+^{d-2}\widetilde{\Upsilon}'[g_+]\\
 &=&\frac{A}{4\,G_N}\left(1+\sum_{k=2}^{K}{k\,c_k\frac{d-2}{d-2k}\, g_+^{k-1}}\right)~.\nonumber 
\eear
This actually coincides with the entropy obtained by other means, like the Wald entropy \cite{Jacobson1993} or integrating the first law \cite{Camanho2011a}. In this way, it was first found in \cite{Myers1988}. The resulting expression for $K=3$ also agrees with the one arising in cubic quasi-topological gravity \cite{Myers2010c}. The general result will presumably coincide in general for that class of theories \cite{Oliva2011}, as they also share the same expressions for the mass and temperature of these solutions. The integration of the first law will trivially yield the same form also for the entropy.

 We fixed the integration constant such that the entropy vanishes when the horizon radius goes to zero. For planar horizons ($g_+=0$) this formula reproduces the proportionality of entropy and area of the black hole horizon, $A=V_{d-2}\;r_+^{d-2}$, whereas it gets corrections for other topologies. Interestingly enough, it has been recently suggested that this expression can be extended towards the interior of the geometry by performing a radial foliation and replacing $g_+$ by $g(r)$; the resulting function, $S(r)$ being interpreted as the information contained inside a given region of the spacetime \cite{Paulos2011}. 

Our entropy formula defines a {\it reference state} ($r_+=0$) for which the entropy vanishes. We can also verify that the radial derivative of the entropy is positive
\begin{equation}
\frac{dS}{dr_+}=\frac{1}{T}\frac{dM}{dr_+}=\frac{(d-2)V_{d-2}}{4\, G_N }\,r_+^{d-3}\;\Upsilon'\left[g_+\right] ~,
\label{entropy}
\end{equation}
for branches free from BD instabilities this implying the positivity of the entropy in that branch of solutions. Negative values for the entropy may however be encountered \cite{Myers1988, Nojiri2001n, Cvetic2002, Neupane2009a, Neupane2009f} in the case of hyperbolic horizons or type (b) spherical solutions, for which such {\it reference state} does not exist. In both cases the zero radius solution does not belong to the same branch where the negative entropies appear. It is actually unclear what is the suitable vacuum solution to be used as a {\it reference state} with hyperbolic topology \cite{Emparan1999}. Generally the horinzonful solution of minimal mass corresponds to a extremal black hole. As the hyperbolic {\it vacuum} ($M=0$) has temperature, the extremal state has been proposed as groundstate \cite{Vanzo1997a,Birmingham1999}.

The black hole solution presented has $\kappa$ as its only parameter, but we can equivalently consider $r_+$. Following the computation of the entropy we may realize that some combination of variations with respect to $r_+$ vanishes,
\be
\frac{(d-2)V_{d-2}}{16\pi G_N}\left(\ \ \partial_+\!\!\left[r_+^{d-1}\Upsilon[g_+]\right]-f'(r_+)\,\partial_+\!\!\left[r^{d-2}\widetilde\Upsilon[g_+]\right]\ \ \right)\delta r_+=0 ~,
\ee
where in this case the derivatives are taken with respect to the horizon radius. This is equivalent to the {\it first law} of thermodynamics, $dM-TdS=0$, the first term being the variation of the mass and the second that of the entropy multiplied by the temperature. To actually check this we have to verify that what we called the mass in \reef{mmass} is actually the total energy of the system (with respect to the reference background). From the above expressions this is also very easy to see as the free energy has the expected expression
\be
F=M-TS\label{free-energy2}
\ee
the expectation value for the energy \reef{Emass} thus being exactly the mass identified before in \reef{mmass}
\be
\langle E\rangle=F+TS=M
\ee
For large mass the relation between the entropy and the mass can be made explicit, being actually equivalent to the relation in the general relativistic or planar case,
\be
S\approx \frac{V_{d-2}}{4 G_N}\left(\frac{16\pi G_N}{(d-2)V_{d-2}} L^2\,M\right)^\frac{d-2}{d-1}
\ee
Notice that this expression is just valid for the EH branch as all other branches do not possess a well-defined high mass limit. 

The entropy is related to the number of microstates compatible with the macrostate of the system. The previous result means that the density of states grows like $\log(\mathcal{N})\sim M^\frac{d-2}{d-1}$, sufficiently slow so as to make the integral defining the partition function, 
\be
\mathcal{Z}=\int{\mathcal{N}(M)\,e^{-\b M}dM},
\ee
converge. This shows that the canonical ensemble in asymptotically AdS space is well behaved, at least for the EH branch, the general case being a bit more involved. In asymptotically flat space the density of black hole states goes as $\log{\mathcal{N}}\sim M^\frac{d-2}{d-3}$ and so the canonical ensemble is pathological. This pathology translates into the thermodynamic instability of these black holes that have negative specific heat. 

In the AdS case, for large $r_+$, we can also approximate
\begin{equation}
M \sim T^{d-1} ~,
\end{equation}
which again coincides with the planar case. Then, $dM/dT > 0$ and the black hole is locally thermodynamically stable; it can be put in equilibrium with a thermal bath. In general, this will not happen for small black holes. This points towards the occurrence of Hawking-Page-like phase transitions, which are interpreted as confinement/deconfinement phase transitions in the corresponding dual conformal plasma \cite{Witten1998,Witten1998a}. These have already been studied in the case of LGB gravity \cite{Nojiri2001j,Cho2002a}. We will further investigate these in the case of general Lovelock gravities for any of the branches of the theory. We will use the preceding formul\ae\ to calculate the free energy (the Euclidean on-shell action) and compare it between the different phases, the lowest value corresponding to the globally stable phase on that branch. This has to be studied in the non-planar case, otherwise scale invariance would only allow for phase transitions at zero temperature. The complicated structure of phase transitions  would then be hidden and any black hole will correspond to the stable highest temperature deconfined phase. 

In chapter \ref{genHP} we will perform the same kind of analysis for a slightly more general class of solutions. In particular, we will be interested in describing transitions between different branches of a given Lovelock theory \cite{Camanho2012,Camanho2013}.


One important feature that will be relevant later on, when discussing classical stability, is that the temperature is proportional to the derivative of the mass (\ref{mass}) with respect to the radius of the black hole horizon,
\begin{equation}
\frac{dM}{dr_+} = \frac{(d-2)V_{d-2}}{4\, G_N }r_+^{d-3}\;\Upsilon'\left[g_+\right]\, T ~.
\label{dMdr}
\end{equation}
The proportionality factor is exactly the radial derivative of the black hole entropy \reef{entropy}. We can see that, as long as  we are in a branch free from Boulware-Deser instabilities, the radial derivatives of the mass and the entropy are positive for $T>0$. On the one hand, this will be important when discussing classical instability as the heat capacity of the black hole reads
\begin{equation}
C=\frac{dM}{dT}=\frac{dM}{dr_+}\frac{dr_+}{dT} ~,
\end{equation}
and then the only factor that can be negative leading to an instability is $dT/dr_+$. 

\begin{figure}
\centering
\includegraphics[width=0.44\textwidth]{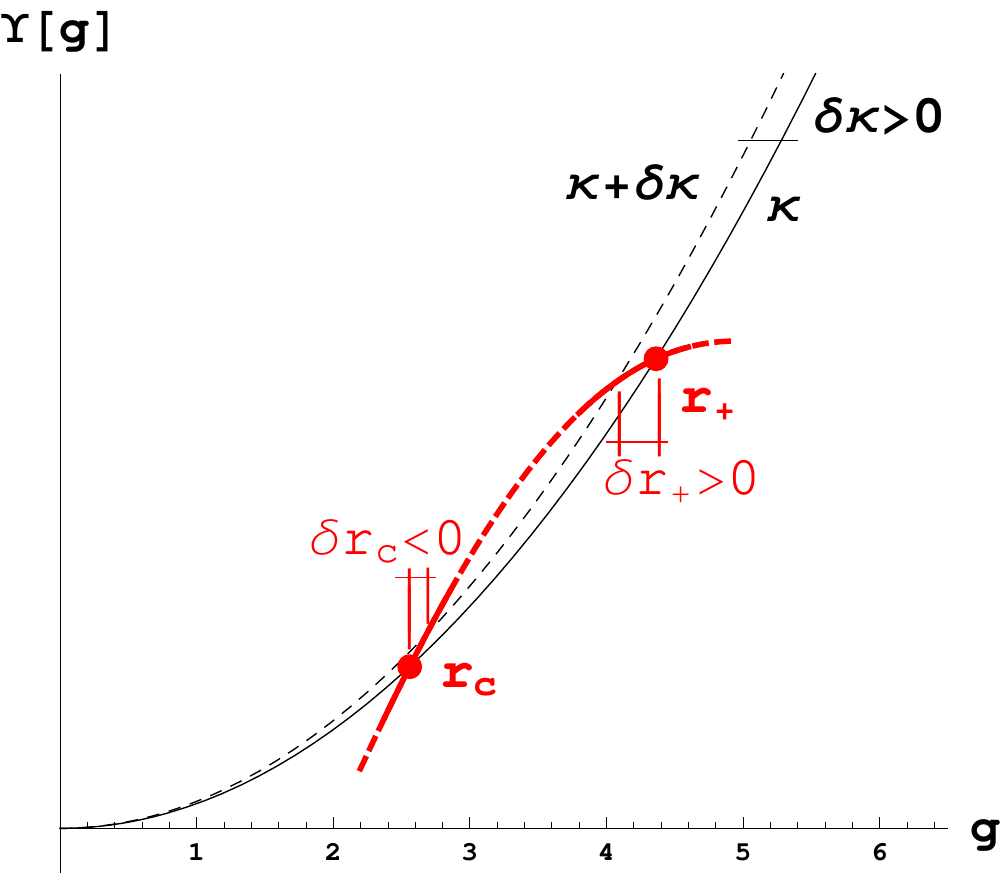}
\caption{Determination of the sign of the temperature for the cosmological and black hole horizon of a dS branch. The cosmological horizon has $T_c\propto d\kappa/dr_c <0$ whereas the event horizon of the black hole has $T_+\propto d\kappa/dr_+ >0$. Recall that ${\rm sign}(\delta g_+)=-{\rm sign}(\delta r_+)$ and the same holds for every horizon.}
\label{signT}
\end{figure}

Another useful application of (\ref{dMdr}) is to determine the sign of the temperature, which is the same as that of the variation of the mass with respect to the horizon radius. One can easily realize (see figure \ref{signT}) that the sign of the temperature depends just on the direction of the change of sign of the function $f$ across the horizon.
As we decrease $r$, if this sign changes from $f<0$ to $f>0$ as, for instance, in the cosmological horizon of any asymptotically dS spacetime, the temperature is negative, whereas it is positive in the opposite case. This is, of course, consistent with temperature being basically the radial derivative of $f$ \reef{Temp}. The black hole horizon, as the largest root of (\ref{rheq}), always corresponds to the latter case, separating an untrapped region ($f>0$) from a trapped one ($f<0$), and as such it always has positive temperature. The inner horizons have alternating signs for the temperature starting from a negative one. Degenerate horizons obviously have zero temperature corresponding to extremal black holes. 

All formul\ae\ presented so far in this chapter are also trivially valid for inner and cosmological horizons as we just used the fact that $f(r_+)=0$. This will also hold for charged solutions \cite{Castro2012a,Castro2013}. Other properties that recently got some attention, \eg the product of inner and outer entropies being independent of the mass, do not have this universal character. It is easy to see that the latter property follows from the simple form of the thermodynamic variables in Einstein-Hilbert gravity and will not hold in general. If we take a derivative with respect to the mass of a product of two horizons from the same black hole, thus corresponding to the same mass, charge, etc. we get
\be
\frac{1}{S_+S_-}\frac{\partial}{\partial M}S_+S_-=\frac{1}{T_+S_+}+\frac{1}{T_-S_-}
\ee
The product of entropies will be independent of the mass when the above derivative vanishes, \ie $T_+S_++T_-S_-=0$, and this will happen in very specific cases within higher curvature theories.  In general we will be just interested in the outermost horizon and we will keep for it the name of $r_+$. 

\subsection{Vacuum horizons and Einstein-Hilbert gravity}

Let us start this subsection by discussing the horizon structure of the vacuum solutions. The general form of the metric function $f$ is in this case
\begin{equation}
f(r)=\sigma-\Lambda\,r^2 ~,
\end{equation}
so it can vanish at $r=\sqrt{\sigma/\Lambda}$, whenever $\sigma$ and $\Lambda$ have the same sign, thus for hyperbolic AdS and spherical dS spacetimes. These horizons are observer dependent features since these spacetimes are maximally symmetric and, thus, any point can be considered as the origin. The dS case is widely known \cite{Gibbons1977a}, this corresponding to the cosmological event horizon.

The AdS case is, however, more obscure as long as the horizon is actually cloaking a finite size region in a similar way as a regular black hole horizon does. The black hole horizon is actually just a deformation of this `vacuum' horizon. This has a problematic interpretation and has led to the proposal that the true ground state for hyperbolic spacetimes is not the massless one, but an extremal negative mass solution \cite{Vanzo1997a, Birmingham1999}. The cosmological horizon of pure dS spacetime has negative temperature, while for the AdS case the temperature is positive as for a regular black hole horizon.  

\begin{figure}
\centering
\includegraphics[width=0.58\textwidth]{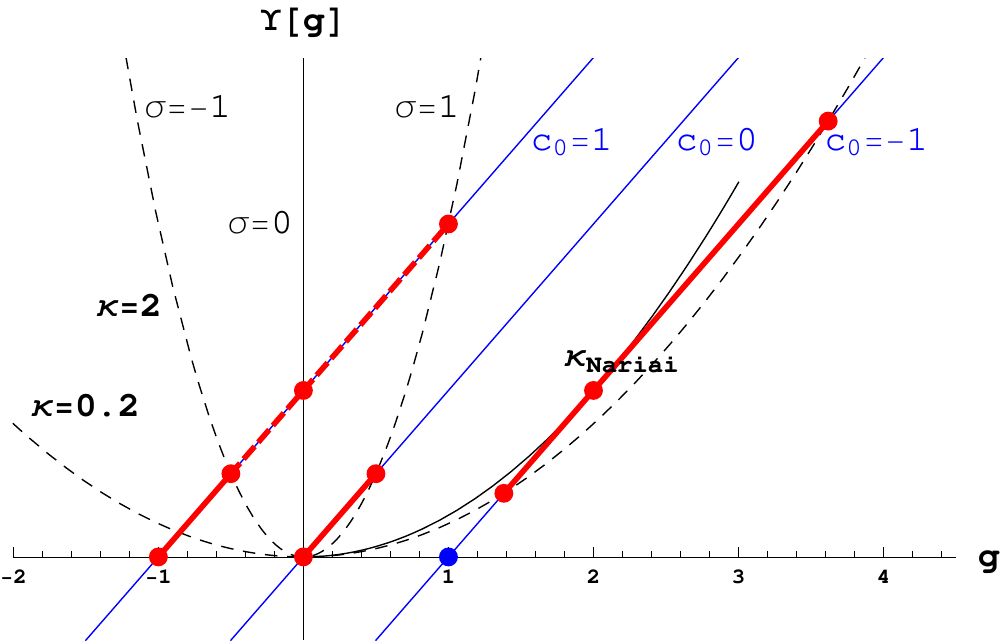} 
\caption{Linear polynomial corresponding to the usual EH branch for negative ($c_0=1$), zero ($c_0=0$) and positive ($c_0=-1$) cosmological constants ($L=1$). The dashed lines are just $\kappa\left(g/\sigma\right)^{\frac{d-1}{2}}$ for $d=5$. The solid black line corresponds to the critical value of the mass, $\kappa_{\rm Nariai}=1/4$, for dS black holes with spherical horizon. The crossing of these lines with the polynomial gives the possible values for $g$ at the horizon and then of $r_+$ (and $r_c$). For $\sigma=1$ and $\kappa>\kappa_{\rm Nariai}$ (or $r_+>r_{\rm Nariai}$), the asymptotically dS branch describes a big crunch spacetime ($f<0$, $\forall r$) without horizons.}
\label{horizonsEH}
\end{figure}

In order to analyze the horizon structure, let us focus on the asymptotically AdS, dS and flat black holes in Einstein-Hilbert gravity with cosmological constant \cite{Birmingham1999}. We include this simple case here for completeness and as an illustration of our method. As clearly depicted in figure \ref{horizonsEH}, the only case accepting all three distinct topologies without exhibiting naked singularities is the asymptotically AdS configuration, the other two cases being well-defined just for spherical topology.
This AdS case, furthermore, has just one horizon for all three topologies. The asymptotically flat spherical black hole has one event horizon as well. The asymptotically dS spherical black hole has in general two horizons: One of them is just the deformation of the cosmological horizon already present in the maximally symmetric solution, while the other corresponds to the black hole. As the mass increases both horizons get closer to each other until, for some critical value of the mass, the so-called {\it Nariai mass},
\begin{equation}
\kappa_{\rm Nariai} = \frac{2L^{d-3}}{d-1}\left(\frac{d-3}{d-1}\right)^{\frac{d-3}{2}} ~,
\end{equation} 
they actually meet (they disappear for masses above that value). The untrapped region, the spacetime as we usually consider it, is comprised between the two horizons and so for this extremal case it seems to disappear. A proper limiting procedure \cite{Ginsparg1983} shows that the geometry remains perfectly regular as $\kappa \rightarrow \kappa_{\rm Nariai}$, and becomes the geometry of the Nariai solution. This space is the direct product of a dS$_2$ and a S$^{d-2}$, both with the same radius. Above this critical mass, though, it describes a {\it big crunch} spacetime.

\begin{figure}
\centering
\includegraphics[width=0.43\textwidth]{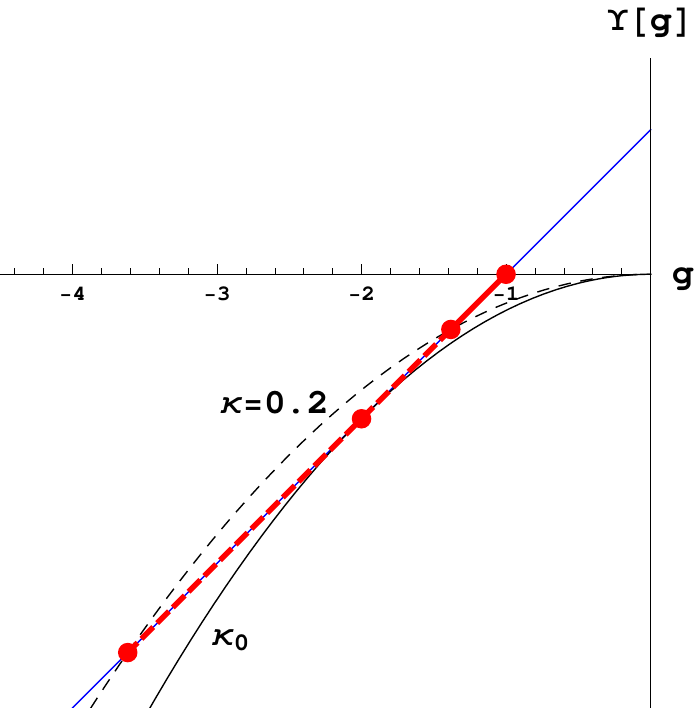} 
\caption{Negative mass hyperbolic black holes in Einstein-Hilbert gravity. The dashed line corresponds to a black hole with an outer and inner horizon (segment in red), while the solid line represents the extremal case, $\kappa_0=-\kappa_{\rm Nariai}$.}
\label{negM}
\end{figure}

In the asymptotically AdS case, a negative mass extremal hyperbolic black hole has been proposed as the ground state in Einstein-Hilbert gravity. The same would apply to any Lovelock theory. Black holes with larger (but negative) mass than this extremal one,
\begin{equation}
\kappa_0=-\kappa_{\rm Nariai}=-\frac{2L^{d-3}}{d-1}\left(\frac{d-3}{d-1}\right)^{\frac{d-3}{d-1}} ~,
\end{equation}
have two horizons, in a way reminiscent of the asymptotically dS spherical black hole with positive mass. The difference being that in the asymptotically dS case the two correspond respectively to the cosmological and black hole horizons, while in the AdS case they are the outer and inner horizons of a black hole (see figure \ref{negM}).
It is worth noticing that for negative masses we are exploring a completely different section of the polynomial than for positive values, the similarities being just due to the extremely simple form of the polynomial in the case under current analysis. In general, positive and negative mass black holes may have dramatically different features.

As we can easily see in this simple example, the existence of just one black hole horizon in all positive mass cases implies that the singularity at the origin is always spacelike, while it is timelike in the negative mass hyperbolic case due to the presence of two black hole horizons that become degenerate in the extremal limit. In the case where the horizon and the singularity coincide, the nature of the latter is null.

\subsection{Black hole entropy at extremality}

As we have seen in the last section, the {\it vacuum} ($M=0$) state with hyperbolic topology has nonzero temperature as it displays a horizon. This is related to the fact that an accelerated observer in AdS would see a horizon with temperature related to its acceleration, in a similar manner as for the Unruh effect in Minkowski space. This makes it difficult to assume this state as groundstate, it cannot be considered at arbitrary temperature as it already has one. The alternative is to consider an extremal state instead. For that we have to make some comments on the special properties of extremal black holes. 

There has been some debate (see \cite{Carroll2009, Horowitz1996g} for some recent discussion) about whether the entropy of extremal black holes corresponds to the usual Bekenstein-Hawking (or Wald in more general cases) or it is simply zero as it seems to indicate the semiclassical calculation \cite{Teitelboim1995, Gibbons1995, Hawking1995}.  Both approaches yield the same result in the non-extremal case, the reason for this discrepancy at extremality being the qualitative difference in the near horizon topology in both cases. In the Euclidean section, it has the topology of $\mathbb{R}^2\times \Sigma_{d-2}$ for non-extremal black holes whereas it is $\mathbb{R}\times S^1\times \Sigma_{d-2}$ in the extremal case, effectively removing the horizon from the geometry. In the first case regularity of the flat factor forces the Euclidean time to be identified such that the period is the inverse of the temperature. This introduces a non-trivial temperature dependence on the solution and thus in the Euclidean on-shell action, $\widehat{\mathcal{I}}$, that it is at the origin of the entropy in this semiclassical picture. The entropy is defined as 
\begin{equation}
S=\left(\beta\frac{\partial}{\partial\beta}-1\right)\widehat{\mathcal{I}} ~.
\end{equation}
In the extremal case the geometry is regular independently of the period $\beta$ and the solution does not depend on this parameter either. The action depends on $\beta$ just because of the integration of the volume of the time circle, $\widehat{\mathcal{I}}$ being just proportional to $\beta$.  The entropy thus vanishes as it can also be explicitly seen from the on-shell action computation. The na\"ive result is 
\begin{equation}
\widehat{\mathcal{I}}=\beta \left(M-\tilde{T}S\right)
\end{equation}
where $\tilde{T}=f'(r_+)/4\pi$ is the usual expression for the black hole temperature. In the non-extremal case, the regularity of the solution at the horizon requires $\beta=\tilde{T}^{-1}$ in such a way that the $\beta$-dependence of the second term cancels and we get
\begin{equation}
\widehat{\mathcal{I}}=\beta M-S
\end{equation}
as expected. The extremal black hole is different as $\beta$ and $\tilde{T}$ are unrelated in that case. Actually $\tilde{T}=0$ yielding $\widehat{\mathcal{I}}^e=\beta M^e$. The entropy would then vanish for the extremal black hole. 

Some other approaches yield the same form of the Bekenstein or Wald entropy but they usually rely on extremal limits of near-extremal solutions, whereas in our case we have assumed that the extremality condition holds {\`a priori}. In case we want to consider any extremal state as ground state for a sector of the theory we will need it to be identified with arbitrary periodicity in Euclidean time. This cannot be done as we take the extremal limit of near-extremal solutions as in that case the temperature would be bound to vanish. We may however include the {\it thermal} extremal states at any temperature with zero entropy as,  as we already explained, in that case the periodicity is not fixed. In what follows we will in general include extremal states in this way even though we will make some comments on the alternative situation in which they are not considered. As we will see the latter case would have a much more problematic interpretation. In particular, the introduction of extremal states as groundstate will avoid the presence of negative entropy states, at least as globally preferred phases, as the free energy of any such state would be bigger than the extremal one
\be
F^e=M^e<M<F=M-TS
\ee
This also happen for the other situations where negative entropies appear. In the case of type (b) spherical branches the analogous r\^ole of the extremal state is played by the vacuum.

\section{Taxonomy of Lovelock black holes}

We will study generic features of maximally symmetric Lovelock black holes in a case by case basis, considering the previously introduced classes of black hole branches (table \ref{cases}). Some work in this direction has already been done considering just the LGB case \cite{Torii2005,Torii2005a}. Let us have a look on the different cases that we can encounter.

\subsection{The Einstein-Hilbert branch}

The Einstein-Hilbert branch is just a deformation of the Schwarzschild-AdS black hole ($c_0=L^{-2}$) and can be identified as the branch crossing $g=0$ with slope $\Upsilon'[0]=1$, exactly as in the Einstein-Hilbert case, and so the slope will be positive for the whole branch. This condition protects this branch from Boulware-Deser-like instabilities. When real, the cosmological constant associated with this branch is negative and so the spacetime is asymptotically AdS. We can proceed with this analysis in an analogous way for asymptotically dS spaces, just by changing the sign of the explicit cosmological constant in the action $c_0\rightarrow -\frac{1}{L^2}$, or for asymptotically flat ones, just by setting $c_0=0$. We include the relevant part of table \ref{cases} above, for the reader convenience.

\begin{table*}
\centering
\begin{tabular}{c||c|c|c|}
asympt. & $\sigma=-1$ & $\sigma=0$ & $\sigma=1$ \\
\hline
\raisebox{9.3ex}{EH}  & \includegraphics[width=0.26\textwidth]{EHh} & \includegraphics[width=0.26\textwidth]{EHf} & \includegraphics[width=0.26\textwidth]{EHs} \\
\hline
\end{tabular}
\caption{Taxonomy of the EH branch black holes.}
\label{EH-case}
\end{table*}

Even though the EH branch is just a deformation of the usual Schwarzschild-AdS black hole, it can be a quite dramatic one. For instance, it may happen that the polynomial has a minimum at $g_{\rm min} < 0$ (if there are several, $g_{\rm min}$ refers to the lowest one in absolute value), such that $\Upsilon[g_{\rm min}] > 0$. A naked singularity would arise for large radius: the solution does not approach AdS asymptotically. This case was first discussed in \cite{Boer2009a,Camanho2010a} for third order Lovelock theory, but the same applies in the general case for a vast region of the space of parameters that will be named, following the aforementioned reference, the {\it excluded region}. In order to avoid the excluded region the value of $\Upsilon[g_{\rm min}]$ at the biggest negative minimum, $g_{\rm min}$, has to be negative. Notice that this does not depend on the topology of the solution, the excluded region being the same for all of them. The sector of the parameter space where this new kind of singularity appears has to be excluded in general, not only because of its nakedness but in reason of the perturbative instability of the corresponding solution \cite{Camanho2013a}, as we will see in section \ref{CCH}. 

For {\it hyperbolic} or {\it planar} topology, as this branch always crosses $g=0$ with positive slope, $\Upsilon'[0]=1$, it has always a horizon hiding the singularity of the geometry that is located (see table \ref{EH-case})
\begin{itemize}
\item
either at $r=0$ [(a) type],\vskip2mm
\item
or at the value corresponding to a maximum of $\Upsilon[g]$ [(b) type].
\end{itemize}
For hyperbolic horizons we have again the possibility of considering negative mass black holes, for masses above a critical value corresponding to the extremal case. This makes no difference with respect to the same situation taking place in an AdS branch and, thus, will be discussed at length below.

The {\it spherical} case is quite more involved. For high enough mass, the existence of the horizon is ensured, but this is not the case in general. For the (a) type EH branch the existence of the horizon can be elucidated by analyzing (\ref{rheq}) in the limit of small mass $g_+\rightarrow\infty$,
\be
\Upsilon[g_+]\approx c_K g_+^K\approx \kappa g_+^{\frac{d-1}{2}}
\ee
For the horizon to exist in this limit, we need the right hand side of the equation to be bigger than the left hand side. This is ensured for $d>2K+1$ as in this case the biggest power in the left hand side would be smaller than $(d-1)/2$. The existence of a horizon in the small mass limit ensures the existence of (at least) one for all values of the mass, simply due to the continuity of $\Upsilon[g]$.

The case $d=2K+1$ is {\it critical}. There will be a minimal mass ($\kappa_{\rm crit} \leq c_K$) below which a naked singularity appears. In principle, for high enough orders of the Lovelock polynomial, more than one horizon can exist but for the critical case, at some point, all of them disappear. The number of black hole horizons determines the type of singularity situated at $r=0$, space or timelike. For $d>2K+1$ we will always have an odd number of horizons (taking into account possible degeneracies), since the (spacelike) singularity is in the trapped region of the spacetime. For $d=2K+1$, the number of horizons depends on the value of the mass. For masses above $c_K$ the number is odd  and at least one horizon will always exist, whereas for masses below this critical mass the number of horizons will change to an even quantity, and will actually disappear at some point. In any case the minimal mass horizonful solution always corresponds to a zero temperature state with a gap to the actual vacuum. This is similar to what happens for quasi-topological black holes with the difference of the number of horizons being always even. The zero temperature state is reached as two of these horizons merge in this case.

The case where the EH branch ends up at a maximum for some positive value of $g=g_\star$ ($r=r_\star$), or (b) type branch, is even simpler. There is a critical value of the mass for which $r_+=r_\star$. Below that mass a naked singularity appears. This is very similar to the situation described in the previous paragraph, the only difference being that in that case the radius of the singularity is zero, $r_\star=0$. Also, for type (b) black holes, the temperature diverges as we approach the radius of the singularity. 

\begin{figure}
\centering
\includegraphics[width=0.41\textwidth]{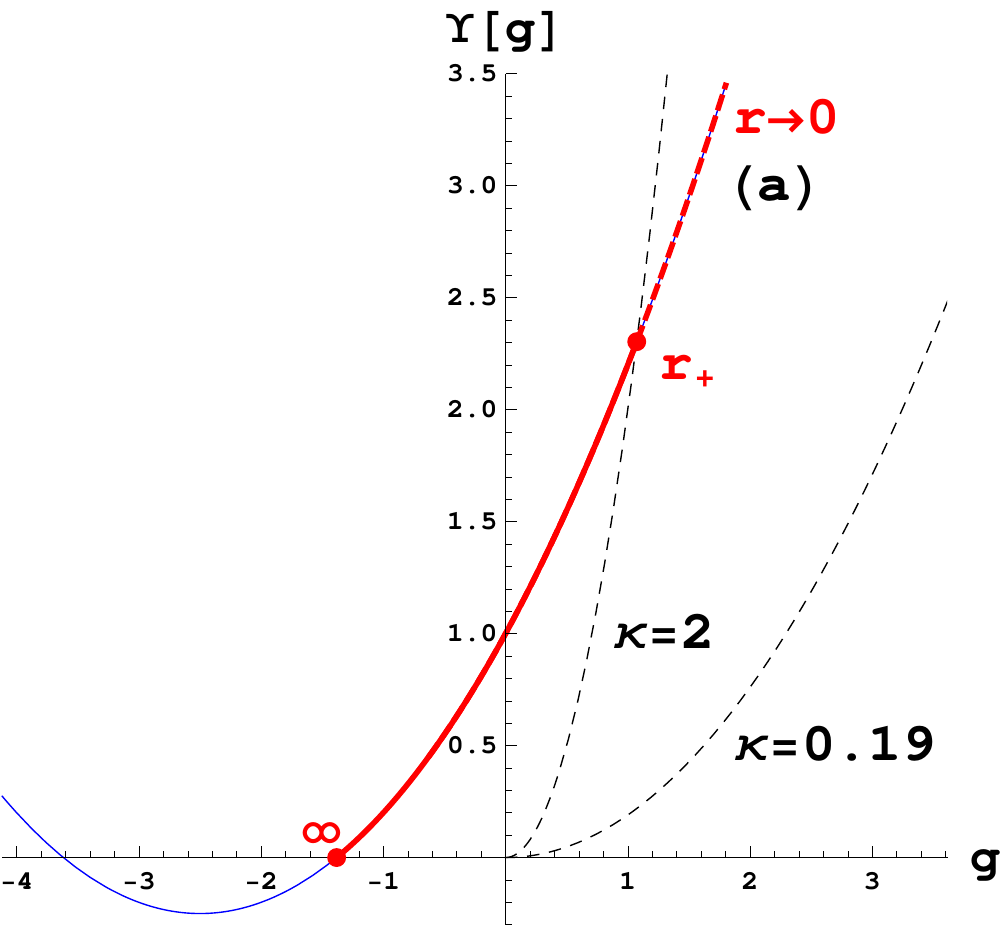} ~~\includegraphics[width=0.41\textwidth]{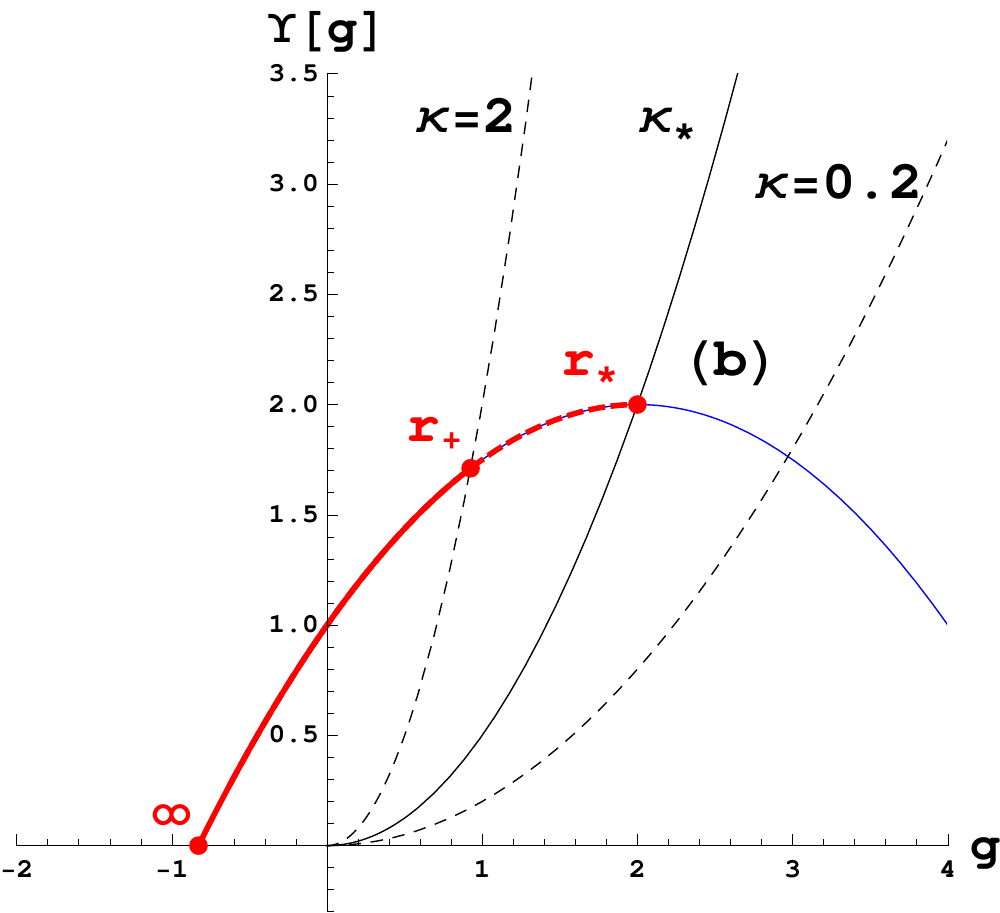} 
\caption{EH branch in GB gravity  in 5 dimensions for $\lambda=1/5$ and $\lambda=-1/4$ respectively ($L=1$). The dashed curves correspond to the $\kappa\left( g/\sigma \right)^{\frac{d-1}{2}}$ for spherical topology ($\sigma=1$). The singularity becomes naked for $\kappa\leq \lambda=0.2$ in the first case and for $\kappa\leq \kappa_{\star}=0.5$ in the second case. For higher dimensions, the second figure would be qualitatively the same with just a different value of the critical mass. The first one, instead, changes as long as the horizon exists for all positive values of $\kappa$ in that case. For the (b) type branches the singularity is always located at $r_{\star}=\sqrt{-2\lambda}\,L$.}
\label{relevantGB}
\end{figure}

The simplest example of Lovelock theory is GB gravity \cite{Cai2002}, where we have just two branches, one of them suffering from BD instabilities. The remaining branch is thus an EH branch. For $\lambda>0$ this branch is of the (a) type, extending all the way up to a singularity situated at $r=0$. For $\lambda<0$, instead, this branch has a maximum at positive values of $g$ (see figure \ref{relevantGB}).
This is a singularity at a finite value of $r$ that may or may not be naked depending on the value of the mass. The mass for which the horizon coincides with the singularity is
\begin{equation}
\kappa_{\star}=\frac{1}{2} (1-4\lambda) (-2\lambda L^2)^{\frac{d-3}{2}} ~.
\end{equation}
For bigger masses we have a well-defined horizon while below this bound the singularity becomes naked.

Another intriguing possibility, that cannot be observed within the simple setting of GB gravity, is the would be appearance of several black hole horizons. For this to happen we need inflection points in $\Upsilon[g]$, and so the minimal example would be the cubic Lovelock theory. In the critical $d=7$ case, we have the possibility of obtaining two black hole horizons for some regions of the space of parameters, while this number is three in higher dimensions. One remarkable thing worth noticing here is that, as couples of horizons appear or disappear when we vary the value of the mass, the value of (the biggest horizon) $r_+$ may change discontinuously. Thus, the temperature as a function of the mass also varies discontinuously when crossing the values of $\kappa$ for which the outermost couple of horizons appear or disappear. One of the sides of the discontinuity has zero temperature (since the black hole is extremal for such critical mass) while the other has finite temperature. We must recall that the inner horizons are in general unstable \cite{Matzner1979,Poisson1990,Brady1995,Dafermos2003} and this possibly means that we should not trust our solution behind the outermost inner horizon. We may interpret these extremal states as black hole ground states, each one for a given range of masses. These seem naively accessible by evaporation and, thus, point towards a violation of the third law of black hole dynamics \cite{Torii2006}. They are in general unstable solutions \cite{Anderson1995a,Marolf2010}, though (see also section \ref{extremalBH}). At low temperatures, this particular branch will have several possible black hole masses and transitions might occur among them and the thermal vacuum. As for temperatures close to zero the free energy essentially coincides with the mass, the globally preferred solution is the less massive one: the vacuum. We will comment more on this later on.

\subsection{AdS (other than EH) branches}

The second class of branches that we describe in what follows are asymptotically AdS black holes different from the EH branch. The latter will be included just for the discussion of negative mass solutions since the analysis is exactly the same. 

Consider first the positive mass solutions, for which the AdS branches always end at a maximum of the polynomial (see table \ref{AdS-case}).
\begin{table*}
\centering
\begin{tabular}{c||c|c|c|}
asympt. & $\sigma=-1$ & $\sigma=0$ & $\sigma=1$ \\
\hline
\raisebox{9.3ex}{AdS}  & \includegraphics[width=0.26\textwidth]{AdSh2} & \includegraphics[width=0.26\textwidth]{AdSf} & \includegraphics[width=0.26\textwidth]{AdSs} \\
\hline
\end{tabular}
\caption{Taxonomy of the AdS (other than EH) branches.}
\label{AdS-case}
\end{table*}
As before, the existence of a horizon cloaking the singularity fixes the topology of such branches: As the considered sections of the polynomial run over negative values of $g$, horizons exist just for $\sigma=-1$. On the other cases, the solutions describe a spacetime with a timelike naked singularity. The condition for the existence of a horizon sets an upper bound on the mass, $\kappa<\kappa_{max}$, $\kappa_{max}$ corresponding to the critical value for which the radius of the horizon coincide with that of the singularity ($r_+=r_{max}$), {\it i.e.},
\begin{equation}
\kappa_{max}=r_{max}^{d-1}\;\Upsilon[g_{max}]~.
\end{equation}

When we encountered a naked singularity with positive mass in the Einstein-Hilbert theory, it corresponded to the low mass limit of a multi-horizon black hole (below a given mass, the two horizons merge and disappear altogether, leaving the singularity naked). The case here is somehow different as the black hole horizon cannot degenerate as we increase the mass approaching the critical value. Thereby, the solution infinitesimally close to the critical one has nonzero temperature, diverging as we approach the bound. This is more reminiscent of the low mass limit of Schwarzschild black holes (ultimately leading to a regular geometry) than of the usual naked singularities in Einstein-Hilbert gravity. It is also similar to what happens for type (b) spherical black holes.

\begin{figure}
\centering
 \includegraphics[width=0.55\textwidth]{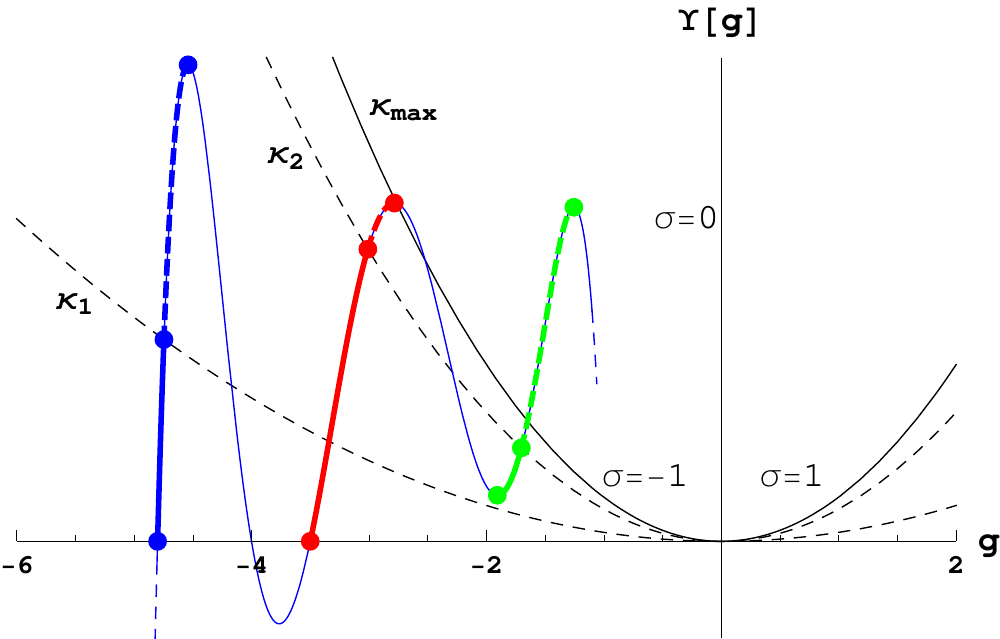}
\caption{Qualitative representation of (positive mass) AdS branches. The blue and red branches will be referred to as type (b1) and (b2) respectively, when considered for negative masses. The green one is an excluded branch. There are asymptotically AdS massive black holes for $\kappa<\kappa_{max}$ (red branch), otherwise the geometry displays a naked singularity.}
\label{horizonsAdS}
\end{figure}

For negative mass solutions, the analysis of the existence of black hole horizons and its number is more involved. The main qualitative feature is the possibility of having a minimum of the polynomial associated to the branch under analysis (see blue and red branches in figure \ref{horizonsAdS} for instance).
We will refer to the case without such minimum (blue branch) as type (b1) solution and as type (b2) for the other one (red branch). The structure of horizons and the type of singularity will differ in both types of branches. 

\begin{figure}
\centering
\includegraphics[width=0.41\textwidth]{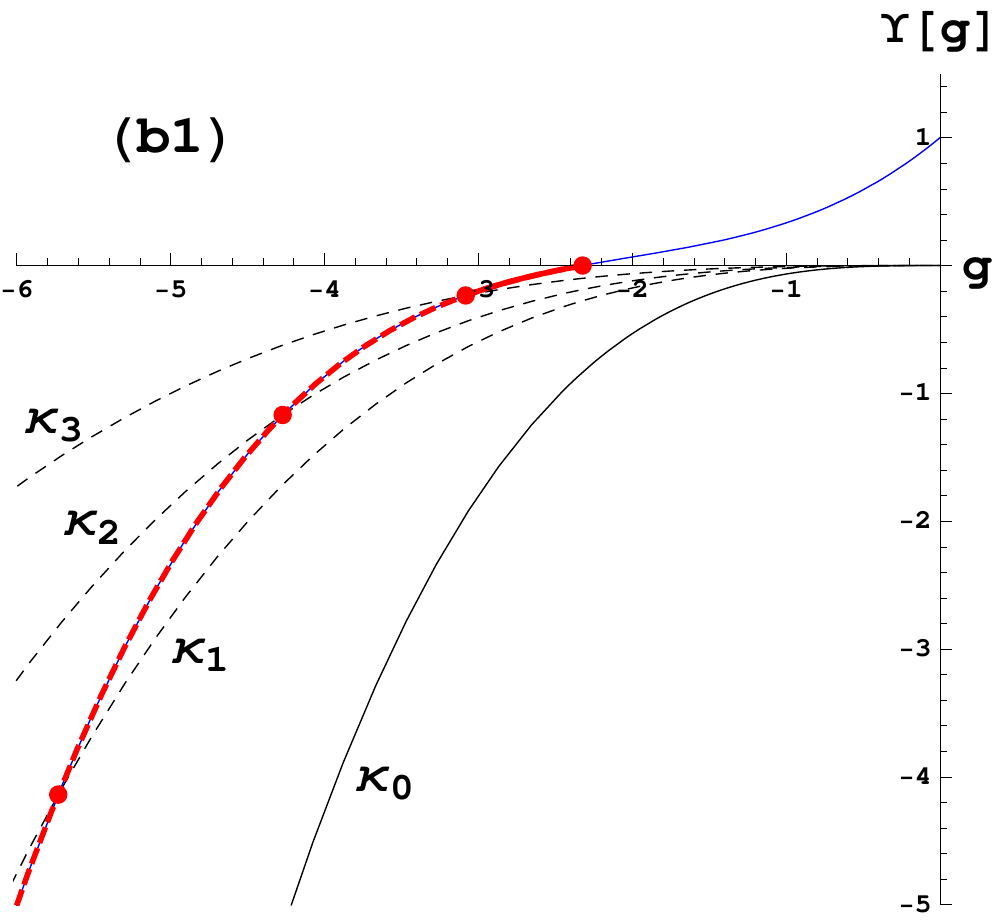} 
\caption{EH branch in the cubic theory in 7 dimensions for $\lambda=0.4$ and $\mu=0.2$ ($L=1$). The dashed curves correspond to $\kappa\left( g_+/\sigma \right)^{\frac{d-1}{2}}$ with $\kappa_0=-\mu/3=-2/30$, $\kappa_1=-0.022$, $\kappa_2=\lambda=-0.015$, and $\kappa_3=-0.008$ ($\sigma=-1$). For masses above $\kappa_0$ we have one horizon with no distinction between positive and negative masses. For $\kappa \leq \kappa_0$ there is a naked singularity at $r=0$. No extremal state exists.}
\label{nohyperbolicground}
\end{figure}

When $d=2K+1$ and the branch we are considering is of (b1) type, there is a minimal mass for which, instead of an extremal regular spacetime with a degenerate horizon, we have a naked singularity (see figure \ref{nohyperbolicground}).
The temperature also vanishes asymptotically as we decrease the mass. We will not comment further on these kind of solutions as they are gravitationally unstable against perturbations \cite{Camanho2013b}.  

\begin{figure}
\centering
\includegraphics[width=0.43\textwidth]{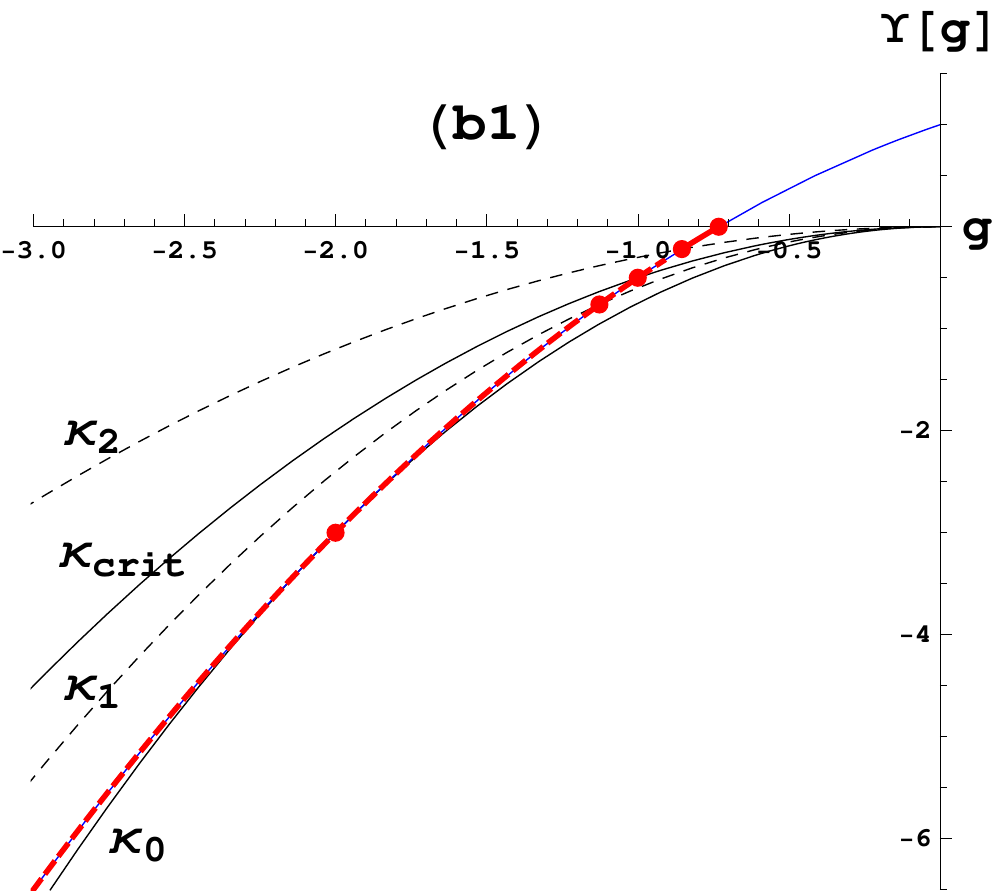} ~~\includegraphics[width=0.43\textwidth]{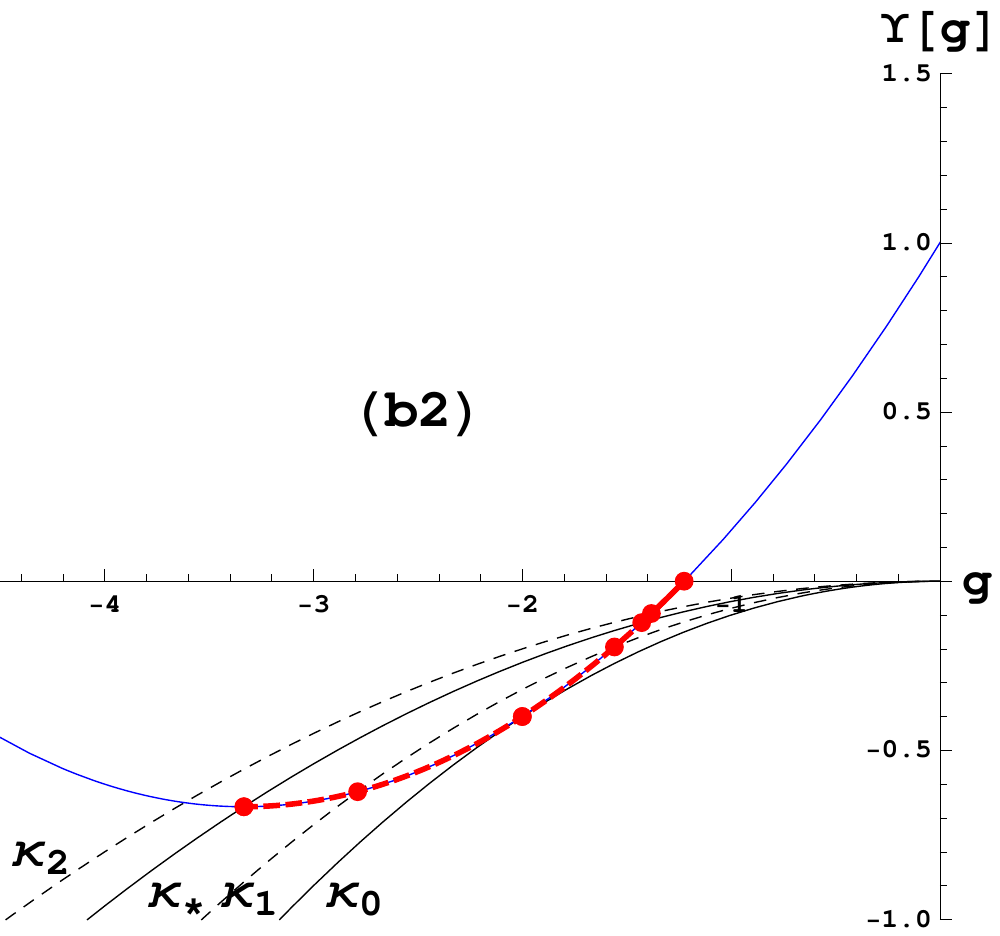} 
\caption{EH branch in GB gravity  in 5 dimensions for $\lambda=-1/2$ and $\lambda=0.15$ respectively ($L=1$). The dashed curves correspond to $\kappa\left( g_+/\sigma \right)^{\frac{d-1}{2}}$ with $\kappa_0=-0.75$, $\kappa_1=-0.6$, $\kappa_{\rm crit}=\lambda=-0.5$, and $\kappa_2=-0.3$ (left) and $\kappa_0=-0.1$, $\kappa_1=-0.08$, $\kappa_\star=-0.06$ and $\kappa_2=-0.05$ (right) ($\sigma=-1$). In both cases we have two horizons for $\kappa_0<\kappa<\kappa_{\star}$ and one for $\kappa_\star<\kappa$. For $\kappa=\kappa_0$ we have a degenerate horizon. The singularity becomes naked for $\kappa\leq \lambda-1/4$ in both cases. In higher dimensions, the behavior is qualitatively the same. For the (b2) type branches the singularity is always located at $r_{\star}=L\sqrt{-2\lambda}$, while it is at the origin in the (b1) case.}
\label{negGB}
\end{figure}

In case we have a well-defined extremal negative mass black hole, we always have at least two horizons for (b1) type branches and $d>2K+1$, as we depart from extremality. For (b2) type branches, however, the inner horizon disappears when its radius coincides with the radius of the singularity, changing its nature from timelike to spacelike, or viceversa. There is a critical mass for which this happens. This is irrelevant for an outside observer who cannot extract information from the inner horizon. Figure \ref{negGB} shows both kinds of solutions in the simplest case of GB gravity, for (b1) type ($\lambda<0$) and (b2) type ($\lambda>0$), respectively.   

Both classes of branches may have several horizons in the presence of inflection points. The (b1) type will always have an even number, except when $\kappa\geq c_K$ where the smallest horizon disappears. Thus, the singularity at $r=0$ is timelike below this critical mass and spacelike above it. The same change of behavior for the singularity appears in the other type of branches with the critical mass being set by the minimum, $\kappa_{\star}$. In the same way as described for spherical black holes in the EH branch, the couples of horizons appearing or disappearing as a function of the mass translate into discontinuous changes on the temperature. The possibility of having several extremal solutions in one branch amounts to several ground states for different ranges of (negative) masses, with possible transitions among them. At zero temperature, though, one does not have a thermal vacuum to compare with, contrary to what happens at finite temperature. The negative mass black holes are the preferred phase in that regime, namely the lowest mass one among them.

\subsection{dS branches}

The existence of event horizons will again set a series of constraints both on the admitted topologies as well as in the possible values for the mass parameter. For hyperbolic or flat horizons, the metric function $f$ is always negative and the solution describes a big crunch. For $\sigma=1$ the solution may still describe a big crunch if we consider a high enough mass, above the Nariai mass. Slightly below it, at least two horizons exist, the biggest one being the cosmological horizon. Notice that there may be several Nariai masses.
\begin{table*}
\centering
\begin{tabular}{c||c|c|c|}
asympt. & $\sigma=-1$ & $\sigma=0$ & $\sigma=1$ \\
\hline
\raisebox{9.3ex}{dS}  & \includegraphics[width=0.26\textwidth]{dSh} & \includegraphics[width=0.26\textwidth]{dSf} & \includegraphics[width=0.26\textwidth]{dSs} \\
\hline
\end{tabular}
\caption{Taxonomy of the dS branches.}
\label{dS-case}
\end{table*}

\begin{figure}
\centering
 \includegraphics[width=0.53\textwidth]{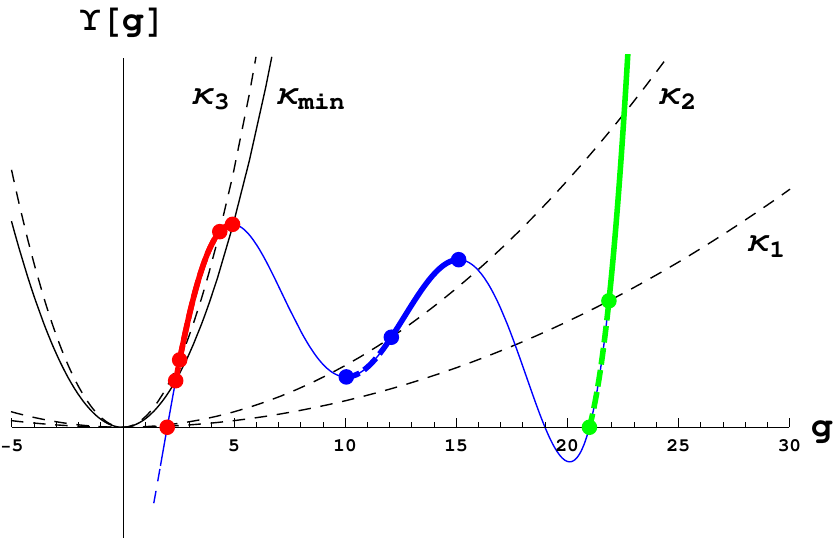} \qquad  \includegraphics[width=0.28\textwidth]{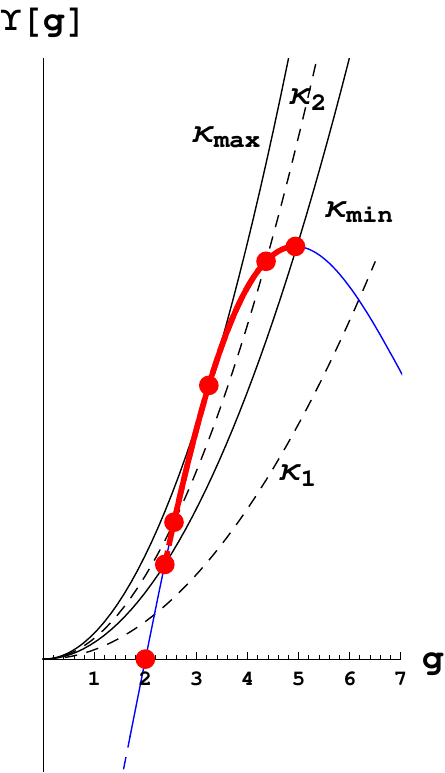}
\caption{Generic features of dS branches are captured in these figures. The different sections of the polynomial qualitatively represent (b) type (in red), excluded (in blue) and (a) type (in green) dS branches. The right figure zooms on the (b) type dS branch showing that for $\kappa_2$, $\kappa_{\rm min} < \kappa_2 < \kappa_{\rm max}$, there are asymptotically dS massive black holes, with outer and inner horizons. At $\kappa_{\max}$ the black hole becomes extremal. For $\kappa$ not in this range, the red branch displays a naked singularity. Even though the (a) type branch does not seem to have a black hole horizon (in the left figure, {\it e.g.}, the green dots correspond to the cosmological horizon), it always exists for $d>2K+1$ and, for some range of masses, also in the $2K+1$ dimensional case.}
\label{horizonsdS}
\end{figure}

Like the EH branch, the dS branches can end up at a maximum, (b) type, or extend all the way to $r=0$ (a) type; red and green branches on figure \ref{horizonsdS} respectively.
For (b) type branches, there is a critical mass, $\kappa_{\rm min}$, for which the outermost black hole radius coincides with the radius of the singularity, $r_+=r_\star$. Below that mass, the horizon disappears. A naked singularity will always show up for sufficiently low masses.

For (a) type branches, the same happens if $d=2K+1$ just replacing the critical mass by $\kappa_{\star}\equiv c_K$. Again, we cannot avoid naked singularities as we approach arbitrarily low masses. Therefore, the existence of horizons sets, in this case, two bounds for the mass. Above the upper bound the geometry displays a spacelike singularity, whereas below the lower bound it has a naked timelike singularity. For $d>2K+1$, instead, the black hole horizon always exists, all the way down to zero mass. Thus, solely the upper Nariai bound for the mass exists.  

Another situation that we may encounter, for $d=2K+1$, is the absence of a Nariai solution. Below a certain critical mass we are faced with a naked singularity, the only remaining horizon being the cosmological one. This is the symmetric situation to the non-existence of extremal negative mass black hole discussed earlier for AdS branches. Mirroring that case, here we also find solutions with a large number of horizons. Their variation, as a function of $\kappa$, may translate into discontinuities in the temperature as a function of the mass. The discussion regarding how to interpret this phenomenon is the same as before.

\section{Heat capacity and local thermodynamic stability}

The details of the solutions and the behavior of their associated thermodynamic quantities strongly depend on the particular case under consideration. Therefore, a general thermodynamic analysis is cumbersome. However, once we have described the qualitative features of the different branches of solutions, much and very interesting information can be extracted, most of it arising as universal features of these black hole solutions.  

For positive mass black holes, the sign of the heat capacity on a BD stable branch, because of (\ref{dMdr}),  depends just on
\begin{equation}
\frac{dT}{dr_+}=-\frac{g_+}{2\pi}\left[(d-2)-\frac{d-1}{2}\frac{\Upsilon[g_+]}{g_+\Upsilon'[g_+]}\left(1+2g_+\frac{\Upsilon''[g_+]}{\Upsilon'[g_+]}\right)\right] ~,
\label{dTdr}
\end{equation}
where the second term (in brackets) seems to be related to the potential felt by perturbations in the shear channel \cite{Camanho2012}. We did not manage to check classical stability in full generality, but in the regimes of {\it high} and {\it low} masses. This should not be confused with the stability analysis focusing on perturbations of the black hole solutions that will be treated in the next chapter. Earlier relevant works on this include \cite{Dotti2005b,Gleiser2005,Beroiz2007,Takahashi2009h,Takahashi2009b,Takahashi2010e,Takahashi2010d}).

We will just consider solutions possessing event horizons, since they are the only ones with associated thermodynamic variables. Then, taking into account our earlier discussion (see table \ref{cases}), we will have to deal with hyperbolic AdS branches; hyperbolic, flat and spherical EH branches and spherical dS branches, classified in different subclasses. 

\subsection{Black holes in the EH branch}

The simplest case to analyze is that of toroidal or planar black holes in the EH branch. The thermodynamic variables, in this case, do not receive any correction from the higher curvature terms in the action and the expression reduces to the usual formula of Einstein-Hilbert gravity,
\begin{equation}
\frac{dT}{dr_+}=\frac{d-1}{4\pi L^2} ~.
\end{equation}
This expression is positive. Therefore, these black holes are locally thermodynamically stable for all values of the mass. 

For the EH branch, the only one admitting all three topologies, the situation is exactly the same for $\sigma=\pm 1$ in the large mass limit. The value of $g$ at the horizon, $g_+=\sigma/r_+^2$, asymptotically approaches zero from both sides, and the formulas reproduce the planar case. Einstein-Hilbert gravity captures the universal thermodynamic description of Lovelock black holes with large enough mass. It does not capture other features in this regime. For instance, those related to the stability and causality preserving properties of the solutions \cite{Camanho2010a,Boer2009a,Camanho2010d}. EH branch's black holes in this regime are then always stable.

This is also the case for a special class of (maximally degenerated) Lovelock theories whose analysis is not considered in the present article. Those theories admit a single (EH-)branch of black holes \cite{Crisostomo2000}. The results there also coincide with the generic analysis for spherical black holes presented here.

It is worth recalling here that for sufficiently large mass, just the EH branch admits black holes, the other branches describing geometries with naked singularities or big crunch spacetimes.

\begin{figure}
\centering
\includegraphics[width=0.46\textwidth]{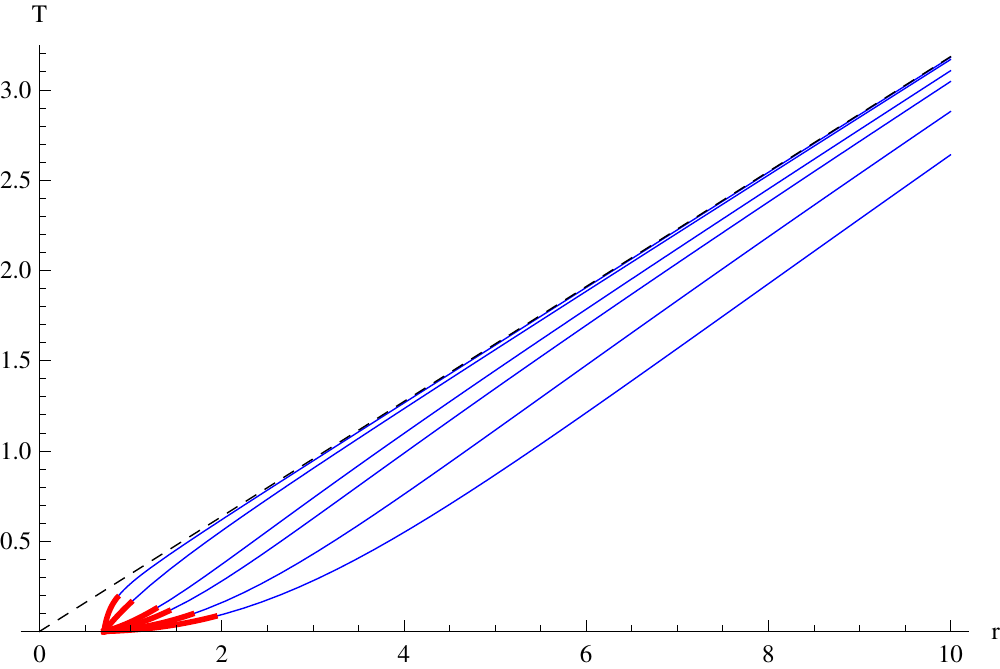}\qquad \includegraphics[width=0.46\textwidth]{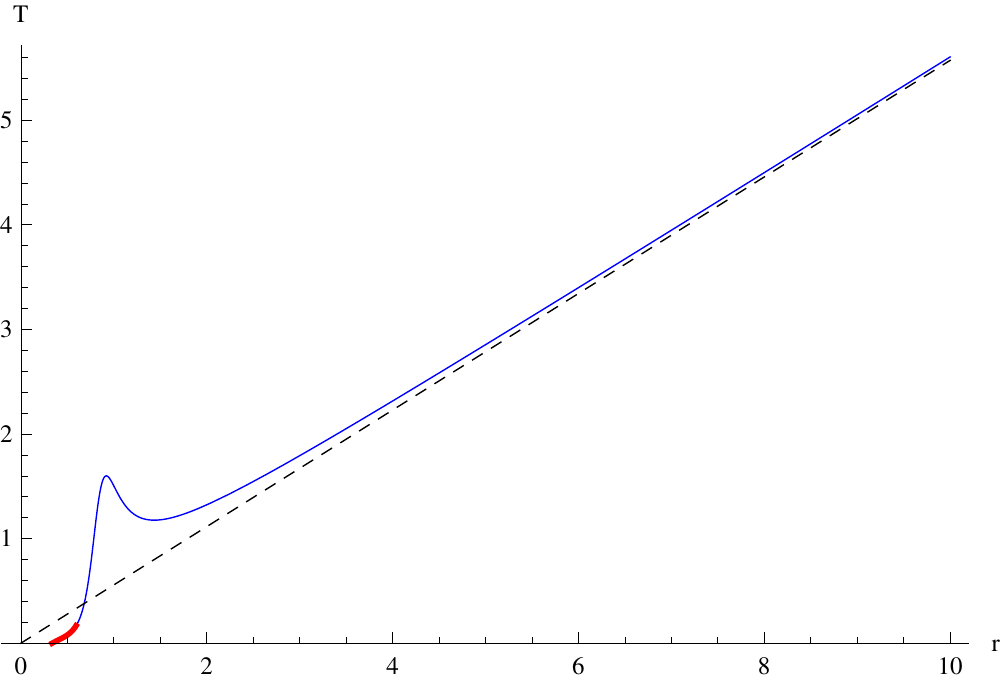} 
\caption{Temperature versus horizon radius (equivalently versus black hole mass, $\kappa$) for hyperbolic black holes ($L=1$). The first figure corresponds to GB gravity in $d=5$ for $\lambda=-10,-5,-2,-1,0,0.2$ (from bottom to top). The behavior is qualitatively the same in higher dimensions. Such black holes are stable for all values of the mass (even negative, in red). The black dashed line corresponds to the planar case to wich all curves asymptote. The second figure corresponds to an example of intermediate unstable phase in cubic Lovelock theory in $d=8$ with $\lambda=0.65$ and $\mu=0.5$.}
\label{stabGBh}
\end{figure}

As we decrease the mass, particular features of the different topologies pop up and we need to consider them separately. Small mass hyperbolic black holes correspond to a smooth deformation of the vacuum. In that regime, the second term inside the brackets of (\ref{dTdr}) becomes negligible, $\Upsilon[g_+] \approx \Upsilon[\Lambda] = 0$, and the expression approaches
\begin{equation}
\frac{dT}{dr_+}\approx \frac{d-2}{2\pi}\,(-g_+) ~.
\end{equation}
Therefore, as $g_+<0$, the low mass hyperbolic black holes are also stable. Notice, though, that even if both extrema of the spectrum for hyperbolic black holes in the EH branch are stable, one may encounter unstable intermediate phases. This is not the case in GB gravity (contrary to what is stated in \cite{Neupane2004}; the negative specific heat found there corresponds to inner horizons and, thus, does not indicate any instability of the system), where we find that hyperbolic black holes are always locally thermodynamically stable, as can be seen in figure \ref{stabGBh} (left), even for negative masses above the extremal one (depicted in red in the figure).
The cubic theory, in turn, as shown in figure \ref{stabGBh} (right), already displays intermediate mass hyperbolic black holes which are locally thermodynamically unstable.

The extremal hyperbolic black hole can be shown to be always stable from this point of view. It has zero temperature and all black holes with higher masses have positive temperature. Thereby, the heat capacity has to be positive, close enough to this state. In some cases, for $d=2K+1$, the extremal negative mass black hole does not exist. In that situation we may consider, in principle, infinitesimally small black holes, $r_+\rightarrow 0$, with temperatures approaching zero asymptotically. The {\it singular} `zero size black hole', again, fixes a bound in the mass, and black holes close to that bound are stable, in the same way as the ones close to the extremal black hole.

For spherical black holes we have to distinguish between different cases. In the (a) type, the small black hole limit, $r_+\rightarrow0$, may correspond respectively to finite or zero mass for $d=2K+1$ and $d>2K+1$. If the branch is of (b) type, there is a lower bound for the mass of the black hole for which the temperature diverges. When $d>2K+1$, in the low mass regime,
\begin{equation}
\frac{dT}{dr_+}\approx -\frac{d-2K-1}{4\pi\,K}\,g_+ < 0 ~.
\end{equation}
Thereby, small spherical black holes in the EH branch are thermodynamically unstable, in exactly the same way as in the usual Einstein-Hilbert gravity. The situation changes for $d=2K+1$, where the temperature vanishes asymptotically. Therefore, small black holes are stable in odd dimension for the highest order Lovelock gravity (in particular, we need $c_K>0$ for type (a) EH or dS branches to exist), whereas they are unstable in all other cases \cite{Cai2004}.

\begin{figure}
\centering
\includegraphics[width=0.45\textwidth]{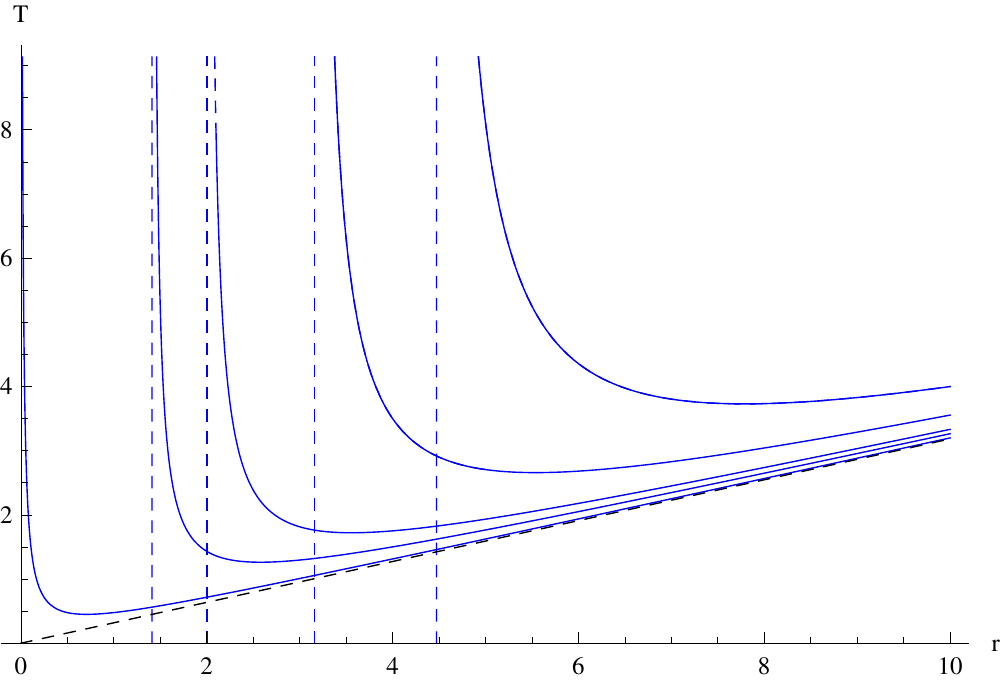}\qquad \includegraphics[width=0.45\textwidth]{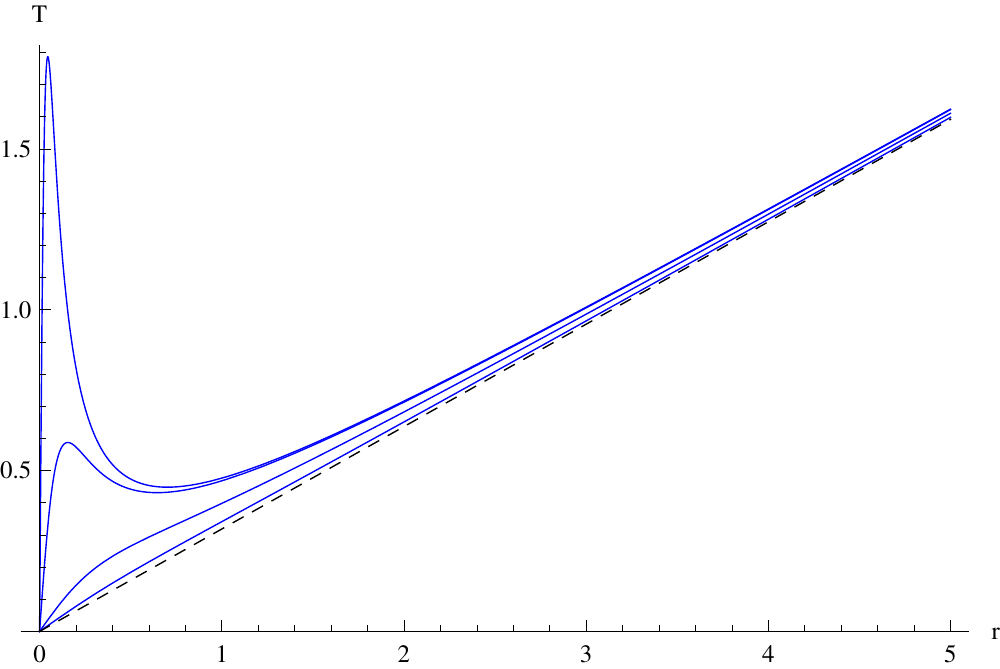} 
\caption{Temperature versus horizon radius (equivalently, mass) for spherical black holes in $d=5$ GB gravity ($L=1$). The first figure corresponds to negative values of the GB coupling, $\lambda=-10,-5,-2,-1,0$ (from top to bottom), whereas the second considers positive values, $\lambda=0.001,0.01,0.1,0.2$ (from top to bottom). The dashed blue lines indicate the value of the mass, $\kappa_{\rm min}$, for which the temperature diverges. We observe the appearance of a new stable phase for small positive values of $\lambda$ and even the disappearance of the unstable region for high enough $\lambda$. This stable region of small black holes disappears in higher dimensions, for all positive values of $\lambda$, the qualitative behavior being similar to the $\lambda=0$ case.}
\label{stabGBs}
\end{figure}

For (b) type branches something similar happens as we approach the minimal mass, $\kappa_{\rm min}$, set by the local maximum of the polynomial $\Upsilon[g]$. At this critical mass the temperature diverges, but in this case as we lower the mass. The heat capacity is negative, as we can also infer from (\ref{Upsmax}), taking into account that $g_+$ has a positive value. Black holes close enough to the minimal mass one are then unstable. This kind of behavior can be seen for instance in GB gravity (see figure \ref{stabGBs}).  

\begin{figure}
\centering
\includegraphics[width=0.47\textwidth]{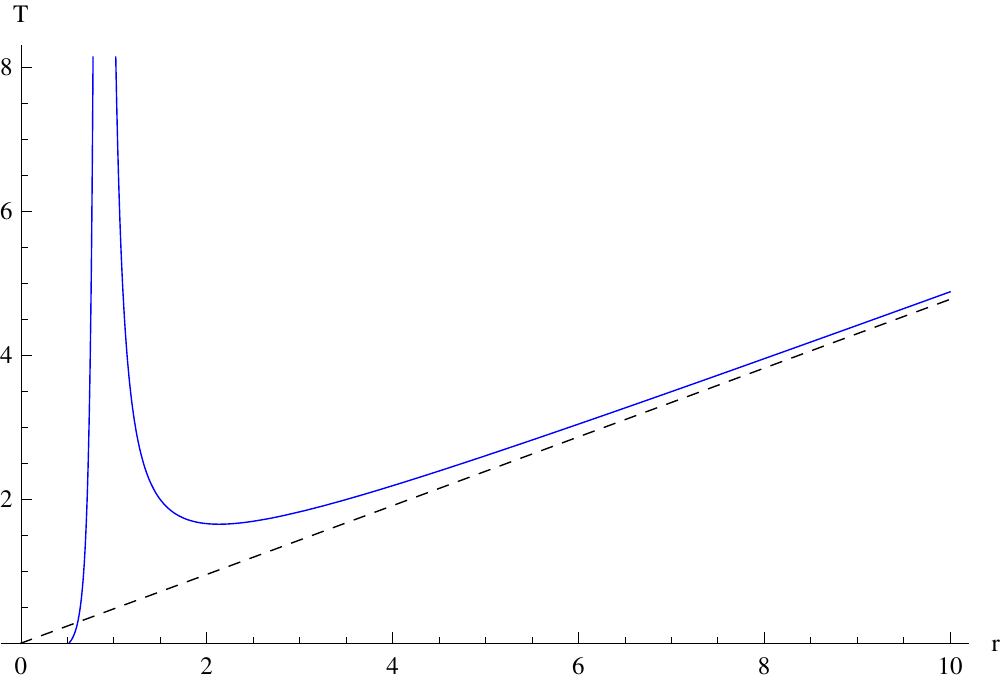}~~ \includegraphics[width=0.47\textwidth]{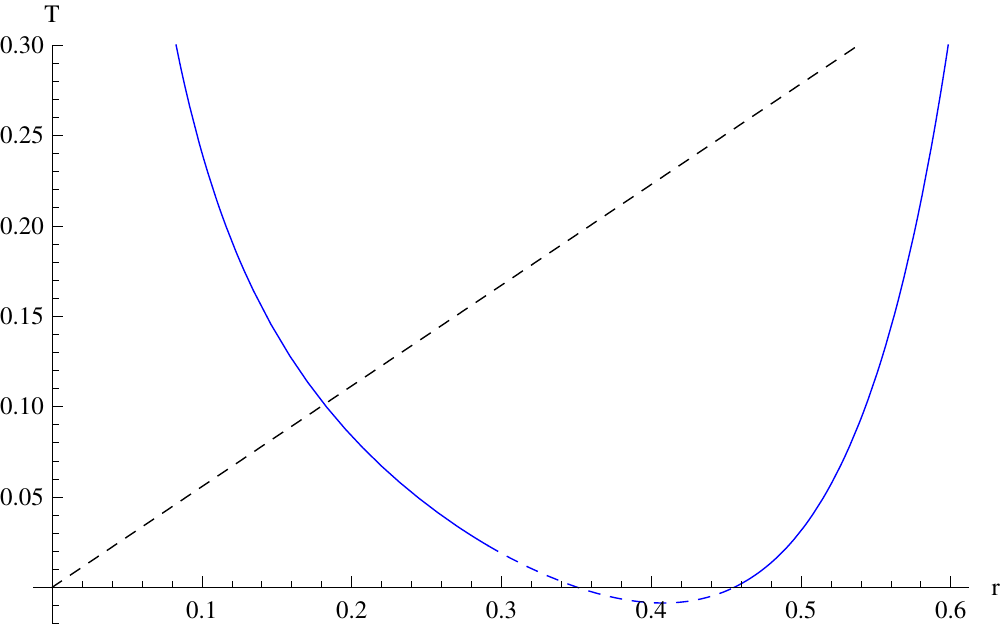} 
\caption{Temperature versus horizon radius (equivalently, mass) for spherical black holes in $d=7$ cubic Lovelock gravity with $\lambda=-0.746$, $\mu=0.56$ and $L=1$ (left). The black hole reaches zero temperature for finite (positive) mass. For $d=8$ (right), the shape of the curve is qualitatively the same except in the low radius region where a new branch of black holes with diverging temperature appears. We just show a zoom of the left bottom corner. We can also verify the existence of the temperature (and radius) jump commented in the main text. The dashed blue line corresponds to inner horizons, one of them becoming outer for masses below the extremal one.}
\label{stable}
\end{figure}

There is one further possibility when the polynomial is such that it allows for more than one black hole horizon. Then, either for (b) type as well as (a) type solutions (in $d=2K+1$ dimensions), we may still have an event horizon cloaking the singularity for some range of masses below the na\"ive $\kappa_{\rm min}$. As we lower the mass further we encounter at least one extremal black hole, which is stable in the same way as the extremal hyperbolic black hole of negative mass.  One such case is shown in figure \ref{stable}, that corresponds to the cubic polynomial plotted in figure \ref{horizons}.
In general, we may also have unstable regions in between the two stable ones, and also for masses below the extremal one if $d > 2K + 1$ (see figure \ref{stable}, right). This situation does not decisively depend on the dimensionality of spacetime. This case has features of the previously discussed spherical black holes but its behavior is similar to the (negative mass) near extremal hyperbolic black holes.

\subsection{Hyperbolic black holes in the AdS branches}

Most of the discussion on hyperbolic black holes in the EH branch also applies, on general grounds, to the AdS branches. The only difference being that, in general, there is a maximal mass for which the temperature diverges. Thus, close enough to that point the heat capacity has necessarily to be positive and the black hole thermodynamically stable. This can be seen directly from (\ref{dTdr}), as the heat capacity close to the maximum approaches
\begin{equation}
\frac{dT}{dr_+}\approx\frac{d-1}{2\pi}\,\frac{g_+ \Upsilon[g_+]\,\Upsilon''[g_+]}{\Upsilon'[g_+]^2} ~,
\label{Upsmax}
\end{equation}
diverging as well when we reach the critical mass. $\Upsilon[g_+]$ is positive due to the positivity of the mass. The second derivative $\Upsilon''[g_+]$ is negative as we are close to a maximum, but $g_+$ is negative as well. The plot of temperature versus horizon radius will be in general qualitatively similar to that corresponding to the EH branch, with the difference that the temperature diverges at some finite value of $r_+$. 

\subsection{Spherical black holes in the dS branches}

The low mass regime of the spherical solutions corresponding to dS branches is very similar to that of the EH branch. The high mass regime, instead, is very different. These black holes may increase their mass until they reach a maximal (so-called {\it Nariai}) mass, which is set by the shape of the polynomial. This is an extremal state with zero temperature and, as we reach it from lower mass configurations, the system is thermodynamically unstable close to it.

\begin{figure}
\centering
 \includegraphics[width=0.47\textwidth]{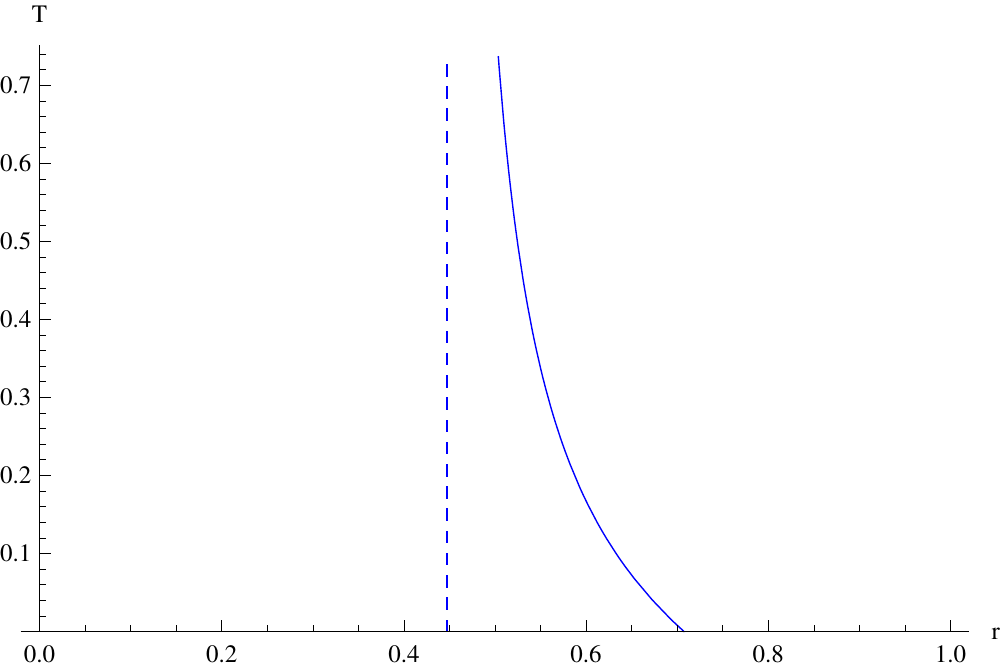}\quad \includegraphics[width=0.47\textwidth]{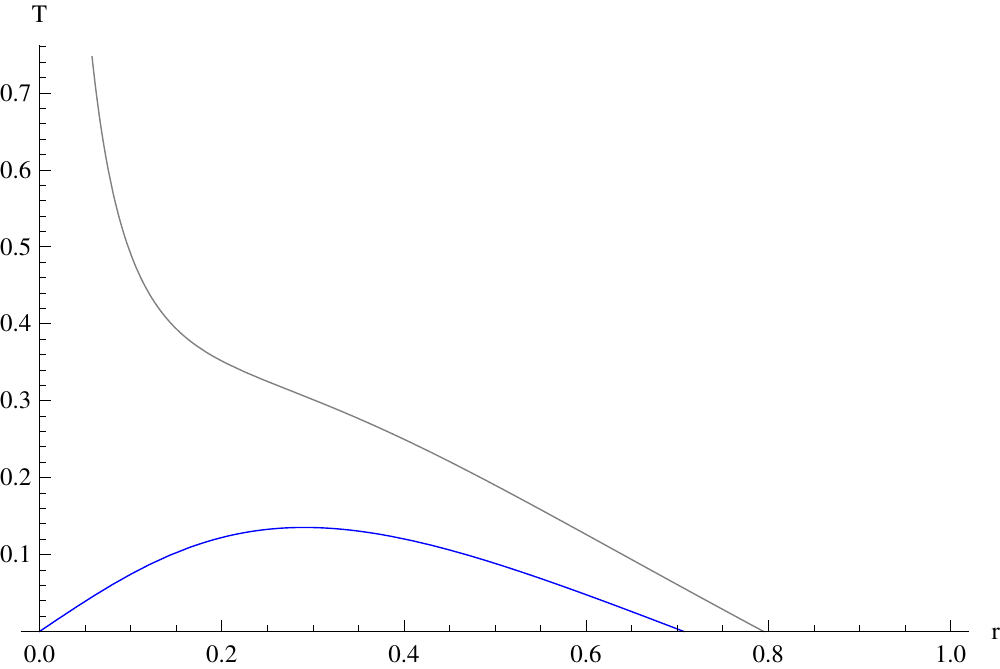} 
\caption{Temperature versus horizon radius (equivalently, mass) for spherical black holes in $d=5$ GB gravity with positive cosmological constant, $L^2=-1$ (in blue). The first figure, $\lambda=-0.1$, corresponds to a (b) type dS branch. We can identify the zero temperature state in the high mass regime with the Nariai solution. The temperature diverges as we approach the lower bound, $\kappa_{\rm min}$, indicated by the dashed blue line. For $\lambda=0.1$, in the second figure, a spherical (a) type dS branch arises. We may identify again the extremal state with maximal mass with the Nariai solution, for which the temperature goes to zero, as well as the radius of the black hole horizon. The `zero size black hole' has finite mass in this case. For higher dimensions the stable region of small black holes disappears (gray line corresponds to $d=6$), the spherical black holes being unstable as their (b) type counterparts. The only difference is that the temperature diverges in the zero mass limit.}
\label{dSbranch}
\end{figure}

We may construct dS branches of (a) and (b) types using GB gravity with positive cosmological constant (setting $c_0=-L^{-2}$). In this case, the (b) type branch (figure \ref{dSbranch}, left) is unstable for all allowed values of the mass, since we are considering $d=5$ (which is $d=2K+1$ in this case), while a stable region of small black holes appears for (a) type branches (figure \ref{dSbranch}, right) \cite{Cai2004a}.
This stable region disappears in higher dimensions. In general, spherical dS branches may have some stable intermediate region, but they are unstable (or even non-existent) for high enough temperature.

Let us summarize the main results of this section. Hyperbolic black holes generically have two stable domains, at low and high temperatures. For intermediate temperatures, these solutions may have more than one possible mass, some of them unstable. The only difference of AdS branches with respect to the EH case is that the high temperature regime has a maximal finite mass in the former. 

The case of spherical black holes exhibit quite distinct features. From the thermodynamic point of view, we may distinguish those situations where there is, or there is not, a minimal temperature, $T_{\rm min}$, for the black holes to exist. In the former case, we do not reach any extremal black hole, neither in the low mass, nor in the high mass regimes. In the case of the EH branch, for $T < T_{\rm min}$, only the thermal vacuum may exist whereas, for high enough temperatures, a black hole may exist with two very different masses, one close to $\kappa_{\rm min}$ (which is unstable) and one very high (stable). For intermediate temperatures close to $T_{\rm min}$, we may in principle encounter several stable and/or unstable black holes. In the latter case, instead, black hole solutions exist for the whole range of temperatures, except in the case where a dS branch reaches an extremal state at low as well as at high masses. For the EH branch, in this situation, we have two stable phases again, one in the low and one in the high temperature regimes. At intermediate temperatures, the black hole may have several possible masses, some of them unstable. 
 
\section{Hawking-Page-like phase transitions}
\label{HP1}

The existence of unstable phases, as well as the several possible black hole solutions at the same temperature, suggest the occurrence of Hawking-Page-like phase transitions, as already observed in the case of LGB gravity \cite{Cvetic2002,Nojiri2001j}. These phase transitions should be also relevant when studying the physics of the dual CFT plasma. In order to analyze this we need to discuss the global stability of the solutions. Any system in thermal equilibrium with an infinite heat reservoir (and thus at constant temperature) will be described by the canonical ensemble, whose relevant thermodynamic potential is the Helmholtz free energy, $F$. The preferred, and so globally stable, solution is the one that minimizes $F$. For instance, the free energy of the black hole solution calculated in (\ref{free-energy}), is the free energy with respect to the vacuum solution, except for the hyperbolic case where the finite ground state free energy (its mass) must be subtracted. Therefore, the sign of the free energy determines which solution is globally preferred at any given temperature, the appropriate black hole (if several are possible) or a thermal bath for the groundstate ({\it vacuum} or extremal).

The general analysis is, again, hard and not very enlightening. We will just concentrate in showing general features of these black hole solutions without entering into the details of the different cases. We will consider the same regimes analyzed for the local stability, as there we can easily find the expression for the free energy. In the planar case the analysis is very simple since the free energy reads
\begin{equation}
F=-\frac{V_{d-2}}{16\pi G}\frac{r_+^{d-1}}{L^2} ~.
\end{equation}
The black hole is then always the preferred solution, as indicated by the negative sign of the free energy, and no phase transitions occur. This will be the situation in the large mass limit of the other topologies in the EH branch. As for large enough $r_+$ the free energy may be as large as one wants, ambiguities on the reference background do not matter in this limit.   

\subsection{Spherical black holes}

For spherical black holes, we will restrict our discussion to the two most generic situations. For the EH branch, we will consider separately the case of having a stable low temperature phase (as in $d=5$ GB gravity with positive $\lambda$), and the case where a minimal temperature is needed for black hole solutions to exist. The second case is the analogue of the usual situation in Einstein-Hilbert gravity. At low temperatures, the thermal vacuum can be considered as the globally stable solution whereas, for higher temperatures, two or more black hole solutions are possible. For high enough temperature just two of them remain. The small one has always positive free energy. For an (a) type branch,
\begin{equation}
F=\frac{(d-2)V_{d-2}}{16\pi G}\frac{r_+^{d-2K-1}}{d-2K}c_K ~,
\label{freesmall}
\end{equation}
whereas for the (b) type case the temperature diverges as we approach the maximum $g_+\rightarrow g_\star$, and the free energy is $F\approx -T S$. Then, for positive entropy, the small black hole solution has negative free energy and is stable against the vacuum. However, it is not the minimum of the free energy since the big black hole has always a lower one. This is quite easy to see by realizing that the small black hole entropy goes to a constant as we approach $g_\star$ whereas the entropy grows indefinitely for big black holes, since they approach the planar limit. Thus, the small black holes are not just locally but also globally unstable. 

The big black holes have, in general, negative free energy. We have then a Hawking-Page-like phase transition, from the thermal vacuum at low temperatures to big black holes at high temperatures. The difference with respect to the Einstein-Hilbert case is that we may have several black holes at intermediate temperatures, with either sign of the free energy. For ranges of temperature where several black holes have negative free energy, transitions among them may happen, the globally preferred solution being the one with the lowest free energy. This would be an example of a new kind of phase transition, different from the Hawking-Page one, where one of the phases is always the thermal vacuum. 

If the EH branch has stable low temperature black holes, {\it i.e.}, for (a) type in $d=2K+1$, these are globally unstable as indicated by their positive free energy that asymptotes a constant when $r_+ \to 0$, 
\begin{equation}
F=\frac{(d-2)V_{d-2}}{16\pi G}c_K~,
\end{equation}
actually to the mass as both the entropy and temperature vanish in this limit.
This is exactly equal to the earlier formula (\ref{freesmall}) for the given dimension. The same happens for the would be extremal black holes that one may encounter in the EH branch. In the limit of low temperatures, the free energy coincides with the mass and, as such states have positive mass, they are globally unstable. The globally preferred phase is the thermal vacuum which is the minimal mass solution. Then, again, one has the same kind of transition described in the previous paragraph.

Another situation we did not comment at length is the possibility of having negative entropy for the spherical black hole with critical mass, $\kappa_{\rm min}$. This happens already in the simplest possible case of GB gravity, for negative $\lambda$, where there is a maximum in the EH branch situated at $g_{\star}=-1/2\lambda$. As pathological as it may seem, the consequence of this from the global stability point of view is clear. Again, the globally preferred solution is the big black hole as before, and the discussion goes through. This is quite general: a negative entropy state necessarily has bigger free energy than the vacuum (characterized by minimal mass and vanishing entropy).

For dS branches the situation at low mass is exactly the same as for spherical solutions in the EH branch. At low temperatures the free energy approaches the value of the mass and the globally stable phase is always thermal vacuum. For (a) type branches and $d >2K+1$, no high temperature black hole exists, thereby the preferred phase in that regime would be trivially the thermal vacuum and no Hawking-Page-like phase transition seems to occur (see, for instance, \cite{Cai2004a}, for the GB case). In any other situation ({\it e.g.}, (b) type branches) with positive entropy, the globally stable solution would be the near-critical black hole approaching the maximum of the polynomial. Therefore, these branches seem to display phase transitions, even though high temperature black holes are locally thermodynamically unstable.

The inclusion of the extremal {\it Nariai space} with arbitrary temperature and zero entropy does not change these conclusions. As it has higher mass than the vacuum it is always globally (and locally) unstable

\subsection{Hyperbolic black holes}

For the hyperbolic black holes one may compute the free energy at the high and low temperature regimes as before. As the maximally symmetric space has temperature in this case it is not clear how to use it as a ground state. Instead, we will consider the extremal negative mass black hole as the reference state --with vanishing entropy-- given that it can be identified with any temperature \cite{Hawking1995}\footnote{This has been disputed by some authors (see \cite{Anderson1995a} for instance) in reason of the semiclassical instability of these solutions.}, as explained earlier. Otherwise, the analysis would become trivial with just one or more black hole solutions, no matter the value of the temperature. No Hawking-Page-like phase transitions would occur in that case, just the possibility of transitions among black holes of different masses at intermediate temperatures.

For the EH branch in the high temperature regime we just have one possible black hole solution. It has negative free energy as it approaches the planar limit and so it is globally preferred. The same happens for the AdS branches, that end up at a maximum of the polynomial. As we approach the critical mass, $\kappa_{\rm max}$, the free energy approaches $F\approx-TS$ that is arbitrarily negative for positive entropy. For negative entropy at the maximum, which corresponds to the biggest possible black hole in the AdS branch, every black hole has negative entropy. In this case the globally preferred phase is the reference state, since it has zero entropy and minimal mass. 

For lower temperatures we have to consider black holes close to the extremal one. In the zero temperature limit only these extremal states matter and their free energy is simply given by their mass. Then, the globally preferred phase in that limit is the lowest mass state. For slightly higher temperatures one expects that the globally preferred solution is still described by the same minimal mass extremal state (identified with finite temperature) or the corresponding black hole solution that is just a smooth deformation of it. It is not hard to elucidate in general which of both solutions has the lowest free energy. To first order in the temperature the free energy is given by $F\approx M^e-S^e\ T$ where $S^e$ is in this case the entropy as we approach zero temperature, being zero for the extremal vacuum and non-vanishing for the non-extremal black hole. For positive entropy states the globally stable phase at low temperatures is then the black hole. We do not necessarily have a phase transition in this case, we have a black hole at low and high temperatures.  At intermediate  temperatures however, we might encounter transitions among extremal or non-extremal solutions that share the same temperature. 

\section{Discussion}

In this and previous chapters, we presented a novel approach to deal with the full classification and description of black hole solutions with constant curvature horizons in Lovelock gravity. Our proposal allows to treat the generic case where the whole set of Lovelock coupling constants is arbitrary, contrary to most existing studies in the literature where the analysis is restricted to particular cases. Most of these cases, moreover, correspond to degenerate vacua of Lovelock gravity, while our approach is valid in general and is most useful in the non-degenerate case.

We discussed the main features of all possible configurations, focusing in the neutral case. In particular, we have established a recipe to scrutinize the number of horizons and their evolution with the mass, something that we expect to be useful to visualize and gain intuition in physical processes involving black holes in such theories: evaporation, mass accretion and appearance of naked singularities \cite{Camanho2013b}. We will comment more on this on the next chapter. The analysis of charged black holes and even cosmological solutions can be performed in a very similar manner. The same happens for some more general classes of higher curvature gravities described recently \cite{Oliva2011} that share the form of the black hole solutions (and consequently their thermodynamic properties) with Lovelock theories. Most of the results of this chapter are also of direct application there. In particular quasi-topological share some crucial properties with Lovelock's in their critical dimension, $d=2K+1$, despite their higher curvature order.

We presented some general features of Lovelock black holes' thermodynamics, analyzed their local and global stability and the possible existence of phase transitions. Even if these solutions show some seemingly pathological features, such as negative values for the entropy, these are avoided if we restrict ourselves to the globally preferred phase.

For asymptotically AdS solutions (either in the EH or AdS branches), global stability in the high temperature regime always selects the biggest black hole, being the one with biggest entropy. If we apply the same criterion to all possible solutions, regardless of the branch to which they belong, the selected solution is always the one approaching the planar limit, since it is the only one that has arbitrarily big entropy. The comparison of solutions belonging to different branches is not really allowed though, as they have different asymptotics. The usual Euclidean prescription says that we must compare all solutions {\it with the same boundary conditions} what certainly includes the asymptotics. However the existence of {\it bubble} solutions \cite{Gravanis2010a, wormholes, Garraffo2008b} separating regions corresponding to different vacua suggest the possible existence of mixed solutions and transitions between branches. In that context, the high temperature phase would naively always correspond to the universal planar limit. This kind of branch transitions will be the focus of chapter \ref{genHP} where we will show that the situation is not as simple.

The usual Einstein-Hilbert gravity admits in principle topological solutions displaying naked singularities. These may arise as a result of a bad choice of topology or as associated to negative mass, below the extremal one for hyperbolic horizons. The latter are just a special case of {\it trans-extremal} solutions, where the values of the parameters are chosen in such a way that, an otherwise well-defined black hole with positive temperature, is taken {\it beyond} the extremal state. Example of this are also Reissner-N\"ordstrom black holes above the critical charge. In principle, all these situations are ruled out by the cosmic censorship conjecture that states that naked singularities do not form in the evolution of generic initial conditions. For instance, the evaporation process for black holes with several horizons should stop at the extremal state as it has zero temperature and this avoids the formation of trans-extremal solutions in that case.

The situation in generic Lovelock theories of gravity is rather different. For this wide family, there are several situations that suggest a possible violation of the cosmic censorship conjecture, as we have seen analyzing the case of static uncharged black holes. In addition to the cases pointed out in the previous paragraph, new kinds of naked singularities appear, some of them which naively seem to be formed in the evolution of these black holes. This can happen for the otherwise well-behaved Einstein-Hilbert branch, and it certainly happens generically on the extra {\it higher order} (A)dS branches. Most of these naked singularities arise because the branch of interest ends up at a maximum or a positive (for positive mass) minimum of the Lovelock polynomial. The latter corresponds to a complex cosmological constant associated with this particular branch. The former, in turn, appear in a variety of cases. They imply an upper mass bound for hyperbolic black holes of AdS branches as well as a lower bound for some spherical black holes in the EH or dS branches.

The other possibility for naked singularities to appear is just the spherical case in the maximal $d=2K+1$ Lovelock theory, for the type (a) EH or dS branches. 
In those cases, we again find a naked singularity for masses below some critical value. Any other possible naked singularity may be considered in the same class as those appearing in Einstein-Hilbert gravity. 

As we think of the evolution of the black holes studied in this thesis, we realize that naked singularities seem easy to form, at least naively. Consider for instance the evaporation of spherical black holes. For (b) type EH or dS branches these black holes always reach a critical mass where the horizon coincides with the singularity. At that point the temperature diverges but a finite mass naked singularity remains. The naked singularity inevitably forms. We emphasized the word {\it naively} before since the present analysis just considers the thermodynamic stability of the solutions. These solutions are locally and globally unstable in that sense, however they are still valid solutions that may form under evolution of generic spherically symmetric initial conditions. In the next chapter we will perform a more detailed analysis in order to elucidate whether naked singularities may form or not in these theories \cite{Camanho2013b}.

We have fixed, throughout this thesis, the values of the cosmological constant and the Newton constant appearing in the lagrangian to their customary values in AdS Lovelock gravities. It is worthwhile mentioning that a straightforward generalization of this work amounts to studying the case of dS Lovelock theories (note that there may be also AdS vacua  in this case), as well as theories where the Newton constant has negative sign. On the one hand, we shall mention that this sign flip was already considered in the context of three dimensional topologically massive gravity, where it was found that a negative Newton constant is useful to render otherwise negative energy modes harmless for the stability about flat space \cite{Deser1982a}. Furthermore, in higher dimensions, even though $G_N < 0$, the generic structure of branches discussed in this chapter will remain, and there will always be solutions corresponding to well-defined gravities with positive Newton's constant.

Lovelock theories have the remarkable feature that lots of physically relevant information is encoded in the characteristic polynomial $\Upsilon[g]$. Boulware-Deser-like instabilities, for instance, can be simply written as $\Upsilon'[\Lambda] < 0$, which has a beautiful CFT counterpart telling us that the central charge, $C_T$, has to be positive. Now, $\Upsilon'[\Lambda]$ can be thought of as the asymptotic value of the quantity $\Upsilon'[g]$ that is meaningful in the interior of the geometry, and has to be positive all along the corresponding branch. Naked singularities taking place at extremal points of the polynomial are suggestive of the fact that $\Upsilon'[g]$ should be a meaningful entry of the holographic dictionary (see \cite{Paulos2011} for related ideas) that does not exist in the case of Einstein-Hilbert gravity. The relevance of Lovelock and more general higher curvature gravity theories in the context of the AdS/CFT correspondence will be the topic of the second part of the thesis.

\chapter{\bfseries\itshape Metric perturbations and stability}
\chaptermark{Metric perturbations and stability}
\label{chp:bhstability}

\vspace{.6cm}

\begin{quotation}
\flushright
{\it ``The most incomprehensible thing about the universe\\ is that it is comprehensible''}\\

\vspace{.3cm}

Albert Einstein
\end{quotation}

\vspace{3cm}

\noindent Perturbative analysis is a very powerful tool to investigate the dynamic response of a system against small disturbances. It is particularly important for the case of gravity due to their strong non-linear character. This is already true for general relativity but even more for Lovelock or other higher curvature theories which are much more complicated systems. The equations of motion on any of these theories are very hard to solve analytically and exact solutions are known just on very special circumstances. Perturbative analysis of exact solutions plays a crucial r\^ole in many physical situations and opens a window into the intricate dynamics of gravity in four and higher dimensions. 

For our purposes, the most important examples of application of perturbative analysis in the context of gravity are the studies of black holes. Such an investigation was first systematically done for the Schwarzschild black hole by Regge and Wheeler \cite{Regge1957} in 1957 and completed some 13 years later by Zerilli \cite{Zerilli1970} and Teukolsky \cite{Teukolsky1972}. 
This is a fundamental question at many levels as unstable solutions are less likeky to form through any physical process and even when they do they will certainly not remain on that state for very long. This has many implications, from the dynamics of black holes to the determination of the final fate of gravitational collapse. 

Apart from the stability issue, perturbative studies also tells us a lot about
basic properties of black hole solutions. For instance the study of stationary perturbations of a stationary black hole solution provides a criterion for uniqueness and the search for new solutions.

The stability of higher dimensional black holes in EH theory has been intensively studied (for a review see for instance \cite{Ishibashi2011a}), higher dimensional Schwarzschild black holes being stable for any type of perturbations. For static spherically symmetric Lovelock black holes, the analysis is not as straightforward. This is due to the complicated form of the equations of motion, even in the linearized case, but also because of the form of the solutions themselves, the existence of branches, etc.
Stability analyses under all type of perturbations have been performed a few years ago, though, in the case of LGB gravity  \cite{Dotti2005a,Dotti2005b,Gleiser2005,Neupane2004,Beroiz2007,Konoplya2008}. More recently, the problem was tackled in the realm of Lovelock theory by Takahashi and Soda in a series of papers
\cite{Takahashi2009h,Takahashi2009b,Takahashi2010g,Takahashi2010e}, also for the charged case \cite{Takahashi2011,Takahashi2012}, including some connections with the AdS/CFT correspondence \cite{Takahashi2011a}.

Perturbative analysis can be formulated in the language of master equations for gauge invariant gravitational perturbations, a much simpler and intuitive approach.  In the context of Lovelock theories of gravity, generic master equations have been found in \cite{Takahashi2010g}. They cannot hide the intricacy of the theory, though. Thereby, except for some restricted efforts, most of the work has been performed in the context of LGB gravity. In this chapter, we will make use of these master equations, or rather the effective potentials they define, in order to analyze the stability of black hole solutions in a particular regime, that of high momentum gravitons. This restricted analysis will greatly simplify the computations; nonetheless, it will still be general enough to uncover some very interesting features of Lovelock black hole solutions. All the computations will be performed analytically and will be useful in order to gain general intuition about gravitational instabilities in Lovelock gravities.

The most familiar instabilities occurring in Lovelock theory are of so-called Boulware-Deser (BD) type \cite{Boulware1985a}. They affect the vacua of the theory. For any Lovelock gravity, it is possible to define a formal polynomial $\Upsilon[\Lambda]$, whose coefficients are the gravitational couplings, and whose (real) roots are nothing but the plausible cosmological constant of the different vacua of the theory. A vacuum with cosmological constant $\Lambda_\star$ such that $\Upsilon'[\Lambda_\star]$ is negative, would host ghostly graviton excitations with the wrong sign of the kinetic term \cite{Camanho2013b}. If we insist in constructing a black hole solution for that vacuum, it is easy to see that the resulting configuration is unstable. Interestingly enough, there is a holographic counterpart of this result, which has to do with the unitarity of the dual CFT \cite{Camanho2010d}.

We will push the analysis of Lovelock black hole linear stability a step forward, aimed at understanding how and if the cosmic censorship conjecture \cite{Penrose} holds in these theories. There are earlier works considering its status, and there seems to be a consensus towards the idea that naked singularities can be produced via gravitational collapse in Lovelock theory. For instance, the case of LGB gravity without cosmological constant has been analyzed, both for the spherically symmetric gravitational collapse of a null dust fluid that generalizes Vaidya's solution \cite{Maeda2005} and for that of a perfect fluid dust cloud \cite{Maeda2006}. Some particular cases of Lovelock gravity have been considered as well; namely, so-called dimensionally continued gravity \cite{NozawaMaeda} and cubic Lovelock theory \cite{DehghaniF}. More recently, a broader (still greatly restricted) analysis has been tackled by several authors arriving essentially to similar conclusions \cite{Rudra,OhashiSJ2011,Zhou_etal,OhashiSJ2012}. A rough upshot of the most distinctive aspects that these papers share is:
\begin{quotation}
\noindent
{\bf (i)} The critical case --maximal higher curvature term of degree $K$ in $d_{\star} = 2K+1$ spacetime dimensions-- is qualitatively different from the $d > d_{\star}$ case.\newline\vskip-3mm
\noindent
{\bf (ii)} There is a lower bound for the mass, in the critical case, below which spherical collapse leads to a naked singularity.\newline\vskip-3mm
\noindent
{\bf (iii)} The situation smoothens for $d > d_{\star}$, where cosmic censorship seems to be respected (to the extent it was challenged). This holds, in particular, for general relativity in higher dimensions \cite{Goswami}.
\end{quotation}
In these articles, cosmic censorship is a statement about the causal structure of spacetime, as due to the dynamical collapse. It is reasonable to question, however, whether the endpoint of the collapse, whenever it leads to a naked singularity, is stable. If it is not, the corresponding violation to the cosmic censorship hypothesis becomes dubious \cite{Joshi}. A stability analysis in the framework of gravitational collapse is cumbersome, though.

In this chapter, we will tackle this problem from a different perspective. We will show that {\bf (i)} and {\bf (ii)} can be understood rather easily from a direct analysis of the spectrum of spherically symmetric static solutions, developed in \cite{Camanho2011a}. We naively focus on the possible endpoint states of the gravitational collapse by relying in Birkhoff's theorem, which is valid in this context \cite{Zegers2005}, and show that this leads to the same results. This allows us to extend the analysis, by the same token, to the case of planar and hyperbolic geometries. It was already seen, in the case of static planar black holes, that the Lovelock couplings have to obey certain constraints in order to avoid instabilities \cite{Camanho2010d}. For non-planar black holes, the situation is more involved but we will still be able to describe some important qualitative features of their instabilities, specially concerning their relation with the cosmic censorship hypothesis. In this way, far from being discarded as pathological, instabilities are given a very precise physical significance.

We will then discuss {\bf (iii)}. It is easy to see that the statement is wrong or, better, non-generic. It certainly applies in the cases discussed in the literature, but it is not true in a generic Lovelock theory. This is connected to the existence of two kinds of solutions, dubbed type (a) and type (b) in previous chapters. While type (a) solutions are those extending all the way to the singularity, at $r=0$, as it is customarily taken for granted, type (b) solutions reach a finite radius singularity due to the existence of a critical point of the characteristic polynomial $\Upsilon$ alluded to above. This polynomial is entirely dictated by the Lovelock couplings and, as such, type (b) solutions cannot be discarded. They are pretty generic and their existence challenges the cosmic censorship hypothesis \cite{Camanho2011a}. We will nevertheless show that perturbative analysis provides valuable information on this respect, supporting its validity. We are of course aware of the many subtleties  linked to the cosmic censorship conjecture. In that respect, our study is not conclusive but should be taken as a significant piece of evidence.

\section{Graviton potentials}
\label{GRpots}

Throughout this chapter we will be considering the same black hole solutions of the preceeding sections as described by the metric form \reef{bhansatz} and the corresponding black hole polynomial \reef{eqg} defining $f$. We add a generic perturbation of the metric, $h_{ab}$, with fixed frequency, $\omega$, and momentum, $q$ in a fixed direction. These fluctuations split into three channels according to their polarization relative to the momentum, namely the tensor, shear and sound channels \cite{Policastro2002}  (or equivalently helicity/spin two, one and zero respectively). The equations of motion for these dynamical degrees of freedom, $\phi_h(r)$, the subindex $h=0, 1, 2$ indicating the corresponding helicity, can be recast as Schr\"odinger type equations \cite{Takahashi2010g}, that in the large momentum limit they reduce to a very simple form\footnote{In the notation of \cite{Takahashi2010g}, we must identify $\gamma_i \equiv \gamma = q^2 L^2$, and our potentials are related to theirs, $U_i \to - V_i/\gamma\,\Lambda$, as $\gamma \to \infty$.}
\begin{equation}
\label{Scheq}
-\hbar^2\, \partial^2_y \Psi_{\! h}+U_{\! h}(y)\,\Psi_{\! h}=\alpha^2\,\Psi_{\! h}~,\qquad\qquad \hbar\equiv\frac{1}{L q}\rightarrow 0~,
\end{equation}
where $\alpha = \omega^2/q^2$, $y$ is a dimensionless tortoise coordinate defined as $dy/dr=\sqrt{-\Lambda}/f(r)$, and $\Psi_{\! h}(y) = B_h(y)\,\phi_h(y)$, where $B_h(y)$ are functions of the metric whose specific expression can be found in \cite{Takahashi2010g}. For a regular solution $\Psi_{\! h}(y)$, the metric perturbation $\phi_h(y)$ blows up as $B_h(y)$ approaches zero. In such case, it could not be considered any longer as a perturbation and, in that sense, the linearized analysis would be spoiled. We then need to make sure that the function $B_h(y)$ is non-vanishing. In our regime of interest, the effective potentials $U_{\! h}$, can be determined as the speed of large momentum gravitons in constant $y$ slices,
\begin{equation}
U_{\! h}(y) = \begin{cases} {\bf c}_h^2 (y) & \qquad y < 0 ~, \cr
                      + \infty & \qquad y = 0 ~,
        \end{cases} 
\end{equation}  
$y=0$ being the boundary of the spacetime. The explicit computation for the helicity two graviton and details on the other helicities are given on appendix \ref{PertEq}

For a generic Lovelock theory, in terms of the original radial variable, one finds for the tensor, shear and sound channels, respectively \cite{Camanho2010a}:
\bear
{\bf c}_2^2(r) & = & \frac{L^2 f(r)}{(d-4)\,r^2} \frac{\mathcal{C}^{(2)}_d[g,r]}{\mathcal{C}^{(1)}_d[g,r]} ~, \nonumber\\ [0.6em]
{\bf c}_1^2(r) & = & \frac{L^2 f(r)}{(d-3)\,r^2} \frac{\mathcal{C}^{(1)}_d[g,r]}{\mathcal{C}^{(0)}_d[g,r]} ~, \label{pots}\\ [0.6em]
{\bf c}_0^2(r) & = & \frac{L^2 f(r)}{(d-2)\,r^2} \left(\frac{2\,\mathcal{C}^{(1)}_d[g,r]}{\mathcal{C}^{(0)}_d[g,r]}-\frac{\mathcal{C}^{(2)}_d[g,r]}{\mathcal{C}^{(1)}_d[g,r]}\right) ~, \nonumber
\label{GRpotentials}
\eear
where $\mathcal{C}^{(k)}_d[g,r]$ are functionals involving up to $k$th-order derivatives of $g$ defined in \reef{Cpots}. Notice also that there is a quite simple relation between the three potentials,
\be
(d-2) {\bf c}_0^2(r)-2(d-3){\bf c}_1^2(r)+(d-4){\bf c}_2^2(r)=0
\label{potsconstraint}
\ee
in such a way that any of the three can be written as a combination of the other two.

These expressions are valid in general, also for charged black holes. In the uncharged case however the black hole equation \reef{eqg} is simpler and this allows us to make a simplifying change of variables. Instead of $r$ we take $\Upsilon$ as independent variable (the relation is one-to-one in absence of charge) and we define $x\equiv\log L^2\Upsilon$ and $F\equiv\log L^2\Upsilon'$ so that $r\partial_r=-(d-1)\partial_x$  yielding for the potentials
\begin{eqnarray}
{\bf c}_2^2(x) & = & \frac{(d-1)L^2 \,f}{(d-4)\,r^2}\frac{\left(\frac{d-3}{d-1}-F'(x)\right)\left(\frac{d-4}{d-1}-F'(x)\right)+F''(x)}{\left(\frac{d-3}{d-1}-F'(x)\right)} ~, \nonumber\\[0.6em]
{\bf c}_1^2(x) & = & \frac{(d-1)L^2 \,f}{(d-3)\,r^2}\left(\frac{d-3}{d-1}-F'(x)\right) ~, \label{potentialsF}\\ [0.6em]
{\bf c}_0^2(x) & = & \frac{(d-1)L^2 \,f}{(d-2)\,r^2}\frac{\left(\frac{d-3}{d-1}-F'(x)\right)\left(\frac{d-2}{d-1}-F'(x)\right)-F''(x)}{\left(\frac{d-3}{d-1}-F'(x)\right)} ~.\nonumber 
\end{eqnarray}
In this way the potentials can be thought as functions of the metric function $g$ only, everything can be written in terms of the Lovelock polynomial $\Upsilon$ and its derivatives, \eg $F'=\Upsilon[g]\Upsilon''[g]/\Upsilon'[g]^2$. This makes it very easy to analyze the potentials for the different branches of solutions (corresponding to different ranges of $g$) and different values of the mass that controls the place where the solution {\it ends}, \ie the horizon $g=g_+$.

\section{Black hole instabilities}
\label{bhstab}

It has been found in \cite{Dotti2005,Gleiser2005,Buchel2010a} that, for certain values of the LGB coupling, some effective potentials might develop negative values close to the horizon. The r\^ole of $\hbar$ in the Schr\"odinger-like equations (\ref{Scheq}) is played by $1/q$, in such a way that taking sufficiently large spatial momentum (small $\hbar$), we can make an infinitesimally small well to support a negative {\it energy} ($\alpha^2$) state in the effective potential. Going back to the original fields, this translates into an exponentially growing and therefore unstable mode \cite{Myers2007}. Thereby, in order to analyze the stability of black holes in our regime of interest we will then  just be concerned with the sign of \reef{potentialsF}.

The simplest potential is the one corresponding to the shear mode. Remarkably, it has exactly the same expression as the denominators of the other two potentials. The absence of instabilities in the shear channel is then linked to the condition needed to ensure the validity of the linear analysis \cite{Takahashi2009h}, since the $B_i(r)$ functions are proportional to some power of $\Upsilon'[g]\, {\bf c}_1^2(x)$, and $\Upsilon'[g]>0$ in order to guarantee that the branch of interest is free from BD-instabilities. Thus, if we approach ${\bf c}_1^2(x) \approx 0$, the linear analysis of perturbations simply break down. There is a further reason for the shear potential to be positive. It is related to the coefficient appearing in front of the kinetic term for the gravitons and, thereby, has to be positive for the sake of unitarity. Either way, this constraint can be seen as redundant since, due to (\ref{potsconstraint}), the positiveness of the shear potential is guaranteed, as far as the tensor and sound potentials are positive. Henceforth we can restrict the stability analysis to these channels.    

We will be particularly interested in black holes with planar horizons. This is due to their simplicity but also as this case is the most relevant in the context of holography. It describes the would be dual gauge theory in Minkowski space. Within this framework, graviton potentials will be important to probe several important aspects of the duality, ranging from causality to hydrodynamic properties of plasmas. Also, the planar limit corresponds to the high mass regime of the other two topologies and provides very important information about them. We restrict ourselves to the EH branch as it is the only one displaying a horizon in that case. 

We can show that a negative potential well can be found at the horizon in a general Lovelock theory\footnote{Some work in this direction has been done recently in \cite{Takahashi2009,Takahashi2009a} under some generic circumstances.}. Such negative values of $\mathbf{c}_h^2$ develop whenever the slope of the effective potential at the horizon is negative, and so we must require $\partial_x \mathbf{c}_h^2\leq 0$ there. Remember that $x=0$ corresponds to the horizon -- where $\mathbf{c}_h^2$ vanishes --  and $x=-\infty$ to the boundary. That analysis is particularly simple for planar black holes as then we have $g=0$ at the horizon, and it is straightforward to make use of $\Upsilon(g)=L^{-2}e^{x}$ to find the derivatives of $g(x)$ there, {\it e.g.}, $\partial_x g|_{g=0} = \left(\Upsilon[g]/\Upsilon'[g]\right)|_{g=0} = c_0/c_1 = L^{-2}$. In this way, we can relate the values of the derivatives of $F(x)$ at $x=0$ with the coefficients of the polynomial $\Upsilon(g)$. In particular,
\begin{equation}
F'(0)=2\lambda ~, \qquad \qquad F''(0)= 2(\mu +\lambda(1-4\lambda)) ~,
\end{equation}
where, again, $\lambda=c_2/L^2$, and we have defined $\mu = 3 c_3/L^4$. Using these results, we expand the graviton effective potentials close to the horizon to get
\begin{eqnarray}
{\bf c}_2^2(x) & \approx & -x\frac{d-1}{d-4}\frac{\left(\frac{d-3}{d-1}-2\lambda\right)\left(\frac{d-4}{d-1}-2\lambda\right)+2(\mu +\lambda(1-4\lambda))}{\left(\frac{d-3}{d-1}-2\lambda\right)}+\mathcal O\left(x^2\right) ~, \nonumber\\[0.7em]
{\bf c}_1^2(x) & \approx & -x\frac{d-1}{d-3}\left(\frac{d-3}{d-1}-2\lambda\right)+\mathcal O\left(x^2\right) ~, \label{horexp}\\[0.7em]
{\bf c}_0^2(x) & \approx & -x\frac{d-1}{d-2}\frac{\left(\frac{d-3}{d-1}-2\lambda\right)\left(\frac{d-2}{d-1}-2\lambda\right)-2(\mu +\lambda(1-4\lambda))}{\left(\frac{d-3}{d-1}-2\lambda\right)}+\mathcal O\left(x^2\right)\nonumber ~.
\end{eqnarray}
Stability requires that the potentials are positive. This gives rise to the following constraints for the Lovelock couplings:
\begin{eqnarray}
& &(d-3)-2 \lambda (d-1) > 0 \label{inscons1} ~, \\ [0.7em]
& & (d-3)(d-4)-2\lambda (d-1)(d-6)-4\lambda^2 (d-1)^2+2\mu (d-1)^2 \geq 0 \label{inscons2} ~, \\ [0.7em]
& & (d-2)(d-3)-6\lambda (d-1)(d-2)+12\lambda^2 (d-1)^2-2\mu (d-1)^2 \geq 0 \label{inscons0} ~.
\end{eqnarray}
These inequalities represent the constraints from the  shear, tensor and sound channels respectively. They are required to hold, otherwise the black brane solution is unstable. For $\mu=0$, these results match those for LGB gravity derived in \cite{Buchel2010a}. I that case the helicity two constraint becomes irrelevant in six or higher dimensions whereas in five it isolates the {\it apex} $\lambda=1/4$ from the rest of the stable range,
\bear
{\text d=5:}&\quad& -\frac18\leq\lambda\leq\frac18 \quad ; \quad \lambda=\frac14\\[0.8em]
{\text d\geq 6:}&\quad& -\frac{d-6+\sqrt{5 d^2-40 d+84}}{4 (d-1)} \leq \lambda \leq \frac{-(d-6)+\sqrt{5 d^2-40 d+84}}{4 (d-1)}\ \ \ \ \
\eear
It is important to notice that, in spite of considering a completely general Lovelock theory with as many terms as we wish, these stability constraints involve just the lowest two Lovelock couplings, $\lambda$ and $\mu$. 

\begin{figure}
\centering
\includegraphics[width=0.9\textwidth]{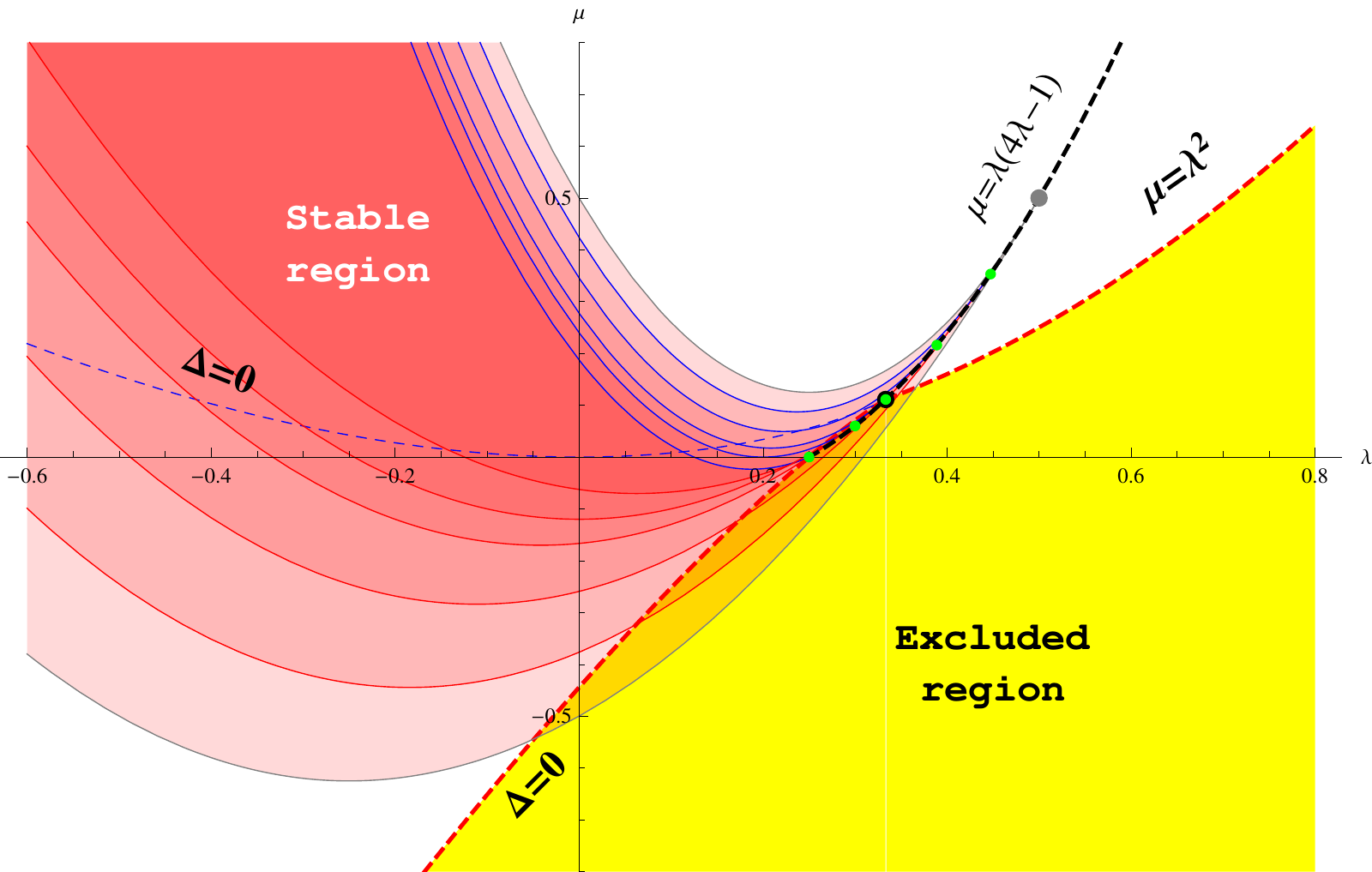}
\caption{Stability wedge (red regions) as defined by (\ref{stabwedge1}, in red) and (\ref{stabwedge2}, in blue) for $d=5,6,7,10,20$. For $d=5,6$ just the points on the $\mu=0$ axis have physical significance and bound the LGB coupling. The gray curves that bound the faintest red region correspond to infinite dimensionality  whereas the yellow area represents the excluded region as discussed in chapter \ref{chp:LLbh}. The dimensionality increases from the innermost to the outermost curves. A given dimension stability wedge always contains those of lower dimensions. The dashed black curve describes the locus of the apex (\ref{apex}, green dots) that goes inside the {\it excluded region} (yellow area) for $d=6,7$. This apex coincides with the maximally degenerate point of cubic Lovelock (black dot) for $d=7$ and with the gray dot as $d\to \infty$. Except for $d=5$ there is always a part of the stability that has to be discarded as it belongs to the excluded region. The dashed blue and red curves are nothing but the degenerate locus $\Delta=0$ where of the possible cosmological constants coincide.   }
\label{wedged}
\end{figure}

The first constraint comes from the shear mode and is irrelevant once the remaining ones are taken into account. The stability constraints define a new allowed {\it stability wedge} in parameter space (see figure \ref{wedged}):
\begin{eqnarray}
& & \mu \,\geq\, - \frac{(d-3)(d-4)-2\lambda (d-1)(d-6)-4\lambda^2 (d-1)^2}{2(d-1)^2} ~, \label{stabwedge1} \\ [0.7em]
& & \mu \,\leq\, \frac{(d-2)(d-3)-6\lambda (d-1)(d-2)+12\lambda^2 (d-1)^2}{2(d-1)^2} ~,
\label{stabwedge2}
\end{eqnarray}
with apex at
\begin{equation}
\lambda = \lambda_c = \frac{d-3}{2(d-1)} ~, \qquad \qquad \mu=\frac{(d-3)(d-5)}{2(d-1)^2} ~,
\label{apex}
\end{equation}  
or, equivalently, on the intersection of the $\lambda=\lambda_c$ line with $\mu = \lambda (4\lambda - 1)$. In general, for five and seven dimensions, the apex coincides with the point of maximal symmetry ($\lambda=1/4$ and $(\lambda,\mu)=(1/3,1/9)$ respectively), where the polynomial $\Upsilon[g]$ has a single maximally degenerate root. This is also the Chern-Simons point of the LGB or cubic Lovelock theory where it actually becomes a Chern-Simons theory for the AdS group (see, for example, \cite{Zanelli2005}). In higher dimensions the apex will have the same values of $\lambda$ and $\mu$ as the Chern-Simons point but all the other coefficients are unconstrained.

The above constraints will be extensively analyzed in chapter \ref{LLfate} in the context of holographic plasmas. The gauge/gravity correspondence allows for the description of such strongly coupled fluids in terms of black holes in AdS, in such a way that the above instabilities are interpreted as instabilities of the dual plasma. The constraints found will then be important to restrict the values of the Lovelock couplings that describe stable plasmas and the possible values of transport coefficients for those fluids, namely the shear viscosity \cite{Camanho2010d}. In particular the above constraints rule out negative values of the shear viscosity, that also represents an instability in the dual picture.

In addition to horizon stability we may encounter negative potential wells in the bulk of the spacetime as examples found in \cite{Camanho2010d}. For planar black holes this will be relevant when discussing more stringent constraints on the parameters and the shear viscosity to entropy density. We will come back to this on section \ref{bulkins}. 

For non-planar topology the situation is much more involved. In addition to the Lovelock couplings the instabilities will in general depend on the other parameter of the solutions, the mass or the radius of the black hole. Its value controls the range of values for $g$ that correspond to the untrapped region of the spacetime where instabilities may be found. For spherical black holes this range decreases as we increase the mass and the opposite happens in the hyperbolic case. The planar limit is just in between in such a way that when the planar case is unstable all spherical black holes in the EH branch are unstable as well. Hyperbolic black holes are always stable if we decrease sufficiently the mass, and they are stable for any mass in the EH branch when the planar limit is stable. In any case, all vacuum solutions are always stable as $F'=F''=0$
and so are black holes on any branch sufficiently close to them. Despite the complexity of the generic case we may still use the stability analysis to shed light on some interesting aspects of Lovelock theories such as evaporation of black holes or the status of the cosmic censorship conjecture on these theories.

\section{Stability of new uncharged extremal black holes}
\sectionmark{New uncharged extremal black holes}
\label{extremalBH}

In section \ref{singhor} a new type of spherically symmetric extremal black holes has been described in Lovelock gravities without charges \cite{Camanho2011a}. This result follows from the possibility for such black holes of having more than one horizon in general, depending on the shape of the polynomial, and couples of black hole horizons appearing or disappearing depending on the values of the mass parameter $\kappa$. We can argue that the third law of thermodynamics protects these black holes to become extremal by evaporation. It can be easily shown that the black hole would spend an infinite amount of time to reach the extremal mass. The real problem is the possibility of approaching these extremal solutions from lower masses. It would be possible to throw matter at a non-extremal solution in the exact amount so that a new degenerate horizon appears, this being an obvious violation of the third law.

This surprising behavior also represents a puzzle in other ways. It amounts to a discontinuous change on the thermodynamic variables associated with the horizon.  Moreover, we know that inner horizons are unstable \cite{Anderson1995a,Marolf2010}, so that we cannot trust our solution behind any such horizon that presumably becomes a null-spacelike singularity. Then the appearance of the new degenerate horizon would create a bigger singularity that suddenly {\it swallows} some previously untrapped region of the spacetime. 
We will see that all these problematic properties are ruled out once we take into consideration the possible instabilities of these spacetimes.  

This possibility of appearance and disappearance of horizons is associated with particular points in the polynomial that verify the following relation
\begin{equation}
\Upsilon'[g_+^e]=\frac{d-1}{2\,g_+^e}\Upsilon[g_+^e]\quad,
\label{extremalpoint}
\end{equation}
which sets the black hole temperature \reef{Temp} to zero. For masses slightly above and below that point the number of horizons differs by two and the thermodynamic variables have a discontinuity (see figure \ref{extremalBH} for a specific example). Two consecutive outer and inner horizons merge and disappear. In some cases below the critical mass the central singularity might even become naked. This however does not necessarily imply a violation of the cosmic censorship hypothesis as an evaporating black hole would need an infinity amount of time to reach zero temperature. 
\begin{figure}
\centering
\includegraphics[width=0.52\textwidth]{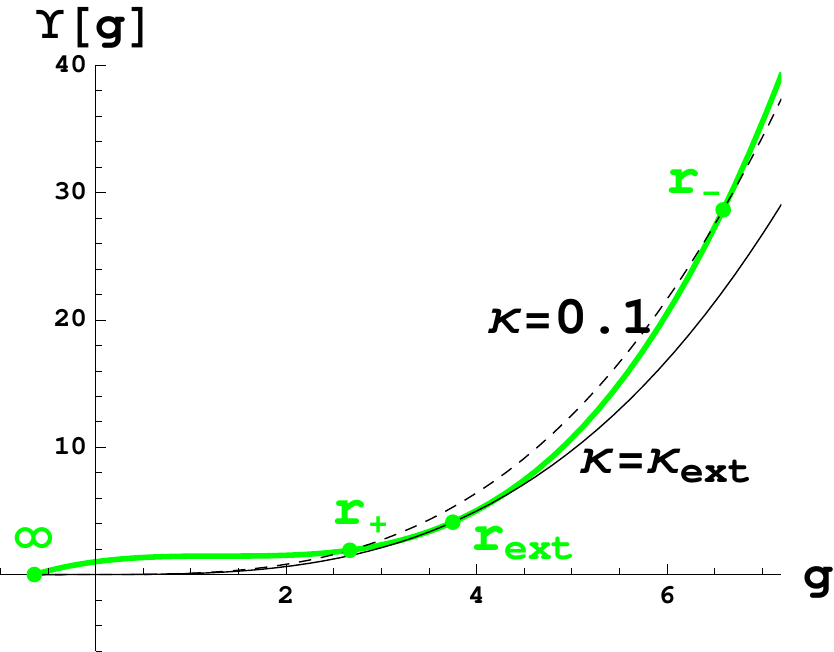}
\caption{Cubic polynomial for $L=1$, $\lambda=-0.746$ and $\mu=0.56$. The dashed lines are just $\kappa\left(g/\sigma\right)^{\frac{d-1}{2}}$ for $d=7$ and the indicated values of $\sigma$ and $\kappa$ (in units of $L$). The crossing of these lines with the polynomial give the possible values for $g$ at the horizon and then of $r_+$. We observe the occurrence of two (outer and inner, respectively) horizons, $r_+$ and $r_-$ which disappear for $\kappa<\kappa_{ext}$ leaving a naked singularity. For $d=8$ we find a similar behavior with one further inner horizon, three in total, with no naked singularity no matter the value of the mass.}
\label{extremalBHf}
\end{figure}

Consider $\Xi[g] := \Upsilon[g] - \kappa_e\,g^{\frac{d-1}{2}}$. It is fairly easy to see that $\Xi[g_+^e]$ and $\Xi'[g_+^e]$ vanish, {\it i.e.}, $g_+^e$ is an {\it extremal horizon}, and $\Xi''[g_+^e] > 0$ (see Figure \ref{extremalBHf}), which implies\footnote{Let us further point out that there is nothing \`a priori preventing $\Xi''[g_+^e] = 0$. However, ${\bf c}_1^2(x_+^e)$ will vanish in such case, which means that the other two potentials diverge with opposite signs, the instability being unavoidable.}

\begin{equation}
\Upsilon''[g_+^e]>\frac{(d-1)(d-3)}{4\,(g_+^e)^2}\Upsilon[g_+^e]= \frac{d-3}{d-1}\frac{\Upsilon'[g_+^e]^2}{\Upsilon[g_+^e]} ~.
\end{equation}
Both conditions make no reference to the specific value of the mass but rather express a property of the polynomial for some particular value of $g$.  Moreover, taking into account that we are considering positive mass solutions, $\Upsilon[g_+^e]>0$, we obtain the condition
\begin{equation}
F'(x_+^e)=\frac{\Upsilon''[g_+^e]\Upsilon[g_+^e]}{\Upsilon'[g_+^e]^2}>\frac{d-3}{d-1}~.
\label{extremalcondition}
\end{equation}
Remarkably this is exactly equivalent to the violation of one of the stability conditions
\begin{equation}
{\bf c}_1^2(x_+^e)<0 \quad.
\end{equation}
but also to having a negative coefficient for the graviton kinetic term in the black hole background and thus to unitarity. As we reduce the mass approaching the extremal value the black hole becomes unstable, earlier than any jump in the black hole radius. The solution is unstable also for any lower mass as the value of $g$ for the extremal point $g^e$ would belong to the untrapped region in that case.  The puzzle posed by these solutions is then solved, the pathological behavior being related to non-unitarity of the gravitons and instabilities of the black hole. Before any pathological behavior is encountered an instability sets in driving the system somewhere else. By the same token, this also forbids the possibility of the previously described  violations of the third law of thermodynamics. 

In the case where several extremal points (or masses) exist the relevant one is always the one corresponding to the lowest value of $g^e$ (biggest radius) and the associate mass. Notice that in general the instability is triggered for masses slightly above that extremal one. 


A final comment is in order. The previous analysis can be performed for other extremal states that appeared in the classification of Lovelock black hole solutions \cite{Camanho2011a}. These are (i) asymptotically dS spherical black holes at the Nariai mass(es) and (ii) asymptotically AdS extremal hyperbolic black holes with negative mass. They are also present in general relativity (which might seem dangerous under the light of the previous paragraph!), and it is easy to show that they respect the stability constraints at the extremal point. The value of $F'(x_+^e)$ is not bigger but lower than the critical value of $(d-3)/(d-1)$ in these situations as either (i) $\Xi''[g_+^e] < 0$ (see footnote 3), or (ii) $\Upsilon[g_+^e]>0$ due to the negative mass. We then do not expect instabilities for such backgrounds, though a generic stability analysis for the tensor and sound channels is not straightforward. This supports the consideration of some of these extremal solutions as groundstates \cite{Vanzo1997a}.

\section{The cosmic censorship hypothesis and stability}
\sectionmark{The cosmic censor}
\label{CCH}

The result of previous section is also relevant, as we shall see, for the discussion of the cosmic censorship hypothesis in Lovelock theories. For the minimal dimension $d=2K+1$ at any given order $K$ it may happen that the two merging horizons are the only ones of the solution and thus the singularity behind them would then become naked. It is hard to imagine any physical process that would reduce the black hole mass below the extremal threshold, nevertheless, even if it existed, we have just shown that the instability sets in before the singularity becomes naked, in fact before the black hole becomes extremal.

We also need to analyze the case of low mass black holes, as long as they can \`a priori be created directly by collapse. In order to complete the proof of their instability, we have to address the critical case, $d = d_\star$. These are naked singularities at $r=0$ (or $g \to \infty$), corresponding to type (a) branches, that take place when the highest Lovelock coupling is positive, $c_K > 0$, in the limit of small mass, $\kappa \to c_K$ (in units of $L$). But these instabilities, for $g \to \infty$, have been already observed in \cite{Takahashi2010e}, without any reference to the cosmic censorship hypothesis. In this limit, taking $F[g] := F(x[g])$,
\begin{equation}
F'[g] \approx \frac{K-1}{K} + \frac{A_K}{g^2} ~, \qquad F''[g] \approx - \frac{2 A_K}{K g^2} ~,
\end{equation}
to leading order, for some constant $A_K$ that is not relevant for our discussion. These expressions translate into the following potentials near $r = 0$:
\begin{equation}
{\bf c}_2^2(r) \approx \frac{3 L^2\,f}{(2K-3)\,r^2} ~, \qquad {\bf c}_0^2(r) \approx - \frac{3 L^2\,f}{(2K-1)\,r^2} ~.
\end{equation} 
the constraint (\ref{potsconstraint}) implying ${\bf c}_1^2(r) \approx 0$. The sound mode is unstable in accordance with \cite{Takahashi2010e}. Again, this is the case not only for the would be naked singularity but also for black holes whose mass is bigger but close enough to $c_K$. In summary, we have just shown that these solutions are unstable, a strong indication that, contrary to the claim of many papers in the literature \cite{Maeda2005,Maeda2006,NozawaMaeda,DehghaniF,Rudra,OhashiSJ2011,Zhou_etal,OhashiSJ2012}, they cannot be the end point of gravitational collapse under generic circumstances, neither be reached by any physical process.

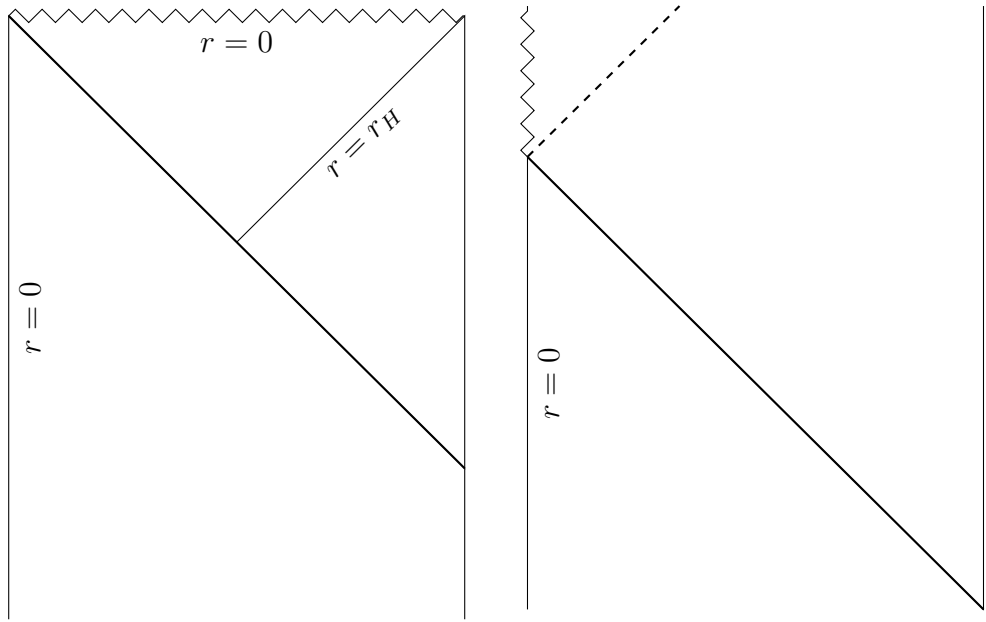
\begin{figure}
\centering
\begin{tikzpicture}
\node (I)    at ( 3,0)   {};
\node (II)   at (-3,0)   {};
\node (III)  at (0, 1.2) {};
\node (IV)   at (0,-3)   {};

\path  
  (II) +(90:3)  coordinate (IItop)
       +(-90:5) coordinate (IIbot)
       +(0:3)   coordinate (IIright)
       +(180:0) coordinate (IIleft)
       ;
\draw (IIbot) --
				node[midway, below, sloped] {$r=0$}
			(IItop) -- (IIright);

\path 
   (I) +(90:3)  coordinate (Itop)
       +(-90:3) coordinate (Ibot)
       +(180:3) coordinate (Ileft)
       +(0:0)   coordinate (Iright)
       ;
\draw  (Ileft) -- 
				node[midway, below, sloped] {$r=r_H$}
			 (Itop) -- (Iright) -- (Ibot) -- (Ileft) -- cycle;

\draw[decorate,decoration=zigzag] (IItop) -- (Itop)
      node[midway, below, inner sep=2mm] {$r=0$};
\draw (Ibot) -- (3,-5);
\draw[thick] (Ibot) -- (IItop);


\end{tikzpicture} 
\quad
\begin{tikzpicture}
\node (Ib)    at ( 12,0) {};
\node (IIb)   at (6,0)   {};
\node (IIIb)  at (9, 3)  {};
\node (IVb)   at (9,-3)  {};

\path  
  (IIb) +(90:3)  coordinate  (IIbtop)
       +(-90:3) coordinate   (IIbbot)
       +(0:3)   coordinate   (IIbright)
       +(180:0) coordinate   (IIbleft)
       ;
			
\path 
   (Ib) +(90:3)  coordinate (Ibtop)
       +(-90:3) coordinate (Ibbot)
       +(180:3) coordinate (Ibleft)
       +(90:5)   coordinate (Ibright)
       ;

\draw (IIbbot) -- 
					node[midway, below, sloped] {$r=0$}
			(IIbtop) -- (Ibbot) -- (12,5);

\draw[decorate,decoration=zigzag] (IIbtop) -- (6,5);
\draw[dashed,thick] (IIbtop) -- (8,5);
\draw[thick] (Ibbot) -- (IIbtop);


\end{tikzpicture} 
\caption{Penrose diagrams for the collapse of a shell of radiation (thick line) to a black hole (left) and a naked singularity (right). In the case of the naked singularity the radiation has no obstacle to scape {\it across} (or bouncing on) the singularity, the hypothetical trajectory corresponding to the dashed line.}
\label{collPenrose}
\end{figure}

For matter collapsing to a regular black hole with a horizon, the formation of the latter constitutes a critical moment for the matter contained within it. Think of a spherically symmetric configuration. The causal properties of event horizons force all the matter to end up at the central singularity, no matter the details of the matter distribution, and it cannot scape from there as causally it would imply travelling backwards in time. This is radically different if the end point is a naked singularity. As this solution is not provided with an event horizon, matter may in principle scape from the singularity without violating any causal structure and the final configuration may be much more sensitive to the details of the configuration under collapse. The Penrose diagrams of both such processes are schematically depicted in figure \ref{collPenrose}. It makes then sense to analyze the stability of such hypothetical solutions in order to assess whether or not they represent good candidates for endpoints of gravitational collapse. 

We have just shown the singular solutions to be unstable, such that, most probably, any small departure from spherical symmetry would imply that such naked singularity would not be formed. Even in the spherically symmetric case the singularity might just be {\it spurious}, mathematically just a result of all the matter ending up at the same point at the same time due to the rigid symmetry imposed. Furthermore, once it reaches the singularity, matter may still bounce back, even in presence of other non-gravitational interactions. This is even clearer if we distort slightly the matter configuration in such a way that we avoid this {\it coincidence problem}. Different parts of the matter configuration will arrive at different times at slightly different points in such a way that the singularity is not formed. \eg if we provide some angular momentum the centrifugal barrier would also do the job, at least in some cases. 

\begin{SCfigure}
\centering
\includegraphics[width=0.47\textwidth]{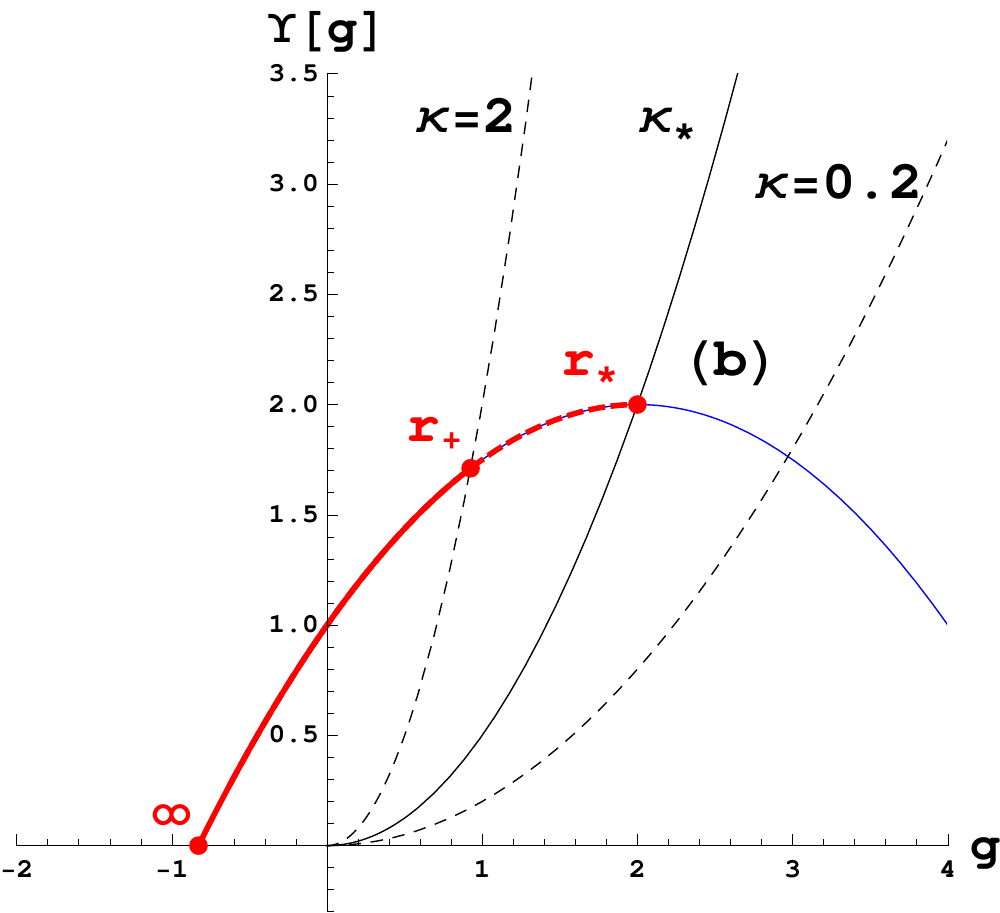} 
\caption{Spherical black holes in the EH branch of the type (b). The singularity is naked for $\kappa \leq \kappa_{\star} = 0.5$ (in $L$ units). In the limiting case, the event horizon shrinks to the singularity point $r_\star$, $g_+ = g_\star := g(r_\star)$.}
\label{relevantGB2}
\end{SCfigure}

There is however a substantially different kind of naked singularity that belongs to the realm of Lovelock theory (possibly to other higher curvature gravities as well). It is quite generic and can be formed through seemingly plausible physical processes. It may occur for the otherwise well-behaved EH branch, that is the main focus of most papers, as well as for the {\it higher curvature} AdS/dS branches. They arise due to the fact that the branch of interest ends up at a maximum of the Lovelock characteristic polynomial, a finite radius, $r_\star = r^{-1}(g_\star)$, with $\Upsilon'[g_\star] = 0$, before an event horizon is encountered,\footnote{Another type of singularity is that associated with positive minima of $\Upsilon$. They would correspond to a branch with a complex cosmological constant. This singularity can never be avoided by any choice of mass parameter and is part of what we have called, in the case of the EH branch, the excluded region.} a so-called type (b) branch (also AdS branches fit into this behavior) in the classification of \cite{Camanho2011a}; see Figure \ref{relevantGB2}.

The above discussions make clear that the stability of these solutions is an essential point to be analyzed in order to unravel the status of the cosmic censorship hypothesis in these theories. In fact, we will see that these instabilities show up whenever a violation of the cosmic censor may take place. It is enough to probe the solution close to the singularity in order to show its instability. We just need to analyze the behavior of the potentials for values of $g$ close to the critical one, $g_\star$. Given that it is a local maximum of $\Upsilon$, we can approximate the polynomial by
\begin{equation}
\Upsilon[g] \approx \Upsilon[g_\star] + \frac12 \Upsilon''[g_\star]\, (g-g_\star)^2 ~,
\end{equation}
and use this to compute the leading contribution to the derivatives of $F[g] := F(x[g])$, that diverge in the vicinity of the singularity,
\begin{equation}
F'[g] = \frac{\Upsilon[g] \Upsilon''[g]}{\Upsilon'[g]^2} \approx \frac{\Upsilon[g_\star]}{\Upsilon''[g_\star]\,(g-g_\star)^2} \to - \infty ~, \qquad F''[g] \approx -2 F'[g]^2 ~,
\end{equation}
the negative sign in $-\infty$ coming from the fact that $g_\star$ is a maximum of $\Upsilon$. Consequently, the tensor and sound potentials diverge with opposite sign,
\begin{equation}
{\bf c}_2^2 \approx \frac{(d-1)L^2\,f_\star}{(d-4)\, r^2}F'~, \qquad {\bf c}_0^2 \approx \frac{(d-1)L^2\,f_\star}{(d-2)\, r^2}(-3F') ~.
\end{equation}
this being a clear sign of unavoidable instability. Again, since we have expanded around $r_\star$, not only the naked singularity is unstable but any black hole solution whose horizon is close enough to it, $r_+ \simeq r_\star$; {\it i.e.}, $\kappa \gtrsim \kappa_\star$. Hence, the black hole solutions cannot be continuously connected to the singular solution by any physical process since the instability would inevitably show up with increasing importance before the singularity is attained, driving the system somewhere else. This is an important point, as long as the threshold between type (b) black holes and naked singularities is not an extremal solution. 
As the horizon approaches the singularity, given that $F'$ diverges in that limit, the instability is increasingly important and dramatic when we get there.

\section{Instabilities and black hole evaporation}

In our quest to qualitatively understand in which situations Lovelock black holes become unstable we pass on to the case of evaporating black holes. This case is relevant as in previous analysis the instabilities under study seem to appear whenever the black hole horizon approaches {\it too much} the singularity. 
This is for instance what happens for type (b) spherical solutions. As the black hole evaporates it looses mass approaching the critical value (for which the temperature diverges) in finite time. This would leave behind a naked singularity but we already proved that neither the extremal solution nor the naked singularity can be reached, the solution becomes unstable as the horizon gets {\it close} to the singularity.

The same happens for the other cases analyzed throughout this chapter and the same will also hold for other classes of evaporating spherical black holes. We have also seen that for type (a) branches in $d=2K+1$ the black hole solution becomes unstable as well before reaching the $r_+=0$ state, that in this case is extremal. The black hole would spend an infinite amount of time to become extremal but just a finite amount to reach the instability. Hence, the instability seem to play a r\^ole in the evaporation process of black holes in Lovelock gravities, at least for spherical topology.  

The remaining case to be understood is that of type (a) solutions, either on the EH or dS branches, for dimensions bigger than the critical, $d>2K+1$. As the mass of the black hole shrinks to zero the horizon also shrinks approaching the central singularity. In this case the singularity can never become naked but still instabilities show up before the black hole shrinks to zero size. In \cite{Takahashi2009a, Takahashi2010e} it has been observed that when all possible Lovelock couplings are turned on -- the highest one being positive, $c_K>0$ -- the instability always shows up, in even and odd dimensions. Here we will generalize that analysis. 

Expanding the tensor and scalar potentials around $g\rightarrow\infty$ we simply get
\begin{equation}
{\bf c}_2^2\approx\frac{(d-3K-1)L^2\,f}{K(d-4)\,r^2}\quad , \qquad
{\bf c}_0^2\approx\frac{(d-K-1)L^2\,f}{K(d-2)\,r^2}~.
\end{equation} 
Then any spherical branch under the above conditions will present a tensor instability in the small mass limit for $2K+1< d<3K+1$, where $K$ is the order of the Lovelock theory. Einstein-Hilbert is a special case from this perspective. The $d=3K+1$ is special and we need to go to the following order in the tensor potential. In that case whether the black hole is stable or not will depend on the actual values of $(c_{K-1}/c_K)^2$ and $c_{K-2}/c_K$. The case of Einstein-Hilbert gravity is stable in any number of dimensions.

Summarizing, for generic Lovelock theory we have encountered instabilities at the end of the evaporation process of any type (a) spherical black hole in dimensions lower than $3K+1$, whereas for type (b) the instability appears for any spacetime dimensionality, regardless of the asymptotics. The latter dimensionality, $d=3K+1$, has to be considered more carefully. Whether this instability is pointing towards an inconsistency of the theory that forces us to constrain $K \leq\left[\frac{d-1}{3}\right]$ or, else, it can be ascribed to a physically sensible process taking place during the black hole evaporation has yet to be answered.

\section{Discussion}

Perturbative analysis of exact solutions appears to be an extremely useful tool to get a deeper understanding of the dynamics of gravity in four and higher dimensions, specially for Lovelock theories. In this chapter we have presented and analyzed the graviton potentials for generic Lovelock gravities in some particularly simple regime. Still, despite the simplicity of the  approach adopted, the results are general enough to provide some valuable information about the behavior of black holes in these higher curvature theories. 

Lovelock solutions display several seemingly pathological features, ranging from naked singularities to violations of the third law or discontinuous changes on the horizon radius -- and consequently also on their associated thermodynamic variables. We have shown that all the puzzling properties of these solutions are ruled out once their stability is considered. Before any naked singularity may show up the corresponding solution becomes unstable, this applying as well for black holes undergoing any physical process as for the collapse of any kind of matter. Therefore, naked singularities cannot be formed under the evolution of generic initial conditions. The existence of such solutions would require extremely {\it fine-tuned} initial data for the gravitational collapse. Any perturbed set of initial conditions will not end up with the formation of a naked singularity due to the instability. To the extent we could challenge it, the cosmic censorship hypothesis seems at work in Lovelock theory. This seems relevant from the point of view of the uses of Lovelock gravities in the framework of the AdS/CFT correspondence. 

The cosmic censorship hypothesis can be saved in Lovelock gravities, at least for now. The same happens also for the third law of thermodynamics, the solutions that would lead to the  violations described in this chapter being also unstable. As a consequence, the thermodynamic quantities for the remaining (stable) ranges of parameters are always continuous.

These results provide a new insight into these graviton instabilities, too often taken as pathological. Much on the contrary, they seem to acquire physical significance and play an important r\^ole, healing Lovelock theory from what otherwise would lead to an ill-defined behavior. Their analysis is crucial for the understanding of the dynamics of black holes in Lovelock theory and possibly other theories containing higher powers of the curvature. In particular, we have shown that Lovelock black holes with spherical symmetry generically become unstable as they evaporate. In most cases there is a mass gap between the lightest stable black hole and the corresponding maximally symmetric vacuum. The instability may be a hint pointing to the existence of new solutions that fill this gap. The only possibilities we foresee for such hypothetical states are {\it stars} made either of regular matter or {\it gravitational hair}.\footnote{See \cite{Dias2011} for an example of the class of solutions we are referring to.} In some cases we might need to break spherical symmetry and provide some angular momentum, as the perturbations considered do.

We have focused mostly on spherical symmetry and the EH branch in this chapter. The results can be straightforwardly extended to AdS/dS branches. In the former case, hyperbolic symmetry is the relevant one, and many results mirror those of spherical black holes in dS branches. For hyperbolic AdS branches the stability of the black hole solutions provides an upper bound on the gravitational mass coming, again, from maxima of the characteristic polynomial. The bound should not be of fundamental nature, thus we expect more massive solutions in the form of hairy black holes, the {\it dressing} of the horizon providing the extra required energy. The fact that the instability is restricted to a finite (radius) spacetime region seems to support this hypothesis. This would be the region filled by our matter configuration or, better, our lump of gravitational energy or {\it geon}, the rest of the spacetime remaining untouched. A more detailed analysis is necessary in order to confirm or disprove this intuition.

\begin{figure}
\centering
\includegraphics[width=0.45\textwidth]{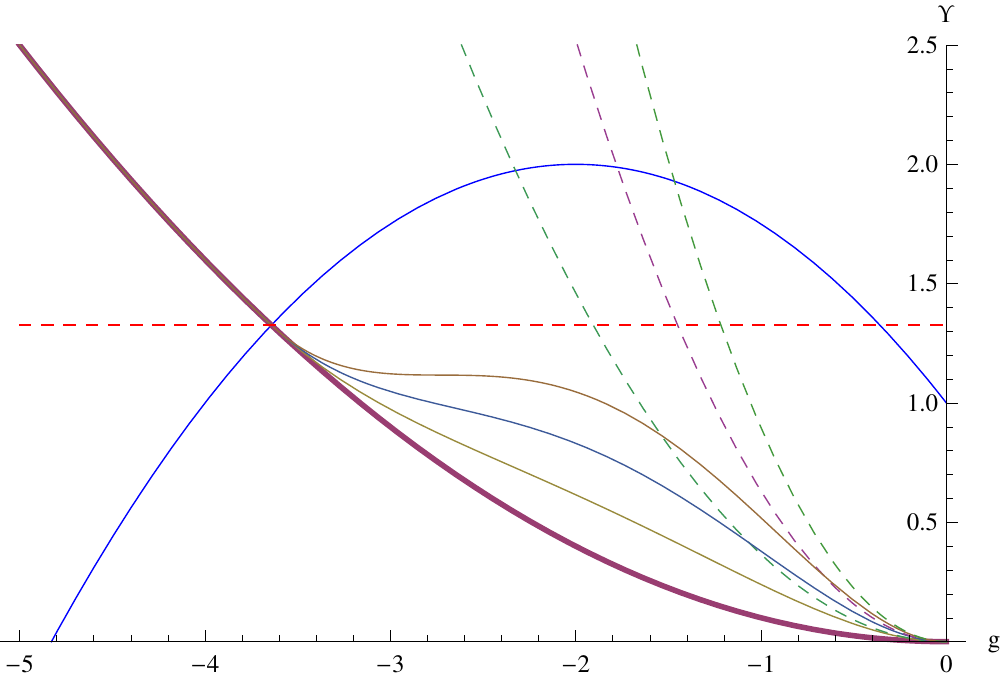} ~~ \includegraphics[width=0.45\textwidth]{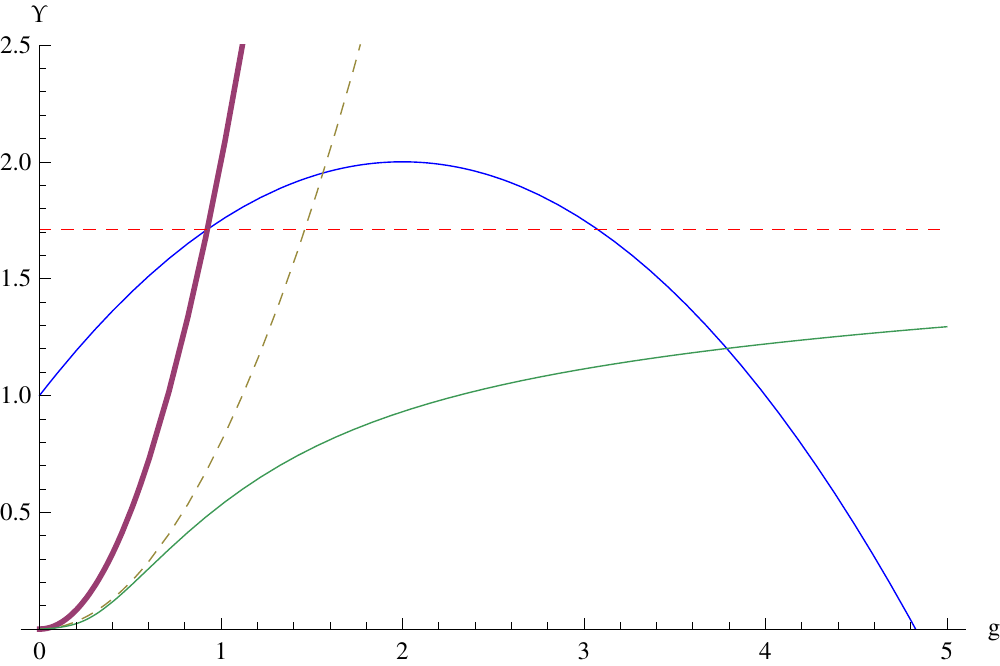}
\caption{On the left hairy black hole configurations respecting the stability bound (horizontal line) for an AdS branch. As we can see the asymptotic mass (corresponding to the dashed lines) in each case would violate the bound but some mass remains outside the horizon. On the right we considered the case of a type (b) EH branch with spherical symmetry. In this situation we may have either a black hole (in purple; for masses higher than the lower stability bound) or a star (in green; below that threshold).  }
\label{stars}
\end{figure}

Coming back to the spherical case, the instability is associated with a particular value of $g$, $g_\ast < g_\star$, above which some of the potentials become negative. The threshold of the instability is then the mass $\kappa_\ast$ for which the radius of the horizon is $r_\ast = g_\ast^{-1/2}$. For lower values of the mass, we can translate the stability constraints into a bound on the amount of matter that can be contained in a hypersphere of radius $r$. For any quantity of matter contained in such a hypersphere, with mass $\kappa(r)$, the radius should be bigger than the would be unstable region for the same mass. For a continuous distribution of matter, this has to be verified for all values of the radius down to $r=0$ so that the configuration is stable. The equation for the metric function $g$ should be schematically modified \cite{Paulos2011} in that case to
\begin{equation}
\Upsilon[g(r)] = \frac{\kappa(r)}{r^{d-1}} \leq \Upsilon[g_\ast] ~.
\label{denssity}
\end{equation}
Of course, to fully address this issue we would need to consider adding matter fields that in principle may change the form of the stability criteria. The characteristic polynomial plays the r\^ole of an {\it effective density} and its value at the threshold of stability can be interpreted as the maximal one so that the instability is avoided; see Figure \ref{stars}.
%
%
We could have played the same game with the value of $g$ for the singularity $g_\star$, and we get another, less stringent, bound on $\Upsilon$ so that the naked singularity is not formed, $\Upsilon[g]<\Upsilon[g_\star]$. Notice that the bounds set by stability are always finite whereas the one that avoid the nakedness of the singularity is infinite in the case of $r=0$ singularities. The polynomial $\Upsilon[g_\star]$ diverges in that case and any finite density matter configuration would be regular. 

Even in the case these matter configurations are not possible, there is another possible endpoint for gravitational collapse. The collapsing matter may just disperse again as it is not trapped by any event horizon. In case the symmetry is relaxed the matter will not collapse exactly to one point in such a way that the density may remain finite through the evolution. Then it may disperse back to infinity or form some type of bound state. To be conclusive in this point it seems unavoidable to explore the inclusion of matter.

It would be interesting to investigate the status in Lovelock theory of the nonlinear instability of AdS that was found in general relativity under small generic perturbations \cite{Bizon2011}. Such an instability is triggered by a mode mixing that produces energy diffusion from low to high frequencies. This mechanism can be understood as a result of the interaction of geons that end up producing a small black hole \cite{Dias2011}. Given the existence of a mass gap in the Lovelock black hole spectrum, it is natural to wonder whether AdS does not display turbulent instabilities when small generic perturbations are considered within the higher curvature dynamics. The only piece of information on this direction that we are aware of is the analysis of Choptuik's critical phenomenon \cite{Choptuik} in the case of LGB theory without cosmological constant \cite{Golod}.

A final comment is in order. There is a different notion of instability that may supersede that unravelled by perturbative analysis in this chapter. Black hole solutions are subjected to the laws of thermodynamics and, as such, they may become unstable at the threshold of a (for instance, Hawking-Page) phase transition. In a thermal bath, it has been recently shown, indeed, that Lovelock theory admits generalized phase transitions allowing for jumps between the different branches \cite{Camanho2012,Camanho2013}. This phenomenon, that will be described in detail in the next chapter, further enriches the rules of the game and make it cumbersome and interesting to reinstate a basic question with which we would like to finish this section: are higher curvature gravities like Lovelock theory fully consistent?

\chapter{\bfseries\itshape Bubbles and new phase transitions}
\chaptermark{Bubbles and new phase transitions}
\label{genHP}

\vspace{.6cm}

\begin{quotation}
\flushright
{\it ``Nothing has such power to broaden the mind\\ as the ability to investigate systematically and truly\\ all that comes under thy observation in life.''}\\

\vspace{.3cm}

Marcus Aurelius
\end{quotation}

\vspace{3cm}


\noindent Phase transitions between two competing vacua of a given theory are a quite common phenomenon in physics. They occur when some parameter of the system is varied so that the (free) energy of the initial vacuum becomes greater than the other. If the energy barrier between the two is big enough, the system may stay in the false vacuum for some time (metastability), and proceed to decay via quantum tunneling or, at finite temperature, jump over the wall due to a thermal kick. The decay of metastable systems usually proceeds by nucleation of bubbles of true vacuum inside the false vacuum. In field theories at zero temperature, this was first studied by Coleman in his classic paper \cite{Coleman1977}. There, he introduced Euclidean methods for computing the probability of the quantum nucleation of a bubble, whose dynamics, after nucleation, may be followed classically. In the first (tree-level) semiclassical approximation, the probability of bubble nucleation is given by
\begin{equation}
P \propto e^{-\widehat{\mathcal{I}}} ~,
\label{prob}
\end{equation}
where $\widehat{\mathcal{I}}$ is the Euclidean action of the system evaluated at the appropriate solution; in this case, the instanton. It is a time-dependent solution which, in the simplest case of a particle in a potential, starts and ends its trajectory at the bottom of the false vacuum (given that the potential in the Euclidean section is the negative of its Lorentzian counterpart, it is a local maximum).
\begin{figure}
\centering
\includegraphics[width=13cm]{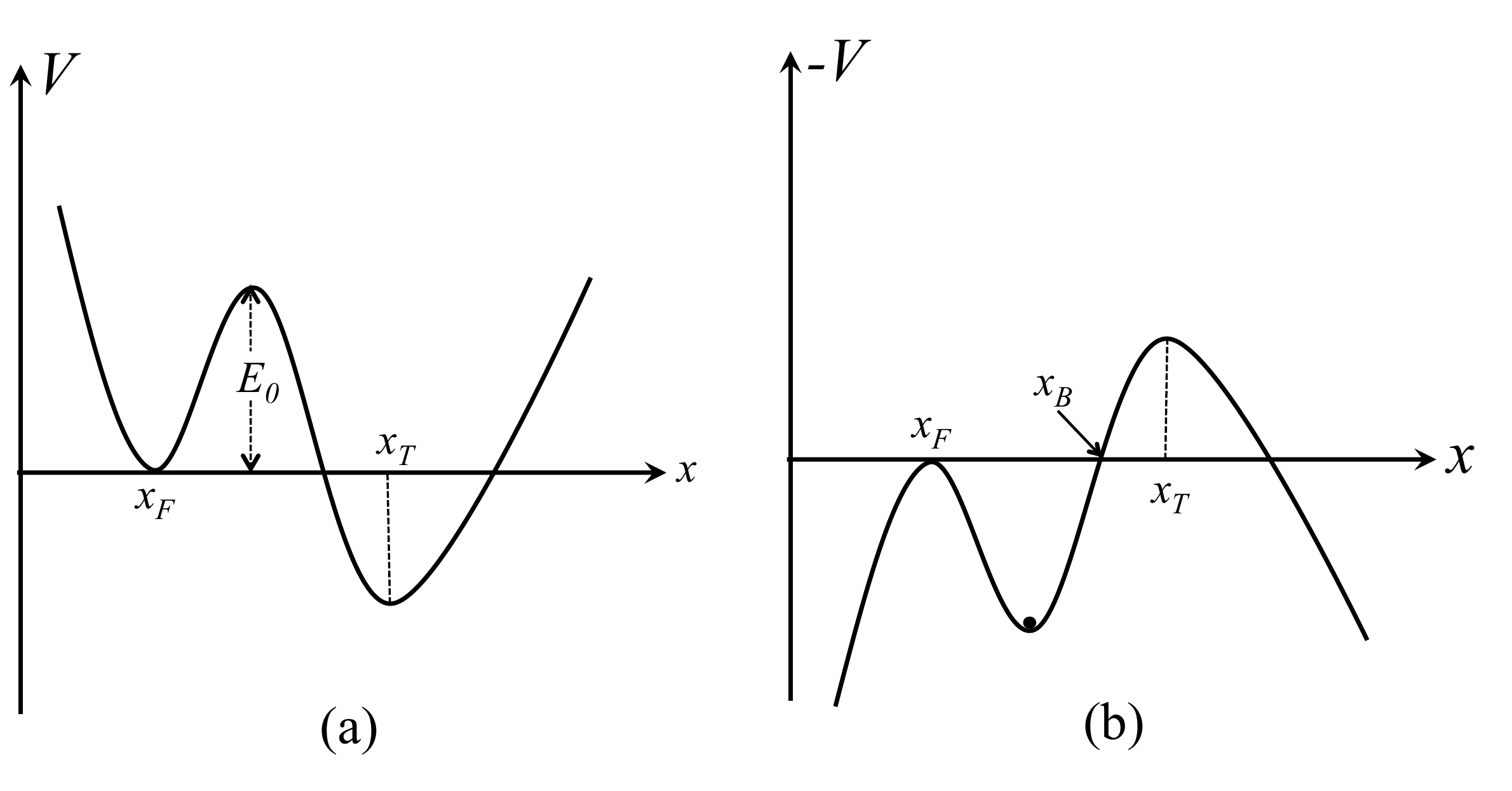}
\caption{In (a) we depict a particle in a potential with two local minima.At $x_F$, the {\it false vacuum}, while the true vacuum is at $x=x_T$. There is a potential barrier of height $E_0$ between the two vacua. In (b) the Euclidean counterpart of the same system. The potential is now $-V$.}
\label{pot}
\end{figure}
This is the point $x=x_F$ in figure \ref{pot}(b), where the particle starts its trajectory, then bounces at $x=x_B$, and finally gets back to $x_F$ in infinite time. The work of Coleman generalizes this mechanism to a scalar field theory. The instanton corresponds to  a scalar field configuration with  SO(4) symmetry. The technique was then further generalized by Linde, who considered a scalar field at finite temperature \cite{Linde1981}. In this case, the probability of nucleation is still given by (\ref{prob}), but the Euclidean action has to be evaluated using a different classical configuration, which has  SO(3) symmetry. In the mechanical example of figure \ref{pot}, if the temperature is not high enough, this is the solution oscillating inside the Euclidean well, between two points in $(x_F,x_B)$. The period  $\beta$ of that solution is identified with the inverse temperature, $\beta=1/T$. This period has a minimum for small oscillations at the bottom of the well. For temperatures higher than that, one may use the static solution with the particle at the bottom of the well, which, of course, has any periodicity. It is easy to see that for this case, (\ref{prob}) gives precisely the Boltzmann factor $e^{-\beta E_0}$ one expects for the probability of the particle to jump over the barrier. This is the {\it sphaleron} or {\it thermalon} \cite{Klinkhamer1984, Gomberoff2004}. 

Gravitational instantons where first discussed by Coleman-De Luccia in \cite{Coleman1980}, where a scalar field with a potential interacts with a dynamical metric. Now, the different vacua correspond to solutions with different cosmological constants. The false vacuum decays by nucleating an expanding bubble of true vacuum. Later, Brown and Teitelboim\cite{Brown1987,Brown1988}  found an analogue instanton when gravity was coupled to an electromagnetic $3$-form potential and its sources, electrically charged membranes.
In that case, there are infinitely many false vacua, and the decay may proceed many times, changing the (positive) cosmological constant at each step. The authors showed that this mechanism could relax the cosmological constant, providing a possible mechanism for understanding the {\it cosmological constant problem}  \cite{Weinberg1989}. For finite temperatures, this same physical system may also decay. Now, a thermalon solution controls the decay rate, and, interestingly enough, the decay of a pure de Sitter geometry,  turns out to leave a black hole behind\cite{Gomberoff2004}. In this chapter we show an analogue process that occurs in higher-curvature theories of gravity. In general, this theories contain degenerate vacua even in the absence of matter. Furthermore, one vacuum may decay into the other by nucleating bubbles made of nothing but gravity itself \cite{Camanho2012,Camanho2013a,Camanho2013}.


Due to the non-linearity of the equations of motion, these theories generally admit more than one maximally symmetric solution, $R_{\mu \nu \alpha\beta}=\Lambda_i(g_{\mu\alpha}g_{\nu\beta}-g_{\mu\beta}g_{\nu\alpha})$; (A)dS vacua with effective cosmological constants $\Lambda_{i}$, whose values are determined by a polynomial equation \cite{Boulware1985a}, 
\vskip-2mm
\begin{equation}
\Upsilon [\Lambda] \equiv \sum_{k=0}^{K}c_{k}\,\Lambda^{k} = c_{K}\prod_{i=1}^{K}\left( \Lambda -\Lambda _{i}\right) =0 ~.
\end{equation}
$K$ being the highest power of curvature (without derivatives) in the field equations. $c_0=1/L^2$ and $c_1=1$ give canonically normalized cosmological and EH terms, $c_{k\geq 2}$ are the LGB and higher order couplings (see chapters \ref{chp:LLgravity} and \ref{chp:LLbh} for more details).

Any vacua is \`a priori suitable in order to define boundary conditions for the gravity theory we are interested in; {\it i.e.} we can define sectors of the theory as classes of solutions that asymptote to a given vacuum \cite{Camanho2011a}. In that way, each branch has associated static solutions, representing either black holes or naked singularities,
\vskip-2mm
\begin{equation}
ds^{2}=-f(r)\,dt^{2}+\frac{dr^{2}}{g(r)}+r^{2}\ d\Omega_{d-2}^{2} ~, \qquad \qquad f,g \xrightarrow{r\rightarrow \infty} -\Lambda_i r^2 ~,
\label{bh2}
\end{equation}
and other solutions with the same asymptotics. The main motivation of the present work is that of studying transitions between different branches of solutions. This is important in order to investigate whether a new type of instability involving non-perturbative solutions occurs in the theory. This new kind of phase transitions have been recently investigated in the context of LGB \cite{Camanho2012} and Lovelock gravities \cite{Camanho2013}. 

\section{Higher order free particle}

The existence of branch transitions in higher curvature gravity theories is a concrete expression of the multivaluedness problem of these theories. In general the  canonical momenta, $\pi_{ij}$, are not invertible functions of the velocities, $\dot{g}^{ij}$ \cite{Teitelboim1987}. An analogous situation may be illustrated by means of a simple one-dimensional example \cite{Henneaux1987b}. 

Consider a free particle lagrangian containing higher powers of velocities,
\vskip-1mm
\begin{equation}
L(\dot{x})=\frac{1}{2}\dot{x}^2-\frac13\dot{x}^3+\frac1{17}\dot{x}^4
\label{paction}
\end{equation}
In the Hamiltonian formulation the equation of motion just implies the constancy of the conjugate momentum, $\frac{d}{dt}p=0$. However, being this multivalued (also the hamiltonian), the solution is not unique. Fixing boundary conditions $x(t_{1,2})=x_{1,2}$, an obvious solution would be constant speed 
$
\dot{x}=(x_2-x_1)/(t_2-t_1)\equiv v
$
but we may also have jumping solutions with constant momentum and the same mean velocity. Obviously for that to happen at least one of the degenerate velocities has to be bigger than $v$, the other one being smaller.

\begin{figure}
\centering
\includegraphics[width=.6\textwidth]{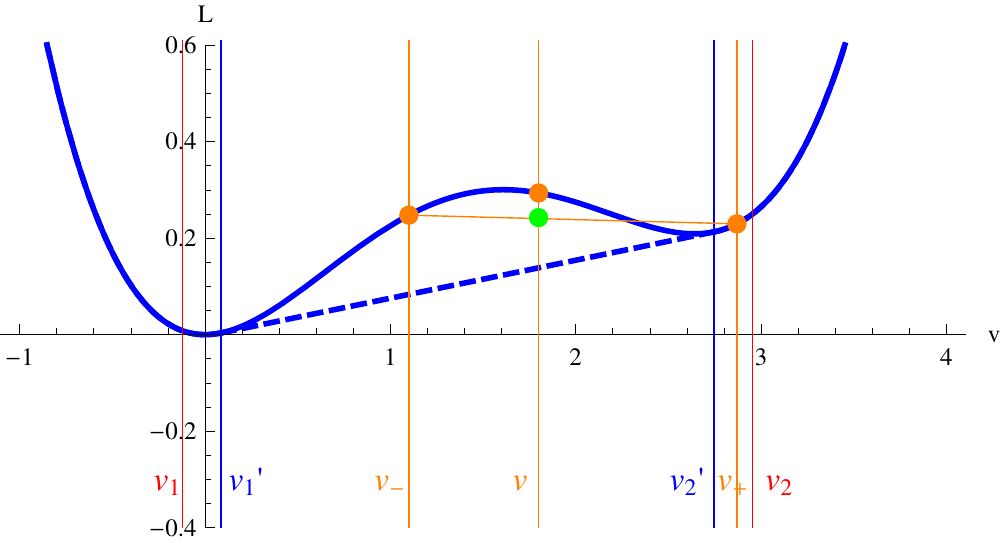}  \\
\includegraphics[width=.6\textwidth]{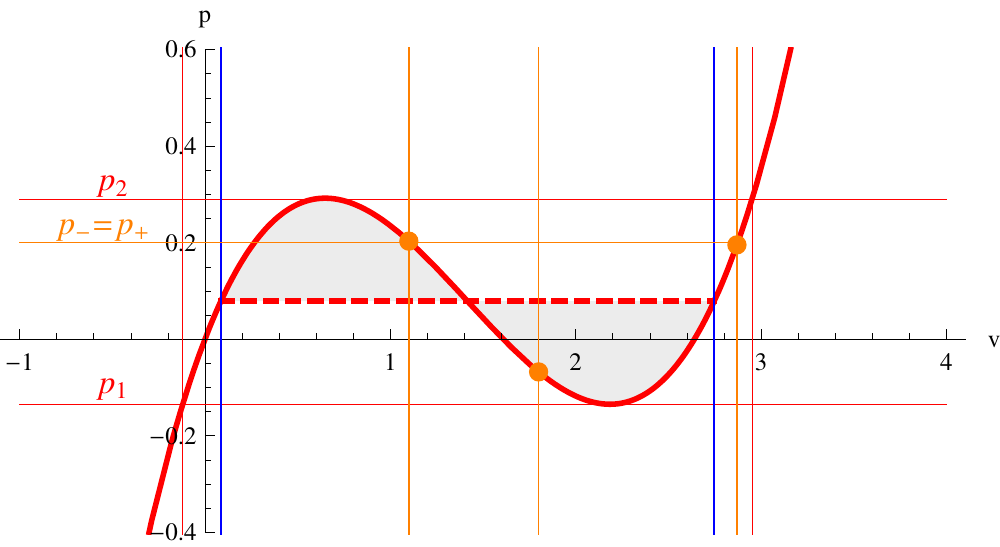} 
\caption{Lagrangian and momentum for the action (\ref{paction}). For the same mean velocity $v$, the action is lower for jumps between $v_\pm$ (big dot) than for constant speed, the minimum action corresponding to the value on the dashed line ({\it effective} lagrangian).}
\label{L-v}       
\end{figure}

\begin{figure}
\centering
\includegraphics[width=.6\textwidth]{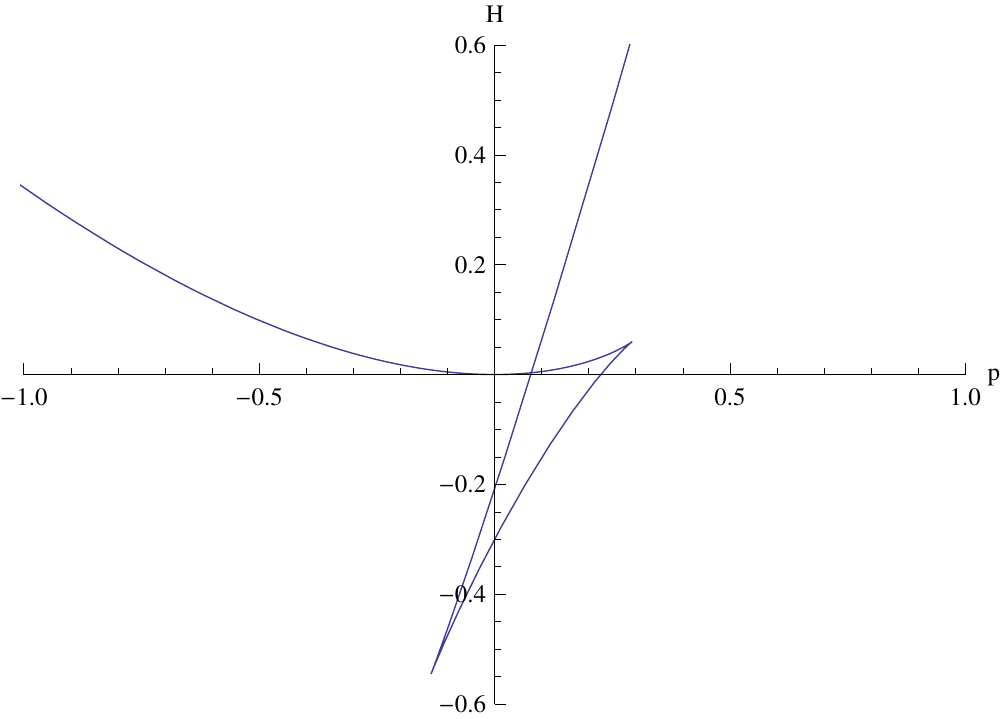} 
\caption{Hamiltonian as a function of the momentum for the action (\ref{paction}). Multivalued momenta correspond to the swallowtail part of the curve whereas those corresponding to the {\it effective} lagrangian are the ones without that part of the curve, just the two upper branches. The crossing correspond to the {jump} depicted in Figure \ref{L-v} as a dashed line}
\label{H-p}       
\end{figure}

In our example, for mean velocities in the range $(v_1,v_2)$, corresponding to multivalued momentum, the solutions are infinitely degenerate. The jumps may occur at any time and unboundedly in number, as long as the mean velocity is the same. This degeneracy is lifted however when the value of the action is taken into account. The minimal action path is the na\"ive one for mean velocities outside the range $(v_1',v_2')$ whereas in that range it corresponds to arbitrary jumps between the two extremal velocities. One can actually compare the action for both kinds of trajectories directly in the figure, we have depicted one example. In the top figure \ref{L-v} we have plotted the lagrangian as a function of the velocity. When this is constant, $\dot{x}=v$, the action will just be the lagrangian $L(v)$ multiplied by the time span of the trajectory. On the other hand, for a jumping trajectory, the mean lagrangian can be found as the intersect of the straight line joining the values of the lagrangian for both values of the velocity, $\dot{x}=v_\pm$, and $\dot{x}=v$,
\be
\bar{L}(v;v_\pm)=L(v_-)+\frac{v-v_-}{v_+-v_-}L(v_+)
\ee
 Notice in the figure the action corresponding to $v$ is bigger than the one corresponding to jumps between $v_{\pm}$ (green dot). This mean lagrangian is minimal for the lowest lying such straight line and it is easy to see that this corresponds to the dashed line for mean velocities, $v_1'<v<v_2'$. Outside that range the momentum is a convex function of the velocity and any line joining to points on opposite sides of a given velocity will yield necessarily a higher action. The effective lagrangian (dashed line) is actually also a convex function of the velocities and the effective momentum dependence corresponds to the Maxwell construction. In that case the dashed line corresponds to jumps conserving both the energy and the momentum, \ie the crossing point on figure \ref{H-p}. Also the lowest energy state does not generically correspond to $\dot{x}=0$ even classically what has been referred to as a {\it time crystal} \cite{Shapere2012,Wilczek2012}. The quantum mechanical version of the model is well defined \cite{Henneaux1987b}.

In the presence of a potential, the na\"ive choice of continuous velocities runs into problems as we would hit the degenerate points $\frac{d}{d\dot{x}}p=0$ where the acceleration is not well defined. In the above effective approach however the momentum is a monotonous function of the velocity except at the {\it transition} that correspond to conservation of not only momentum but energy.


\section{Generalized Hawking-Page transitions}

In the context of general relativity in asymptotically AdS spacetimes, the Hawking-Page phase transition \cite{HP} is the realization that above certain temperature the dominant saddle in the gravitational partition function comes from a black hole, whereas for lower temperatures it corresponds to the thermal vacuum. The {\it classical} solution is the one with least Euclidean action among those with a smooth Euclidean section. In the case of Lovelock gravities the occurrence of these transitions was analyzed in chapter \ref{chp:thermo}. 

When one deals with higher curvature gravity however there is a crucial difference that has been overlooked in the literature. In addition to the usual continuous and differentiable metrics (\ref{bh2}), one may construct distributional metrics by gluing two solutions corresponding to different branches across a spherical shell or {\it bubble} \cite{wormholes,Garraffo2008b}. The resulting solution will be continuous at the bubble --with discontinuous derivatives, even in absence of matter. The higher curvature terms can be thought of as a sort of matter source for the Einstein tensor. The existence of such {\it jump} metrics, as for the one-dimensional example, is due to the multivaluedness of momenta in the theory.

In the gravitational context, continuity of momenta is equivalent to the junction conditions that need to be imposed on the bubble. In the EH case, Israel junction conditions \cite{Israel1967}, being linear in velocities, also imply the continuity of derivatives of the metric. The generalization of these conditions for higher curvature gravity contain higher powers of velocities, thus allowing for more general situations.

Static bubble configurations, when they exist, have a smooth Euclidean continuation. It is then possible to calculate the value of the action and compare it to all other solutions with the same asymptotics and temperature. This analysis has been performed for the LGB action \cite{Camanho2012} for unstable boundary conditions \cite{Boulware1985a}. The result suggests a possible resolution of the instability through bubble nucleation.


The phenomenon described here is quite general. It occurs also for general Lovelock gravities \cite{Camanho2013} and presumably for more general classes of theories. One may even think of the possibility of having different gravity theories on different sides of the bubble. This has a straightforward physical interpretation if we consider the higher order terms as sourced by other fields that vary across the bubble. For masses above $m^2>\|\Lambda_{\pm}\|$ a bubble made of these fields will be well approximated by a thin wall and we may integrate out the fields for the purpose of discussing the thermodynamics. If those fields have several possible vacuum expectation values leading to different theories we may construct interpolating solutions in essentially the same way discussed in this chapter. In this case the energy carried by the bubble can be interpreted as the energy of the fields we have integrated out.

\section{Junction conditions}

We are interested in configurations involving a (timelike, $n^2=1$) junction surface dividing two regions, each of them given by a solution corresponding to a different branch of the Lovelock action. To this end, we will split the geometry as $\mathcal{M} = \mathcal{M}_-\, \cup\, ( \Sigma \times \xi )\, \cup\, \mathcal{M}_+$, where $\mathcal{M}_\mp$  denote the interior/exterior regions respectively, $\Sigma$ is the codimension one junction (hyper)surface, and $\xi \in [0,1]$ is a real parameter used to interpolate both regions. The matching condition can be obtained from an action principle constructed by means of two auxiliary quantities \cite{Gravanis2009}
\begin{equation}
E^a = \xi\, e^a_+ + (1-\xi)\, e^a_- ~, \qquad \Omega^{ab} = \xi\, \omega^{ab}_+ + (1-\xi)\, \omega^{ab}_- ~,
\end{equation}
and the associated generalized curvatures
\begin{equation}
\mathcal{R}_\xi^{ab} = d \Omega^{ab} + \Omega_{~c}^{a} \wedge \Omega^{cb} = \xi\, R^{ab}_+ + (1-\xi)\, R^{ab}_- - \xi (1-\xi)\, ({\omega_+} - {\omega_-})_{~c}^{a} \wedge ({\omega_+} - {\omega_-})^{cb} ~,
\label{genR}
\end{equation}
and $\mathfrak{F}^{ab} = \delta\Omega^{ab} + \mathcal{R}^{ab}$, where $\delta$ is the exterior derivative on the convex simplex in the space of connections, $\delta = d\xi\,\frac{\partial}{\partial \xi}$. That is, $\delta\Omega^{ab} = d\xi \wedge (\omega^{ab}_+ - \omega^{ab}_-)$. These curvatures are used to construct the secondary characteristic classes \cite{ChernSimons}. Their use in the context of Chern--Simons gravities has been considered in \cite{Mora2006}.

In the bulk regions $\xi$ takes a fixed value, $\xi=0,1$ respectively in $\mathcal{M}_\mp$, whereas it runs from 0 to 1 in $\Sigma$ and has to be integrated over. The so-called {\it secondary} action is obtained by substituting $e^a$ by $E^a$ and $R^{ab}$ by $\mathfrak{F}^{ab}$ in (\ref{LLaction}),
\begin{equation}
\widetilde{\mathcal{L}}_k = \e_{a_1 a_2 \cdots a_d}\mathfrak{F}^{a_{1}a_{2}} \wedge \cdots \wedge \mathfrak{F}^{a_{2k-1}a_{2k}} \wedge E^{a_{2k+1} \cdots a_{d}} ~,
\label{LLbulkjunc}
\end{equation}
We can readily expand the secondary action in powers of $\delta\Omega$, taking into account that $\delta\Omega \wedge \delta\Omega = 0$. The leading term corresponds to the bulk integrals on $\mathcal{M}_\mp$ and contributes to the standard Lovelock action (\ref{LLaction}), while the first order term captures the integral along $\xi$ on $\Sigma$, 
\begin{equation}
\widetilde{\mathcal{Q}}_k = - k \int_0^1\! d\xi\ (\omega_+ - \omega_-)^{a_{1}a_{2}} \wedge \mathcal{R}^{a_{3}a_{4}} \wedge \cdots \wedge \mathcal{R}^{a_{2k-1}a_{2k}} \wedge E^{a_{2k+1}\cdots a_d}\e_{a_{1} \cdots a_{d}} ~,
\label{matchaction}
\end{equation}
where the minus sign has been chosen so that it coincides with the sign in section \ref{bdyterms}. If we further impose the continuity of the metric such that $E^a=e^a$ for some choice of vielbein, we may also define $\theta^{ab}_\pm = \omega^{ab}_\pm - \omega^{ab}_0$. From this $\omega^{ab}_+ - \omega^{ab}_-$ can be replaced by $\theta^{ab}_+ - \theta^{ab}_-$, which is zero unless one of the indices lies along the normal direction. Thus, the only relevant contributions to $\mathfrak{R}^{ab}$ are those whose two indices are tangent to $\Sigma$. Now,
\begin{equation}
\mathcal{R}^{ab} = R_0^{ab} + {\Theta}_{~c}^{a} \wedge {\Theta}^{cb} + \ldots ~,
\label{genRd}
\end{equation}
where $R_0^{ab}$ is the intrinsic curvature of $\Sigma$ computed from $\omega_0^{ab}$, ${\Theta}^{ab} \equiv \xi\, {\theta}_+^{ab} + (1-\xi)\, {\theta}_-^{ab}$, and the dots amount to terms having at least one index in the normal direction. Recalling the relation between the second fundamental form and the extrinsic curvature \reef{extrinsiK}, we can finally write
\begin{equation}
\widetilde{\mathcal{Q}}_k =  2 k \int_0^1\! d\xi (K_+ - K_-)^{A_1} \wedge \mathfrak{R}_{\rm j}^{A_{2}A_{3}} \wedge \cdots \wedge \mathfrak{R}_{\rm j}^{A_{2k-2}A_{2k-1}} \wedge e^{A_{2k}\cdots A_d}\e_{A_{1} \cdots A_{d}} ~,
\label{matchactionbis}
\end{equation}
where the junction generalized curvature $\mathfrak{R}_{\rm j}^{AB}(\xi)$, is given by $\mathfrak{R}_{\rm j}^{AB}(\xi) \equiv R_0^{AB} - (\xi\, K_+^A + (1-\xi)\, K_-^A) \wedge (\xi\, K_+^B + (1-\xi)\, K_-^B)$ (we can recover the boundary generalized curvature \reef{intrinsicaction} by switching-off $K^{a}_+ \to 0$). We can surprisingly split $\widetilde{\mathcal{Q}}_k = \mathcal{Q}^{-}_k - \mathcal{Q}^{+}_k$, where
\begin{equation}
\mathcal{Q}_k^{\pm} = - 2 k \int_0^1\!\! d\xi \ K^{A_1}_{\pm} \wedge \mathfrak{R}_{0\pm}^{A_2 A_3}(\xi) \wedge \cdots \wedge\mathfrak{R}_{0\pm}^{A_{2k-2} A_{2k-1}}(\xi) \wedge e^{A_{2k}\cdots A_d}\e_{A_{1} \cdots A_{d}} ~,
\label{twosideaction}
\end{equation}
with $\mathfrak{R}_{0\pm}^{AB}(\xi) \equiv R_0^{AB} - \xi^2\, K_\pm^A \wedge K_\pm^B$, either term coming from each boundary region. That is, the junction acts as a two-sided boundary. All in all, the $k^{\text{th}}$ contribution to the total action can be written in terms of quantities coming from both sides,
\begin{equation}
\mathcal{I}_k = \left(\int_\mathcal{M_-}\!\!\!\mathcal{L}_k^{-} - \int_\Sigma \mathcal{Q}_k^{-}\right) + \left(\int_\mathcal{M_+}\!\!\!\mathcal{L}_k^{+}+\int_\Sigma\mathcal{Q}_k^{+}-\int_{\partial \mathcal{M}}\!\!\!\mathcal{Q}_k^{+}\right) ~,
\label{splitaction}
\end{equation}
the relative plus sign in the second term coming from the reverse orientation of the surface in that region. The last term is irrelevant for the purposes of the present letter and will not be considered in the following, it just makes the whole outer contribution to vanish if we take the junction surface to infinity. The infinitesimal variation of the boundary action with respect to $\omega_{\pm}$ gives two terms. One is a total derivative. The other cancels with the total derivative term coming from the bulk. The surface contribution to the equations of motion is just given by the variation with respect to the frame, that correspond on each side to the canonical momentum at the surface. As explained in section \ref{bdyterms}, the junction conditions amount just to continuity of the momenta across the hypersurface $\Sigma$ defined in \reef{momenta},
\begin{equation}
\pi^+_{AB} = \pi^-_{AB} ~.
\label{junction}
\end{equation}
In the particular case we are interested in, given that all the forms involved in the above expression are diagonal, we should only care on those components.

\subsection{Thermalon configuration}

\begin{SCfigure}
\centering
\includegraphics[width=.45\textwidth]{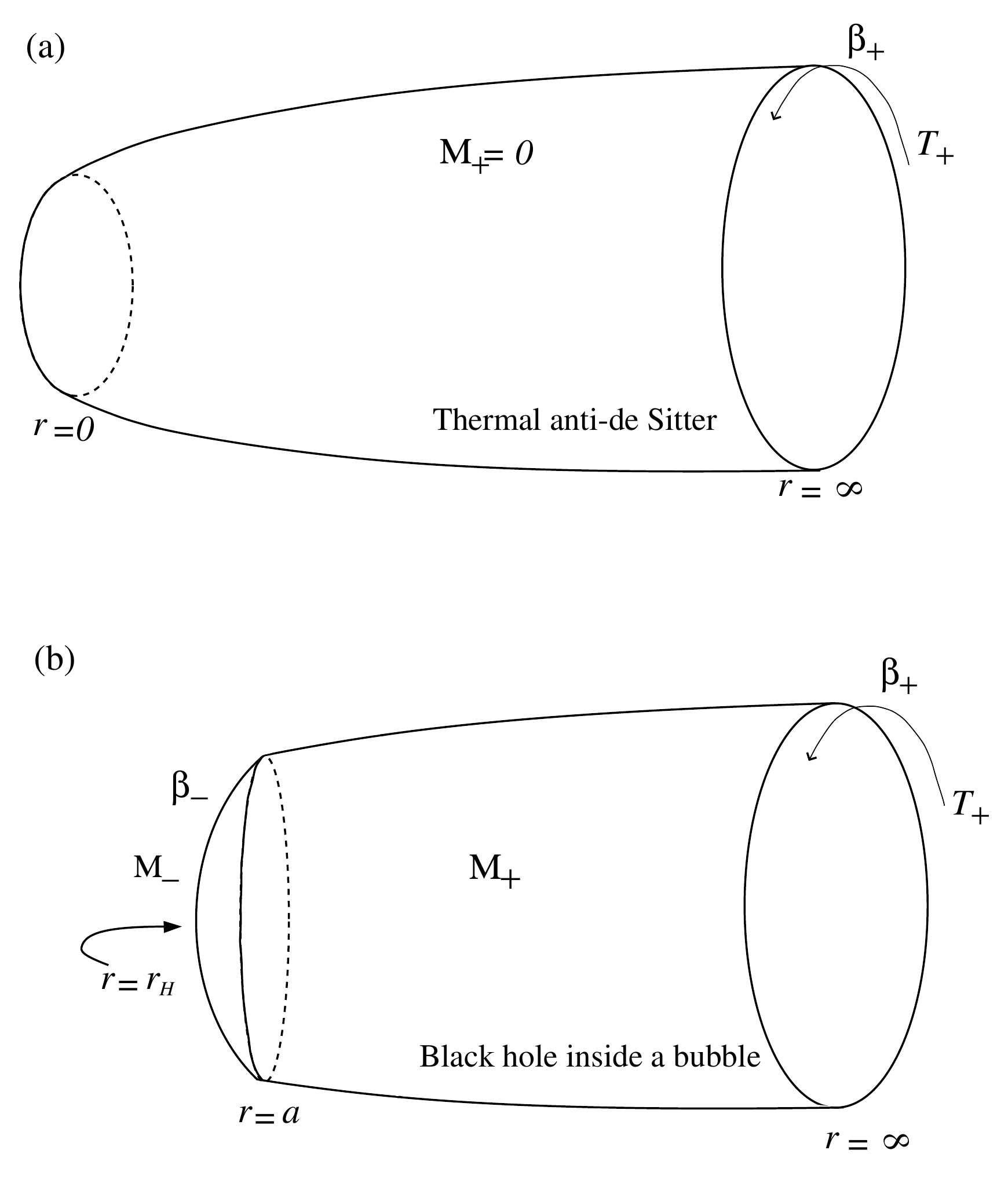}
\caption{Euclidean section of a {\it thermalon} configuration; a {\it bubble} pops out in empty AdS space with cosmological constant $\Lambda_+$, hosting a black hole belonging to a different branch of the Lovelock theory characterized by $\Lambda_-$.}
\label{thermalon}
\end{SCfigure}

Let us be more detailed in the kind of configurations we are interested in. They correspond to a {\it bubble}, whose outer region asymptotes AdS with a cosmological constant $\Lambda_{+}$, while the inner region corresponds to another branch solution characterized by the effective cosmological constant $\Lambda_-$ (see figure \ref{thermalon}).
The junction conditions will constrain the possible combinations of parameters that will describe the solutions of interest, instantons and thermalons. Across the junction, the vielbein has to be continuous. We thus consider an Euclidean section of the form
\begin{equation}
ds^2 = f_{\pm}(r) dt^2 + \frac{dr^2}{f_{\pm}(r)} + r^2\ d\Omega_{\sigma,d-2}^2 ~,
\label{Eucsection}
\end{equation}
where the $\pm$ signs denote the outer/inner regions. The junction is conveniently described by the parametric equations
\begin{equation}
r = a(\tau) ~, \qquad t_\pm = T_{\pm}(\tau) ~,
\label{parequations}
\end{equation}
with an induced metric of the form
\begin{equation}
ds^2 = d\tau^2 + a(\tau)^2\ d\Sigma_{d-2,\sigma}^{2} ~,
\label{indmetric}
\end{equation}
where we have made the choice
\begin{equation}
f_{\pm}(a)\,\dot{T}^2_{\pm}+ \frac{\dot{a}^2}{f_{\pm}(a)} = 1 ~, \qquad \forall \tau ~.
\label{RTcond}
\end{equation}
The function $a(\tau)$ appears in the induced metric and so it has to be the same seen from both sides, as long as the induced metric itself has to be continuous. The thermalon is characterized by a static configuration, $\dot{a} = \ddot{a} = 0$ ({\it i.e.}, $a(\tau) = a_\star$, the location of the thermalon), that, in view of (\ref{RTcond}), translates into
\begin{equation}
\sqrt{f_{-}(a_\star)}\ T_{-} = \sqrt{f_{+}(a_\star)}\ T_{+} = \tau ~,
\label{Tglue}
\end{equation}
which means that the physical length of the circle in Euclidean time is the same as seen from both sides. This matching condition will eventually allow us to determine the temperature of the solution. In fact, regularity of the Euclidean section (see figure \ref{thermalon}) demands the inner solution is a black hole and fixes its temperature. Once the periodicity in Euclidean time is fixed to attain a regular horizon, the periodicity in the outer time will be determined through
\begin{equation}
\sqrt{f_{-}(a_\star)}\ \beta_{-} = \sqrt{f_{+}(a_\star)}\ \beta_{+}\equiv \beta_0 ~,
\label{matchbetas}
\end{equation}
where $\beta_{-}$ is the usual inverse Hawking temperature of the inner solution while $\beta_{+}$ is the one seen by an observer at infinity, that will be different, in general, from the Hawking temperature corresponding to the black hole of the given mass on either branch. $\beta_0$ corresponds to the periodicity in the new variable $\tau$.

Consider the induced vielbein basis $e^{\tau} = d\tau$ and $e^{i} = a(\tau)\ \tilde{e}^{\varphi_i}$, which is intrinsic to $\Sigma$. If we call $x^\mu_\pm(\zeta^a)$ the coordinates of the embedding of $\Sigma$ with intrinsic coordinates $\zeta^A = (\tau,\varphi_i)$ on $\mathcal{M}_\pm$, the unit normal vectors read, in Lorentzian signature,
\begin{equation}
n_\pm^\mu = (\frac{\dot{a}}{f_\pm(a)},\sqrt{\dot{a}^2+f_\pm(a)},0,\cdots,0) ~.
\label{normalvectors}
\end{equation}
The extrinsic curvature is then given by
\begin{equation}
K_{AB} = -n_{\pm\mu} \left(\frac{\partial^2 x_\pm^\mu}{\partial\zeta^A\partial\zeta^B}+\Gamma^{\mu}_{\ \alpha\beta}\, \frac{\partial x_\pm^\alpha}{\partial\zeta^A}\, \frac{\partial x_\pm^\beta}{\partial\zeta^B}\right) ~,
\label{Kab}
\end{equation}
yielding
\begin{equation}
{K_\pm}^\tau_{\ \tau} = \frac{\ddot{a} + \frac12 f'_\pm(a)}{\sqrt{\dot{a}^2+f_\pm(a)}} ~, \qquad  {K_\pm}^{\varphi_i}_{\ \varphi_j} = \frac{\sqrt{\dot{a}^2+f_\pm(a)}}{a}\ \delta^i_{\ j} ~.
\end{equation}
The intrinsic curvature, in turn, is given by
\begin{equation}
R^{i\tau}_0=\frac{\ddot{a}}{a}\,e^{i}\wedge e^{\tau} ~, \qquad R^{ij}_0 = \frac{\sigma+\dot{a}^2}{a^2}\,e^{i}\wedge e^{j} ~.
\label{R0ab}
\end{equation}
Notice that it is trivially the same as seen from either side as it is calculated from the induced metric and this is continuous. The vielbein basis is continuous but the spin connection is not. In that case it should be possible to write the surface term as the difference of (\ref{intrinsicaction}) seen from each of the bulk regions. It is clear that all the components aligned along the normal direction of this intrinsic spin connection are zero in the same way as it happens for the corresponding intrinsic curvature. In the Euclidean signature we may just change the signs of the squared velocity and its acceleration as $(\dot{a}^2,\ddot{a})\ \to \ (-\dot{a}^2_E,-\ddot{a}_E)$.

\subsection{Junction conditions and bubble dynamics}

The junction conditions (\ref{junction}) for the configurations of interest have just diagonal components related by a conservation equation (Bianchi identity) that constrains them in such a way that only the $\tau\tau$ component matters. The rest are related to that one as \cite{Davis2003,wormholes}
\begin{equation}
\frac{d}{d\tau}\left(a^{d-2}\, \pi^\pm_{\tau\tau}\right) = (d-2)\, a^2 \dot{a}\, \pi^\pm_{\varphi_i\varphi_i} ~, \qquad \forall i ~,
\label{Bianchi}
\end{equation}
in such a way that if $\pi^\pm_{\tau\tau}$ verifies (\ref{junction}), all the components automatically do. This is reminiscent of the analogous field equations for cosmological solutions studied in section \ref{LLcosmo} Thus, we just need to compute $\pi^\pm_{\tau\tau}$. This junction condition involves just the {\it angular} components of both the intrinsic and extrinsic curvatures; the expression reduces to
\begin{equation}
\Pi^\pm_{(\tau\tau)} \equiv \frac{\sqrt{\dot{a}^2+f_\pm(a)}}{a} \int_0^1\! d\xi\ \Upsilon'\left[\frac{\sigma-\xi^2\,f_\pm(a)+ (1 - \xi^2)\,\dot{a}^2}{a^2}\right] ~,
\end{equation}
where we avoid the inclusion of some (irrelevant for our discussion) factors, and the polynomial $\Upsilon$ is again seen to play a central r\^ole. In the future we will also avoid the use of indices, $\Pi^{\pm}$ referring to the $(\tau\tau)$-component. If we define $\widetilde{\Pi}=\Pi^+-\Pi^-$, because of \reef{Bianchi} the junction equations can be written as $\widetilde{\Pi}=\partial_\tau\widetilde{\Pi}=0$. Notice that the dimensionality of spacetime is somehow irrelevant in this expression. In the case of LGB, this expression reduces to the one in \cite{Garraffo2008b}. If we conveniently introduce the functions\footnote{Not to be confused with the previously used notation for the horizon value of $g$, $1/r_+^2$}
\begin{equation}
g_\pm \equiv g_\pm(a) = \frac{\sigma-f_{\pm}(a)}{a^2} ~, \qquad H \equiv H(a,\dot{a}) = \frac{\sigma + \dot{a}^2}{a^2} ~,
\end{equation}
we can rewrite
\begin{equation}
\Pi^\pm\,[g_\pm, H] = \sqrt{H-g_\pm}\int_0^1\! d\xi\ \Upsilon'\left[\xi^2\,g_\pm + (1 - \xi^2)\,H \right] ~,
\label{Sjunc}
\end{equation}
where it becomes clear that all the information about the branches is contained in $g_\pm$, that are implicitly given by (\ref{eqg}). Notice that $g_\pm \leq H$ in order to have a real value for $\Pi^\pm$ (which tantamount to $f_\pm(a) \geq 0$ in order to have a real Euclidean boundary action). This implies that the static bubble corresponding to the {\it thermalon} necessarily forms at $a_\star \geq \max(r_{H+},r_{H-})$, outside the would be black hole horizons corresponding to both branches. Moreover, this reality condition is also necessary for the equation to yield real values of the velocity, at least for some region of the spacetime.  

The difference in canonical momenta $\widetilde{\Pi}$ may be rewritten in a couple more useful ways as
\begin{equation}
\widetilde{\Pi}=\int^1_0 d\xi \left(\sqrt{H-g_+}-\sqrt{H-g_-}\right)\Upsilon'\left[H-\left(\xi\sqrt{H-g_+}+(1-\xi)\sqrt{H-g_-}\right)^2\right]
\label{totaljunc}
\end{equation}
that is the variation of \reef{matchactionbis}, or changing the integration variable
\be
\widetilde{\Pi}=\int_{\sqrt{H-g_-}}^{\sqrt{H-g_+}}{\!\!\!\!dx\,\Upsilon'[H-x^2]}
\label{changePi}
\ee
that is more compact and easier to manipulate.

We can interpret $\widetilde{\Pi}(\dot{a},a) = 0$ as a conservation equation with a non-canonical kinetic term. We can make this statement more precise by noticing that
\begin{equation}
\Pi_+^2 = \Pi_-^2 \qquad \Longleftrightarrow \qquad \prod_{i=1}^{K-1}\left(\frac12\dot{a}^2+V_i(a)\right) = 0 ~,
\end{equation}
where we have to take into account that some of the {\it roots} or potentials, $V_i(a)$, correspond to the actual junction conditions (\ref{junction}), while some other may correspond to the reversal orientation, $\Pi^+ = - \Pi^-$, that corresponds to gluing two interiors or two exteriors ({\it wormhole}). Besides some roots may be discarded in case they yield imaginary momenta, $g_\pm>H$.  This reality condition amounts to
\be
\dot{a}^2\geq -f_{\pm} \qquad ; \quad \dot{a}^2_E\leq f_\pm
\ee
in the Lorentzian and Euclidean sections respectively. Remark that in the Euclidean version the brane may just propagate outside the horizon whereas in the Lorentzian case the bubble may cross it. This leads to some seemingly pathological situations, namely the destruction of the horizon by the bubble leading to a violation of the cosmic censorship hypothesis.

Roots of $\widetilde{\Pi}=0$ and $\Pi^++\Pi^-=0$ may just join where $\Pi^+=\Pi^-=0$. The same also happens for solutions yielding real and imaginary values of momenta, as this happens for instance for $H=g_-$ or equivalently $\dot{a}^2=-f_-$ something that is possible  just inside the horizon or for negative values of $\dot{a}^2$, forbidden region of the potential. 
%
%
In these particular points the potential ceases to be a solution of $\Pi^+-\Pi^-=0$ to become a solution of $\Pi^++\Pi^-=0$. This is generic feature of any Lovelock theory as $\sqrt{H-g_-}$ does not change sign at $H=g_-$, as it can be seen in the figure, whereas $\Pi^-$ changes sign there. When this happens inside the horizon it is unclear what happens to the bubble. It cannot go further but it cannot turn back as it would be travelling backwards in time. We will comment more on this later on. 

\begin{SCfigure}
\includegraphics[width=.45\textwidth]{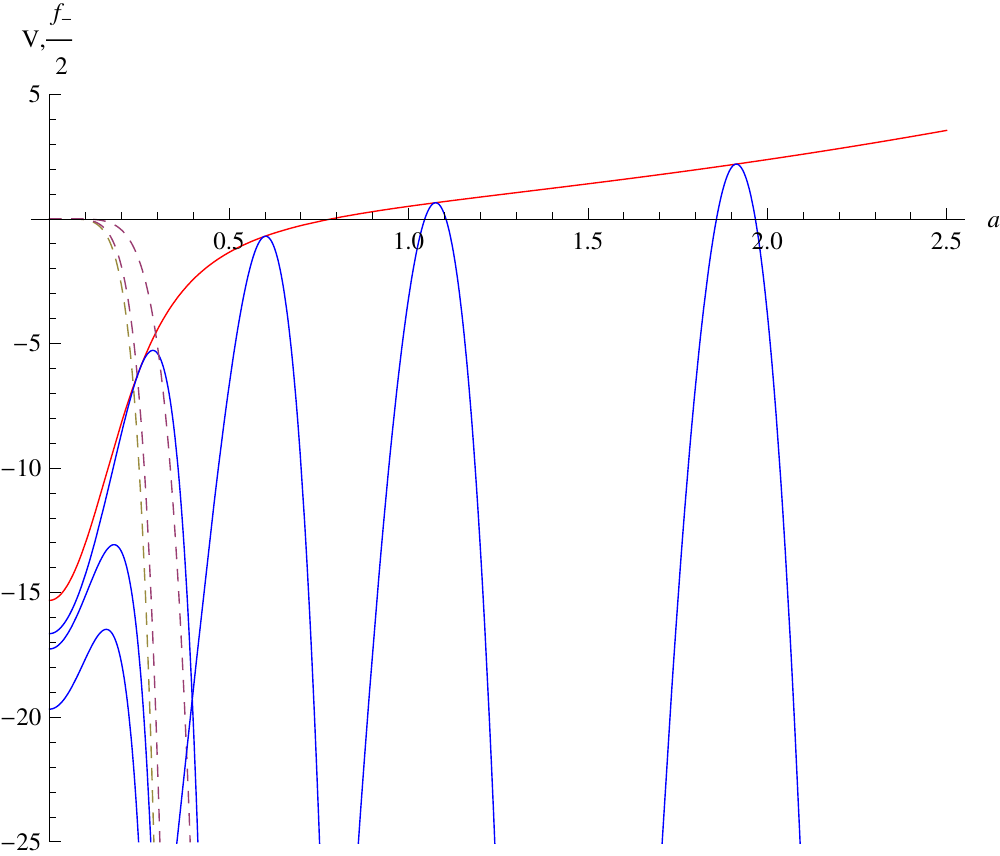}
\caption{Potential (in blue) of a brane gluing spherically symmetric stable and unstable branches of LGB gravity with parameters $\lambda=0.001$, $L=1$, $\kappa_-= 1$ and $\kappa_+=2,2.4,7,10^2,10^3,10^4$ from left to right. In red $f_-/2$, metric function corresponding to the stable branch. The blue curve slides up the red one as we increase $\kappa_+$. The dashed curve corresponds to the asymptotic behavior \reef{asymptH} for the three lower values of $\kappa_+$.}
\label{rootchange}
\end{SCfigure}

See figure \ref{rootchange} for specific examples of these facts. The potential becomes unphysical beyond the point where it meets $f_-/2$. Notice that we can get to the origin if we reduce $\kappa_+$ or, for fixed $\kappa_+$ if we increase $\kappa_-$, actually for $\kappa_+\leq 4\kappa_-$.

Once a particular potential is chosen, say $V_j(a)$, we can use (\ref{Bianchi}) to determine the corresponding acceleration equation that governs the dynamics of the bubble
\begin{equation}
\ddot{a} = -V_j'(a) ~.
\label{dynbubble}
\end{equation}
Notice that this dynamics may be difficult to determine, in general, since for generic $K$ it might be impossible to have an explicit expression for the potentials, and, on top of that, several of them may provide a suitable dynamics for the bubble. In all the expressions the r\^oles of the two branches can be exchanged yielding the same dynamics of the bubble. As there is no matter on the bubble, the corresponding equations are blind to which is the inner/outer solution, the bubble behaves in exactly the same way.

There are two limiting cases of the junction conditions that are of special interest. On the one hand we are interested in the static configurations ({\it thermalons}) and their stability. For that it is enough to consider the slow limit of (\ref{junction}), in a double expansion about $a_\star$ and $\dot{a}=0$, 
\begin{equation}
\widetilde{\Pi}\approx \widetilde{\Pi}^\star + \frac{\partial \widetilde{\Pi}^{\,\star}}{\partial H}\ \frac{\dot{a}^2}{a_\star^2} + \frac{\partial \widetilde{\Pi}^{\,\star}}{\partial a}\ (a-a_\star)+\frac12 \frac{\partial^2 \widetilde{\Pi}^{\,\star}}{\partial a^2}\ (a-a_\star)^2 ~,
\end{equation}
where the upper star means that a quantity is being evaluated after the replacement $H \to H_\star \equiv \frac{\sigma}{a_\star^2}$ and $g_\pm^\star\rightarrow g_\pm(a_\star)$. Besides the two conditions
\begin{equation}
\widetilde{\Pi}^\star = \frac{\partial \widetilde{\Pi}^\star}{\partial a} = 0 ~,
\end{equation}
the junction condition (\ref{junction}) at $a = a_\star$ adopts the canonical form of an energy constraint for an auxiliary harmonic system described by $a(\tau)$,
\begin{equation}
\frac12 \dot{a}^2 + V_{\rm eff}(a) = 0 ~,
\end{equation}
where
\begin{equation}
V_{\rm eff}(a) = \frac12 k\, (a-a_\star)^2 ~, \qquad k = \frac{a_\star^2}{2} \left( \frac{\partial \widetilde{\Pi}^{\,\star}}{\partial H} \right)^{-1}\! \frac{\partial^2 \widetilde{\Pi}^{\,\star}}{\partial a^2}
\label{potstability}
\end{equation}
provided the denominator is non-vanishing, which is equivalent to the potential being a smooth function of the radius at $a_\star$. That factor vanishes when two potentials merge, thus becoming complex beyond that point. The bubble cannot go inside the region of complex potential, analogously to the case of branch singularities in the cosmological context (see section \ref{LLcosmo}), it will turn back along the other {\it root} that merges with the one it was following. 

The other interesting limiting case corresponds to the speed of the brane becoming very large, since it will be relevant to discuss the asymptotic behavior of the running away bubble. As $a\rightarrow\infty$, the behavior of $H$ has to be given by a power law, and so it can either diverge or asymptote to a constant. We can verify that one of the solutions diverges as $H\sim a^{d-1}$ whereas the rest ($K-2$) are asymptotically constant. We can prove that the maximum number of solutions is $K-1$ in general just by squaring each side of the $\Pi^+=\Pi^-$ equation in order to obtain a polynomial equation on $H$. Even if the na\"ive degree of such equation is $2K-1$ it is easy to show that the first non-vanishing coefficient of such equation is order $K-1$ in $H$. 

In the limit of large speeds (\ref{totaljunc}) can be evaluated directly using
\begin{eqnarray}
H-\left(\xi\sqrt{H-g_+}+(1-\xi)\sqrt{H-g_-}\right)^2&\approx & \xi g_+ + (1-\xi)g_-+\frac{1}{4H}\xi(1-\xi)\left(g_+ - g_-\right)^2\nonumber\\
\left(\sqrt{H-g_+}-\sqrt{H-g_-}\right)&\approx &\frac{1}{2\sqrt{H}}\left(g_- -g_+\right)\left(1+\frac1{4H}(g_-+g_+)\right)\nonumber
\end{eqnarray}
so that we can verify that 
\begin{equation}
\widetilde{\Pi} \approx -\frac{1}{2\sqrt{H}}\left[\left(\Upsilon[g_+]-\Upsilon[g_-]\right)-\frac{1}{2H}\left(\int_{g_-}^{g_+} \!\!\!\!\!dx\, \Upsilon[x]-\left(g_+\Upsilon[g_+]-g_-\Upsilon[g_-]\right)\right)\right]
\end{equation}
to next-to-leading order in $1/H$, where we have used the change of variable $x=\xi g_++(1-\xi)g_-$ thus solving
\begin{equation}
H\approx\frac{1}{2(\kappa_+-\kappa_-)}\left[a^{d-1}\int_{g_-}^{g_+} \!\!dx\, \Upsilon[x]-(g_+\kappa_+-g_-\kappa_-)\right]
\label{highH}
\end{equation}
In the lightlike limit ($H\rightarrow\infty$),  we get a consistent equation for $H$ either if $a\rightarrow\infty$ and $H\sim a^{d-1}$ or $\kappa_+\rightarrow \kappa_-$ which coincides with the result of \cite{Gravanis2009}. In the former case the asymptotic behavior of the potential can be just read as
\begin{equation}
H\approx\frac{a^{d-1}}{2(\kappa_+-\kappa_-)}\int_{\Lambda_-}^{\Lambda_+} \!\!dx\, \Upsilon[x]
\label{asymptH}
\end{equation}
and this solution is {\it physical} in the sense of being solution of $\widetilde{\Pi}=0$ as opposed to $\Pi^++\Pi^-=0$ roots. The asymptotic solution \reef{asymptH} is useful to analyze in which cases it is possible for the brane to reach infinity. For that, one has to also analyze the sign of the asymptotically constant solutions, when present, and also determine along which branch is propagating the bubble. The constant roots are the solutions of 
\be
\widetilde{\Pi}[\Lambda_\pm,H]=0
\ee
that is degree $K-2$ in $H$ as the leading power is proportional to $\Upsilon[g_+]-\Upsilon[g_-]\sim 1/a^{d-1}$ and vanishes as we approach the asymptotic region. In that limit the topology of the horizon becomes irrelevant.


 In case the bubble may run away to infinity, it asymptotically approaches the speed of light. This can be seen for instance from \reef{asymptH}, as $\dot{a}^2$ grows faster than the function $f$. the radial speed is then
\begin{equation}
\frac{dr}{dt}=\frac{\dot{a}}{\sqrt{f+\dot{a}^2}} f \rightarrow f
\end{equation}
exactly the same as for a null geodesic in that background. 
In that way, as AdS space has a timelike boundary, the bubble gets there in a finite time and as such it can be interpreted as a change of boundary conditions for the theory, \ie as a jump from one branch of solutions to the other. The time from a position $a_0$ far away to the boundary is actually
\begin{equation}
\Delta \tau=\int_{a_0}^\infty \frac{da}{\dot{a}}\sim \int_{a_0}^\infty \frac{da}{a^{\frac{d+1}{2}}}\sim\frac{1}{\, a_0^{\frac{d-1}{2}}}<\infty
\end{equation}
and is also finite for the asymptotically constant values of $H$, although the asymptotic speed is a fraction of the speed of light in that case.

In the case of LGB gravity there is just one possible potential that determines the dynamics of the brane separating two solutions belonging to the two different branches. We will follow the same notation as in \reef{GBbranches}, the $(-)$ branch being the stable one. This effective potential can be simply written as 
\begin{equation}
V(a)=a^{d+1}\frac{\Delta\left[g(3+2\lambda g)^2\right]}{24\lambda\;\Delta\kappa}+\frac{\sigma}{2}
\label{GBpots1}
\end{equation}
It verifies all of the general properties described above. 
For instance, as the integral of $\Upsilon[g]$ between $\Lambda_-$ and $\Lambda_+$ is always positive the bubble can scape to infinity as long as $\kappa_+>\kappa_-$, \ie the mass of the unstable branch has to be higher than the stable one. Remember that any of the two can be in principle on either side of the brane and the dynamics is exactly the same. The tension depends on the position and there is no pressure term as it would depend on the side that corresponds to each branch. The bubble has tension despite the fact that it contains no matter. 

Depending on the values of the masses the stable branch may have a horizon and the naked singularity of the unstable solution may be located at a higher or lower radial position. Moreover, depending on the choice of external and internal solutions and the position of the bubble, the (dynamical) metric may display the horizon in some situations. 

 Quite generally (\eg see figure \ref{rootchange} for low enough $\kappa_+$) the horizon is accessible in the sense that the potential is finite and negative at that position. In case the stable branch is the inner one the metric may display a horizon but, as the bubble may cross that point, it can be destroyed. It will necessarily undergo subsequent collapse until it reaches the naked singularity that becomes visible. For that we just need the radius of the horizon to be larger than the singularity, otherwise the latter would become naked even before the bubble reaches the horizon. In the opposite case, with the unstable branch inside, the contraction of the brane would instead create a horizon.
\begin{SCfigure}
\centering
\includegraphics[width=.45\textwidth]{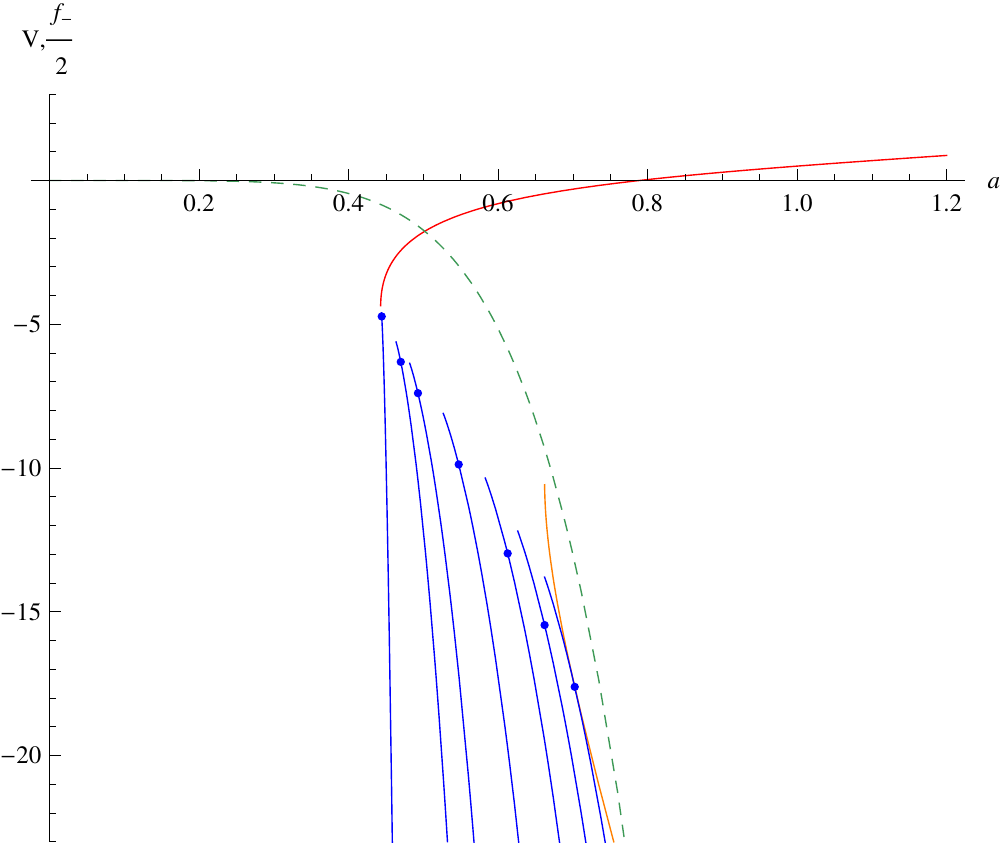}
\caption{Potential (in blue) of a brane gluing spherically symmetric stable and unstable branches of LGB gravity with parameters $\lambda=-0.01$, $L=1$, $\kappa_-=1$ and $\kappa_+= 1.01,1.2,1.4,2,3,4,5$ from left to right. In red $f_-/2$, metric function corresponding to the stable branch and in orange the one corresponding to the unstable branch, $f_+/2$, for $\kappa_+=5$. Singularities are located where the curves end, the potential ending when we find the outermost of them. Before that we encounter a point where $V=f_+/2$ beyond which the potential is unphysical. The dashed curve corresponds to the asymptotic behavior \reef{asymptH} for $\kappa_+=5$. }
\label{braneHorizon}
\end{SCfigure}
In fact, even if these examples might seem fine-tuned we will provide specific examples. The case of a expanding looks a bit more natural, at least with the unstable branch outside. In that case one may think of a bubble popping out from a naked singularity and creating a horizon, leaving a regular black hole behind. This points towards a possible instability of the BD-unstable branch, namely their naked singularity metrics, decaying through the formation of this kind of bubbles. Remember that these solutions are also perturbatively unstable. We will analyze this issue, although in a different manner in the next sections.

\subsection{Bubbles, horizons and the cosmic censor}

The destruction of the horizon by the bubble might seem unnatural but seems to be what really happens. In order to verify that this is indeed the case one has to be careful and check that nothing goes wrong at the horizon or inside it. In particular, in order for the metric to be continuous, the existence of a common induced metric on the junction surface is not enough. We also have to ensure that the change of variables between the coordinate frames on both sides is regular. 

In the vicinity of the junction ($\rho=0$) we might write the metric in terms of  a coordinates set adapted to the surface
\be
ds^2\approx -d\tau^2+d\rho^2+a^2(\tau) d\Sigma_{\sigma,d-2}^2
\ee
the same in both sides and $\rho$ being the normal. We then get constraints on the coordinate functions of the brane $T_\pm(\tau,\rho)$ and $a_\pm(\tau,\rho)$ (such that $T_\pm(\tau,0)\equiv T_\pm(\tau)$ and $a_\pm(\tau,0)\equiv a(\tau)$). One is the Lorentzian analogue of the previously found \reef{RTcond} and additionally,
\be
-f_\pm {T'}_{\!\!\!\pm}^2+\frac{{a'}_{\!\!\!\pm}^2}{f_\pm}=1 \qquad \qquad; \quad\qquad 
-f_\pm {T'}_{\!\!\!\pm}\dot{T}_\pm+\frac{{a'}_{\!\!\!\pm}\dot{a}_\pm}{f_\pm}=0
\ee
where primes indicate derivatives with respect to the new variable $\rho$. Remark that the radial variables are not equal in this case even though they  are at the junction. 

In this way we may write the change of variables as 
\bear
dt_\pm&=&\dot{T}_\pm\, d\tau+\frac{\dot{a}}{f_\pm}d\rho \nn\\
dr_\pm&=&\dot{a}\, d\tau + f_\pm \dot{T}_\pm d\rho 
\label{dtaudrho}
\eear
that corresponds to a boost in the $(t_\pm,r_\pm)$-plane. This can be easily verified using the orthonormal frame \reef{vierbh}
\be
\left(
\begin{array}{c}
e^0_\pm \\
e^1_\pm
\end{array}
\right)=U_\pm \left(
\begin{array}{c}
d\tau \\
d\rho
\end{array}
\right) \qquad \quad U_\pm=\left(
\begin{array}{cc}
\sqrt{f_\pm}\dot{T}_\pm & \frac{\dot{a}}{\sqrt{f_\pm}} \\
\frac{\dot{a}}{\sqrt{f_\pm}} & \sqrt{f_\pm}\dot{T}_\pm 
\end{array}
\right) ~.
\ee
the boost matrix having unit determinant with an inverse obtained just by changing the sign of $\dot{a}$. Remember that $\dot{T}_\pm$ can be written in terms of $\dot{a}$ using \reef{RTcond}, being invariant under that change. The change of variables between inner and outer coordinates corresponds thus to a composition of boosts $e_+=U_+U^{-1}_-e_-$, a transformation that does not change the  causal structure, \ie null geodesics are continuous across the bubble. 

We can now address the question of the behavior of the bubble as we cross the horizon. In that case we have to change $\sqrt{f_-}=i\sqrt{\|f_-\|}$, as $f_-$ goes negative, in such a way that the timelike and spacelike vielbeine exchange their r\^oles\footnote{Inside the horizon 
\be
\hat{e}^0=-\frac{dr}{\sqrt{\|f\|}} \qquad \qquad ; \qquad \quad \hat{e}^1=\sqrt{\|f\|}dt
\ee
} preserving the orientation ($e^0_-\wedge e^1_-=\hat{e}_-^0\wedge \hat{e}_-^1$)
\be
e^0_-=i\hat{e}^1_- \qquad\qquad ; \qquad \quad e^1_-=i\hat{e}_-^0 ~.
\ee
Consequently the change of variables behind the horizon
\be
\left(
\begin{array}{c}
\hat{e}^0_- \\
\hat{e}^1_-
\end{array}
\right)=\hat{U}_- \left(
\begin{array}{c}
d\tau \\
d\rho
\end{array}
\right) \qquad \quad \hat{U}_-=\frac{1}{\sqrt{\|f_-\|}}\left(
\begin{array}{cc}
-\dot{a} &-\sqrt{f_-+\dot{a}^2}  \\
-\sqrt{f_-+\dot{a}^2} & -\dot{a} 
\end{array}
\right) ~.
\label{changebeyond}
\ee
 again corresponds to a boost and is completely well defined. The change is actually continuous across the horizon, the bubble being able to cross it. After all there is nothing special -- locally -- about that point. Besides, once it gets to the horizon the causal structure makes it impossible for the bubble to get back (see figure \ref{collPenrose} for the Penrose diagram of the process) as it would be travelling backwards in time. This can also be seen from  \reef{changebeyond} as the diagonal components of the $\hat{U}_-$ matrix have to be positive and bigger than one. This is also true for $U_\pm$ outside the horizon. 

It is remarkable that, even though the original horizon is destroyed by the bubble the causal structure inherited by it makes it impossible for the bubble to go back. In this case we cannot blame the symmetry of the solutions considered as we did in chapter \ref{chp:bhstability}. Once the horizon is crossed, its (previous) presence forces the bubble to actually reach the central singularity, regardless of the details of the collapse.  The singularity unavoidably becomes naked, this being a clear violation of the cosmic censorship hypothesis. Nonetheless, this violation might be a marginal one if, due to the instabilities suffered by those singular solutions (see chapter \ref{chp:bhstability} for details), the energy leaks the singularity, thus leaving regular pure AdS$_+$ behind. This process is allowed by the causal structure of the spacetime. Notice that the bubble itself may also emit part of its mass, $M+-M_-$, to infinity before it actually reaches the singularity. However the naked singularity will still form as the mass contained in the inner solution, $M_-$, cannot scape until the bubble reaches the center.

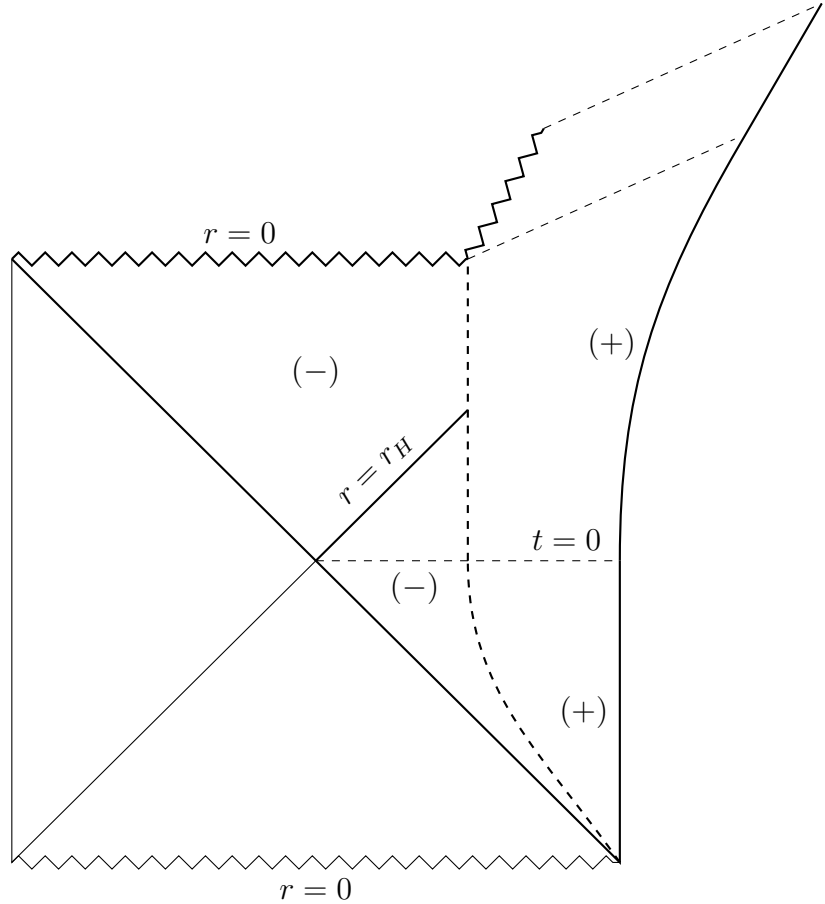
\begin{figure}
\centering

\begin{tikzpicture}
\node (I)    at ( 4,0)   {};
\node (II)   at (-4,0)   {};
\node (III)  at (0, 2.5) {$(-)$};
\node (IV)   at (0,-2.5) {};
\node (boosted) at (45:8) {};

\path  
  (II) +(90:4)  coordinate (IItop)
       +(-90:4) coordinate (IIbot)
       +(0:4)   coordinate (IIright)
       +(180:0) coordinate (IIleft)
       ;
\draw[thick] (IItop) -- (IIright);
\draw (IIright) -- (IIbot) --  (IItop);

\path 
   (I) +(90:4)  coordinate (Itop)
       +(-90:4) coordinate (Ibot)
       +(180:4) coordinate (Ileft)
       +(0:0)   coordinate (Iright)
       ;
			
\path 
	 ($(IItop)!.75!(Itop)$) coordinate (boostBL)
							+(60:2) coordinate (boostTL)
							;
\path
	 (boosted)  +(-120:.2) coordinate (boostBR)
							+(60:2) coordinate (boostTR)
							;
\draw[thick]  (Iright) --
		 node[midway,left] {$(+)$}
		 (Ibot) -- (Ileft) --
		 node[midway, above, sloped] {$r=r_H$}
				($(Ileft)!.5!(Itop)$);

\draw[decorate,decoration=zigzag,thick] (IItop) -- (boostBL) 
      node[midway, above, inner sep=2mm] {$r=0$};

\draw[decorate,decoration=zigzag] (IIbot) -- (Ibot)
			node[midway, below, inner sep=2mm] {$r=0$};
			
\draw[dashed,thick] 
    ($(Ileft)!.5!(Iright)$) to[out=90, in=-90, looseness=1.5] (boostBL);

\draw[dashed,thick] 
    (Ibot) to[out=125, in=-90, looseness=1] ($(Ileft)!.5!(Iright)$);

\draw[thick]
    (Iright) to[out=90, in=-120, looseness=1]
		node[midway,left] {$(+)$}
		(boosted) -- (boostTR);
		
\draw[decorate,decoration=zigzag,thick] 
    (boostBL) to[out=60, in=-120, looseness=1] (boostTL);
		
\draw[dashed] 
    (boostBL) -- (boosted);
		
\draw[dashed] 
    (boostTL) -- (boostTR);
		
\draw[dashed] 
    (Ileft) --
		node[below] {$(-)$}
		($(Ileft)!.65!(Iright)$) -- 
		node[above] {$t=0$}
		(Iright);

\end{tikzpicture}

\caption{Schematic Penrose diagram for a {\it nearly} static bubble that undergoes collapse from $t=0$ onwards. The thick dashed line represents the trajectory of the bubble with inner black hole and outer naked singularity geometries. The spacelike singularity becomes timelike as the bubble reaches $r=0$. The resulting spacetime is a (boosted) naked singularity, dashed lines representing constant $t_+$ slices. The lowest one corresponds to the Cauchy horizon introduced by the timelike singularity. The upper and right wedges represent the physical region of the spacetime corresponding to a black hole formed by collapse whereas the rest corresponds to the complementary white hole region.}
\end{figure}

\section{Thermalons and thermodynamics}

Bubble configurations when static are subject to the same discussion of chapter \ref{chp:thermo}, concerning the thermodynamics aspects of black holes. In very much the same way, the new solutions will be characterized by the same thermodynamic variables. 

In this section we will compute the value of the Euclidean on-shell action for these {\it thermalons}. Such static solutions will exist in some cases as it can be seen in figure \ref{rootchange} for the case of LGB gravity with positive $\lambda$ coupling. Being static, they trivially have a smooth Euclidean section whenever they display a horizon in the inner solution covering the singularity. We have to choose the periodicity in Euclidean time accordingly, so that the full configuration is smooth. Once the inner periodicity is fixed to attain a regular horizon, the periodicity in the outer time will be determined through \reef{matchbetas}, thus fixing the temperature. 

As for black holes, the Euclidean on-shell action is in general divergent and needs to be regularized, either by background substraction or by other means. We will measure the free energy with respect to the specific solution used as groundstate, usually the maximally symmetric one. In order to simplify the discussion we will calculate the on-shell action in terms of three parameters, $a_\star$ and $\beta_\pm$, the first being the equilibrium position of the bubble and the others the periodicity in Euclidean time in the inner and outer regions. It is important to keep in mind that these two variables are not independent from each other. We will use $+$ to denote the outer region and $-$ for the inner one. This is more general but it will be consistent with the notation of the LGB branches when we undertake that analysis.

Unlike the computation of the Hawking-Page effect described in section \ref{chp:thermo}, where the fields are continuous, here we have to consider the contribution to the action arising on the bubble, when writing the Euclidean action in the form \reef{splitaction},
\be
\widehat{\mathcal{I}}=\widehat{\mathcal{I}}_-+\widehat{\mathcal{I}}_{\Sigma}+\widehat{\mathcal{I}}_+   ~.
\ee
The outer piece includes all the boundary terms at infinity necessary to both have a well defined variational principle.  It regularizes its divergence by subtracting the background $M_+ = 0$ with the same periodicity at infinity, yielding
\be
\widehat{\mathcal{I}}_+(a_\star,\beta_+)= \beta_+\frac{(d-2)V_{d-2}}{16\pi G_N}\left. \partial_r\left[r^d\,\widetilde\Upsilon[g_+]\right]\right|_{a_\star} ~,
\label{Iout}
\ee
that has the same expression as for the usual black holes \reef{Eaction}  evaluated at the position of the bubble instead of the horizon. The term $\widehat{\mathcal{I}}_-$, in turn, is integrated from the horizon to the location of the bubble yielding two terms of the same form, one evaluated on the bubble and one at the horizon. 
\be
\widehat{\mathcal{I}}_-(a_\star,\beta_-)= \beta_-\frac{(d-2)V_{d-2}}{16\pi G_N}\left(\left. \partial_r\left[r^d\,\widetilde\Upsilon[g_-]\right]\right|_{r_H}-\left. \partial_r\left[r^d\,\widetilde\Upsilon[g_-]\right]\right|_{a_\star} \right)~.
\label{Iin}
\ee
where $r_H$ is the radius of the horizon, if any, or zero.
Finally, $\widehat{\mathcal{I}}_\Sigma$ is given by
\be
\widehat{\mathcal{I}}_\Sigma=-\widehat{\mathcal{I}}_\partial^-(a_\star,\beta_0)+\widehat{\mathcal{I}}_\partial^+(a_\star,\beta_0)
\ee
where the periodicity in Euclidean time is inherited from the bulk regions $\beta_0=\sqrt{f_\pm(a)}\beta_\pm$. Then, we can collect all the contributions that depend upon the location of the bubble, 
\bear
\widehat{\mathcal{I}}_{bub}(a_\star,\beta_0)&=&\beta_0\frac{(d-2)V_{d-2}}{16\pi G_N}\left(\frac1{\sqrt{f}} \partial_r\left.\left[r^d\, \tilde\Upsilon[g]\right]\right|_{a_\star}\right.\nn\\
&+&2\left.\left.\partial_r\left.\left[ r^{d-2} \sqrt{f}\;\int_0^1 dt\, \tilde\Upsilon'\left[(1-t^2)g_H+t^2\, g\right]\right]\right|_{a_\star}\right)\right|_-^+
\label{Ibrane2}
\eear
in the static case, otherwise we cannot perform the $\tau$ integration, and where $\left.\mathcal{F}(g)\right|^+_-=\mathcal{F}(g_+)-\mathcal{F}(g_-)$, indicates the difference between the terms evaluated on both sides of the brane. The rest can be consequently called 
\be
\widehat{\mathcal{I}}_{bh} = \widehat{\mathcal{I}}-\widehat{\mathcal{I}}_{bub} = \beta_-M_--S~,
\ee
the usual contribution from the inner black hole. Trivially the surface term vanishes when the same solution is taken in both sides of the junction.

The contribution from the bubble can be greatly simplified making use of the relations between the two characteristic polynomials $\Upsilon[g]$ and $\widetilde\Upsilon[g]$, as in \reef{twopoly}, and also some properties obtained by integrating by parts expressions of the type
\be
\int_0^1dt P[t^2g+(1-t^2)H_\star]= P[g]+2(H_\star-g)\int_0^1dt\, t^2 P'[t^2g+(1-t^2)H_\star]
\label{byparts}
\ee

Remarkably enough, a neat result comes out after a quite lengthy calculation, once the junction conditions are imposed
\begin{equation}
\widehat{\mathcal{I}}_{bub}=\frac{(d-2)V_{d-2}}{16\pi G_N}\left(\beta_+\kappa_+-\beta_-\kappa_-\right)=\beta_+M_+-\beta_-M_-
\end{equation}
which is the exact value needed to correct the on-shell action in such a way that the thermodynamic interpretation is safely preserved. In fact, because of this
contribution, the total action takes the form
\begin{equation}
\widehat{\mathcal{I}}=\widehat{\mathcal{I}}_{bh}+\widehat{\mathcal{I}}_{bub}=\beta_+M_+ -S ~.
\label{treintaytres}
\end{equation}
That is, the brane contributes as mass (it carries the mass difference between the two solutions) but not as entropy that comes solely from the horizon. From the Hamiltonian point of view this is naturally understood as follows. The canonical action vanishes in this case, the only possible contributions coming from boundary terms both at infinity and at the horizon, yielding respectively $\beta_+M_+$ and the
entropy, which are nothing but the total charges of the solution. 

For the above result we have just imposed the first junction condition, $\widetilde{\Pi}[g_\pm^\star,H_\star]=0$, where $H_\star=\sigma/a^2_\star$ is the static value of $H$. The other junction condition corresponds to the radial derivative of the first one in such a way that
\begin{equation}
\partial_a \widetilde{\Pi}[g_\pm^\star,H_\star]=\partial_{g_+} \widetilde{\Pi}[g_\pm^\star,H_\star] \,\left.g'_+\right|_\star+\partial_{g_-} \widetilde{\Pi}[g_\pm,H_\star] \,\left.g'_-\right|_\star + \partial_H \widetilde{\Pi}[g_\pm,H_\star] \,H_\star'=0 ~.
\end{equation}
The $H$ derivative of $\widetilde{\Pi}$ plays another fundamental r\^ole as it gives the overall normalization of the effective potential (see discussion above). Using \reef{changePi} we can show
\be
\partial_{g_\pm} \widetilde{\Pi}[g_\pm,H_\star]
= \frac{-1}{2\sqrt{H_\star-g_\pm^\star}}\Upsilon'[g_\pm^\star]~.
\ee
Also for the other derivative after a bit of massage we get
\be
\partial_H \left(\Pi^+-\Pi^-\right) =-\frac12\int_{\sqrt{H_\star-g_-}}^{\sqrt{H_\star-g_+}}{\!\!\!\! dx\,\frac{\Upsilon'[H_\star-x^2]}{x^2}}
\ee
Then, using the black hole equation \reef{eqg}, we can write the second junction condition as 
\be
\frac{\kappa_+}{\sqrt{f_+}}-\frac{\kappa_-}{\sqrt{f_-}}=\frac{4\sigma}{d-1}a^{d-4}\partial_H \widetilde{\Pi}[g_\pm,H_\star]
\ee
Notice that the left hand side of this expression is proportional to the contribution of the bubble \reef{Ibrane2} to the on-shell action (or equivalently the free energy).

This junction condition is also very important to ensure a consistent thermodynamic picture. We have seen in the previous paragraphs that the  on-shell action  adopts the expected form from the semi-classical approach \reef{treintaytres} where the different quantities correspond to the expected ones in the solution, \ie
\begin{equation}
\beta=\beta_+ \quad , \qquad M=M_+ ~,
\end{equation}
 the entropy being unchanged as coming from the inner black hole horizon. In order to show the consistency of the thermodynamics in addition to this the relations between this quantities and the corresponding thermodynamic potentials should be also the correct ones, as seen in \ref{chp:thermo}. It is enough to show that assuming the above values for mass and entropy we recover the correct temperature as
\begin{equation}
T=\frac{dM}{dS}=\frac{dM_+}{da_\star}\frac{da_\star}{dr_H}\frac{dr_H}{dS}
\end{equation}
where we have to use the implicit relation between the two parameters involved, $a_\star$ and $r_H$
\begin{equation}
\kappa_-(a_\star)=r_H^{d-1}\Upsilon\left[\frac{\sigma}{r_H^2}\right]~.
\end{equation}
Equivalently we can use the property
\begin{equation}
dS=\beta_- dM_-
\end{equation}
so that the inverse temperature is
\begin{equation}
\beta=\frac{dS}{dM}=\beta_-\frac{dM_-}{dM_+}~.
\end{equation}
In order for this to be equal to $\beta_+$ as required we must verify
\begin{equation}
\beta_+ dM_+=\beta_- dM_- =dS
\label{new1law}
\end{equation}
where the differential is taken with respect to the parameter $a_\star$ or equivalently $r_H$ or $T$. This implies that the first law of thermodynamics holds true not only for the inner black hole ($\beta_-$ and $M_-$) but for the whole configuration ($\beta_+$ and $M_+$). This can be easily proven using both junction conditions. First we derive the first one with respect to $a_\star$. For that we have to take into account that, in contrast with the differentiation of the potential with respect to $a$ (not $a_\star$) from which we obtained the second junction condition, the mass parameters $\kappa_\pm$ depend now on the variable. The expression itself depends on $\kappa_\pm$ through $g_\pm^\star$ and so one gets
\begin{equation}
\Upsilon[g_\pm^\star]=\frac{\kappa_\pm(a_\star)}{a_\star^{d-1}} \quad \Rightarrow \qquad \left(g_\pm^\star\right)'=\frac{-(d-1)\kappa_\pm+a_\star\kappa'_\pm}{a^d_\star \Upsilon'[g_\pm^\star]}
\end{equation}
The final result is the same as obtained for the second junction condition with an extra term proportional to the derivatives of the mass parameters, $\kappa_\pm$,
\begin{equation}
\frac{\kappa_++a\kappa'_+}{\sqrt{f_+}}-\frac{\kappa_++a\kappa'_+}{\sqrt{f_-}}=\frac{4\sigma}{d-1}a^{d-4}\partial_H \widetilde{\Pi}[g_\pm^\star,H_\star]~.
\end{equation}
Then, the term proportional to the mass parameter cancels the right hand side of the equation and we get just
\begin{equation}
\frac{\kappa'_+}{\sqrt{f_+}}-\frac{\kappa'_+}{\sqrt{f_-}}=0
\end{equation}
that multiplied by $\beta_0$ yields \reef{new1law} that is exactly what we were looking for.

We have shown that the first law of thermodynamics holds for the whole bubble configuration whereas the thermodynamic parameters verify the expected relations, this proving the consistency of the thermodynamic approach. Having computed the free energy of this geometry, we can now compare it to the rest of the solutions sharing the same boundary conditions and temperature and thus identify the preferred or classical configuration as the one of lower free energy. We can also analyze the local thermodynamic stability of the solutions as we did for the black hole solutions in chapter \ref{chp:thermo}.

\subsection{LGB thermalons}

Before analyzing the free energy of the thermalon configurations and the occurrence of phase transitions let us describe a bit in more detail the configurations we are dealing with in the simplest case of LGB gravity.  In this particular theory just the EH branch may display a horizon for planar or spherical black holes, it being  the only possible inner branch. The other, accordingly outer branch, has long been known to be unstable {\it \`a la} Boulware-Deser, its vacuum contains ghosts in the sense that the kinetic term for gravitons has the wrong sign.  We will be interested in the thermodynamics of the system in the case of {\it wrong} boundary conditions, \ie setting the asymptotics corresponding to this ill defined branch, with some topology and some temperature. For these topologies the thermodynamics corresponding to EH branch asymptotics is unchanged from what has been discussed in chapter \ref{chp:thermo}. This may shed some light on the long-standing question of the fate of the ghosty branch in this theory (see \cite{Charmousis2008a} for some related ideas).

In the hyperbolic case the naked singularity of the BD-unstable branch may be covered by a event horizon but just for positive LGB coefficient, $\lambda$, and masses below a certain threshold. This is the only situation in which we may have thermalons with the two possible asymptotics. This will be the only situation in which we may have the opportunity of describing transitions between branches in both ways.  Even though these present some puzzling features and the interpretation of the phenomena is not as clear as in the planar or spherical case, we will comment on them in the next sections.

In this case the potential has a simple expression \reef{GBpots1} and in order to define a thermalon configuration we have to find mass parameters $\kappa_\pm$ such that a equilibrium position, $a_\star$, exists. We have to solve, $V(a_\star;g_\pm^\star)=V'(a_\star;g_\pm^\star)=0$, where the only real variable is $a_\star$ and the extra two unknowns $g_\pm^\star$ are given implicitly by \reef{eqg}, once chosen the branches such as in \reef{GBbranches}
so that the inner solution corresponds to the stable branch and the outer one to the unstable.

We can write the potential in a more suitable way by using \reef{eqg} in order to reduce the order of the polynomial in $g$,
\begin{equation}
V(a_\star)=a_\star^{d+1}\frac{\left.[(1-4\lambda)g^\star+4(2+\lambda g)\frac{\kappa}{a^{d-1}}]\right|^+_-}{24\lambda(\kappa_+-\kappa_-)}+\frac{\sigma}{2}
\label{V2}
\end{equation}
and its $a$-derivative
\begin{equation}
V'(a_\star)=a_\star^{d}\frac{\left.[(1-4\lambda)(d+1)g+(d-17+2(d-5)\lambda g)\frac{\kappa}{a^{d-1}}]\right|^+_-}{24\lambda(\kappa_+-\kappa_-)}
\label{dV}
\end{equation}
It is then possible use these two equations to find the mass parameters of both solutions in terms of $g_\pm^\star$ and $a_\star$ and combine these expressions with the usual ones from \reef{eqg} and get a couple of equations 
\begin{equation}
\Upsilon[g_\pm^\star]=1+g_\pm^\star+\lambda \left(g_\pm^\star\right)^2=\frac{1-4\lambda}{4\lambda}\frac{a_\star^2(d-1)(3+2\lambda g_\mp^\star)+4\lambda(d+1)\sigma}{a_\star^2(d-1)+2(d-5)\lambda\sigma}
\end{equation}
that can then be solved for $g_\pm^\star=g_\pm(a_\star)$. The general expressions are not very enlightening but they simplify quite a lot in the planar case as the $a_\star$-dependence completely drops out. The bubble radius just remains as the overall scale.

Due to the symmetry of the equations we have several solutions. The relevant solution for our purposes is
\begin{equation}
g_{\pm}^\star=\frac{-3+4\lambda\mp\sqrt{3(1-4\lambda)(3+4\lambda)}}{4\lambda}
\label{thermsol}
\end{equation}
as we have to choose the values that correspond to the inner ($-$) branch being stable and the outer ($+$) one being unstable. In particular both $g_\pm^\star$ cannot be equal.

If we plot these solutions  as well as the mass parameters as functions of $\lambda$ (see figure \reef{planargs}) we can readily see that for negative values of the coupling the outer mass, corresponding to the unstable solution, is negative. Besides, the values of $g_\pm^\star$ are positive for negative $\lambda$ which corresponds to bubble being formed inside the black hole horizon, $g_-=0$. There are no static or thermalon configurations for negative values of the coupling. This feature is not specific of planar black holes, no thermalons exist for negative $\lambda$ on any topology. also for the other topologies. For this reason we will restrict our analysis for positive values of the LGB coupling. Remark that for $\lambda<0$ the bubble would separate AdS and dS branches. 

\begin{SCfigure}
	\centering
		\includegraphics[width=0.45\textwidth]{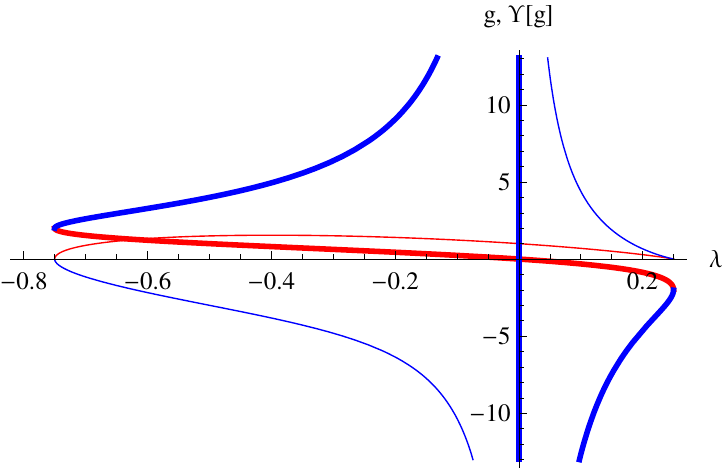}
			\caption{$g_{\pm}^\star$ and $\Upsilon[g_\pm^\star]\sim\kappa_\pm$ (thick and thin lines respectively) as a function of $\lambda$, red and blue lines corresponding the inner, $g_-$, and outer, $g_+$, solutions respectively.}
		\label{planargs}
\end{SCfigure}

\begin{SCfigure}
	\centering
		\includegraphics[width=0.45\textwidth]{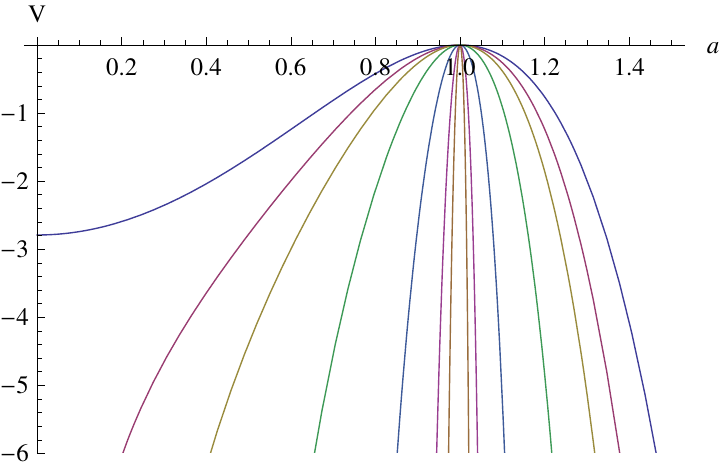}
			\caption{Bubble potential in LGB gravity for $\lambda=0.1$, $a_\star=1$ and $d=5,6,7,10,20,50,100$ (from top to bottom).}
		\label{planarpotential}
\end{SCfigure}

We can plot the potential with the recently found values for the different parameters involved in the configuration, $\kappa_\pm$ and the equilibrium position $a_\star$ as showed in figure \ref{planarpotential}, the qualitative behavior of the potential being the same for all values of $a_\star$ and $\lambda$ (positive). 
Notice the fundamental difference between five dimensions and higher that the potential at zero radius goes to a constant.

\section[Generalized Hawking-Page transitions in Lovelock gravity]{Generalized Hawking-Page transitions in Lovelock theories of gravity%
  \sectionmark{Generalized HP transitions in Lovelock gravity}
}
\sectionmark{Generalized HP transitions in Lovelock gravity}

Having proven the consistency of the thermodynamic picture in the case of the static bubble by deriving  (\ref{treintaytres}) and (\ref{new1law}), we are ready to address the question of the global (and local) thermodynamic stability of that configuration. This amounts to analyzing the free energy associated to that solution, the analysis being analogous to that of section \ref{HP1}. The only difference is that we now need to include the thermalons discussed above. This will allow for the study of the formation of these new configurations and the consequences for the dynamics of the system. 

In the same way as for the usual Hawking-Page phase transition we can identify the thermodynamically preferred or classical configuration as the background of least on-shell action among all those with a smooth Euclidean continuation and the same period in imaginary time. 

The first example that can be trivially analyzed is that of planar black holes in general Lovelock gravity. In this case the only branch displaying a horizon, as already discussed in the previous sections, is the EH one, all the others being unstable and displaying naked singularities. Because of that, the only possible smooth Euclidean metrics correspond to the vacua of the theory and bubble configurations with the inner region corresponding to the EH branch. Then, for any choice of asymptotics (different from EH), the configurations  we will be comparing are corresponding thermal vacuum and, when available, the horizonful bubble configuration with the same temperature.

The junction conditions simplify a lot in this case, the free energy corresponding just to the black hole with no bubble contribution. The presence of the junction plays no r\^ole except for the change on the temperature according to \reef{matchbetas}. The free energy of the thermal vacuum is zero, as we have taken it as reference background, and that of the thermalon yields
\be
F_+=\frac{\mathcal{I}_E}{\beta_+}=\frac{\beta_-}{\beta_+}F_-=-\frac{M_+}{d-2} ~,
\ee
which is always negative. This implies that the preferable classical solution is, when available, the thermalon. One of them if there are several with the same asymptotics. For any branch of solutions such that the planar thermalon with the inner EH branch exists, the bubble will always form. Remember that $\beta_- M_-=\beta_+ M_+$ and the inner mass has to be positive in order to have a horizon. 

In the LGB case the thermalon exists just for positive $\lambda$, the equilibrium position being unstable. It will therefore eventually expand in such a way that it engulfs the whole spacetime in finite proper time, thus changing its asymptotic behavior (and the associated thermodynamic variables) \cite{Camanho2012}. 

\begin{figure}
\centering
\includegraphics[width=0.6\textwidth]{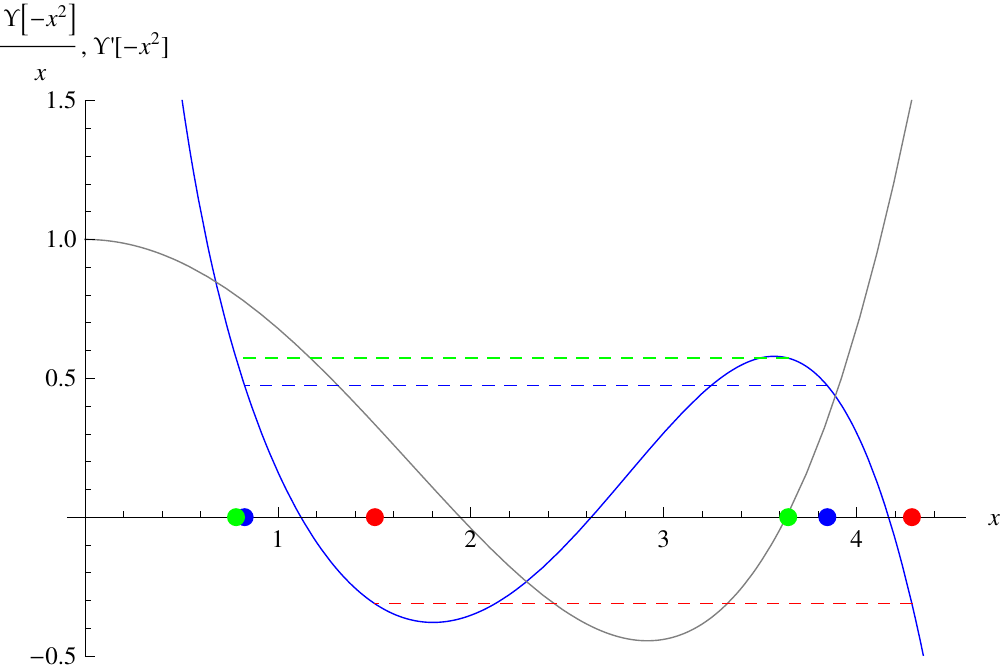}
\caption{$\Upsilon[-x^2]/x$ (solid blue line) and $\Upsilon'[-x^2]$ (in gray) for cubic Lovelock theory with parameters $L=1$, $\lambda=0.17$ and $\mu=0.02$. The couples of points with the same color correspond to each of the three solutions of the static junction conditions \reef{staticjunk} in that case. The area below the gray curve between each couple of points vanishes according to the second equation.}
\label{CubicPlanar}
\end{figure}

In the planar case the problem of the existence of these kind of configurations translates into finding couples $g_\pm^\star$ that verify the static junction equations, that can be written as
\be
\left.\frac{\Upsilon[g^\star]}{\sqrt{-g^\star}}\right|_-^+=0 \qquad \qquad ; \qquad \quad \int_{\sqrt{-g_-^\star}}^{\sqrt{-g_+^\star}}{dx\,\Upsilon'[-x^2]}=0
\label{staticjunk}
\ee
and subsequently as polynomial equations for $g_\pm^\star$. The solutions for any Lovelock theory can be plot, a specific example for cubic Lovelock theory being depicted in figure \ref{CubicPlanar}. Even though the independent  variable $x=\sqrt{-g}$ is not the usual one for the polynomial, we can easily recognize the three vacua, $\Upsilon=0$, and associated branches. These correspond to the three ranges with definite sign for $\Upsilon'$, positive for the branches that are stable ({\it \`a la} Boulware-Deser) and negative for the intermediate unstable one. The EH branch is the one closest to the vertical axis and then we have two higher curvature branches, one unstable and another stable. For this particular choice of couplings, we have three possible equilibrium configurations connecting the same two branches. One of the points on each pair corresponds to the EH branch and its couple to the other stable branch. The blue and green pairs correspond to positive mass solutions whereas the red one has negative mass, both in the outer and inner regions, as indicated by the positive and negative respective values of the polynomial. In the latter case (red dots), the inner region does not display a horizon and we have to discard this solution. Unlike the LGB case analyzed before there is no thermalon connecting the unstable branch to the EH one for this particular choice of parameters. In case we fix such {\it sick} boundary conditions the system cannot scape via bubble nucleation. Being unstable, it would evolve by some other means the endpoint of such instability remaining unclear.

The analysis of cubic Lovelock theory is particularly interesting in general as, for the range of couplings displaying three real vacua, it is the minimal example where we may analyze the occurrence of bubble transitions between two stable vacua, as in our example. The roots of \reef{staticjunk} depend however on the values of the coefficients $\lambda$ and $\mu$ and in general we may have also bubbles connecting any of the stable branches with the intermediate unstable one. Bubbles between two higher curvature branches will not have a smooth Euclidean section however, unless the inner region corresponds to the vacuum, something not allowed in general. 

We can also analyze the stability of the potential at the equilibrium point by means of \reef{potstability}, that up to irrelevant positive factors is
\be
V_{\rm eff}\sim\left(\frac{\partial \widetilde\Pi}{\partial H}\right)^{-1}\left(\frac{\partial^2 \widetilde\Pi}{\partial a^2}\right)\sim \frac{\left(\frac{1}{\sqrt{-g_+}\Upsilon'[g_+]}-\frac{1}{\sqrt{-g_-}\Upsilon'[g_-]}\right)}{\int_{\sqrt{g_-}}^{\sqrt{g_+}}{{\! dx\, \frac{\Upsilon'[-x^2]}{x^2}}}}
\label{thermstab}
\ee
This expression is positive for stable equilibria and negative in the opposite case. The above expression is negative for LGB gravity with $\lambda>0$, as expected for unstable equilibria.

\begin{SCfigure}
\centering
\includegraphics[width=0.45\textwidth]{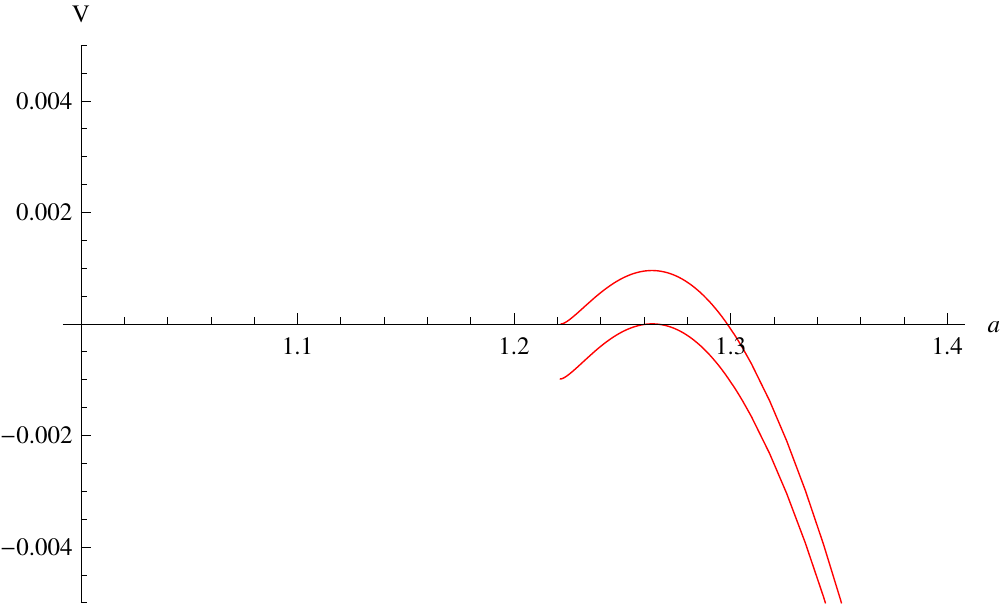}
\caption{Bubble potentials for cubic Lovelock theory with $\lambda=0.17$ and $\mu=0.02$. We just show the relevant branch of the potential connected to the physical equilibrium points, those with positive mass. The upper potential corresponds to the stable configuration (in green in figure \ref{CubicPlanar}) while the other is the unstable one (in blue also in figure \ref{CubicPlanar}). There is a minimum although it is hard to see in the plot.}
\label{cubicPots}
\end{SCfigure}

In the cubic case we have two possible potentials and the equilibrium points can, in principle, be associated to any of them. This seems much more involved but one may still use \reef{thermstab} in order to study the stability of those static points. For our specific example we find that the local potential is positive for the red and green pairs whereas it is negative in the remaining case. Thus we have one possible stable bubble (green) and one unstable (blue) (see figure \ref{cubicPots}).  

We can get even more information from the asymptotic behavior of the potential. For the branch approximated by \reef{asymptH}, we need to first realize that the outer mass is always bigger than the inner one due to the first condition in \reef{staticjunk}. We can compute the integral of the polynomial between the two stable vacua
\be
\int_{\Lambda_{EH}}^{\Lambda_+}{dx\,\Upsilon[x]}\approx -11<0
\label{polintegral}
\ee
and from \reef{asymptH} we can then realize that $H$ is asymptotically negative. The bubble cannot reach infinity along this branch of the potential. It can however reach the boundary following the branch that asymptotes to a constant $H\approx 0.76$, the potential being  asymptotically negative along. This is actually the potential the two positive mass equilibrium points correspond to, as it can be seen in figure \ref{cubicPots}. One of the bubbles, the unstable one may reach infinity by expansion while the other is fixed on its position unless it can {\it tunnel} across the barrier to subsequently expand. The point to the left where the potentials end corresponds to a naked singularity of the outer solution that appears before we even reach the horizon, situated at the origin of the plot. 

We have analyzed a very particular example but the same analysis can be performed in general, in cubic or any higher order Lovelock theory. The possible situations one may encounter are extremely varied. As already seen in the cubic case, we may have stable or unstable equilibrium points whose number may change as we vary the values of the couplings, or $a_\star$ for non-planar topology. These equilibria may in principle correspond to any of the $K-1$ bubble potentials of the theory, and each of these potentials may have any sign at infinity. This will determine whether the bubble may reach the boundary or not, thus the possibility of a change of branch. Also, depending on the couplings, the branches connected by the static configurations may change, all being connected to the EH branch, some being connected or none being connected. Also for two given branches we may have several static configurations connecting them. All of them have to be compared in order to decide which is the globally stable phase. It might even happen that no static configuration exists. 

Depending on the characteristics of the globally preferred phase for given asymptotics, always a thermalon when it exists, the fate of the system may be very different.  The junction conditions considered above determine not only the equilibrium configuration but also, in Lorentzian signature, the effective potential felt by the bubble and, consequently, its subsequent dynamics. 

If no thermalon connecting our choice of boundary conditions with the EH branch exists the system will remain on the only possible solution, pure AdS$_+$. This is the case for the unstable asymptotics of the cubic example depicted in figure \ref{cubicPots}. On the contrary, when that static configuration exists it will form but whether it changes the asymptotics or not will depend on the form of the potential. If the bubble is unstable and no potential barrier appears on its way to the boundary, we will have a change of branch. This is for instance what happens for LGB gravity with positive $\lambda$. For a stable bubble the situation is also quite interesting. In a sense this stable configuration provides a regular black hole to a branch of solutions that naively had none. The bubble being frozen at the equilibrium position, this situation is very similar to the Hawking-Page transitions studied in section \ref{HP1}, the system remaining in the black hole phase. In this case however, when the asymptotic potential allows the bubble to reach the boundary, there might be a non-vanishing probability for the bubble to {\it tunnel} to that region. In that case the boundary conditions may eventually change and we again have a branch transition. This is the case of the unstable configuration found in our cubic example (see figure \ref{cubicPots}).

In case the bubble collapses instead of expand, it will generally lead to the destruction of the horizon and the consequent formation of a naked singularity. We will not comment more on this here. By the arguments given in chapter \ref{chp:bhstability} and those outlined in the previous sections, we will just assume the system then comes back to the {\it initial} phase, pure AdS$_+$. This is analogous to the formation of bubbles in a fluid. When these expand the phase transition to the gas phase proceeds while, when it collapses, the system remains liquid.


The results of this section are more general than just the planar case as it also corresponds to the high mass limit of the other two topologies. 
Even though the equations are much more involved in case of spherical or hyperbolic symmetry, the analysis follows in the same way. In the next sections we concentrate in the case of LGB gravity with $\sigma=\pm 1$.

\subsection{Spherical LGB bubbles}

In the case of LGB gravity the thermalon configurations being described here are just relevant for unstable or {\it ghosty} boundary conditions, that of the unstable or ($+$) branch of solutions, and positive LGB coupling. Again, we have to compare the free energy of the thermalon configurations found in the preceding sections with the corresponding thermal vacuum. From the Euclidean point of view this is the only other smooth metric with that unstable boundary conditions. 

\begin{SCfigure}
\includegraphics[width=0.45\textwidth]{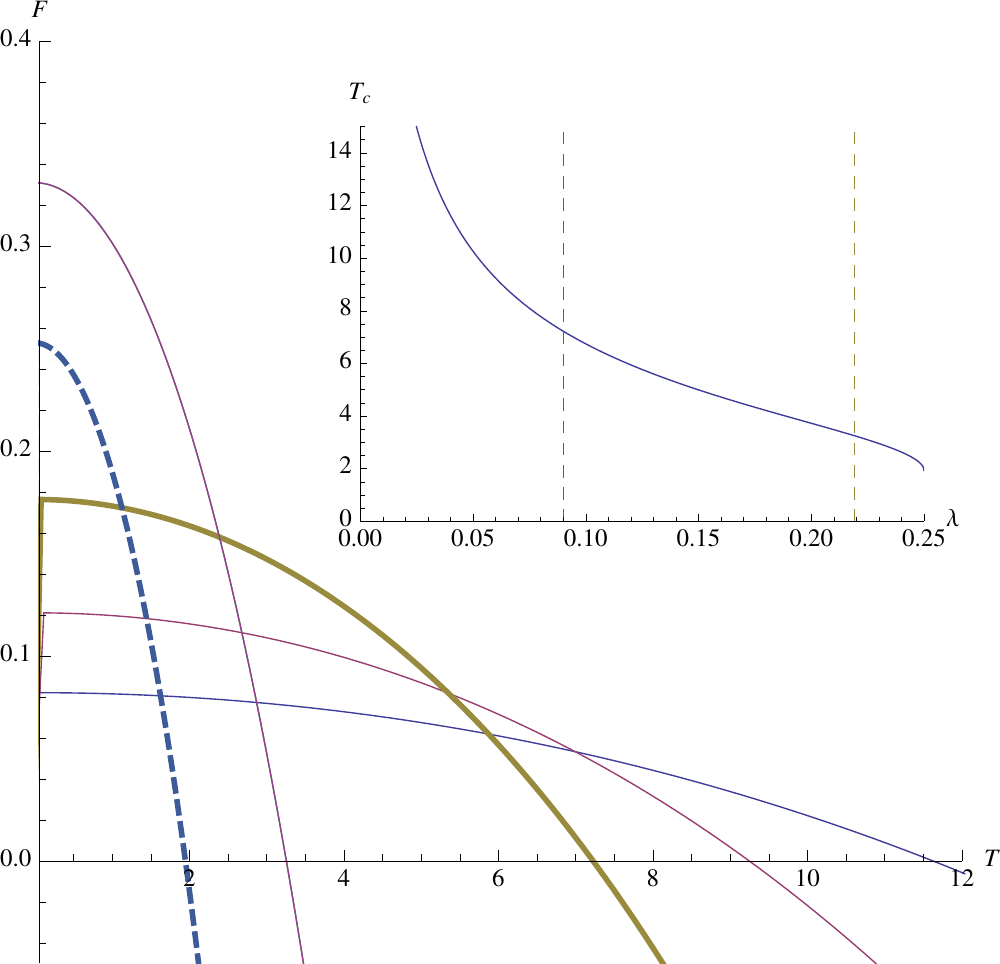}
\caption{Free energy versus temperature in 5d for $\lambda = 0.04, 0.06, 0.09$ (positivity bound), $0.219$ (maximal $F(T=0)$), and $\lambda \to 1/4$ (from right to left). The $\lambda$ dependence of the critical temperature is displayed in a separate box.}
\label{CEGGFig2}
\end{SCfigure}

The resulting phase diagram is similar to the usual Hawking-Page phase transition. For any value of the LGB coupling, the free energy, $F$, as a function of the temperature $1/\beta_+$ displays a critical temperature above which it becomes negative and, thus, the phase transition occurs (see figure \ref{CEGGFig2}). 
If the free energy is positive, however, the system is metastable. It decays by nucleating bubbles with a probability given, in the semiclassical approximation by $e^{-\beta_+ F}$. Therefore, after enough time, the system will always end up in the stable, EH black hole solution.  This is reminiscent of the HP transition, except for the fact that, here, the thermal AdS vacuum decays into a black hole {\it belonging to a different branch}. $T_c(\lambda)$ is monotonically decreasing, the phase transition becoming increasingly unlikely the more we come closer to the EH -- {\it classical} -- limit. In this sense, it is a quantum mechanical phenomenon.

\subsection{Hyperbolic LGB bubbles}

Asymptotically AdS spacetimes with hyperbolic topology are an interesting playground for checking various facts about the recently discovered type of phase transitions. Despite some unclear features of the thermodynamics of these spacetimes, they allow in particular for the discussion of transitions between two asymptotically AdS branches both having regular horizonful black hole solutions. In particular, for the simplest case of LGB theory with $\lambda>0$, both, the EH and the ghosty branch black holes, may have horizons depending on the value of the mass. Even more, even though the range of masses for the ghosty branch is bounded from above (and below) this branch has black hole solutions for all possible temperatures. Besides, although one of the branches is unstable we would like to investigate if the discovered transition protects the theory against this instability in the sense that the unstable branch would never be the preferred phase. This is actually the case for LGB theory with planar or spherical symmetry. Despite the fact that the vacuum is preferred for some range of temperatures in the spherical case, the thermalon may be formed with small but finite probability in that case leading to a change of branch. 

This case is much richer, not only due to the possibility of transitions in both directions, but also because the spectrum of configurations gets richer. The existence of extremal black holes in both branches for specific values of the mass, leads to the corresponding extremal thermalons. These extremal configurations can be considered as qualitatively different from their non-extremal counterparts as regularity of the horizon in the Euclidean section does not fix their temperature. As a limit of non-extremal black holes the extremal solutions necessarily have zero temperature and the entropy corresponding to the Wald formula. Quite the opposite, {\it ab initio} extremal configurations may be identified with any temperature and zero entropy (see chapter \ref{chp:thermo} for details). The same happens for the bubble configurations even though the limiting temperature is not zero. In addition to the non-extremal bubbles seen so far there will be extremal solutions where the horizon of the inner black hole is degenerate. 

For the direct transition, the one leading to the EH branch we will fix the BD-unstable asymptotics and compare the free energy of all possible configurations at the same temperature. These are in principle four families of solutions, extremal and non-extremal black holes and the corresponding bubble counterparts. The same will happen for the reverse transition.  

Let me first comment briefly on the thermodynamics of the unstable branch black holes. These are unstable not only {\it \`a la} Boulware-Deser but also thermodynamically.  Their specific heat is negative and they also have negative entropy. 
Due to this fact their free energy will always be bigger than the one corresponding to the extremal black hole that coincides with its mass. This extremal solution corresponds to the black hole with mass saturating the upper bound, $\kappa=\lambda$ in five dimensions or the one corresponding to a degenerate horizon in higher dimensions. In the 5d case the degenerate horizon coincides with the singularity at $r=0$ but this does not represent a problem as that point is at the end of an infinite throat and should be removed from the geometry. The near horizon geometry can be written as 
\begin{equation}
ds^2\sim 2\lambda d\rho^2+e^{2\rho}\left(-\frac{dt^2}{2\lambda}+d\Sigma_{-1,3}^2\right)
\end{equation}
where we have explicitly removed the point $r=0$ considering a change of variables as $\rho=Log r$. In higher dimensions the horizon is at finite radius and the geometry completely regular. We will concentrate in the five dimensional case in what follows.
Thus, this particular solution is a smooth geometry interpolating between $AdS_5$ at spatial infinity and the previous geometry as we approach $r=0$. Also, it has zero specific heat and entropy and can be considered with any periodicity in Euclidean time and thus it is a well motivated candidate as groundstate of the theory for the sector we are considering. Also as the black holes are unstable all of them have higher free energy than the extremal one and thus the chosen groundstate is always preferred in a semiclassical basis. The only relevant configuration for the next step of the analysis will then be the extremal black hole.

We will have to compare the free energy of our thermalon configurations to that of this groundstate. The free energy of the non-extremal configuration has the usual expression $F=M-TS$ and the one corresponding to the extremal once again corresponds to a constant, thus implying a vanishing entropy. However the extremal free energy does not coincide with the mass in this case. In the extremal case the bubble is situated exactly at the inner horizon, $f_-(a_\star)=0$, in such a way that the rescaling of the temperature just cancels the zero of $f'$ at the horizon,
\be
\frac{\sqrt{f_+}}{\sqrt{f_-}}\frac{f'_-}{4\pi} \to \tilde{T}_+^e ~.
\ee
This is the temperature corresponding to the black dot in the figure where the non-extremal bubble curve ends. The resulting on-shell action is then,
\begin{equation}
\widehat{\mathcal{I}}^e=\beta_+ (M_+^e-\tilde{T}_+^e S^e)
\end{equation}
Again attending at the usual semiclassical picture the entropy of the solution vanishes and the free energy corresponds to an {\it effective} mass that picks some contribution proportional to the Wald entropy. 
\begin{equation}
F_{e,b}=M_+^e-\tilde{T}_+^e S^e
\end{equation}
Here the word extremal does not necessarily mean zero temperature but it is rather a question about the topology of the near horizon region.  The curves for the free energy as a function of the temperature for the different configurations is shown in figure \reef{directH} for several values of the coupling $\lambda$. 
\begin{figure}
	\centering
		\includegraphics[width=0.4\textwidth]{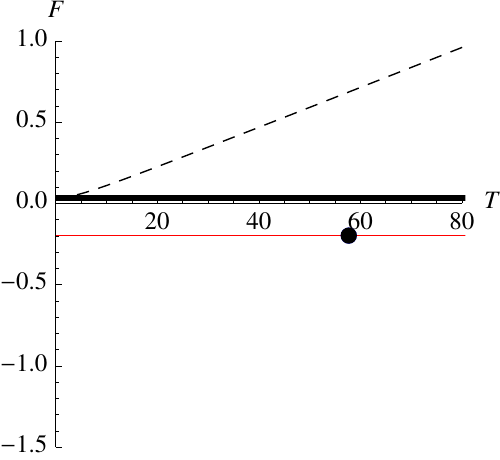}\quad\includegraphics[width=0.4\textwidth]{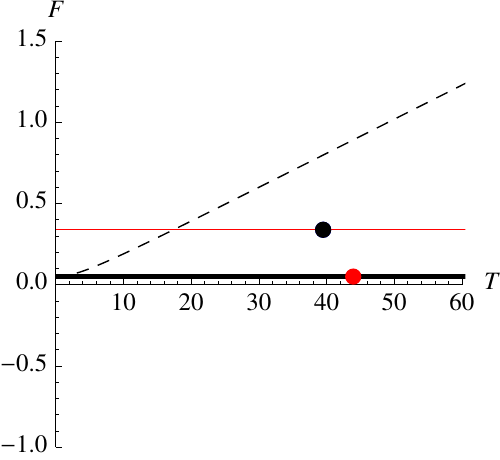}\\[1em]
		\includegraphics[width=0.4\textwidth]{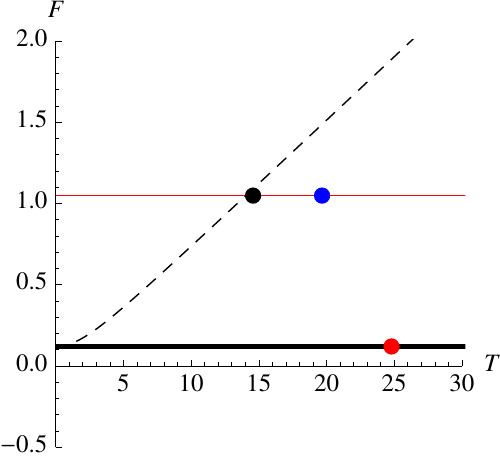}\quad\includegraphics[width=0.4\textwidth]{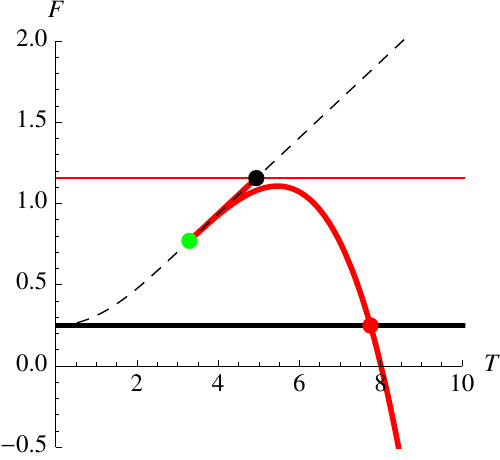}
			\caption{Free energy versus temperature for $\lambda=0.035, 0.05, 0.12, 0.249$. The dashed black curve corresponds to the unstable branch of black holes while the thick black one represents its extremal free energy. The thick red curve corresponds to the bubble with the same asymptotics and its extremal limit is indicated by the black dot. The thin red curve corresponds to the free energy of such extremal state when vanishing entropy is assumed. The dots indicate the locus of different kinds of possible phase transitions even though just some of them may take place in each case. The black dot when relevant represents a phase transition from a extremal to a non-extremal bubble and the red one a transition from the extremal black hole to the non-extremal bubble. The green dot indicate the minimal temperature for the existence of non-extremal bubbles when these are divided in two branches one thermodynamically stable and one unstable. }
		\label{directH}
\end{figure}

For the discussion of the phase diagram we will assume that the extremal configurations are present at any temperature with zero entropy. At the end of this section we will comment on the alternative approach not considering them but as limiting cases  of their non-extremal counterparts. 

\begin{figure}
	\centering
		\includegraphics[width=0.45\textwidth]{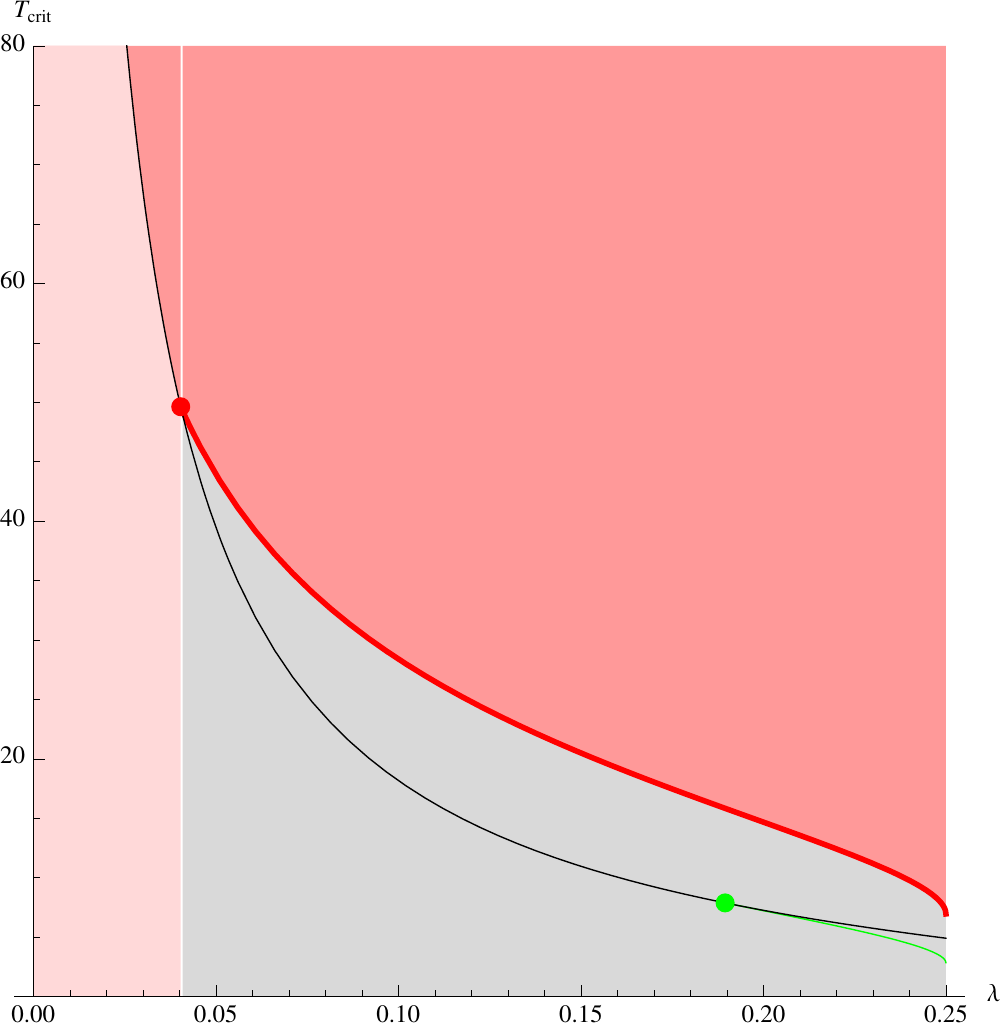}
			\caption{In red the bubble states, light for the extremal, and in gray the extremal (BD-unstable) black hole. The red dot indicate the value of $\lambda$ for which the transition between the extremal unstable black hole and the non-extremal bubble (red line) begins to exist. The red line corresponds to the bubble extremal limit and the red one to the minimal temperature of these configurations when this does not correspond to the extremal one.}
		\label{directHT}
\end{figure}
We have different behavior for different ranges of the LGB coupling. For small values, $\lambda<\lambda_{crit}\approx 0.0404$, the bubble is always the stable phase with a transition between the extremal configuration at low temperatures and the non-extremal one preferred as we increase the temperature. For values of $\lambda$ above that critical value the low temperature phase is the extremal black hole instead, our hypothetical vacuum, the non-extremal bubble being again the stable phase at high temperature. 

Another interesting feature is that for values of $\lambda$ bigger $1/12$ negative entropies appear for the bubble, notice the positive slope of the non-extremal curve in the third and fourth graphics. This negative entropy states however are never the preferred phase of the system and can safely be discarded as unphysical. 

This however would not happen if we consider the alternative approach to the extremal states. In that case these are just limiting cases of the non-extremal configurations already included as endpoints of the corresponding curves (black dot for the extremal bubble). We have to consider the same graphics of figure \reef{directH} but discarding the thick black and thin red curves corresponding to the extremal states. The stable phase at low temperature would then always be the black hole with bubble formation at high temperature. This situation is puzzling as the low temperature phase would in that case have negative specific heat and entropy. The negative entropy bubble states would also be the preferred phase for some range of the temperature in that case. We cannot just discard them as unphysical in this case as we would not have any available metric at low temperatures. The inclusion of the extremal states seems to solve the problem of negative entropy configurations, and reduces it to a thermodynamic instability.

In the region of the $\lambda-T$ phase diagram of figure \reef{directHT} coloured in gray, the bubble configurations have both lower free energy than the {\it vacuum}, this being the thermodynamically preferred phase. Still the probability of bubble formation is nonzero being proportional to the exponential of the difference of the actions of both solutions. Thus after enough time a bubble will form leading again the system to the EH branch. 


For the inverse transition we proceed in the same way but setting the opposite asymptotics. The results are depicted in figure \reef{inverseH}. Strikingly, due to several physical constraints, for low values of the LGB coupling bubbles, either extremal or non-extremal, do not exist, the critical value being $\lambda=\frac{1}{24}(7-\sqrt{13})\approx 0.14$ where the two extremal states degenerate. They are shown in the figure (dashed red and thin red respectively for non-extremal and extremal bubbles) but they would be formed inside the outer horizon. Viable bubble geometries appear above the critical $\lambda$  value represented by the thick red line and they also have a well-defined extremal extension (thin red line). 

For low enough values of the LGB coupling the black hole states are the only available configurations. Below $\lambda=1/12$ the non-extremal solutions are stable and have positive entropy and as a consequence have lower free energy than the extremal {\it vacuum}. The non-extremal black hole is the stable phase of the system for all temperatures. For higher values of $\lambda$, negative entropy non-extremal states appear at low temperatures but the extremal configuration has lower free energy in that case. We can again safely remove the unphysical states. The low temperature phase corresponds to the extremal black holes with a transition to the non-extremal ones at high temperature. 

The stable low  and high temperature phases are still the same above the critical $\lambda=\frac{1}{24}(7-\sqrt{13})$ where the bubble configurations appear. These can be divided in two branches, one stable that merges with the black hole curve at zero temperature and one unstable close to the extremal state that is irrelevant as it has higher free energy than the former. Any of these bubbles have higher free energy than the extremal black holes for all temperatures. Still and as for the transition in the other sense, the probability of a bubble being formed is nonzero and thus after enough time a bubble the BD-unstable phase would form. This would in principle drive the system to that pathological branch but it is also unstable to the formation of bubbles of the EH branch. The final fate of the system seems to be a chaotic situation with bubbles of both phases popping up everywhere. This seemingly problematic situation is avoided for $\lambda<\frac{1}{24}(7-\sqrt{13})$.

In case we discard the free energy curves corresponding to the extremal states, thin red and black lines, we would encounter some problematic behavior for even lower values of $\lambda$. For $\lambda > 1/12$ negative entropy black holes appear and they would be the preferred phase for low temperatures. Above $\lambda=\frac{1}{24}(7-\sqrt{13})$ the bubbles appear with also negative entropy and lower values of the free energy for some range of temperatures, between the lower thin red and green curves of figure \reef{inverseHT}. 

We can summarize saying that for values of the LGB coupling below $\lambda=\frac{1}{24}(7-\sqrt{13})$ the BD-unstable branch of hyperbolic black holes is always driven to the EH one, this being stable against the formation of bubbles. For this scheme the EH branch is protected and the only transition that takes place is a change from the extremal black hole at low temperatures to the non-extremal at higher ones. This transition only occurs when negative entropy states are in the spectrum, namely close to the extremal configuration. For higher values of $\lambda$ the situation is chaotic with bubbles of the opposite branch popping for any choice of the boundary conditions, the interpretation of this being unclear.

Moreover, the transition mechanism presented here provides a possible resolution of the instabilities found for the {\it ghosty} branch of LGB gravity. For spherical and planar topology it is always driven to the EH branch via bubble nucleation. We have just verified that the same happens for hyperbolic spacetimes  as long as the LGB coupling is below $\lambda=\frac{1}{24}(7-\sqrt{13})$. This will presumably be also valid in the cubic case for moderate values of $\mu$.

\begin{figure}
	\centering	\includegraphics[width=0.34\textwidth]{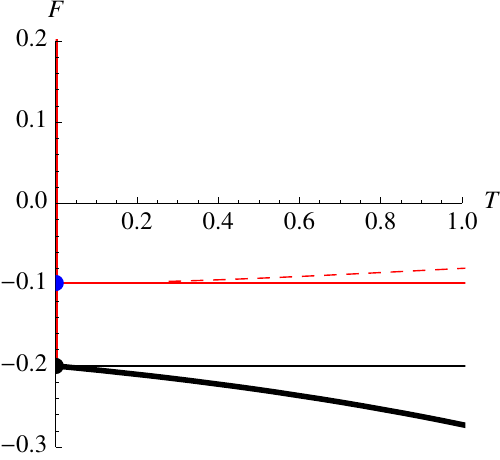}\qquad\includegraphics[width=0.34\textwidth]{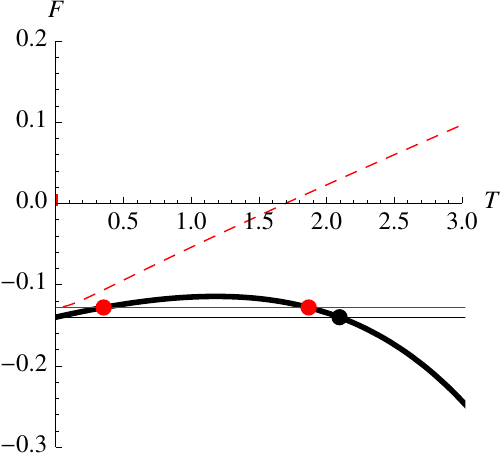}\\	\includegraphics[width=0.34\textwidth]{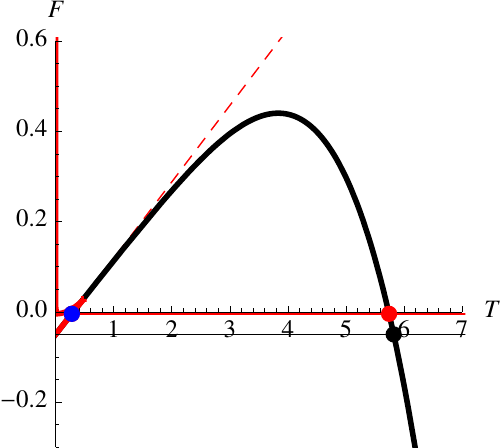}\qquad\includegraphics[width=0.34\textwidth]{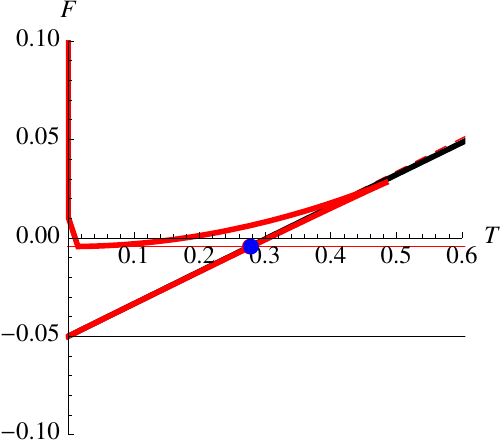} \\
	\includegraphics[width=0.34\textwidth]{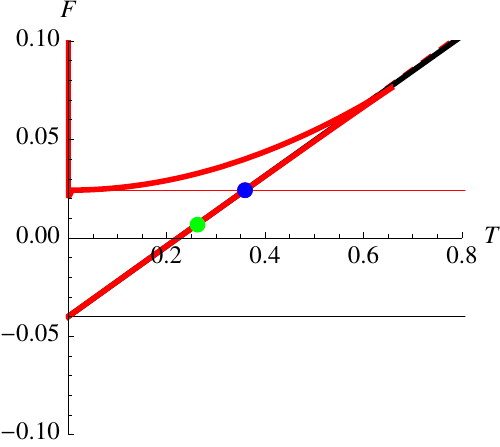}\qquad\includegraphics[width=0.34\textwidth]{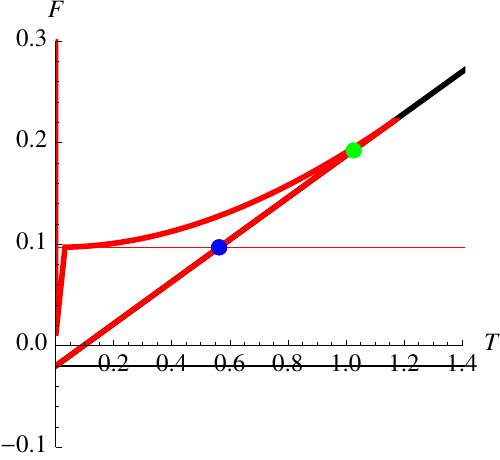}
			\caption{Free energy versus temperature for $\lambda=0.05,0.11,0.2 \text{(and zoom)}, 0.21, 0.23$. In red the bubble states and in black the black holes in the stable branch. The thin black line corresponds to the extremal geometry once assumed the vanishing of the entropy whereas the green line corresponds to the usual Wald entropy for the same geometry. The dashed red lines correspond to the extension for $a<1/\sqrt{2}$ of $M_+-T_-S$. These are not viable bubble solutions in the same way as its extremal extension represented by the thin red line. Acceptable bubble geometries appear for $\lambda\geq\frac{1}{24}\left(7-\sqrt{13}\right)$ represented by the thick red line and they also have a well-defined extremal extension. The red dots indicate the locus of extremal bubble to black hole transitions and viceversa whereas the blue and the green ones indicate non-extremal extremal bubble transitions and bubble black hole transitions respectively. Below the green dot when there the black line is below the red one and so we have a low temperature black hole phase as well as a high temperature one. }
		\label{inverseH}
\end{figure}

\begin{figure}
	\centering
		\includegraphics[width=0.45\textwidth]{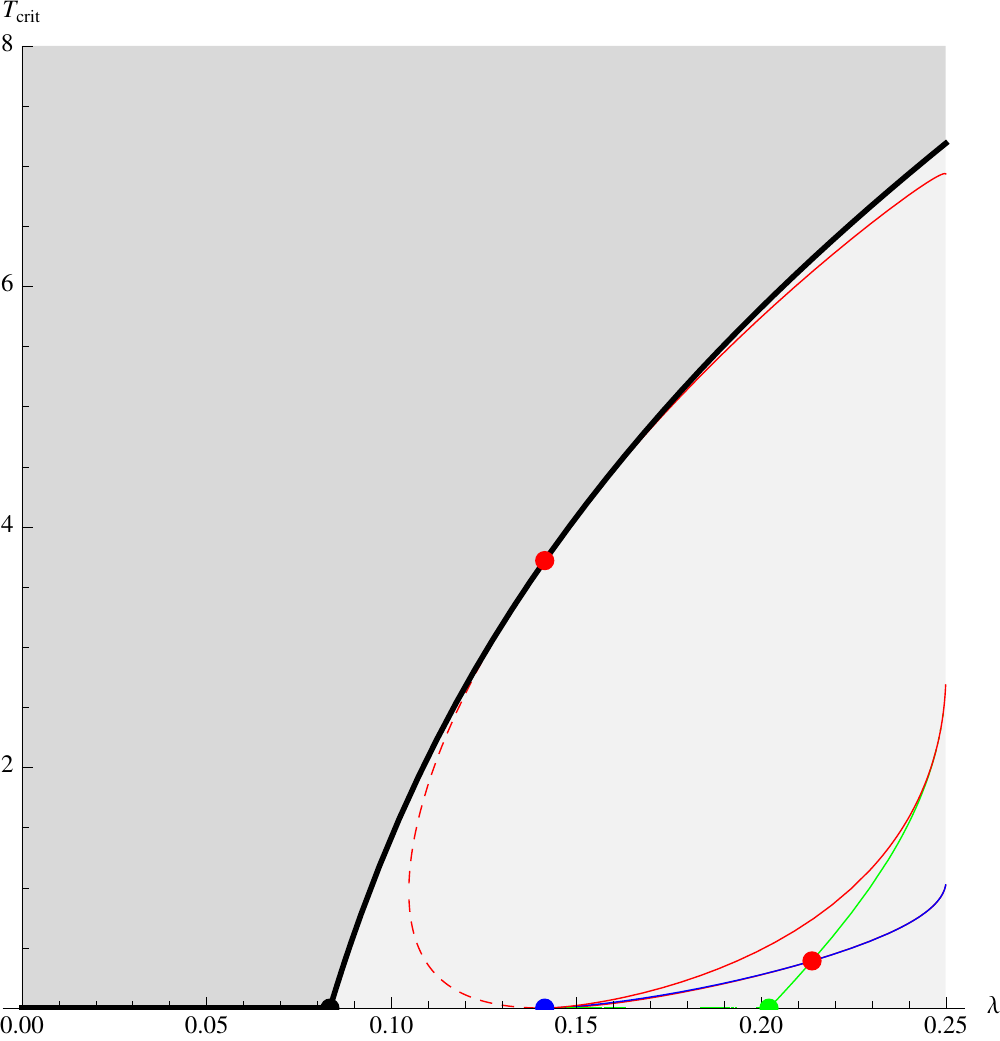}
			\caption{In red the bubble states and in gray the black holes, light for the extremal configurations. Below the red dot the red curve is below the blue one and they reverse passed that point.  }
		\label{inverseHT}
\end{figure}

\section{Discussion}

Throughout this chapter, we have broaden the scope of our analysis of Lovelock theories and their solutions. In particular, we have included the possibility of distributional solutions for which the spin connection is discontinuous at some given junction (hyper)surface. In the absence of matter in the {\it bubble}, it glues two solutions that correspond to different branches of the same theory, \ie to different asymptotics. From the Hamiltonian point of view the existence of such configurations is allowed by the multivaluedness of the canonical momenta. 

We have also proven that it is possible to generalize the thermodynamic notions  
usually applied to black holes to these new solutions. Under certain mild regularity assumptions, static bubble configurations can be assigned temperature, mass, entropy and free energy, all verifying the expected thermodynamic relations. Having proven the consistency of the thermodynamic picture, we have then  analyzed local and global stability of our system in this generalized context and the occurrence of phase transitions. We have restricted our attention mainly to the LGB case even though the same kind of transitions occur also in the general case, as it can be explicitly shown for planar symmetry. This is a novel mechanism for phase transitions that is a distinctive feature of higher curvature theories of gravity. Specifically, phase transitions among the different branches of the theory are driven by this mechanism. Mimicking the thermalon configuration \cite{Gomberoff2004}, a bubble separating two regions of different cosmological constants pops out, generically hosting a black hole.

In the context of LGB gravity, this configuration is thermodynamically preferred above some critical temperature. The corresponding phase transition can be interpreted as a generalized HP transition for the higher-curvature branches, driving the system towards the EH branch. This happens even for the hyperbolic case in which the reverse transition is also possible. For the EH-asymptotics the usual black hole is always the preferred phase. Below some critical $\lambda$ thermalon solutions do not even exist in this sector of the theory. 

The junction conditions do not just determine the existence of the static configurations but also their dynamics. In the LGB case, the bubble configuration, being unstable, dynamically changes the asymptotic cosmological constant, transitioning towards the stable horizonful branch of solutions, the only one usually considered as relevant. This is then a natural mechanism for the system to select the general relativistic vacuum among all possible ones. We are aware of the fact that the vacuum $\Lambda_+$ in the LGB theory exhibits ghosts. The phenomenon presented in this chapter, however, takes place in general Lovelock gravity as well, where there may be further healthy vacua than the one connected to the EH action \cite{Camanho2010d}. The selection of the EH branch among all the stable ones is not as universal as one may naively think, however, as not all branches of solutions are connected to the EH one by a thermalon. Thus, choosing such boundary conditions the asymptotics cannot change due to the mechanism presented here. Besides, even when a suitable static configuration exists, the dynamics of the bubble might not allow the bubble to scape to infinity, \eg being in stable equilibrium, so that it cannot change the asymptotics either. In that case, bubble configurations might represent new phases for higher curvature vacua, in some situations very similar to a regular black hole. Remember that in many cases these branches do not have smooth black hole solutions.

Usually, AdS spacetimes are considered to have perfectly reflective boundaries so that anything bounces back in finite proper time. This is not the case here. As the theory has several possible asymptotics, all  \`a priori equally valid, we might choose any of them but we have to allow for asymptotics changing solutions such as those described in this chapter. Through the evolution of the bubble, these also change the temperature and the mass of the solution, as it also happens in the case of {\it quenches} (see \cite{Buchel2013a} for an example in the context of holography). We might think of the thermalon mechanism as a kind of {\it thermodynamic quench} induced by the system, not externally. 

\begin{figure}
\centering
\includegraphics[width=0.65\textwidth]{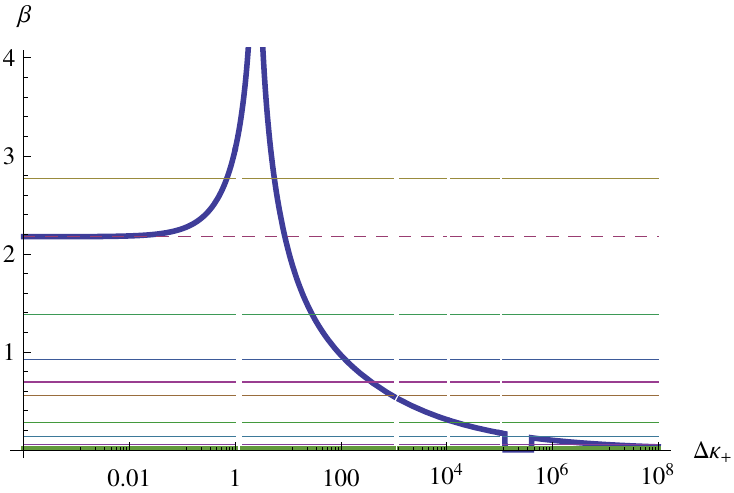}
\caption{Oscillation period for a bubble in 5d LGB gravity for $\lambda=0.1$, $\kappa_-=\kappa_-^{th}$ and $\kappa_+=\kappa_+^{th}+\Delta\kappa_+$, $\kappa_\pm^{th}$ being the mass parameters corresponding to $a_\star=1$. The horizontal lines, from top to bottom, correspond to $\beta_-/n$ for $n=1,2,3,4,5,10,20,50$. In this specific example we have an infinite tower of smooth manifolds with increasing $n$ and $\kappa_+$.}
\label{ISpectrum}
\end{figure}

Following the same approach that lead us to consider the thermalon configurations discussed so far, one can include in the analysis more complex solutions. For instance, in the LGB case with positive $\lambda$ (see figure \ref{rootchange}), for values of $\kappa_+$ slightly above the thermalon value giving the equilibrium point, the Euclidean trajectory would be oscillatory, between a maximal and a minimal value of the radius $a_\pm$. The same happens in general for any unstable bubble. Instead of picking the static value, we may tune the value of $\kappa_+$ so that the period of the oscillation
\be
\beta=2\int_{a_-}^{a_+}\frac{\dot{T}_-}{\sqrt{2 V(a)}}da
\ee
{\it fits} in the periodicity $\beta_-$ given by the regularity at the horizon, an integer number of times. We can plot this periodicity as a function of $\kappa_+$ (see figure \ref{ISpectrum}) and identify the values that verify 
\be
\beta_-=n\beta
\ee
for some integer $n$ as the values of the mass parameter that yield smooth Euclidean manifolds. The picture is similar to the thermalon of figure \ref{thermalon} with a wiggly line as junction. The above condition is such that the trajectory is closed. These new solutions have to be considered as well in the canonical ensemble and may lead to more general phase transitions than those previously discussed. 

Lovelock gravities, as well as any other higher curvature theory, have several branches of asymptotically (A)dS solutions that might admit an interpretation as different phases of the dual field theory. Phase transitions among these are driven by the mechanism described in the present chapter. 

From the holographic point of view this looks like a confinement/deconfinement phase transition in a dual CFT, involving an effective change in the microscopic properties, \eg the central charges, both phases being strongly coupled. Whether a phenomenon like this takes place in a 4d or higher dimensional CFTs, particularly within the framework of the fluid/gravity correspondence --where both phases might be characterized by different transport coefficients--, or it is overtaken by higher curvature corrections, is an open question at this point.





\thispagestyle{empty}


\part{HIGHER CURVATURE GRAVITY \ {\bfseries\itshape \&}\ \ HOLOGRAPHY}

\chapter{\bfseries \itshape AdS/CFT and Lovelock gravity}
\chaptermark{AdS/CFT and Lovelock gravity}
\label{chp:AdSCFT}

\vspace{.6cm}

\begin{quotation}
\flushright
{\it ``But although all our knowledge begins with experience,\\
it does not follow that it arises from experience.''}\\

\vspace{.3cm}

Immanuel Kant
\end{quotation}

\vspace{3cm}

\noindent Near the end of the XIX$^{\rm th}$ century, physics was thought to be an {\it almost} finished discipline. For the scientific community at the time, Newtonian mechanics on the one hand and Maxwell's theory of electromagnetism on the other would describe perfectly any physical process. They could not be more mistaken. During the last years of the century and the first decades of the next, some new experimental evidence would require a complete revision of our scientific paradigm. This revolution is the origin of Einstein's special theory of relativity and also of quantum mechanics. Moreover, the quest for reconciling the apparent inconsistencies between these two theories motivated the birth of quantum field theory (QFT), the basis of our current understanding of particle physics.

Furthermore, Newtonian gravity was neither compatible with the new paradigm of special relativity, the solution of this problem being on the origin of Einstein's formulation of general relativity. We already talked about general relativity and its higher curvature generalization in the first part of this thesis. QFT and general relativity are the two pillars of contemporary theoretical physics, key to our current knowledge of the Universe. Both theories have been successfully tested to extremely high accuracy and are thought to faithfully describe any physical process within the corresponding regimes of applicability. The Higgs boson was the only missing particle of the standard model of particle physics, the last discovered building block of our understanding of the most fundamental aspects of the physical world. Is this description of physics {\it complete} as our XIX$^{\rm th}$ century fellows thought?

QFT ad general relativity are two conceptually very different theories and apply to almost complementary regimes. However, both theories are contradictory in situations for which both gravitational and quantum mechanical effects are relevant. In other words, despite all its successes, general relativity, or any classical modification of it such as Lovelock theory, can not be a {\it complete} theory of gravity. In order to understand phenomena at energy scales beyond the Planck mass, $M_{Pl}=\sqrt{\frac{\hbar c}{G_N}}$ (or distances smaller than the Planck length), a full quantum theory of Gravity is needed. Moreover, the new theory must approach the precedent theories in the respective regimes of validity, while solving their incompatibilities. The search for a viable theory of Quantum Gravity is one of the most important problems of our time, where by viable we mean theoretically but also experimentally consistent with the Universe we live in. 

One of the key points of the disagreement between the two frameworks is the r\^ole played by the spacetime on each of them. From the QFT point of view this four dimensional manifold is just a fixed background, the arena where all physical processes happen. In contrast, from the GR perspective gravity is just another dynamical field and, as such, there is no \`a priori reason for it not to be quantized. Nevertheless it cannot be quantized in the same way as all the other fields. Einstein-Hilbert, or any other gravity action, describes the dynamics of a spin two particle but renormalizability constrains the possible spin in the spectrum to be one or less. Moreover, the Weinberg-Witten theorem \cite{Weinberg1980} also restricts the spin of particles if they are charged under Poincar\'e covariant currents. Gauge bosons and gravitons scape the assumptions of the theorem because of gauge symmetry, but this also forces the spacetime to be dynamical. Furthermore, this result also forbids the existence of composite gravitons (see \cite{Loebbert2008} for a recent discussion on the topic). One cannot get a massless spin two bound state in a theory with a local stress-energy tensor, showing that one cannot start with a local theory in Minkowksi space and generate Einstein gravity as an emergent phenomenon, thus eliminating the last possibility of consistently describe spin two particles in a fixed Minkowsky background.

One of the candidate quantum theory of gravity is string theory, even though it was born in a quite different context. In fact, it was first proposed in the sixties  as a theory of strong interactions, mainly because of the way the meson spectrum is organized in Regge trajectories. We now describe the strong interaction through Quantum Chromodynamics (QCD) although there is still much to be understood, mostly due to the fact that it is strongly coupled at low energies. Most tools used in QFT are only suited for the peturbative regime.

It was after the rise of QCD that it was recognized that the spectrum of closed string theories always includes a spin 2 massless excitation with the correct properties to describe a graviton, the quantum of the gravitational interaction. String theory was the first consistent  candidate for a complete quantum description of gravity. Its main advantage is that softer interactions between {\it stringy} objects would help bridle the divergences encountered in previous quantization approaches using point-like gravitons.

The early connection to strong interactions was premonitory. As we will describe string theory has offered many new and deep  perspectives into QFT and its connection to gravity. Gerard 't Hooft was one of the scientists that most contributed to the understanding of that connection. Looking for new ways of simplifying computations in QCD, he considered $SU(N)$ instead $SU(3)$ gauge theory with a very large number of colors, $N$ \cite{Hooft1974}. Perhaps the understanding of this simplified theory would shed some light on QCD or we can try to approach the physical value $N = 3$ with a perturbative expansion in $1/N$. Using his famous double line representation, 't Hooft found that Feynman diagrams may be organized in a double expansion in terms of the genus $g$ of the two dimensional surface defined by the diagrams and the 't Hooft coupling\footnote{Even though we use the same symbol $\lambda$ for the 't Hooft and the LGB couplings, it will be easy to know which one we are referring to as they never appear in the same context.} $\lambda=g_{\rm YM}^2 N$ as
\be
\sum_{g=0}^{\infty}\left(g_{\rm YM}^{2}\right)^{2g-2}\sum_{h=2}^{\infty}C_{g,h}\lambda^h ~,
\ee
strongly reminiscent of the perturbative expansion of string theory with string coupling
\be
g_s\sim g_{\rm YM}^2~.
\ee
The 't Hooft limit corresponds to large $N$, $N\to\infty$, keeping the 't Hooft coupling $\lambda$ fixed. In this regime, only planar ($g = 0$) Feynman diagrams  contribute and the putative string theory becomes effectively free. 

The other contribution of 't Hooft, that of {\it holography} \cite{Stephens1994,Susskind1995}, is not less fundamental for the aim of the present essay. Motivated by the work of Bekenstein and Hawking on the entropy of black holes and its scaling with the area of the system, 't Hooft and later Susskind proposed the {\it holographic principle}. According to this proposal, the
maximum entropy in a spacetime volume is proportional to its surface area. This is far smaller than the entropy of any local QFT on the same space,
even with some UV cut-off, as such a theory would have an entropy which scales with the volume. Hence, for a string theory or any quantum theory of gravity to be dual to a QFT, the field theory would have to be formulated on one lower dimension, so that the degrees of freedom can in principle be matched. The idea of the gravitational theory as effectively described in a lower dimensional space is also reinforced by the fact that it cannot have local gauge invariant observables. The notion of locality is tightly tied to the notion of coordinate frame and in a diffeomorphism invariant theory local quantities are thus not gauge invariant, observables having to lie at the boundary of the spacetime. One example of this is the gravitational energy not being {\it localized} anywhere in spacetime. Just the total mass, integrated over the spacetime hence not dependent on positions, is a well-defined gauge invariant observable.

\section{The Maldacena conjecture}

The Weinberg-Witten theorem effectively forbids the notion of a spin two particle in a local QFT with Poincar\'e invariance. However the holographic principle readily offers an elegant way out that immediately solves some of the apparent inconsistencies. The graviton does not appear in the QFT but it has to live in a higher dimensional spacetime. 

As it has been argued, the large $N$ limit of SU($N$) gauge theories is expected to be described in terms of strings. Naively, for $d = 4$ Yang Mills theory one would expect to get a bosonic string theory in four dimensions. However, the bosonic string is not consistent quantum mechanically in $d = 4$ and this is not correct. The reason for this inconsistency is that the classical Polyakov action,
\be
I\sim \int{d^D x\sqrt{-g}g^{\a\b}\partial_\a X\partial_\b X}~,
\ee
has a Weyl symmetry $g_{\a\b}\rightarrow\Omega g_{\a\b}$ which is not a symmetry quantum mechanically, it has an anomaly. In the quantum theory, under a change of the metric of the form $g_{\a\b}=e^\phi\hat{g}_{\a\b}$ the partition function,

\be
e^{-I_{\rm eff}}=\int{\cD[X]\cD[b,c]e^{-I[X, g]-I[b,c,g]}}~,
\ee
changes as \cite{Maldacena2003b}
\be
I_{\rm eff}(g) -I_{\rm eff} (\hat{g})=\frac{26-D}{48\pi} \int{d^D x\sqrt{-g}\left( \frac{1}{2}(\hat{\nabla}\phi)^2+\hat{R}^{(2)}\phi+\mu^2e^\phi \right)}~,
\ee
where $\hat{R}^{(2)}$ is the Ricci scalar on the worldsheet of the string.

This action for $\phi$ is called {\it Liouville action}. Even though the initial classical action for the conformal factor in the metric was zero, a non-trivial action was generated quantum mechanically. This new dynamic field can be thought of as an extra dimension of spacetime in the context of string theory. Being interested in gauge theories in four dimensions, we then need to look for five dimensional strings. We need to specify the space where these strings move. It should have four dimensional Poincar\'e symmetry, so the metric has the form

\be
ds^2=\o^2(z)(ds^2_{1+3}+dz^2)~,
\ee
where we have used the reparametrization invariance to set the coefficient of $dz^2$ equal to that of the 4 dimensional metric.

Consider now a conformal field theory, e.g. $\cN = 4$  Super Yang Mills. In this case the action is also invariant under conformal coordinate transformations such that $g = \Omega^2(x)g$, and $\Omega$ is any function. Then, a scale transformation $x\rightarrow\lambda x$ should be a symmetry. However, string theory has a scale, the string length $l_s$. Then, the only way that our string theory could be symmetric under the scale transformation is that this corresponds to an isometry of the background. This implies that the extra dimension must also change under this scaling, $z\rightarrow \lambda z$ and that the form of the warp factor is $\o (z) = b/z$. So we end up dealing with 5 dimensional anti-de Sitter space
\be
ds^2 = \frac{b^2}{z^2}(dx^2_{1+3} + dz^2).
\ee

In 1997 Maldacena \cite{Maldacena1998} provided a precise realization of the duality between gauge theory and gravity, allowing for a detailed understanding of the facts examined above. This work, together with \cite{Gubser1998} and \cite{Witten1998} where a precise dictionary between these naively very disparate theories was constructed, laid the foundations of what later became  the gauge/gravity duality, or AdS/CFT in its most popular incarnation (see also \cite{Aharony1999}). 

The AdS/CFT correspondence is a relation between a conformal field theory (CFT) in $d$ dimensions and a gravity theory in ($d+1$)-dimensional anti-de Sitter background. The first hint of this duality is the degree of symmetry in both sides of the correspondence. The group of isometries of $AdS_{d+1}$ space is SO($2,d$), which coincides with the conformal group in $d$ dimensions. 
In fact it can be shown  that the group of isometries of AdS space acts on its boundary as the conformal group acting on Minkowski space. This is one of the reasons why the CFT is usually said to live on the boundary of AdS. The extra dimension is related to the energy scale in such a way that it is sometimes said that the AdS/CFT correspondence {\it geometrizes} the renormalization group flow of the QFT.


In the context of string theory, gauge theories describe the low energy dynamics of open strings living on several superposed D-branes. For example, the field theory on $N$ D3-branes is 4d $\cN = 4$ U($N$) Yang Mills at low energies. These open string have the possibility of colliding and merging to form a closed string, not bound to the brane any more. The bulk spacetime where this closed strings move is warped due to the presence of the D-branes. In fact, as Polchinski proved \cite{Polchinski1995}, D-branes are actually the same object as p-branes, which are solutions of IIB supergravity in ten dimensions. The metric for one of this solutions extended in three spatial dimensions is

\be
\label{pbrane}
ds^2=H^{-1/2}(r)(-dt^2+ d\vec{x}_3^2+H^{1/2}(r)(dr^2+r^2 d\Omega_5^2)~,
\ee
where $H=1+b^4/r^4$ and $b^4=4\pi g_s N l_s^4$. The directions $d\vec{x}_3$ are those  parallel to the brane, and $r$ is the distance from the brane. From the point of view of a distant observer, any excitation near the brane (the horizon, $r=0$) has very small energy,

\be
E_{\infty}(r)=\sqrt{-g_{00}}{E_{r}}\sim rE_{r}\rightarrow 0~.
\ee.

Then, we conclude that at low energies only these excitations will survive. The near horizon geometry, $r\rightarrow0$ due to the D3-branes is

\be
ds^2\approx \frac{b^2}{z^2}(-dt^2+d\vec{x}_{3}^2+dz^2)+b^2 d\Omega_5^2 ~,
\ee
where we have defined $z=b^2/r$. This is the metric of $AdS_5\times S^5$, both factor spaces having the same radius $b$. The low energy dynamics of closed strings is determined by the usual action of general relativity (with some extra fields), then, we have gravity in an $AdS_5\times S^5$ background at low energies. The AdS/CFT correspondence conjectures that, in the low energy regime, the two alternative descriptions should be equivalent. So $\cN = 4$ U($N$) Yang Mills is equivalent to gravity in $AdS_5\times S^5$ \cite{Aharony2000}. This relation can be understood as an example of the {\it open-close} string duality.  In this context, gravity (or more precisely type IIB supergravity) is the low energy limit of (type IIB) superstring theory, which is consistent in ten dimensions. 
 
Since the gauge theory describes the dynamics of open strings whereas gravity describes closed strings moving in the bulk, we can identify in the usual {\it stringy} way
\be
g^2_{\rm YM}=4\pi g_s ~,
\ee
in agreement with the generic expectations of 't Hooft's $1/N$ expansion. Beyond that leading SUGRA approximation, $g_s$ and $\alpha'$ correspond to $1/N$ and $1/\lambda$ corrections respectively.

The relation between the two theories is a duality. The gravity description is a good approximation to string theory if the radius of the space is much larger than $l_s$, since $l_s$ is the intrinsic size of the graviton. The radius of AdS space (\ref{pbrane}) is related to the 't Hooft coupling such that this implies
\be
\label{coupling}
\frac{b}{l_s}\sim \lambda^{1/4}\gg 1
\ee
We also require that string corrections are small so $g_s\rightarrow 0$, thus we need to consider the large $N$ limit of the theory. 

In the field theory side, the large $N$ theory has effective coupling $\lambda=g_{\rm YM}^2 N$ such that when this coupling is small, $\lambda\ll 1$, we have a well-defined perturbative field theory, whereas in the opposite strongly coupled regime (the 't Hooft limit) of the gauge theory we expect strings. This is the same regime for which the classical gravitational description is valid.

Then, in the region where the 't Hooft coupling is very small, $\lambda\ll 1$, the gauge theory description is weakly coupled and the other (gravity) is strongly coupled. The opposite is true when this parameter is large. This is what makes the conjecture so interesting but also so hard prove or refute. In the specific example being discussed here, supersymmetry ensures that there are some quantities that are independent of the coupling.  Computing these on both sides, and checking that they agree, we have checks of the duality \cite{Aharony2000}. Some of the most convincing tests have been recently reviewed in \cite{Klebanov2008a}.

The AdS/CFT correspondence is by now a well-established non-perturbative duality of paramount importance. After several years of research, it has overpassed dozens of checks and has been applied in a plethora of systems that go far beyond the large $N$ limit of $\mathcal{N} = 4$ super Yang-Mills theory in four dimensions, originally portrayed by Maldacena \cite{Maldacena1998}.  The strongest version of the conjecture is that this correspondence is valid for any value of $g_s$ and $N$, even if we are only able to make calculations in certain limits. In fact, there have been found many non-trivial examples of finite $N$  and $\lambda$ corrections that agree between AdS and CFT theories. The pursuit of a interpolation between the respective regimes where the string or gauge theory descriptions are reliable 
has delivered many insightful developments and ideas, {\it e.g.} integrability of planar $N = 4$ SYM (see \cite{Beisert2011} for a review) and also recently the recovery of the bootstrap program for CFTs \cite{El-Showk2012}. 

The correspondence has also been generalized to a plethora of other examples, distilled in a similar manner from other brane configurations in string theory. The known dual couples are impressively varied, with different dimensionality, global and gauge symmetries or particle content. However, the relevant closed string or (super)gravity theory dual to physical QCD remains elusive. The advent of the gauge/gravity correspondence has motivated important achievements not only for QFT and gravity but also in mathematics and in other seemingly unrelated branches of physics, from Condensed Matter to hydrodynamics.

\subsection{Correlation functions}

The dynamics of any CFT is defined by the correlation functions of the gauge invariant local operators in its spectrum. The dynamical information of AdS/CFT duality can then be realized as a relation between the generating functional for CFT correlators and the string theory partition function with appropriate boundary conditions on the AdS boundary \cite{Gubser1998b,Witten1998}. 
In the case of a scalar field, $\phi(z, \mathbf x)$ , we have
\be
\left\langle \exp\left(i\int_\Sigma dx^\m \phi_0(\mathbf x)\,\mathcal{O}(\mathbf x)\right)\right\rangle_{\rm CFT}=\mathcal{Z}_{\rm string}[\phi(0,\mathbf x)=\phi_0(\mathbf x)]~,
\ee
where the asymptotic value (or boundary condition) of the scalar field in AdS acts as a source for the dual CFT scalar operator $\mathcal{O}$. Moreover, the hamiltonian is realized in the CFT as the dilatation operator in such a way that energy in AdS corresponds to conformal dimension in the field theory, these being related as 
\be
m^2=\Delta(\Delta-d)~.
\ee
The same can be also generalized to operators with spin. In the CFT side the spectrum will always contain many operators with different spins and conformal dimension. One of them is {\it universal} in the sense that it is present for any CFT and that it has some very specific properties. This operator is the stress-energy tensor and it is realized in the dual gravitational picture by the graviton. Restricting to purely gravitational theories will then amount to the analysis of correlators of the stress-energy tensor\footnote{\S\ This truncation, to pure gravity in the bulk, is not always self-consistent, as demonstrated for instance in \cite{Camanho2013c}}. In that case we can compute the generating function as
\begin{equation}
\left\langle \exp \bigg( \int\!d\mathbf x ~\eta^{ab}(\mathbf x)\; T_{ab}(\mathbf x) \bigg) \right\rangle_{\!\rm CFT} = \mathcal{Z} \left[ g_{\mu\nu} \right] \approx \exp \left( - \mathcal{\widehat{I}}[g_{\mu\nu}] \right)  ~,
\end{equation}
where $\mathcal{Z} \left[ g_{\mu\nu} \right]$ is the partition function of quantum gravity  --the same we used to discuss thermodynamic properties of black holes-- and $g_{\mu\nu} = g_{\mu\nu}(z,\mathbf x)$ such that $g_{ab}(0,\mathbf x) = \eta_{ab}(\mathbf x)$. From this expression, correlators of the stress-energy tensor can be obtained by performing functional derivatives of the gravity action with respect to the boundary metric. This, in turn, is simply given by considering gravitational fluctuations around an asymptotically AdS configuration of the theory. The bulk metric acts as a source for the stress-energy tensor in the boundary (and viceversa).

\subsection{Confinement/deconfinement phase transition}

The gravity (or string theory) partition function is in general computed by integrating over bulk metrics, $g$, that induce a given metric $h$ on the boundary. This is calculated in the same way we described at the end of section \ref{EQG}, in the framework of Euclidean Quantum Gravity. Again, in the stationary phase approximation, one can compute $\mathcal{Z}(h)$ by finding a solution $g$ for the Einstein equations with the required boundary behavior, and setting $\mathcal{Z}(h) \approx \exp\left(-\widehat{\mathcal{I}}[g]\right)$. If there are several solutions which are asymptotic to the given boundary, the leading contribution comes from the solution with the lowest action and we have phase transitions when other solution dominates over the previous one. The only difference with the Euclidean Quantum Gravity case is that here we have to consider solutions of $AdS_5\times S^5$, not just $AdS_5$. In Einstein's theory, any asymptotically $AdS_5$ solution with an extra $S^5$ factor is a consistent solution for the corresponding field equations in ten dimensions. In the higher curvature case this is not as simple.

We consider $\cN=4$ Super Yang Mills theory at finite temperature on a spatial manifold $S^3$ or $\IR^3$. We must compute the partition function on $S^1\times S^3$ with respective radius $\b$ and $b$. By conformal invariance only the ratio $\b/b$ matters. This can be regarded as a dimensionless temperature on the CFT side, $T_\text{CFT}=b/\b=bT$. 

It has been shown (see citation in \cite{Witten1998a}) that the $\cN=4$ theory on $S^1\times S^3$ has a phase transition as a function of $\b/b$ in the planar limit. The large $\b/b$ phase has some properties in common with the usual large $\b$ (small $T$) phase of confining gauge theories, while the small $\b/b$ phase is analogous to a deconfining phase. In the large $N$ limit, a criterion for confinement is whether the free energy (after subtracting a constant from the ground state energy) is of order one, reflecting the contributions of singlet color hadrons, or of order $N^2$, reflecting the contributions of gluons \cite{Witten1998a}. 

We can compute the partition function of this theory from the gravity side using the correspondence. In the case of Einstein-Hilbert gravity in five dimensions,  considering the extra $Vol(S^5)\sim b^5$ factor arising from the volume of the 5-sphere, for the solution containing a black hole we find $F/T\sim b^8 \sim N^2$ as expected for a deconfining phase. In contrast, the low temperature phase we have a trivial $N^2$ contribution coming from the spacetime volume. The excitations (just thermal radiation) have no dependence on $N$ and give a contribution of order one, as in a confining phase. 

Thus, the Hawking-Page phase transition we found in the analysis of the canonical ensemble of AdS space corresponds to a confinement/deconfinement phase transition in the $\cN=4$ Super Yang Mills theory living on its boundary, for Einstein-Hilbert gravity and, by analogy, also for Lovelock theories in AdS.
For non-conformal theories, such as that described by the Klebanov-Strassler model \cite{Klebanov2000}, the confinement/deconfinement transition occurs at finite temperature also in $\mathbb{R}^3$. In that context the associated QGP might be affected by hydrodynamic effects that may dynamically shift the temperature at which the phase transition occurs \cite{Buchel2013b}. The viscous terms in a relativistic fluid result in reducing the effective pressure, thus facilitating the nucleation of bubbles of a stable phase. This is known as {\it cavitation}. The effect is particularly pronounced in the vicinity of (weak) first-order phase transitions. We can use the holographic correspondence to study cavitation in strongly coupled non-conformal plasmas close to their confinement phase transition. An example is presented in Annex \ref{cavitation} for the case of planar cascading gauge theory. While in this particular model the shift of the deconfinement temperature due to cavitation does not exceed 5\%, we speculate that cavitation might be important near the QCD critical point.

\subsection{Effective conformal theories and local AdS dynamics}

We can take an alternative point of view on the AdS/CFT correspondence and use the above prescription in order to effectively {\it define} and study CFTs in higher dimensions. We just need to write down a gravitational action with any number of matter fields. The AdS partition function will then provide correlation functions for the dual operators that automatically satisfy conformal symmetry. These will however be subject to other physical constraints such as unitarity, causality, crossing, etc. in such a way that these translate into constraints on the spectrum and possible interaction terms. This {\it effective conformal theory} (see \cite{Fitzpatrick2011c} and references therein) would describe the low-lying spectrum of the dilatation operator in a CFT, equivalent to low energy spectrum in AdS. Such an effective theory is useful when the spectrum contains a hierarchy in the dimension of operators, in such a way that we can perform an expansion for large {\it gap}. These criteria ensure that there is a regime where the dilatation operator is modified perturbatively. Local interactions in AdS thus provide a very efficient way of organizing perturbations of the dilatation operator respecting conformal invariance, much as Minkowski space naturally describes Lorentz invariant perturbations of a QFT hamiltonian. This approach  might also be useful to understand which CFTs have a well-behaved AdS description. 

This perspective may also provide some interesting insights into higher dimensional CFTs. It is indeed unclear whether there are non-trivial higher dimensional CFTs. 
In general it is thought that interacting CFTs do not exist in dimensions higher than six. Due to dimensionality any interaction term is irrelevant so that there is no obvious way to define interacting CFTs. There seems to be an avenue for their formulation in terms of $p$-forms with $p \geq 2$ (so-called generalized gerbe theories) \cite{Witten2007c}. Also from the bootstrap perspective, it is not at all obvious that there is no solution. It would be good to understand this better and the study of gravity in AdS opens an interesting window that might shed some light on the issue. 

In the remaining chapters of this thesis, we will investigate the uses of the  framework of the gauge/gravity duality in the case of Lovelock theories and extract some of its consequences. This provides interesting information both from a fundamental as well as from a more {\it effective} perspective. Higher-curvature corrections to the Einstein-Hilbert action appear as next to leading orders in the low-energy expansion of string theory, \eg the Lanczos-Gauss-Bonnet term \cite{Zwiebach1985}, being in any case perturvative. In the context of holography, generic higher curvature terms may correspond both to finite $N$ or $\lambda$ corrections. See for instance \cite{Kats2009,Buchel2009} for a recent discussion regarding the LGB term. Even if, in the case of LGB gravity, its {\it stringy} origin allows for establishing a detailed holographic dictionary, this does not happen for higher order Lovelock theories. Still Lovelock gravity in seven dimensions might help to understand the r\^ole of cubic terms in the duality and provide information about more general six dimensional CFTs. In turn higher order Lovelock terms, may help understand the elusive higher dimensional CFTs, although there is no explicit string construction for them, neither for AdS spaces in dimension higher than seven. These theories constitute perfect toy models to test our ideas about the holographic duality with the addition of finite higher curvature corrections. These introduce some of the characteristic features of higher curvature gravities, such as several branches, new types of singularities and more complex dynamics, without problematic higher derivative degrees of freedom. 

From the CFT point of view, any stringy realization of the gauge/gravity duality will yield correlation functions for the stress-energy tensor analogous to those arising for Einstein-Hilbert gravity, with the possibility of some small correction due to higher order terms. The analysis of the Lovelock family will in turn allow for the exploration of much more general CFTs of arbitrary dimensionality and beyond the perturbative regime. All results are consistent and compelling enough to pursue our investigations.  

\section{Lovelock theories and holography}

Lovelock theory has a rich structure of AdS vacua. Correspondingly, the AdS/CFT correspondence tells us that there should be an analogously complex structure of dual CFTs. We know very little about these higher-dimensional CFTs. Indeed, at least for the case of supersymmetric CFTs, one naively expects to have at most six dimensional non-trivial unitary conformal field theories, based on the algebraic construction by Nahm \cite{Nahm1978}. For higher-dimensional CFTs, the general expectation is that they are not Lagrangian theories and their constituent degrees of freedom are not gauge fields but possibly self-dual p-forms (in the case of even spacetime dimensions) \cite{Witten2007c}.  These theories should have a stress tensor and conformal symmetry will constrain their $2$- and $3$-point functions as we shall presently describe.

\subsection{CFT unitarity and 2-point functions}

Consider a CFT$_{d-1}$. The leading singularity of the $2$-point function in any number of dimensions is fully characterized by the central charge $C_T$ \cite{Osborn1994}
\begin{equation}
\langle T_{ab}(\mathbf x)\, T_{cd}(\mathbf 0)\rangle = \frac{C_T}{\mathbf x^{2(d-1)}}\;\mathcal{I}_{ab,cd}(\mathbf x) ~,
\label{TTcorrelator}
\end{equation}
where
\begin{equation}
\mathcal{I}_{ab,cd}(\mathbf x) = \frac12 \left( I_{ac}(\mathbf x)\, I_{bd}(\mathbf x) + I_{ad}(\mathbf x)\, I_{bc}(\mathbf x) - \frac1{d-1}\, \eta_{ab}\, \eta_{cd} \right) ~,
\end{equation}
whereas $I_{ab}(\mathbf x) = \eta_{ab} - 2\,{x_a\, x_b/\mathbf x^2}$. This structure is completely dictated by conformal symmetry. For instance, $C_T$ is proportional in a CFT$_4$ to the standard central charge $c$ that multiplies the (Weyl)$^2$ term in the trace anomaly, $C_T = 40\, c/\pi^4$.

The holographic computation of $C_T$ was performed in \cite{Buchel2010a} for LGB in various dimensions and in \cite{Camanho2010d} for Lovelock theory. Since $C_T$ appears in the general $2$-point function of the stress tensor, it is sufficient to consider a particular set of components. Following \cite{Buchel2010a} we consider the correlator $\langle T_{xy}(\mathbf x)\,T_{xy}(\mathbf 0)\rangle$.  According to the AdS/CFT dictionary, it is sufficient to take a metric fluctuation $h_{xy}(z,\mathbf x) = L^2/z^2 \;\phi(z,\mathbf x)$ about empty AdS with cosmological constant $\Lambda_\star$. Expanding the Lovelock action to quadratic order in $\phi$, and evaluating it on-shell,
\begin{equation}
\mathcal{I}_{\rm quad} = \frac{\Upsilon'[\Lambda_\star]}{2 (- \Lambda_\star)^{d/2}} \int\!d{\bf x} ~z^{2-d} \left(\phi\,\partial_z \phi\right) ~.
\label{Iquad}
\end{equation}
Imposing the boundary conditions $\phi(0,\mathbf x) = \hat\phi(\mathbf x)$, the full bulk solution reads
\begin{equation}
\phi(z,\mathbf x) = \frac{d}{d-2} \frac{\Gamma[d]}{\pi^{\frac{d-1}2} \Gamma\left[\frac{d-1}2\right]} \int\!d{\bf y} ~\frac{z^{d-1}}{(z^2 + |\mathbf x - \mathbf y|^2)^{d-1}} \mathcal{I}_{ab,cd}(\mathbf x - \mathbf y) \;\hat\phi(\mathbf y) ~.
\label{fbulk}
\end{equation}
Plugging this expression into $\mathcal{I}_{\rm quad}$, we obtain
\begin{equation}
\mathcal{I}_{\rm quad} = \frac{C_T}{2} \int\!d{\bf x} \int\!d{\bf y} ~\frac{\hat\phi(\mathbf x)\;\mathcal{I}_{ab,cd}(\mathbf x - \mathbf y)\;\hat\phi(\mathbf y)}{|\mathbf x - \mathbf y|^{2(d-1)}} ~,
\label{Iquadfinal}
\end{equation}
where $C_T$ is the central charge of the dual CFT$_{d-1}$,
\begin{equation}
C_T = \frac{d}{d-2} \frac{\Gamma[d]}{\pi^{\frac{d-1}2} \Gamma\left[\frac{d-1}2\right]} \frac{\Upsilon'[\Lambda_\star]}{(- \Lambda_\star)^{d/2}} ~.
\end{equation}
It is easy to see that this had to be the result as the computation basically corresponds to that of Einstein-Hilbert gravity with the overall constant $\Upsilon'[\Lambda_\star]$. The upshot of this computation in an AdS vacuum, $\Lambda_\star < 0$, is thought-provoking:
\begin{equation}
C_T > 0 \qquad\Longleftrightarrow\qquad \Upsilon'[\Lambda_\star] > 0 ~.
\end{equation}
The latter inequality, in the gravity side, corresponded to the generalized BD condition preventing ghost gravitons in the branch corresponding to the AdS vacuum with cosmological constant $\Lambda_\star$. Thereby, unitarity of the CFT and the absence of ghosts gravitons in AdS, seem to be the two faces of the same holographic coin. The positivity of $\Upsilon'[\Lambda]$ has been proven to be true for the EH branch of solutions in any Lovelock theory \cite{Camanho2010a}. Thus, this branch is protected from BD instabilities and the corresponding dual theory is unitary. 

\subsection{3-point function and conformal collider physics}
\label{3pcorr}

The form of the $3$-point function of the stress-tensor in a $(d-1)$-dimensional conformal field theory is highly constrained. In \cite{Osborn1994,Erdmenger1997}, it was shown that it can always be written in the form
\be
\langle T_{ab}(\mathbf x)\,T_{cd}(\mathbf y)\,T_{ef}(\mathbf z)\rangle=
\frac {\left( \mathcal A\, \mathcal I^{(1)}_{ab,cd,ef}+\mathcal B\, \mathcal I^{(2)}_{ab,cd,ef}+\mathcal C\, \mathcal I^{(3)}_{ab,cd,ef}\right)}{(|\mathbf x - \mathbf y|\,|\mathbf y - \mathbf z|\,|\mathbf z - \mathbf x|)^{d-1}}
\ee
where the form of the tensor structures $\mathcal I^{(i)}_{ab,cd,ef}$ will be irrelevant for us here. Energy conservation also implies a relation between the central charge $C_T$ appearing in the $2$-point function, and the parameters $\mathcal A,\mathcal B,\mathcal C$, namely
\be
C_T=\frac{ \pi^{\frac{d-1}{2}}}{\Gamma\left[\frac{d-1}2\right]}\,\frac{(d-2)(d+1)\mathcal{A}-2\mathcal{B}-4 d \, \mathcal{C}}{(d-1)(d+1)}~.
\ee
Since we have already computed $C_T$ in the previous section, we are left with two independent parameters to be calculated.

A convenient parametrization of the $3$-point function of the stress-tensor was introduced in \cite{Hofman2008}. Consider a {\it gedanken} collision experiment in an arbitrary CFT$_d$ that mimics the framework developed in \cite{Basham1978,Basham1978a} for e$^+$--e$^-$ annihilation in QCD. We would like to measure the total energy flux per unit angle deposited in calorimeters distributed around the collision region,
\begin{equation}
{\cal E}(\mathbf n) = \lim_{r \to \infty} r^{d-3}\! \int_{-\infty}^\infty\! dt\; n^i\, T^0_{~i}(t, r\, \mathbf n) ~,
\label{etheta}
\end{equation}
the unit vector $\mathbf n$ pointing towards the actual direction of measure. The expectation value of the energy on a state created by a given local gauge invariant operator $\mathcal{O}$,
\begin{equation}
\langle {\cal E}(\hat{n}) \rangle_\mathcal{O} = \frac{\langle 0| \mathcal{O}^\dagger {\cal E}(\hat{n}) \mathcal{O} |0 \rangle}{\langle 0| \mathcal{O}^\dagger \mathcal{O} |0 \rangle} ~,
\label{vevenergy}
\end{equation}
is written in terms of 2- and 3-point functions in the CFT. There is a natural operator of this sort to be considered, that CFTs in any spacetime dimension possess, which is the stress-energy tensor, $\mathcal{O} = \epsilon_{ij}\, T_{ij}$. For such operators, $\langle {\cal E}(\hat{n}) \rangle_\mathcal{O}$ is given in terms of 2- and 3-point correlators of $T_{\mu\nu}$.

Using the fact that $\epsilon_{ij}$ is a symmetric and traceless polarization tensor with purely spatial indices, $O(d-2)$ rotational symmetry allows to write \cite{Boer2009, Camanho2010a}
\be
\langle \mathcal E(\mathbf n)\rangle_{\e T}=\frac{E}{\Omega_{d-3}}\left[1+t_2 \left(\frac{n_i n_j \epsilon^*_{ik}\epsilon_{jk}}{\epsilon^*_{ik}\epsilon_{ik}}-\frac 1{d-2}\right)
+t_4 \left(\frac{|n_i n_j\epsilon_{ij}|^2}{\epsilon^*_{ik}\epsilon_{ik}}-\frac 2{d(d-2)}\right)\right]~,
\ee
with $E$ the total energy of the insertion, and $\Omega_{d-3}$ the volume of a unit $(d-3)$-sphere. The energy flux is almost completely fixed by symmetry up to coefficients $t_2$ and $t_4$. 

The existence of minus signs in the above expression leads to interesting constraints on the parameters $t_2$ and $t_4$, by demanding that the measured energy flux should be positive for any direction $\mathbf n$ and polarization $\epsilon_{ij}$. In spite of the fact that the positivity of the energy flux is not self-evident, there are field theoretic arguments supporting this claim \cite{Hofman2008,Hofman2009} (see also \cite{Zhiboedov2013} for a more recent discussion). Furthermore, interestingly enough, it was recently argued that this condition is equivalent to unitarity of the corresponding CFT \cite{Kulaxizi2011}. This condition looks physically reasonable and holds in all known examples, though we are not aware of its general proof.

\begin{figure}
\centering
\includegraphics[width=0.67\textwidth]{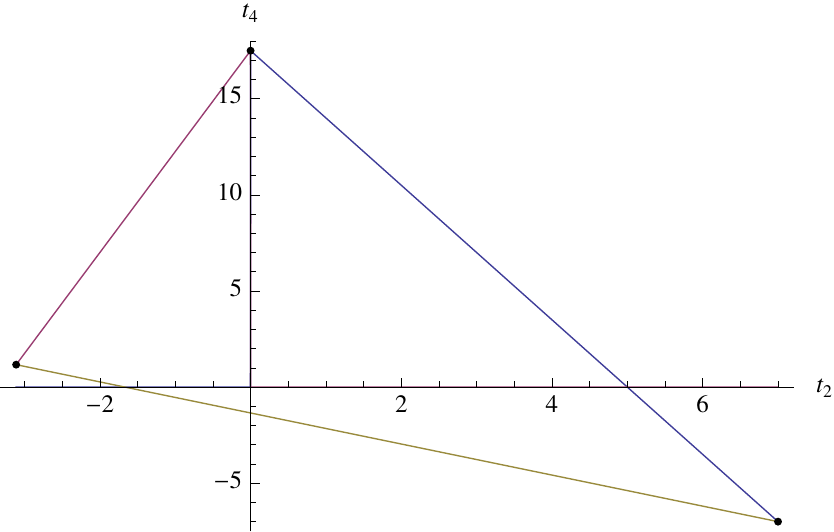}
\caption{Constraints (\ref{tensorboundd})--(\ref{scalarboundd}) restrict the values of $t_2$ and $t_4$ to the interior of the depicted triangle (in $d=7$).}
\label{triangle}
\end{figure}

In the example under study here we can thus analyze the conditions imposed by demanding positivity of the deposited energy irrespective of the calorimeter angular position. These depend on the different polarizations $\epsilon_{ij}$,
\begin{eqnarray}
{\rm tensor:}\qquad & & 1 - \frac{1}{d-2}\, t_2 - \frac{2}{d(d-2)}\, t_4 \geq 0 ~,\label{tensorboundd} \\ [0.7em]
{\rm vector:}\qquad & & \left( 1 - \frac{1}{d-2}\, t_2 - \frac{2}{d(d-2)}\, t_4 \right) + \frac{1}{2}\, t_2 \geq 0 ~, \label{vectorboundd} \\ [0.7em]
{\rm scalar:}\qquad & & \left( 1 - \frac{1}{d-2}\, t_2 - \frac{2}{d(d-2)}\, t_4 \right) + \frac{d-3}{d-2} \left( t_2 + t_4 \right) \geq 0 ~. \label{scalarboundd}
\end{eqnarray}
The three expressions come from the splitting of $\epsilon_{ij}$ into tensor, vector and scalar components with respect to rotations in the hyperplane perpendicular to $\mathbf n$. In a way very similar to the splitting of graviton polarizations for obvious reasons. These constraints restrict the possible values of $t_2$ and $t_4$ for any CFT to lie inside a triangle (see figure \ref{triangle}) whose sides are given by (\ref{tensorboundd}), (\ref{vectorboundd}) and (\ref{scalarboundd}). The vertices of the triangle are $(-\frac{2 (d-3) d}{d^2-5 d+4},\frac{d}{d-1})$, $(0,\frac{d(d-2)}{2})$ and $(d,-d)$. Each of the constraints in saturated in a free theory with no antisymmetric tensor fields, no fermions or no scalars respectively \cite{Hofman2008,Boer2009}. It is a straightforward consequence to see that the helicity one contribution is not restrictive for $t_4 < \frac{d}{d-1}$. In particular, of course, it is so for $t_4 = 0$.

The presence of a non-trivial $t_4$ seems to be linked to the absence of supersymmetry \cite{Hofman2008,Boer2009}. In particular, for any supersymmetric CFT $t_4$ vanishes and the above constraints translate into bounds that $t_2$ must obey: $t_2 \leq d-2$ (tensor), $t_2 \geq - 2(d-2)/(d-4)$ (vector) and $t_2 \geq - (d-2)/(d-4)$ (scalar). The restriction imposed by the vector polarization, as it happens in 4d, is less restrictive than the scalar one and, thus, irrelevant. Summarizing, positive energy constraints impose the following restriction on $t_2$ for any SCFT,
\begin{equation}
- \frac{d-2}{d-4} \leq t_2 \leq d-2 ~.
\label{t2general}
\end{equation}

We could also analyze 1-point functions for states created by other operators of lower spin, scalars or conserved currents, that have a much simpler form. For the scalar, as all directions are equivalent we will have just the isotropic term
\be
\langle \mathcal E(\mathbf n)\rangle_\mathcal{O}=\frac{E}{\Omega_{d-3}}~,
\label{1pscalar}
\ee
whereas in the spin one case, $\mathcal{O}=\e_i j_i$, there is just one extra contribution parametrized by $a_2$. In this case
$O(d-2)$ symmetry and the energy conservation condition constrain the form of the 1-point function to
\be
\langle \mathcal E(\mathbf n)\rangle_{\e j}=\frac{E}{\Omega_{d-3}}\left[1+a_2 \left(\frac{n_i n_j \epsilon^*_{i}\epsilon_{j}}{\epsilon^*_{i}\epsilon_{i}}-\frac 1{d-2}\right)\right]~.
\label{1pvector}
\ee
These energy functions are determined in terms of 2-point functions of scalars/currents and 3-point functions of two scalar/currents with the stress-energy tensor.


For conformal field theories with a weakly curved gravitational dual, it is possible to compute $t_2$ and $t_4$ holographically. Also $a_2$ if we include a vector field, otherwise this coefficient is trivially zero. This was first done in detail for LGB theory in various dimensions \cite{Buchel2010a} and later applied to quasi-topological gravity in \cite{Myers2010d}. 
The calculation proceeds by considering the vacuum AdS solution perturbed by a shock wave, which corresponds holographically to a $T_{--}$ insertion. By adding a transverse metric fluctuation, one reads off the interaction vertex from the action, and from that one obtains $t_2$ and $t_4$.

Shock wave backgrounds were also considered by Hofman \cite{Hofman2009} in the context of five-dimensional LGB theory and was to arbitrary higher-dimensions in \cite{Camanho2010} There it was found that in the presence of the shock wave there is the possibility for causality violation in the dual field theory. This places bounds on the parameters of the theory, which precisely match those portrayed in (\ref{tensorboundd} - \ref{scalarboundd}). This will be the focus of the next chapter. 

We would now like to generalize this story to higher Lovelock theories. We consider, along the lines of \cite{Camanho2010}, a helicity two perturbation $\phi(u,v,r)$ in the shock wave background 
\begin{equation}
ds^2_{{\rm AdS},sw}=\frac{N_{\#}^2\,L^2}{z^2}\left(-du\,dv+dx^i\,dx^i+2\,\epsilon\,\phi\,dx^2\,dx^3+dz^2\right)+f(u)\,\varpi(x^{a},z)\,du^2~,
\label{dsSW}
\end{equation}
where $z=L^2/r$ is the Poincar\'e coordinate and $u,v=x^0\pm x^{d-1}$ are light-cone coordinates. This amounts to choosing just one non-vanishing component of the polarization tensor, $\epsilon_{23}\neq0$. Leading contributions (in the high momentum limit) to the equations of motion come from the exterior derivative of the perturbation of the spin connection.\footnote{The $3$-point function will be determined by these kinds of terms; actually, terms in $\partial_v^2\phi$ since the vertex has to be of the form $\phi\,\partial_v^2\phi\,\partial_i\partial_j\varpi$, as $\varpi\sim h_{uu}$.} We get,
\begin{eqnarray}
& & d(\delta \omega^{02})  \approx \frac{\epsilon\; z^2}{L^2 N^2_{\#}}\, \left[ \partial_v^2 \phi\; e^1 \wedge e^2 + \left( \partial_u \partial_v \phi + \frac{z^2}{L^2 N^2_{\#}}\, f(u) \,\varpi(x^a, z) \, \partial_v^2 \phi \right)\, e^0 \wedge e^2 \right]  ~, \nonumber \\ [0,7em]
& & d(\delta \omega^{12}) \approx \frac{\epsilon\; z^2}{L^2 N^2_{\#}}\, \left[ \left( \partial_u \partial_v \phi + \frac{z^2}{L^2 N^2_{\#}}\, f(u) \,\varpi(x^a, z) \, \partial_v^2 \phi \right)\, e^1\wedge e^2 + \left( \cdots \right)\, e^0\wedge e^2 \right] ~,  \nonumber
\end{eqnarray}
the ellipsis being used in the second expression since the corresponding term does not contribute to the equations of motion. The components with index 3 instead of 2 are the only remaining non-vanishing ones, and they are obtained just by changing\footnote{$dx^2\rightarrow -dx^3$, $dx^3\rightarrow dx^2$, $\phi\rightarrow-\phi$ is a symmetry of the background as well as of the vielbein basis.} $\phi\rightarrow- \phi$. The other ingredient we need is the curvature 2-form of the background metric, that can be written as
\begin{equation}
R^{ab} = \Lambda (e^a\wedge e^b + f(u) X^{ab}) ~,
\end{equation}
where $\Lambda=-\frac{1}{L^2 N_{\#}^2}$ and $X^{ab}$ is an antisymmetric 2-form accounting for the contribution of the shock wave. The relevant component here is 
\begin{eqnarray}
& & X^{1a} = \frac{z^2}{L^2 N^2_{\#}}\, \left[\left(2\varpi+z\partial_z \varpi-z^2 \partial_a^2\varpi\right)\; e^0\wedge e^a+\left(\cdots\right)\; e^0\wedge e^z+\left(\cdots\right)\; e^0\wedge e^b\right]  ~,  \nonumber \\ [0.7em]
& & X^{1(d-1)} = -\left(\frac{z}{L N_{\#}}\right)^2\, \left[(3z\partial_z
\varpi+z^2 \partial_z^2\varpi)\; e^0 \wedge
e^{d-1}+\left(\cdots\right)\; e^{0}\wedge e^{a}\right] ~, \nonumber
\end{eqnarray}
with $a,b\neq 0,1,d-1$, and $b\neq a$. The relevant equation of motion is given by $\delta \mathcal{E}_3\wedge e^3 =0$, where
\begin{eqnarray}
& &\!\!\!\! \delta \mathcal{E}_3\wedge e^3 = \sum_{k=1}^{K}{}k\,a_k\; \epsilon_{3 f_1 \cdots f_{d-1}} d(\delta \omega^{f_1 f_2})\wedge R^{f_3 \cdots f_{2k}}\wedge e^{f_{2k+1}\cdots\, f_{d-1} \,3} \nonumber \\ [0.9em]
& &  = 4(-1)^{d-1} \frac{\epsilon\; z^2}{L^2 N^2_{\#}}\, \sum_{k=1}^{K}{}k\,a_k\,\Lambda^{k-1}\left[(d-3)!\left( \partial_u \partial_v \phi + \frac{z^2}{L^2 N^2_{\#}}\, f(u) \,\varpi \, \partial_v^2 \phi \right) \right.\nonumber\\ [0.9em]
& &  - \left. (k-1) (d-5)!\; \frac{z^2}{L^2 N^2_{\#}}\, f(u)\left(-4\varpi-2z\partial_z\varpi+z^2(\partial_2^2\varpi+\partial_3^2\varpi)\right)\partial_v^2\phi\right] ~.
\label{eq3}
\end{eqnarray} 
We made use of the equation of motion for the shock wave profile
\begin{equation}
2(d-3)\varpi+(d-6)z\partial_z \varpi-z^2(\partial_a\partial^a\varpi+\partial_z^2 \varpi)=0~,\qquad a=2,\ldots d-2\; .
\label{SWeq}
\end{equation}
This equation admits several solutions of the type
\begin{equation}
\varpi =\alpha_0 \frac{z^\alpha}{\left(z^2+(x^2-x^2_0)^2+\cdots+(x^n-x^n_0)\right)^\beta} ~,
\end{equation}
where $n$ is the number of transverse coordinates. The relevant solution for our discussion here is the one given by $\alpha=d-3$ and $\beta=d-2$. This shock wave profile has been argued \cite{Hofman2008} to be the dual field configuration to $\mathcal{E}(\mathbf n)$ provided $x_0^i=\frac{n^i}{1+n^{d-2}}$ and $f(u)=\delta(u)$. 

From (\ref{eq3}), we shall focus on those terms proportional to $\partial_{v}^2\phi$. The shock wave interaction term thus gives the following contribution to the equations of motion,
\begin{eqnarray}
\left[\Upsilon'[\Lambda]\, \varpi+\frac{\Lambda\,\Upsilon''[\Lambda]\,}{(d-3)(d-4)}\left(4\varpi+2z\partial_z \varpi-z^2(\partial_2^2\varpi+\partial_3^2\varpi)\right)\right]\partial_v^2\phi~.
\end{eqnarray}
There are extra non vanishing components of the form
\begin{eqnarray*}
& & \delta \mathcal{E}_{i}\wedge e^2\sim \partial_3\partial_i\varpi \partial^2_{v}\phi ~, \\ [1em]
& & \delta \mathcal{E}_{d-1}\wedge e^2\sim \frac1{x^{d-1}}\left(\partial_3\varpi +x^{d-1} \partial_3\partial_{d-1}\varpi\right)\partial^2_{v}\phi ~, \\ [1em]
& & \delta \mathcal{E}_{i\neq2,3}\wedge e^{j\neq2,3}\sim \partial_2\partial_3\varpi \partial^2_{v}\phi \delta_i^j ~.
\end{eqnarray*}
However, these are irrelevant for computing the 3-point function we are interested in. The tensor channel mixes with other modes but they would only affect other correlators irrelevant for our discussion.

The $3$-point function follows from evaluating the effective action for the field $\phi$ on-shell, on a particular solution which depends on all coordinates, including $x_2,x_3$ \cite{Buchel2010a}. The cubic interaction vertex of $\phi$ with the shock wave appearing in the action will be essentially the one in the equation of motion determined above. Up to an overall factor, the cubic vertex is then
\be
\mathcal{I}^{(3)}\sim C_T \int d^d x \sqrt{-g} \,\phi\, \partial^2_v \phi\, \varpi\,\left(1- \frac{\Lambda\,\Upsilon''(\Lambda)}{\Upsilon'(\Lambda)}\frac{T_2}{(d-3)(d-4)}\right)~,
\ee
where 
\be
T_2=\frac{z^2(\partial_2^2 \varpi+\partial_3^2 \varpi)-2 z \partial_z \varpi-4 \varpi}{\varpi}~.
\ee
This is nothing but the same $T_2$ appearing in \cite{Buchel2010a}. Indeed, following that paper the relevant graviton profile is
\begin{equation}
\phi(u=0,v,x^a,z)\sim e^{-iEv}\,\delta^{d-3}(x^a)\,\delta(z-1)~,
\end{equation}
so that we need to impose $x^a=0$ and $z=1$ yielding
\be
T_2=2(d-1)(d-2)\left(\frac{n_2^2+n_3^2}{2}-\frac1{d-2}\right)~,
\ee
and we therefore read off
\be
t_2=-\frac{2(d-1)(d-2)}{(d-3)(d-4)} \frac{\Lambda \Upsilon''[\Lambda]}{\Upsilon'[\Lambda]}~, \quad\qquad t_4=0~. \label{t2t4}
\ee
This expression reproduces previous results in LGB gravity \cite{Camanho2010,Buchel2010a}, and is exactly the same as conjectured by de Boer, Kulaxizi and Parnachev in \cite{Boer2010}. Using these results together with the expression for $C_T$, we find expressions for the usual $3$-point function parameters $\mathcal A,\mathcal B, \mathcal C$ in Appendix \ref{3point}.

In the next chapter we will analyze the constraints positivity of energy correlators poses in the gravitational theory and its interpretation in that context.

\section{Shock waves in higher curvature gravity}
\label{SW}

Shock waves as those considered in the previous section are very interesting backgrounds for several reasons. They have been considered as suitable testbed for numerical relativity studies with applications to heavy ion collisions and many other holographic settings. In the next chapter they will be used to probe the causality properties of Lovelock theories. These solutions have a very particular structure that makes them exact solutions of any gravity theory. In fact, Horowitz and Itzhaki \cite{Horowitz1999a} showed that these solutions are not corrected once higher curvature corrections are included. They have also been shown to correspond to exact solutions of string theory \cite{Horowitz1990}. We will rephrase their argument here for completeness, following closely the approach of \cite{Frolov2005}. 

We may write the ansatz for the shock wave in asymptotically AdS space \reef{dsSW} as 
\begin{equation}
g_{\mu\nu}=\hat{g}_{\mu\nu} + F(u, x^i)l_\mu l_\nu  ~,
\end{equation}
with $\hat{g}_{\m\n}$ the AdS metric and $l^\mu=\delta^\mu_v$ a null Killing vector. Hatted quantities correspond to those arising from the pure AdS background, $\hat{g}_{\mu\nu}$, whereas extra contributions are denoted  as checked. 

We will be considering AdS in null Poincar\'e coordinates, $x^{\mu}=(u,v,x^i,z)$. We can write its connection in terms of a new vector $V^\mu=\partial_\m \log z$,
\begin{equation}
\hat{\Gamma}^{\mu}_{\ \alpha \beta}=-\left[V_\alpha\delta^{\mu}_\beta+V_\beta\delta^{\mu}_\alpha-V^{\mu}\hat{g}_{\alpha\beta}\right] ~.
\end{equation}
while the total connection,
\begin{equation}
\Gamma^\mu_{\ \alpha\beta}=\hat{\Gamma}^\mu_{\ \alpha\beta}+\gamma^\m_{\ \alpha\beta}   ~,
\end{equation}
has an extra term completely orthogonal to the Killing vector, $\gamma^\m_{\ \a\b}l^\a=\gamma^\m_{\ \a\b}l_\m=0$, in such a way that it does not contribute to the covariant derivative of $l^\m$,
\begin{equation}
\nabla_\alpha l_\beta= \hat{\nabla}_\alpha l_\beta=-2V_{[\alpha} l_{\beta]}
\label{killing}
\end{equation}
and the Killing condition is the same for the full metric as for the pure AdS one. 

The Riemann tensor can be written as
\begin{equation}
R_{\mu\nu\alpha\beta}=-\left(g_{\mu\alpha}g_{\nu\beta}-g_{\mu\beta}g_{\nu\alpha}\right)+\check{R}_{\mu\nu\alpha\beta}
\end{equation}
where the extra term has a very simple expression in terms of $l^\mu$
\begin{equation}
\check{R}_{\mu\nu\alpha\beta}=l_{[\mu}K_{\nu][\alpha}l_{\beta]}
\end{equation}
where $K_{\mu\nu}$ is a symmetric tensor made from $F$, its derivatives and $V^{\mu}$ verifying
\begin{equation}
K_{\mu\nu}l^{\nu}=0
\end{equation}
The last property just follows from $(\nabla_{\alpha}\nabla_{\beta}-\nabla_{\beta}\nabla_{\beta})l_\nu=R_{\mu\nu\alpha\beta}\, l^{\mu}$ that because of the previous properties has to be equal to $(\hat{\nabla}_{\alpha}\hat{\nabla}_{\beta}-\hat{\nabla}_{\beta}\hat{\nabla}_{\beta})l_\nu=\hat{R}_{\mu\nu\alpha\beta}\, l^{\mu}$ in such a way that $\check{R}_{\mu\nu\alpha\beta}\, l^{\mu}=0$. 

This form of expressing the Riemann tensor is particularly useful to discuss the equations of motion for this type of backgrounds. The AdS background is a solution of Einstein's equations with cosmological constant
\begin{equation}
R_{\mu\nu}=-(d-1)g_{\mu\nu}
\end{equation}
in such a way that in order for the shock wave metric to be also a solution it must verify
\begin{equation}
K^\mu_{\ \ \mu}=0
\label{SWeq2}
\end{equation}
which is a linear equation for $F$, precisely \reef{SWeq} with the notation of the previous section. Once this equation is imposed $\check{R}_{\m\n\a\b}$ corresponds to the Weyl tensor.

We can now proceed to discuss higher curvature corrections to that equation. For that we have to verify that, once the equation of motion is taken ito account any (2,0)-tensor formed solely from the metric and its covariant derivatives is necessarily zero. It may also be proportional to the metric but this will just change the value of the AdS radius, that we always scale back to one.

First we will consider scalars made just from the metric and the Riemann tensor. Their contribution can be organized in powers of the extra piece $\check{R}$ of the Riemann  that contains two $l^\mu$ vectors. These have to be contracted with either another $l^\mu$ or $K_{\mu\nu}$ to form a scalar. In any case these terms necessarily vanish. 

In order to include covariant derivatives we will make use of the fact that $l^\mu$ is a null Killing vector orthogonal to all the tensors that can be constructed from the metric -- except the metric itself. We will denote by $S_n$ a given scalar with $n$ covariant derivatives. The degree $n$ of a given scalar can be reduced following two steps. 

If the given term contains $\nabla_{\alpha}l_\beta$ we can eliminate the covariant derivative by using (\ref{killing}) until we eliminate all covariant derivatives acting on it. The resulting scalar necessarily has lower degree than the original one.

For a general scalar, $l^\mu$ has to be contracted either with one of the available tensors $l^\mu$, $K_{\mu\nu}$ or any of their covariant derivatives. We can again make use of the Killing equation in order to commute the vector $l^\mu$ with the covariant derivatives without increasing the degree of the scalar, using
\begin{equation}
l_\alpha \nabla_\beta(\cdots)=\nabla_\beta[l_\alpha(\cdots)]+2V_{[\beta}l_{\alpha]}(\cdots)
\label{commute}
\end{equation}
then we can reduce all possible scalars to two possible kinds of terms depending on where the $l^\mu$ is contracted. We may have either
\begin{equation}
\nabla\cdots\nabla[l^\alpha\nabla_\alpha(\cdots)] \qquad \text{or} \qquad \nabla\cdots\nabla[l^\alpha T_{\ldots \,\alpha \ldots}(\cdots)]
\end{equation}
contracted with some other tensor with the necessary number of indices to construct the scalar. The first term vanishes because the quantity inside the square brackets is just the Lie derivative with respect to a constant Killing vector that of course vanishes. The second term is also zero as any of the available tensors is orthogonal to $l^\mu$.

In this way we can reduce the degree of the initial scalar until we have just scalars without any covariant derivative which, as discussed before, necessarily vanish.
Any scalar made from the metric, the Riemann tensor and its covariant derivatives, containing at least one power of $\check{R}$, is necessarily zero. This is actually an off-shell property as we did not need to impose (\ref{SWeq2}) on any step. 

We can also verify that the only (0,2)-tensor containing at least one power of $\check{R}$  that is non-vanishing off-shell is the Ricci tensor. In this kind of terms we have at least two $l^\mu$ factors that can either be contracted as discussed in the previous section or free. If it is not contracted with anything we can again use the property (\ref{commute}) to put it in front as a global factor. If it is contracted we can use the same steps discussed above to reduce the degree of the tensor until it either has no covariant derivatives, in which case if the $l^\mu$ is contracted necessarily vanishes, or the Killing vector is no longer contracted in which case we can again put it as a global factor. We can perform the same type operations for the other $l^\mu$ in such a way that the only non-vanishing terms will be of the form $l_\alpha l_\beta$ times some scalar made from everything else. This scalar however cannot contain any further $l^\mu$ and thus just terms linear in $\check{R}$ contribute.

There are two such terms we can form, one is the Ricci tensor,
\begin{equation}
\check{R}_{\mu\nu}=-\frac14 K^\alpha_{\ \ \alpha}l_{\mu}l_\nu~,
\end{equation}
and, once this is set to zero, we are left just with
\begin{equation}
\nabla^\alpha\nabla^\beta\check{R}_{\alpha\mu\beta\nu}=\nabla^2 \check{R}_{\mu\nu}-\nabla^\alpha\nabla_{\mu}\check{R}_{\alpha\nu}
\end{equation}
and its covariant derivatives. All of them necessarily vanish because of the Bianchi identity and the fact that $\check{R}_{\mu\nu}=0$ on-shell. As a consequence, the Einstein equation for the shock wave \reef{SWeq2} is not corrected by higher order terms. Any extra gravitational term in the action will just give a contribution either proportional to this, to the metric or just vanish. The same reasoning applies for shock waves about any of the vacua of the full higher curvature gravity, we just have a different effective Newton's and cosmological constants. 

There are a couple of useful properties of these background metrics that will be used later on. By using the Bianchi identity we can show
\begin{equation}
\square \check{R}_{\mu\nu\alpha\beta}\sim \check{R}_{\mu\nu\alpha\beta}+ \mathcal{O}(\check{R}^2)
\label{Rprop1}
\end{equation}
and, once the $\check{R}_{\mu\nu}$ is also set to vanish, also
\begin{equation}
\nabla^\mu \check{R}_{\mu\nu\alpha\beta}=0~.
\label{Rprop2}
\end{equation}
We will also make extensive use of the fact that commuting two covariant derivatives just produces terms with an extra Riemann tensor. The terms of the Riemann containing just the metric will contract some of the remaining indices whereas the rest will be higher order in $\check{R}$. This will be important in the following as the most interesting terms will just be linear in this tensor.

\subsection{Probing shock waves in general theories of gravity}

We will now analyze which terms contribute to the equation of motion for perturbations in the large momentum regime. This will be relevant for discussing which such terms contribute to the class of computation discussed in section \ref{3pcorr} for the 3-point function. For that we will have to restrict to two derivative equations of motion for perturbations about the shock wave background (and AdS), at least up to dimension six terms. There is also the possibility of considering more than two derivatives acting on the perturbation as $\nabla^n \delta R$ for the spin two case. These contributions in general change the number of degrees of freedom of the
theory and will not be considered here.

We will see that, to that order, just the 3-point function matters, in the sense that the only terms that enter the equations of motion for perturbations about the shock wave are those contributing to\footnote{\S\ See \cite{Camanho2013c} for a more explicit derivation of the connection to 3-point functions.} $t_{2}$ and $t_4$. These are the only free parameters in the energy flux 1-point function in a state created by the stress-energy tensor and also parametrize the 3-point functions of this operator. This is also consistent with a field redefinition analysis of the independent terms in the action. 

Before considering gravitons, it is instructive to analyze what happens for bosonic fields of lower spin. If we consider a minimally coupled scalar in this background, its equation of motion will just be
\begin{equation}
\square\phi=0~,
\label{Soem}
\end{equation}
where we consider $\phi$ as a small perturbation in such a way that we do not need to add backreaction. In the large momentum limit considering just terms yielding two derivative equations of motion for $\phi$ in the action we have to consider contributions of the form
\begin{equation}
\mathcal{H}^{\mu\nu}\nabla_\mu\nabla_\nu\phi~,
\end{equation}
where the tensor $\mathcal{H}$ is just made from the background metric, Riemann and covariant derivatives. However we already saw that there is no (2,0)-tensor  not vanishing on-shell except for the metric. 

Something similar happens for a vector field $A_\mu$. Maxwell's equations ,
\begin{equation}
\nabla^\mu F_{\mu\nu}=0~,
\end{equation}
may be corrected in the two derivative high momentum regime by terms of the form
\begin{equation}
\mathcal{H}^{\mu\nu\alpha\beta}\nabla_\nu F_{\alpha\beta}~,
\end{equation}
where again $\mathcal{H}$ is a background tensor. In principle, applying the same procedure discussed in the previous section, this tensor can be at most quadratic in $\check{R}$ as it has four indices and each of them can accommodate one factor $l^\mu$. Therefore, in the quadratic case $\mathcal{H}_{\mu\nu\alpha\beta}\sim l_\mu l_\nu l_\alpha l_\beta$ but $F_{\alpha\beta}$ is antisymmetric in its two indices in such a way that $l^{\alpha}l^\beta\nabla F_{\alpha\beta}=0$ and the term above vanishes. Consequently, we are left with terms linear in $\check{R}$. Taking into account \reef{Rprop1}, \reef{Rprop2} and the fact that, at the linear level in $\check{R}$, covariant derivatives commute, we just have to consider tensors
\begin{equation}
\nabla_{\alpha_1}\cdots \nabla_{\alpha_n}\check{R}_{\beta_1 \cdots \beta_4}
\end{equation}
with no indices contracted. Otherwise the term is either zero or can be reduced to another term of the above type. The number of free indices is four so that we cannot include any covariant derivative. The only extra contribution to the equation of motion of the vector field is then
\begin{equation}
\check{R}^{\mu\nu\alpha\beta}\nabla_\nu F_{\alpha\beta}
\end{equation}
that is the high momentum limit of $\nabla_\nu(\check{R}^{\mu\nu\alpha\beta}F_{\alpha\beta})$ that arises for instance from a $\check{R}^{\mu\nu\alpha\beta}F_{\mu\nu}F_{\alpha\beta}$ term in the action.

In both cases the possible contributions to the perturbation equations are in one to one correspondence with the possible structures in the corresponding energy flux 1-point functions as in \reef{1pscalar} and \reef{1pvector}. In the vector case the coefficient of the extra term is proportional to the $a_2$ constant appearing in the energy 1-point function of a state created by a conserved current. If we parametrize the equation of motion for the perturbation as 
\begin{equation}
\nabla^{\mu}F_{\mu\nu} - \alpha_2\, \check{R}_{\nu}^{\ \ \mu\alpha\beta}\nabla_\mu F_{\alpha\beta} = 0
\end{equation}
we can actually compute the value of the $a_2$ coefficient in a similar manner as for $t_{2,4}$, this yielding
\begin{equation}
a_2 = (d-1)(d-2) \alpha_2
\end{equation}

The expectation is that the same should happen also for graviton 3-point functions. One should be able to find three possible structures corresponding to the Einstein-Hilbert (isotropic) contribution, the one appearing in Lovelock theories and another one, as parametrized by $t_2$ and $t_4$ respectively.

Again, in the high momentum limit at two derivative level, the possible extra contributions are of the form,
\begin{equation}
\mathcal{H}^{\mu\nu\alpha\beta\rho\sigma}\;\delta R_{\alpha\beta\rho\sigma}
\end{equation}
We have six slots in $\mathcal{H}$ in order to have up to six $l^\mu$ vectors so in principle we have to consider up to cubic terms in $\check{R}$. The cubic terms however do not give any contribution similarly to what happens with quadratic terms for vector perturbations. The contraction $l^\mu l^\nu \delta R_{\mu\nu\alpha\beta}$ vanishes because of the antisymmetry of the first pair of indices. 

The Einstein-Hilbert contribution is 
\begin{equation}
\delta R_{\mu\nu}\sim \square h_{\mu\nu}~,
\end{equation}
when we choose the transverse traceless gauge, $\nabla^\mu h_{\mu\nu}=h^\alpha_{\ \ \alpha}=0$. This is the only possible $\mathcal{O}(\check{R}^0)$ term. Then the full equation will correspond to that plus higher order terms in such a way that we can trade $\delta R_{\mu\nu}$ by those higher order terms.

The linear terms are also quite simple. In this case we have just
\begin{equation}
\check{R}_{(\mu}^{\ \ \rho\alpha\beta}\delta R_{\nu)\rho\alpha\beta} \qquad \text{or} \qquad (\nabla\nabla \check{R})^{\mu\nu\alpha\beta\rho\sigma}\;\delta R_{\alpha\beta\rho\sigma} ~. 
\end{equation}
We could also include terms in $\delta R_{\mu\nu}$ but these would reduce to quadratic terms when the equation of motion is imposed. Also the covariant derivative indices in the second term are symmetric as the antisymmetric part reduces to the first term plus a quadratic contribution. There is only one independent way of contracting the indices in the second kind of term
\begin{equation}
\left[\nabla_{(\mu}\nabla_{\nu)}\check{R}^{\alpha\beta\rho\sigma}\right]\delta R_{\alpha\beta\rho\sigma}
\end{equation}
all other possible terms are related to this one via the Bianchi identity and the symmetries of $\delta R$ plus maybe some quadratic terms.

These two terms exactly correspond to the two extra (in addition to the isotropic one from the Einstein-Hilbert term) possible structures appearing in the 3-point function of gravitons. The coefficients of both terms contribute to $t_2$ while the second is the only one yielding $t_4$. This can be easily checked computing the energy 1-point function for a $h_{12}$ perturbation as in \ref{3pcorr} to obtain
\begin{eqnarray}
\check{R}_{(\mu}^{\ \ \rho\alpha\beta}\delta R_{\nu)\rho\alpha\beta} & \sim & \left(\frac{n_1^2+n_2^2}{2}-\frac1{d-2}\right)\equiv T_2 \\
\left[\nabla_{(\mu}\nabla_{\nu)}\check{R}^{\alpha\beta\rho\sigma}\right]\delta R_{\alpha\beta\rho\sigma} & \sim & \left(2n_1^2 n_2^2-\frac2{d(d-2)}\right) + \#  T_2
\end{eqnarray}
that exactly correspond to the $t_{2,4}$ structures in the energy flux 1-point function. If we parametrize the equation of motion for perturbations as 
\begin{equation}
\delta R_{\mu\nu} + 2 \gamma_2 \; \check{R}_{(\mu}^{\ \ \rho\alpha\beta}\delta R_{\nu)\rho\alpha\beta} + 2 \gamma_4 \; [\nabla_{(\mu}\nabla_{\nu)}\check{R}^{\alpha\beta\rho\sigma}]\delta R_{\alpha\beta\rho\sigma}=0
\end{equation}
in five dimensions the relation to the energy 1-point function parameters is
\begin{equation}
t_2=24(36\gamma_4 -\gamma_2) \qquad  \qquad t_4=-1440\gamma_4
\end{equation}
%


The terms leading to contributions on the energy 1-point function are also in one to one correspondence to the independent contributions to the effective action under field redefinitions. In flat space we know that (at least up to cubic order {\it i.e} dimension six terms) the coefficients of terms containing the Ricci tensor are ambiguous in the sense that they can be fixed freely by a general field redefinition. In AdS the analysis can be performed in exactly the same way if we write the action in terms of 
\begin{equation}
\check{R}_{\mu\nu\alpha\beta}=R_{\mu\nu\alpha\beta}+2g_{\mu[\alpha}g_{\beta]\nu}
\end{equation}
and organize all terms depending on the order in $\check{R}$. Then on Einstein-shell
\begin{equation}
\check{R}_{\mu\nu}=0
\end{equation}
(or more generally $\check{R}_{\mu\nu}\sim \mathcal{O}(\check{R}^2)$) and $\check{R}_{\m\n\a\b}$ becomes the Weyl tensor. We can then use a general field redefinition
\begin{equation}
\tilde{g}_{\mu\nu}\rightarrow g_{\mu\nu}+\alpha_1\; \check{R}_{\mu\nu}+\alpha_2\; \check{R}\; g_{\mu\nu} + \mathcal{O}(\check{R}^2)
\end{equation}
and the analysis carries over in exactly the same way as for asymptotically flat case. 

Then, up to dimension six we have just four possible independent contributions \cite{Metsaev1987d}, the Einstein-Hilbert term, a quadratic contribution that can be chosen as the LGB term (which vanish in $d=4$), and two cubic terms. One of the cubic lagrangians terms can also be taken to be of Lovelock type while for the other we may choose 
\begin{equation}
\mathcal{I}_3=R^{\mu\nu}_{\ \ \alpha\beta}R^{\alpha\beta}_{\ \ \rho\sigma}R^{\rho\sigma}_{\ \ \mu\nu}+\ldots
\end{equation}
where the dots stand for terms containing the Ricci in such a way that the equation of motion for perturbations about the shock wave background is second order, although this might not be possible. Both Lovelock combinations contribute in the same way, just to $t_2$. Whereas the $t_4$ structure comes solely from $\mathcal{I}_3$. 

The independent order six contributions to the action are already accounted for by the terms contributing to $t_{2}$ and $t_4$. This seems to indicate that in the interaction of the shock wave with other fields is encoded in just 2- and 3-point correlators of the dual CFT. Nonetheless, higher order terms (of dimension higher than six) may still contribute to the equations of motion as
higher derivative terms involving new degrees of freedom. These contributions have not been considered in the present analysis.
\chapter{\bfseries\itshape Causality and positivity of energy}
\chaptermark{Causality and positivity of energy}
\label{chp:LLcausality}

\vspace{.6cm}

\begin{quotation}
\flushright
{\it ``The whole problem with the world\\ is that fools and fanatics are always so certain of themselves,\\ and wiser people so full of doubts.''}\\

\vspace{.3cm}

Bertrand Russell
\end{quotation}

\vspace{3cm}

\noindent 
Despite all of the accumulating evidence in favor of the AdS/CFT correspondence, a novel direction was recently explored by Hofman and Maldacena \cite{Hofman2008}. As discussed in the last chapter, they studied a {\it gedanken} collider physics setup in the context of conformal field theories. They focused on the case of 4d CFTs and found a number of constraints for their central charges\footnote{The conformal anomaly of a four-dimensional CFT can be obtained by computing the trace of the stress-energy tensor in a curved spacetime \cite{Birrell}
\begin{equation}
\langle T^\mu{}_\mu \rangle_{\rm CFT} = \frac{c}{16\pi^2} I_4 - \frac{a}{16\pi^2} E_4 ~,
\label{vevTmunu}
\end{equation}
where $c$ and $a$ are the central charges, and $E_4$ and $I_4$ correspond to the four-dimensional Euler density and the square of the Weyl curvature.} by demanding that the energy measured in calorimeters of a collider physics experiment be positive. They found, for instance, that any 4d $\mathcal{N}=1$ supersymmetric CFT must have central charges within the window, $1/2 \leq a/c \leq 3/2$, the bounds being saturated by free theories with only chiral supermultiplets (lower bound) or only vector supermultiplets (upper bound) \cite{Hofman2008}. Since the computation of $\langle T^\mu_{~\mu} \rangle$ in a state generated by the stress-energy tensor is given by 3-point correlators of $T$, and pure Einstein-Hilbert gravity is well-known to yield $a = c$ \cite{Henningson1998,Henningson2000}, the gravity dual of a theory with $a \neq c$ should contain higher (at least quadratic) curvature corrections.

In a seemingly different context, Brigante {\it et al.} \cite{Brigante2008,Brigante2008a} explored the addition of a LGB term in the gravity side of the AdS/CFT correspondence and showed that, in the background of a black hole, the coefficient of this term, $\lambda$, is bounded from above, $\lambda \leq 9/100$, in order to preserve causality at the boundary.\footnote{Indeed, this is more general since any other curvature squared term can be reduced to LGB by field redefinitions disregarding higher powers of the curvature. See for instance \cite{Brigante2008}.} If this bound is disregarded, boundary perturbations would propagate at superluminal velocities. A natural question is immediately raised as to whether these quadratic curvature corrections arise in the string theory framework. The answer was given in the affirmative by Kats and Petrov \cite{Kats2009}, and further explored more recently by Buchel {\it et al.} \cite{Buchel2009}. Both papers focus on string theory compactifications that are relevant in the context of 4d SCFTs.

The somehow striking result came when Hofman and Maldacena realized that the upper bound on $\lambda$ was nothing but, through holographic renormalization, the lower bound on $a/c$. The matching is exact. This seems to provide a deep connection between two central concepts such as causality and positivity of the energy in both sides of the AdS/CFT correspondence. Besides, these results provided an irrefutable evidence against the so-called KSS bound \cite{Kovtun2005} for $\eta/s$ in quantum relativistic theories, $\eta/s \geq \frac{1}{4\pi}$. This is due to the fact that the value for $\eta/s$ is corrected in presence of a LGB correction to $\eta/s = \frac{1}{4\pi} (1 - 4 \lambda)$. Since the upper bound for $\lambda$ is positive, the shear viscosity to entropy density ratio, for such a SCFT, would be lower than the KSS value. Higher order Lovelock terms do not affect the value of $\eta/s$ \cite{Shu2009a}. However, for positive cubic couplings, $\lambda$ can surpass the upper bound obtained in the pure LGB case. This will push down the would be lower bound for $\eta/s$. We will come back to this in the next chapter, where we will analyze additional constraints in the context of Lovelock gravity and its implications for transport coefficients such as the shear viscosity to entropy density.  

In another paper, Buchel and Myers \cite{Buchel2009a} dug further into the constraints imposed by causality in the holographic description of hydrodynamics. Their paper deals with a black hole background, in which they explicitly relate the value of $\lambda$ to the difference between central charges of the dual CFT. They also showed that a lower bound for the LGB coupling, $\lambda \geq - 7/36$ comes out due to causality constraints, and that it corresponds precisely to the upper bound $a/c \leq 3/2$ of \cite{Hofman2008}. 

It was later pointed out by Hofman \cite{Hofman2009}, that bounds resulting from causality constraints should not be a feature of thermal CFTs. The relation between causality and positivity must lie at a more fundamental level and, as such, should show up at zero temperature. Indeed, by means of an ingenuous computation using shock waves, he proved that the upper bound on $\lambda$ also appears as a causality requirement imposed on a scattering process involving a graviton and a shock wave. He scrutinized deeper in the relation between causality and positive energy, and showed that there are indeed several bounds resulting from the different helicities both of the stress-energy tensor in the CFT side as well as of the metric perturbations in the gravity side.

This perfect match, both qualitative and quantitative, is encouraging and presents new puzzles. The addition of higher curvature corrections in the gravity side has a quantum mechanical nature, thus exploring the holographic principle thoroughly beyond the semiclassical level. It is immediate to ask whether this extends to CFTs in dimensions different than four. A natural candidate to deal with is 6d, since we know that there is a well-studied system in M-theory that corresponds to a $(2,0)$ SCFT in 6d \cite{Witten1995c,Strominger1996d}. If there exists a SCFT in 6d with large central charges but whose difference cannot be neglected in the 't Hooft limit, the gravity dual shall contain terms quadratic in the curvature. This is due to the fact that these differences appear in the 3-point function of the stress-energy tensor and, in the gravity side, this operator is sourced by a 3-graviton vertex. 

The relation between causality and positivity of the energy in 6d CFTs was studied  by de Boer, Kulaxizi and Parnachev \cite{Boer2009}. These authors courageously performed the holographic renormalization computation that allows to relate the central charges of the CFT with the LGB coefficient. This case is more complicated than its 4d counterpart since the CFT has three central charges though the positive energy conditions constrain two independent combinations thereof. They studied causality violation in the gravity side and showed that, again, $\lambda$ is bounded from above, $\lambda \leq 3/16$, which further reduces the value of $\eta/s$ in the corresponding plasma. They also showed that this bound is precisely the one arising in the CFT side from positivity of the energy arguments. These latter constraints also lead to a lower bound for the LGB coupling, $\lambda \geq - 5/16$, that arises from considering excitations with a different polarization. This was confirmed in subsequent papers in which the relation between causality and energy positivity was also generalized to arbitrary dimensions, first in the context of LGB gravity \cite{Buchel2010a,Camanho2010} and later for arbitrary Lovelock theories \cite{Boer2009a,Camanho2010a}

We generalize all the expressions for an arbitrary higher dimensional AdS/CFT dual pair. This is not, \`a priori, guaranteed to have any meaning, but it is tempting to explore this possibility and, as we will show, it leads to interesting results. On higher dimensions, though, the holographic renormalization computation of the CFT central charges is missing. Its difficulty increases heavily with spacetime dimensionality and an alternative avenue was explored in \cite{Buchel2010a}. This computation was generalized in \cite{Camanho2010d} for arbitrary Lovelock theories on any dimension, yielding the results  already presented in section \ref{3pcorr} for $t_{2}$ and $t_4$. It is indeed unclear whether there are non-trivial higher dimensional CFTs. It is still possible to argue within the conformal collider physics setup, on general grounds, that there should be bounds due to positivity of the energy conditions in these conjectural theories. The formulas we obtain for higher $d$ match these expectations. All the expressions are extended smoothly and meaningfully, as we discuss below. This may provide evidence supporting the possibility that AdS/CFT is not necessarily related to string theory.

An interesting puzzle indeed has to do with the string theory origin of quadratic curvature corrections as those of LGB gravity. These curvature corrections may appear in type II string theory due to $\alpha'$ corrections to the DBI action of probe D-branes \cite{Buchel2009}. The natural context in the 6d case, however, is to see their emergence in M-theory. Even though corrections of this sort are known to exist due to the presence of wrapped M5-branes \cite{Bachas1999}, it is not straightforward to see how they would extend to our case. They will presumably emerge\footnote{We thank Juan Maldacena for his comments about this issue.} from $A_{k-1}$ singularities produced in M-theory by a $\IZ_k$ orbifold of the AdS$_7 \times$ S$^4$ background \cite{Gaiotto2009d}. Indeed, thinking of the $S^4$ as an $S^3$ fibered on $S^1$, modding out by $\IZ_k \subset U(1) \subset SU(2)_{\rm L}$, where $SU(2)_L$ acts on the left on the 3-sphere, after Kaluza-Klein reduction along the $U(1)$ circle, leads to $k$ D6-branes in type IIA string theory (see, for instance, the discussion in \cite{Edelstein2001}). Hence, the $\alpha'$ corrected DBI terms extensively discussed in \cite{Buchel2009} should extend smoothly to our case. 


A natural problem that immediately arises in this context is what happens in the case of higher curvature corrections? Whereas in the quadratic case any combination in the lagrangian can be written as the LGB term by means of field redefinitions, this is not the case for cubic or higher contributions. Indeed, for instance, there are two independent terms that can be written at third order \cite{Metsaev1987}: the cubic Lovelock and a Riemann$^3$ combination. These two are very different in nature. While the cubic Lovelock term is the only third order curvature contribution leading to second order wave equations for the metric, in any spacetime dimensions ($d \geq 7$), the Riemann$^3$ contribution leads to higher order equations of motion and, consequently, to ghosts. Furthermore, both terms lead to a different tensorial structure for the 3-graviton vertex \cite{Metsaev1987}, as we have seen in the precedent chapter. This means that their inclusion should affect differently the corresponding tensorial contributions to the relevant 3-point functions of the stress-energy tensor. These tensorial structure is parametrized by two numbers, $t_2$ and $t_4$ \cite{Hofman2008} (that are related to the central charges of the CFTs). In the case of Lovelock we have shown that $t_4=0$, which is presumably related to the fact that the putative dual CFTs are SCFTs, in accordance with the intuition coming from previous analysis in flat space.

We will study in this chapter the whole family of higher order Lovelock gravities. We will use the AdS/CFT framework to scrutinize these theories in arbitrary dimensions in regard of the possible occurrence of causality violation. We shall address the more general case and explicitly work out the third order Lovelock gravity in arbitrary dimension. Results for higher order theories are contained in our formulas though some extra work is needed to extract them explicitly. The third order case has taught us that it is a subtle issue to determine which is the branch of solutions connected to Einstein-Hilbert and, thus, presumably stable. We also undertake the computation of gravitons colliding shock waves to seek for causality violation processes, and it seems clear to us that the identification of the EH branch is of utmost importance. From the point of view of a black hole background, the relevant question is whether one can find an asymptotically AdS black hole with a well-defined horizon. This is a delicate problem that can be amusingly cast in terms of a purely algebraic setup as has been shown in chapter \ref{chp:LLbh}. We show that the black hole computation and the shock wave one fully agree in Lovelock theory. The shock wave computation has an additional advantage as it makes sense also for non-EH branches of the theory, despite the fact that their black holes do not have a horizon. 

We find the values of the quadratic and cubic Lovelock couplings that preserve causality in the boundary. They define an unbounded (from below) region, whose intersection with the $\mu = 0$ axis reproduces all the results previously obtained for LGB. The region keeps its qualitative shape for arbitrary dimension but grows in width. As it happens in the quadratic case \cite{Camanho2010}, one of the boundaries asymptotically approaches a curve that serves as the limit of the so-called excluded region where there are no well-defined black hole solutions. This is expected since that curve plays the r\^ole of an obstruction to the growth of the allowed region. The more striking behavior happens for the other boundary, which remains at finite distance when nothing seems to prevent it from growing.

\section{Holographic causality and black holes}

We will be interested in analyzed causality properties of black hole backgrounds in Lovelock theory. Naively this seems quite a trivial question. Lovelock theory is locally Lorentz invariant, thus the metric provides local light-cones inside which timelike particles must propagate. This should be sufficient for the theory to be causal (see however \cite{Adams2006} for counterexamples in the context of  QFT). As we will see the issue is not as simple though. First of all gravitons in higher curvature gravity do not propagate according to the background metric but instead feel an {\sl effective metric} related to their equations of motion which is not simple to analyze. Moreover, we will be interested in a different kind of causality, not related to the local propagation in the bulk. With the holographic picture in mind we will analyze the possibility of having trajectories arising from the boundary of AdS and coming back to it. This can be interpreted as bulk disturbances created by local operators in the boundary CFT and we expect microcausality violation in this theory if there exists a bouncing graviton travelling faster than light from the point of view of the boundary theory. 


This can be explicitly seen by writing the equation for perturbations in the background of a black hole \reef{Scheq} in an alternative form as
\begin{equation}
\widetilde g_{\rm eff}^{\mu\nu} \widetilde\nabla_\mu \widetilde\nabla_\nu \Psi = 0 ~,
\end{equation}
where $\widetilde\nabla$ is a covariant derivative with respect to the effective geometry given by $\widetilde g_{\rm eff}^{\mu\nu} = \Omega^2\; g_{\rm eff}^{\mu\nu}$ with
\begin{equation}
g^{\rm eff}_{\mu\nu}dx^\mu dx^\nu=N_{\#}^2 f(r)\left(-dt+\frac{1}{{\bf c}_h^2}\,dx^i dx^i\right)+\frac{dr^2}{f(r)} ~,
\label{geff}
\end{equation}
and $\Omega^2=\mathbf{c}_h^2$ is a Weyl factor whose specific form will be irrelevant for the present discussion. We have rescaled the time variable, $t\to N_{\#}t$, and accordingly the graviton velocities, ${\bf c}_h^2\to {\bf c}_h^2/N_{\#}^2$, in such a way that the latter are asymptotically unity. The rescaling factor is related to the effective cosmological constant of the branch under consideration as 
\be
N_{\#}^2=1/(-L^2\Lambda)~.
\label{Nsost}
\ee
In the large momentum limit, a localized wave packet should follow a null geodesic, $x^\mu(s)$, in this effective geometry (\ref{geff}), this following from standard geometrical optics arguments. If we consider a wave packet with definite momentum
\begin{equation}
\phi=e^{i\Theta(t,r,z)}\phi_{\rm env}(t,r,z) ~,
\end{equation}
where $\Theta$ is a rapidly varying phase and $\phi_{\rm env}$ denotes an {\it almost constant} envelope, to leading order we find
\begin{equation}
\frac{dx^\mu}{ds}\frac{dx^\nu}{ds}\; g^{eff}_{\mu\nu}=0 ~,
\end{equation}
We have to identify $\frac{dx^\mu}{ds}=g^{\mu\nu}_{eff}\; k_\nu = g^{\mu\nu}_{eff}\,\nabla_\nu\Theta$. As our effective background is symmetric under translations in the $t$ and $z$ directions, we can interpret $\omega$ and $q$ as conserved quantities associated with the corresponding Killing vectors,
\begin{equation}
\omega = k_t = \frac{dt}{ds} N_{\#}^2\, f ~, \qquad q = k_z = \frac{dz}{ds} N_{\#}^2\, \frac{f}{{\bf c}_h^2} ~.
\end{equation}
Rescaling the affine parameter as $\tilde{s} = q s/N_{\#}$ (we assume $q\neq0$), we get the following radial equation of motion
\begin{equation}
\left(\frac{dr}{d\tilde{s}}\right)^2=\alpha^2-{\bf c}_h^2 ~, \qquad \alpha\equiv\frac{\omega}{q} ~.
\end{equation}
This equation describes a particle of energy $\alpha^2$ moving in a potential given by $c_2^2$. Remember that this potentials are always asymptotically one and approach zero at the horizon. In most cases the potential is monotonic and thus the graviton inevitably falls into the black hole. Nevertheless, in case there is a maximum in ${\bf c}_h^2$, geodesics starting from the boundary can find its way back to the boundary, with turning point $\alpha^2 = {\bf c}_h^2(r_{\rm turn})$. For a null bouncing geodesic starting and ending at the boundary, we then have
\begin{eqnarray}
\Delta t (\alpha) & = & 2\int_{r_{\rm turn}(\alpha)}^\infty{\frac{\dot{t}}{\dot{r}}\; dr} = \frac{2}{N_{\#}} \int_{r_{\rm turn}( \alpha)}^\infty{\frac{\alpha}{f(r)\sqrt{\alpha^2-{\bf c}_h^2(r)}}\; dr} ~, \nonumber \\ [0,5em]
\Delta z (\alpha) & = & 2\int_{r_{\rm turn}(\alpha)}^\infty{\frac{\dot{z}}{\dot{r}}\; dr} = \frac{2}{N_{\#}} \int_{r_{\rm turn}( \alpha)}^\infty{\frac{{\bf c}_h^2}{f(r)\sqrt{\alpha^2-{\bf c}_h^2(r)}}\; dr} ~,
\label{deltaintegrals}
\end{eqnarray}
where dots indicate derivatives with respect to $\tilde{s}$. Then, as the energy $\alpha$ approaches the value of the speed at the maximum, $\alpha\rightarrow c_{2,{\rm max}}$ ($r_{\rm turn} \rightarrow r_{\rm max}$), the  denominator of the integrand in both expressions diverges and the integrals (\ref{deltaintegrals}) are dominated by contributions from the region near the maximum. Thus, in such a limit we have
\begin{equation}
\frac{\Delta z}{\Delta t} \rightarrow {\bf c}_{h,{\rm max}} > 1 ~.
\end{equation}
These geodesics spend a long time near the maximum, travelling with a speed bigger than one. Interpreting this as originating from local operators in the boundary CFT, the hypothetical dual field theory is not causal if there exists a bouncing geodesic with $\frac{\Delta t}{\Delta z} > 1$, as in this case. Further discussion on this point can be found in \cite{Brigante2008a,Boer2009}. Actually, it can be shown that the superluminal graviton propagation corresponds to superluminal propagation of metastable quasiparticles in the boundary CFT with $\frac{\Delta z}{\Delta t}$ identified as the group velocity of the quasiparticles.

Generically, a graviton wave packet will fall into the black brane very quickly but, precisely when the local speed of graviton can exceed unity, the potential in \reef{Scheq} develops a local minimum that may support long-lived metastable states. Their lifetime is determined by the tunneling rate through the barrier which separates the minimum from the horizon. For very large $q$ we are approaching the classical limit of the effective Schr\"odinger problem and an associated metastable state has lifetime parametrically larger than any other timescale. 

In order to avert causality violation, we must demand these effective potentials \reef{potentialsF} to be always smaller than one \cite{Brigante2008,Brigante2008a}. In particular, at the boundary we have $c_h^2=1$ and there we must demand\footnote{Recall that the boundary corresponds to $x_b=-\infty$ and the horizon to $x_h=0$.} $\partial_x {\bf c}_h^2 \leq 0$, as $x \to -\infty$. Using $\partial_x g=\Upsilon[g]/\Upsilon'[g]$, we get the following constraints:
\begin{eqnarray}
{\rm Tensor:}\qquad & & \Upsilon'[\Lambda]+\frac{2 (d-1)}{(d-3)(d-4)}\,\Lambda\Upsilon''[\Lambda]\geq 0~.\nonumber\\[0.7em]
{\rm Vector:}\qquad & & \Upsilon'[\Lambda]-\frac{ (d-1)}{(d-3)}\,\Lambda\Upsilon''[\Lambda]\geq 0~.\\[0.7em]
{\rm Scalar:}\qquad & & \Upsilon'[\Lambda]-\frac{2 (d-1)}{(d-3)}\,\Lambda\Upsilon''[\Lambda]\geq 0~.\nonumber
\label{genconstraints}
\end{eqnarray}
These can be rewritten in terms of the dual CFT parameters, using the expressions for $t_2, t_4$ in \reef{t2t4}. The constraints become:
\bear
{\rm Tensor:} & & \qquad  1-\frac{1}{d-2}\, t_2 \geq 0 ~. \nonumber \\ [0.7em]
{\rm Vector:} & & \qquad  1+\frac{d-4}{2(d-2)}\, t_2 \geq 0 ~. \\ [0.7em]
{\rm Scalar:} & & \qquad  1+\frac{d-4}{d-2}\, t_2 \geq 0 ~. \nonumber
\eear
Hence precisely matching the constraints (\ref{tensorboundd}--\ref{scalarboundd}) coming from positivity of energy in the dual conformal field theory (with $t_4=0$).

We shall analyze now the would be restrictions imposed on the Lovelock couplings to avoid superluminal propagation of signals at the boundary. The above expressions are valid for arbitrary higher order Lovelock theory and higher dimensional spacetimes. As we saw earlier, though, there could be additional restrictions coming from the stability of the AdS vacuum solution or, further, the existence of a black hole with a well-defined horizon in such vacuum. Instead of analyzing these expressions right away, let us discuss an alternative computation, first introduced in \cite{Hofman2009}, given by the scattering of gravitons and shock waves.

\section{Gravitons and shock waves}

The previous computations are carried on a black hole background. As such, they are adequate in the context of thermal CFTs. As pointed out in \cite{Hofman2009}, one would expect to be able to perform a similar computation in a zero temperature background. The violation of unitarity driven by a value of the LGB coupling outside the allowed range, is not an artefact of the finite temperature. An adequate background to perform a computation that is independent of the temperature is given by a pp-wave. In particular, it is easier to consider the simplest case, provided by shock waves \cite{Hofman2009}. As we already mentioned, they are not subjected to higher derivative corrections \cite{Horowitz1999a}. As such, AdS shock waves are exact solutions in string theory.

We shall thus consider shock wave backgrounds in Lovelock gravity. We will study the scattering of a graviton with a shock wave in AdS. This computation, originally introduced by Hofman in the case of LGB theory in 5d \cite{Hofman2009}, and extended by us to arbitrary higher dimensional LGB gravity \cite{Camanho2010}, can also be generalized to general $d$ spacetime dimensions in the case of higher order Lovelock dynamics \cite{Camanho2010a}. This process is, in a sense, the gravity dual of the energy 1-point function in the CFT \cite{Hofman2008}. We will see that causality violation is again the source of a constraint on the value of $\lambda$. For forbidden values of this coupling, a graviton that is emitted from the boundary comes back and lands outside its light cone. 


It is more convenient to work in Poincar\'e coordinates, $z = 1/r$. We insist in performing all computations in the vielbein formalism since it is significantly simpler than the usual tensorial setup. We define light-cone coordinates\footnote{We then have to change the tangent space metric to $\eta_{00} = \eta_{11} = 0$, $\eta_{01} = \eta_{10} = - \frac12$, $\eta_{AB}=\text{diag}(1,1,\cdots,1), {\scriptstyle A,B = 2, \ldots, 6}$.} $u = t + x^{d-1}$ and $v = t - x^{d-1}$, and consider a shock wave propagating on AdS along the radial direction,
\begin{equation}
ds^2_{\rm AdS,sw} = ds^2_{\rm AdS} + f(u)\, \varpi(x^a,z)\, du^2 ~.
\end{equation}
We should think of $f(u)$ as a distribution with support in $u = 0$, which we will finally identify as a Dirac delta function. As we did in the previous section, we consider graviton perturbations, $h_{\m\n}$, splitted in different helicity channel which we keep infinitesimal
\begin{equation}
d\tilde{s}^2_{\rm AdS,sw} = \frac{N_{\#}^2}{L^2}\, \frac{-du dv + dx^i dx^i + h_{\mu\nu} \, dx^\m dx^\n+ L^4 dz^2}{z^2} + f(u)\, \varpi(x^a,z)\, du^2 ~.
\end{equation}

The relevant shock wave solution is $\varpi = \alpha\, N_{\#}^2\, z^{d-3}$. The procedure is almost the same as in the previous section, just a bit more complicated since the symmetry of the background is lower than in the black hole solution. As before, since we are only interested in the high momentum limit, we keep only contributions of the sort $\partial^2_v\phi$, $\partial_u\partial_v\phi$ and $\partial_u^2\phi$. The explicit computation can be found in the appendix \ref{SWpert}.

We will compute the time delay, $\Delta v$, due to the collision of our perturbation with the shock wave, in order to analyze the occurrence (or not) of causality violation from the boundary point of view. For that it will be important  to make certain that the delay due to free propagation in AdS is negligible compared to the shock wave contribution. We will then need to consider the large momentum regime for our perturbation, in accordance with the analogous computation performed in the Lovelock black hole background. In this limit, the free propagation of a localized wave packet can be well approximated by geodesic motion in AdS that then yields
\begin{equation}
\Delta v^{\rm free} = 2\sqrt{\frac{P_u}{P_v}}\;z_\star ~,
\end{equation}
where $z_\star$ is the radial position of the collision point. We neglected the graviton motion in the transverse directions, thereby we see that we need $P_v \gg P_u$ (that also implies $P_v \gg P_z$). We only keep contributions of the sort $\partial^2_v\phi$ and $\partial_u\partial_v\phi$ in the equations of motion, as shown earlier. The latter, even if subdominant, has to be kept to provide the dynamics of the graviton outside the locus of the shock wave. 

\begin{SCfigure}
\centering
\includegraphics[width=0.4\textwidth]{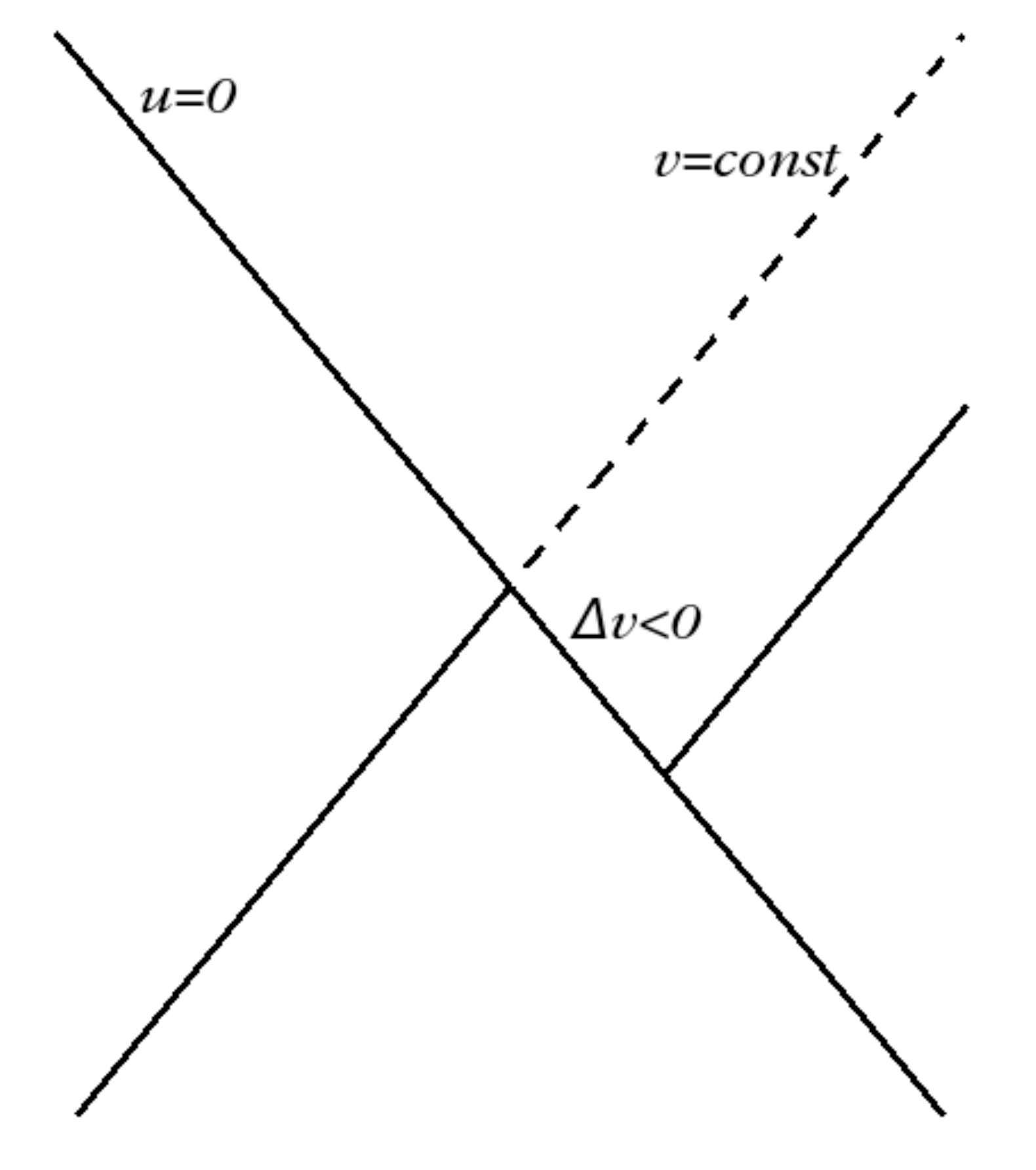}\caption{The line $u=0$ corresponds to the shock wave while the line $v=const.$ corresponds to the graviton. After the collision, if $\Delta v<0$, the particle lands outside its light-cone.}\label{shockwave}
\end{SCfigure}

The resulting equations of motion involve just one component of the perturbation for each helicity channel and all three have a very similar structure. 
\bear
{\rm Tensor:}\qquad & & \partial_u\partial_v\phi+\alpha f(u)L^2 z^{d-1}\left(1-\frac{1}{d-2}t_2\right)\partial_v^2\phi=0 ~,\nonumber\\
{\rm Vector:}\qquad & & \partial_u\partial_v\phi+\alpha f(u)L^2 z^{d-1}\left(1+\frac{d-4}{2(d-2)}t_2\right)\partial_v^2\phi=0 ~,\\
{\rm Scalar:}\qquad & & \partial_u\partial_v\phi+\alpha f(u)L^2 z^{d-1}\left(1+\frac{d-4}{d-2}t_2\right)\partial_v^2\phi=0 ~.\nonumber
\end{eqnarray}
In the large momentum limit and taking the shock wave profile to be a delta function, $f(u) = \delta(u)$, the equation of motion reduces to the usual wave equation $\partial_u \partial_v \phi = 0$ outside $u = 0$. Then, we can consider a wave packet moving with definite momentum on both sides of the shock wave. We can find a matching condition just by integrating the corresponding equation of motion over the discontinuity
\begin{equation}
\phi_> = \phi_<\, e^{-i P_v\, \alpha\,\, z^{d-1}\mathcal{N}_h} ~,
\label{matching}
\end{equation}
where we used $P_v = - i \partial_v$. We can find the shift in the momentum in the $z$-direction acting with $P_z = - i \partial_z$,
\begin{equation}
P_z^> = P_z^< - (d-1) P_v\, \alpha\, \delta(u)\, z^{d-2}\; \mathcal{N}_h ~.
\end{equation}
If we consider a particle going inside AdS, $P_z > 0$, and if we want it to come back to the boundary after the collision we need
\begin{equation}
P_v\, \alpha\; \mathcal{N}_h > 0 ~.
\end{equation}
But we know that $\alpha > 0$ (since the black hole has positive mass) and $P_v = - \frac12 P^u < 0$ (since $P^u = P^0 + P^6$ must be positive for the energy to be so); then we need  
\begin{equation}
\mathcal{N}_h< 0 ~.
\end{equation}
When this happens the graviton can make its way back to the boundary and, as we can read from (\ref{matching}), it comes back shifted in the $v$-direction a negative amount (see figure \ref{shockwave})
%
\begin{equation}
\Delta v = \alpha\, z^{d-1}\, \mathcal{N}_h~.
\end{equation}
The graviton lands, at the boundary, outside its light-cone. This is an explicit break up of causality. We conclude that the theory violates causality unless $\mathcal{N}_h\geq0$ what amounts exactly to the same constraints found in the black hole case and arising from positivity of the energy correlators on the dual CFT. Then we have shown how the calculation involving bouncing gravitons in a black hole background and those scattering shock waves lead exactly to the same result for completely general Lovelock gravities in arbitrary dimensions.



The first non-trivial case corresponds, as always, to the case of LGB gravity. We can now verify that the lower and upper bounds of $\lambda$ come from, respectively, helicity zero and helicity two perturbations or, conversely, positivity of the energy in the dual CFT in the same channels. This would lead us to a formula valid for any dimension $d \geq 5$ (below 5d, the LGB term either is a total derivative or it identically vanishes),
\begin{equation}
-\frac{(d-3) (3 d-1)}{4 (d+1)^2} \leq \lambda \leq \frac{(d-4) (d-3) \left(d^2-3 d+8\right)}{4 \left(d^2-5 d+10\right)^2} ~.
\label{uplambd}
\end{equation}
%
\begin{figure}
\centering
\includegraphics[width=0.67\textwidth]{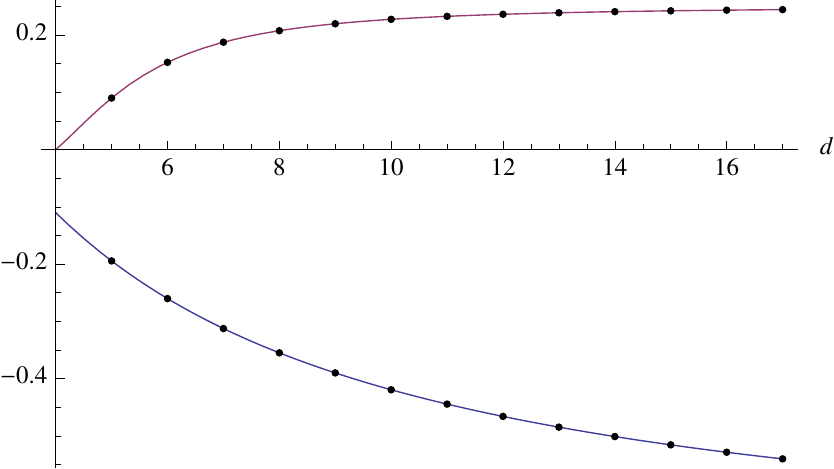}\caption{Upper and lower bound values for $\lambda$. The upper curve corresponds to helicity two modes, while the lower one is due to helicity zero perturbations. The window monotonically increases an asymptotically approaches its maximum range $-3/4 \leq \lambda \leq 1/4$ for infinite dimensional theories.}
\label{bounds}
\end{figure}
\noindent
This formula reproduces earlier results \cite{Brigante2008,Brigante2008a,Boer2009,Ge2009} for the upper bound, and generalizes the lower bound to arbitrary dimensions \cite{Camanho2010,Buchel2010a}.

There are several comments we would like to make about this result. Besides its remarkable simplicity and smoothness, we see that $\lambda_{\rm max}$ asymptotically approaches $1/4$, when $d \to \infty$. This may be expected. One can show that $\lambda_{\rm max}$ is a monotonically increasing function, but there is an obstruction precisely at $\lambda = 1/4$, as we discussed around (\ref{discr}). It is more striking what happens to the lower bound. There is no critical negative value of $\lambda$, at least manifestly. Thus, naively one might expect that $\lambda_{\rm min} \to - \infty$ in the infinite dimensional limit. However, we obtain $\lambda_{\rm min} \to - 3/4$ (see figure \ref{bounds}). We think that this asymptotic behavior calls for a deeper understanding.

One of the main consequences of a positive $\lambda$ is the violation of the so-called KSS bound for the shear viscosity to entropy density ratio \cite{Kovtun2005}. As pointed out in \cite{Brigante2008},
\begin{equation}
\frac{\eta}{s} = \frac{1}{4\pi} \left( 1 - 2 \frac{d-1}{d-3}\, \lambda \right) ~,
\end{equation}
for a CFT plasma dual to a LGB theory. We see that the maximal violation of the KSS bound happens for conjectural 8d CFTs, the minimum value of $\eta/s$ asymptotically approaching the ratio $\eta/s = 1/8\pi$. Whether there exist higher dimensional CFTs with a finite temperature regime admitting a hydrodynamical description with such low values of $\eta/s$ is, of course, an open problem. A warning remark is however worth at this point. The low energy effective gravity action used in these computations is strictly valid in the region of large central charges when their relative differences are very small. Thus, finite values of the GB coupling, $\lambda \sim 1$, are not fully reliable.
\vskip3mm

\begin{figure}
\centering
\includegraphics[width=0.67\textwidth]{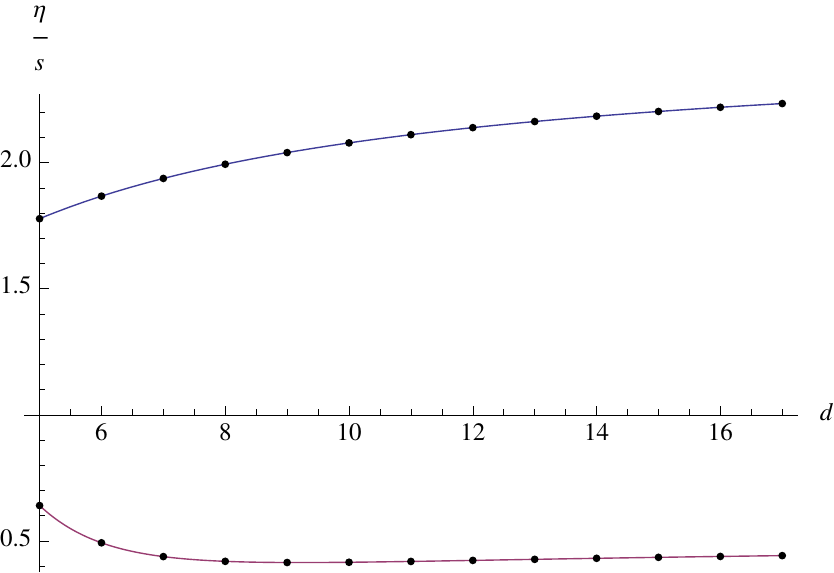}\caption{Upper and lower values for $\eta/s$. The upper curve corresponds to helicity zero, while the lower one is due to helicity two perturbations. The window asymptotically approaches its maximum range $1/2 \leq 4\pi\,\eta/s \leq 5/2$ for infinite dimensions.}
\label{eta-s-vs-d}
\end{figure}

The very existence of a negative lower bound for $\lambda$ seems to imply, naively, that there is an upper bound for $\eta/s$ in strongly coupled CFTs (see figure \ref{eta-s-vs-d}). It is well-known that $\eta/s \to \infty$ is a generic feature in weakly coupled theories but, to the best of our knowledge, there is no \`a priori reason that tells us why the strongly coupled value should be $1/4\pi$ or differ by a factor of order 1. This seems to be a possible interpretation of our result: no matter the dimensionality of a CFT, its strongly coupled plasma will have a very small shear viscosity to entropy density ratio. To put this conclusion on firmer grounds, however, one should study more carefully the effect of higher curvature corrections and have a deeper understanding on the nature of higher dimensional CFTs and the effect of terms with powers higher than two. This will be the focus of the next chapter, where we will be also concerned with additional constraints coming from stability of planar black holes. These represent the gravitational dual of the plasmas with the above shear viscosity and entropy  density, the instability of the black hole thus implying the instability of the fluid.

\section{Causality constraints in cubic Lovelock theory}

We would like to show, for definiteness, the results corresponding to third order Lovelock theory. In this case, the analysis of the previously discussed causality constraints reduce to studying the sign of the following set of polynomials ($x= L^2 \Lambda)$):
\begin{eqnarray}
\mathcal{N}_2 & \sim & 1+\frac{2(d^2-5d+10)}{(d-3)(d-4)}\lambda\, x +\frac{d^2-3d+8}{(d-3)(d-4)}\mu\, x^2 ~, \\ [0.5em]
\mathcal{N}_1 & \sim & 1-\frac{4}{(d-3)}\lambda\, x-\frac{d+1}{(d-3)}\mu\, x^2 ~, \\ [0.5em]
\mathcal{N}_0 & \sim & 1-\frac{2(d+1)}{(d-3)}\lambda\, x-\frac{3d-1}{(d-3)}\mu\, x^2 ~,
\end{eqnarray}
where we have taken into account that the first derivative of the polynomial (in the denominator of $t_2$) is positive for BD-stable branches and in particular for the EH one. 
\hskip-1mm
\begin{figure}
\centering
\includegraphics[width=0.78\textwidth]{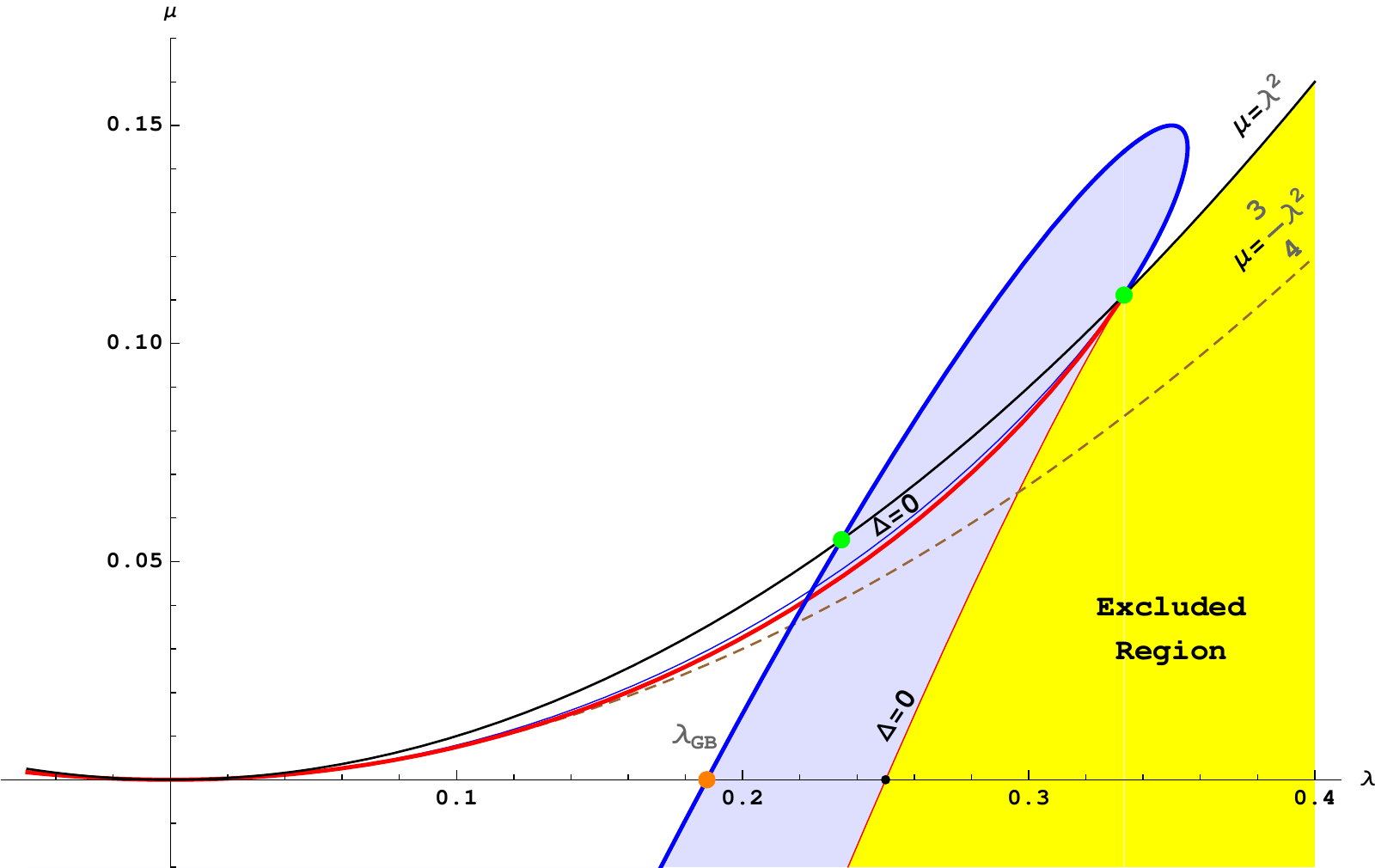}\caption{The blue shaded region correspond to negative values of $\mathcal{N}_2$ for the branch of solutions with a horizon, and therefore causality violation occurs there. The red and blue thick curves correspond to $\mu = \frac{1}{243} \left[ 189 \lambda + 4 \left(-8 \pm (2 - 9 \lambda)\, \sqrt{16 - 45 \lambda} \right) \right]$. The blue one provides the restriction for the relevant solution branch. The thin blue and red lines are nothing but the singular locus $\Delta = 0$. We can observe that this region includes the LGB case ($\lambda_{\rm GB} = 3/16,~ \mu = 0$) (depicted as an orange point) \cite{Boer2009}. If $\mu = \lambda^2$, contrary to what is stated in  \cite{Ge2009a}, there is a forbidden region (limited by the green points) $\frac{19}{81} \leq \lambda \leq \frac{1}{3}$. }
\label{constraintN2}
\end{figure}
The constant $x$, as we discussed earlier, is one of the solutions of
\begin{equation}
1+x+\lambda\, x^2+\frac{\mu}{3}\,x^3 = 0 ~.
\end{equation}
To proceed, we have to find the simultaneous solutions for each of the polynomials and the previous equation, that is where each of the polynomials can change sign. The simplest way of doing this is to find the solutions for each of the polynomials and substitute them into the equation.
This procedure gives several curves $\mu=\mu(\lambda)$ for each polynomial. The most clear way that we find to present these results is by presenting the following figures which illustrate the situation. The analytic expression of the curves can be easily obtained and, indeed, are discussed in the captions. We will discuss in the text their main qualitative features.

In figure \ref{constraintN2} we display both the region of the coupling space excluded by the arguments of the preceding section and (in blue) the sector of the $(\lambda,\mu)$-plane that leads to causality violation in the 7d Lovelock theory. Projected in the $\lambda$ axis we reobtain the upper bound corresponding to LGB gravity and obtained in \cite{Boer2009,Camanho2010}. Even if, from the point of view of AdS/CFT, the trustable region is close to the origin, {\it i.e.}, for small values of $\lambda$ and $\mu$, it is amusing to see that the forbidden region has a structure that seems to be meaningful, even if this is not fully justified. Notice that, had we restricted to third order Lovelock theory with $\mu = \lambda^2$, we would have obtained a higher upper bound for $\lambda$ (this corrects a misleading statement in \cite{Ge2009a}). Indeed, notice that there is a disconnected (from the Einstein-Hilbert theory) region since some finite positive values of $\lambda$ are prevented by causality.

\begin{figure}
\centering\includegraphics[width=0.78\textwidth]{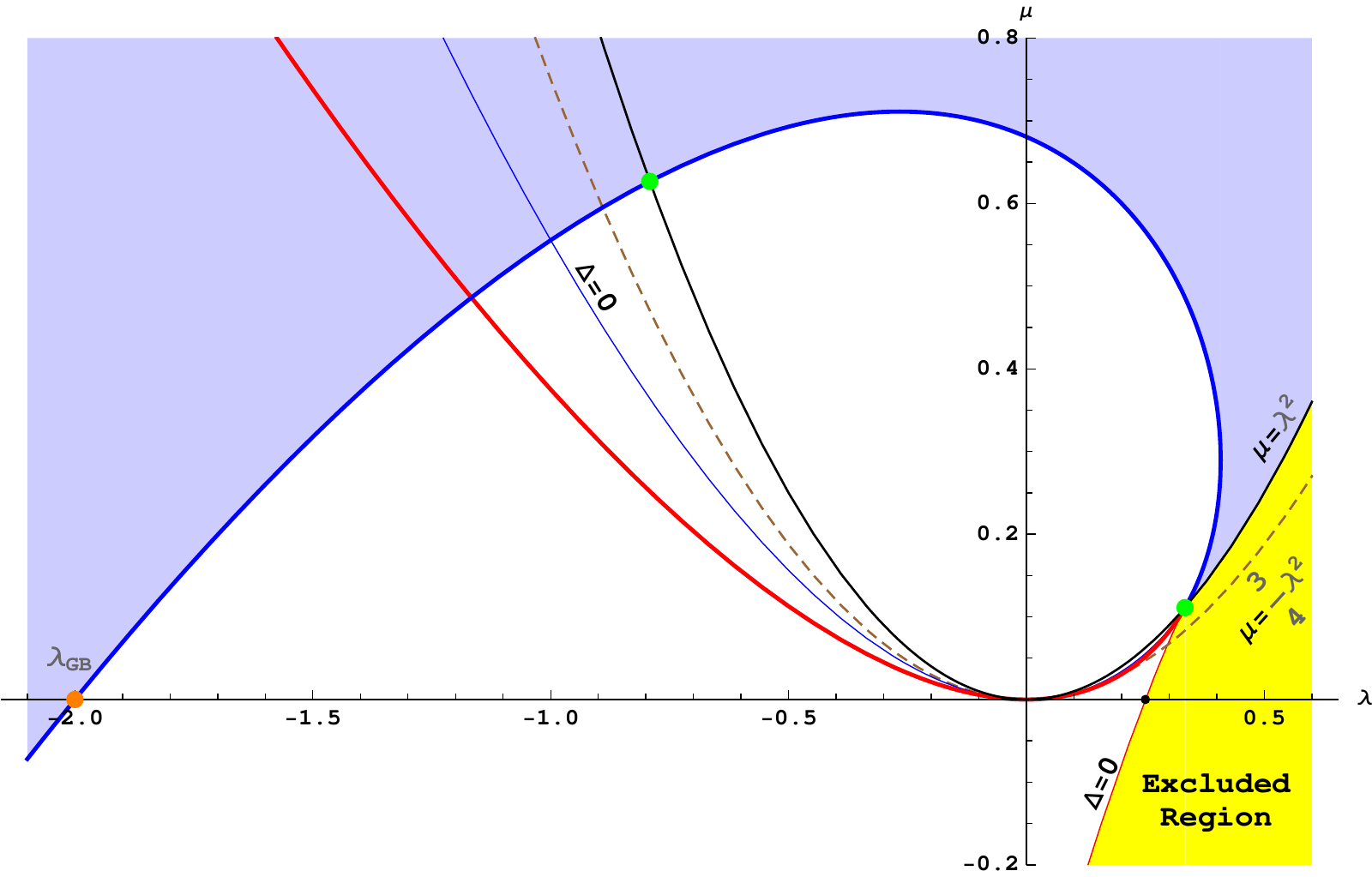}
\caption{The helicity 1 polynomial restrict the parameters to lie inside the region between the blue curve and the excluded region. The shaded region correspond again to negative values of $\mathcal{N}_1$ for the EH branch and thus to causality violation. The equation of the curve is $\mu = \frac{1}{144} \left[ -18 \lambda + 49 \pm (7 + 6 \lambda) \sqrt{49 - 120 \lambda} \right]$. For LGB gravity we found, $\lambda>-2$.}
\label{constraintN1}
\end{figure}

We represented the situation corresponding to helicity one gravitons colliding a shock wave or, alternatively, helicity one perturbations of the AdS black hole background in figure \ref{constraintN1}.\hskip-1mm
We see that it gives a complementary restriction in the coupling space. Again, it intersects the $\lambda$ axis at the point $-2$, as obtained in \cite{Camanho2010}. We will not extend much in the discussion of this curve for reasons that will be clear immediately.

Let us finally explore the causality violation induced by helicity zero gravitons.\hskip-1mm
\begin{figure}
\centering
\includegraphics[width=0.78\textwidth]{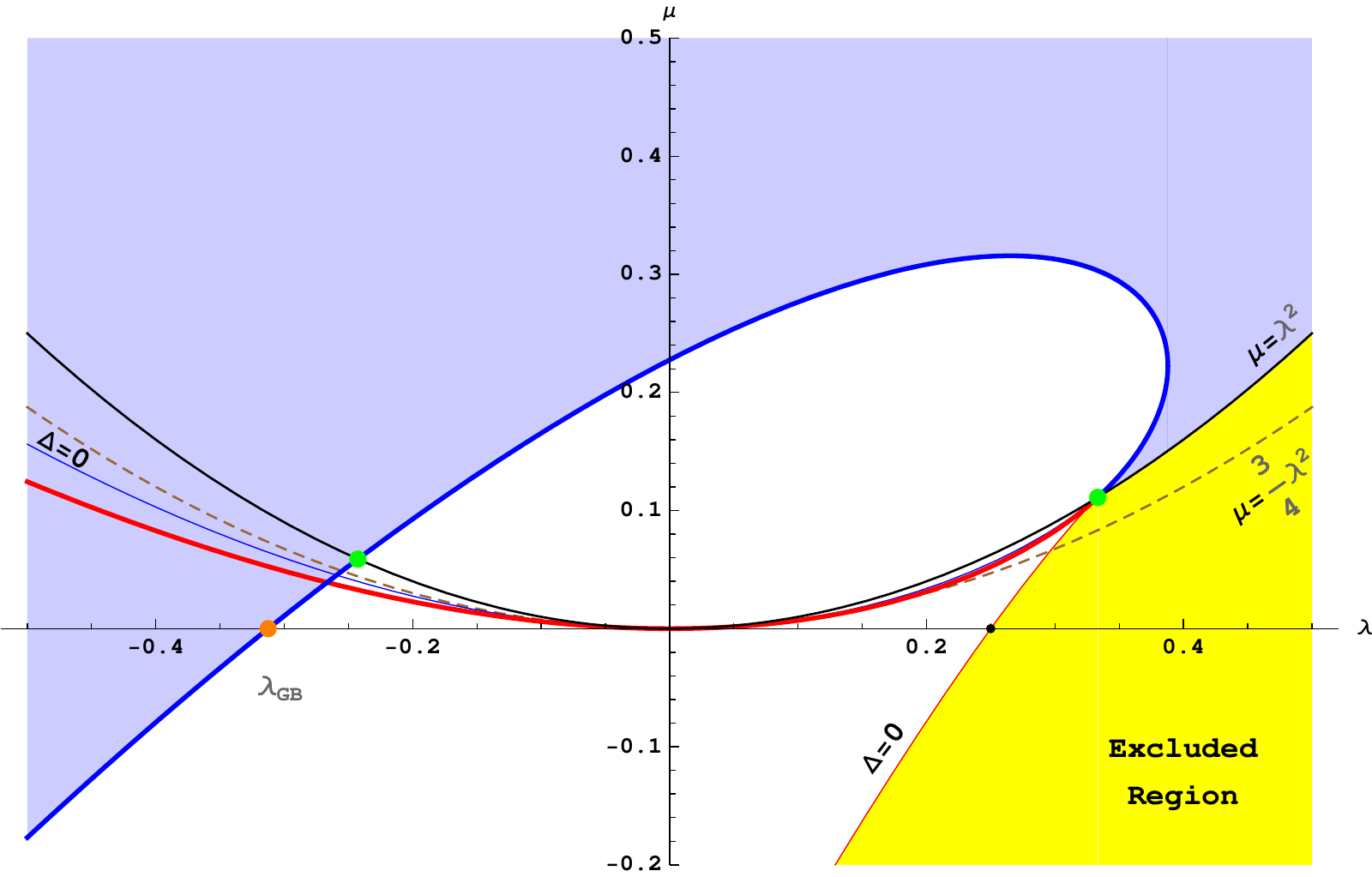}
\caption{As in the helicity 1 case, the helicity 0 constraint restrict the parameters to lie in a strip between the blue line and the excluded region, but it is always more stringent. For LGB gravity, $\lambda>-\frac{5}{16}$. The equation of the curve is $\mu = \frac{1}{1125} \left[ 315 \lambda + 128 \pm 4 (4 + 15 \lambda) \sqrt{64 - 165 \lambda} \right]$.}
\label{constraintN0}
\end{figure}
A very important feature is the fact that this curve imposes more severe restrictions than those forced by the helicity one case. Indeed, the curve fully overcomes the helicity one constraints in such a way that they are finally irrelevant. This has been seen earlier in LGB \cite{Hofman2009,Camanho2010}. Going to the conformal collider physics setup, this result seems to be related to the vanishing of $t_4$ or more precisely to the constraint \reef{potsconstraint}. This equation in particular implies that whenever both the helicity two and zero potentials are causal (less than one), the helicity one also is. Something similar happened also for the stability constraints of chapter \ref{chp:bhstability}. Finally, the intersection with the $\lambda$ axis reproduces the result for LGB \cite{Boer2009,Camanho2010,Buchel2010a}.

Now, the three curves should be put altogether to determine the region in the coupling space that is allowed by all helicities.\hskip-1mm
\begin{figure}
\centering
\includegraphics[width=0.78\textwidth]{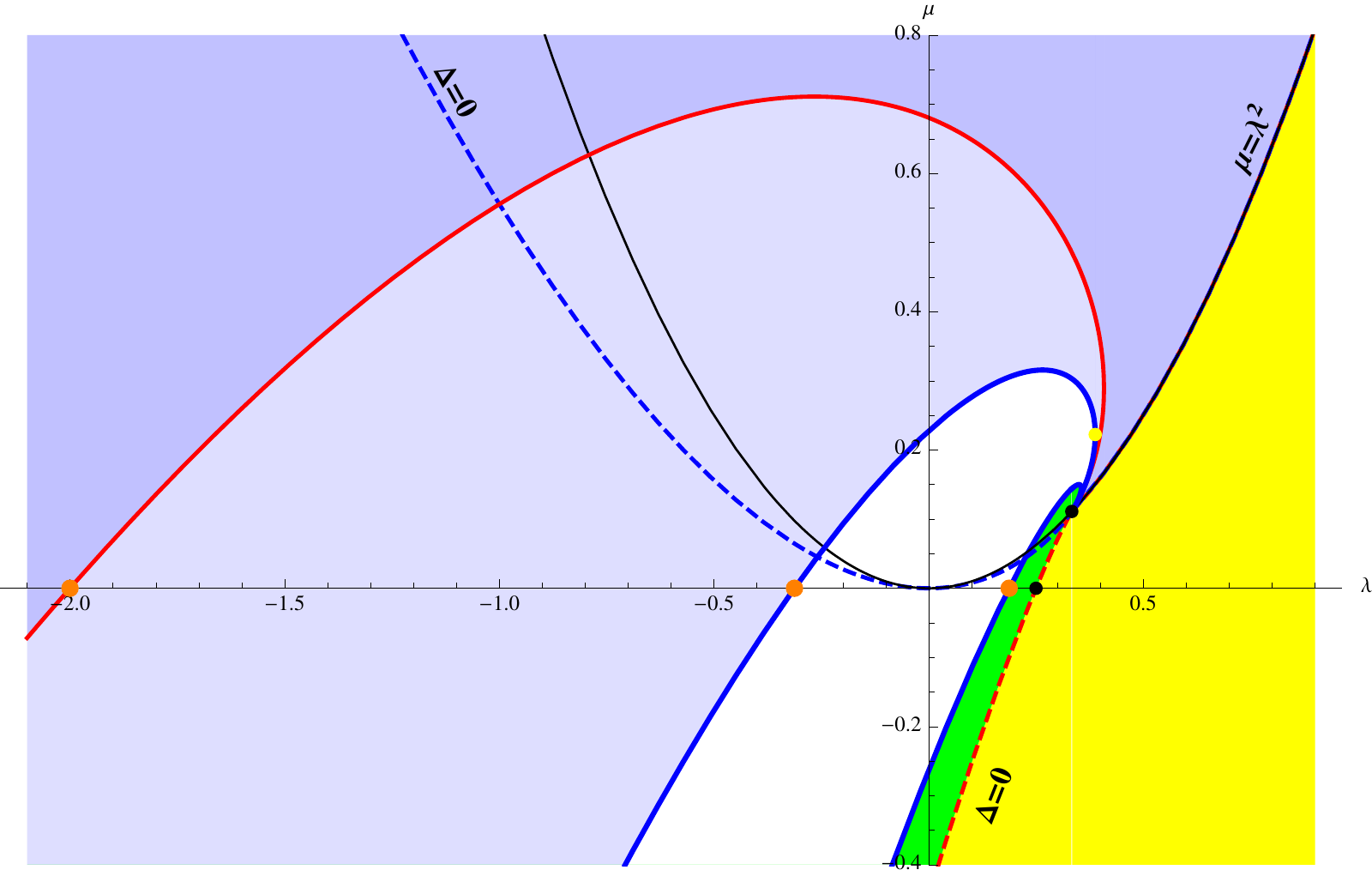}\caption{As the helicity 0 constraint is always more stringent than the helicity 1's, the allowed region of parameters is contained between the curves corresponding to helicity 0 and 2 (red and blue lines respectively). Notice that the maximum value for $\lambda$ is raised from the GB case to the value $64/165$ (yellow point). Strikingly enough this gives a negative value for the shear viscosity to entropy density ratio.}
\label{allowed-region}
\end{figure}
As mentioned, the helicity one curve ends up being irrelevant. The coupling space of the third order Lovelock theory becomes, as we see in figure \ref{allowed-region}, quite restricted. Several questions immediately arise and we will try to pose some of them and answer quite a few in the discussion. It should be pointed out that there is a subregion of the allowed sector with $\lambda$ greater than $1/3$ which immediately leads to negative values of $\eta/s$ \cite{Boer2009a}. We will see in the next chapter that once again stability solves the mystery ruling out the region where $\eta/s$ dips below zero.

To finish this section, it is tempting to use our expressions for third order Lovelock theory which are valid for arbitrary higher dimensional spacetimes.\hskip-1mm
\begin{figure}
\centering
\includegraphics[width=0.78\textwidth]{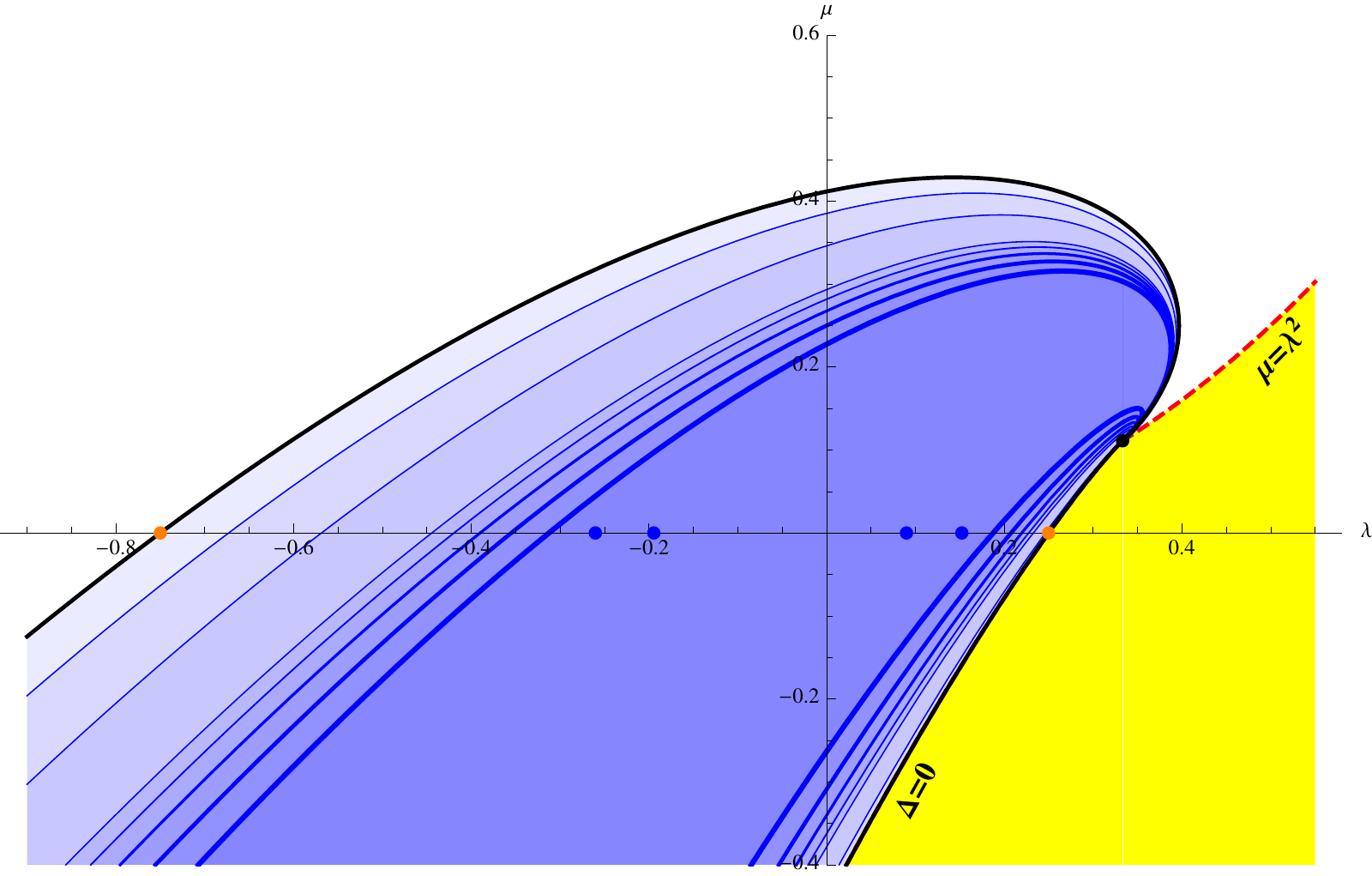}\caption{Allowed region for the GB and cubic Lovelock couplings for arbitrary dimension. The inner region is the one depicted in figure \ref{allowed-region} for $d = 7$. Towards the exterior, we depicted with a decreasing blue shadow successively, those for $d = 8, 9, 10, 11, 20, 50$ and $\infty$. Blue points denote the boundaries of $d = 5$ and $6$, where the cubic Lovelock term identically vanishes. The orange points are the asymptotic values for GB gravity found in \cite{Camanho2010}. Notice that one of the branches that bound the allowed region tend to the singular locus $\Delta = 0$ when $d \to \infty$.}
\label{allowed-d}
\end{figure}
It is not difficult to compute the allowed regions for different dimensionalities. This is represented in figure \ref{allowed-d}. It should be pointed out that the subregion which, being part of the allowed sector, leads to negative values of the shear viscosity disappears for 11d (and higher). Indeed, the maximum allowed values of $\lambda$ are, respectively, $1369/3519$, $147/377$ and $2209/5655$, for 8d, 9d and 10d, which in all the cases exceed the critical value leading to a vanishing shear viscosity (respectively, $5/14$, $6/16$ and $7/18$).\hskip-1mm
\begin{figure}
\centering
\includegraphics[width=0.78\textwidth]{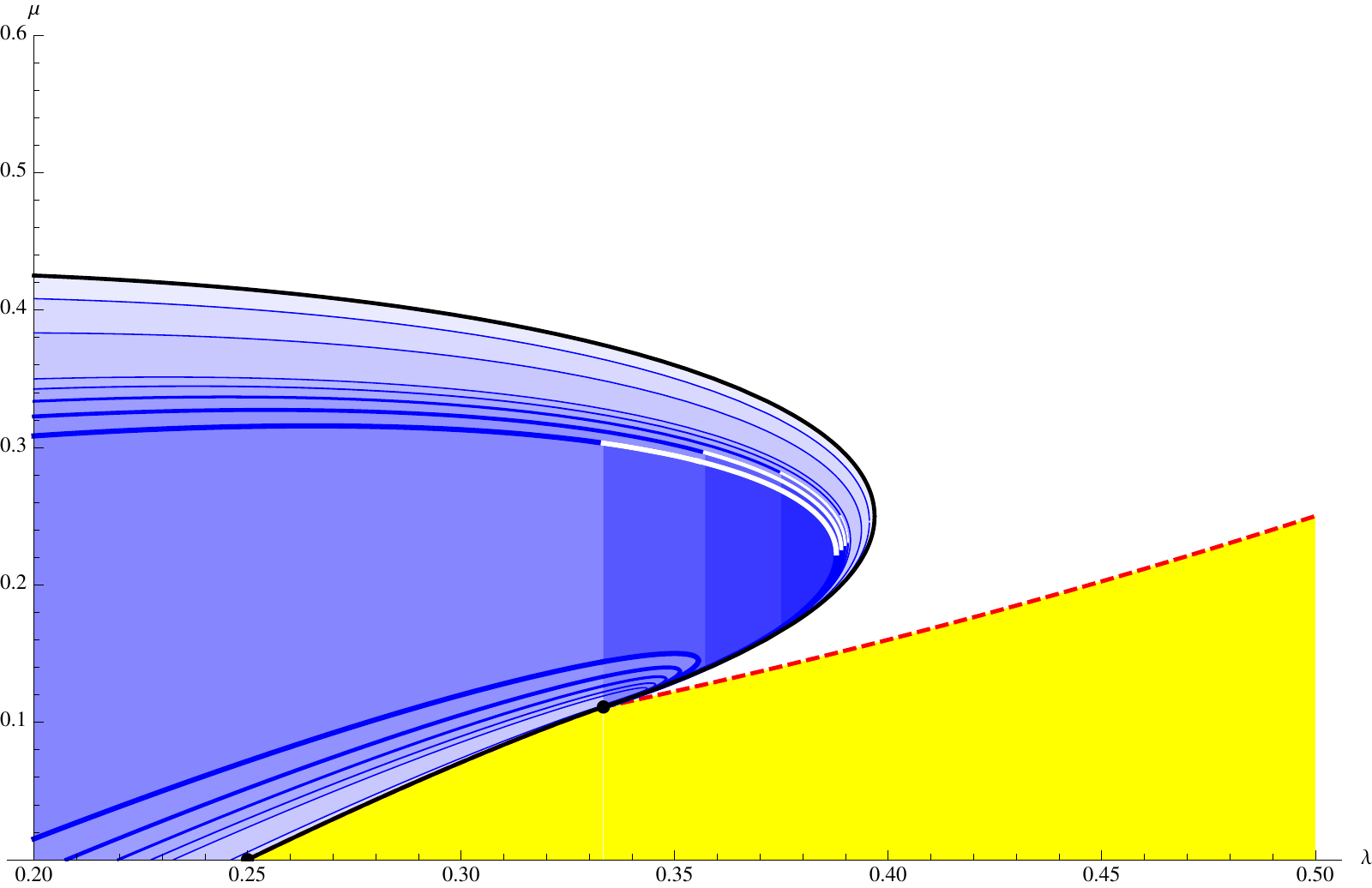}\caption{We have shadowed the region of the allowed space of couplings leading to unphysical negative values of $\eta/s$. It can be seen that the maximum $\lambda$ increases with the spacetime dimensionality, but the critical value leading to a vanishing $\eta/s$ does it faster. From 11d on, this problem disappears.}
\label{allowed-d-zoom}
\end{figure}
As stated, 11d is the first dimensionality at which the maximum value for $\lambda$, $169/432$, is lower than the critical one, $8/20$. Whether this is a numerical coincidence or it is pointing out that something special happens to (super)gravity in 11d is presently unclear.

\section{Causality for higher curvature vacua}

The causality constraints discussed above in the context of third order Lovelock gravity are those arising for the EH branch of the theory. It is the only one with black hole solutions displaying a horizon in the planar case, thus the only one for which the black hole computation of previous sections is valid. Nonetheless, one may still wonder which constraints imposes positivity of the energy for the CFTs defined by these vacua. Notice that the values of $t_2$  changes whereas $t_4$ is still zero. The shock wave scattering provides the corresponding  violation of causality for all branches.  

Causality constraints can be rephrased as constraints in the characteristic polynomial evaluated on the specific vacuum under consideration. For the EH branch it is expected that for any Lovelock theory of arbitrary degree and dimension there is always a region containing the Einstein-Hilbert point where causality (and stability) is respected. These constraints can be rewritten for BD-stable AdS vacua as
\begin{equation}
-\frac{(d-3)(d-4)}{2(d-1)}\leq C[\Lambda_\star]\equiv\frac{\Lambda_\star\Upsilon''[\Lambda_\star]}{\Upsilon'[\Lambda_\star]}\leq\frac{(d-3)}{2(d-1)}
\end{equation}
where $\Lambda_\star$ is the effective cosmological constant of the vacuum being analyzed. The upper and lower bounds correspond  to the zero helicity or scalar channel and the tensor or helicity two channel respectively. The helicity one condition is always less constraining than the zero one and will be ignored. Remark that this bounds depend on the dimension whereas every other expression involved in the problem does not (once fixed the degree of the polynomial). The upper bound is itself bounded from above as
\begin{equation}
\frac{(d-3)}{2(d-1)}<\frac12~,
\end{equation}
while the lower bound is unbounded. Furthermore, we may write the polynomial in terms of its roots,
\begin{equation}
\Upsilon[\Lambda]=c_K\prod_{k=0}^K(\Lambda-\Lambda_k) ~.
\label{pol2}
\end{equation}
From the reality of the Lovelock coefficients (or equivalently of the lagrangian) we know that the roots are either real or come as conjugate dual pairs. The only further constraint we will consider is that it has a well-defined (AdS-)EH limit, \ie  
\begin{equation}
\Upsilon'[g]>0\quad,\qquad \forall g\in[\Lambda_\text{EH},0)
\end{equation}
This defines effectively the EH vacuum. 

From the equivalent polynomial \reef{pol2} we can very easily express $C[\Lambda_i]$ for each vacuum as
\begin{equation}
C[\Lambda_i]=2\sum_{j\neq i}\frac{1}{1-\frac{\Lambda_j}{\Lambda_i}}
\label{C}
\end{equation}
Also the Boulware-Deser stability of such vacuum can be very easily assessed in this way as the first derivative of the polynomial there yields
\begin{equation}
\Upsilon'[\Lambda_i]=c_K \prod_{j\neq i}(\Lambda_i-\Lambda_j)
\end{equation}
Starting from the stable EH-vacuum the AdS roots follow a series of stable/unstable vacua where complex conjugate pairs of roots do not matter. The same happens for the dS vacua.

For $\Lambda_i<0$ we can classify the contributions to \reef{C} according to their sign
\begin{equation}
1-\frac{\Lambda_j}{\Lambda_i}> 0 \qquad\Leftrightarrow\quad \Lambda_i\leq \Lambda_j 
\end{equation}
and the reverse for the opposite sign. This is also valid for the complex conjugate pairs where $\Lambda_j$ has to be replaced by its real part. 

Using this it is very easy to show that in cubic Lovelock gravity, the minimal example with two BD-stable AdS-vacua, just the EH-vacua may be causal. In the case in which both such vacua exist the characteristic polynomial displays three real roots 
\begin{equation}
\Lambda_\text{HC}<\Lambda_{u}<\Lambda_\text{EH}<0
\end{equation}
where the subindices stand for higher curvature, unstable and Einstein-Hilbert respectively. If we want to analyze causality for the higher curvature vacuum $\Lambda_{HC}$ we have just to analyze two terms, both positive and bounded from below
\begin{equation}
C[\Lambda_\text{HC}]=\frac{2}{1-\frac{\Lambda_u}{\Lambda_\text{HC}}}+\frac{2}{1-\frac{\Lambda_\text{EH}}{\Lambda_\text{HC}}}>4
\end{equation}
in such a way that $C[\Lambda_{HC}]$ automatically violates the helicity zero (also the helicity one) constraint. The same happens for the smaller root of the polynomial when real. Assuming there is at least one real AdS root (for instance the EH one) and that the rest has bigger real part we can easily show 
\begin{equation}
C[\Lambda_\text{smaller}]>2
\end{equation}
In case there is any smaller root -- with more negative real part -- this would yield a negative contribution to $C[\Lambda_i]$. In absolute value, this term may be as big as one wants, the corresponding (unstable) vacuum just has to approach sufficiently the one we are analyzing. Thus, one negative term alone may compensate any positive contribution from the rest of the roots and bring the value of $C[\Lambda_i]$ to the interval allowed by causality.

The minimal example for which we may have a BD-stable and causal AdS vacuum in addition to the EH one is quartic Lovelock, we just need $c_4>0$ in the region of parameter space where the four roots are real and negative.

\section{Discussion}

We have used the AdS/CFT framework to scrutinize higher order Lovelock theory in arbitrary dimensions with the focus on possible causality violation. We have formally addressed the more general case and explicitly worked out third order Lovelock gravity. Results for higher order theories are contained in our formulas though some extra work is needed to extract them explicitly. Provided with the values of $t_2$ and $t_4$ appearing in the Maldacena-Hofman parametrization of the $3$-point function of the stress tensor \cite{Hofman2008}, we have checked the complete equivalence between the constraints arising from the positivity of energy bounds conjectured in that paper and those coming from causality. These results constitute a proof of previous conjectures in the literature \cite{Boer2010,Camanho2010a}.

We do this computation in two different ways and  end up showing that they are bound to give the same result. We have first computed all polarization linear perturbations of the black hole AdS solution in Lovelock theory, and subsequently studied the collision of gravitons and shock waves in AdS in the framework of this theory. We found a region in the coupling space where the theory does not violate causality (at least in a way that can be detected by this kind of computations\footnote{\S\ Tighter constraints may arise in case we change the shock wave profile (see for instance \cite{Camanho2013c}).}). This generalizes the intervals found for $\lambda$ in the case of LGB theories \cite{Buchel2009a,Hofman2009,Boer2009,Camanho2010}. Contrary to those intervals, that are now just given by the intersection of the allowed sector and the $\lambda$ axis, the region is unbounded. Arbitrary large (simultaneous) negative values for $\lambda$ and $\mu$ are allowed. The meaning of this result is unclear since this happens in a region of couplings where the computations are not trustable. However, the regular pattern that we found as a function of the spacetime dimensionality is too nice to be discarded so easily. We have seen that the allowed region increases with spacetime dimensionality, as it happened with the LGB coupling \cite{Camanho2010}. In the infinite dimensional limit, one of the boundaries of the allowed region (roughly speaking, the one giving the upper bound; rigorously speaking, the one due to helicity two gravitons), asymptotically approaches the zero discriminant boundary in the coupling space of the theory. This is expected since the region beyond that curve is excluded. The other curve, arising from scalar gravitons, has a less natural behavior that calls for a deeper explanation. There is no restriction for this curve to be asymptotically larger, but it approaches a finite curve, that of course crosses the $\lambda$ axis at $\lambda = -3/4$, in accordance with \cite{Camanho2010}. Thus, moderate (non-negative) values of $\lambda$ and $\mu$ are forbidden even for would be infinite dimensional theories. Whether this is an artefact of our approach or there is a deeper reason for this behavior, is unclear to us at this  point.

It is intriguing that classical Lovelock gravity in AdS space can capture highly non-trivial aspects of conformal field theory physics. While Einstein gravity is known to describe the universal spin-2 sector of a large class of strongly coupled CFTs, no dual to Lovelock gravity (beyond GB) has been found, nor have the Lovelock terms with finite coefficients appeared in string theory. While higher powers of curvature generically appear in string theory as $\alpha'$ corrections to the supergravity action, these corrections are necessarily perturbative. In Lovelock theory, we have dimensionful parameters appearing in the lagrangian which do not appear to be protected from quantum corrections by any symmetry. That from the classical theory we can obtain specific numerical predictions on which values of the parameters should or not be allowed by causality and that these restrictions should agree with analogous restrictions in a hypothetical dual CFTs is highly suggestive. We obtain what seem to be splendid checks of the AdS/CFT correspondence, the results being smooth functions of the spacetime dimensionality.

One possible hint is that Lovelock theories all satisfy $t_4=0$, which is required for supersymmetric CFTs \cite{Kulaxizi2010}. For free field theories, $t_4=0$ is obtained by picking the appropriate supersymmetric matter content. In Lovelock gravity, the $t_2$ and $t_4$ parameters can be obtained in the boundary expansions of the effective potentials \reef{pots}. The $t_4=0$ result should imply some relationship between these. Indeed, it is easy to see that from \reef{pots} we have
\be
(d-4)\, \mathbf{c}_2^2(r) + (d-2)\, \mathbf{c}_0^2(r) - 2 (d-3)\, \mathbf{c}_1^2(r) = 0 ~,
\label{holsusy}
\ee
which is actually valid at any radius $r$ and not just at the boundary, where it indeed implies $t_4=0$.  Beyond the boundary expansion, the relationship written above should impose constraints on higher $n$-point functions of the theory. It is tempting to conjecture that these are the holographic duals of some sort of supersymmetric constraint.

The papers that originally dealt with causality violation did it under the spell of searching possible violations of the KSS bound \cite{Brigante2008,Brigante2008a}. They found that the addition of a LGB term in the gravity lagrangian corrects the universal value of the shear viscosity to entropy density ratio as
\begin{equation}
\frac{\eta}{s} = \frac{1}{4\pi} \left( 1 - 4 \lambda \right) ~.
\label{etavsslambda}
\end{equation}
The addition of further higher order Lovelock terms does not manifestly contribute to this quotient \cite{Shu2009a}. The possible addition of a quadratic curvature correction with positive $\lambda$ is enough to argue that the KSS bound is not universally valid. The appearance of an upper bound for $\lambda$ due to causality conservation naively seems to lower the KSS bound down to a new one. Since (\ref{etavsslambda}) is not affected by higher order Lovelock terms, this has led some authors to speculate about the nature of this newly found lower bound. However, it is not clear if the new bound should be taken seriously: when $\lambda$ approaches the maximal allowed value, the central charges of the dual CFT explore a regime (roughly $(a - c)/c \sim \mathcal{O}(1)$) where the gravity description is \` a priori untrustable. Furthermore, let us point out that the fact that higher order Lovelock couplings do not enter (\ref{etavsslambda}) does not mean that they are irrelevant for this problem. Indeed, we have seen that even an infinitesimal positive value of $\mu$ would raise the upper bound for $\lambda$ thus affecting the putative lower bound for $\eta/s$. Even more weird, there are values of $\mu$ leading to too high values of $\lambda$ in the sense that $\eta/s$ becomes negative. We will see in the next chapter that the mechanism allowing us to disregard this unphysical behavior is actually the stability of the plasma, dual to the stability of planar black holes studied in chapter \ref{chp:bhstability}. These additional constraints will further restrict the possible values of the Lovelock coefficients being crucial for the discussion of bounds on $\eta/s$.

Previously, concerning the LGB case, it was raised the issue about the string theory origin of quadratic curvature corrections. It seems clear after \cite{Kats2009,Buchel2009,Gaiotto2009d} that these terms may possibly arise from D-brane probes or $A_{N-1}$ singularities in M-theory. The situation with the cubic curvature corrections is more delicate in this respect. Besides, the very fact that the results generalize smoothly both to higher order Lovelock theory and to higher spacetime dimensions, seems to suggest that these computations may not necessarily rely in the framework of string theory.

Our results for arbitrary dimensions are compatible with the condition $t_4 = 0$. This is the expected value of $t_4$ for a supersymmetric CFT. Cubic terms are expected to accomplish supersymmetry breaking. However, this is not the case for the Lovelock combination. As for $t_2$, the result of this chapter is unexpected. It is well-known that the third order Lovelock term endows a 3-graviton vertex structure that does not contribute to the 3-point function \cite{Metsaev1987}. However, this result is strictly valid for flat space-times while in AdS, as we see, it is not true. The introduction of the third order Lovelock lagrangian enters non-trivially in the discussion of causality violation. It would be interesting to further clarify how this term contributes to $t_2$ by studying directly the 3-graviton vertex in AdS spacetimes.

\chapter{\bfseries\itshape Holographic plasmas and the fate of the viscosity bound}
\chaptermark{The fate of the viscosity bound}
\label{LLfate}

\vspace{.6cm}

\begin{quotation}
\flushright
{\it ``A poet once said, `The whole universe is in a glass of wine'. \\
We will probably never know in what sense he meant that,\\
for poets do not write to be understood. But it is true that if we\\
look at a glass of wine closely enough we see the entire universe.''}\\

\vspace{.3cm}

Richard Feynman
\end{quotation}

\vspace{3cm}


\noindent
The AdS/CFT correspondence has provided a tool to study hydrodynamical aspects of quantum field theories at strong coupling. This was particularly timely due to the advent of experiments that prompted the exploration of QCD in a region of phase space where it displays such behavior. One of the most striking predictions of AdS/CFT had to do with the shear viscosity to entropy density ratio, $\eta/s$, of strongly coupled plasmas. Policastro, Son and Starinets computed this ratio in the canonical example of $\cN = 4$ $SU(N)$ supersymmetric Yang-Mills (SYM) plasma, in
the planar ('t Hooft) limit and for infinitely large 't Hooft coupling $\lambda=g_{\rm YM}^2\,N\to \infty$ \cite{Policastro2001a}, finding
\be
\frac{\eta}{s}=\frac{1}{4\pi}
\ee
Interestingly enough, it was found that there is a universal value for this ratio, $\eta/s = 1/4\pi$, for theories whose gravity dual is governed by the Einstein--Hilbert action, regardless of the matter content, the number of supersymmetries, the existence or not of a conformal symmetry, and even the spacetime dimensionality. On the other hand, all measured values for this ratio in any quantum relativistic system are above this value. This led to speculations that $\eta/s \geq 1/4\pi$ might be an exact statement in quantum relativistic systems, the so-called KSS bound conjecture \cite{Kovtun2005}. Indeed, it is possible to provide a hand waving argument \cite{Son2007}, relying on a quasiparticle description of the plasma, that links the KSS bound to the Heisenberg uncertainty principle. This argument is questionable, though, since there are convincing hints supporting a non-quasiparticle description of strongly interacting plasmas.

The KSS bound conjecture has been thoroughly scrutinized for many years (see, for example, \cite{Dobado2008} for a recent review). It turned out to be the case, however, that when quantum corrections are included in the gravitational action, under the form of curvature squared terms, the $\eta/s$ ratio can be smaller than the KSS bound \cite{Brigante2008,Kats2009}. In particular, there are string theory constructions where this is the case \cite{Kats2009,Buchel2009}. In these situations the violation of the viscosity bound can be traced back to the inequality between the central charges of the theory, namely to $c-a>0$, quite generic in superconformal gauge theories with unequal central charges \cite{Buchel2009}. Moreover, to ensure reliable computations, higher curvature corrections have to be regarded as being {\it small}. As a result, any violation of the KSS bound realized in string models of the duality is necessarily perturbative. Nonetheless, the question of the existence of any such bound persisted. Because of the universality property of the shear viscosity to entropy density  in the supergravity approximation, any finite violation of the KSS bound has to be studied in an {\it effective} model of AdS/CFT, rather than a particular realization of the correspondence in string theory.

For the particular combination given by the LGB term, one may consider finite values of the coefficient, as the gravitational theory is then two-derivative. As such, it effectively defines via the AdS/CFT correspondence a dual conformal plasma. For this we can compute 2- and 3-point function as discussed in chapter \ref{chp:AdSCFT} and also transport coefficients such as $\eta/s$. In this case the ratio is modified to \cite{Brustein2009a,Shu2010}
\be
\frac{\eta}{s}=\frac{1}{4\pi}\left(1-\frac{2(d-1)}{d-3}\lambda\right)~,
\label{etas}
\ee
$\lambda$ being the appropriately normalized LGB coupling, and the KSS bound is violated whenever $\lambda$ is positive \cite{Brigante2008,Kats2009}. Up to field redefinitions, for very small $\lambda$ the gravitational model with a LGB correction is equivalent to the example of Kats and Petrov \cite{Kats2009} embedded in string theory. For that we have to identify
\be
\frac{c-a}{c}\sim \lambda
\ee
that now can be taken however to finite values.

Even if \`a priori the addition of a LGB term would lead to an arbitrary violation of the KSS bound, it turns out that the causality constraints studied in the previous section arise preventing the possibility of going to arbitrarily low values of $\eta/s$ \cite{Brigante2008a}. This was actually the original motivation of their discovery. In the case of 4d CFTs, for instance, this imposes the constraint $\lambda \leq 9/100$, which reduces the minimum value of $\eta/s$ by a factor of $16/25$.

The study of a possible bound is interesting both from a theoretical standpoint as well as from a phenomenological one -- it has been found experimentally that the quark-gluon plasma created in relativistic heavy ion collisions appears to have a very low shear viscosity to entropy density ratio \cite{Romatschke2007b}. A similar result was also found recently in a radically different context: that of strongly correlated ultracold atomic Fermi gasses in the so-called {\it unitarity} limit \cite{Schafer2009}. Both systems have measured values of $\eta/s$ compatible with $1/4\pi$. On the other hand, from a theoretical standpoint, the causality constraints coming from the behavior of the geometry near the boundary, that rule the attainability of lower values of $\eta/s$, have a beautiful holographic dual in the CFT side: they arise from positivity of energy conditions \cite{Hofman2008,Buchel2009e,Hofman2009}. The perfect matching of these two quite fundamental restrictions on a physically sensible theory constitutes a striking check of the AdS/CFT correspondence. 

Whether this is still valid for higher-dimensional CFTs became a natural question that was subsequently answered in a series of papers \cite{Boer2010a,Camanho2010,Buchel2010a}. Summarizing, it turned out that LGB theories lead to a violation of the KSS bound in any spacetime dimensionality, and that causality constraints exactly match positivity of energy bounds on the CFT side.  The fact that this matching is valid regardless of the dimensionality is puzzling and seems to provide clues on possible non-stringy versions of AdS/CFT. Our knowledge of higher-dimensional CFTs is, however, too poor yet to push these arguments forward. In higher dimensions one has in general the choice of including other Lovelock terms in the action. These are higher order in curvature that still lead to second order equations of motion \cite{Lovelock1971}. A first step in the analysis of these theories was done in \cite{Boer2010,Camanho2010a}. There it was found that causality and positivity of energy constraints still match perfectly. Still, something new happens here. Since there are now more free parameters available, it turns out that causality alone cannot prevent the $\eta/s$ ratio from becoming arbitrarily small (or even negative).

In this chapter we analyze in detail the fate of the $\eta/s$ ratio in AdS/CFT restricted to the case of Lovelock theories. We will focus in the cases of LGB, cubic and quartic Lovelock gravities in arbitrary dimension, even though general results are contained in our formulas.  We will study restrictions coming from positivity of energy correlators of the CFT, as explained in the previous chapter. 
We then consider graviton fluctuations about the black hole background for a general Lovelock theory. Near the boundary, demanding causality of graviton propagation leads to bounds which exactly match those of energy flux positivity. Close to the horizon, in turn, we have shown that unstable modes can appear if the parameters of the theory are not tuned properly. This leads to another set of bounds which, for the LGB gravity or cubic Lovelock theory in seven dimensions, prevent the $\eta/s$ ratio from becoming arbitrarily small. In the case of higher order Lovelock theories, a new phenomenon occurs\footnote{This phenomenon already takes place for the cubic Lovelock theory. It leads to a less stringent (thus irrelevant) constraint in $d=7$, but is crucial  to prevent arbitrarily small values of $\eta/s$ in $8 \leq d \leq 10$ theories. When $d>10$, causality at the boundary is enough to discard this possibility in cubic Lovelock gravity \cite{Camanho2010a}.}: causality violation and plasma instabilities may arise in the interior of the geometry. They should be related to analogous unhealthy properties arising in the CFT plasma far from the UV, whose identification and full characterization is left as an open problem. Causality violations and instabilities should have, however, very different physical interpretations: while we may expect that the former correspond to some illness of the boundary theory, the latter should be related to either plasma instabilities or to issues such as not being expanding around the true vacuum\footnote{We thank Rob Myers for clarifying comments on these points.}. Anyway, interestingly enough, this shows that causality violation is not necessarily a phenomenon arising in the UV and it is not generically dual to positivity of energy in the CFT.

The attentive reader has already realized that higher order Lovelock gravities imply higher-dimensional theories. There is a second more subtle point that should be presently raised. The analysis performed in this chapter deals with higher order curvature corrections whose coupling constants do not need to be necessarily small. Both conditions are unnatural within the string theory realm. It seems however worth inspecting thoroughly these theories based on the very fact that they lead to {\it reasonable} results.  That is, even for these presumably non-stringy setups we obtain what seem to be splendid checks of the AdS/CFT correspondence.

The last section is devoted to the analysis of the $\eta/s$ ratio in higher order Lovelock theories. The first part of that section presents analytic and numerical results for the minimum ratio in any dimension up to quartic order theory. These are obtained by considering the full stability and causality constraints. Our results are relatively smooth functions of the spacetime dimensionality. We also show that for any given Lovelock theory of order $K$ there is always a lower bound on $\eta/s$. It is impossible to get arbitrarily close to $\eta/s=0$, even in the case where the causality constraints would not be taken into account as proposed in \cite{Buchel2010b}.

In the second part we consider a limited analysis of arbitrary Lovelock theories in any dimension. Picking a specific curve through parameter space, the properties of which are carefully discussed in Appendix \ref{appendixB}, we show that there seems not to be a dimension independent viscosity bound. Within our framework, we argue that a strongly coupled ideal fluid could be achieved in the strict limit $d \to \infty$, when a correspondingly infinite number of Lovelock terms are taken into account. This is somehow reminiscent of the so-called {\it species problem}, although it is not clear to us whether there is a more rigorous way in which these two setups are related.

\section{Shear viscosity and finite temperature instabilities}

It has been shown in \cite{Brustein2009a,Shu2010} that the shear viscosity to entropy ratio of a generic Lovelock theory is given by
\be
\frac{\eta}s = \frac{1}{4\pi}\left(1-\frac{\lambda}{\lambda_c}\right) ~,
\label{etaeq}
\ee
where we have defined the critical value $\lambda_c$ where the ratio vanishes
\be
\lambda_c \equiv \frac{d-3}{2 (d-1)} ~.
\label{lambdac}
\ee
It is interesting to stress that the ratio only depends on the quadratic Lovelock coupling, $c_2$. Generically one is not free to take an arbitrary $\lambda$. There are causality constraints which restrict the range of values where the theory is well defined \cite{Brigante2008,Brigante2008a}. These restrictions match the restrictions on $t_2$ and $t_4$, and these necessarily involve all the parameters in the theory. The maximum value $\lambda_{(0)}$ that may be achieved depends on the theory in question. In this way it has been found \cite{Boer2010,Camanho2010a} that causality constraints alone cannot prevent one from reaching (and even overpassing) $\eta/s=0$. In this section we show that as one approaches small values of $\eta/s$, the black hole becomes unstable and the linear approximation breaks down. This places an effective set of constraints in the parameters of the theory, saving the ratio above from ever becoming {\it too small}. 

\subsection{Plasma instabilities}

It has been already discussed in chapter \ref{chp:bhstability} that particular values of the Lovelock couplings lead to instabilities for planar Lovelock black holes. Within the gauge/gravity framework, black holes provide the dual description of a finite temperature plasma. For certain values of the Lovelock couplings coupling, some effective potentials might develop negative values close to the horizon, this indicating an instability of the plasma \cite{Myers2007}. In equation (\ref{Scheq}) we see that the r\^ole of $\hbar$ is played by $1/q$. By taking sufficiently large spatial momentum (small $\hbar$), we can make an infinitesimally small (negative energy) well to support a negative energy state in the effective Schr\"odinger problem. Going back to the original fields, this translates into an exponentially growing and therefore unstable mode \cite{Myers2007}. 

At the horizon the effective potentials vanish in such a way that in order to avoid instabilities we must demand the first radial derivative to be positive. This gives rise to the constraints (\ref{inscons1} -- \ref{inscons0}) that we write here as a remainder, 
\begin{eqnarray}
& \lambda < \lambda_c ~, \nonumber\\ [0.7em]
& (d-3)(d-4)-2\lambda (d-1)(d-6)-4\lambda^2 (d-1)^2+2\mu (d-1)^2 \geq 0 ~, \\ [0.7em]
& (d-2)(d-3)-6\lambda (d-1)(d-2)+12\lambda^2 (d-1)^2-2\mu (d-1)^2 \geq 0  ~. \nonumber
\end{eqnarray}
These inequalities represent the constraints from the  shear, tensor and sound channels respectively. They are required to hold, otherwise the CFT plasma is unstable. For $\mu=0$, these results match those for LGB gravity derived in \cite{Buchel2010a}. It is important to notice that, in spite of considering a completely general Lovelock theory with as many terms as we wish, these stability constraints involve just the lowest two Lovelock couplings, $\lambda$ and $\mu$. 

The first constraint immediately rules out the region of the parameter space where the shear viscosity to entropy ratio becomes negative, since $\lambda < \lambda_c$. It is interesting to point out that the shear channel stability condition is also needed to ensure the validity of the linear analysis \cite{Takahashi2009} and unitarity (see section \ref{bhstab}). Then, we can also relate the would be negative values of the shear viscosity -- that would lead to a manifest instability given its r\^ole as a damping coefficient in the sound channel (a negative value then amplifies the sound mode) --, to the accompanying break down of the perturbative approach, at the horizon, and unitarity there. This restriction is also implied by the other two constraints together. In fact, as already mentioned, the shear channel constraint is always irrelevant once the other two are taken into account. The same happens for the causality constraints, as already explained.  We may restrict our analysis to the tensor and sound channels only. 

The tensor and scalar constraints define a new allowed {\it stability wedge} (\ref{stabwedge1}
--\ref{stabwedge2}) in the parameter space with apex at
\begin{equation}
\lambda = \lambda_c = \frac{d-3}{2(d-1)} ~, \qquad \qquad \mu=\frac{(d-3)(d-5)}{2(d-1)^2} ~,
\end{equation}  
or, equivalently, on the intersection of the $\eta/s=0$ line with $\mu = \lambda (4\lambda - 1)$. Then we can approach, from the inside of this stable region, a unique point (in the cubic case; a (hyper-)line in higher order cases) where $\eta/s=0$, and this is exactly at the apex of such region, where the stability constraints coming from each helicity meet.

\begin{figure}
\centering
		\includegraphics[width=123mm]{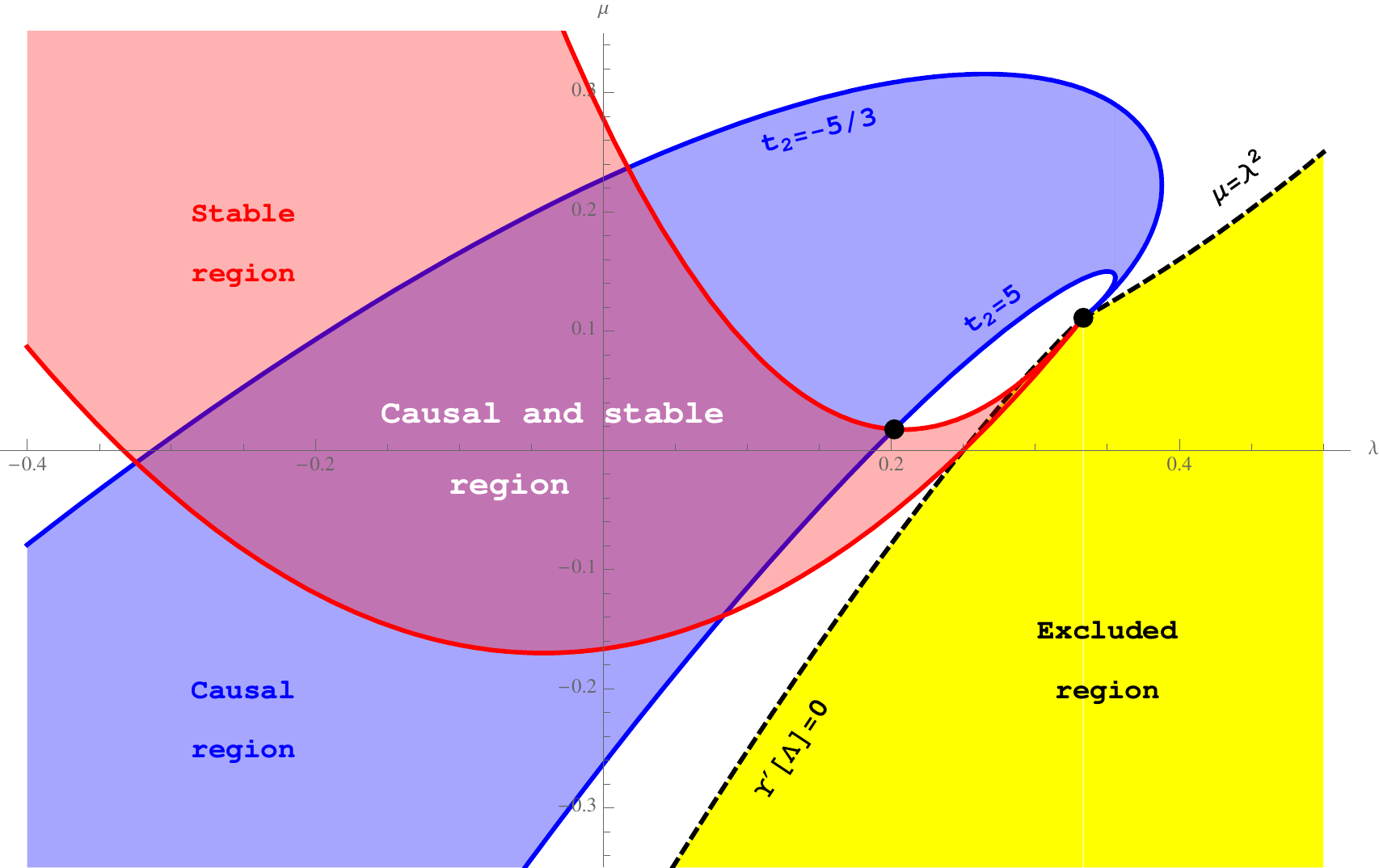}
  \caption{The allowed region by causality and stability for cubic Lovelock theory in $d=7$. The black points are the maximally degenerated point $(1/3,1/9)$ (also the Chern-Simons point in this case) and the intersection of the helicity zero stability constraint and the helicity two causality constraint $\sim(0.20,0.017)$, that gives the lowest value for $\eta/s$.}
{\label{StableCausal}}
\end{figure}

For $d=7$, the apex coincides with the point of maximal symmetry $(\lambda,\mu)=(1/3,1/9)$, where the polynomial $\Upsilon[g]$ has a single maximally degenerate root. This is also the Chern-Simons point of the $d=7$ Lovelock theory. We plot the different regions in the space of parameters in figure \ref{StableCausal}.\!\!\!
The minimum value for $\eta/s$ in this case is attained when the upper (helicity zero) stability curve intersects the lower (helicity two) causality curve, which happens at
\begin{equation}
\lambda \simeq 0.202042 ~, \qquad \qquad \mu \simeq 0.0175986 ~,
\end{equation}
giving a minimum value
\be
\frac{\eta}s \simeq 0.393874 \times \frac{1}{4\pi} < 0.4375 \times \frac{1}{4\pi} ~,
\ee
where the latter one is the correction to the KSS bound coming from LGB alone (this corresponds to the value of $\lambda$ at which the lower blue curve intersects the axis, $\lambda = 3/16$). This seems to be the end of the story in $d=7$, at least in the context of Lovelock gravities. The existence of plasma instabilities (they originate in the behavior of the gravitational potential close to the horizon) sets a definite (positive) lower bound on cubic Lovelock theory. However, as we will see, the situation becomes more involved in higher dimensions, where the restrictions discussed so far are not enough to prevent $\eta/s$ from becoming arbitrarily close to zero.

\section{Bulk causality and stability}
\label{bulkins}

\subsection{The cubic theory in higher dimensions}

Our results imply that the $\eta/s$ ratio can never be too small for cubic Lovelock theory in $d=7$ by a combination of two constraints -- preservation of causality, holographically dual to positivity of energy; and stability of the black brane solution which can be broadly identified with the stability of the thermal plasma. Of course the obvious question is whether similar results are valid in higher order theories. As it turns out, these two constraints are indeed enough, but not in the simple fashion we have described so far -- we must require causality and stability everywhere in the bulk, and not just from a simple near boundary or near horizon analysis.

\begin{figure}
\centering
\includegraphics[width=0.5\textwidth]{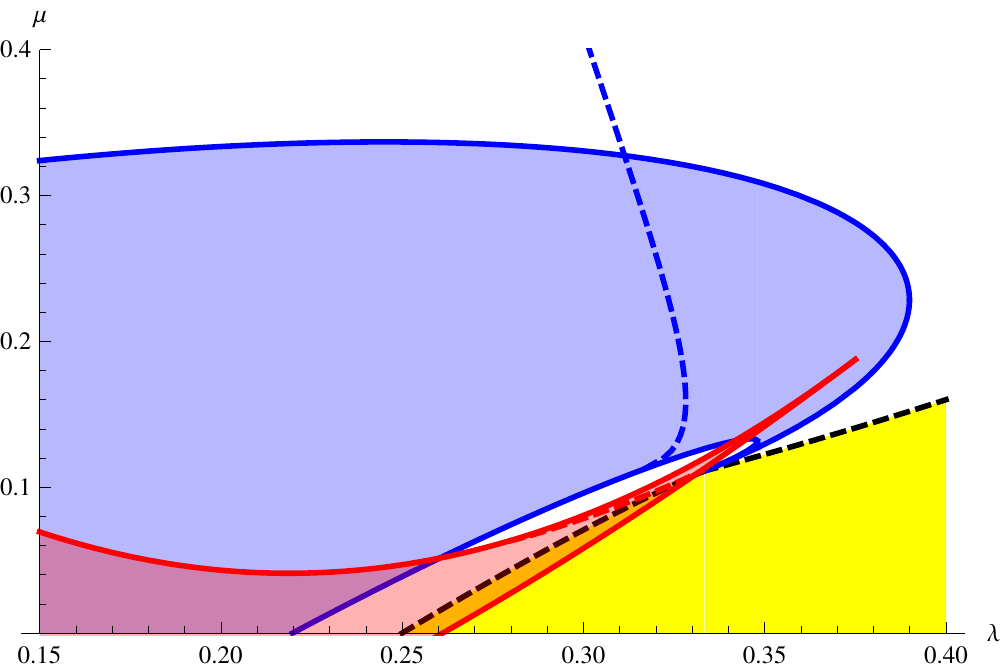}\includegraphics[width=0.5\textwidth]{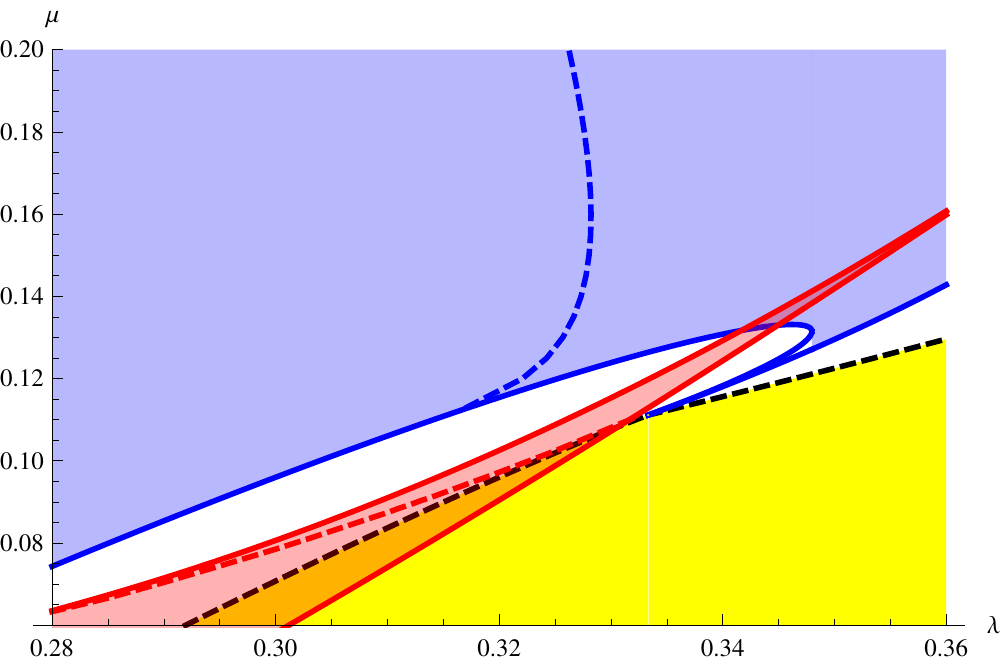}\caption{Causality and stability regions in $d=9$ cubic Lovelock theory (and zoom). The thick lines correspond to causality at the boundary and stability at the horizon, respectively in blue and red, while the dashed lines correspond to causality (in the tensor channel) and stability (in the sound channel) in the full geometry.}
{\label{StableCausal-9d}}
\end{figure}
\begin{figure}
\centering
\includegraphics[width=0.5\textwidth]{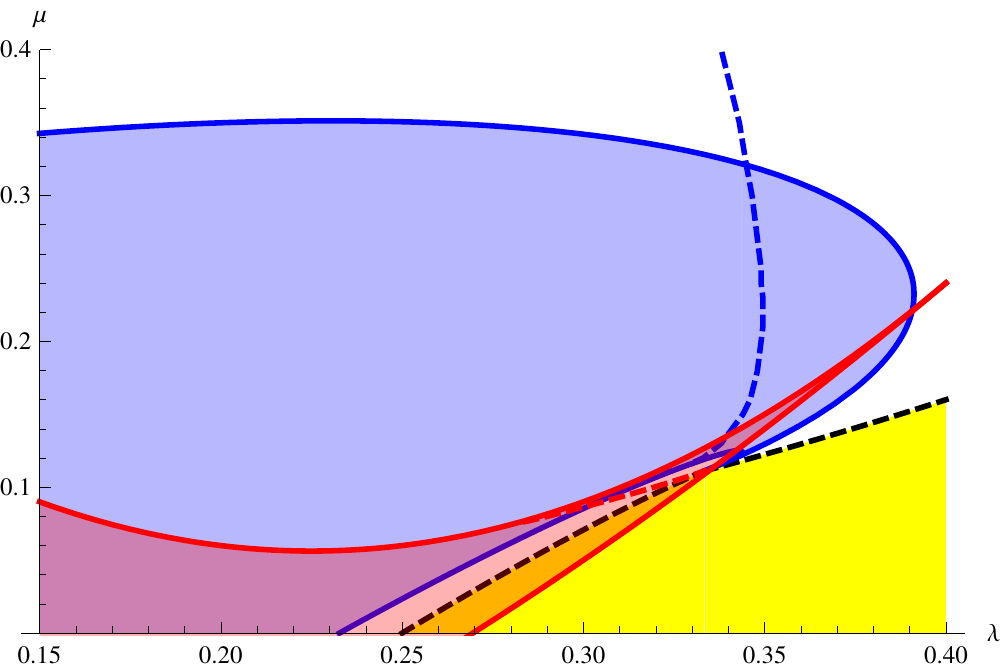}\includegraphics[width=0.5\textwidth]{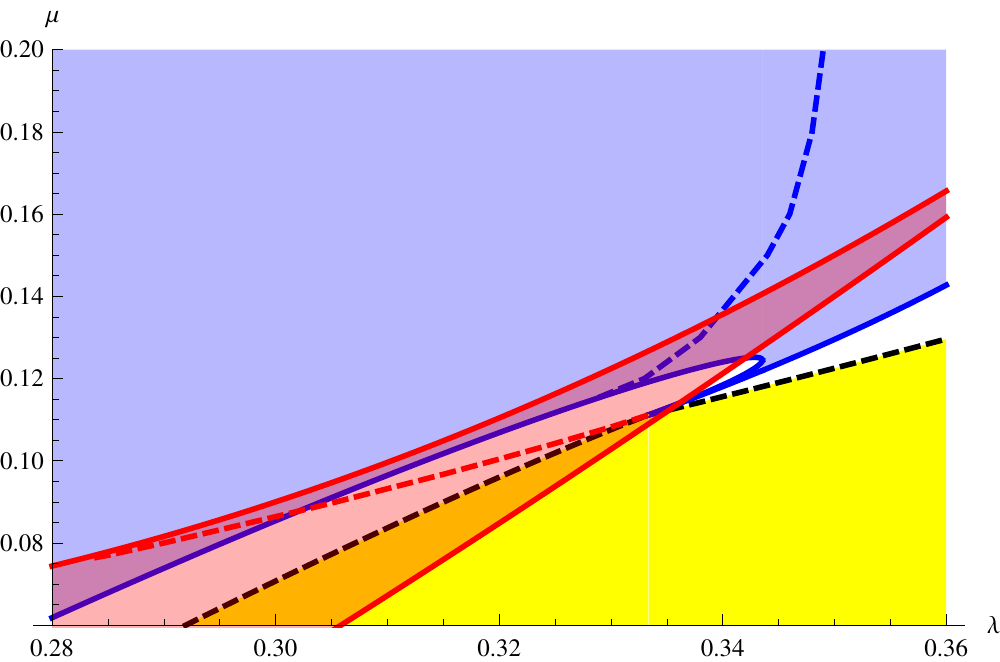}\caption{Causality and stability regions in $d=11$ cubic Lovelock theory (and zoom). The thick lines correspond to causality at the boundary and stability at the horizon, respectively in blue and red, while the dashed lines correspond to causality (in the tensor channel) and stability (in the sound channel) in the full geometry.}
{\label{StableCausal-11d}}
\end{figure}

This can be seen already in the simplest case of cubic Lovelock theory in higher dimensions. In figures \ref{StableCausal-9d} and \ref{StableCausal-11d} we show the relevant part of the causal and stable regions for $d = 9$ and $d = 11$, as determined by the near horizon and near boundary expansions of the effective potentials. There are also a few dashed lines plotted whose meaning we will explain in a moment. It is apparent that by going to higher dimensions, a disconnected region close to the apex of the stability region is now causal as well. This means that such a region satisfies all our previous constraints, and yet it will have an arbitrarily low shear viscosity. This is because, as we have seen earlier, the apex of the stability region actually lives on the surface $\eta=0$ determined by the stability constraint in the shear channel. This point lies inside the causality allowed region for $d \leq 10$ \cite{Camanho2010a}.

\begin{figure}
\centering
\includegraphics[width=0.47\textwidth]{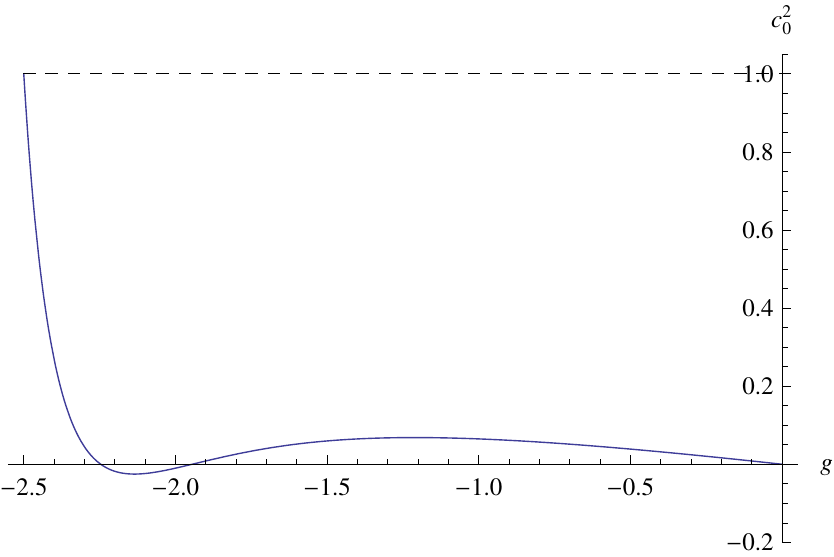}~~\includegraphics[width=0.47\textwidth]{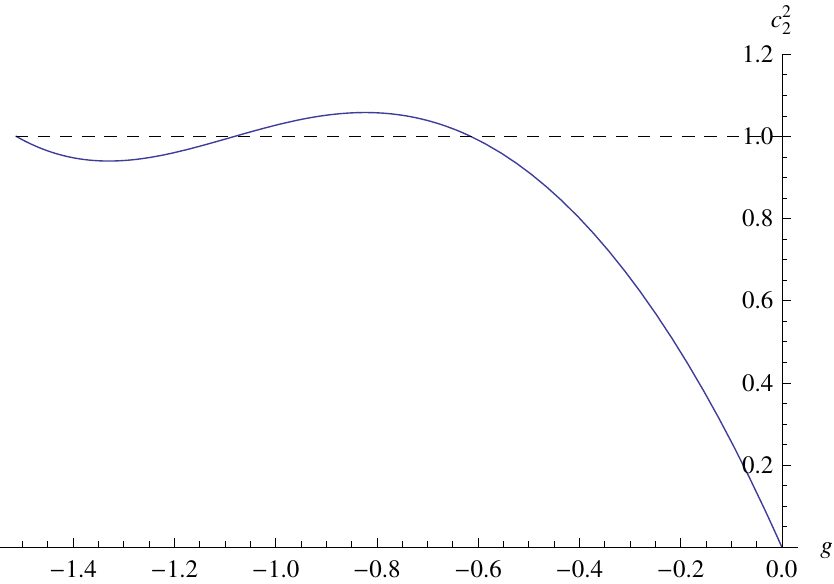}\caption{These figures illustrate the appearance of bulk instabilities (on the left) and bulk causality violation (on the right) for the case of cubic Lovelock theory in ten dimensions. They correspond, respectively, to the sound potential for $\lambda=0.33$ and $\mu=0.108$ and to the tensor potential for $\lambda=0.35$ and $\mu=0.25$. The plot of the potential ends at the boundary where the function $g$ takes the value of the cosmological constant, respectively $\Lambda=-2.5$ and $\Lambda=-1.512$.}
{\label{bulkinst}}
\end{figure}

While in principle there is nothing wrong with having a small ratio, this is not the end of the story. As we have mentioned before, by considering the full effective potentials we can see that they develop causality problems and/or instabilities, but now in the interior of the black hole geometry (see figures \ref{bulkinst}).\!\!\!\!
In the left figure, even though the potential initially rises close to the horizon ($g=0$), it dips into negative values. In the same way, in the figure on the right, the potential goes below one close to the boundary and then makes a hump above this value, leading to causality violation. Such features cannot be fully seen in a perturbative  analysis, although their presence is easily guessed at. Consider for instance moving along the curve where the causality bound is saturated in the tensor channel. This means that the effective potential is of the form
\be
V_{\rm eff} = 1 + 0 \times e^{x} + v_4\, e^{2x} + \ldots
\ee
It is clear then that along this curve we will see causality violation if $v_4$ ever becomes positive. Beyond the point in parameter space where this occurs, there will be a competition between the $e^x$ and $e^{2x}$ terms in the potential expansion which will determine the shape of the causality curve from that point onwards. In general it is clear that to determine this curve one must consider the full effective potential, since all terms in the near-boundary expansion become of the same order. Analogous statements hold for the instability analysis near the horizon. These curves are determined numerically and shown as dashed lines in figures \ref{StableCausal-9d} and \ref{StableCausal-11d}. We can see, in particular, that the sound mode instability curve neatly cuts off the region where $\eta/s$ might become too small, thereby imposing an effective lower bound on this ratio. 

Going to higher order Lovelock theories, one may wonder if this general reasoning will hold. Could the addition of further couplings allow us to escape even the full causality and stability bounds? This does not seem to be so. Evidence for this conclusion comes from an analysis of quartic Lovelock theory, to which we now turn. 

\subsection{Quartic Lovelock theory}
\label{love4}

Quartic theory corresponds to the introduction of an extra nonzero parameter, $c_4 \equiv \frac{\nu}{4}L^6$. The theory exists in $d\geq 9$, and examining it requires solving for quartic polynomials. The analysis of the parameter space of this theory is made in a somewhat similar manner to the cubic case, and we will not go into great detail here. There are resemblances and differences to the cubic theory. For instance, as in the lower order cases the boundary of the excluded region is just the surface where the cosmological constant associated with the EH branch becomes degenerate. However, this surface ends at one of the triply degenerated lines (three cosmological constants equal) and we need another surface to actually close the excluded region. This also happens in the cubic case where the excluded region is closed by $\mu=\lambda^2$, which is the line along which all three solutions become degenerate at some point inside the geometry.\!\!\!
\begin{figure}
\centering
\includegraphics[width=0.76\textwidth]{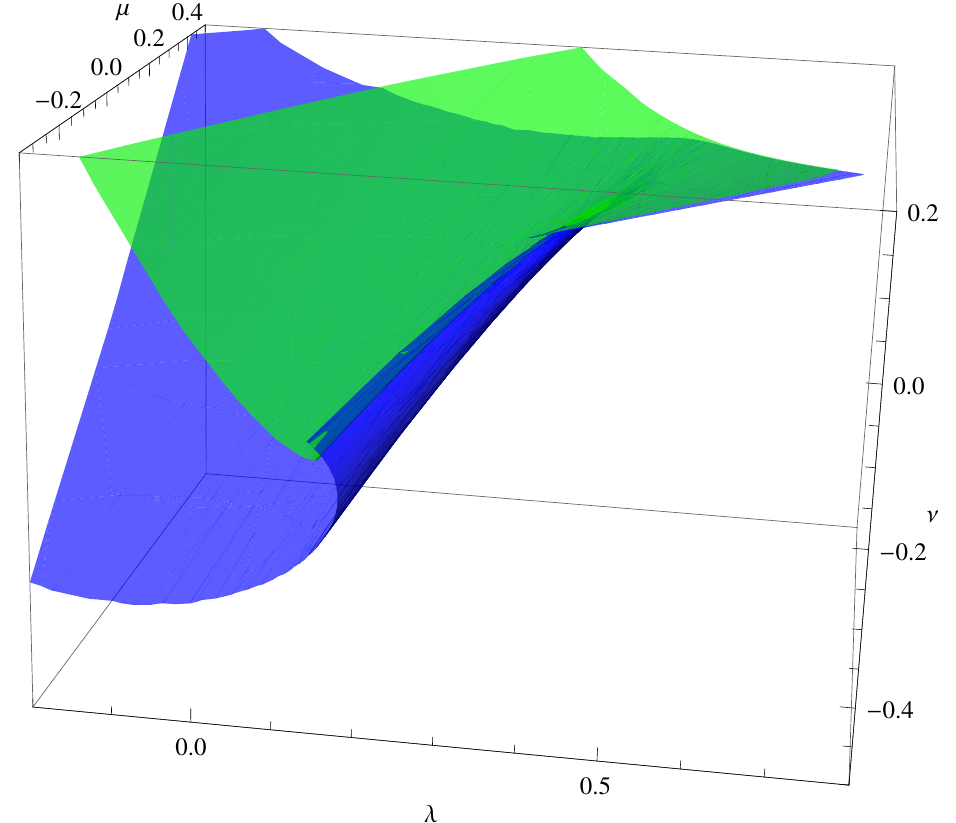}\caption{Surfaces delimiting the (boundary) causal region for the helicity zero and two modes in $d=9$ dimensions; respectively $t_2=-7/5$ (blue) and $t_2=7$ (green). We omit the excluded region boundary surface for the sake of clarity.}
{\label{LL4-causality}}
\end{figure}
In the quartic case the situation is analogous and the excluded region is closed by the surface swept by the line where three solutions degenerate at some value of the radius. This surface cuts off the other surfaces bounding the causal regions for the different helicities: green for tensor channel and dark blue for the sound. As always, the causal region is contained between the tensor and sound surfaces (figure \ref{LL4-causality}). Besides, we also realize that the values of $\lambda$ are not bounded (neither from below nor from above) just due to causality -- by increasing $\nu$ it is possible to go to arbitrarily high $\lambda$.

\begin{figure}
\centering
\includegraphics[width=0.76\textwidth]{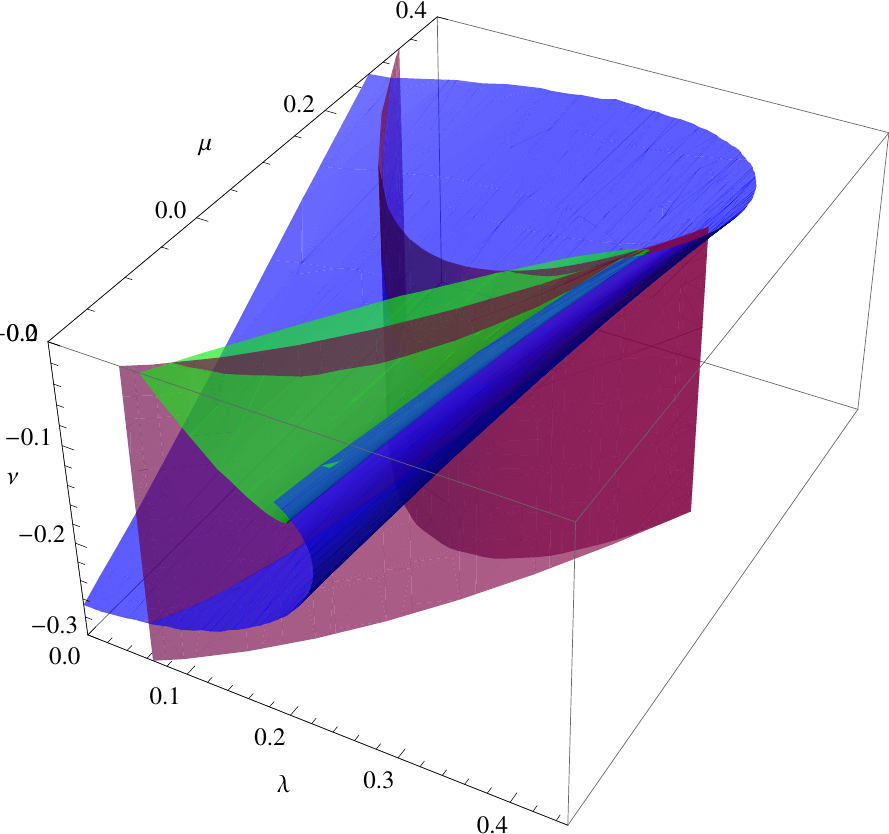}\caption{Stability close to the horizon (in purple) and (boundary) causality surfaces for $d=9$. It can be observed that there is a channel connecting the two disconnected regions found in cubic Lovelock theory (see figure \ref{StableCausal-9d}).}
{\label{LL4-causalstable}}
\end{figure}

\begin{figure}
\centering
\includegraphics[width=0.43\textwidth]{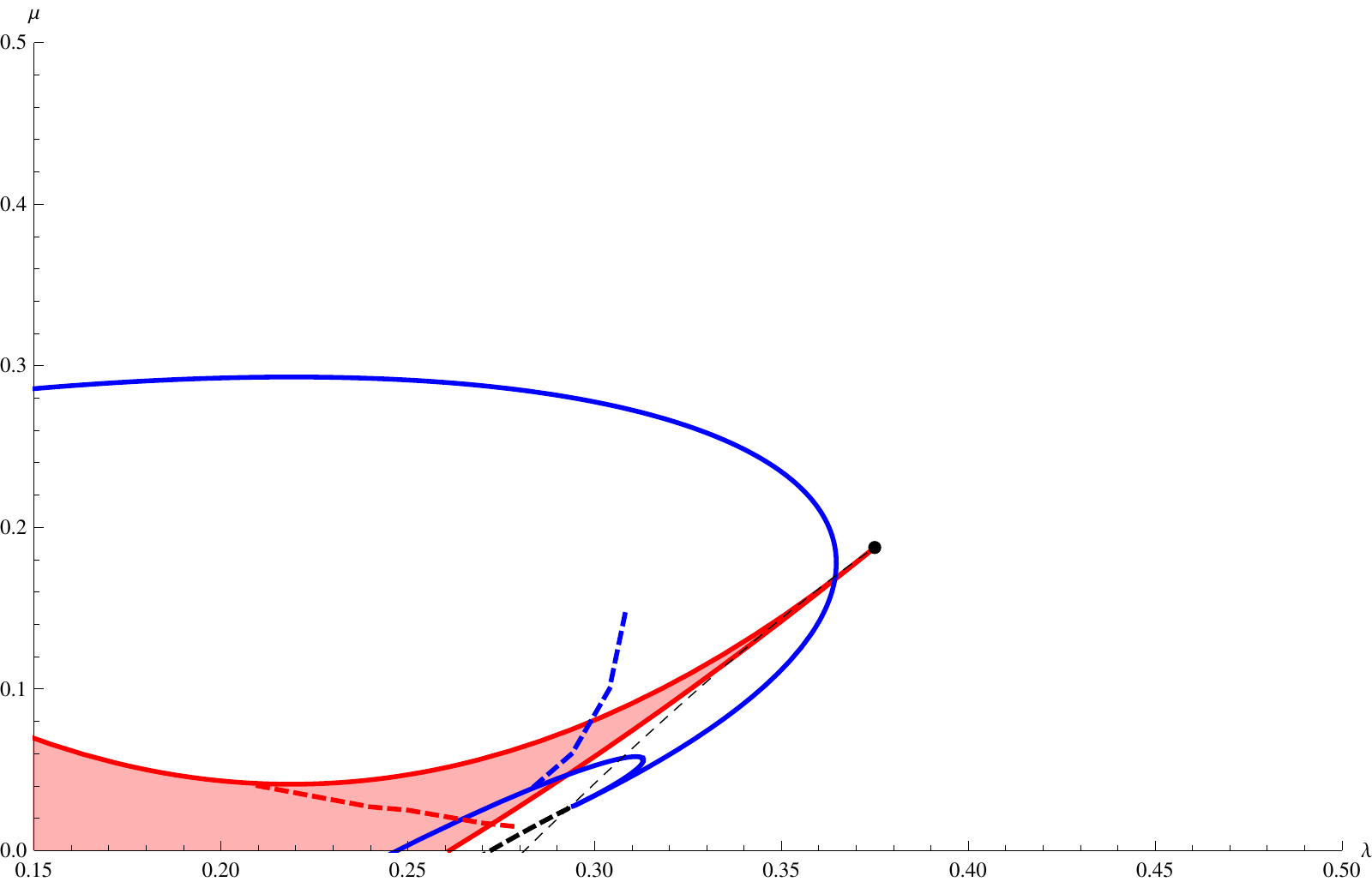} \qquad \includegraphics[width=0.43\textwidth]{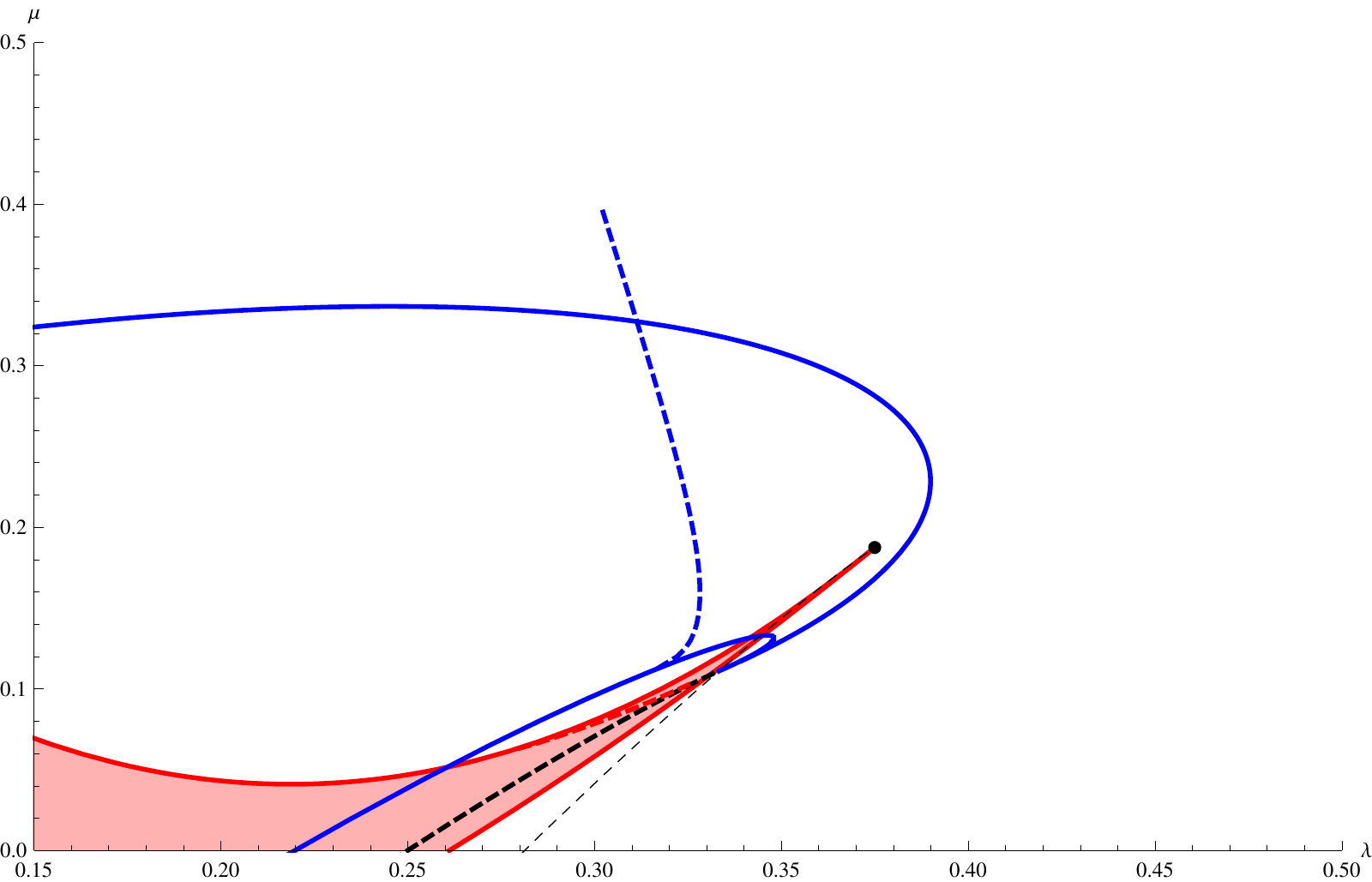} \\
\includegraphics[width=0.43\textwidth]{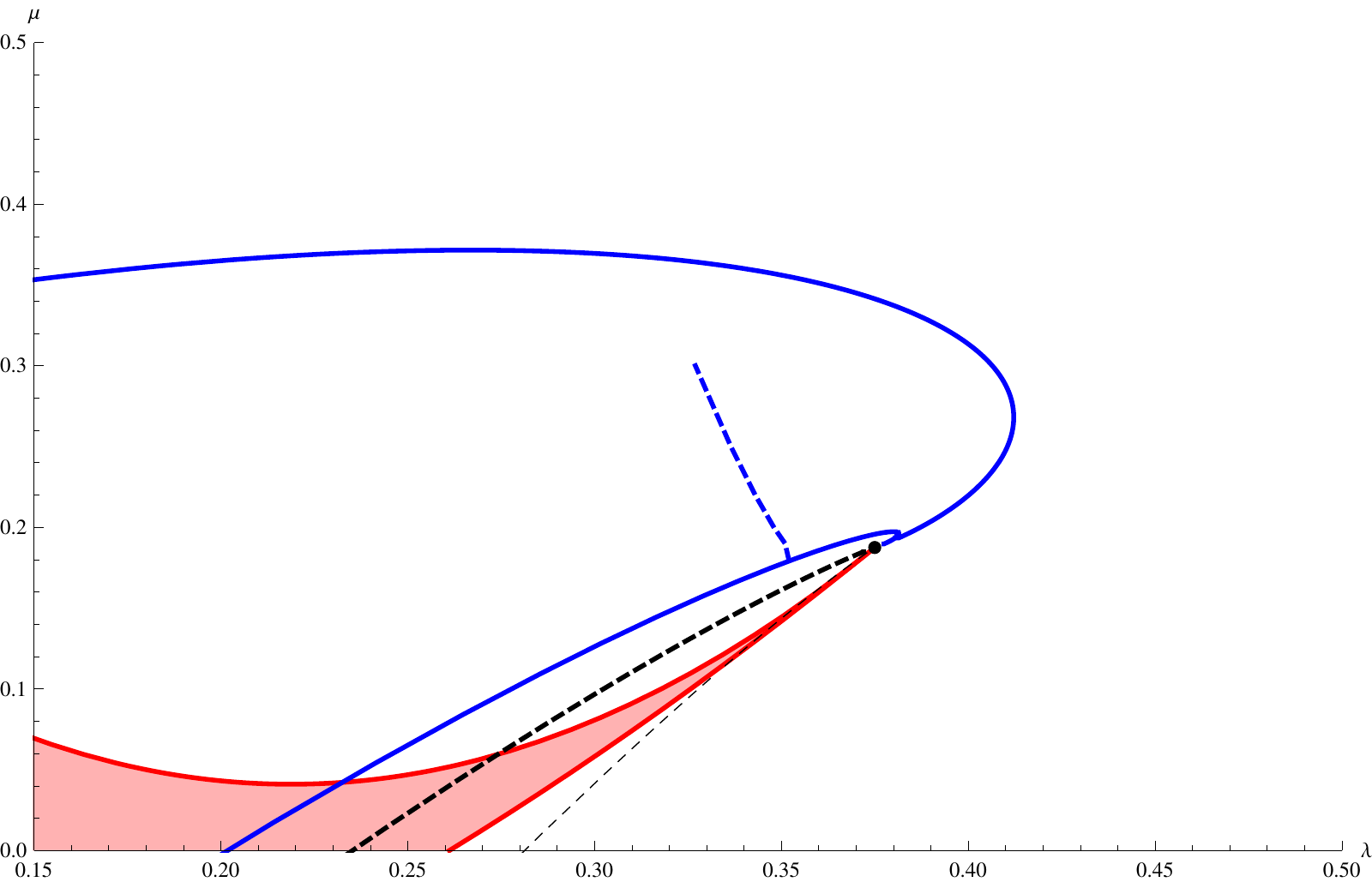}
\caption{Different slices for the quartic case in 9d. Top left corresponds to $\nu=-0.02$, top right $\nu=0$ and below $\nu=1/64$. The black thin dashed line is an imaginary line connecting the different end points of the causality constraints. The thick black line is the boundary of the excluded region. The dashed red and blue lines are the full sound stability and tensor causality curves respectively.}
{\label{slices9d}}
\end{figure}

This situation drastically changes once we consider stability and causality constraints (see figure \ref{LL4-causalstable}).\!\!\!
When horizon stability is imposed, the allowed region of parameters is further constrained by the purple surfaces; notice, in particular, that it includes a channel connecting the Einstein-Hilbert point with the $\eta/s=0$ line (the apex of the stability region, in purple). We are therefore in a situation analogous to the one in the cubic theory, and we expect the full stability and causality constraints to rescue us. This is indeed the case. In figure \ref{slices9d} we show the causality and stability constraints taken at different $\nu$ slicings. As $\nu$ increases, the causality curves move up in the plot until at some point they block the path to the critical point where $\eta/s=0$. For lower values of $\nu$, instead, we cannot reach this point due to the full sound stability constraint. 

We have actually done this analysis for various $\nu$ and have reached the same conclusions. These results add further evidence that for every dimension, stability and causality place an effective lower bound on the value of the shear viscosity to entropy ratio.

\subsection{Expansions at $\eta/s=0$}

We would now like to present a general argument that shows one should expect instabilities to appear generically as the $\eta/s=0$ point is approached. The argument is related to the fact that in the section 4, there is something which might invalidate the analysis. This is when the perturbative expansion on the radius, close to the horizon, breaks down. Notice that the tensor and sound mode effective potentials, $c_2^2$ and $c_0^2$ in \reef{potentialsF} have $(d-3)/(d-1)-F'(x)$ in the denominator. At the horizon we have,
\be
\frac{d-3}{d-1} - F'(x) \simeq 2 (\lambda_c-\lambda) ~.
\ee
This means that the near-horizon expansion breaks down whenever $\lambda\simeq \lambda_c$, precisely where $\eta/s\simeq 0$. Close enough to this point the previous approximation is no longer valid and we must treat this case separately. This is true for any Lovelock theory in any dimension, and it is therefore of interest to see if there are any general statements one can make. 

Let us then investigate the behavior of the effective potentials given in \reef{pots} at the horizon. First notice that the black hole polynomial equation (\ref{eqg}) fixes
\begin{equation}
g(x) \simeq \frac{x}{L^2} + \mathcal O(x^2) ~,
\end{equation}
and we can expand
\begin{equation}
\left( \frac{d-3}{d-1} - F'(x) \right) \approx \left( \frac{d-3}{d-1} - F'(0) \right) - \sum^{}_{n=1} \frac1{n!}\, F^{(n+1)}(0)\; x^{n} ~.
\label{denominator}
\end{equation}
The expansion is such that each term in the above is controlled by a different Lovelock parameter.
For instance $F'(0)=2\lambda$, and so the first term vanishes whenever $\lambda = \lambda_c$. When the following derivatives $F^{(j)}(0)$ vanish, $\forall j<n$, we can actually write
\begin{equation}
F^{(n)}(0) = \Upsilon^{(n+1)}[0] - \prod_{i=1}^{n} \frac{d-2i-1}{d-1} ~,
\label{denomina}
\end{equation}
and so we can set the following term to zero by choosing
\begin{equation}
(c_{n+1})_c = \frac{\Upsilon^{(n)}[0]}{n!} = \frac{1}{n!} \prod_{i=1}^{n-1} \frac{d-2i-1}{d-1} ~.
\end{equation}
In particular,
\begin{equation}
\mu_c =  \frac12 \frac{(d-3)(d-5)}{(d-1)^2} ~, \qquad \nu_c = \frac1{6} \frac{(d-3)(d-5)(d-7)}{(d-1)^3} ~, \qquad \ldots
\end{equation}

For a given (odd) dimensional Lovelock theory, if we set all $K=\frac{d-1}2$ couplings to these particular values, we will be at the so-called AdS Chern-Simons point \cite{Zanelli2005}. Now, notice that (\ref{denominator}) is actually the denominator of the helicity two and helicity zero potentials (and proportional to that of the helicity one mode). The leading correction to those potentials close to the horizon, when we set $F'(0)=2\lambda_c$ and $F^{(j<n)}(0)=0$, is then
\begin{eqnarray}
c_2^2(x) & \approx & - \frac{d-1}{d-4}\; \frac{n-1}{L^2\,\Lambda} + \mathcal O(x) ~, \\ [0.7em]
c_1^2(x) & \approx & - \frac{d-1}{d-3}\; \frac{F^{(n)}(0)}{(n-1)!\,L^2\,\Lambda}\; x^n\; (1 + \mathcal O(x)) ~, \label{horexp2} \\ [0.7em]
c_0^2(x) & \approx & \frac{d-1}{d-2} \; \frac{n-1}{L^2\,\Lambda} + \mathcal O(x) ~.
\end{eqnarray}
The effective potentials have therefore discontinuous limits at the horizon as one approaches $\lambda = \lambda_c$. Since $\Lambda<0$, the sound channel potential tends to a negative constant at the horizon as the critical value $\lambda_c$ is approached\footnote{A caveat to this conclusion is the case where one takes the full function $F'(x)-(d-3)/(d-1)$ to zero, what actually corresponds to the Chern-Simons point. However, it is easy to see from their definition that in this case one of the $c_2^2$ and $c_0^2$ potentials becomes negative.}. We conclude that {\it for any given Lovelock theory, it is impossible to get arbitrarily close to $\eta/s=0$ without running into an instability.} This means that, at least within Lovelock theories of gravity, it seems to be impossible to obtain arbitrarily small values for the $\eta/s$ ratio at any given dimensionality. 

These arguments say nothing about the value of the minimum itself. It seems clear that for any Lovelock theory there will be a minimum, but it is also reasonable to expect that for higher dimensionalities, the existence of extra free parameters would allow a lower value to be reached. For instance, looking naively into the horizon expansion of the effective potential
\be
V_{eff} =\alpha x+\beta x^2+\ldots
\ee
it seems that one could move towards higher values of $\lambda$ staying inside the stability region by simply picking parameters such that $\alpha$, $\beta$, etc. are always negative. In practice, however, we necessarily run into difficulties with this reasoning because of the breakdown of the perturbative expansion as we approach the critical value $\lambda_c$.\footnote{In other words, even though the first term in the horizon expansion might lead to a positive effective potential, the higher order terms might reverse this tendency, and they are becoming more and more important as $\lambda\to \lambda_c$.} In general, the precise interplay between these two issues must be determined by numerics, on a case by case basis.

\section{The $\eta/s$ ratio in higher order Lovelock theories}

\subsection{The minimum ratio in quartic theory}

\begin{figure}
\centering
\includegraphics[width=0.5\textwidth]{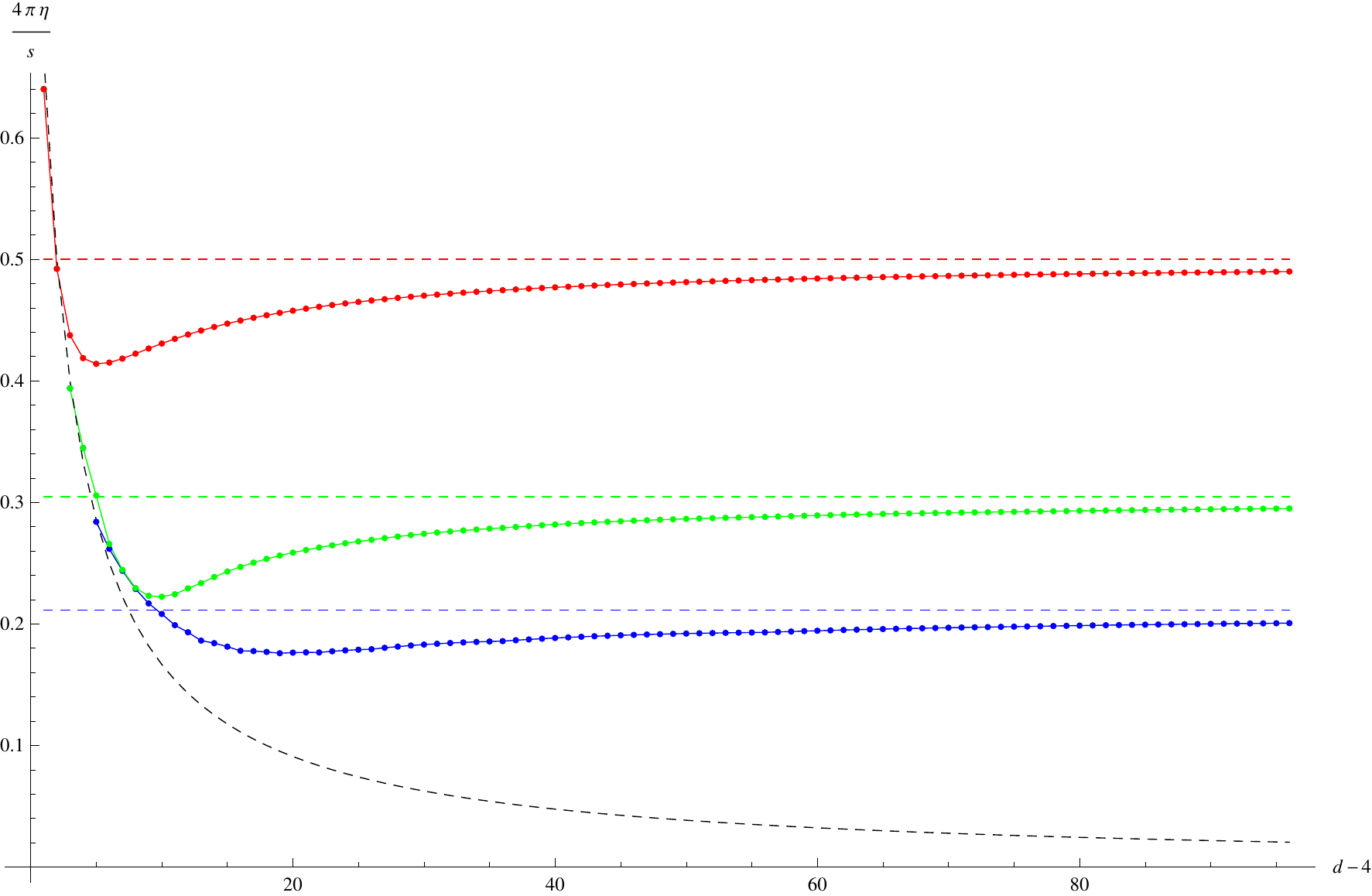}\qquad\includegraphics[width=0.43\textwidth]{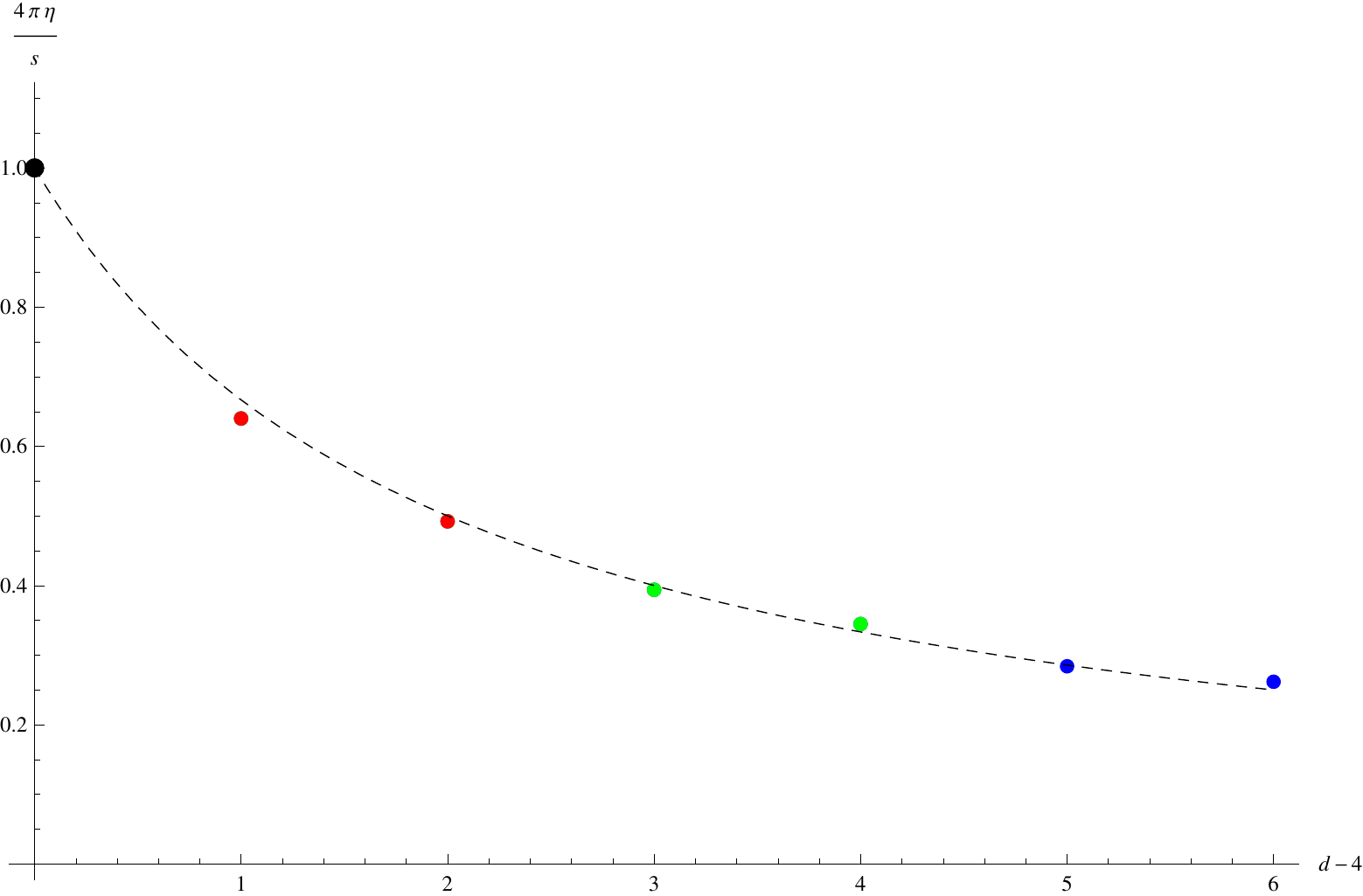}
\caption{ The dots correspond to the numerical (some analytic) bounds for LGB (red), cubic (green) and quartic (blue) Lovelock theories, and the coloured dashed lines correspond to the estimated asymptotic value on each case.  The dashed black line corresponds to the curve $4\pi\eta/s = 2/(d-2)$ that nicely fits (see the zoomed figure at the right) the first seven absolute bounds (including the unit value for $d=4$) and has a tantalizing behavior, $4\pi\eta/s \sim 2/d \sim 1/K_{\rm max}$, for $d \to \infty$, where $K_{\rm max}$ is the integer part of $(d-1)/2$.}
{\label{boundslambda}}
\end{figure}

In LGB gravity and cubic Lovelock theory, the parameter space is sufficiently small that it can be explored easily. In particular, one can find the full numerical causality and instability curves which determine the allowed regions in parameter space. As one goes to higher orders this analysis becomes more and more baroque, as the hypersurfaces in parameter space delimiting the causal and stable regions are more complicated and cannot be visualized in general. However, as was clear from section \ref{love4} we have succeeded in studying in detail quartic Lovelock theory in various dimensions. For each theory and dimensionality we have found the minimum value for the shear viscosity to entropy ratio, and plotted the results in figure \ref{boundslambda}.\!\!\!
The first thing we can notice is the remarkable smoothness either for each separated value of $K$ as well as extending from one value $K$ to $K+1$. There is nothing preventing the actual values to jump sharply given the discreteness in the number of couplings involved in these theories. Interestingly, we find that there is a minimum as a function of dimension for any given order. The critical dimension for these minima and the corresponding values of the minimum $\eta/s$ for each theory are quoted in table \ref{tab:a}. 
\begin{table}
	\centering
		\begin{tabular}{c|c|c}
			$K$ & $d_{min}$ & $(4\pi\eta/s)_{min}$ \\
			\hline
			\hline			
			2 & 9 & $219/529\approx 0.414$\\
			3 & 14 & $0.222$\\
			4 & 23 & $0.176$
		\end{tabular}
\caption{Critical dimension ($d_{min}$) for which the minimum value of the viscosity to entropy density ratio is attained for each order in Lovelock ($K$) theory, as well as the corresponding actual minimum value of the quantity $(4\pi\eta/s)_{min}$.}
\label{tab:a}
\end{table}
Overall, the minimum $\eta/s$ ratio seems to be decreasing rapidly as one increases the number of couplings.

It is interesting to comment on a very simple function that nicely fits the minimum value of $\eta/s$ for $d < 11$. It reads
\begin{equation}
\frac{\eta}{s} \simeq \frac{1}{4\pi}\,\frac{2}{d-2} ~.
\label{fitformula}
\end{equation}
It is the simplest curve that smoothly interpolates these points. It also has a nice asymptotic behavior, as we will discuss shortly. However, it is important to stress that this is not a (dimension dependent) bound for $\eta/s$. Indeed, this is already clear in the most relevant case given by $d=5$, where $2/3$ is actually (slightly) greater than $16/25$ (see the right zoomed figure \ref{boundslambda}). The expression in (\ref{fitformula}), though, approximately captures the dependence of a dimensional dependent novel bound for $\eta/s$ arising in Lovelock theories.

\subsection{No dimension-independent $\eta/s$ bound}

In higher order Lovelock theories, the parameter space becomes too large and we must resort to a more limited analysis. We have chosen to concentrate on a particular line through parameter space, parametrized by the single parameter $\Lambda$:
\begin{equation}
\Upsilon[g]=1+g+\lambda g^2 +\ldots=\left(1-\frac{g}{\Lambda}\right)\left(1-\frac{g}{\tilde{\Lambda}}\right)^{K-1} ~.
\label{maxdeg}
\end{equation}
All other parameters can be determined from $\Lambda$ ($\tilde \Lambda$ is fixed by the overall normalization). The curve has the property that it starts at the Einstein-Hilbert point and ends at the maximally degenerate point (MDP) of the theory, at $\Lambda=-K$ where the polynomial has a single degenerate root. A detailed analysis of the stability and causality properties of this curve is performed in appendix \reef{appendixB}. We have found that for generic values of $K$ and $d$ there is an interval of $\Lambda$ values where there is instability at the horizon\footnote{Except for $K<7$, where it disappears for high enough $d$. Whenever the unstable interval is not present the whole curve is stable.}. This is the range of values where our curve goes outside the stability wedge. Beyond this interval, our curve returns to the stability wedge, but nevertheless there are still instabilities -- no longer at the horizon but in the bulk. These instabilities persist all the way to the end of the curve, where it reaches the maximally degenerate point.

From these results it is clear that generically, the maximum stable and causal point one may reach along this curve is the first crossing point with the stability wedge. This crossing point can be found analytically, though the results are not particularly enlightening. However, there is a considerable simplification at very large dimensionality. To first approximation the crossing point is then determined by solving the algebraic equation,
\be
K^2 (3+2 \Lambda)+2 K \Lambda(4+3\Lambda)+\Lambda^2(6+6 \Lambda+\Lambda^2)=0 ~.
\ee
Solving this equation, and using \reef{lambdamu}, we can find the value of $\lambda$ at the crossing point. However, this value actually decreases as $K$ increases. This leads to a minimum $\eta/s$ ratio which increases as we include more couplings. Notice there is no contradiction here: recall we are moving along a particular curve in parameter space, which is simply not the optimal one in terms of finding the minimum possible value for this ratio.

Although the curve we have chosen is not optimal in this sense, it is still useful as we can use it as a base point for exploring nearby regions of parameter space. In particular, consider moving along the curve towards the maximally degenerate point. For high enough $d$, it is shown in Appendix \ref{appendixB} that close to this point there is only an instability in the sound channel, \ie all other channels are causal and stable\footnote{The other relevant constraint is actually the tensor causality one but the causality violating region can be made as small as one wishes just by increasing the dimensionality.}. We now perturb our curve by considering a small perturbation $\delta \lambda\, g^2$ or $\delta \mu\, g^3$ to \reef{maxdeg}.
The resulting potentials can then be easily found numerically.

\begin{figure}
\centering
\includegraphics[width=0.45\textwidth]{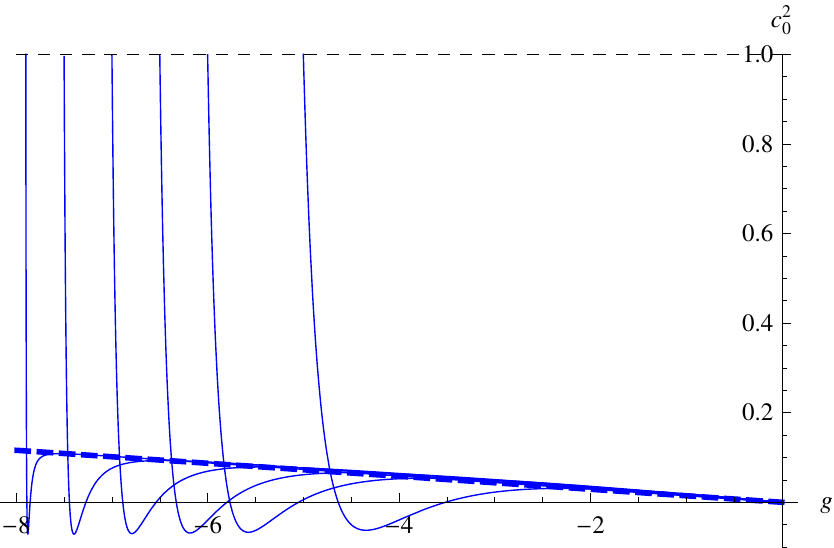}~~\includegraphics[width=0.45\textwidth]{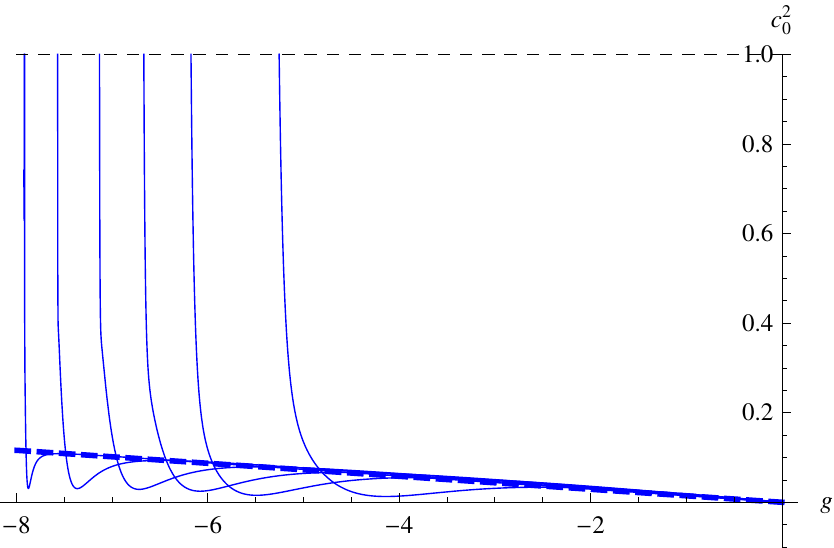}
\caption{On the left, the helicity zero potential for $K=8$, $d=100$ and $\Lambda=-8$ (dashed), $-7.9,-7.5,-7,-6.5,-6,-5$ along our curve. On the right the same curve slightly perturbed by a $\delta \lambda g^2$ term in the polynomial. The precise values of $\delta \lambda$ are $\delta \lambda=0, 1.8\cdot 10^{-18}, 6\cdot 10^{-13}, 1.9\cdot 10^{-10}, 6\cdot 10^{-9}, 8\cdot 10^{-8}, 3\cdot 10^{-6}$.}
{\label{stab0CSpointM}}
\end{figure}

In figure \ref{stab0CSpointM} we show the spin-0 effective potential for $K=8$, for various $\Lambda$ before and after adding a small fluctuation. The dashed curve represents the limiting curve for $K=8$ along our particular trajectory in parameter space ($\Lambda=-8$). On the left-hand figures, we see how although at the MDP itself the potential does not present any instabilities, it is unstable all the way up to that point along our curve, \ie there are negative potential wells in the bulk. On the right-hand side figure we see how adding just a small perturbation is capable of lifting the instability. This sensitiveness of the potential to fluctuations is due to the fact that the maximally degenerate point is special. At this point we have $\Upsilon^{(n)}(\Lambda)=0$, with $n=0,\ldots, K-1$, which means that the perturbative boundary expansion of the potentials breaks down. Accordingly there is a discontinuous limit at that point. Indeed, taking the $\Lambda\to -K$ limit along our original curve we find that the effective potentials become (assuming $d\neq 2K+1$):
\begin{eqnarray}
c_2^2 & = &-\frac{d-3K-1}{K^2(d-4)} g(x) ~, \nonumber\\[0.9em]
c_1^2 & = &-\frac{d-2K-1}{K^2(d-3)} g(x) ~, \\[0.9em]
c_0^2 & = &-\frac{d-K-1}{K^2(d-2)}  g(x) ~. \nonumber
\end{eqnarray}
Since $g(x)\in [\Lambda,0]$ the above are actually always positive for sufficiently large dimensionality\footnote{Remarkably the critical dimensionality, $d>3K+1$ is exactly the same found in chapter \ref{chp:bhstability}, required for arbitrarily small black holes to be stable.}. In particular, although the $c_0$ potential always has a negative region for  $\Lambda$ larger than but close to $-K$, it is actually tending discontinuously to a perfectly reasonable potential. This is clear from figure \ref{stab0CSpointM}.  

Our results then indicate that it is reasonable to expect that sufficiently close to the MDP there should be a trajectory in parameter space which is both stable and causal. We have shown this is certainly true up to $K=8$. If this result holds for higher $K$, then it is always possible to reach sufficiently close to the MDP and have well behaved effective potentials. As such, the theory seems to be well defined at that point.

At the MDP we have
\begin{equation}
\lambda_{\mbox{\tiny MDP}}=\frac{K-1}{2K}
\end{equation}
and therefore a value of the viscosity to entropy ratio for very large dimensionality becomes 
\begin{equation}
\left(\frac{\eta}{s}\right)_{\mbox{\tiny MDP}}=\frac{1}{4\pi}\frac{1}{K}~.
\end{equation}
By taking high enough $K$, this can be made arbitrarily small. Besides, since the maximum $K$, $K_{\rm max}$, is proportional to $d/2$ in the large $d$ limit, we obtain the asymptotic behavior already suggested by the formula (\ref{fitformula}) (see also figure \ref{boundslambda}).

To summarize, we have given evidence that it is possible to find a point in parameter space, sufficiently close to the MDP of a particular Lovelock theory, such that no stability or causality issues occur for high enough $d$. In particular we have checked that it is possible to do this up to $K=8$. We conjecture that this can always be achieved for any $K$, and we therefore come to the conclusion that {\em the viscosity to entropy ratio can be made arbitrarily small by considering a Lovelock theory of high enough order}. In other words, there is no bound for the $\eta/s$ ratio which is independent of the dimensionality, at least in the class of Lovelock theories of gravity. This goes in line with our expectations that adding more and more couplings, or free parameters in the lagrangian, it should be possible to reach lower and lower values for the shear viscosity to entropy ratio. 

It is interesting to mention at this point that the link between causality violation and the viscosity bound does not apply for thermal theories undergoing a low temperature phase transition. In \cite{Buchel2010b} Buchel and Cremonini analyze this issue and made an important remark. The shear viscosity is one of the coupling coefficients of the effective hydrodynamic description of the theory. As such it applies to the regime of lowest momenta and frequency, \ie $\omega,\|{\bf k}\|<< T$,
where the r\^ole of the temperature may also be played by by any other microscopic scale of the plasma -- actually the smallest one. On exactly the opposite regime, microcausality is determined by the propagation of the high energy modes
regime, \ie, for $\omega,\|{\bf k}\| >> T$. A link between both kinds of features is only possible if the same phase of the theory extends over the entire range of the energy scales. In other words, there must not be any phase transitions in the plasma, as it happens for dual plasma of LGB gravity. As it is conformal just zero temperature phase transitions are allowed and we are always in the deconfined high temperature phase. The only free parameter of the model is the LGB coupling, which determines both the shear viscosity ratio and its causality properties. Hence the link between the two is in a sense a mere coincidence. Any constraint on $\lambda$ will translate into an analogous restriction on $\eta/s$ without any necessary deeper connection. 

In the holographic setting UV and IR properties of the field theory correspond to geometrically distinct parts in AdS, the asymptotic and near-horizon regions respectively. In this sense it is clear that causality, found at the boundary,  corresponds to the UV whereas stability, mostly related to the horizon, is a IR property. We have actually verified that causality plays a secondary r\^ole on the discussion of the possible bounds on $\eta/s$. Stability alone already prevents this ratio from becoming arbitrarily small at any fixed dimensionality, at least in the context of Lovelock gravities.  

In the specific model of \cite{Buchel2010b} the LGB term is sourced by the VEV of some irrelevant operator that condensates in the IR below some critical temperature associated with the spontaneous breaking of some global U(1) symmetry. Geometrically, the broken phase is characterized by a scalar field with a non-trivial profile,  vanishing at infinity but finite at the horizon, the effective LGB coupling being proportional to that field. This effective coupling thus affects the value of the shear viscosity without changing the causality properties of the CFT. In a sense, we are lifting the causality restrictions but the stability ones are expected to still hold. Stability still ensures the existence of a bound on $\eta/s$, a bound that reduces its value as we increase the dimensionality and the order of the Lovelock theory accordingly. 

\section{Discussion}

We have devoted this chapter to scrutinize the finite temperature phase of the CFT in the context of Lovelock theories. In particular we have considered in detail the restrictions imposed by causality and stability on the shear viscosity to entropy ratio of the dual plasma. 

An important part of our analysis was the study of these effective potentials, searching for causality violation or instabilities in the large momentum limit which might rule out regions of the parameter space. Notice, however, that while the former condition is a {\it bona-fide} constraint on the theory, since it would lead to causality, the second seems to hold only for the validity of the black brane solution, and is not a fundamental restriction. We have found that causality violation can occur deep inside the bulk, and therefore this effect cannot be captured by a boundary expansion. This means that, contrary to previous expectations, causality violation is not necessarily a UV feature of the CFTs related to positivity of energy restrictions\footnote{\S\ Another example of this can be found in \cite{Camanho2013c}. In that example we again recover the positivity constraints from the asymptotic limit of a more general set.}. Holographically this means that there are interesting constraints on the parameters of the CFT which cannot be seen in perturbation theory. The field-theoretic origin of such constraints remains a mystery.\footnote{It would be of definite interest to study whether these constraints are related or not to unitarity restrictions that were very recently shown to arise in the computation of $2$-point functions of the stress-energy tensor at finite 
temperature and large energy and momenta \cite{Kulaxizi2011}.} On the gravity side, these restrictions were essential to prevent an arbitrarily small $\eta/s$ ratio for a fixed theory, as seen for the concrete examples of cubic and quartic Lovelock theories.

For higher order theories, we have resorted to the study of a particular curve through parameter space. We chose this curve on the basis of simplicity and the capability of analytic treatment. The upshot of our analysis is that it seems likely that there is some curve leading arbitrarily close to the maximally degenerate point of the theory. The reader might be suspicious about this -- after all, at this point the central charge of the dual conformal field theory is vanishing, and so is the kinetic term of the graviton fluctuations around the AdS black hole. That is why we emphasize that one is really considering a curve leading to this point, but causality constraints keep us at a safe distance where everything is well defined. 

In all cases we considered in detail, the minimum value for the ratio was obtained well away from the maximally degenerate point. In any case, we hope our work stimulates study of the properties of the theory around this particular point. In the case $d = 2 K + 1$, these theories display symmetry enhancement becoming gauge theories of the Chern-Simons group: they have local AdS symmetry whereas all other theories with $d > 2 K + 1$ have local Lorenz invariance \cite{Zanelli2005}. Contrary to what is suggested by the fact that $C_T$ vanishes in the dual conformal field theory, there are some hints pointing towards the existence of interesting physics on these theories. For instance, the thermodynamics of their black holes displays a qualitative difference to that of generic AdS Lovelock black holes: the temperature grows linearly with the horizon radius, and the specific heat is a continuous, monotonically increasing and positive function of $r_+$ \cite{Crisostomo2000}. Thus, Chern-Simons black holes can reach thermal equilibrium with a heat bath at any temperature and they are stable under thermal fluctuations. There is also a mass gap between the massless black hole and AdS spacetime, as it happens in the case of $d=3$ \cite{Banados1992a}.

Our results indicate that, in the case of very large $d$, it should be possible to reach a value of the $\eta/s$ ratio of {\em at least}
\be
\frac{\eta}s=\frac{1}{4\pi} \frac{1}{K}
\ee
in the $K$th order theory. Intuitively, this is simply telling us that the more parameters we have, the lower the ratio can become. This is reminiscent of the species counter-argument to a minimum bound on $\eta/s$ \cite{Cohen2007a,Cherman2008}. Of course, in Lovelock gravity one needs to increase the number of dimensions in order to have more parameters. Then it would seem that for all practical purposes there is a bound on $\eta/s$ for any finite dimensionality. This is certainly a possibility, and it is definitely true for Lovelock theory -- a combination of stability and causality ensures it -- even if we disregard the causality constraints {\it \`a la} Buchel and Cremonini \cite{Buchel2010b}. However, even for fixed dimensionality it seems that it might be possible to have other interesting higher curvature corrections with finite coefficients \cite{Oliva2010,Oliva2010a,Myers2010c,Oliva2011}. These quasi-topological gravities have exact black hole solutions of a similar nature to those of Lovelock theory, including their thermodynamic features. However, hydrodynamic as well as causality and stability properties, are quite different \cite{Myers2010d}. While there is much work to be done understanding these theories, it seems plausible that the addition of high order quasi-topological terms would lead to a smaller $\eta/s$ ratio. 

The stability wedge in all Lovelock theories has a non-trivial shape in the ($\lambda$,$\mu$)-plane but is otherwise a cylinder in the remaining couplings. This means, in particular, that there is a lower bound for the cubic Lovelock parameter $\mu$. It results from instabilities of the black brane solution and, as such, should be related to properties of the CFT plasma rather than of the zero temperature CFT. The precise meaning of this bound in the field theory side and its consequences in the computation of physical quantities such as transport coefficients or thermodynamic variables, as well as the absence of a higher bound for this parameter, are interesting avenues for further research.

\begin{figure}
\centering
\includegraphics[width=0.67\textwidth]{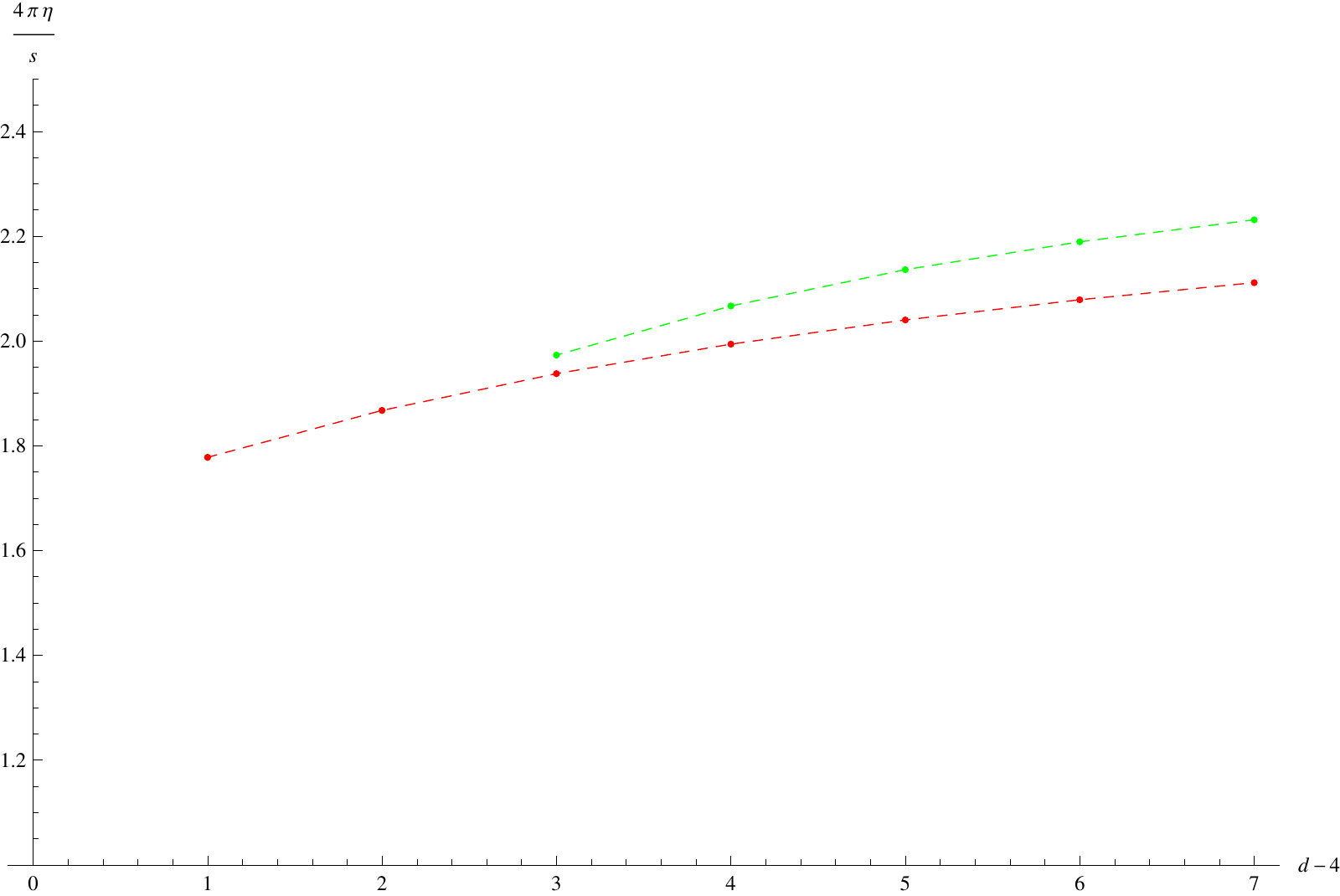}
\caption{ The dots correspond to the numerical (some analytic) $\eta/s$ `upper bounds' for $K=2,3,4$ (red, green and blue respectively).}
{\label{boundslambda-low}}
\end{figure}

As discussed earlier in the case of LGB \cite{Camanho2010} and cubic Lovelock theory \cite{Camanho2010a}, it is tempting to say something about the existence of a lower bound on $\lambda$ (see figure \ref{boundslambda-low}), that implies an upper bound for $\eta/s$. It is due to the causality constraint in the former and the stability constraint on the latter. This seems to be in line with the expectation that the shear viscosity of a strongly coupled system cannot be too large. On this respect, it is interesting to mention that this bound disappears in Lovelock theories of quartic or higher order -- {\it i.e.}, for $d \geq 9$ (see figure \ref{boundslambda-low}) --, since no expected problems as causality violation or instabilities of the sort discussed in this thesis arise in that direction. This might be hinting towards the existence of pathologies of a different nature that we have been unable to characterize.

To conclude, we hope our work stimulates further research into holographic studies of the $\eta/s$ ratio in particular, and the dynamics of Lovelock theories of gravity and holography in general. It might be useful to explore further the effective potentials. These relatively simple functions seem to encode a great deal of information about the hypothetical dual CFT, information which we are only beginning to extract.



\chapter{\bfseries \itshape Summary and conclusions}
\chaptermark{Summary and conclusions}
\label{summary}

\vspace{.6cm}

\begin{quotation}
\flushright
{\it ``Who is more humble? The scientist who looks\\
at the universe with an open mind\\ and accepts whatever the universe has to teach us,\\
or somebody who says everything in this book\\
must be considered the literal truth\\ 
and never mind the fallibility of all the human beings involved?''}\\

\vspace{.3cm}

Carl Sagan
\end{quotation}

\vspace{3cm}

\noindent Lovelock theory is the natural extension of general relativity to higher dimensions. It may also be thought of as a toy model for ghost-free higher curvature gravity. These gravity theories capture some of the defining features of higher curvature gravities, namely the existence of more than one (A)dS vacuum and intricate dynamics, more general black hole solutions and instabilities; while avoiding some of their problems. In particular, Lovelock gravities yield second order field equations so that they can be considered beyond the perturbative regime and are free of higher derivative ghosts. This provides an appealing arena to explore different gravitational and holographic aspects of higher curvature gravity.

Most of the vacua of the theory support black holes that display interesting features. Besides, black holes with maximally symmetric horizons are subject to a version of Birkhoff's theorem and these solutions can be found analytically. Most efforts in the literature have been devoted however to one particular branch of solutions, often restricted to a specific combination of the Lovelock couplings. The branch usually chosen for the analysis is the so-called EH branch, as it actually reduces to the general relativistic solution as we turn off the higher order couplings. In this thesis we have presented some tools that allow for the description of Lovelock black holes for arbitrary values of the whole set of couplings, dimensionality and order of the theory. Despite the fact of the solution being implicit, it is possible to extract most of the relevant information and discuss all possible cases in the general situation: analyze the number of horizons, the thermodynamic stability of the solution, phase transitions, etc. Furthermore, this approach has been generalized to the case of charged and cosmological solutions, and also to the so-called {\it quasi-topological} gravities. These theories share the same functional form of the black hole solutions with the Lovelock family while being lower dimensional.

Our method is very useful to gain intuition about physical processes involving black holes. One can easily visualize the evolution of the position and number of horizons as the mass of the solution varies, this providing crucial information about, for instance, the possible appearance of naked singularities or the violation of the third law of thermodynamics. We have seen that the rigid symmetry imposed on the solution naively allows such problematic behavior which is avoided once the stability of the solution is taken into consideration. 

We have analyzed the propagation of perturbations through black hole and naked singularity spacetimes in the limit of high momentum and frequency. This restricted regime has the advantage of being very simple while still powerful enough to test the stability of the solutions of interest. Besides, the instability seems to appear first at high energy. We have probed in this way the solutions and found that an instability always shows up as we approach the spaces with naked singularities, \ie  the solution becomes unstable before the horizon of the black hole disappears, its fate being still unclear. The naked singularity solution is also unstable in this very same way and it cannot be formed by collapse either. This also happens for the cases for which one naively expected the appearance of an (uncharged) extremal black holes and sudden {\it jumps} on the position of the horizon, thus effectively preserving the third law of thermodynamics in Lovelock gravities, at least for this class of solutions. 

The same instabilities appear also in the evaporation process of spherical black holes, for a wide range of situations. In fact, as the mass of the black hole is reduced, the solutions can just avoid becoming unstable when they display a zero radius singularity (as opposed to a branch singularity) in dimensions greater or equal than $3K+1$. Einstein-Hilbert seems to be special in this respect as it is the only Lovelock theory that is stable for every dimension in which it can be defined. The 3-dimensional case is special as Einstein-Hilbert gravity becomes  Chern-Simons theory for the AdS group, being topological. The global picture that arises from this analysis seems to point to these instabilities as having some crucial r\^ole in the dynamics of the theory. Rather than being a pathology of the theory as it has sometimes been referred to, instabilities avoid the appearance of truly pathological behavior. Moreover, they generically impose an upper bound on the {\it effective density}, $M/R^{d-1}$, of any object, bound that depends on the specific branch of the theory we are considering. This seems to naively forbid the existence of point-like entities in most cases.

We have verified that Hawking-Page phase transitions generally occur in Lovelock theories. The key difference with the general relativistic case is the possibility of more than one thermodynamically stable black hole for some ranges of the temperature in such a way that the transition may also happen between two black holes of different mass. By analogy with the Einstein-Hilbert case, the transition from a thermal vacuum to a black hole may be identified as a confinement/deconfinement phase transition of the dual CFT plasma, thus transitions between different black holes should correspond to distinct phases of the field theory deconfined plasma.   

Another seemingly pathological feature of Lovelock black holes is that they may have negative values for the entropy. Nonetheless, this problematic states are avoided if we restrict ourselves to the globally preferred phase at every temperature. For that, in some cases, \eg hyperbolic black holes, it has been crucial to consider in our analysis the possibility of extremal configurations with arbitrary temperature and zero entropy, in accordance with the semi-classical approach. This is important as we need to have a well-defined {\it groundstate} with minimal mass for any temperature in such a way that negative entropy states always have higher free energy and are thus unstable. This might also shed some light on the r\^ole of the extremal configurations and to whether they should be considered with arbitrary temperature or just as limiting cases of their non-extremal counterparts. 

Lovelock gravities display a family of vacua and corresponding black hole solutions that point towards the possibility of an intricate phase diagram. In order to study the phase diagram of the full theory, not of a particular branch as discussed previously, we then need to be able to describe phase transitions between different branches of the theory. This could also be envisaged as a possible endpoint of the instabilities mentioned above. Indeed, this was one of the original motivations for our analysis of these transitions. We have seen, though, that instabilities and   phase transitions are correlated just in very particular situations. 

In order to describe transitions among the different branches of the theory, we have broaden the class of solutions to be considered. Keeping the same symmetry requirements, we allow for solutions that, being continuous, have discontinuous first derivative of the metric (equivalently discontinuous spin connection). The existence of these solutions is due to the multivaluedness of the momenta in Lovelock gravities. At any point, the {\it velocity} may jump as long as the momentum is continuous, this continuity of the canonical momenta across the resulting junction being due to the absence of matter on that surface. Nonetheless, this does not mean that the mass parameters of the inner, $M_-$ and outer, $M_+$, solutions are the same, the bubble effectively carrying energy $M_+-M_-$. For static configurations, either stable or unstable, in addition to the mass we may also assign these {\it thermalons} with temperature and entropy. We have shown that the contribution of the boundary terms on the {\it bubble} is such that the free energy of the whole spacetime takes the expected form $F=M_+-T_+S$. The mass and temperature, $M_+$ and $T_+$, correspond to the outer solution while the entropy $S$ remains untouched as coming from the inner black hole horizon. We have also checked that the usual relations between the thermodynamic variables and the first law of thermodynamics are verified, thus effectively proving the consistency of the thermodynamic picture for these generalized configurations. 

The only necessary assumption for all this to work is the existence of a horizon for the inner region of the bubble. As simple as it may seem, this imposes severe restrictions on the kind of configurations that can be considered. For instance,  the inner solution for planar topology has to necessarily correspond to the EH branch. This in particular implies that the thermodynamics of the EH branch (identified by its asymptotics) is not changed by the existence of these new solutions and that any branch transition (under some assumptions) will necessarily lead to this branch. Actually the {\it thermalon} is always the thermodynamically preferred phase and thus a bubble necessarily pops up hosting a black hole in its interior. In some cases however, depending on the choice of asymptotics, the thermalon with a regular EH black hole inside does not exist and the transition cannot happen. The non-planar cases are more involved and generically display one (or several) finite temperature phase transitions. We have analyzed the simplest case of LGB gravity in great detail. The hyperbolic case is special as it also allows for thermalons with EH asymptotics, but just above some critical value of the LGB coupling. 

The junction conditions used in order to find the static bubble configurations also determine the dynamics of the bubble, namely the stability of its equilibrium position. In the LGB case, the bubble, being unstable, will eventually expand reaching the boundary in finite time, thus changing the asymptotics and charges of the solution. This is reminiscent of what happens in the case of {\it quenches} where the energy and temperature of the system change as a result of the work done by a sudden variation of the boundary conditions. The thermalon mechanism may be thought of as a thermodynamically induced {\it quench} as opposed to an external action over the system. In the case of LGB gravity with {\it ghosty} boundary conditions, in the spherical case the thermal vacuum is the globally stable phase at low temperature whereas there is a critical temperature for which the free energy becomes negative indicating the phase transition. Still, for positive values of the free energy, the bubble solution exists and may be formed with some finite probability, the {\it ghosty} vacuum is a metastable state. In any case, the system is necessarily driven to the EH branch. 

The most interesting situation is that of hyperbolic topology, where we have the direct transition (to the EH branch) with {\it ghosty} asymptotics, but also the reverse one with EH asymptotics. The situation for the direct transition is very similar to the spherical and planar cases the system necessarily transitioning to EH asymptotics. 
On the contrary, in the reverse case with the well behaved asymptotics, the system is stable for moderate values of the LGB as there is no thermalon configuration to mediate the transition. The spacetime is thus necessarily driven to the EH branch and it remains there. This is then both a natural mechanism to select the EH vacuum and to also avoid the instability of the {\it ghosty} branch. For higher values of $\lambda$, however, static bubbles exist and, even though the black hole is the globally preferred phase, they may form with some probability. The result is a system in which bubbles of both phases form in a chaotic way  at any temperature. In this case the inclusion of the extremal states at any temperature seems to be crucial in order to have a well-defined ensemble. 

In the context of the AdS/CFT duality black holes correspond to deconfined phases of the plasma whereas the confined phase is the thermal vacuum. The generalized Hawking-Page phase  transitions mentioned in the previous paragraph seem to correspond to confinement/deconfinement transitions involving a further change in the microscopic properties of the plasma -- these are associated with the asymptotics of the solution.

In the case of more general Lovelock theories the situation may change quite a bit. For instance, the equilibrium point for some thermalons may be stable instead of unstable, in such a way that the resulting phase is more similar to a regular black hole than to the thermalons of LGB gravity. The system may stay in that phase if there is no other thermalon to take it somewhere else. From the holographic point of view these stable thermalons may provide some higher curvature branches with a deconfined phase.  

Lovelock theories have the remarkable feature that lots of physically relevant information is encoded in the characteristic polynomial $\Upsilon[g]$. Boulware-Deser-like instabilities, for instance, can be simply written as $\Upsilon'[\Lambda] < 0$, which has a beautiful CFT counterpart telling us that the central charge, $C_T$, has to be positive. We have computed the central charge $C_T$, making the above statement concrete, as well as the two extra parameters, $t_2$ and $t_4$, involved in 3-point functions of the stress-energy tensor. $t_4$ is actually vanishing for the particular case of Lovelock theories. This might be related to some kind of supersymmetry constraint on the dual CFT, although the supersymmetric version of general Lovelock theories is unknown. We have also discussed the possible terms that may contribute to the computation of $t_2$ and $t_4$ in the context of general higher curvature gravity. Restricting to terms yielding 2-derivative equations of motion in the shock wave background, there is a one-to-one correspondence between the possible structures appearing in the energy flux 1-point function and the contributions from higher curvature terms. This is also consistent with the number of field redefinition invariant combinations that can be written in the action. 

Another very interesting correspondence that has been observed using Lovelock theory is the one that relates holographic causality in the bulk of AdS with positivity of the energy flux expectation values. Both physical requirements are very different in nature, yet they imply the same set of constraints on $t_2$ (for $t_4=0$). Lacking a definitive proof of the positivity of energy fluxes, this may also be considered as indirect evidence in the restricted context of Lovelock holography. Moreover, this has led to the discovery of these conditions being required by unitarity of the corresponding CFT \cite{Kulaxizi2011}. Furthermore, even though the connection was first found for the case of black hole backgrounds, this is not a feature of thermal CFTs. We have in fact performed a similar computation using a shock wave background and found a complete agreement with the previous sets of constraints. This second computation is also simpler in the sense that it makes clear that the problem arises from the 3-graviton vertex. Notice that, contrary to the instabilities found before, causality violation is a {\it bona fide} pathology and it should be discarded as unphysical. The instabilities, in turn, might just be pointing to the existence of new solutions or to other issues such as not being expanding about the true vacuum. In the case of Lovelock gravity stability is just necessary for the validity of the black hole solution and is not a fundamental restriction. 

The above physical constraints restrict the possible values of the Lovelock couplings that describe a causal CFT with positive energy fluxes. Notice that this connection could have never been discovered in the perturbative regime as we need $t_2$ or $t_4$ of order one for the theory to ever violate causality. Besides, in the supergravity approximation, $t_2\sim \alpha'$. All this, and the fact that all formul\ae\ seem to extend meaningfully and smoothly to arbitrary dimension and order of Lovelock gravity, seem to entail the applicability of the AdS/CFT correspondence beyond the context of critical string theory. Although the simplest Lovelock invariant has appeared in the context of string realizations of the duality, higher order Lovelock terms do not show up in that context, nor higher curvature corrections with finite coefficients. Do strongly coupled CFTs necessarily have equal central changes in four dimensions? The answer to this question does not seem to be easy and we have not found any hint pointing in any clear direction. 

When dealing with higher order Lovelock gravities, contrary to previous expectations we have also found causality violation far from the boundary of the spacetime, thus showing that this particular pathology is not necessarily linked to the UV properties of the field theory, neither to positivity of the energy flux expectation values alone. It would be interesting to study if this effect could be traced back to positivity of energy flux higher point functions.

Finally, we have analyzed the constraints imposed by causality/positivity to other relevant quantities of the CFT, namely transport coefficients such as the shear viscosity to entropy density ratio. This ratio gets corrected from the universal KSS value $\eta/s=1/4\pi$ by the LGB coupling in such a way that any positive value of $\lambda$ would entail a violation of the celebrated bound. Besides, an upper bound on $\lambda$ automatically translates into a lower bound in $\eta/s$. Even though higher order Lovelock terms do not appear explicitly in the corrected expression for $\eta/s$, they actually affect the value of the  bound. 
They add some freedom that consequently allows $\lambda$ to take higher values, thus lowering the bound. We have discussed the existence of a bound in that context and realized that causality alone cannot prevent the ratio to take arbitrarily low values, even negative in some cases. Besides, one can construct effective holographic models where the LGB term is sourced by a VEV of another field that condenses in the IR, \ie it has a finite value just close to the horizon being zero asymptotically. In this way, the addition of the LGB correction does not affect the asymptotic causal properties of the theory while changing the value of the shear viscosity. In a sense we can lift in this manner the causality restriction on $\eta/s$.

  The restrictions that avoid unphysical values of the shear viscosity arise once again from stability. After all, negative shear viscosity also corresponds to an instability of the system. Using the stability analysis in the high momentum regime, we have shown that a nonzero bound on the shear viscosity to entropy density actually exists for any dimensionality $d$, the minimum value of $\eta/s$ approaching zero as we go to higher and higher dimensions. Our results indicate that for very large $d$ it should be possible to reach a value of $4\pi\eta/s$ of at least $1/K$. Up to ten dimensions (\ie quartic order) it can be shown that the lowest bound for each dimensions roughly follows a curve given by $4\pi\eta/s\approx 1/(d-2)$.

Along this thesis I have tried to present Lovelock theories as an interesting playground for testing our ideas about gravity and the holographic duality. This provided interesting information about viable modifications of Einstein-Hilbert gravity and, by contrast, also shed some light on general relativity in  four or higher dimensions. Lovelock gravities have many interesting features, a rich dynamics and effects that cannot be observed in that simpler case. This would also allow us to look at Einstein-Hilbert gravity under new light, putting it in a much broader context and maybe understanding what is so special about it. Our investigations led to deep connections between seemingly unrelated concepts. These unsuspected insights have the most celebrated instance in the exact equivalence between causality in AdS and positivity of the energy in the CFT, but we have uncovered more subtle examples. Instabilities play a central r\^ole in this context being related to the preservation of the cosmic censorship hypothesis and the third law of black hole mechanics in Lovelock gravity. Stability seems to be relevant for the evaporation process of black holes as well, and for the existence of a bound for $\eta/s$. The diversity of situations for which instabilities appear to have a relevant r\^ole is quite vast and deserves further investigation. Also, it would be interesting to analyze further implications of them in the context of AdS/CFT, particularly on the possible values of other transport coefficients of the dual plasma.

 After this long journey, Lovelock theories emerge as a complex and interesting set of theories, consistent as far as we can tell. Still, there are plenty of open avenues for further research, many unresolved issues. In higher dimensions there is a enormous increase in the variety of rotating solutions, black rings, black saturns, multi black rings, etc. One interesting question to be answered in the context of Lovelock gravity would be to what extent this rich structure survives. There is a long way to go though as we do not even know the exact form of the simplest rotating solutions. Another insufficiently explored issue is that of the dimensional reduction of Lovelock theories, the relations under compactification between theories of different order -- namely with Einstein-Hilbert gravity -- and the stimulating possibility of these theories undergoing self-driven compactification. In exploring ways to simplify and get further insight into the dynamics of gravity, some attention has been recently drawn to the large $d$ limit of gravity. Lovelock gravity provide the most obvious extension also in this respect as they offer a second parameter $K$, the order of the theory, that can scale as fast as the dimension. In the holographic context the notion of entanglement entropy has also recently attracted a lot of attention and Lovelock gravity seem to be a suitable testbed in that case as well. Finally, as it was already mentioned in the main text, Lovelock theories, through the AdS/CFT correspondence, open a window into the realm of higher dimensional CFTs, which we do not know much about. 

We hope that the tools and information obtained through our analysis will be valuable for tackling these and other interesting questions, also when dealing with higher curvature corrections of generic type.



\appendix
\chapter{\bfseries\itshape Master equations for high momentum perturbations}
\chaptermark{Master equations}
\label{PertEq}

\vskip1.5cm

\section{Black hole perturbations}
\label{BHpert}

We shall consider perturbations of the metric around the black hole solution obtained in the previous sections  along a given direction parallel to the boundary (say, $x^{d-1} \equiv z$) and propagating towards the interior of the geometry. Using the direction $z$ as an axis of symmetry, we can classify the perturbations in helicity representations of the rotation group around it. It is convenient to analyze each case separately. The linear order contribution to the equations of motion (\ref{eqlambda}) can be written as
\begin{eqnarray}
\delta\mathcal{E}_a & = & \epsilon_{a f_1 \cdots f_{d-1}}\; \sum_{k=0}^{[\frac{d-1}{2}]} c_k \left[ k\, \delta R^{f_1 f_2}\wedge R^{f_3 \cdots f_{2k}} \wedge e^{f_{2k+1} \cdots f_{d-1}} \right. \nonumber\\ [0.5em]
& & \left. \qquad +\, (d-2k-1)\, R^{f_1 \cdots f_{2k}} \wedge e^{f_{2k+1} \cdots f_{d-2}} \wedge \delta e^{f_{d-1}} \right] = 0 ~.
\label{lineareq}
\end{eqnarray}

\subsection*{Helicity two perturbation}

The easiest case is the one with higher helicity. For symmetry reasons we can choose the helicity two perturbation, $h_{\m\n}(t,r,z)dx^\m dx^\n$, simply as\footnote{In principle, one should consider $h_{ij}$, with $i < j = 2, \ldots, 6$, but their equations are all decoupled and give rise to the same answer. The remaining helicity two components are $h_{ii} - h_{jj}$, with $i,j = 2, \ldots, 6$, but since diagonal components can be made all equal by rotation, $h_{ii}= 0$.} $h_{23}(t,r,z)\, dx^2 dx^3$. Since we will consider small perturbations, it is convenient to include an infinitesimal parameter $\epsilon$, $h_{23}(t,r,z) = \epsilon\, \phi(t,r,z)$. This can be readily included in the vielbein as $\tilde{e}^a = e^a +\epsilon\,\delta e^a$ \cite{Camanho2010},
\begin{eqnarray}
& & \tilde{e}\,^0 = \sqrt{f}\, dt ~, \qquad\quad \tilde{e}^1 = \frac{1}{\sqrt{f}}\,  dr  ~, \qquad \tilde{e}^F = \frac{r}{L}\,  dx^F ~, \quad {\scriptstyle F = 4 \ldots d-1} ~, \nonumber \\ [0.5em]
& & \tilde{e}^2 = \frac{r}{L}\,\left( 1 + \frac{\epsilon}{2}\,\phi \right) \left( dx^2 + dx^3 \right)  ~, \qquad \tilde{e}^3 = \frac{r}{L}\,\left( 1 - \frac{\epsilon}{2}\,\phi \right) \left( dx^2 - dx^3 \right) ~.
\label{vierbh-h2}
\end{eqnarray}
From the torsionless condition we can now calculate the first order corrections to the spin-connection, $\tilde{\omega}^a_{~b}=\omega^a_{~b} + \epsilon\,\delta \omega^a_{~b}$, and from them those to the curvature 2-form,
\begin{equation}
\delta R^a_{~b}= d(\delta \omega^a_{~b})+\delta \omega^a_{~c} \wedge \omega^c_{~b} + \omega^a_{~c} \wedge \delta \omega^c_{~b} ~.
\label{deltaRab}
\end{equation}
Consider now the Fourier transform of the perturbation,
\begin{equation}
\phi(t,r,z) = \mathop\int \frac{d\omega}{2\pi}\; \frac{dq}{2\pi}\; \hat\phi(r;k)\; e^{- i \omega t + i q z} ~, \qquad k = (\omega, 0, \ldots, 0, q) ~.
\label{Fourt}
\end{equation}
For the sake of clarity, we omit the $k$ dependence in $\hat\phi$ in what follows.

The first thing we have to realize in order to carry out this calculation is that the relevant contributions to order $\omega^2$, $q^2$ and $\omega\, q$, come from derivatives along the directions $e^0$ and $e^{d-1}$. In the simplest case, for helicity two perturbations, these contributions have only an effect on the expressions of $\delta \omega^{02}$, $\delta \omega^{03}$, $\delta \omega^{(d-1)2}$ and $\delta \omega^{(d-1)3}$. Since we are at the linearized level, we conclude that the only non-trivial contributions to order $\omega^2$, $q^2$, $\omega\, q$ come from their exterior derivative (the second term in (\ref{lineareq}) is also irrelevant),
\begin{eqnarray}
& & \delta R^{02} \approx d (\delta \omega^{02}) = - \frac{\omega^2}{2 f}\;\phi\; e^0 \wedge e^2 + \frac{\omega\, q L}{2 r \sqrt{f}}\; \phi\; e^{d-1}\wedge e^2 ~, \label{spcon-1} \\ [0,5em]
& & \delta R^{(d-1)2} \approx d (\delta \omega^{(d-1)2}) = - \frac{q^2 L^2}{2 r^2}\; \phi\; e^{d-1} \wedge e^2 - \frac{\omega\, q L}{2 r \sqrt{f}}\; \phi\; e^0\wedge e^2 ~,
\label{spcon-2}
\end{eqnarray}
where the symbol $\approx$ refers to those contributions relevant to compute the propagation speed of a boundary perturbation. There are analogue expressions (with the opposite sign) for the components with a leg along the $e^3$ direction. Notice that the cosmological constant term is irrelevant for our purposes.

Recalling from (\ref{riemannbh}) that the curvature 2-form of the black-hole spacetime is proportional to $e^a \wedge e^b$, it is immediate to see that the only non-vanishing contributions are those with $d (\delta \omega^{ab})$ behaving equally,
\begin{equation}
d (\delta \omega^{02}) \approx - \frac{\omega^2}{2 f}\;\phi\; e^0 \wedge e^2 ~, \qquad d (\delta \omega^{(d-1)2}) \approx - \frac{q^2 L^2}{2 r^2}\; \phi\; e^{d-1} \wedge e^2 ~,
\end{equation}
(and, of course, an analogous expression for $d (\delta \omega^{03})$ and $d (\delta \omega^{(d-1)3})$). Some of the equations of motion are trivially satisfied. Indeed, notice that $\delta \mathcal{E}^{(d)}_a \approx 0$ trivially, unless  $a = 2$ or $3$. This is due to cancellations of contributions coming from $d(\delta \omega^{02})$ and $d(\delta \omega^{03})$ (similarly, $d(\delta \omega^{(d-1)2})$ and $d(\delta \omega^{(d-1)3})$). Moreover, by symmetry, the two non-trivial equations just differ by a global sign. We must then focus on a single component of (\ref{lineareq}), say, $\delta \mathcal{E}^{(d)}_3 = 0$. The contribution to first order from each Lovelock term can be written as
\begin{eqnarray}
\delta \mathcal{E}^{(d)}_3 & = & 2 \left[ \epsilon_{302 f_3 \cdots f_{d-1}}\; \sum_{k=1}^{[\frac{d-1}{2}]} k\;c_k\; d(\delta\omega^{02}) \wedge R^{f_3 \cdots f_{2k}} \wedge e^{f_{2k+1} \cdots f_{d-1}} \right. \nonumber \\
& &  +\,\left. \epsilon_{3(d-1)2f_3 \cdots f_{d-1}}\; \sum_{k=1}^{[\frac{d-1}{2}]} k\;c_k\; d(\delta\omega^{(d-1)2}) \wedge R^{f_3 \cdots f_{2k}} \wedge e^{f_{2k+1} \cdots f_{d-1}} \right]=0 ~.\quad
\end{eqnarray}
To proceed, the only thing to worry about is where the $0$ and $1$ indices are, since depending on them the curvature 2-form components change (\ref{riemannbh}). Notice that the first (second) line gives the $\omega^2$ ($q^2$) contribution. The former, for instance, can be nicely rewritten in terms of $g(r)$,
\begin{equation}
\delta\mathcal{E}_{3,\,\omega}=-\frac{\omega^2}{f}\,\phi\,\sum_{k=0}^{K}{}k\,c_k\; \left(r(g^{k-1})'+(d-3)g^{k-1}\right)(d-4)! ~,
\end{equation}
while the latter reads
\begin{eqnarray}
\delta\mathcal{E}_{3,\,q} & = & \frac{q^2 L^2}{r^2}\,\phi\,\sum_{k=0}^{K}{}k\,c_k\;\left(r^2(g^{k-1})'' \right. \nonumber \\
& & \left. \qquad +\;2 (d-3) r (g^{k-1})'+(d-3)(d-4) g^{k-1} \right)(d-5)! ~.
\end{eqnarray}
Here we assume that $c_{k>K} = 0$. It is convenient to define the following functionals, $\mathcal{C}^{(k)}_d[f,r]$, involving up to $k$th-order derivatives of $f$ (or $g$):
\begin{eqnarray}
{C}^{(0)}_d[g,r] & = & \Upsilon' [g]~, \\ [0.9em]
{C}^{(1)}_d[g,r] & = & \left(r \frac{d}{dr}+(d-3)\right)\Upsilon' [g] ~,\\ [0.6em]
{C}^{(2)}_d[g,r] & = & \left(r \frac{d}{dr}+(d-3)\right)\left(r \frac{d}{dr}+(d-4)\right)\Upsilon' [g]~.
\label{Cpots}
\end{eqnarray}
Then, the speed of the helicity two graviton can be written as (see \cite{Camanho2010} for details)
\begin{equation}
{\bf c}_2^2(r)\equiv \frac{q^2}{\omega^2}= \frac{L^2 f}{(d-4)r^2} \frac{\mathcal{C}^{(2)}_d[g,r]}{\mathcal{C}^{(1)}_d[g,r]} ~,
\end{equation}
%

\subsection*{Helicity one perturbation}

It is a bit more involved but the same can be done for the other polarizations. In order to choose an helicity one perturbation, we can proceed with a gauge fixing ($h_{0a}=0$) and, by symmetry arguments, just turn on the components $h_{12}(t,r,z) = \epsilon\, \phi(t,r,z)$ and $h_{26}(t,r,z) = \epsilon\, \psi(t,r,z)$ (recall that the direction $1$ is the radial one and $d-1$ is the propagation $z$). The perturbation, thus, can be parametrized as, 
\begin{eqnarray}
& & \tilde{e}\,^0 = \sqrt{f}\, dt ~, \qquad\quad \tilde{e}^1 = \frac{1}{\sqrt{f}}\,  dr + \frac{r}{L}\, \epsilon\,\phi\, dx^2 ~, \nonumber \\ [0,7em]
& & \tilde{e}^2 = \frac{r}{L}\, \left( dx^2 + \epsilon\,\psi dz\right) ~, \qquad \tilde{e}^B = \frac{r}{L}\,  dx^B ~, \quad {\scriptstyle B = 3, 4, \ldots d-1} ~.
\label{vierbh-h1}
\end{eqnarray}
Proceeding as in the previous case we get a set of algebraic equations that sets one of the components, $h_{2(d-1)}$, to zero. The remaining equation yields directly the speed of the graviton as in the previous case
\begin{equation}
{\bf c}_1^2=\frac{L^2 f}{(d-3)\,r^2} \frac{\mathcal{C}^{(1)}_d[g,r]}{\mathcal{C}^{(0)}_d[g,r]} ~.
\end{equation}
%

\subsection*{Helicity zero perturbation}

We can proceed in the same way as we did for the other two types of perturbations. The helicity zero perturbation is, anyway, a bit more involved. After gauge fixing, we still need to turn on several components $h_{11} = \psi(t,r,z)$, $h_{22} = h_{33} = h_{44} = h_{55} = \xi(t,r,z)$, $h_{16} = \phi(t,r,z)$, and $h_{66} = \varphi(t,r,z)$ (as before, we will call their Fourier transforms respectively $\hat\psi(r)$, $\hat\xi(r)$, $\hat\phi(r)$ and $\hat\varphi(r)$. The vielbein for this case reads,
\begin{eqnarray}
& & \tilde{e}\,^0 =\sqrt{f}\, dt ~, \qquad\quad \tilde{e}^1 = \frac{1}{\sqrt{f}}\, \left( 1 + \frac{\epsilon}{2}\,\psi \right) dr + \frac{r}{L}\, \epsilon\,\phi\, dz ~, \qquad\\
& & \tilde{e}^{d-1} = \frac{r}{L}\, \left( 1 + \frac{\epsilon}{2}\,\varphi \right) dz ~, \qquad \tilde{e}^B = \frac{r}{L}\, \left( 1 + \frac{\epsilon}{2}\,\xi \right) dx^B ~, \quad {\scriptstyle B = 2, 3, \ldots d-2} ~.
\label{vierbh-h0}
\end{eqnarray}
One of the components, $\phi$, is set to zero by the equations of motion, while other two can be written in terms of only one degree of freedom
\begin{equation}
\psi = -\frac{\mathcal{C}^{(1)}_d[g,r]}{\mathcal{C}^{(0)}_d[g,r]}\,\xi ~, \qquad
\varphi = \left(\frac{q^2}{\omega^2}\frac{L^2 f(r)}{r^2}\frac{\mathcal{C}^{(1)}_d[g,r]}{\mathcal{C}^{(0)}_d[g,r]}-(d-3)\right)\,\xi ~.
\end{equation}
When we substitute these expressions into the equations of motion, only one of them remains linearly independent and gives the speed of the helicity zero graviton
\begin{equation}
{\bf c}_0^2 = \frac{L^2 f}{(d-2)\,r^2}\left( \frac{2\,\mathcal{C}^{(1)}_d[f,r]}{\mathcal{C}^{(0)}_d[f,r]} - \frac{\mathcal{C}^{(2)}_d[f,r]}{\mathcal{C}^{(1)}_d[f,r]}\right) ~.
\end{equation}
The above expressions are valid for arbitrary higher order Lovelock theory and higher dimensional spacetimes. Instead of analyzing these expressions right away, let us discuss an alternative computation, first introduced in \cite{Hofman2009}, given by the scattering of gravitons and shock waves.

\section{Shock waves and gravitons}
\label{SWpert}

It is more convenient to work in Poincar\'e coordinates, $z = 1/r$. We insist in performing all computations in the formalism used in the previous section since it is significantly simpler than the usual tensorial setup. We define light-cone coordinates\footnote{We then have to change the tangent space metric to $\eta_{00} = \eta_{11} = 0$, $\eta_{01} = \eta_{10} = - \frac12$, $\eta_{AB}=\text{diag}(1,1,\cdots,1), {\scriptstyle A,B = 2, \ldots, 6}$.} $u = t + x^6$ and $v = t - x^6$, and consider a shock wave propagating on AdS along the radial direction,
\begin{equation}
ds^2_{\rm AdS,sw} = ds^2_{\rm AdS} + f(u)\, \varpi(x^a,z)\, du^2 ~.
\end{equation}
We should think of $f(u)$ as a distribution with support in $u = 0$, which we will finally identify as a Dirac delta function. As we did in the previous section, we consider an helicity two graviton perturbation, $h_{23}\, dx^2 dx^3$, which we keep infinitesimal $h_{23} = \epsilon\, \phi$,
\begin{equation}
d\tilde{s}^2_{\rm AdS,sw} = \frac{N_{\#}^2}{L^2}\, \frac{-du dv + dx^i dx^i +2 \epsilon \phi \, dx^2dx^3+ L^4 dz^2}{z^2} + f(u)\, \varpi(x^a,z)\, du^2 ~.
\end{equation}
The calculation is very similar to the one in the previous section. We just have to modify the vielbein considered as the input. In this case,
\begin{eqnarray}
& & \tilde{e}\,^0 = \frac{N_{\#}}{L z}\, du ~, \qquad\quad \tilde{e}^1 = \frac{N_{\#}}{L z}\,dv-\frac{L z }{N_{\#}} f(u)\, \varpi(x^a,z)\,du ~, \nonumber \\ [0,5em]
& & \tilde{e}^2 = \frac{N_{\#}}{\sqrt{2} L z} \left(1+\frac{\epsilon}{2}  \phi \right)(dx^2+dx^3) ~, \qquad \tilde{e}^3 = \frac{N_{\#}}{\sqrt{2} L z} \left(1-\frac{\epsilon}{2} \phi \right)(dx^2-dx^3) ~, \nonumber \\ [0,5em]
& & \tilde{e}\,^K =\frac{N_{\#}}{L z}\, dx^K ~, \quad {\scriptstyle K = 4, 5, \ldots d-2} ~, \qquad \tilde{e}^{d-1} = \frac{N_{\#}L}{z}\, dz ~.
\label{viersw-h2}
\end{eqnarray}
The constant $N_{\#}$ (\ref{Nsost}) is related to the radius of the AdS space, and the perturbation depends only on $(u,v,z)$ as before. The shock wave is parametrized by the function $\varpi(x^a,z)$.

Introducing (\ref{viersw-h2}) into the equations of motion for the background ($\epsilon \to 0$) we get an equation for $N_{\#}$ yielding the already known  possible values, and the equation for the shock wave propagating on AdS \reef{SWeq}.
There are several possible solutions for this equation. The one we are going to consider is
\begin{equation}
\varpi = \alpha\, N_{\#}^2\, z^{d-3} ~,
\end{equation}
which, as discussed in \cite{Hofman2009}, can be obtained from the black hole background by boosting the solution while keeping its energy constant. The normalization constant $\alpha$ is proportional to the energy density and, as such, must be positive if the solution has a positive mass.

For perturbations propagating on top of these backgrounds, the procedure is almost the same as in the previous section, just a bit more complicated since the symmetry of the background is lower than in the black hole solution. As before, since we are only interested in the high momentum limit, we keep only contributions of the sort $\partial^2_v\phi$, $\partial_u\partial_v\phi$ and $\partial_u^2\phi$. These contributions come again from the exterior derivative of the perturbation of the spin connection. In the helicity two case
\begin{eqnarray}
& & d(\delta \omega^{02}) = \frac{L^2\, z^2}{N_{\#}^2}\, \left[ \partial_v^2 \phi\; e^1 \wedge e^2 + \left( \partial_u \partial_v \phi + \alpha f(u) L^2 z^{d-1}\, \partial_v^2 \phi \right)\, e^0 \wedge e^2 \right] ~, \label{dw02} \\ [0,7em]
& & d(\delta \omega^{12}) = \frac{L^2\, z^2}{N_{\#}^2}\, \left[ \left( \partial_u \partial_v \phi + \alpha f(u) L^2 z^{d-1}\, \partial_v^2\phi \right)\, e^1\wedge e^2 + \left( \cdots \right)\, e^0\wedge e^2 \right] ~, \label{dw12}
\end{eqnarray}
the ellipsis being used in the second expression since the corresponding term does not contribute to the equations of motion. The components with index 3 instead of 2 are the only remaining ones non-vanishing, and they are obtained changing $\phi\rightarrow- \phi$. The other thing we need is the curvature 2-form of the background metric, that can be written as
\begin{equation}
R^{ab} = \Lambda (e^a\wedge e^b + X^{ab}) ~,
\label{RabXab}
\end{equation}
where $\Lambda=-\frac{1}{L^2 N_{\#}^2}$ and $X^{ab}$ is an antisymmetric 2-form accounting for the contribution of the shock wave 
\begin{eqnarray}
& & X^{1a} = (d-1)\, \alpha\,  f(u)\, L^2\, z^{d-1}\; e^0\wedge e^a ~,\qquad a\neq 0,d-1 ~, \nonumber \\ [0.5em]
& & X^{1(d-1)} = - [(d-2)^2 - 1]\, \alpha\, f(u)\, L^2\, z^{d-1}\; e^0 \wedge e^{d-1} ~. \nonumber
\end{eqnarray}
Now, the relevant equation is, as before, given by a single component of (\ref{lineareq}),
\begin{eqnarray}
\delta \mathcal{E}_3 &=&\sum_{k=1}^{K}{}k\,c_k\, \epsilon_{3 f_1 \cdots f_{d-1}} d(\delta \omega^{f_1 f_2})\wedge R^{f_3 \cdots f_{2k}}\wedge e^{f_{2k+1}\cdots\, f_{d-1}}\nonumber\\
&=&2\sum_{k=1}^{K}{}k\,c_k\, \left[\epsilon_{302f_3 \cdots f_{d-1}} d(\delta\omega^{02})+\epsilon_{312f_3 \cdots f_{d-1}}d(\delta\omega^{12})\right]\wedge R^{f_3 \cdots f_{2k}} \wedge e^{f_{2k+1}\cdots\,f_{d-1}}\nonumber\\
&=& \frac{4 L^2 z^2}{N_{\#}^2}\sum_{k=1}^{K}{}k\,c_k\,\Lambda^{k-1}\left[(d-3)!\left(\partial_u\partial_v\phi+ \alpha f(u) L^2 z^{d-1} \partial_v^2 \phi \right) - (k-1) \times \right.\nonumber\\
&&\left.\times (d-5)!\left((d-5)(d-1)-[(d-2)^2-1]\right)\alpha f(u)L^2 z^{d-1}\partial_v^2\phi\right] dVol ~,
\end{eqnarray} 
since the diagonal part of $d(\delta\omega^{ab})$ contributes to all the diagonal parts of $R^{ab}$ and the off-diagonal part of $d(\delta\omega^{ab})$ contributes to the off-diagonal part of one of the $R^{ab}$. Collecting terms type $\partial_u\partial_v\phi$ and $\partial_v^2\phi$ and simplifying constant factors (not depending on the degree of the Lovelock term, $k$), $\delta \mathcal{E}_3=0$ implies
\begin{equation}
\Upsilon'[\Lambda]\left[\partial_u\partial_v\phi+\left(1+\frac{2(d-1)\Lambda\Upsilon''[\Lambda]}{(d-3)(d-4)\Upsilon'[\Lambda]}\right)\alpha f(u)L^2 z^{d-1}\partial_v^2\phi\right]=0 ~.
\end{equation}
As it should be expected we get $\Upsilon'[\Lambda]$ as an overall constant related to the unitarity properties of these perturbations and the constant factor multiplying the $\partial_v^2\phi$ term can be written in terms of $t_2$ defined in \reef{t2t4} as
\be
\mathcal{N}_2=1+\frac{2(d-1)\Lambda\Upsilon''[\Lambda]}{(d-3)(d-4)\Upsilon'[\Lambda]}=1-\frac{1}{d-2}t_2
\ee
The equations of motion for the other two helicities have exactly the same form with
\bear
\mathcal{N}_1=1+\frac{d-4}{2(d-2)}t_2 \qquad\qquad ; \qquad\quad \mathcal{N}_0=1+\frac{d-4}{d-2}t_2
\eear
As discussed in chapter \ref{chp:LLcausality} this equations will be relevant when discussing holographic causality properties of this spacetimes in the context of the AdS/CFT correspondence. 

\chapter{\bfseries\itshape 3-point function parameters}
\chaptermark{3-point function parameters}
\label{3point}

Conformal symmetry is powerful enough to determine the form of the $3$-point function for the stress-energy tensor up to five constants and this are further constrained by conservation laws allowing us to reduce the number of independent parameters to three, $\mathcal{A}$, $\mathcal{B}$ and $\mathcal{C}$. Further one finds that Ward identities relate the 2- and 3-point functions and so $C_T$ can be expressed in terms of these three constants.
\begin{equation}
C_T=\frac{\Omega_{d-2}}{2}\frac{(d-2)(d+1)\mathcal{A}-2\mathcal{B}-4d\mathcal{C}}{(d-1)(d+1)}
\end{equation}
where $\Omega_{d-2}$ is the area of the unit $(d-2)$-sphere. Also as the energy flux 1-point function is just a quotient of a 3- and a 2-point function $t_2$ and $t_4$ should also be writeable in terms of the three parameters $\mathcal{A}$, $\mathcal{B}$ and $\mathcal{C}$. This was done in \cite{Buchel2010a} yielding
\begin{eqnarray}
t_2&=&\frac{2d}{d-1}\frac{(d-3)(d+1)d\mathcal{A}+3(d-1)^2\mathcal{B}-4(d-1)(2d-1)\mathcal{C}}{(d-2)(d+1)\mathcal{A}-2\mathcal{B}-4d\mathcal{C}} ~, \nonumber \\ [0.9em]
t_4&=&-\frac{d}{d-1}\frac{(d+1)(2(d-1)^2-3(d-1)-3)\mathcal{A}+2(d-1)^2(d+1)\mathcal{B}-4(d-1)d(d+1)\mathcal{C}}{(d-2)(d+1)\mathcal{A}-2\mathcal{B}-4d\mathcal{C}}~,\nonumber
\end{eqnarray}
Then the scalar, vector and tensor constraints coming from from the positivity of energy can also be translated into
\begin{eqnarray}
-\frac{d}{d-1}\frac{(d-3)(d+1)\mathcal{A}+2(d-1)\mathcal{B}-4(d-1)\mathcal{C}}{(d-2)(d+1)\mathcal{A}-2\mathcal{B}-4d\mathcal{C}}&\geq& 0 ~,\nonumber\\[0.9em]
d\frac{(d-3)(d+1)\mathcal{A}+(3d-5)\mathcal{B}-8(d-1)\mathcal{C}}{(d-2)(d+1)\mathcal{A}-2\mathcal{B}-4d\mathcal{C}}&\geq& 0 ~,\\[0.9em]
-2d^2(d-2)\frac{\mathcal{B}-2\mathcal{C}}{(d-2)(d+1)\mathcal{A}-2\mathcal{B}-4d\mathcal{C}}&\geq& 0 ~.\nonumber
\end{eqnarray} 
constraints that can be identified with those coming from causality in the boundary theory in order to obtain expressions for $\mathcal A, \mathcal B,\mathcal C$. However in order to unambiguously determine the normalization of these coefficients we can just obtain them in terms of $C_T, t_2, t_4$ and plug the expressions for these into them. We obtain then
\begin{eqnarray}
\mathcal{A} &=& - \frac{(d-1)^3}{(d-2)^3}\, \frac{\Gamma[d]}{\pi^{d-1}}\, \frac{1}{(-\Lambda)^{d/2}} \left(\frac{2  d \Lambda  \Upsilon''[\Lambda]}{(d-4)^2}+ \Upsilon'[\Lambda]\right) ~, \nonumber \\[0.9em]
\mathcal{B} &=& -\frac{  (d-1) }{(d-2)^3 }\frac{\Gamma[d]}{\pi^{d-1} } \frac{1}{(-\Lambda)^{d/2}}\left(\frac{ (d-1) d \left(d^2-4 d + 6\right) \Lambda  \Upsilon''[\Lambda]}{(d-4)^2}+ \left(d^3-4
d^2+5 d-1\right) \Upsilon'[\Lambda]\right) ~,\nonumber\\[0.9em]
\mathcal{C} &=& -\frac{(d-1)^2 }{2(d-2)^3}\frac{\Gamma[d]}{\pi^{d-1}} \frac{1}{(-\Lambda)^{d/2}}\left(\frac{ \left(d^3-3 d^2+3 d-4\right) \Lambda  \Upsilon''[\Lambda]}{(d-4)^2}+\frac{1}{2}  (2 (d-3) d+3) \Upsilon'[\Lambda]\right) ~. \nonumber
\end{eqnarray}
%

\chapter{\bfseries\itshape A curve through the parameter space of Lovelock gravities}
\chaptermark{A curve through parameter space}
\label{appendixB}

\begin{figure}
\centering
\includegraphics[width=0.8\textwidth]{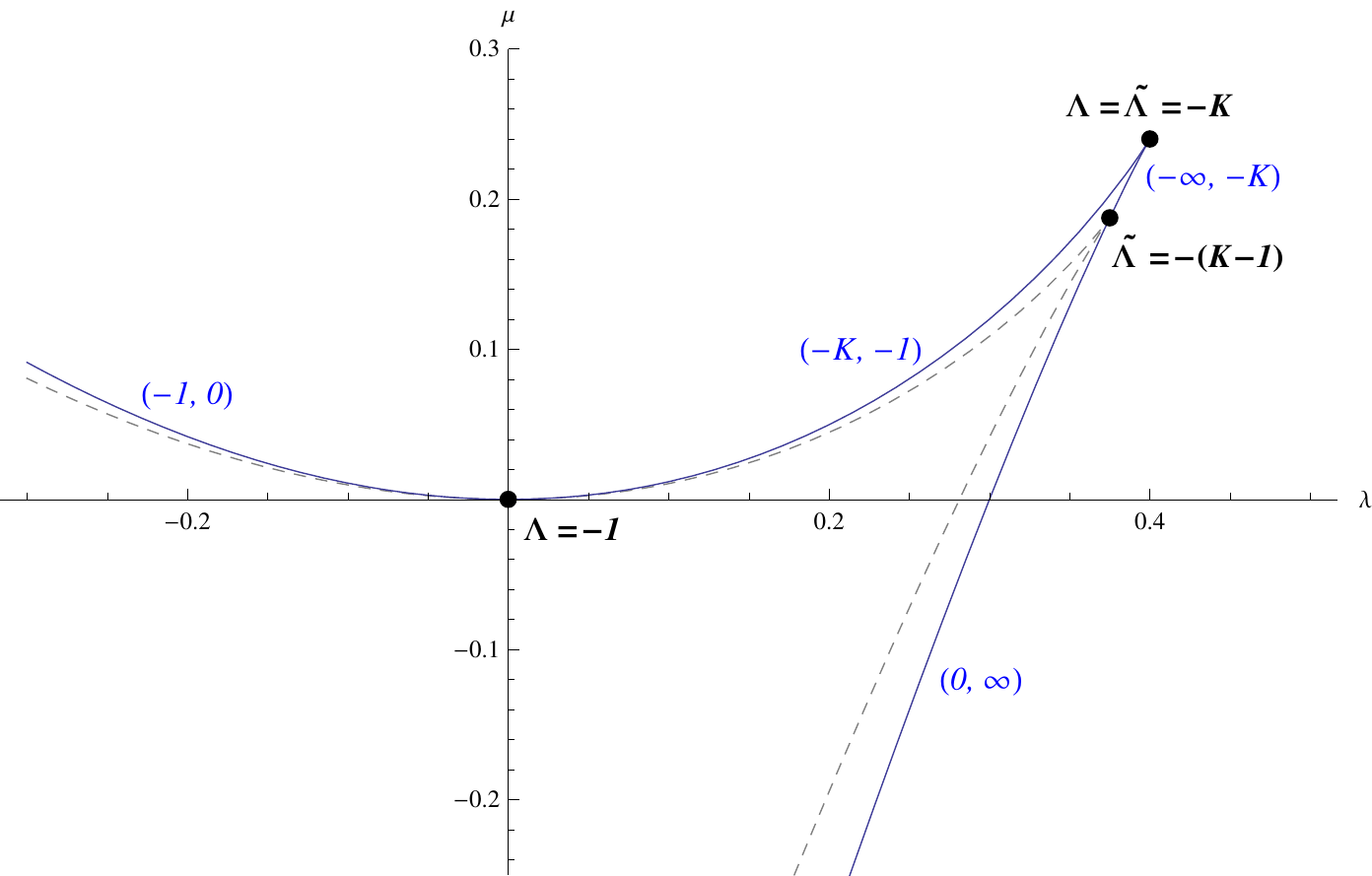}\caption{Projection of the points corresponding to (\ref{maxdegb}) in the ($\lambda$,$\mu$)-plane for $K=5$. The dashed line correspond to the projection of the $(K-1)$th order trajectory, and in blue the corresponding interval of $\Lambda$ values for each part of the curve is indicated.}
{\label{trajectory}}
\end{figure}

Exploring the full parameter space for a generic Lovelock theory is fairly intractable. As such, we choose to consider a particular curve through parameter space, simple enough that we can determine where along it a given theory preserves causality and stability. This curve corresponds to such a choice of parameters that the defining polynomial of the $K^{th}$ order theory becomes (we set $L=1$ here for simplicity)
\begin{equation}
\Upsilon[g]=1+g+\lambda g^2 +\ldots=\left(1-\frac{g}{\Lambda}\right)\left(1-\frac{g}{\tilde{\Lambda}}\right)^{K-1} ~,
\label{maxdegb}
\end{equation}
where there is just one free parameter as
\begin{equation}
\tilde{\Lambda}=-\frac{(K-1)\Lambda}{1+\Lambda} ~,
\label{cosmolambda}
\end{equation}
to ensure that the coefficient of order $g$ on the polynomial (the Einstein-Hilbert term) is actually one. All the Lovelock coefficients can be written in terms of this parameter,
\begin{eqnarray}
\lambda&=&\frac{(1+\Lambda)((K-2)\Lambda-K)}{2(K-1)\Lambda^2} ~, \nonumber \\[0.9em]
\mu&=&\frac{(1+\Lambda)^2(K-2)((K-3)\Lambda-2K)}{2(K-1)^2\Lambda^3} ~, \qquad \ldots
\label{lambdamu}
\end{eqnarray}
We call this from now on the {\sl maximally degenerated trajectory} (MDT). In figure \ref{trajectory} we show the projection of our curve on the ($\lambda$,$\mu$)-plane for the particular case $K=5$. For the range $\Lambda\in[-K,0]$, the cosmological constant, corresponding to the root with minimum absolute value of $\Upsilon=0$, is given by $\Lambda$, and beyond that by $\tilde \Lambda$. The curve parametrized by $\Lambda$ connects Einstein-Hilbert gravity ($\Lambda=-1$) with the maximally degenerated Lovelock theory ($\Lambda=\tilde{\Lambda}=-K$), henceforth denoted by the maximally degenerate point (MDP). 
Given this curve, we want to see where instabilities or causality violation may occur. We will restrict ourselves to the range $\Lambda\in(-K,-1)$ section for any value $K$, {\it i.e.}, to the part of the trajectory connecting the Einstein-Hilbert point with the MDP.
We are looking for solutions of $c_i^2(r)$ equals zero or one for the stability or the causality analysis respectively. These constraints can be written as polynomial equations in $g$, and the appearance of new roots is signalled by a change of sign of the corresponding discriminant. An important point worth noticing here is that we are just interested in roots, $g_{\star}$, in the interval of allowed values for the function $g$ 
\begin{equation}
g_{\star}\in[\Lambda,0] ~,
\label{grange}
\end{equation}
where $\Lambda$ is the closest to zero root of $\Upsilon[\Lambda]=0$. The zeros of the discriminant separate the theory space spanned by $(d,K)$ into distinct regions. By examining the properties of curve for a particular pair $(d,K)$ we are guaranteed to have found the generic behavior of the curve within the region containing that point.

\section{Stability analysis}

All three potentials are bound to yield zero at the horizon and one at the boundary. Thus, any root of $c_i^2[g]=0$ coming from outside the interval \reef{grange} has to enter it through the horizon ($g=0$) and in that case the horizon (which is always a zero for the equation) becomes a degenerate root. This is exactly what we studied when analyzing stability at the horizon -- we look for the vanishing of the first coefficient in the horizon expansion. The same phenomenon occurs at the boundary for the causality analysis. Conversely, if any root leaves the interval between the boundary and the horizon one will also see a degenerate root appearing. The only other option for an instability to show up is as a new double root appearing in the bulk. In any case, it is clear that  a study of the discriminants of the polynomial equations tells us exactly when these double roots occur. By carefully studying the regions in parameter space where the discriminant changes sign, one can learn where the stable and/or causal regions lie in parameter space. While this program is very involved in the generic case, the particular trajectory in parameter space we have chosen is simple enough so that this analysis of stability and causality can be done for all three potentials.

\subsection{Stability at the horizon}

The horizon stability analysis can be performed by combining the expressions (\ref{lambdamu}) for the Lovelock couplings along the MDT,  with the stability constraints found in (\ref{stabwedge1})-(\ref{stabwedge2}). For each constraint we get a quartic equation on $\Lambda$ and we can study the number of real roots by examining the corresponding discriminants. For the lower boundary of the stability wedge, corresponding to the tensor channel, we get 
\begin{eqnarray}
\Delta_{s2}^{(h)}&=&-16 (d-1)^6 (K-1)^4 K^2 (d-3 K-1) (d-2 K-1)\times\nonumber\\ [0.7em]
&& \left[ (d (162 d-1529)+3768) K^4-(d (3 d (45 d-391)+1796)+4044) K^3\right.\\ [0.7em]
& & +(d (d (3 d (9d-56)-530)+3576)-504) K^2\nonumber\\ [0.7em]
&& \left. -4 (d-1) (d (d (2 d-21)-16)+300) K+8 (d-1)^2 ((d-2) d-12) \right] ~, \nonumber
\end{eqnarray}
where $s$ stands for stability and the number to the corresponding helicity of the channel considered. The $h$ indicates that we are just considering stability at the horizon, for the moment. 

For general values of $K$ this discriminant vanishes for $d=2K+1$ and $d=3K+1$. For $d<3K+1$ there is an interval of values for $\Lambda$ around $\Lambda=-K$ for which the tensor potential becomes unstable. This is simply due to the fact that the end point of the MDT is outside the stability wedge for $d<3K+1$. Then, by considering high enough $d$ for fixed $K$, the tensor channel becomes stable along our curve, close to the MDP.

\begin{figure}
\centering
\includegraphics[width=0.6\textwidth]{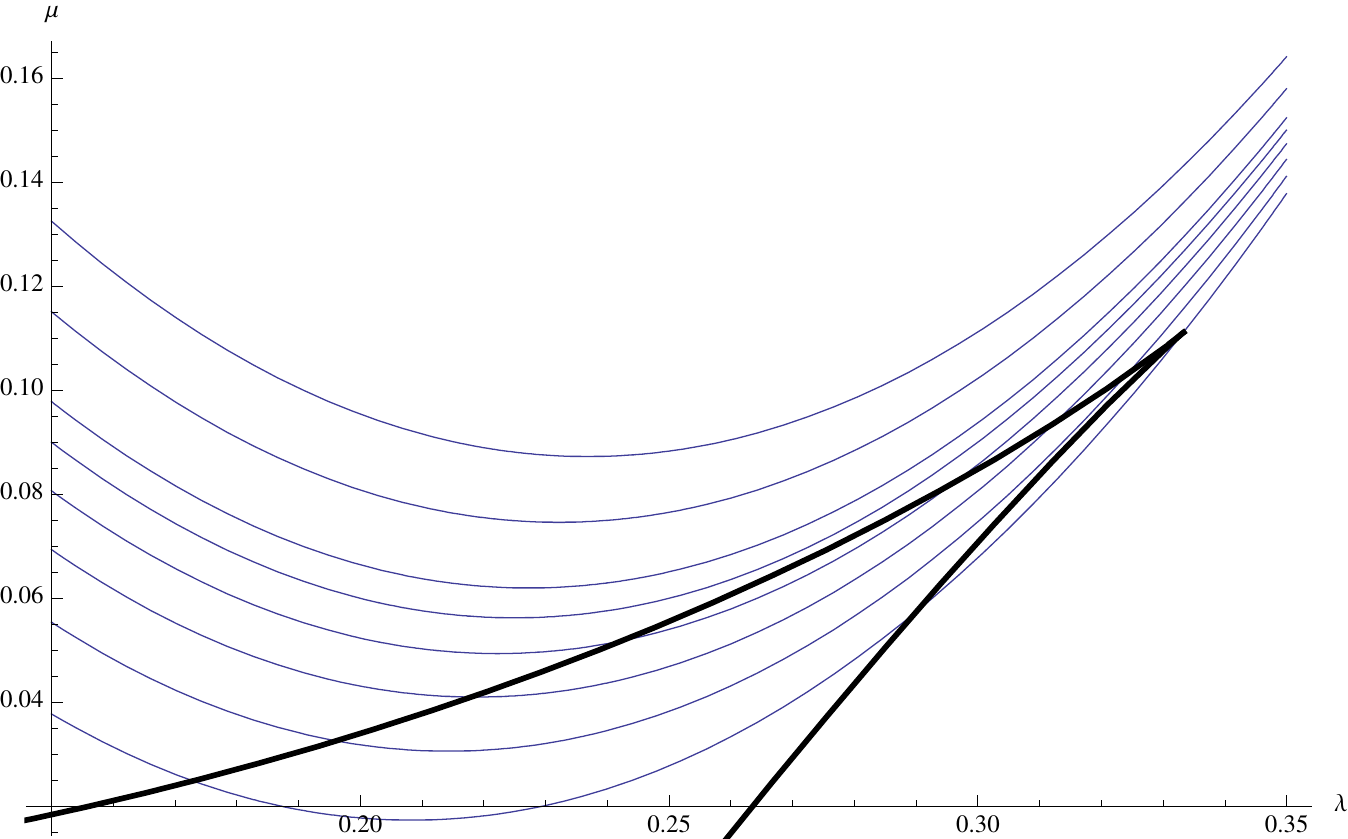}\caption{$K=3$ trajectory (in black) and $d=7,8,9,10,11,12,15,20$ helicity 0 stability constraint at the horizon (in blue). The unstable region is above the blue line. All the roots are in the figure in this case.}
{\label{s0h_n3}}
\end{figure}

\begin{figure}
\centering
\includegraphics[width=0.6\textwidth]{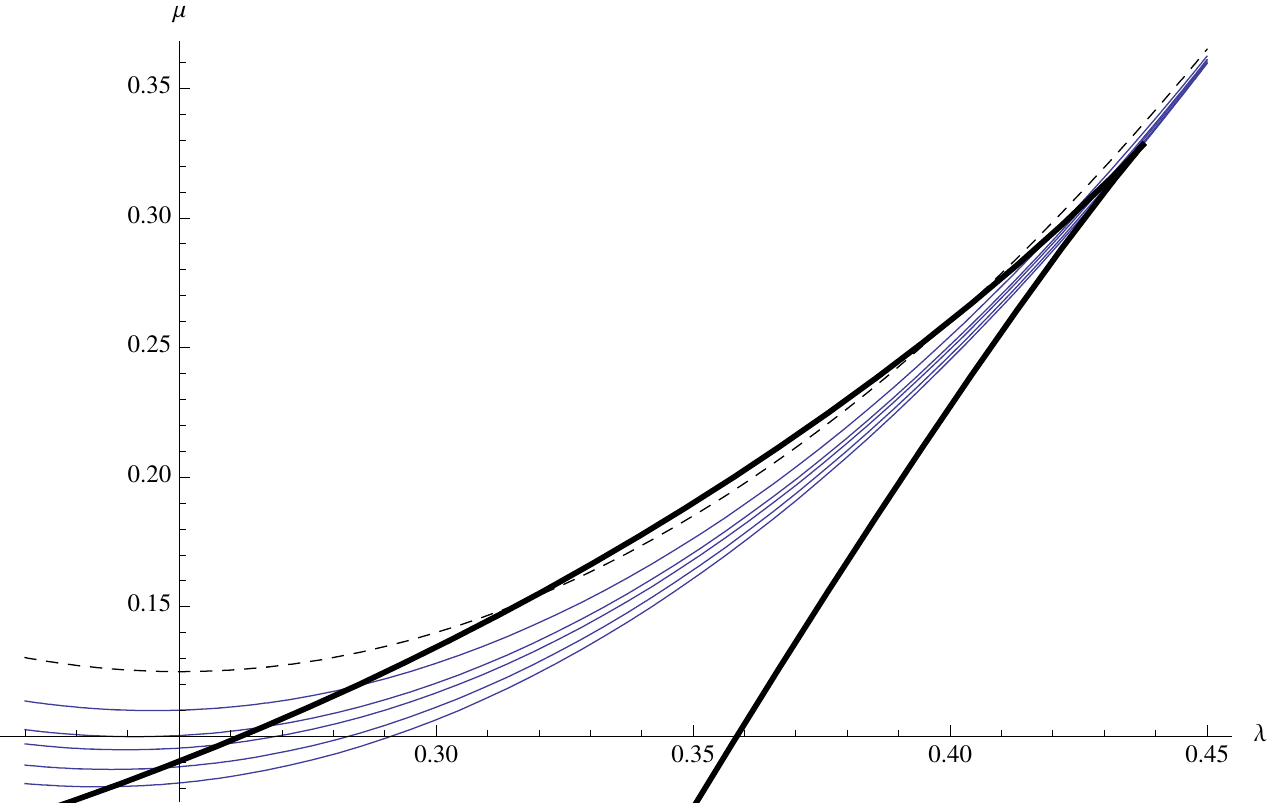}\caption{$K=8$ trajectory (in black) and $d=17, 20, 25, 30, 50,\infty $ helicity 0 stability constraint at the horizon (in blue and dashed black for $d\rightarrow\infty$). The unstable region is above the blue line. All the roots are in the figure in this case.}
{\label{s0h_n8}}
\end{figure}

We can proceed in a similar manner with the upper boundary of the na\"ive stability region, given by the sound channel. The discriminant in this case is 
\begin{eqnarray}
\Delta_{s0}^{(h)}&=&-432 (d-2) (d-1)^6 (d-2 K-1) (d-K-1) (d-K) (K-1)^4 K^2 \times\\ [0.7em]
& \times&\!\!\!\!\left(((K-8) K+8) d^2-2 (K ((K-13) K+6)+8) d+K (20-K (23K+2))+8\right)\nonumber
\end{eqnarray}
and it yields zero whenever $d=2K+1$ and
\begin{equation}
d_{\pm}=\frac{K^3-13 K^2+6 K \pm \sqrt{(K-2)^2 K^2 \left(K^2+K-1\right)}+8}{K^2-8 K+8}~.
\label{stabd}
\end{equation}
The relevant root for the part of the curve we are interested in is actually $d_-$. For $d<d_-$ there is a unstable interval of $\Lambda$ values strictly inside $\Lambda\in(-\Lambda,-1)$ as can be seen in figure \ref{s0h_n3}. For $d=2K+1$ the MDP is at one of the edges of this interval. In the plot we can also check that for $d>d_-$ this unstable interval disappears leaving a perfectly stable trajectory until it reaches the MDP. The situation changes dramatically as we increase the order of the Lovelock theory though. This is true for $K<7$ -- for $K\geq7$ the denominator of $d_-$ changes sign and this critical value becomes negative and thus irrelevant. In other words, for $K\geq7$ the unstable interval remains for every dimension. These statements are illustrated by figures \ref{s0h_n3} and \ref{s0h_n8}.


\subsection{Full stability}

For $K>7$ it would appear that the sound channel is stable beyond the forbidden interval. This turns out not to be the case. As one goes beyond this interval the instability moves deeper in the bulk, so that there is always an instability in this channel. The full stability analysis is performed as follows. First notice that the two functions involved in the effective potentials, $F'(x)$ and $F''(x)$, have particularly simple expressions along the MDT:
\begin{eqnarray}
F'(x(g)) & = & \frac{(K-1) (g-\Lambda ) (\Lambda +1) \left(2 \Lambda ^2+K \left(-\Lambda
   ^2+g \Lambda +\Lambda +g\right)\right)}{\left(\Lambda ^2+K
   \left(-\Lambda ^2+g \Lambda +g\right)\right)^2}~,\\[0.9em]
F''(x(g)) & = & \frac{2 (K-1) (g-\Lambda ) \Lambda ^2 (\Lambda +1) (K+\Lambda )^2 ((K-1)
   \Lambda +g (\Lambda +1))}{\left(\Lambda ^2+K \left(-\Lambda ^2+g
   \Lambda +g\right)\right)^4}~.
\end{eqnarray}
We have removed the explicit dependence on the derivatives of the function $g(r)$ by using the polynomial equation (\ref{eqg}). For instance one gets:
\be
g'(r)=-\frac{d-1}{r}\frac{\Upsilon[g]}{\Upsilon'[g]} ~.
\ee
In this way we can analyze the potentials just knowing the range of values $g$ can take and avoiding solving the polynomial equation explicitly. 
In order to complete the stability analysis, now in the bulk of the geometry, we have to check if there is any value of $g\in[\Lambda,0]$ for which any of the potentials takes negative values. The appearance of any instability will be signalled with a change of sign of the discriminant of the corresponding polynomial equation $c_i^2=0$\footnote{This equation is not really polynomial but rational. However the denominator of the expression doesn't play any relevant r\^ole in this discussion and so we will restrict ourselves to the polynomial equation given by the numerator}. This is where a minimum of the potential touches the zero axis yielding a new double root of the equation. It is then enough to look for zeros of the discriminant and check if there is instabilities in the different disconnected regions. 

The full discriminant in the helicity zero case can be nicely written in terms of the corresponding discriminant of the horizon condition,
\begin{equation}
\Delta_{s0}=\frac{\Lambda ^{12} (K+\Lambda )^{12}}{(d-1)^{12} (\Lambda +1)^{12}}\Delta_{s0}^{(h)} ~,
\end{equation}
and then the appearance of instabilities in the bulk is controlled by the same discriminant analyzed in the precedent section. The $\Lambda$ dependent factor vanishes just at the maximally symmetric point as $\Lambda=0$ is never possible in the present case. From the previous analysis we know that there are two naively stable regions for $d<d_-$ and just a full naively stable trajectory as we go to higher dimensions. We have to consider separately the allowed values for $\Lambda$ on both sides of the unstable interval. 

\begin{figure}
\centering
\includegraphics[width=0.45\textwidth]{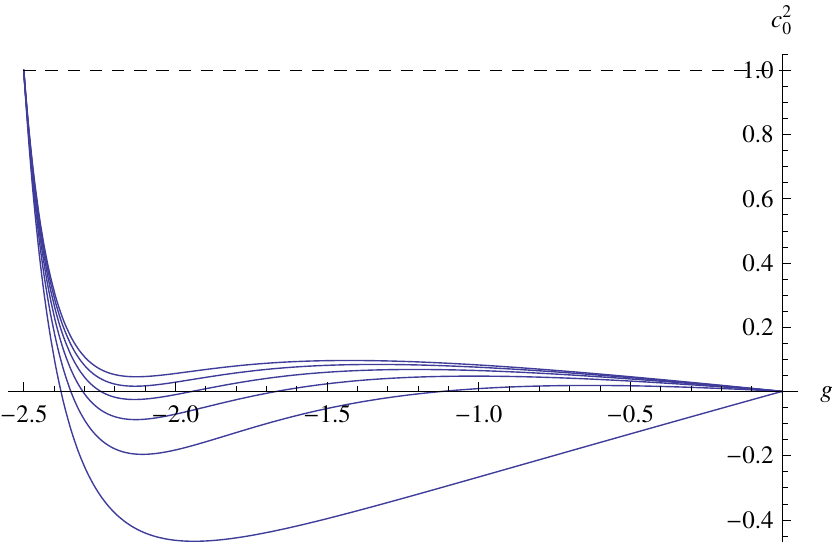}~~\includegraphics[width=0.45\textwidth]{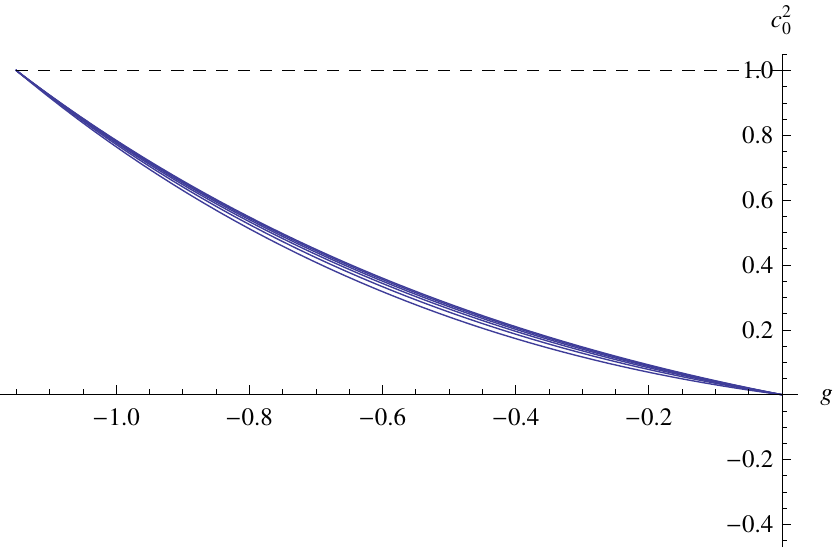}
\caption{Helicity zero potential for $K=3$ and $d=7,8,9,10,11,12$ to the left and to the right of the forbidden interval on $\Lambda$, $\Lambda=-2.5$ and $\Lambda=-1.15$ respectively.}
{\label{fullstab0_lowd}}
\end{figure}

\begin{figure}
\centering
\includegraphics[width=0.45\textwidth]{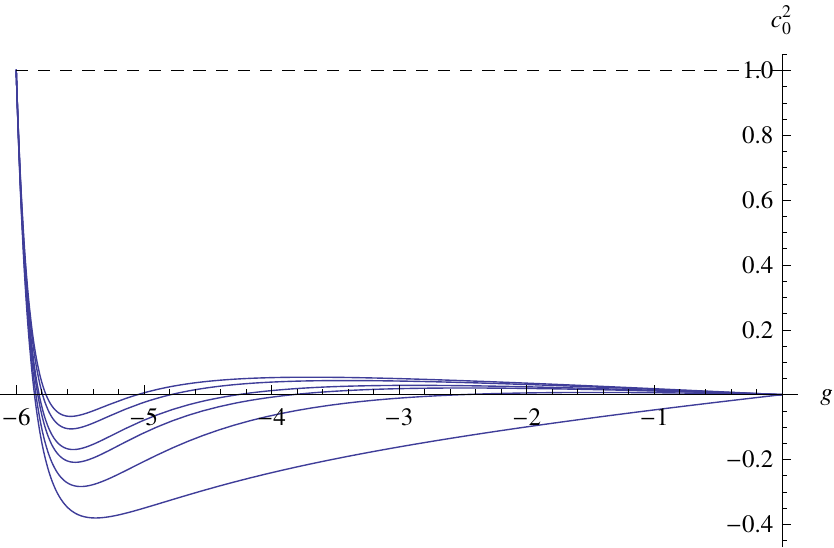}~~\includegraphics[width=0.45\textwidth]{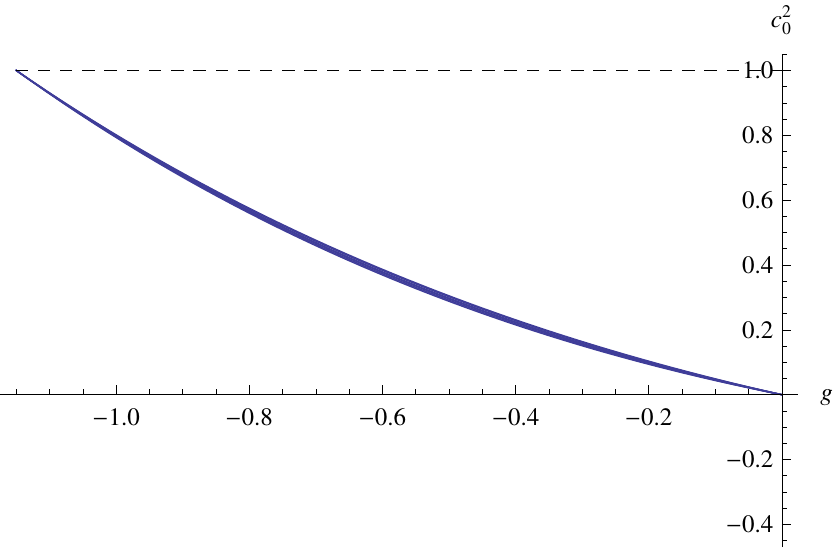}
\caption{Helicity zero potential for $K=8$ and $d=17,20,25,30,50,100$ to the left and to the right of the forbidden interval on $\Lambda$, $\Lambda=-2.5$ and $\Lambda=-1.15$ respectively.}
{\label{fullstab0_highd}}
\end{figure}

For $K<7$ and any value of $\Lambda$ in between the forbidden interval and the MDP there is an interval of values for $g$ where the potential goes below zero. This instability disappears for $d>d_{-}$, at the same dimension as the na\"ive unstable interval does (see left figure \ref{fullstab0_lowd}). Notice there's no instability either on the other allowed region as can be seen on the right hand side of figure \ref{fullstab0_lowd}.
For higher values of $K$ the instability in the allowed interval for $\Lambda$ close to the MDP never disappears. No matter how high the dimensionality some values of $g$ lead to negative values of the sound potential (see figure \ref{fullstab0_highd}). As one enters the stability wedge the instability simply moves away from the horizon and out towards the boundary of the geometry. 

\begin{figure}
\centering
\includegraphics[width=0.6\textwidth]{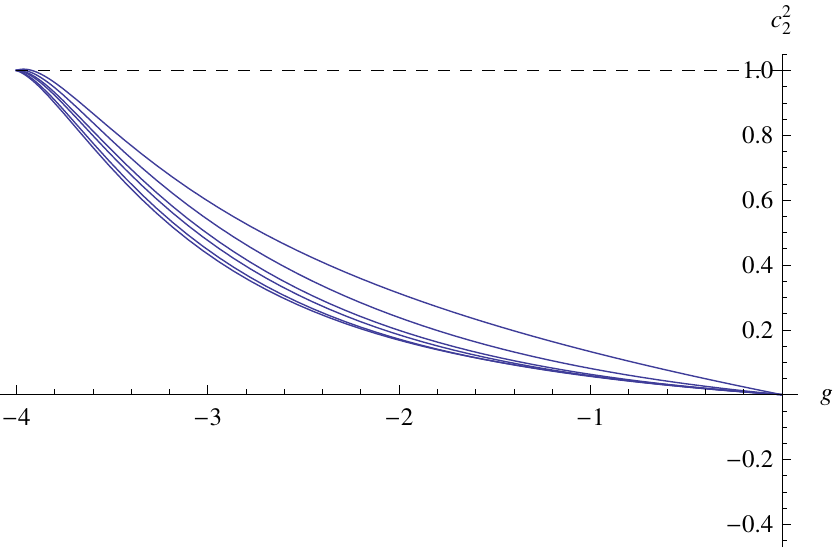}
\caption{Helicity two potential for $K=8$ and $d=17,20,25,30,50,100$ and cosmological constant $\Lambda=-4$. We find causality violation, but no instabilities.}
{\label{fullstab2}}
\end{figure}

The same analysis can be done in the helicity two channel. There the full discriminant turns out to be
\begin{equation}
\Delta_{s2}=\frac{\Lambda ^{12} (K+\Lambda )^{12}}{(d-1)^{12} (\Lambda +1)^{12}}\Delta_{s2}^{(h)}~.
\end{equation}
In this case we just have one naively stable region for $\Lambda\in(-K,-1)$ and the tensor potential has been found to be stable everywhere there. Then the tensor channel is free of instabilities as long as it is so at the horizon (see figure \ref{fullstab2}).

Even though we haven't made any reference to it yet, we didn't forget the shear potential. The reason for this disregard is that the corresponding stability constraint is always less constraining than the helicity zero and two ones. In fact, it amounts just to $\lambda<\lambda_c$, or equivalently $\eta/s>0$, when the other two constraints are respected (see \reef{potentialsF}).

All the features described in this section for the maximally degenerated trajectory can be easily observed in the cubic Lovelock case as we increase the dimensionality (see figures \ref{LL3}).  
\begin{figure}
\centering
\includegraphics[width=0.5\textwidth]{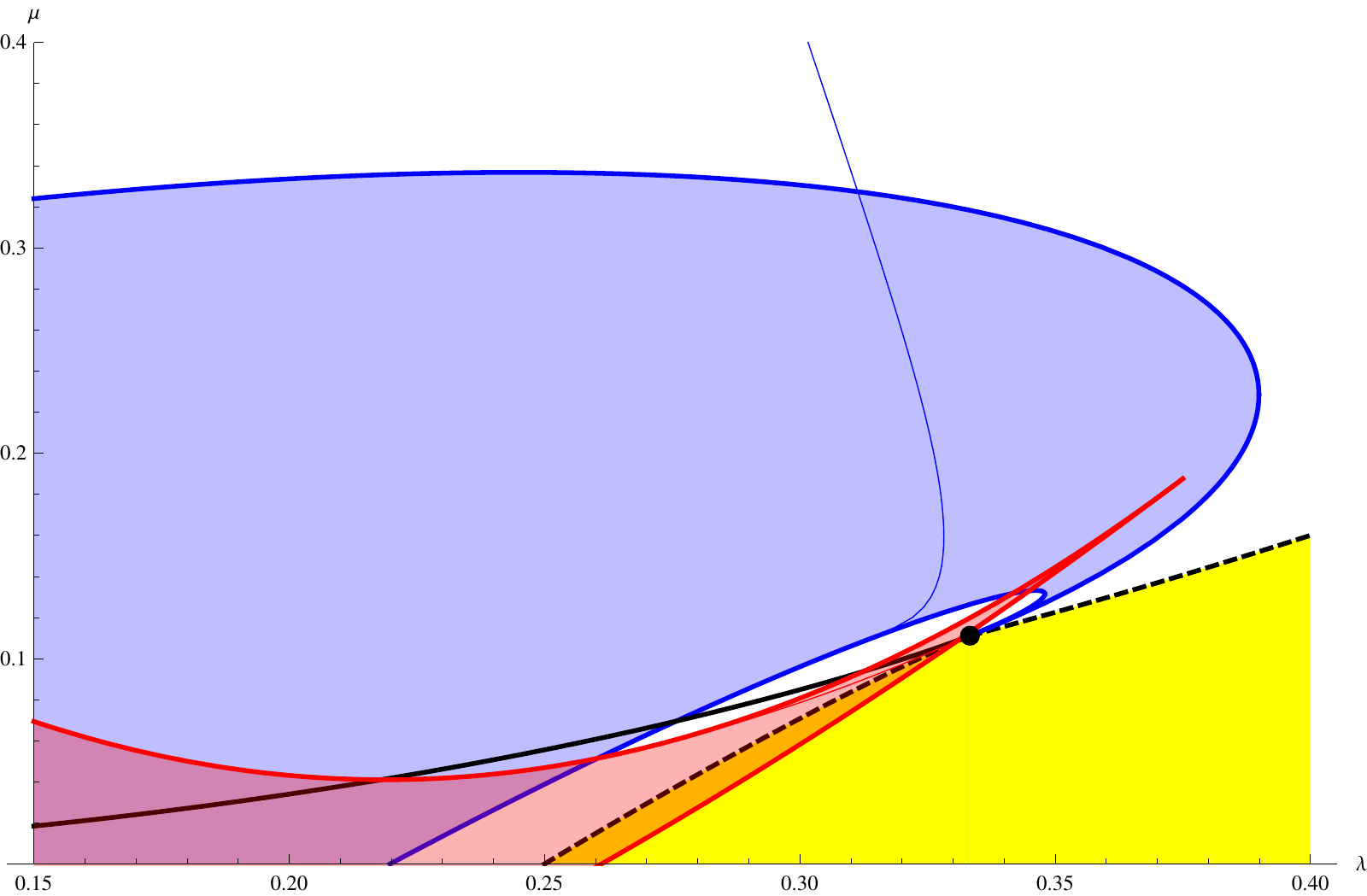}~~\includegraphics[width=0.4\textwidth]{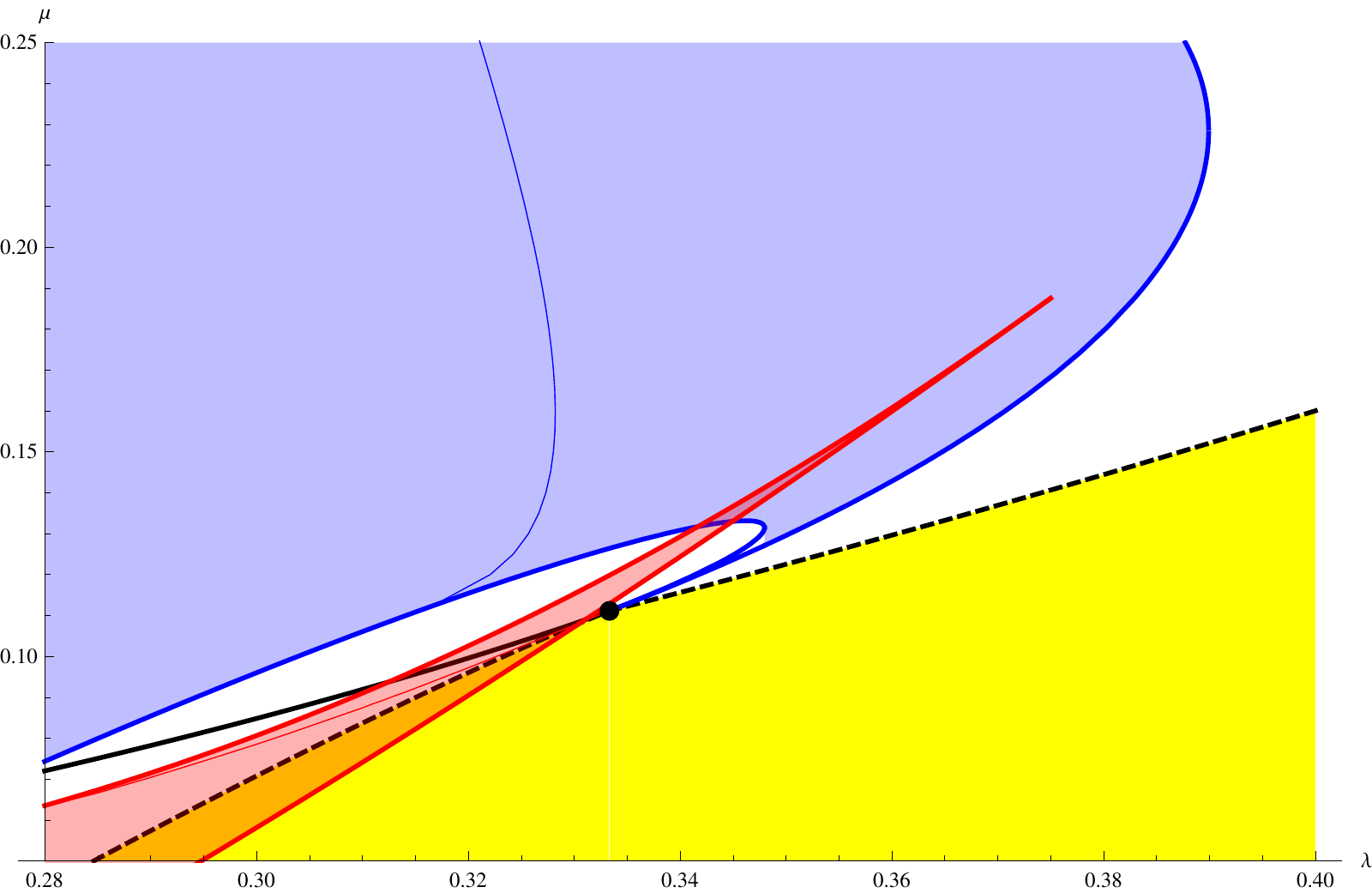}\\[0.5em]
\includegraphics[width=0.5\textwidth]{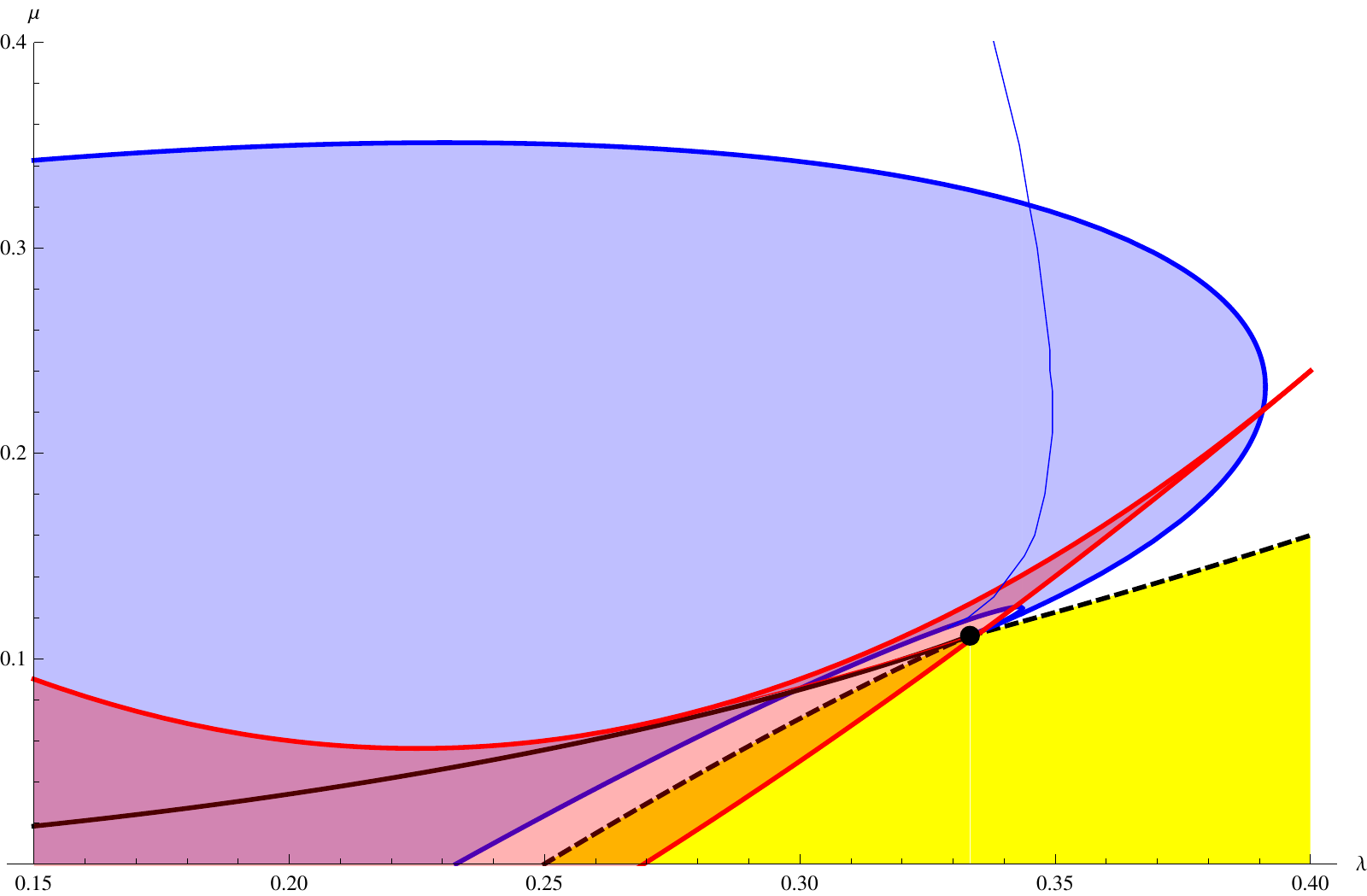}~~\includegraphics[width=0.4\textwidth]{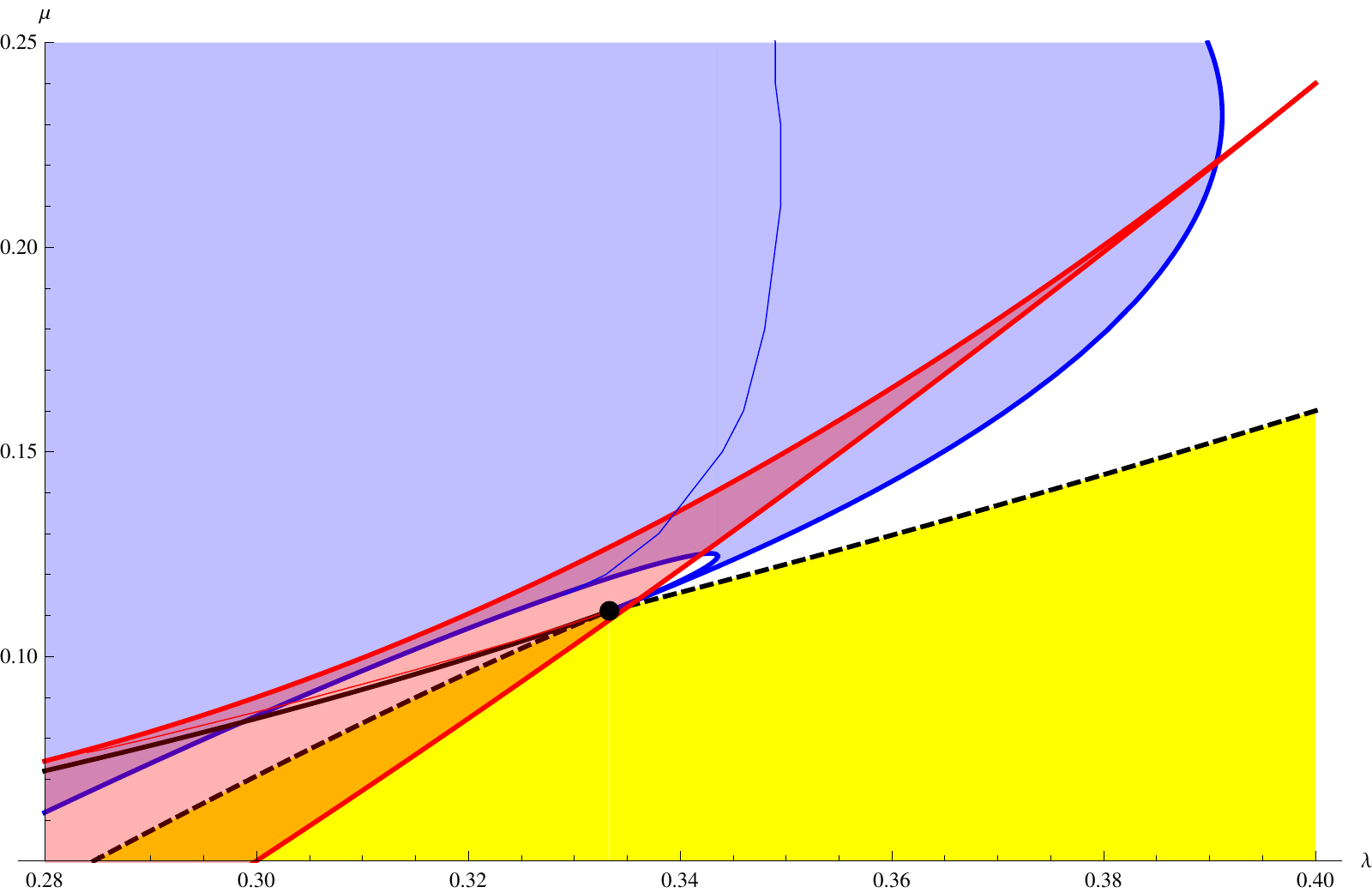}\\[0.5em]
\includegraphics[width=0.5\textwidth]{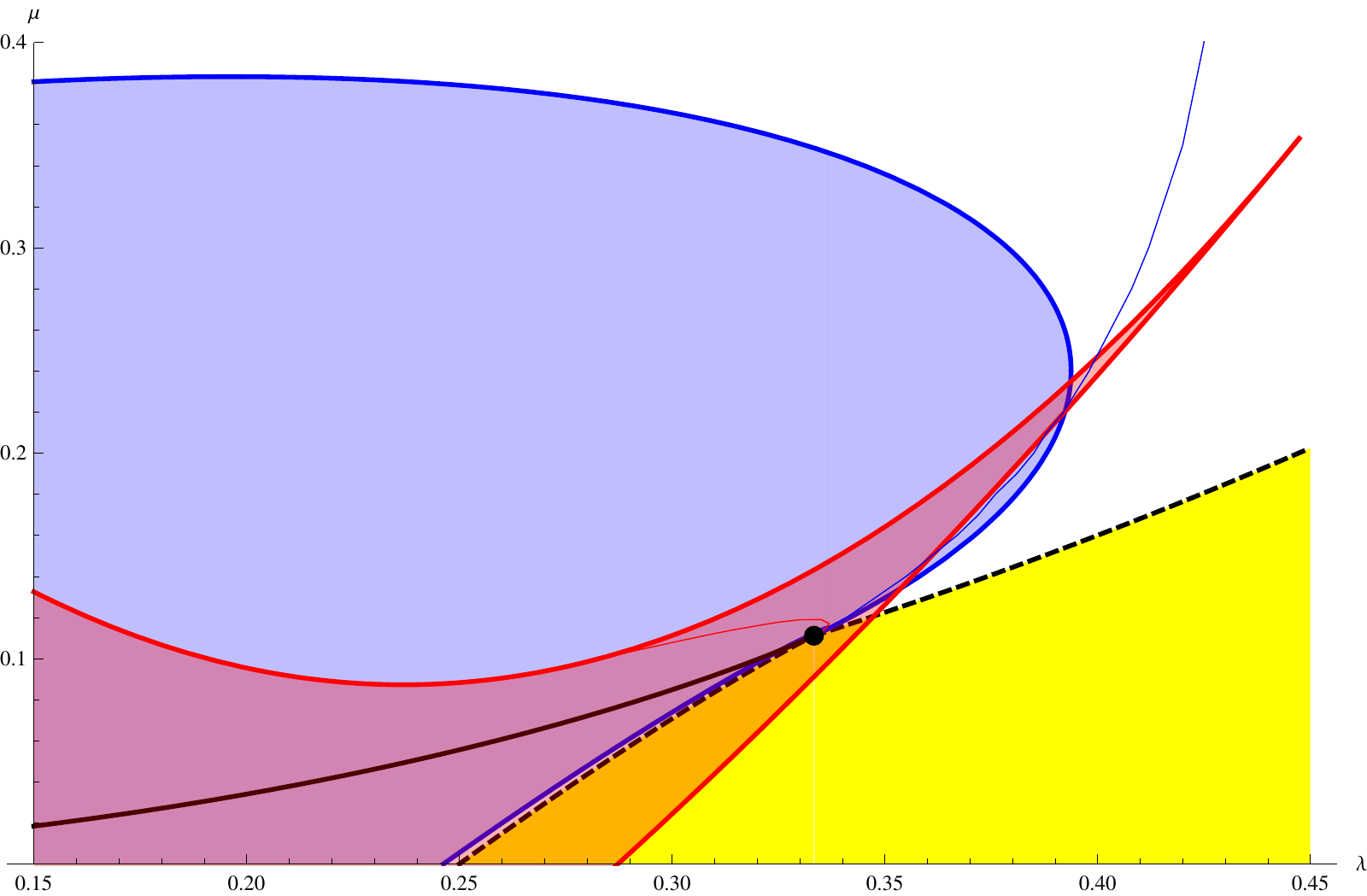}~~\includegraphics[width=0.4\textwidth]{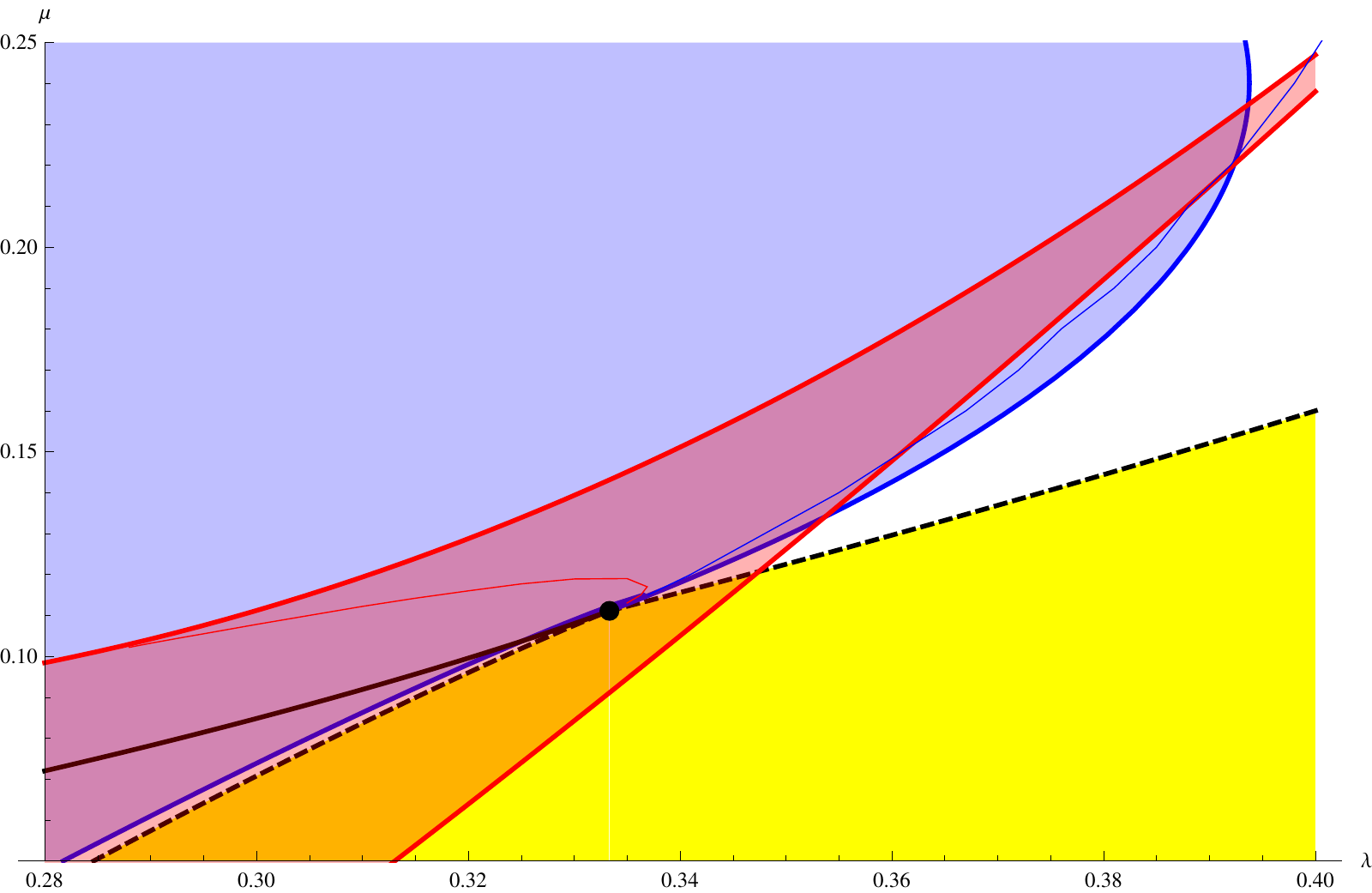}\\
\caption{All the different regions in the third order Lovelock case for $d=9,11,20$ (and zoom), from top to bottom. In black, the maximally degenerated line considered in the `full' causality and stability analysis. For $d<11$ the region of this curve closest to the maximally symmetric point (black dot) is unstable whereas for $d\geq11$ it becomes stable.}
{\label{LL3b}}
\end{figure}

\section{Causality analysis}

Let's now try doing the same kind of analysis with the causality constraints. These are of the form
\begin{equation}
\Upsilon'[\Lambda]+\tau \Lambda\Upsilon''[\Lambda]=0
\end{equation}
For the MDT trajectory, solving this equation yields
\begin{equation}
\Lambda_{\star}=-\frac{K+2 (K-1) \tau }{1+2 (K-1) \tau}.
\end{equation}
Depending on the channel $\tau$ takes different values namely (see (\ref{genconstraints})),
\begin{eqnarray}
\tau_2&=&\frac{2(d-1)}{(d-3)(d-4)}\nonumber\\
\tau_1&=&-\frac{d-1}{d-3}\\
\tau_0&=&-\frac{2(d-1)}{(d-3)}\nonumber
\end{eqnarray}
In order to respect $\Lambda_{\star}<0$ we must have $\tau>-\frac{1}{2(K-1)}$ or $\tau<-\frac{K}{2(K-1)}$.
Then, the crossing point of the MDT with the helicity two constraint is between the Einstein-Hilbert point and the MDP, and the crossing with the other two constraints is before the Einstein-Hilbert point. The helicity two crossing happens at a particular value of the LGB coupling, namely 
\begin{equation}
\lambda=\frac{K-1}{2K}\frac{K(K+2 (K-1) \tau)}{(K+2 (K-1) \tau_2)^2} ~.
\end{equation}
For the helicity two constraint the coefficient $\tau$ goes to zero as $d$ increases, that is, the value of $\lambda$ at the crossing point (the maximal value of $\lambda$ regarding causality) is the value at the  maximally symmetric point with an increasing extra factor that goes to one as $d\rightarrow\infty$. Then the crossing point approaches the MDP as $d$ increases. This is explained by the fact that the helicity two causality reduces to $C_T\geq 0$ for $d\to \infty$, and hence our trajectory can only cross it at the MDP itself.

The full causality analysis is quite more complicated that the stability one and not very enlightening. Here we will simply cite some partial results in the most relevant case, the helicity two one. In that case the equation $c_2^2(r)=1$ can be reduced to a polynomial equation on $g$ with discriminant
\begin{equation}
\Delta_{c2}= \Lambda^{20} (1+\Lambda)^{12} (K+\Lambda)^{12} \left(\Lambda-\Lambda_\star\right) \tilde{\Delta}_{c2}.
\end{equation}
The factor $\tilde{\Delta}_{c2}$ is a complicated expression which has zeroes at points irrelevant for our analysis. In particular, when restricted to the part of the curve connecting the Einstein-Hilbert point with the maximally symmetric point ($\Lambda\in(-K,-1)$) the discriminant has just one relevant  root at $\Lambda=\Lambda_{\star}$, exactly the point determined by demanding causality at the boundary. The situation is analogous for the other two helicities. Then, causality in the MDT reduces to causality at the boundary. In particular, close to the MDP the only constraint comes from the helicity two channel, as has been discussed.

In the same way as for the stability analysis, all the previously referred features can be seen in the third order case (see figures \ref{LL3}). 

\chapter{\bfseries\itshape Cavitation effects on the confinement/deconfinement transition}
\chaptermark{Cavitation effects}
\label{cavitation}

Hydrodynamics is a universal framework to describe strongly coupled systems at energy scales much lower than their characteristic microscopic scales (masses, temperature, etc). The basic hydrodynamic equation is that of the conservation of the stress-energy tensor
\be
\nabla_\mu T^{\mu\nu}=0 \,.
\label{const}
\ee
For  an ideal  relativistic fluid the stress-energy tensor takes the form 
\be
T^{\mu\nu}_{ideal}=\cE\ u^\mu u^\nu +\cP\ \Delta^{\mu\nu}\,,
\label{tideal}
\ee
where $\cE$ and $\cP$ are the energy density and pressure,
$$
\Delta^{\mu\nu}=g^{\mu\nu}+u^\mu u^\nu\,, 
$$
and $u^\mu$ is the fluid four-velocity, normalized so that $u_\mu u^\mu=-1$. The leading viscous corrections are parametrized by the fluid shear $\eta$ and bulk $\zeta$ transport coefficients in the viscous tensor $\Pi^{\mu\nu}$:
\bear
&&T^{\mu\nu}=T^{\mu\nu}_{ideal}+\Pi^{\mu\nu}\,,
\label{ns} \\ [0.7em]
&&\Pi^{\mu\nu}=-\eta\ \s^{\mu\nu}-\zeta\;\nabla u\ \Delta^{\mu\nu}\,,
\eear
where $\nabla u \equiv \nabla_\a u^\a$, and we have adopted 
\be
\sigma^{\mu\nu} = \left(\Delta^{\mu\lambda} \nabla_\lambda u^\nu + \Delta^{\nu\lambda} \nabla_\lambda u^\mu\right)
- \frac{2}{3}\;\nabla u\ \Delta^{\mu\nu}\,.
\ee
Assuming that the relevant microscopic scale is the temperature, leading hydrodynamic approximation is valid provided 
\be
\frac{|\nabla_{\mu} u^{\nu}|}{T}\ll 1\,,
\label{upper}
\ee 
otherwise, higher-order gradients (typically infinitely many of them) must be included \cite{Muronga2002,Muronga2004,Baier2008}.

It is easy to see that viscous terms tend to reduce the pressure \cite{Rajagopal2010}. For example, for fluids comoving in an expanding background such as an FRW metric,
\be
ds_4^2=-dt^2+a(t)^2 (d\vec{x})^2\,,
\ee 
we find 
\be
\s^{\mu\nu}\big|_{\rm FRW}=0\,,\qquad \nabla u\big|_{\rm FRW} =3\frac {\dot{a}}{a}\,,
\ee
resulting in an isotropic effective pressure
\be
\cP^{\rm eff} = \cP - \zeta\;\nabla u\,.
\label{pfrw}
\ee
In the case of a boost invariant fluid expansion, the pressure is no longer isotropic \cite{Muronga2004}:
\be
\cP_{\perp}^{\rm eff}=\cP+\frac{2\eta-3\zeta}{3\tau}\,,\qquad \cP_\xi^{\rm eff}=\cP-\frac{4\eta+3\zeta}{3\tau}\,,
\ee
where $ _\perp$ and $ _\xi$ are the transverse and longitudinal directions of the boost invariant expansion\footnote{Such expansion is conveniently described changing variables from $(t,z)$ to $(\tau,\xi)$: $\tau=\sqrt{t^2-z^2}$, $\xi={\rm arctanh}\frac zt$.}, and $\tau$ is the proper time.  Notice that in this case the {\it spatially averaged} pressure, $\left(\frac 23 P_{\perp}+\frac 13 P_\xi\right)$ still takes the form (\ref{pfrw}). In what follows, we take the trace-averaged form (\ref{pfrw}) as a generic expression for the effective pressure.

Consider now a system which, in thermal equilibrium, can exist in one of the two phases $A$ or $B$. A first-order phase transition between these phases implies the existence of a critical temperature $T_c$, such that $\cP_A> \cP_B$ for $T>T_c$, and $\cP_A< \cP_B$ otherwise. The phase with the higher pressure is thermodynamically favoured, and the transition at $T=T_c$ proceeds through nucleation of bubbles of a stable phase. If the system flows, the relevant pressure determining the stability of a phase is the effective one:
\be
\cP_{A/B}^{\rm eff}=\cP_{A/B}-\zeta_{A/B}\;\nabla u\,.
\ee
Close\footnote{We use the first law of thermodynamics $d\cF=-d\cP=-\cS\ dT$.} to $T_c$,
\be
\cP_{A/B}=\cP_c+\cS_{A/B}\ (T-T_c)+\cO\left((T-T_c)^2\right)\,,
\ee
where $\cS_{A/B}$ are the entropy densities of the corresponding phases. Thus, viscous hydrodynamics effects would shift the transition temperature according to 
\be
\frac{|\delta T_c|}{T_c}\ \sim\ \frac{|\zeta_A-\zeta_B|}{|\cS_A-\cS_B|}\ \frac{|\nabla u|}{T_c} \ \lesssim\ \frac{|\zeta_A-\zeta_B|}{|\cS_A-\cS_B|}\,,
\label{big}
\ee
where the upper bound is enforced from the consistency of truncating hydrodynamics at the first order in the velocity gradients, see (\ref{upper}). Notice that cavitation affects the transition temperature the more weakly the first-order transition is (the smaller the difference between $\cS_{A/B}$ is), and the larger the bulk viscosity difference of the two phases at $T_c$ is. 

Ideally, we would like to evaluate (\ref{big}) for  QCD close to confinement/deconfinement transition. While the recent lattice results  provide a reliable equation of state\footnote{At least at vanishing baryon chemical potential.} \cite{Borsanyi2010}, rather than doing it from first principles, one has to rely on various models to evaluate transport coefficients of gauge theory plasma at strong coupling \cite{Policastro2001a,Buchel2009,Karsch2008,Buchel2008,Buchel2010c}. In what follows we present the first self-consistent estimate of (\ref{big}) for a strongly coupled gauge theory plasma. 

~

\noindent
\textit{Cascading gauge theory.---}
Consider \cite{Klebanov2000} $\cN=1$ four-dimensional supersymmetric $SU(K+P)\times SU(K)$ gauge theory with two chiral superfields $A_1, A_2$ in the $(K+P,\overline{K})$ representation, and two fields $B_1, B_2$ in the $(\overline{K+P},K)$. This gauge theory has two gauge couplings $g_1, g_2$ associated with two gauge group factors,  and a quartic superpotential
\be
W\sim {\rm tr} \left(A_i B_j A_kB_\ell\right)\e^{ik}\e^{j\ell}\,.
\ee
The theory is not conformal, and develops a strong coupling scale $\Lambda$ through dimensional transmutation of the gauge couplings. In the UV/IR it undergoes the {\it cascade} of Seiberg \cite{Seiberg1995} dualities with $K\to K\pm P$. The net result of the duality cascade is that the rank $K$ of the theory becomes dependent on the scale $E$ at which the theory is probed \cite{Buchel2001}:
\be
K\to K_{\rm eff}(E)\approx 2P^2 \ln \frac{E}{\Lambda}\,,\qquad E\gg \Lambda\,.
\ee
While not  QCD, the theory shares some of  the IR features of the latter: when $K$ is an integer multiple of $P$, the cascade ends in the IR with $SU(P)$ supersymmetric Yang-Mills theory  which confines with spontaneous breaking of the chiral symmetry. Cascading gauge theory is always strongly coupled in the UV. In the planar limit and for large 't Hooft coupling of the IR $SU(p)$ factor, the theory is strongly coupled along its full RG flow, and thus can be studies using its holographic dual \cite{Klebanov2000}. We focus on the cascading gauge theory in the regime where the holographic description is reliable.

\begin{figure}[t]
\begin{center}
\includegraphics[width=0.67\textwidth]{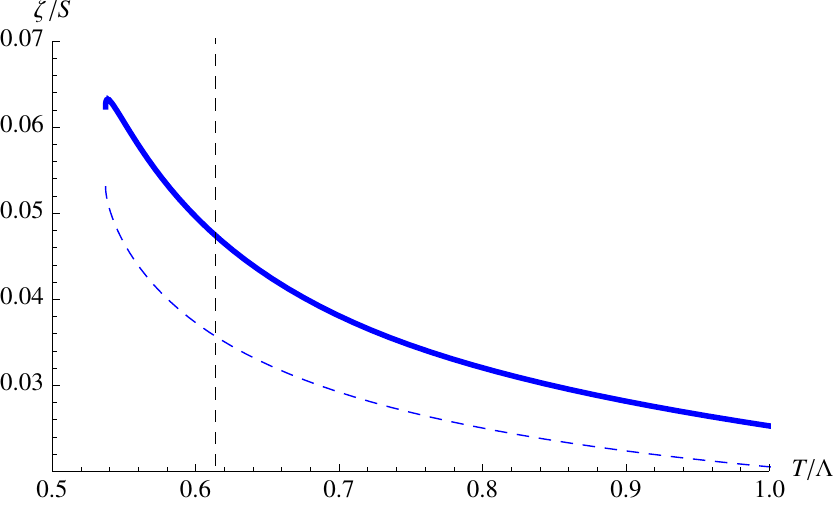}
\end{center}
\caption{The ratio of the bulk viscosity $\zeta$ to the entropy density $\cS$ in cascading gauge theory plasma (solid curve) and the bulk viscosity bound \cite{Buchel2008} (dashed). The dashed vertical line denote the critical temperature $T_c$ of the confinement/deconfinement phase transition.} 
\label{figure1}
\label{fig1}
\end{figure}

Thermodynamics of the cascading gauge theory plasma has been studied extensively in the past \cite{Aharony2007,Buchel2009e,Buchel2010d}: cascading gauge theory plasma simultaneously undergoes (first-order) confinement and the chiral symmetry breaking at $T_c=0.6141111(3) \Lambda$; at a slightly lower temperature $T_{\chi sb}=0.882503(0) T_c$ the deconfined phase becomes unstable towards spontaneous development of a chiral condensate, finally, at $T_u=0.8749(0) T_c$, the deconfined phase of the theory approaches a critical point with a divergent specific heat \cite{Buchel2010c}. The shear viscosity of the plasma is universal for all phases and at all temperatures \cite{Buchel2004a},
\be
\frac{\eta}{\cS}=\frac{1}{4\pi}\,.
\ee 

The bulk viscosity of the theory is technically difficult to compute --- so far it is known only to the fourth order in the high temperature expansion, $\left(\ln\frac T\Lambda\right)^{-1}$ \cite{Buchel2009e}, which is not enough to determine its value at the critical point $T_c$. We use Eling-Oz formula \cite{Eling2011,Buchel2011a} to compute bulk viscosity of the deconfined phase of the cascading gauge theory over all temperature range.
The results are presented in figure \ref{figure1}. We find
\be
\frac{\zeta}{\cS}\bigg|_{T=T_c}=0.04(8)\,.
\label{bigc}
\ee
Besides, the bulk viscosity bound \cite{Buchel2008} is respected all across the phase transition.

We can now address the question whether or not cavitation is expected to affect the temperature of the deconfinement transition in cascading plasma. Here, the phase $A$ of a fluid is the deconfined phase of the plasma, and $B$ is the confined phase. Since in the planar limit both the transport coefficients and the entropy density are suppressed, we obtain combining (\ref{big}) and (\ref{bigc})
\be
\frac{|\delta T_c|}{T_c}\ \lesssim\ \frac{\zeta_A}{\cS_A}=0.04(8)\,.
\ee    

~

\noindent
\textit{Discussion.---}
In this Letter we asked to which extent cavitation in confining  gauge theories affects the critical temperature of the confinement/deconfinement transition. We used the specific example of a cascading gauge theory to argue that in the planar limit and at strong coupling the effect is small. It is reasonable to expect that the result is universal as it reflects the fact that large-$N$ phase transitions are typically strong (as opposite to weak) first-order, and that the bulk viscosity at the critical point remains finite. Some phenomenological models suggest \cite{Karsch2008} that QCD bulk viscosity might diverge at the critical point of the $T - \mu_B$ phase diagram. Since the QCD critical point \cite{Stephanov2005} separates the line of first-order phase transitions (at large chemical potential) from crossovers (at low chemical potential), both of these effects tend to increase $|\delta T_c|/T_c$.


\providecommand{\href}[2]{#2}\begingroup\raggedright\endgroup



\newpage




\mbox{}
\newpage

\thispagestyle{empty}

\begin{center} 

\vskip5cm

\begin{figure}[b]
\centering
\includegraphics[width=0.4\textwidth]{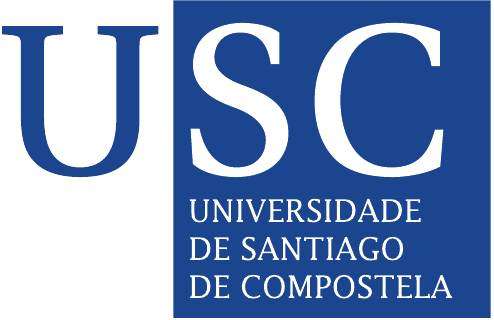}
\end{figure}

\end{center}

\end{document}